\documentclass[
fontsize=12pt,          
numbers=noenddot,      
listof=totoc,        	
bibliography=totoc,  	
headsepline=true,       
footsepline=false, 		
DIV=12                	
]{scrartcl}
\addtokomafont{disposition}{\boldmath}

\usepackage[utf8]{inputenc}
\usepackage[T1]{fontenc}
\usepackage[english]{babel}

\usepackage{enumitem}

\usepackage{graphicx}
\usepackage[table,x11names]{xcolor}

\usepackage{amsmath, amsfonts, amssymb, bbm, bm, mathabx}
\usepackage{csquotes}
\usepackage[citestyle = authoryear,
bibstyle = authoryear, dashed = false, giveninits, 
backend=bibtex, maxnames = 2, maxbibnames = 10, uniquename = false, uniquelist = false]{biblatex} 
\renewbibmacro*{volume+number+eid}{%
	\printfield{volume}%
	\setunit*{\addnbthinspace}%
	\printfield{number}%
	\setunit{\addcomma\space}%
	\printfield{eid}}
\DeclareFieldFormat[article]{number}{\mkbibparens{#1}}

\usepackage[a4paper, left = 2cm, right = 2cm, top = 2cm, bottom = 2cm]{geometry}

\usepackage{booktabs} 	
\usepackage{longtable} 
\usepackage{tabularx}
\usepackage{xltabular}
\usepackage{lscape}
\usepackage{hyperref} 
\usepackage{cleveref}

\usepackage{algorithm}
\usepackage{algpseudocode}

\usepackage{todonotes}
\usepackage{lscape}
\usepackage{nowidow}
\usepackage{placeins}
\usepackage[bf, format = plain]{caption}
\usepackage{subcaption}

\usepackage{fontawesome5}
\usepackage{tikz}
\usetikzlibrary{arrows.meta, positioning, calc, decorations.pathreplacing}

\usepackage{orcidlink}

\DeclareNameAlias{sortname}{family-given}

\renewbibmacro*{publisher+location+date}{%
	\iflistundef{publisher}
	{\setunit*{\addcomma\space}}
	{\setunit*{\addcolon\space}}%
	\printlist{publisher}%
	\setunit*{\addcomma\space}%
	\printlist{location}%
	\usebibmacro{date}%
	\newunit}

\renewbibmacro*{volume+number+eid}{%
	\printfield{volume}%
	\setunit*{\addnbspace}
	\printfield{number}%
	\setunit{\addcomma\space}%
	\printfield{eid}}
\DeclareFieldFormat[article]{number}{\mkbibparens{#1}}

\AtEveryBibitem{\clearlist{language}}
\AtEveryBibitem{\clearfield{month}}
\AtEveryCitekey{\clearfield{month}}

\AtEveryBibitem{%
	\ifentrytype{book}{
		\clearfield{url}%
		\clearfield{urlyear}%
	}{}
	\ifentrytype{article}{
		\clearfield{url}%
		\clearfield{urlyear}%
	}{}
	\ifentrytype{collection}{
		\clearfield{url}%
		\clearfield{urlyear}%
	}{}
	\ifentrytype{incollection}{
		\clearfield{url}%
		\clearfield{urlyear}%
	}{}
}

\DeclareMathOperator{\E}{\mathbb{E}}
\DeclareMathOperator{\Prob}{\mathbb{P}}
\DeclareMathOperator{\Cov}{\mathbb{C}\text{ov}}

\newcommand{\dif}{\mathop{}\!\mathrm{d}}

\title{\Large An Empirical Comparison of Methods for Quantifying the Similarity of Numeric Datasets}

\author{\normalsize Marieke Stolte$^1$\thanks{Corresponding author, e-mail: \texttt{stolte@statistik.tu-dortmund.de}}\orcidlink{https://orcid.org/0009-0002-0711-6789} \and\normalsize Jörg Rahnenführer$^1$\orcidlink{https://orcid.org/0000-0002-8947-440X} \and\normalsize Andrea Bommert$^1$\orcidlink{https://orcid.org/0000-0002-1005-9351}}
\date{\small$^1$Department of Statistics, TU Dortmund University}

\begin{document}
	\maketitle
	\section*{Abstract}
	Methods for quantifying the similarity of datasets are relevant in applications where two or more datasets, or their underlying distributions, need to be compared, ranging from two- and $k$-sample testing to applications in machine learning and synthetic data generation. 
	Many methods for quantifying the similarity of datasets are available from the literature, but due to the lack of neutral comparison studies, it is unclear which method to choose when.  
	Here, 36 methods applicable to continuous data are compared across various scenarios, including two or more datasets drawn from different distributions. 
	Several deviations between datasets are considered, including shift and scale alternatives or differences in higher moments.
	An overall method ranking is established based on the methods' abilities to differentiate between datasets from different distributions, combined with computational aspects. 
	Based on this, concrete decision rules for finding the best method based on characteristics of the datasets are determined. 
	Moreover, combinations of four to six methods are proposed in the two-sample case such that in 90\,\% to 95\,\% of the considered scenarios, at least one of these methods is almost as good as the best method.
	In the multi-sample case, a combination of two to three methods is proposed analogously.

	\paragraph{Keywords:} dataset similarity; two-sample test; multi-sample test; neutral comparison

	\section{Introduction}	
	Quantifying the similarity of datasets is useful in many applications, ranging from classical statistical testing to modern machine learning applications. 
	For example, the similarity between two or more datasets can be used in two- or $k$-sample tests to check whether two or more datasets originate from the same distribution.
	Other examples for using the similarity of datasets include transfer-learning, where it can be used to transfer insights from one learning task to another, or synthetic data generation and simulation studies, where it can be used to compare the generated data to a real-world dataset of interest. 
	
	There are many methods for quantifying the similarity of datasets proposed in the literature for these various applications. 
	Previously, \textcite{stolte_methods_2024} reviewed 118 dataset similarity methods that are non-parametric and applicable to multivariate data and compared them based on a set of theoretical criteria that they proposed. 
 	However, no empirical comparison was conducted there and therefore the methods' performance in detecting differences between datasets was not considered. 
 	Thus, the current study uses that theoretical comparison as a basis to conduct a focused empirical comparison of the most promising methods.
 	The best methods according to the theoretical criteria are compared here with respect to their empirical ability to detect differences between datasets coming from different data-generating mechanisms to identify methods with both good theoretical properties as well as good empirical performance.
 	
	There already exist some simulations on the newer methods or comparisons of methods with parametric or univariate methods, especially for the earlier methods \parencite{baringhaus_new_2004, baringhaus_rigid_2010, zhang_graph-based_2022, friedman_multivariate_1979, rosenbaum_exact_2005, schilling_multivariate_1986, yu_two-sample_2007}. 
	Moreover, there are some limited comparisons of methods in original articles, mostly for the more recently proposed methods \parencite{biswas_distribution-free_2014, biswas_nonparametric_2014, chen_new_2017, chen_weighted_2018, gretton_kernel_2012, hediger_use_2021, huang_kernel_2022, li_measuring_2022, lopez-paz_revisiting_2017, mukherjee_distribution-free_2022, mukhopadhyay_nonparametric_2020, pan_ball_2018, petrie_graph-theoretic_2016,  song_generalized_2021, song_gtestsmulti_2022, szekely_testing_2004, zaremba_b_2022}. 
	These studies show that the considered methods perform well in selected scenarios and in comparison to some limited competitor methods.
	However, the studies are all conducted in the context of proposing new methods and are therefore not considered neutral by the definition of \textcite{boulesteix_plea_2013}. 
	Also, almost exclusively, power comparisons are conducted of asymptotic or permutation / Bootstrap two- or $k$-sample tests, excluding those methods that do not define a statistical test. 
	Moreover, due to the fact that the method comparison is not the main focus in the articles presenting those studies, the considered scenarios and competitor methods are typically very limited. 
	For categorical data, already a comprehensive comparison of such methods was conducted previously based on this lack of neutral comparison studies for dataset similarity methods \parencite{stolte_comparing_2025}.
	Here, for numeric data, a corresponding neutral and comprehensive comparison is performed, building up on the previous studies. 	
	
	The current study, therefore, provides a neutral comparison study of the most promising methods for comparing two or more numeric datasets, selected from the preceding theoretical comparison \parencite{stolte_methods_2024}. 
	In particular, it aims to compare how good the methods are at detecting certain differences between datasets. 
	Note that since the comparison is not limited to two- and $k$-sample tests, this study does not conduct power comparisons. 
	A similar quantity is used instead, which compares the statistic values simulated for datasets drawn from different distributions to those values simulated for datasets drawn from the same distribution. 
	It is not expected to identify a single method that compares best in all scenarios \parencite{strobl_against_2024}. 
	Therefore, the goal is rather to identify groups of methods that act similarly across different simulation scenarios and to determine which deviations between datasets these groups of methods can detect well.
	Based on this, methods from different groups can be combined such that they can together detect a high number of different deviations. 
	Another goal of the study is to find out which methods are computationally feasible and numerically stable for application in practice. 
	Combining both aspects, concrete rules for choosing a good method based on dataset characteristics and deviations of interest are given, and combinations of methods that together cover many possible deviations are proposed.
	
	The remaining manuscript is structured as follows. 
	First, in Section~\ref{sec:setup}, the simulation set-up is described, including the aims, data-generating mechanisms, estimands, methods, performance measures, and software used for the study. 
	Afterwards, in Section~\ref{sec:results}, the results of the study are presented for the two- and for the multi-sample case. 
	In each case, first, computational aspects are discussed, and then an overall method ranking, decision rules for finding the best method depending on the setting, and an overall best method combination are proposed. 
	Last, in Section~\ref{sec:summary}, the results are summarized and discussed.

	\section{Simulation Set-Up}\label{sec:setup}
	The following describes the simulation set-up according to the \textit{ADEMP} (aims, data-generating mechanisms, estimands / other targets, methods, and performance measures) structure \parencite{morris_using_2019}.
	
	\subsection{Aims}
	The aims of the simulation study are to: 
	\begin{enumerate}
		\item Compare dataset similarity measures with respect to their performance in detecting differences of datasets drawn from distributions that differ in certain aspects, and to identify groups of dataset similarity measures that act similarly across different alternatives.
		\item Combine methods such that in combination they are able to detect well a high number of different deviations.
		\item Compare dataset similarity measures with respect to their consumption of computational resources.
	\end{enumerate}
	
	\subsection{Data-Generating Mechanisms}
	In the following, the data-generating mechanisms of the simulation study are explained. 
	The mechanisms considered here are based on those used in the previous study on categorical data \parencite{stolte_comparing_2025}.
	The data-generating mechanisms can be divided into two cases: 
	\begin{enumerate}
		\item Comparison of two datasets
		\begin{enumerate}
			\item[1.1] without a target variable,
			\item[1.2] with a target variable.
		\end{enumerate}
		\item Comparison of four datasets (without a target variable).
	\end{enumerate}
	In each case, multiple true data-generating mechanisms are considered. 
	Based on previous studies, parametric data-generating mechanisms are chosen, which grant full control. 
	This facilitates the interpretation of the results as all aspects of the data-generating mechanism are known and well-understood, and it allows for a very general comparison of the methods that is not restricted to one particular application of interest. 
	Two or more datasets are generated from the same underlying distributions or from different underlying distributions. 
	For the two-sample case, datasets with an additional target variable are created, as well as datasets without such a target variable. 
	For the $k$-sample case, only the case of datasets without a target variable is considered since none of the methods for multiple samples can appropriately consider a target variable in the data. 
	The $k$-sample case has different options for how many and which distributions can differ. 
	The number of possible settings increases with increasing $k$. 
	Here, only $k = 4$ is considered for the $k$-sample case as a compromise between comparing multiple samples but still having a reasonably low number of possible settings.
	For $k = 4$, there are four possible settings for how many distributions differ from each other. 
	These are presented in Table~\ref{tab:groupings} and are considered for case~2 of the data-generating mechanisms.
	The cases 2 a)\,--\,c) have mostly been neglected in the literature previously \parencite[e.g.][]{mukherjee_distribution-free_2022, song_new_2022}. 
	
	\begin{table}[!t]
		\begin{tabular}{lp{12.5cm}}
			\toprule
			Case & Description \\
			\midrule
			a) $3 + 1$ & One distribution differs from the others, which are equal, e.g.\ $F_1 = F_2 = F_3 \ne F_4$ \\
			b) $2 + 2$ & Two groups of two distributions each, where the distributions within the groups are equal but the distributions between the groups differ, e.g.\ $F_1 = F_2 \ne F_3 = F_4$ \\
			c) $2 + 1 + 1$ & Two distributions are equal, and the other two distributions are different from these and each other, e.g.\ $F_1 = F_2 \ne F_3 \ne F_4, F_1\ne F_4$ \\
			d) $1 + 1 + 1 + 1$ & All distributions differ, $F_i \ne F_j, i\ne j\in\{1,\dots,4\}$ \\
			\bottomrule
		\end{tabular}
		\caption{Possible settings for how many distributions differ from each other for $k = 4$ datasets.}\label{tab:groupings}
	\end{table}

	\subsubsection{Numbers of Observations and Variables}
	For each setting of the underlying distributions, different numbers of variables and observations are considered, and the imbalance of the number of observations of different datasets is also varied, as this can impact the method performance \parencite{chen_weighted_2018}.
	The number of variables, observations, and the imbalance are chosen as in the previous study for categorical datasets \parencite{stolte_comparing_2025}.
	The parameter settings are given in Table~\ref{tab:N.p.bal}.
	The number of variables $p$ represents low- to middle-dimensional data since not all methods are intended for high-dimensional data, and many previous studies only considered lower numbers of variables as well \parencite{friedman_multivariate_1979, schilling_multivariate_1986, baringhaus_new_2004, szekely_testing_2004, rosenbaum_exact_2005, baringhaus_rigid_2010, lopez-paz_revisiting_2017, pan_ball_2018,  li_measuring_2022}.
	The overall sample sizes are chosen to cover typical sample sizes and one balanced and unbalanced sample size setting, respectively.
	Moreover, the runtime of many methods increases both in $p$ and in the sample size $N$ such that high values are infeasible in the scope of a simulation study.
	A full factorial design is used, i.e.\ all combinations of $p$, $N$, and the settings for the individual sample sizes are used in each scenario, and in the four-sample case also for each of the settings a)\,--\,d). 
	
	\begin{table}[!b]
		\begin{tabular}{lllp{5cm}}
			\toprule
			Datasets & No.\ Variables & Overall Sample Size & Balance \\
			\midrule
			$k = 2$ & $p \in \{2, 10, 50\}$ & $N \in \{50, 100, 200, 500, 1000\}$ & $n_1 = \pi\cdot N$, $n_2 = (1-\pi)\cdot N$, $\pi\in\{0.2, 0.5\}$ \\
			& &  &  \\
			$k = 4$ & $p \in \{2, 10, 50\}$ & $N \in \{100, 200, 400\}$ & $n_i = \pi_i\cdot N$, $\pi_i= 0.25\forall i$ or $\pi_i= 0.1\cdot i$, $i = 1,\dots,4$\\
			\bottomrule
		\end{tabular}
		\caption{Parameter settings for number of variables, overall sample size, and balance of the individual sample sizes.}\label{tab:N.p.bal}
	\end{table}

	\subsubsection{Generation of Numeric Data}
	The scenarios for generating numeric data are chosen to cover different distributions. 
	The choice of the distributions that are used is based on previous simulation studies to cover a wide range of scenarios considered relevant by different researchers.
	In addition to normal data, log-normal and $\chi^2$ distributions are used to generate skewed data, and $t$ distributions are used to generate heavy-tailed data.
	For each distribution class, a null scenario is chosen in which the distributions of all datasets coincide. 
	For normal data, the null scenarios are multivariate standard normal distributions. 
	For the log-normal data, the null scenarios are log-normal distributions with the parameters chosen such that the means and variances are all one, and all variables are uncorrelated. 
	For the $t$ distributions, the degrees of freedom are set to $3$ as the lowest possible degrees of freedom with existing first and second moments, since for increasing degrees of freedom, the $t$ distribution converges to a normal distribution, which is already included. 
	The means are set to zero and the variances to one. 
	For the $\chi^2$ distributions, the degrees of freedom are set to one for a very skewed distribution. 
	Then, for the alternative scenarios, the distributions of one or three datasets, respectively, are varied with respect to some moments. 
	A full list of the parameter settings can be found in Table \ref{tab:scen.num.no.y} to \ref{tab:scen.num.multi} in Appendix~\ref{app:scen.tabs}.
	
	The consideration of shift and scale alternatives for normal data, i.e.\ deviations in the means or variances, respectively, are very common in the comparison of methods for quantifying dataset similarity \parencite{biswas_distribution-free_2014, biswas_nonparametric_2014, baringhaus_rigid_2010, baringhaus_new_2004, chen_weighted_2018, chen_new_2017, friedman_multivariate_1979, gretton_kernel_2012, huang_kernel_2022, mukherjee_distribution-free_2022, li_measuring_2022, mukhopadhyay_nonparametric_2020, pan_ball_2018, petrie_graph-theoretic_2016, rosenbaum_exact_2005, schilling_multivariate_1986, song_generalized_2021, song_new_2022, szekely_testing_2004}. 
	In some of the previous studies, the correlations are varied, which is done here as well. 
	
	The use of multivariate $t$-distributed data with location changes, or sometimes also scale or the degrees of freedom, is also common \parencite{chen_weighted_2018, huang_kernel_2022, li_measuring_2022, pan_ball_2018, rizzo_disco_2010, song_generalized_2021, song_new_2022}.
	Moreover, often one dataset is generated from a normal distribution and the other from a $t$ distribution \parencite{baringhaus_rigid_2010, biswas_distribution-free_2014, lopez-paz_revisiting_2017, mukhopadhyay_nonparametric_2020, rosenbaum_exact_2005, szekely_testing_2004}.
	The log-normal distribution was also used previously by some studies for including skewed distributions \parencite{friedman_multivariate_1979, chen_new_2017, pan_ball_2018, song_new_2022}.
	
	Some previous studies included other distributions like logistic distributions \parencite{baringhaus_new_2004}, Weibull distributions \parencite{baringhaus_rigid_2010}, or Gaussian mixtures \parencite{biswas_distribution-free_2014}, which are not used here to limit the computing time of the study. 
	Others include special cases motivated by a certain application \parencite[e.g.][]{biswas_distribution-free_2014, huang_kernel_2022, lopez-paz_revisiting_2017, yu_two-sample_2007, zaremba_b_2022}. 
	\textcite{song_new_2022} include also higher moments in addition to the most common settings. 
	The simulations here are therefore based on the scenarios chosen there.
	
	Location alternatives are considered for data generated from normal, log-normal, and $t$ distributions here.
	For location alternatives, the mean vector of the deviating distributions is shifted such that the Euclidean distance of the mean vectors is equal to a certain value $\delta$, i.e.\ each component of the deviating mean vectors is shifted by $\delta/\sqrt{p}$. 
	This ensures that the difference between the datasets is kept constant across different numbers of variables $p$.
	The concrete parameter settings can be found in Tables~\ref{tab:scen.num.no.y} and \ref{tab:scen.num.multi} in Appendix~\ref{app:scen.tabs}.
	
	For the three above-mentioned distribution classes, scale alternatives are also considered.
	Each variable is scaled by a certain factor. 
	To ensure that the volume of level sets of the densities is constant across different numbers of variables $p$, each component is scaled by a factor of $s^{1/p}$. 
	For normal and $t$ distributions, the covariance matrix is transformed accordingly. 
	For log-normal distributions, the data is scaled after generation, since expressing a certain scale is not directly possible by choosing a certain covariance matrix anymore.
	However, this also influences the location; therefore, the scale alternative is simultaneously also affecting the shift in that case.
	The concrete settings are again given in Tables~\ref{tab:scen.num.no.y} and \ref{tab:scen.num.multi} in Appendix~\ref{app:scen.tabs}.
	
	For the normal and $ t$-distributed data, changes in the correlation structure are considered additionally, as these can be varied directly as parameters of the distributions.
	For simplicity, all pairwise correlations are set to the same value $\rho$. 
	This, however, restricts the analysis to positive correlations, as when setting all correlations to the same negative value, the resulting matrices would no longer be positive definite. 
	
	Changes in kurtosis for the multivariate $t$ distributions are induced by varying the degrees of freedom. 
	The dispersion matrix is adjusted such that the variances stay constant. 
	For the $\chi^2$ distribution, each variable is standardized. 
	For changes in the skewness and kurtosis, the degrees of freedom are varied as well.
	
	One additional scenario is that data under the null is generated from a normal distribution, and the deviation is that the data is generated from a $t$ distribution. 
	In previous studies, this was often set to a single $t$ distribution with low degrees of freedom. 
	Here, a decreasing sequence of degrees of freedom is used starting at $30$, which is often claimed to be sufficiently similar to normal \parencite[e.g.][]{kwak_central_2017} and getting more dissimilar from the normal distribution with decreasing degrees of freedom. 
	
	The concrete parameter settings were chosen based on the cited literature, as well as a small pre-simulation study with few iterations, which was used to determine the range of parameter values up to the deviation for which most methods can distinguish well between the datasets. 
	
	For the case of numerical datasets with a target variable, there is one true outcome-generating model and then certain deviations from this. 
	The true outcome-generating model is chosen as a logistic regression model since all methods need a categorical outcome variable. 
	Moreover, the number of classes should be low, as the smallest generated datasets consist of only 25 data points; it has to be ensured that there are observations from all classes in all datasets. 
	Therefore, the logistic model was chosen such that for all chosen distributions of the covariates, a reasonable number of ones and zeroes is expected to be generated even for low sample sizes.
	As deviations of the outcome-generating model, once the signs of all coefficients except for the intercept are switched, once each coefficient is divided by two, and once completely different coefficients are chosen.
	For a discussion of the resulting separating hyperplanes, see \textcite{stolte_comparing_2025}. 
	The concrete settings for the outcome-generating models are given in Table~\ref{tab:scen.num.y} in Appendix~\ref{app:scen.tabs}.
	
	\subsection{Estimands}
	The population quantity of interest is the similarity or, equivalently, distance of the underlying distributions of the datasets. 
	This is estimated by each method.

	\subsection{Methods}
	The most promising methods from the previous review and theoretical comparison \parencite{stolte_methods_2024} are included in this empirical comparison. 
	Methods are selected from the review if any of the following criteria are fulfilled:
	\begin{enumerate}
		\item The method is implemented in R.
		\item The method fulfils at least 11 (i.e.\ more than half of the) criteria in the theoretical comparison, excluding the consistency criteria. 
		\item The method is the best in its subclass in the theoretical comparison, and no other method from this subclass was chosen with the first two criteria.
		\item The method is applicable to numeric data.
	\end{enumerate}
	An overview of all methods that fulfil these criteria is provided in Table~\ref{tab:methods} in Appendix~\ref{app:meth.tabs}.
	These 36 methods are explained in the following.
	Most of these methods are only applicable to two samples. 
	For the methods that are applicable to multiple samples, this is explicitly stated. 
	All methods are used with default parameters based on recommendations from the literature, if available. 
	If no sensible default is available, different options are compared.
	The choices of these are explained in Appendix~\ref{app:meth.pars}. 
	The methods' parameters are chosen by a practitioner with good knowledge of the underlying literature, but without expert knowledge of the methods, could use them.
	\begin{itemize}
		\item Kernel Measure of Multi-Sample Dissimilarity (KMD, \cite{huang_kernel_2022}): The KMD is a kernel-based test, which uses the association between the variables and the sample membership to quantify the dissimilarity of multiple samples. 
		It can be estimated using a similarity graph $\mathcal{G}$, e.g., the $K$-nearest neighbour (NN) graph or the minimum spanning tree (MST) on the pooled sample.
		Denote by $(Z_i,Z_j)\in\mathcal{E}(\mathcal{G})$ that there is an edge in $\mathcal{G}$ connecting two points $Z_i$ and $Z_j$ in the pooled sample.
		Moreover, let $o_i$ be the out-degree of $Z_i$ in $\mathcal{G}$. Then, an estimator for the KMD $\eta$ is defined as
		\[
		\hat{\eta} := \frac{\frac{1}{N} \sum_{i=1}^N \frac{1}{o_i} \sum_{j:(Z_i,Z_j)\in\mathcal{E}(\mathcal{G})} K(\Delta_i, \Delta_j) - \frac{1}{N(N-1)} \sum_{i\ne j} K(\Delta_i, \Delta_j)}{\frac{1}{N}\sum_{i=1}^N K(\Delta_i, \Delta_i) - \frac{1}{N(N-1)} \sum_{i\ne j} K(\Delta_i, \Delta_j)},
		\]
		where $\{Z_1,\dots, Z_N\}$ denotes the pooled sample and $\{\Delta_1,\dots, \Delta_N\}$ the dataset membership of each point and $K(x, y)$ is a positive definite kernel function. Here, the suggested discrete kernel $ K(x, y) := \mathbbm{1}(x = y)$ is used.
		\item Energy statistic \parencite{szekely_testing_2004} and Cramér test \parencite{baringhaus_new_2004}: The $k$-sample Energy statistic is defined as 
		\begin{align*}
			T_\text{Energy} = &\sum_{1\le i<j\le k} \frac{n_i n_j}{n_i + n_j} \left(\frac{2}{n_i n_j}\sum_{u = 1}^{n_i}\sum_{v = 1}^{n_j} \|X^{(i)}_u - X^{(j)}_v\|_2 \right.\\
			& \left.- \frac{1}{n_i^2} \sum_{u = 1}^{n_i}\sum_{v = 1}^{n_i} \|X^{(i)}_u - X^{(i)}_v\|_2 - \frac{1}{n_j^2} \sum_{u = 1}^{n_j}\sum_{v = 1}^{n_j} \|X^{(j)}_u - X^{(j)}_v\|_2\right),
		\end{align*}
		with datasets $X^{(i)}$ consisting of observations $X^{(i)}_u, u = 1,\dots,n_i,i = 1,\dots,k$. For the two-sample case, the energy distance is equivalent to the Cramér test. 
		\item Rank energy statistic by \textcite{deb_multivariate_2021} (DS): The rank energy statistic is a version of the energy statistic that uses multivariate ranks, which are assigned using optimal transport.
		\item Distance-based test by \textcite{baringhaus_rigid_2010} (BF) and Bahr test \parencite{bahr_ein_1996}: The test statistic generalizes the Cramér test statistic by using a continuous function $\varphi$ such that $\varphi(\|x-y\|^2)$ is a negative definite kernel instead of the Euclidean distances. Different examples for $\varphi$ are given, including as special cases the Cramér test \parencite{baringhaus_new_2004}, and the test by \textcite{bahr_ein_1996}.
		\item Distance-based test by \textcite{biswas_nonparametric_2014} (BG2): The statistic uses inter-point distances and is defined as
		\begin{align*}
			T &= ||\hat{\mu}_{D_{F_1}} - \hat{\mu}_{D_{F_2}}||^2_2, \text{ where}\\
			\hat{\mu}_{D_{F_1}} &= \left[\hat{\mu}_{{F_1}{F_1}} = \frac{2}{n_1(n_1 - 1)}\sum_{i=1}^{n_1}\sum_{j=i+1}^{n_1}||X_{i}^{(1)} - X_{j}^{(1)}||,\right.\\
			& \;\;\;\;\left. \hat{\mu}_{{F_1}{F_2}} = \frac{1}{n_1 n_2}\sum_{i=1}^{n_1}\sum_{j=1}^{n_2}||X_{i}^{(1)} - X_{j}^{(2)}||\right]^T,\\
			\hat{\mu}_{D_{F_2}} &= \left[\hat{\mu}_{{F_1}{F_2}} = \frac{1}{n_1 n_2}\sum_{i=1}^{n_1}\sum_{j=1}^{n_2}||X_{i}^{(1)} - X_{j}^{(2)}||,\right.\\
			& \;\;\;\;\left. \hat{\mu}_{{F_2}{F_2}} = \frac{2}{n_2(n_2 - 1)}\sum_{i=1}^{n_2}\sum_{j=i+1}^{n_2}||X_{i}^{(2)} - X_{j}^{(2)}||\right]^T.
		\end{align*}
		\item Distance Components (DISCO) analysis \parencite{rizzo_disco_2010}: It partitions the total dispersion in the samples 
		\[
		T_{\alpha} = \frac{N}{2} \sum_{i,j = 1}^{k} g_{\alpha}(X^{(i)}, X^{(j)}), 
		\]
		into the between-sample measure of dispersion $S_{\alpha}$ and the within-sample measure of dispersion $W_{\alpha}$
		\[
		T_{\alpha} = S_{\alpha} + W_{\alpha}
		\]
		analogous to the variance components in ANOVA, where
		\[
		g_{\alpha}(X^{(i)}, X^{(j)}) = \frac{1}{n_i n_j}\sum_{u = 1}^{n_i}\sum_{v=1}^{n_j} \|X^{(i)}_u - X^{(j)}_v\|_2^{\alpha}, i,j\in\{1, \dots, k\}.
		\]
		The parameter $\alpha\in(0,2)$ has to be chosen in advance. 
		$\alpha= 2$ yields a usual ANOVA. 
		Either an $F$-type statistic or the between-sample measure of dispersion can be used as the test statistic.		
		\item Classifier two-sample test (C2ST, \cite{lopez-paz_revisiting_2017}): The pooled dataset is split into a training and test set, and a classifier is trained on the training set to distinguish between the datasets. 
		The classifier's accuracy on the test set is used as the statistic. 
		For similar datasets, an accuracy close to the accuracy of the naive prediction of the larger dataset is expected, while for different datasets, higher accuracies are expected. 
		A Binomial test can be used to compare the accuracy of the naive prediction.
		The procedure can also be used for multiple samples.
		\item Random forest-based test by \textcite{hediger_use_2021} (HMN): A random forest is trained on the entire pooled dataset to distinguish between the individual datasets. 
		The out-of-bag prediction error is used as a test statistic. 
		For similar datasets, the error should be close to that expected for always predicting the larger dataset. 
		\item Tree-based test by \textcite{yu_two-sample_2007} (YMRZL): A classification tree is trained to distinguish between the datasets using a training dataset that is a subset of the pooled sample. 
		Its classification error on the left-out test set is used as the statistic. 
		For similar datasets, the error should be close to that expected for always predicting the larger sample.
		\item Direction Projection Permutation (DiProPerm) test \parencite{wei_direction-projection-permutation_2016}: A linear classifier is trained to distinguish between the two datasets. 
		All points in the pooled dataset are then projected onto the normal vector of the separating hyperplane defined by that classifier. 
		Then, a univariate two-sample statistic is calculated on these projected values, and a permutation test is performed using that statistic. 
		\item Original edge count test by \textcite{friedman_multivariate_1979} (FR): A graph (originally the minimum spanning tree, MST) is constructed on the pooled sample using an appropriate distance measure (here: Euclidean distance). 
		For calculating the test statistic for a given similarity graph, the number of edges connecting points from different samples is counted.
		The expectation and variance of this edge-count statistic under the null hypothesis of equal distributions are known and can be calculated analytically.
		The standardized edge count is then used as the test statistic. 
		It follows a standard normal distribution asymptotically under the null, given certain assumptions on the size and density of the similarity graph.
		For similar datasets, higher numbers of edges connecting points from different samples are expected. 
		\item Generalized edge count test by \textcite{chen_new_2017} (CF): The Friedman-Rafsky test is generalized to improve the power for detecting both location and scale alternatives. 
		Here, the number of edges connecting points within each of the two samples,  $R_1, R_2,$ respectively, is counted in a similarity graph on the pooled sample.
		The Mahalanobis distance 
		\[
		(R_1 - \E_{H_0}(R_1), R_2 - \E_{H_0}(R_2)) \Cov_{H_0}^{-1}(R) \begin{pmatrix}
			R_1 - \E_{H_0}(R_1)\\ R_2 - \E_{H_0}(R_2)
		\end{pmatrix}
		\] 
		of the vector $R = (R_1, R_2)^T$ is used as the test statistic. Under the null, it is asymptotically $\chi^2_2$-distributed under certain assumptions on the graph. 
		Small values of the statistic indicate the similarity of the datasets.
		\item Weighted edge count test by \textcite{chen_weighted_2018} (CCS): The Friedman-Rafsky test is generalized to improve the power in settings with unequal sample sizes. 
		The weighted statistic is defined as 
		\[
		R_w = \frac{n_1}{N} R_1 + \frac{n_2}{N} R_2,
		\]
		where $R_1, R_2$ are defined as above and $n_i$ denotes the sample size of the $i$-th sample, $i = 1, 2$, $N = n_1 + n_2$. 
		Again, the expectation and standard deviation of $R_w$ can be calculated analytically and are used to define a standardized test statistic that is asymptotically standard normally distributed under the null, given certain assumptions on the graph. 
		Small numbers of edges connecting points within the same sample indicate similar datasets.
		Therefore, small values of $R_w$ or its standardized version indicate similarity. 
		\item Max-type edge count test by \textcite{zhang_graph-based_2022} (ZC): The test is another generalization of the Friedman-Rafsky test. 
		Its test statistic is given by 
		\[
		R_m = \max\{\kappa R_w, |R_1 - R_2|\},
		\]
		where $\kappa >0$ is a parameter that has to be chosen before testing. 
		Again, a standardized version is given by standardizing  $R_w$ and $R_d = |R_1 - R_2|$ with their expectations and standard deviations under the null, with both components asymptotically following a standard normal distribution under the null, given certain assumptions on the graph. 
		Small values of the statistic indicate similarity. 
		\item Multi-sample graph-based test by \textcite{song_new_2022} (SC): Two statistics are proposed, which use the numbers of between- and within-sample edges of a similarity graph on the pooled sample. 
		The first test statistic is given by
		\begin{align*}
			S &= S^W + S^B,\text{ where}\\
			S^W &= \left(R^W - \E_{H_0}(R^W)\right)^{T} \mathbb{C}\text{ov}_{H_0}^{-1}\left(R^W\right) \left(R^W - \E_{H_0}(R^W)\right),\\
			S^B &= \left(R^B - \E_{H_0}(R^B)\right)^{T} \mathbb{C}\text{ov}_{H_0}^{-1}\left(R^B\right) \left(R^B - \E_{H_0}(R^B)\right),
		\end{align*}
		where $R^W$ denotes the vector containing the numbers of within-sample edges for each of the $k$ samples and $R^B$ denotes the vector containing the numbers of between-sample edges for all $k(k-1)$ pairs of different samples. 
		The second statistic is defined as
		\[
		S^A = \left(R^A - \E_{H_0}(R^A)\right)^{T} \mathbb{C}\text{ov}_{H_0}^{-1}\left(R^A\right) \left(R^A - \E_{H_0}(R^A)\right),
		\]
		where $R^A$ denotes the vector of all linearly independent numbers of edges between and within samples, i.e.\ excluding the last pair. 
		\item Test based on the shortest Hamilton path by \textcite{biswas_distribution-free_2014} (BMG): An edge count statistic similar to that of the Friedman-Rafsky test is calculated but using the shortest Hamilton path instead of the MST as the similarity graph.
		\item Cross-Match test by \textcite{rosenbaum_exact_2005}: For the cross-match test, the optimal non-bipartite matching is used as a similarity graph on the pooled sample. 
		The test statistic is calculated analogously to that of the Friedman-Rafsky test. 
		\item Multi-sample Cross-Match statistic by \textcite{petrie_graph-theoretic_2016}: The optimal non-bipartite matching is calculated on the pooled sample, and the overall number of edges connecting points from different samples is calculated. 
		It is standardized by the analytical expectation and standard deviation under the null hypothesis. 
		The standardized statistic is asymptotically standard normally distributed under the null.
		High values of the cross-match statistic indicate similarity between the datasets.  
		\item Multi-sample Mahalanobis Cross-Match (MMCM) statistic \parencite{mukherjee_distribution-free_2022}: The optimal non-bipartite matching is calculated on the pooled sample.
		The numbers of edges $a_{ij}$ connecting points from sample $i$ and sample $j$, $i\ne j \in \{1,\dots,k\}$, is calculated. 
		The Mahalanobis distance of the cross-match vector $A = a_{12}$ in the two-sample case and $A = (a_{12}, a_{13}, a_{23}, a_{24})^T$ in the four-sample case, respectively, is used as the test statistic 
		\[
		\text{MMCM} = (A - \E_{H_0}(A))^T \Cov_{H_0}^{-1}(A) (A - \E_{H_0}(A)),
		\]
		where again expectations and covariances under the null can be calculated analytically. 
		The MMCM statistic asymptotically follows a $\chi^2_{k-1}$-distribution under the null. 
		For similar datasets, low MMCM values are expected. 
		For two samples, this is analytically equivalent to Petrie's test and the \textcite{rosenbaum_exact_2005} cross-match test.
		\item Nearest-neighbour test by \textcite{schilling_multivariate_1986} and \textcite{henze_multivariate_1988} (SH): The $K$-nearest neighbour graph is calculated on the pooled sample. 
		The mean within-sample edge count
		\[
		\text{SH} := L := \frac{1}{KN} (R_1 + R_2)
		\]
		is used as the test statistic.
		The number of nearest neighbours, $K$, has to be chosen.
		\item Nearest-neighbour test by \textcite{barakat_multivariate_1996} (BQS): This method generalizes the Schilling-Henze test such that the parameter $K$ does not have to be chosen. 
		Therefore, the sum of edge counts is calculated, summing up over all possible values of $K$. 
		\item Test by \textcite{mukhopadhyay_nonparametric_2020} (MW): 
		A set of orthogonal functions designed non-parametrically (LP polynomials) is derived by orthonormalizing a collection of functions that are defined as orthonormal polynomials based on mid-distribution transformations. 
		A polynomial kernel of degree two is created based on these functions, which captures the similarity between two data points within the LP-transformed domain.
		The resulting values from the kernel Gram matrix are used as weights for a graph on the pooled sample. 
		Then, the graph's vertices are clustered into $k$ groups with higher connectivity to assess how closely this clustering aligns with the actual memberships of the points in the $k$ datasets.
		This leads to an independence testing problem, which can be addressed by checking if all LP means are equal to zero.
		\item Jeffreys divergence \parencite{jeffreys_invariant_1997}: It is the symmetrized version
		\[
		J(F_1, F_2) = \text{KL}(F_1, F_2) + \text{KL}(F_2, F_1)
		\]
		of the Kullback Leibler (KL) divergence \parencite{kullback_information_1951}
		\[
		\text{KL}(F_1, F_2) := \int \log\left(\frac{f_1(x)}{f_2(x)}\right) f_1(x) \dif x
		\]
		of two distributions $F_1$ and $F_2$.
		\item Wasserstein distance \parencite{vaserstein_markov_1969}: The $q$-Wasserstein distance of two distributions $F_1$ and $F_2$ on $\mathcal{X}$ is defined as 
		\[
		\text{W}(F_1, F_2) := \left(\min_{\pi\in\Pi(F_1, F_2)} \int_{\mathcal{X}\times\mathcal{X}} d_{\mathcal{X}}(x, y)^q \dif \pi(x, y)\right)^{1/q}, 
		\]
		where $d_{\mathcal{X}}$ is the metric that $\mathcal{X}$ is provided with, and 
		\[
		\Pi(F_1, F_2) := \{\pi_{1,2}\in\mathcal{P}(\mathcal{X}\times\mathcal{X})\arrowvert \pi_1 = F_1, \pi_2 = F_2\}
		\]
		is the set of joint distributions over the product space $\mathcal{X}\times\mathcal{X}$ with marginal distributions $F_1$ and $F_2$.
		\item Ball divergence \parencite{pan_ball_2018}: The Ball divergence measures the difference between two probability measures.
		It is defined as the square of the measure difference over a given closed ball collection.
		\item Maximum Mean Discrepancy (MMD) \parencite{gretton_kernel_2006} and block MMD \parencite{zaremba_b-test_2013}: The MMD uses an extension of feature maps as used by other kernel methods like support vector machines to the space of probability distributions by representing each distribution as a so-called mean function $\mu$. 
		This maps each probability distribution to an element in the reproducing kernel Hilbert space (RKHS) corresponding to a chosen kernel function. 
		The MMD then measures the distance of the mean functions measured in the RKHS $\mathcal{H}$
		\[
		\text{MMD}(\mathcal{H}, F_1, F_2) = \|\mu_{F_1} - \mu_{F_2}\|_{\mathcal{H}}.
		\]
		For the block MMD, the estimation is performed block-wise, which decreases the runtime from quadratic in $N$ to linear in $N$ for fixed block sizes. 
		\item Generalized permutation kernel two-sample test (GPK) \parencite{song_generalized_2021}: \textcite{song_generalized_2021} propose another kernel-based test for which they decompose the squared MMD estimator as
		\[
			\widehat{\text{MMD}}^2 = \alpha + \beta - 2\gamma,
		\]
		where
		\begin{align*}
			\alpha &= \frac{1}{n_1(n_1 - 1)}\sum_{i=1}^{n_1}\sum_{\substack{j = 1\\ j\ne i}}^{n_1} K(X_i^{(1)}, X_j^{(1)}),\\
			\beta &= \frac{1}{n_2(n_2 - 1)}\sum_{i=1}^{n_2}\sum_{\substack{j = 1\\ j\ne i}}^{n_2} K(X_i^{(2)}, X_j^{(2)}),\\
			\gamma &= \frac{1}{n_1n_2}\sum_{i=1}^{n_1}\sum_{j = 1}^{n_2} K(X_i^{(1)}, X_j^{(2)}).
		\end{align*}
		Their proposed test statistic is defined as
		\[
			\text{GPK} = (\alpha - \E_{H_0}(\alpha), \beta - \E_{H_0}(\beta)) \Cov^{-1}_{H_0}\left(\begin{pmatrix} \alpha\\ \beta \end{pmatrix}\right) \begin{pmatrix} \alpha - \E_{H_0}(\alpha)\\ \beta - \E_{H_0}(\beta)\end{pmatrix}.
		\]
		It can be decomposed into $\text{GPK} = Z_W^2 + Z_D^2$, where $Z_W$ and $Z_D$ are the standardized versions (with expectation and variance under $H_0$) of
		\begin{align*}
			W &= \frac{n_1}{N}\alpha + \frac{n_2}{N}\beta,\\
			D &= n_1 (n_1 -1) \alpha - n_2(n_2 - 1)\beta.
		\end{align*}
		Based on this, $W$ can be further generalized to
		\[
			W_r = r\frac{n_1}{N}\alpha + \frac{n_2}{N}\beta
		\]
		and $Z_W$ to $Z_{W,r}$. 
		Since the asymptotic distribution of $Z_W = Z_{W, 1}$ is complicated but that of $Z_{W,r}, r\ne 1$, is a standard normal under mild assumptions, fast tests are proposed using a Bonferroni correction for the individual tests based on $Z_D$, $Z_{W,1.2} =: Z_{W1}$ and $Z_{W,0.8} =: Z_{W2}$. 
		\item Characteristic distance by \textcite{li_measuring_2022} (LHZ): It is defined as
		\begin{align*}
			\text{CD} (X, Y) &= \E\left[\| \E\left(\exp\left(i\langle X^{\prime\prime}, X - X^{\prime}\rangle\right)\arrowvert X - X^{\prime}\right)\right.\\
			&\qquad ~~- \left.\E\left(\exp\left(i\langle Y, X - X^{\prime}\rangle\right)\arrowvert X - X^{\prime}\right)\|^2\right]\\
			&\quad + \E\left[\| \E\left(\exp\left(i\langle X, Y - Y^{\prime}\rangle\right)\arrowvert Y-Y^{\prime}\right)\right. \\
			&\qquad \quad ~-\left. \E\left(\exp\left(i\langle Y^{\prime\prime}, Y - Y^{\prime}\rangle\right)\arrowvert Y - Y^{\prime}\right)\|^2\right],
		\end{align*}
		where $X^{\prime}, X^{\prime\prime}$ and $Y^{\prime}, Y^{\prime\prime}$ denote independent copies of $X\sim F_1$ and $Y\sim F_2$, respectively.
		\item Test based on cumulative distribution functions by \textcite{biau_asymptotic_2005} (BG): The $L_1$-distance between the two empirical distributions restricted to a finite partition is calculated.
		For this, a finite partition of the subspace spanned by the two datasets has to be defined, e.g.\ a rectangular partition.
		The $L_1$ distance between the empirical distribution functions restricted to the elements of this partition is calculated.
		\item FS and RI test \parencite{paul_clustering-based_2022}: These tests are distribution-free $k$-sample tests proposed for the high dimension low sample size (HDLSS) setting.
		The tests are based on clustering the pooled sample and comparing the resulting clustering to the true dataset membership via a contingency table.
		If the datasets come from the same distribution, the cluster and dataset membership are independent, while if the datasets come from different distributions, the clustering depends on the true dataset membership. 
		As a clustering algorithm, $K$-means based on the generalized version of the \textit{mean absolute difference of distances (MADD)}
		\[
		\rho_{h,\varphi}(z_i, z_j) = \frac{1}{N-2} \sum_{m\in \{1,\dots, N\}\setminus\{i,j\}} \left| \varphi_{h,\psi}(z_i, z_m) - \varphi_{h,\psi}(z_j, z_m)\right|,
		\]
		is used as proposed by \textcite{sarkar_perfect_2020} for the HDLSS setting. 
		Here, $z_i, i = 1,\dots,N$, denote realizations from the pooled sample and \[\varphi_{h,\psi}(z_i, z_j) = h\left(\frac{1}{p}\sum_{l=1}^p\psi|z_{il} - z_{jl}|\right),\] where $h:\mathbb{R}^{+} \to\mathbb{R}^{+}$ and $\psi:\mathbb{R}^{+} \to\mathbb{R}^{+}$ are continuous and strictly increasing functions. 
		$\psi_1(t) = t^2$, $\psi_2(t) = 1 - \exp(-t)$, $\psi_3(t) = 1 - \exp(-t^2)$, $\psi_4(t) = \log(1 + t)$, and $\psi_5(t) = t$ are considered in combination with $h_1(t) = \sqrt{t}$ and $h_2(t) = t$. 
		The number of clusters has to be chosen in advance. 
		A natural choice is to set the number of clusters to the number of datasets $k$.
		For the RI test, the Rand index of the clustering, which is zero when the clustering is perfect, is used as a test statistic. 
		For the FS test, the generalized Fisher's test statistic for testing for independence in a $k\times\ell$ contingency table is used.
		Modified versions of the tests (MRI, MFS test) are proposed, for which the number of clusters is estimated from the data using the Dunn index.
		Moreover, multi-scale versions of the tests are presented (MSRI, MSFS test), where the respective tests are performed for different numbers of clusters, and the results are aggregated using a Bonferroni adjustment for the individual tests. 
		Lastly, aggregated tests (AFS / ARI test) can be used in the multi-sample setting that perform all pairwise FS / MFS or RI / MRI tests, respectively, on the samples and aggregate the results by taking the minimum test statistic value and applying a multiple testing procedure.
		\item Engineer metric \parencite{rachev_probability_1991}: The $L_q$-\textit{Engineer metric} of two  $p$-dimensional random vectors $X\sim F_1$ and $Y\sim F_2$ is defined as
		\[
		\text{EN}(X, Y; q) = \left[ \sum_{i = 1}^{p} \left\arrowvert \E\left(X_i\right) - \E\left(Y_i\right)\right\arrowvert^q\right]^{\min(q, 1/q)} \text{ with } q> 0.
		\]
		\item Decision tree-based dataset distance by \textcite{ganti_framework_1999} (GGRL): The GGRL requires the datasets to include a target variable. 
		A decision tree is fit to each dataset. 
		The partitions of the sample space induced by these trees are intersected, and the proportions of data points falling into each segment of this so-called greatest common refinement (GCR) are determined for each dataset. 
		The resulting probability vectors $p$ and $q$ are then compared using a difference function $f$, and the results are aggregated using an aggregate function $g$: 
		\[
		\text{GGRL} = g(f(p, q)).
		\]
		Proposed choices for the difference function are the absolute component-wise differences ($f_a$) or the absolute component-wise differences scaled by their means ($f_s$). 
		For the aggregate function, the sum or the maximum is proposed.
		\item Decision tree-based dataset similarity by \textcite{ntoutsi_general_2008} (NKT): 				The same partition as for GGRL is calculated and based on that, the probability densities $P_D(\mathcal{X})$ and $P_D(Y^{(j)},\mathcal{X})$ are estimated for $D \in \{X^{(1)}, X^{(2)}, Z\}$.
		Then, \textcite{ntoutsi_general_2008} consider the similarity index 
\[s(p, q) = \sum_{i} \sqrt{p_i \cdot q_i}\]
for vectors $p$ and $q$, where $(n_r \times n_c)$-matrices are interpreted as $(n_r \cdot n_c)$-dimensional vectors.
		For the conditional distribution, the similarity vector $S(Y|\mathcal{X}) \in \mathbb{R}^{n_r}$ is computed with $S(Y|\mathcal{X})_i = s(\hat{P}_{X^{(1)}}(Y|\mathcal{X})_{i \bullet}, \hat{P}_{X^{(2)}}(Y|\mathcal{X})_{i \bullet})$ and index $i \bullet$ denoting the $i$-th row.
		Based on this, three similarity measures for datasets are proposed:
		\begin{align*}
			\text{NTO1} &= s(\hat{P}_{X^{(1)}}(\mathcal{X}), \hat{P}_{X^{(2)}}(\mathcal{X})),\\
			\text{NTO2} &= s(\hat{P}_{X^{(1)}}(Y, \mathcal{X}), \hat{P}_{X^{(2)}}(Y, \mathcal{X})),\\
			\text{NTO3} &= S(Y|\mathcal{X})^{T} \hat{P}_{Z}(\mathcal{X}).
		\end{align*}
		All three measures have values in the interval $[0, 1]$, where high values correspond to high similarity.
		\item Optimal Transport Dataset Distance \parencite{alvarez-melis_geometric_2020} (OTDD): 
		The OTDD is a distance between datasets that takes into account a target variable included in the datasets. It is defined as 
		\[
		d_{\text{OT}}(X^{(1)}, X^{(2)}) = \min_{\pi\in\Pi(F_1, F_2)} \int_{\mathcal{Z}\times\mathcal{Z}} d_{\mathcal{Z}}(z, z^\prime)^q \dif \pi(z, z^\prime)
		\]
		where $X^{(1)}, X^{(2)}$ denote the two datasets, and $\Pi(F_1, F_2)$ is again the set of joint distributions over the product space $\mathcal{Z}\times\mathcal{Z}$ over the sample space of the pooled sample with marginal distributions $F_1$ and $F_2$, and
		\[
		d_{\mathcal{Z}}(z, z^\prime) := (d_{\mathcal{X}}(x, x^{\prime})^{q^\prime} + W_{q^\prime}(\alpha_y, \alpha_{y^{\prime}})^{q^\prime})^{1/{q^\prime}}.
		\]
		defines a distance of two points $z^{T} = (x^{T}, y)$, and ${z^\prime}^{T} = ({x^\prime}^{T}, y^\prime)$ in the pooled sample.
		$d_{\mathcal{X}}$ defines a distance on the covariate space, e.g., the Euclidean distance, and  $W_{q^\prime}(\alpha_y, \alpha_{y^{\prime}})$ is the $q^\prime$-Wasserstein distance of the distribution of the subset of covariate data with corresponding response value $y$ and the distribution of the subset of covariate data with corresponding response value $y^\prime$. 
		Different speed-ups of the calculation are possible by assuming normal distributions or by using a Sinkhorn approximation of the optimization problems.
		For more details see Appendix~\ref{app:meth.pars}.
	\end{itemize}
	All of these methods are applied in the two-sample case. 
	The KMD, energy statistic, DISCO, C2ST, SC, MW, MMCM, the method of \textcite{petrie_graph-theoretic_2016}, and all versions of the FS and RI test are also applied in the multi-sample case. 
	For the other methods, no generalizations to the multi-sample case are available in the literature.
	For most methods, a permutation test is proposed in addition.
	However, no permutation test is performed here due to the high runtime. 
	More details on the methods and their theoretical properties can be found in \textcite{stolte_methods_2024} and the references therein.  
	
	\subsection{Performance Measures}
\begin{figure}[!tb]
		\begin{center}
			\resizebox{0.82\linewidth}{!}{\input{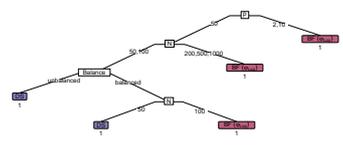}}
		\end{center}
		\caption{Overview of analysis approach. The simulation results in the depicted PESR curves were slightly manipulated for illustration purposes.}\label{fig:workflow}
	\end{figure}

	Two aspects are evaluated here. 	
	First, it is evaluated how well the methods can detect the differences between the distributions that were described in the previous subsections. 
	Second, computational aspects are taken into account.
	For the first evaluation, no classical power comparison can be conducted since not all methods define a test.
	Moreover, such a comparison would not be possible due to the very high runtimes of the many permutation tests. 
	Instead, the methods are compared as in \textcite{stolte_comparing_2025}. 
	To make the performances comparable, an approach inspired by classical power comparisons is performed. 
	It is illustrated in Figure~\ref{fig:workflow}. 

	In short, the null distribution of each test statistic is approximated by the simulated test statistic values in each of the considered null settings (per DGP). 
	From this, the 95\% quantile (for dissimilarity measures), or 5\% quantile (for similarity measures) is calculated as a threshold that is then applied for the corresponding alternative settings.
	The proportion of simulation repetitions in which a more extreme statistic value than this threshold is observed is used to quantify the performance of the method. 
	This \textit{Proportion of Extreme Simulation Repetitions} w.r.t.\ the null threshold is abbreviated as \textit{PESR} in the following.
	It is used to evaluate how well the methods detect the difference in the distributions, and it can be compared between different methods. 
	Note that the determined PESR does not directly equal the testing power since only one specific null situation is considered, see \textcite{stolte_comparing_2025}. 
	
	The number of simulation repetitions is set to $500$ repetitions per scenario as in the previous evaluation on categorical data, to achieve sensibly small Monte Carlo standard errors (MCSE) of the estimated PESR.
	For $500$ iterations, the maximum MCSE is then $\sqrt{0.5(1-0.5) / 500}\approx 0.022$. 
	For a proportion of $0.05$, which corresponds to the PESR in null situations, the MCSE is $\sqrt{0.05(1-0.05) / 500}\approx 0.01$. 
	This is considered sufficiently small here. 
	Using $1000$ iterations would only bring down the MCSEs to $\approx0.015$ and $\approx 0.006$, respectively, but this would double the already high runtime.
	
	In case of computational errors resulting in missing or infinite values of the statistics, the affected repetitions are excluded from the PESR calculation.
	If there are missing values in more than $100$ of the $500$ iterations, the corresponding PESR value is set to missing. 
	This ensures that the calculated PESR values are based on a reasonably high number of repetitions.
	
	For the evaluation of computational aspects, the runtime, memory consumption, and any numerical problems are taken into account. 
	The runtime and memory consumption are measured for selected scenarios only to ensure a fair comparison in a controlled setting. 
	This cannot be ensured in the main simulation study that is run distributed across different compute clusters with differing hardware and software components and varying workloads on the clusters.
	Therefore, the computational cost of the methods is evaluated in an additional benchmark under consistent conditions, but for selected scenarios only.
	Here, the null situations for normal data are chosen. 
	All combinations of $N$ and $p$ as discussed before are used, and datasets are generated once for each combination of $N$ and $p$ for that scenario. 
	On each dataset, each similarity method is then applied once to measure the memory consumption and afterwards at least $10$~times to measure the runtime per method call. 
	For methods with low runtimes, the number of repetitions is increased such that the method is run for at least $1$~second to get stable estimates of the runtime per method call. 
	Each method is called once before starting the benchmark to ensure that all required packages and objects are already loaded at the start of the benchmark, and the results are not distorted by lazy loading.
	The benchmarks are performed on a Lenovo ThinkPad laptop with an AMD Ryzen~5 PRO 4650U processor with six cores and 16\,GB of RAM under Windows~11. 
	Benchmarks are run during the night-time when the laptop is not used for any other work to ensure that the results are not disturbed by other computations.
	
	The results of the PESR comparisons and of the computational aspects are in the end combined for the overall performance assessment of the methods. 
	An overview of the analysis is given in Figure~\ref{fig:workflow}. 
	First, the individual PESR values per combination of the dataset dimension, DGP and method are calculated as described above. 
	Next, the difference of these PESR values to those of an ``ideal'' method, i.e.\ always picking the best method with the highest PESR for each specific deviation, is calculated as shown in the fifth plot on the left in Figure~\ref{fig:workflow}.
	For that, for each parameter setting of the DGP, the highest PESR across all methods is determined and the differences of this value and the PESR value of each method is taken. 
	This is repeated for all parameter settings.
	These differences per method are then averaged across the specific deviation values per scenario. 
	These mean differences to the ``ideal'' method are used to compare the methods. 
	There are different interpretations of these quantities possible. 
	The difference can be seen as quantifying how much worse picking a certain fixed method is on average than making the best possible decision in each situation (ideal method). 
	Moreover, it can be interpreted as the number of times that the ideal method correctly identifies the difference between the datasets but the respective method does not. 
	When this number is then normalized by the number of different deviation values to make the scenarios comparable, as there are, e.g.\ six different shift values but only four different correlation values in the data-generating mechanisms, this again equals the mean differences to the best method. 
	For equally spaced deviation values, it is also proportional to the area between the curves. 
	In the calculation, missing values for the PESR are penalized by setting the corresponding differences to the maximum value of one. 
	
	The mean differences to the ideal method per scenario and method are then presented in heatmaps to enable an overview and comparison between the methods. 
	Moreover, the results regarding runtime, memory, and errors are summarized as follows. 
	The runtimes are measured for different $N$ and $p$ for the normal null scenario. 
	For the $i$-th $N$ and $p$ combination, the median runtimes are scaled to $[0, 1]$: 
	\[
	r_{i,j}^{\text{scaled}} = \frac{r_{i,j} - \min_{j}r_{i,j}}{\max_j r_{i,j} - \min_j r_{i,j}},
	\]
	where $r_{i,j}$ denotes the median runtime of the $j$-th method, $j = 1,\dots,n_{\text{meth}}$, for the $i$-th $N$ and $p$ combination.
	Then, for each method, the median $\operatorname{med}_i r_{i,j}^{\text{scaled}}$ of these scaled values across all $N$ and $p$ combinations is taken. 
	Analogously, the measured memory consumption values are scaled to $[0,1]$ per $N$ and $p$ combination, and the median per method is taken. 
	To quantify the overall error susceptibility of the method, the maximum proportion of missing values due to computational errors is calculated across all scenarios of the simulation. 
	For the methods that can take a target variable into account, all aforementioned calculations are performed only for the scenarios where the OGM is not altered for comparability with the other methods. 
	
	These summaries of the practical applicability criteria are also incorporated in the heatmaps for the method comparisons. 
	Based on the performance quantified by the mean differences in PESR to the ideal method and these practical aspects together, an overall method ranking is established.
	
	In the next step, for each method, it is determined for which deviations it shows an acceptable performance in detecting the differences between datasets.
	For this, methods are considered acceptable if their mean differences to the ideal method for a given scenario are not more than 0.1 above the minimum mean difference to the ideal method across all methods in that scenario.
	For the example shown in the Table in Figure~\ref{fig:workflow} this means that as the smallest mean difference is given as 0.02, all methods with a mean difference of up to 0.12.  
	The cut-off of $0.1$ was chosen by visual assessment of which PESR-curves the authors would still consider as acceptable.
	Again, the acceptable scenarios can be visualized in heatmaps to see how many scenarios are covered by each method.
	Here, a scenario is said to be covered by a method, if the method is acceptable for it. 
	
	To give concrete recommendations for choosing a suitable method, the methods that cover most scenarios are identified depending on the number of variables, observations, and the balance of the sample sizes. 
	This is performed once overall, once for each deviation, once for each DGP, and once for each combination of deviation and DGP, to provide recommendations using different levels of prior knowledge about the data-generating mechanism of interest. 
	In each case, a decision tree is fit to predict the best method based on the dataset dimensions ($N, p$, balance). 
	The tree is always grown until there are only observations of one class left in each leaf node to provide sets of unambiguous rules for finding the best method.
	
	As it is not expected that one method can cover all scenarios, additionally, the best method combination that covers all scenarios, or 95\%, and 90\%, respectively, of the scenarios is identified. 
	These method combination then ensures that in (almost) all considered scenarios one of the selected methods is close to the best-performing one. 
	Finding this method combination can be seen as an instance of the set cover problem, which is an NP-complete discrete optimization problem. 
	As an exhaustive search for the best combination is infeasible, a greedy search is performed, which is known to give the best polynomial time approximation of the solution to the problem \parencite{trevisan_non-approximability_2001}.
	The proposed method combinations could be applied in practice in a multiple testing framework to simultaneously check for various differences between two or more datasets.

	\subsection{Software}
	The simulations for the two-sample case are performed using \texttt{R} version 4.4.0 \parencite{R} on the Linux-HPC-Cluster (LiDO3) at TU Dortmund University. The simulations for the multi-sample case are performed using \texttt{R} version 4.4.2 \parencite{R} on the compute cluster of the Department of Statistics of TU Dortmund University.
	Further analyses and benchmarks are performed using \texttt{R} version 4.5.1 \parencite{R} on a personal computer.
	The implementation of all methods can be found in the \texttt{DataSimilarity}
	package \parencite{DataSimilarity}. 
	The \texttt{bench} package \parencite{bench} is used for measuring runtime and memory consumption. 
	The \texttt{pheatmap} \parencite{pheatmap} and the \texttt{cba} \parencite{cba} packages are used for visualizing and clustering the PESR values of the methods across scenarios. 
	The \texttt{rpart.plot} \parencite{rpart.plot} package is used for visualising the decision rules for finding the best-performing methods for a specific scenario.
	The full \texttt{R} code of the study can be found on Zenodo \parencite{code_zenodo_num}.
	
	\FloatBarrier  
	
	\section{Results}\label{sec:results}
	
	In the following, the results of the simulation study are presented, starting with a pre-selection of the best variant(s) per method. 
	The analysis is then split into the two- and the multi-sample case. 
	In each case, first, the numeric stability, runtime, and memory allocation of the methods are compared to evaluate practical aspects of their computation. 
	Afterwards, the overall method rankings based on the mean PESR differences to the ideal methods are presented, together with an overview the scenarios in which each method is considered acceptable. 
	Based on this, next, the decision rules for finding the method that covers most scenarios based on characteristics of the datasets are presented. 
	Lastly, the results of the greedy search for the method combinations that cover most scenarios are shown.

	\subsection{Pre-Selection of Methods}
	As a first step, a pre-selection is performed in which only the best variants per method are chosen. 
	This is done to reduce the number of methods in the overall comparisons for clarity in the following analyses. 
	The pre-selection focuses on the performance of the methods and ignores the computational aspects since it is assumed that practitioners would be willing to trade slightly higher resource consumption for better performance, and for most of the methods, the resource consumption and computational problems did not differ by much between the variants. 
	The variants are selected as follows. 
	First, the mean differences of the PESR values from those of the ideal method are calculated per combination of the DGP and deviation. 
	The values are averaged over the values per deviation, $N$, $p$, the balance, and in the multi-sample case, also the grouping. 
	Then, per method, only the best variant is selected per deviation, i.e.\ the variant that has the lowest mean difference to the ideal method. 
	In case of (approximate) ties, all tied winner variants are chosen. 
	For the subsequent method comparison, the union of the sets of method variants is taken over all DGP and deviation combinations, i.e.\ the best variants of a method for normal shift / scale / correlation alternatives as well as $t$ distribution shift / scale / correlation / kurtosis alternatives, and so on, are all included in the overall comparison. 
	The corresponding heatmaps of differences to the ideal method that were used in the pre-selection, along with the selected parameter settings for all methods, can be found in Appendix~\ref{app:meth.sel}.
	All following analyses will be performed on the pre-selected method variants only.

	\subsection{Two-Sample Case}
	
	In the following, the methods for the two-sample case are compared. 
	First, their numerical properties, i.e.\ the number of computational errors during the execution of the simulation study, and the runtime and memory consumption in the benchmark setting, are compared. 
	Afterwards, the method ranking according to the mean PESR differences to the ideal method is established, and the scenarios for which each method is deemed acceptable are evaluated. 
	Based on this, decision rules for finding the method that covers most scenarios as well as method combinations that cover most scenarios, are proposed.
	
	\subsubsection{Evaluation of Computational Aspects}

\begin{figure}[!t]
		\centering
		\includegraphics[width=\linewidth]{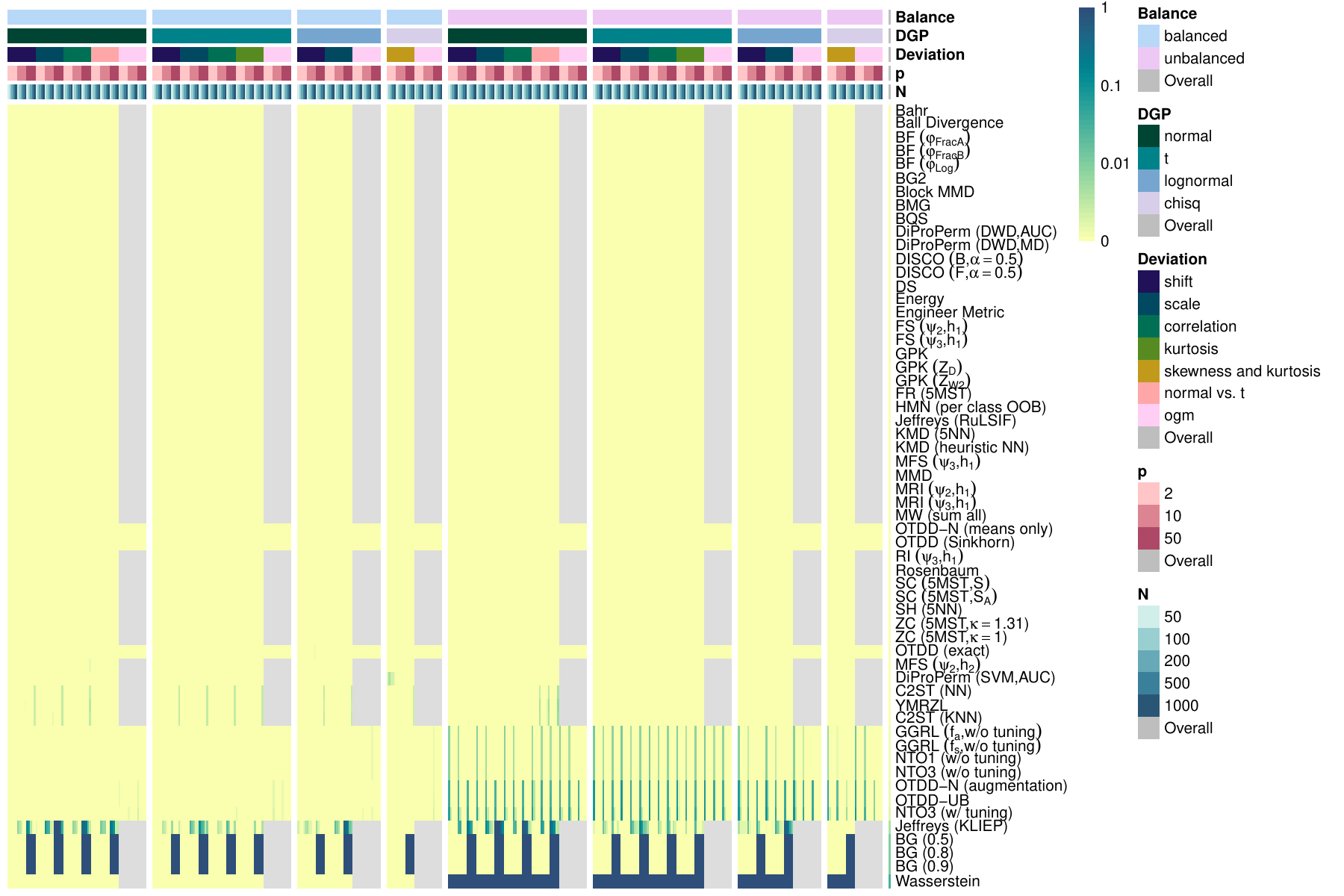}
		\caption{Proportion of iterations per simulation scenario in which computational errors occurred. The total iteration number per scenario is 500. Proportions are shown on a $\log_{10}$-scale with proportions of 0 set to $1/1000$. Grey tiles in the heatmap indicate missing values that occur for methods that do not consider a target variable in scenarios where the deviation is in the OGM. Methods are ordered from top to bottom by increasing row means of the proportions and alphabetically in case of ties.}\label{fig:heat.err.two}
	\end{figure}

	First, the number of computational errors that each method produced while running the simulation study is compared. 
	The respective error proportions per scenario are shown in Figure~\ref{fig:heat.err.two} as a heatmap. 
	For many methods, no computational errors occurred. 
	The methods that encountered errors are the following. 
	
	The BG test is not working for $p = 50$ due to too many cells in the required partition of the sample space. 
	The enumeration of the partition becomes impossible for increasing $p$ due to the RAM limitation of a computer.
	
	The C2ST~(KNN) and C2ST~(NN) encountered one or two iterations with errors in many cases, mostly for $N = 1000, p = 50$, few for $N = 500$ or smaller $p$.
	These errors seem to be related to problems with the file system that occurred while running the study.
	Similarly, the YMRZL ran into errors in up to 3 iterations per scenario, mostly for $N = 1000$, $p = 50$.
	
	The DiProPerm~(SVM, AUC) test shows 1\,--\,5 iterations per scenario in cases with balanced sample sizes for $\chi^2$-distributed data, and $p = 2$, where the test statistic values are missing without any error message. 
	
	For GGRL, 1\,--\,6 iterations with errors per scenario occurred for many scenarios with $N = 50$, where not enough ones were generated in the target variable. 
	Similarly, there are 1\,--\,49 (mostly 1\,--\,4) iterations per scenario for NKT with computational errors, almost all in scenarios with $N = 50$, few with $N = 100$. 
	In these cases, not enough ones were generated for the target variables.
	The higher numbers of iterations with errors mostly occurred for extreme deviations, which lead to lower success probabilities in the logistic OGM.
	
	For Jeffreys' divergence, up to all iterations per scenario are affected in certain scenarios by problems in the density ratio estimation.
	This happens especially for stronger deviations, low $N$ or high $p$ scenarios, but at least some problems occurred in almost all scenarios. 
	The other density ratio estimation methods were numerically more stable in this regard, but showed worse performance in almost all scenarios where both variants were able to produce a result.
	
	The MFS test failed in six iterations for correlation alternatives with balanced sample sizes, normal data, $N = 1000$, $p = 50$. 
	This can happen for specific clustering situations during the calculation of the statistic (e.g.\ if almost all observations are in one cluster and the remaining clusters consist of single observations, then the probability of observing the corresponding contingency table cannot be calculated from the generalized hypergeometric distribution). 
	
	For the OTDD~(augmentation), errors happened in up to 36 iterations per scenario, mostly for $N = 50$, some for $N = 100$. 
	This can happen if exactly one 0 or 1 is generated in one of the datasets.
	The same holds for OTDD~(UB), where up to 38 iterations per scenario are affected, mostly for $N = 50$, sometimes for $N = 100$.
	The OTDD~(exact) encountered only one error in an iteration for log-normal data with a shift of 0.5, balanced sample sizes, $N = 1000$, and $p = 10$. 
	The implementation of the Wasserstein distance does not work for unbalanced sample sizes; therefore, no results could be retrieved in those cases.

\begin{figure}[!t]
		\centering
		\includegraphics[width=\linewidth]{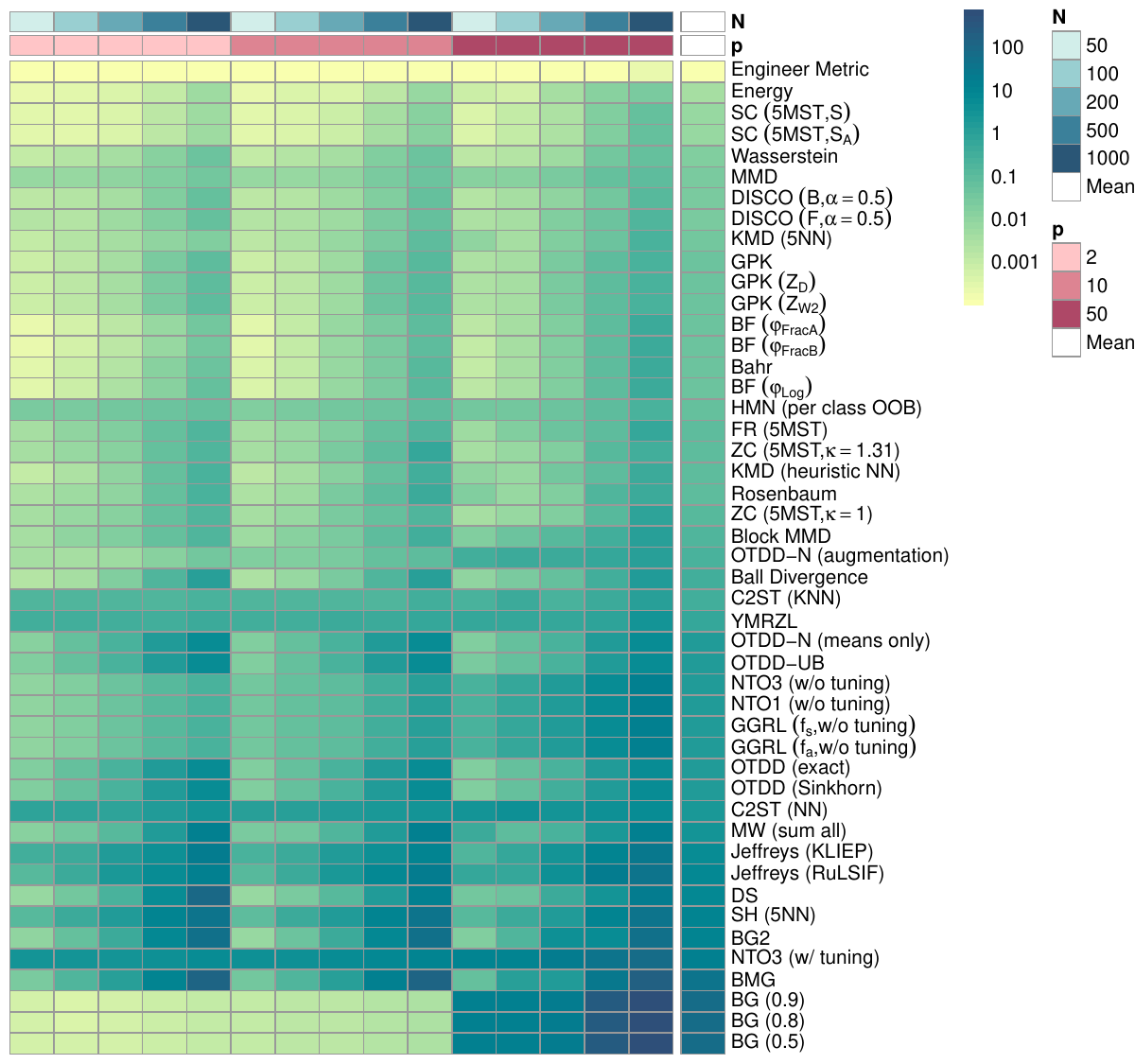}
		\caption{Median runtime in seconds for each method in the benchmark scenarios. The scenarios all consider the null situation for normal data with equal sample sizes, and only $N$ and $p$ are varied. Each method is run at least ten times. If necessary, the number of runs is increased to guarantee that the method is run for at least 1\,s. The runtime is shown on a $\log_{10}$-scale. Methods are ordered from top to bottom by increasing row means and alphabetically in case of ties.}\label{fig:heat.time.two}
	\end{figure}

	Figure~\ref{fig:heat.time.two} shows a comparison of the median runtimes per method across the benchmarking settings. 
	Here, still the untransformed runtimes are shown on a log scale.
	On average, the Energy distance, the SC, and the BG tests are the fastest. 
	It should, however, be noted that the BG has low runtimes for low $p$ but very high runtimes for $p = 50$, where it is no longer possible to calculate the statistic values. 
	Overall, all methods that can handle a target variable are among the slower methods. 
	Moreover, methods that tune some classifiers are particularly slow, as well as the density estimation in Jeffreys or the optimal transport solutions in DS and OTDD. 
	The runtimes of the graph-based methods differ a lot depending on the implementation of the graph that is used. 
	Therefore, some of the graph-based methods are among the fastest (SC), but some others are also among the slowest (SH, BMG).
	Except for BG, which has excessively higher runtimes for $p=50$, which can be attributed to the problems discussed before, the ordering is mostly consistent across the different $N$ and $p$ combinations, with increasing runtimes especially in $N$ but also in $p$. 
	However, there are some methods whose runtimes are almost independent of $N$ and $p$.
	These are the MMD, the HMN, the C2ST, and the YMRZL. 

	\begin{figure}[!t]
		\centering
		\includegraphics[width=\linewidth]{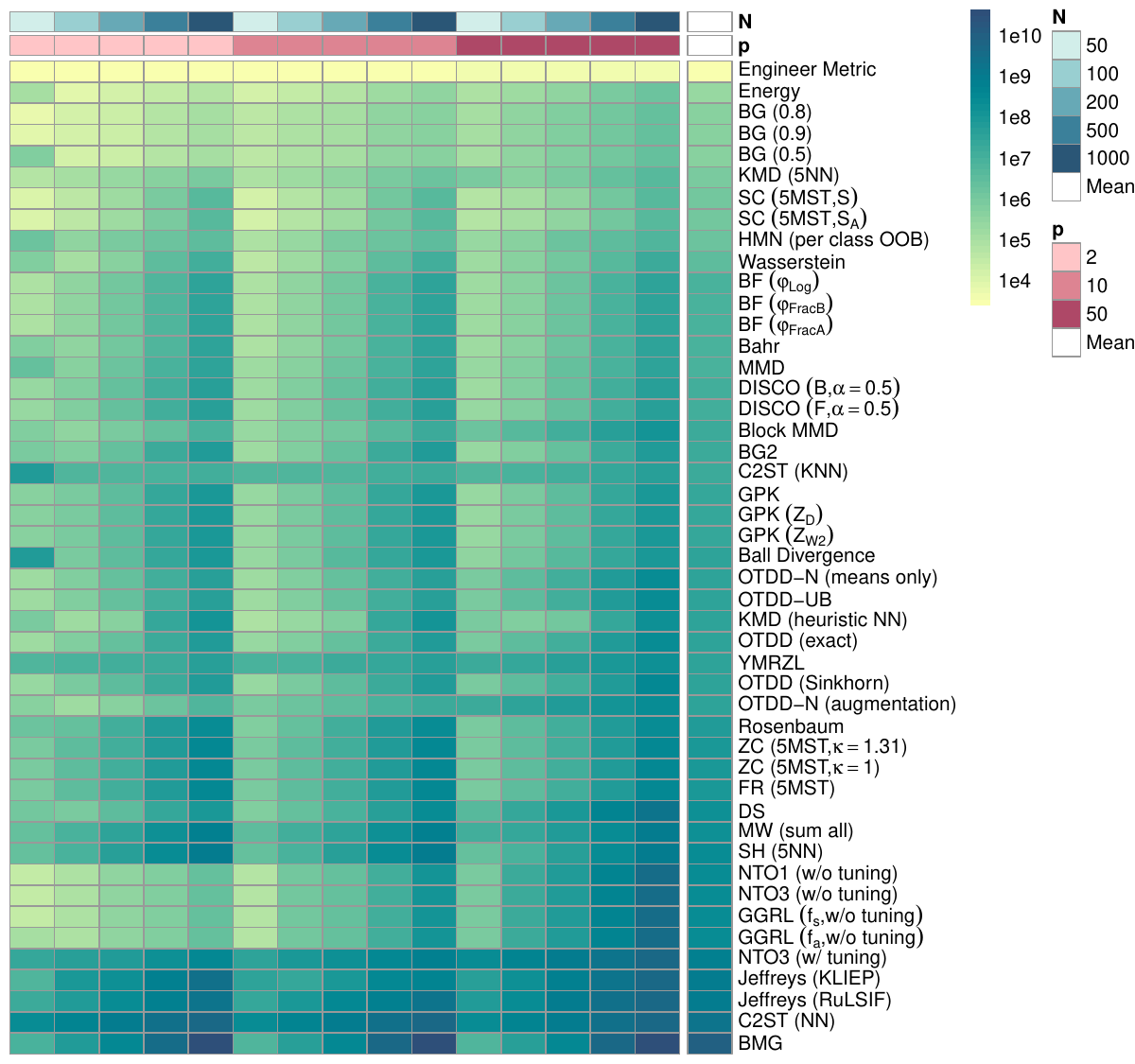}
		\caption{Memory allocation in bytes for each method in the benchmark scenarios. The scenarios all consider the null situation for normal data with equal sample sizes, and only $N$ and $p$ are varied. Each method is run once. The memory consumption is shown on a $\log_{10}$-scale. Methods are ordered from top to bottom by increasing row means and alphabetically in case of ties.}\label{fig:heat.mem.two}
	\end{figure}
	
	The results with regard to the memory consumption are similar (see Figure~\ref{fig:heat.mem.two}). 
	The top three methods with the lowest memory consumption are the Energy distance, the BG, and the SC tests. 
	The BG test shows no increased memory consumption for high $p$, although its calculation is impossible since the available memory would not suffice. 
	There are some changes in the ordering in the midfield compared to the runtime ranking, e.g.\ the GPK test, DISCO, and edge count / cross-match tests are higher in the ranking regarding runtime than regarding memory, while the KMD, NKT, and the BF test are lower. 
	Similar methods, like for the runtime, are ranked in the last places with some changes in the exact ordering. 
	It should be noted that the memory measurements include only allocations made by the \texttt{R} process itself. 
	Therefore, allocations resulting from external code in other languages (e.g.\ \texttt{C}) could not be accounted for in this comparison, which might give an unfair advantage to methods where parts of the calculation are performed in another language. 
	This is, for example, the case for the calculation of the similarity graphs for some of the graph-based methods or for the fitting of the classifiers in classifier-based methods.

	In summary, the classifier-based methods, and methods that can handle a target variable, as well as MW, SH, Jeffreys, BMG, and DS show high resource consumption. 
	The classifier-based methods, methods that can handle target, and especially Jeffreys divergence, also encountered comparably high numbers of computational problems. 
	The BG test and the Wasserstein distance are not applicable for high-dimensional data or unequal sample sizes, respectively, but apart from that, they appear numerically stable and show low resource consumption. 
	The remaining methods can be regarded as good to acceptable with regard to all considered aspects of their practical usability.

	\subsubsection{Overall Method Ranking}	
\begin{figure}[!t]
		\centering
		\includegraphics[width=\linewidth]{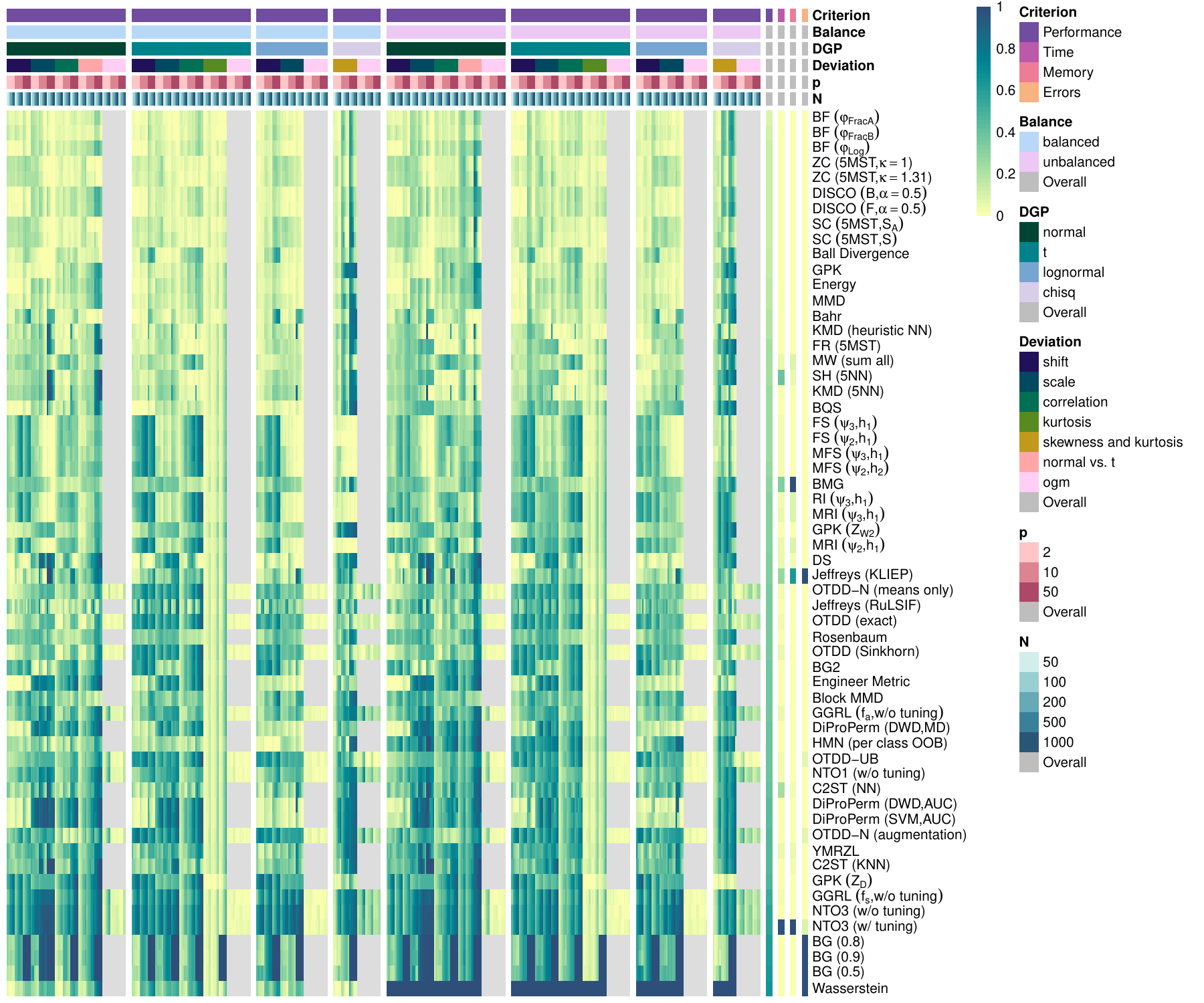}
		\caption{Mean difference to best method per scenario. Grey tiles in the heatmap indicate missing values that occur for methods that do not consider a target variable in scenarios where the deviation is in the OGM. Methods are ordered from top to bottom by increasing mean difference over all scenarios (``Overall Performance''), and in case of ties by decreasing runtime, and decreasing memory consumption, and alphabetically.}\label{fig:heat.diff.best.two}
	\end{figure}

	The overall method ranking based on mean differences of the PESR values per method to the ``ideal'' method is shown in Figure~\ref{fig:heat.diff.best.two}. 
	The mean differences per scenario and overall for each method are shown as a heatmap. 
	It can be seen that for a given method, the mean differences to the best method in terms of performance can vary between scenarios, especially with respect to the deviation, but also with respect to $N$, $p$, and the DGP. 
	Typically, the performances of the methods get better with increasing $N$ or decreasing $p$, and for balanced compared to unbalanced sample sizes. 
	Averaged over all scenarios, the BF gives the best performance with almost no differences between its variants of different choices for the dissimilarity function $\varphi$. 
	The BF is then followed by the ZC~(5MST) and DISCO. 
	Again, the variants for these methods perform very similarly to each other. 
	The Wasserstein distance and the BG test (all variants) are ranked worst, mainly due to the penalization of the scenarios for which they do not give any results.
	Purely in terms of performance when applicable, the C2ST~(KNN), GPK~($Z_D$), GGRL~($f_s$, w/o tuning), and NTO3 are ranked particularly poor in comparison.
	Often, the methods that are ranked best according to their performance are also performing well with respect to computational resource consumption. 
	Exceptions to this are the SH~(5NN), BMG, Jeffreys~(KLIEP), and C2ST~(NN) that require comparatively more computing time. 
	The BMG requires by far the most memory, followed by Jeffreys~(KLIEP).
	Jeffreys~(KLIEP) performs in the middle range overall despite high proportions of errors, i.e.\ if it works, it gives quite good performance.
	The best method that takes the target into account is the OTDD-N~(means only), but it is in the lower half performance-wise.

	In Appendix~\ref{app:diff.best.dev}, the heatmaps of the mean differences to the ``ideal'' method are given per deviation as there are clear differences between some deviations visible from the overall heatmap  (Figure~\ref{fig:heat.diff.best.two}). 
	For shift alternatives (Figure~\ref{fig:heat.diff.best.two.shift} in Appendix~\ref{app:diff.best.dev}), the interpoint distance based methods dominate with DISCO~(B or F) as the best method, followed by BF~($\varphi_{\text{Log}}$), Energy, DS, and the other BF variants. 
	Interestingly, BG2, as the remaining interpoint distance-based method, is among the worst methods. 
	The rank-based Energy distance DS shows a complementary pattern to other methods in that group and might therefore be a valuable addition in method combinations. 
	The best method that takes a target variable into account is the GGRL~($f_a$, w/o tuning), which is again in the lower half of methods. 
	
	For scale alternatives (Figure~\ref{fig:heat.diff.best.two.scale} in Appendix~\ref{app:diff.best.dev}), the Ball divergence, GPK, and the BF variants perform best. 
	Contrary to the shift alternatives, the FS~($\psi_3$ or $\psi_2$,$h_1$) is among the best as well and especially good for larger numbers of variables $p$. 
	The other FS and RI variants also perform considerably better than for the shift alternatives in the comparison. 
	The interpoint distance-based methods, on the other hand, perform worse in the comparison and are mostly ranked in the midfield, except for DISCO, which is again among the best methods. 
	The best method that can take a target variable into account is OTDD-N (means only), but it is already ranked in the lowest third of the methods. 
	Interestingly, some methods are better at detecting scales larger than one than scales smaller than one or vice versa. 
	See for example the PESR curves for the GPK variants in Figure~\ref{fig:pesr.curve.gpk.normal.balanced.scale} in Appendix~\ref{app:diff.best.dev}, where the individual components Z$_{\text{D}}$, Z$_{\text{W1}}$, and Z$_{\text{W2}}$ can only detect one direction and only their combination GPK is then able to detect both directions.
	If one is interested specifically in one of these cases, one could additionally leverage that knowledge. 
	Here, it is assumed that no such knowledge is available, and therefore, methods are preferred that perform well overall, and it is not discriminated between scales larger or smaller than one.
	
	For correlation alternatives (Figure~\ref{fig:heat.diff.best.two.cor} in Appendix~\ref{app:diff.best.dev}), the graph-based methods based on dense graphs dominate with FR~(5MST), SH~(5NN), SC~(5MST), ZC~(5MST) at the top, followed by KMD~(heuristic NN) which also uses a graph but its density depends on the sample size $N$. 
	The interpoint distance-based methods (except for BG2, which is again worse) and the remaining graph-based methods are in the midfield. 
	The best method that can take a target variable into account is GGRL~($f_a$, w/o tuning), which is ranked in the midfield. 
	
	For the comparison of normal and $t$-distributed datasets (Figure~\ref{fig:heat.diff.best.two.normalt} in Appendix~\ref{app:diff.best.dev}), the Bahr test is best, followed by the BF~($\varphi_{\text{FracB}}$) and the Ball divergence. 
	In contrast to other deviations, there is a clear distinction here between the BF variants with the $\varphi_{\text{Log}}$ as the dissimilarity function performing worst. 
	The graph-based tests ZC~(5MST) and SC~(5MST) also perform comparably well.
	The OTDD~(Sinkhorn) and OTDD-N~(means only) are ranked in the lower midfield as the best methods that take a target variable into account. 
		
	For kurtosis alternatives on $t$ distributed data (Figure~\ref{fig:heat.diff.best.two.kurtosis} in Appendix~\ref{app:diff.best.dev}), the performance of all methods is rather bad, especially for low $N$ or $p$, and the differences between the top methods are small.  
	The BF~($\varphi_{\text{FracB}}$) is again the best together with the ZC(5MST,$\kappa = 1$) and the Ball divergence. 
	The other BF, ZC, and SC variants are also among the best methods. 
	It is noteworthy that the GPK~($Z_D$), which was often among the worst methods for other alternatives, is the 5\textsuperscript{th} best method here. 
	The best method that takes a target variable into account is NTO1~(w/o tuning), which is ranked in the lower mid-range. 

	For the skewness and kurtosis alternatives on $\chi^2$-distributed data (Figure~\ref{fig:heat.diff.best.two.skew.kurtosis} in Appendix~\ref{app:diff.best.dev}), the FS variants and GPK~($Z_D$) are best. 
	These were typically among the worst-performing methods for many other alternatives. 
	However, the ZC and SC using the 5MST and the BF variants that were often among the best for other alternatives, are also among the best methods here, together with the Engineer Metric and the RI variants. 
	There are large differences in the methods' performances, especially for larger numbers of variables $p$. 
	The Jeffreys divergence, the BG test, and the Wasserstein distance, which were ranked among the worst in all other settings due to large numbers of errors, or being not applicable for high $p$ or unequal sample sizes, respectively, are here in the mid-range.
	This happens because many of the other methods perform very poorly for this alternative, especially the classifier-based methods, the methods that can handle a target variable, and, curiously, the GPK~($Z_{W2}$).
	However, the exact Sinkhorn, and means only variant of the OTDD, perform quite well in the comparison. 
	Overall, the ranking of methods here differs greatly from the rankings in the previous scenarios.	
	
	For OGM alternatives (Figure~\ref{fig:heat.diff.best.two.ogm} in Appendix~\ref{app:diff.best.dev}), only the methods that can handle datasets that include a target variable are compared. 
	It should be noted that none of the methods perform well for this alternative. 
	The PESR values are generally low, especially for the alternatives where only the sign or the size of the coefficients is altered. 
	Overall, the Sinkhorn approximation of the OTDD performs best, with the exact version performing only slightly worse.
	All selected OTDD variants perform better than the NKT and GGRL variants. 

\begin{figure}[!tb]
		\centering
		\includegraphics[width=\linewidth]{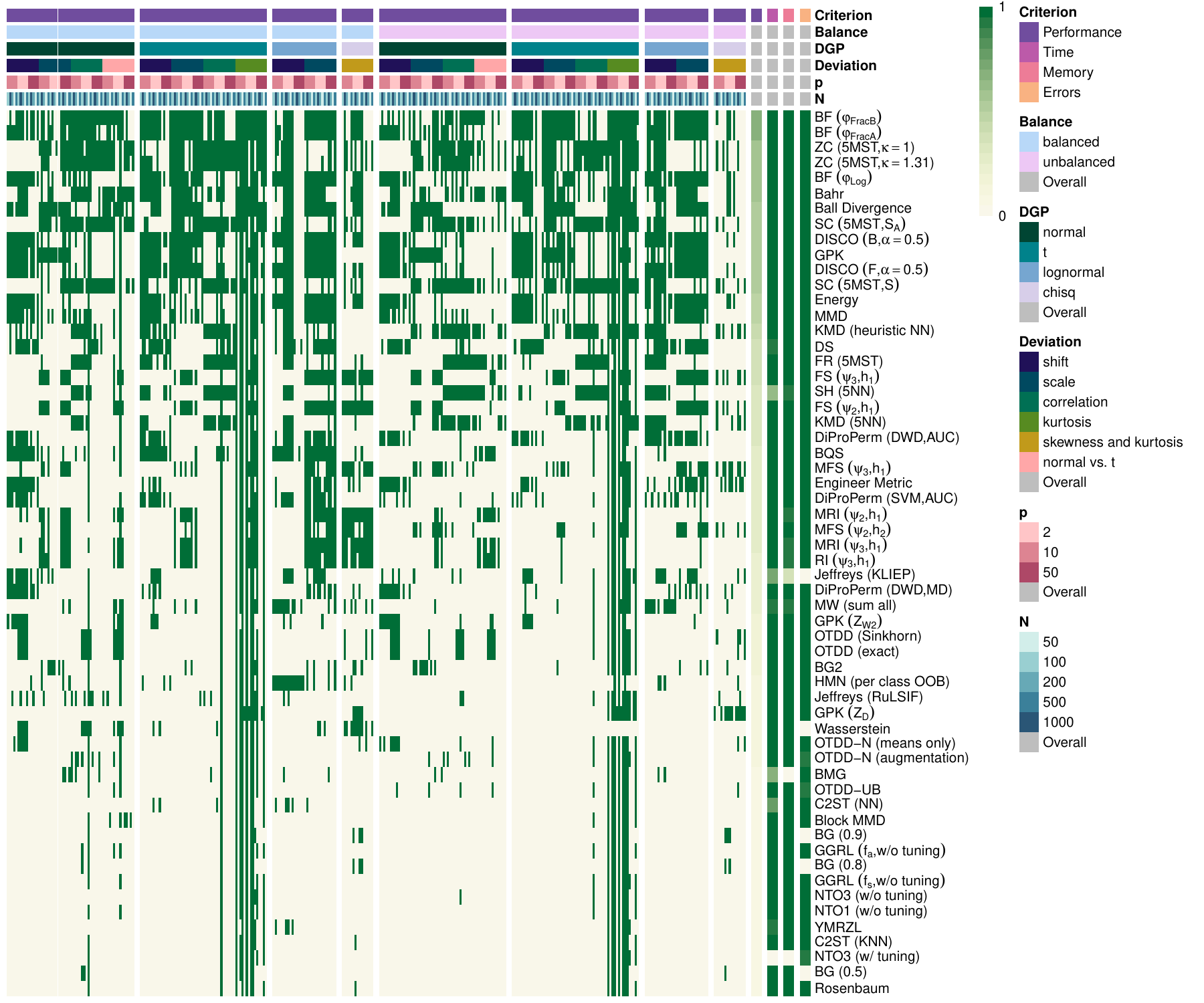}
		\caption{Acceptable methods per scenario. Light beige corresponds to non-acceptable, green corresponds to acceptable. Methods are ordered from top to bottom by increasing proportion of scenarios for which they are acceptable and alphabetically in case of ties.}\label{fig:heat.accept.two}
	\end{figure}

	Figure~\ref{fig:heat.accept.two} shows for which scenarios each method is considered acceptable. 
	The ordering for the top performing methods is similar to that based on the mean differences to the ``ideal'' method (Figure~\ref{fig:heat.diff.best.two}). 
	See Figure~\ref{fig:comp.rank.two} in Appendix~\ref{app:comp.rank} for a visualization of the differences between the rankings according to the two criteria. 
	However, for the mid-range and for the worst-performing methods, there are larger changes in the ordering. 
	These differences are due to the use of the fixed cutpoint for the calculation of the acceptable methods and due to the comparison of the acceptable method that is relative to the best of the methods in the comparison. 
	The most extreme changes between the rankings are observed for the Bahr test, which is ranked sixth according to the number of covered scenarios but only 14$^\text{th}$ according to the mean difference to the ideal method, the DS that is 16$^\text{th}$ instead of 30$^\text{th}$, as well as MW that is ranked 33$^\text{rd}$ instead of 17$^\text{th}$ and BMG that is ranked 44$^\text{th}$ instead of 25$^\text{th}$. 
	Moreover, the DiProPerm variants are ranked much better according to the number of acceptable scenarios than according to the mean difference to the ideal method. 
	These differences can appear when a method is either always performing okay, so ranked in the middle according to the mean difference to the ideal method, but often slightly misses the cut-off for being acceptable, so ranked badly according to the number of acceptable scenarios, or if a method performs very badly in some and okay in other alternatives and therefore has higher mean distance to the ideal method but is still included as acceptable in many scenarios. 
	
	\subsubsection{Decision Rules for Finding the Best Method}
	In the following, the rules for finding the method that covers most scenarios are given. 
	The rules are presented as decision trees where the split rules depend only on the dataset dimensions, i.e.\ $N$, $p$, and on the sample size balance. 
	Without further prior knowledge, the resulting decision tree is complex and consists of many rules and a diverse set of methods in the leaf nodes (see Figure~\ref{fig:tree.num.two} in Appendix~\ref{app:dec.trees}). 
	When incorporating prior knowledge by restricting the scenarios to certain DGPs or deviations, the decision rules get clearer. 
	Especially the restriction to certain deviations leads to clearer rules. 
	This is probably due to the previously observed differences between the deviations with respect to the method performance and ranking. 
	In the following, the decision rules per deviation are discussed. 
	For the rules for each DGP, or for each combination of DGPs and deviations, see Appendix~\ref{app:dec.trees}. 
	
	Figure~\ref{fig:tree.shift.two} shows the decision rules for choosing a method that can detect most shift alternatives.
	For low $p$, the DISCO~(B,$\alpha=0.5$) covers all scenarios and is therefore recommended. 
	For $p = 50$ and high sample sizes ($N = 500, 1000$), it is also recommended except for the case of balanced sample sizes and $N = 1000$, where the HMN is recommended instead. 
	For $p = 50$ and low sample sizes, mostly the DS is recommended, except for the case of $N = 200$ and balanced sample sizes, where the BQS is recommended instead. 
	
	\begin{figure}[!b]
		\centering
		\includegraphics[width=\linewidth]{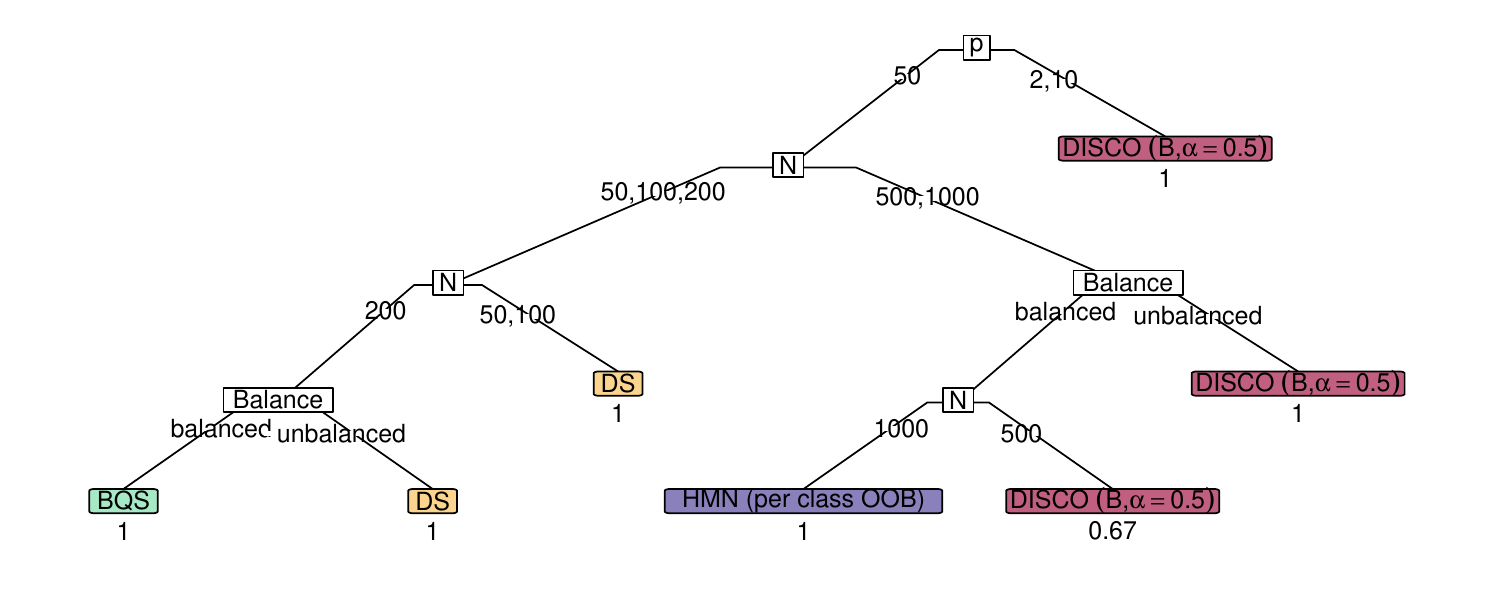}
		\caption{Decision rules for choosing a method that covers most scenarios for two numeric datasets and shift alternatives given only the information about the dataset dimensions. Numbers underneath the leaf nodes give the proportion of scenarios in the respective leaf node that are covered by the chosen method. }\label{fig:tree.shift.two}
	\end{figure}
	
	For scale alternatives, the Ball Divergence is recommended regardless of $N$, $p$, and the sample size balance. 
	Therefore, for this case no decision tree is shown.
	

	
\begin{figure}[!tb]
		\centering
		\includegraphics[width=\linewidth]{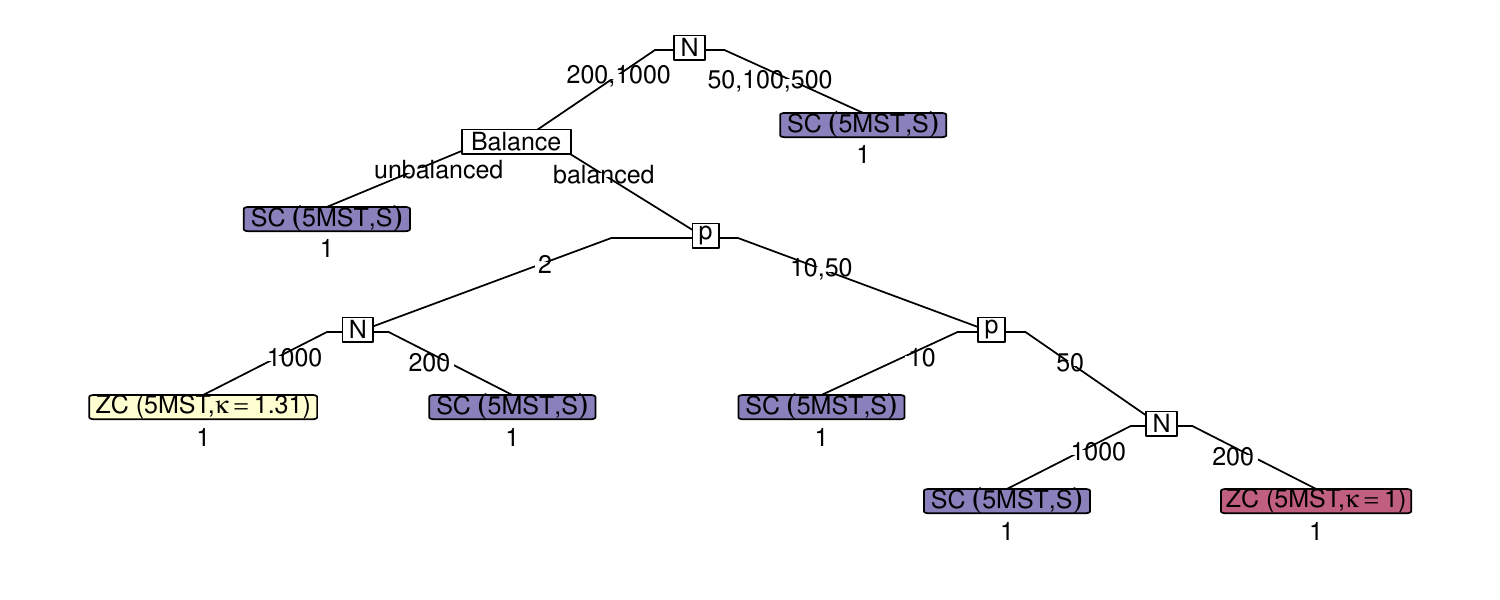}
		\caption{Decision rules for choosing a method that covers most scenarios for two numeric datasets and correlation alternatives given only the information about the dataset dimensions. Numbers underneath the leaf nodes give the proportion of scenarios in the respective leaf node that are covered by the chosen method. }\label{fig:tree.cor.two}
	\end{figure}

	For correlation alternatives (Figure~\ref{fig:tree.cor.two}), mostly the SC~(5MST,S) is recommended, except for $p=2$, $N=1000$, balanced samples sizes, when the ZC~(5MST,$\kappa=1.31$) is recommended instead, or $p = 50$, $N = 200$, and balanced sample sizes, when the ZC~(5MST,$\kappa=1$) is recommended.
	
\begin{figure}[!tb]
		\centering
		\includegraphics[width=\linewidth]{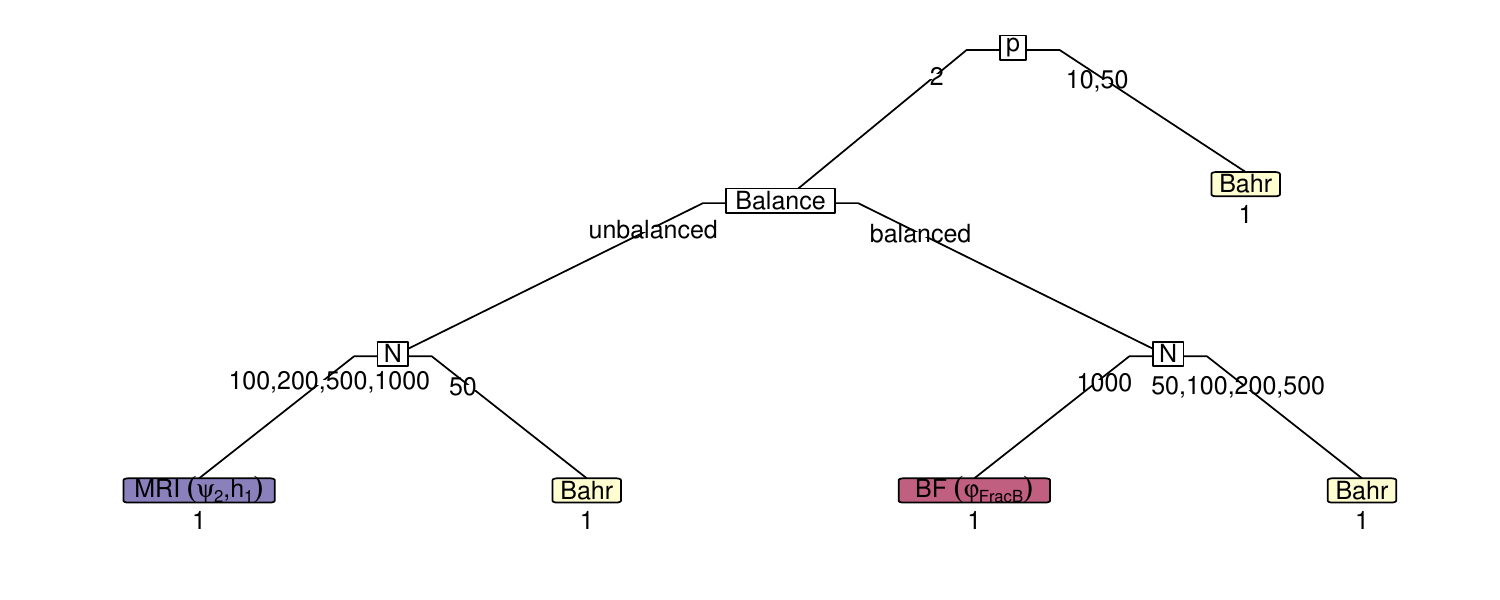}
		\caption{Decision rules for choosing a method that covers most scenarios for two numeric datasets and normal vs.\ $t$ alternatives given only the information about the dataset dimensions. Numbers underneath the leaf nodes give the proportion of scenarios in the respective leaf node that are covered by the chosen method. }\label{fig:tree.normalt.two}
	\end{figure}

	For the comparison of normal and $t$-distributed datasets (Figure~\ref{fig:tree.normalt.two}), mostly the Bahr test is recommended, except for $p = 2$, unbalanced sample sizes and $N > 50$, where the MRI~($\psi_2$, $h_1$) is recommended instead, and for $p = 2$, $N = 1000$ and balanced sample sizes, where the BF~($\varphi_{\text{FracB}}$) is recommended. 
	
	For the kurtosis alternative for $t$-distributed data, the BF~($\varphi_{\text{FracB}}$) is recommended for all $N$, $p$, and sample size balance settings.
	

	
\begin{figure}[!t]
		\centering
		\includegraphics[width=\linewidth]{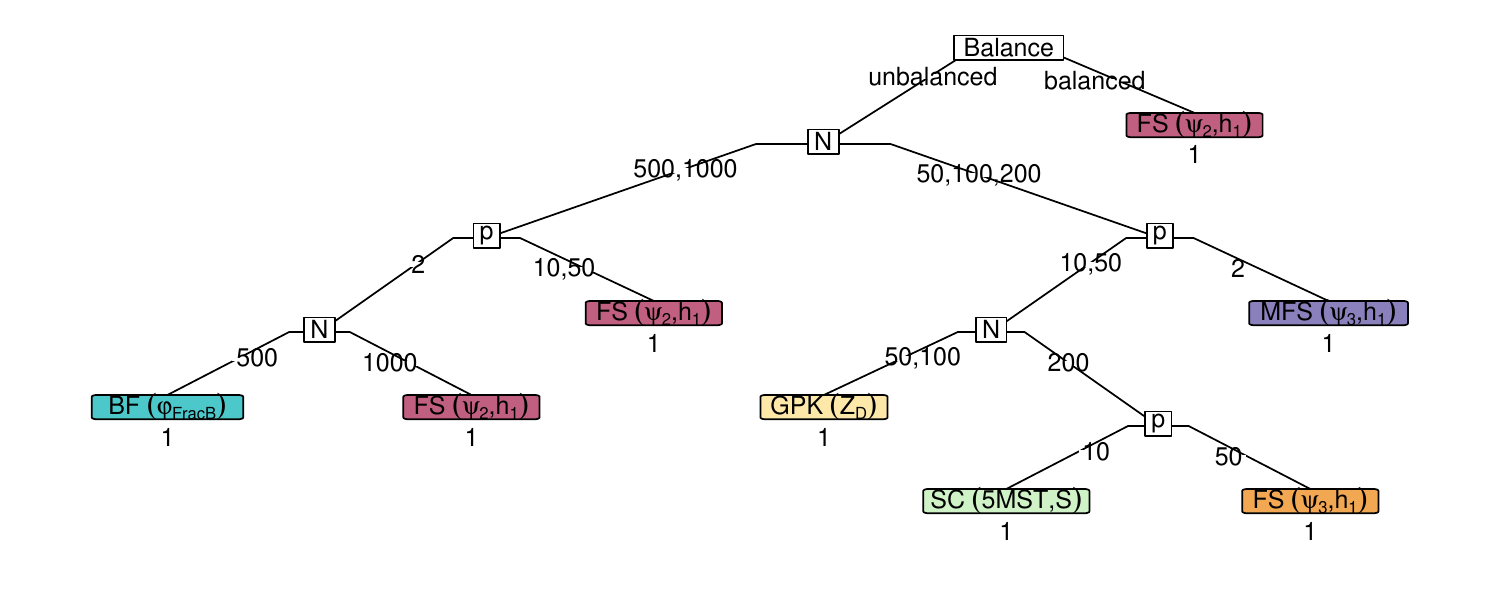}
		\caption{Decision rules for choosing a method that covers most scenarios for two numeric datasets and skewness and kurtosis alternatives given only the information about the dataset dimensions. Numbers underneath the leaf nodes give the proportion of scenarios in the respective leaf node that are covered by the chosen method. }\label{fig:tree.skew.kurtosis.two}
	\end{figure}

	For the skewness and kurtosis alternatives for $\chi^2$ distributed data, the set of recommended methods is more diverse (Figure~\ref{fig:tree.skew.kurtosis.two}). 
	The FS~($\psi_2$,$h_1$) can be recommended for all $N$ and $p$, when the sample sizes are balanced, and for unbalanced sample sizes for $N \ge 500$, except for $p = 2$ and $N = 500$, where the BF~($\varphi_{\text{FracB}}$) is recommended instead. 
	For unbalanced sample sizes and smaller $N$, and $p = 2$, the MFS~($\psi_3$,$h_1$) is recommended.
	For larger $p$, the GPK~($Z_{\text{D}}$), SC~(5MST, S), or FS~($\psi_3$,$h_1$) are recommended.
	
\begin{figure}[!t]
		\centering
		\includegraphics[width=\linewidth]{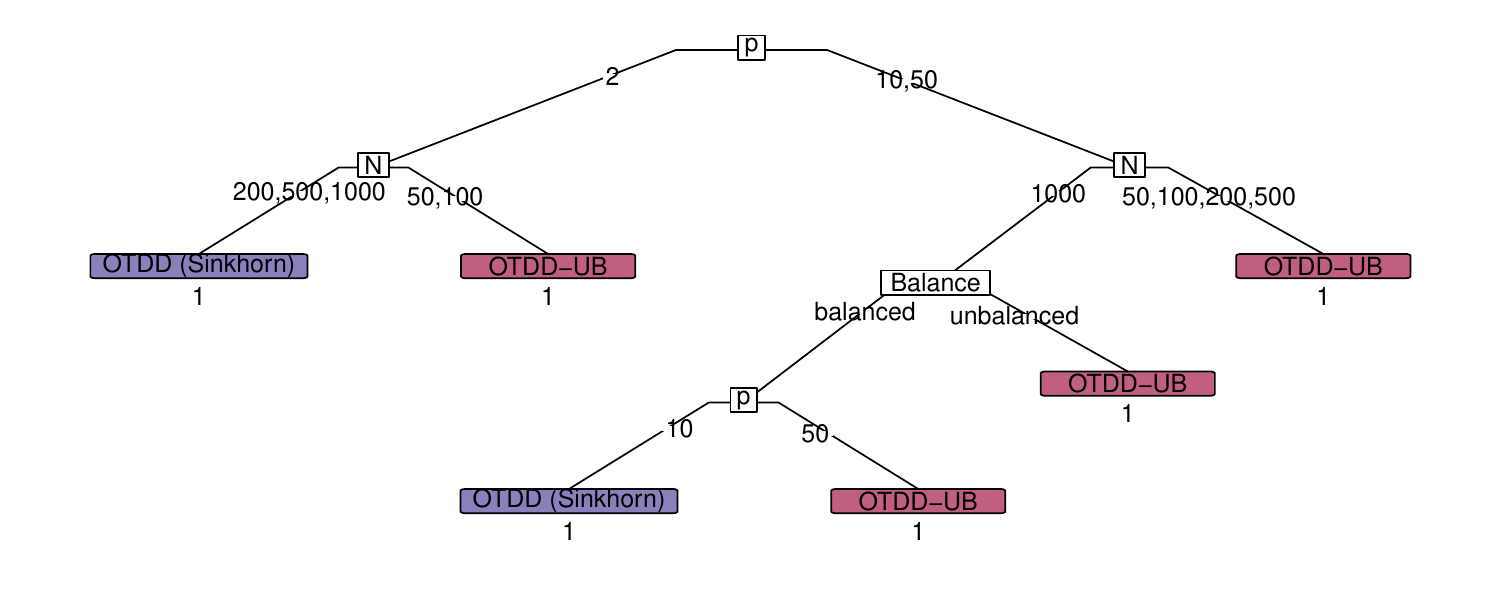}
		\caption{Decision rules for choosing a method that covers most scenarios for two numeric datasets and OGM alternatives given only the information about the dataset dimensions. Numbers underneath the leaf nodes give the proportion of scenarios in the respective leaf node that are covered by the chosen method. }\label{fig:tree.ogm.two}
	\end{figure}

	For detecting OGM alternatives (Figure~\ref{fig:tree.ogm.two}), mostly the OTDD-UB is recommended, except for $p = 2$ and $N>100$, or $p = 10$, $N = 1000$, and balanced sample sizes, where the OTDD~(Sinkhorn) is recommended instead. 
	
	\subsubsection{Best Method Combination}	
	
	The results for the method rankings and the decision rules show that, as expected, there is no single method that can detect all alternatives for all DGPs and dataset dimensions.
	Therefore, the aim is to find combinations of a few methods that can together detect most of the considered differences between two numeric datasets. 
	This means, the combinations of one, two, three,~\dots{} methods that have together the highest proportion of scenarios covered as acceptable are sought after. 
	Such a combination might be applied in practice in a multiple testing framework to simultaneously check for various differences between the distributions.
	Alternatively, if the focus is on quantifying similarity or distance between the distributions rather than on testing, the results according to each method in the combination could be reported to ensure detecting various differences. 
	
	\begin{figure}[!b]
		\centering
		\includegraphics[width=\linewidth]{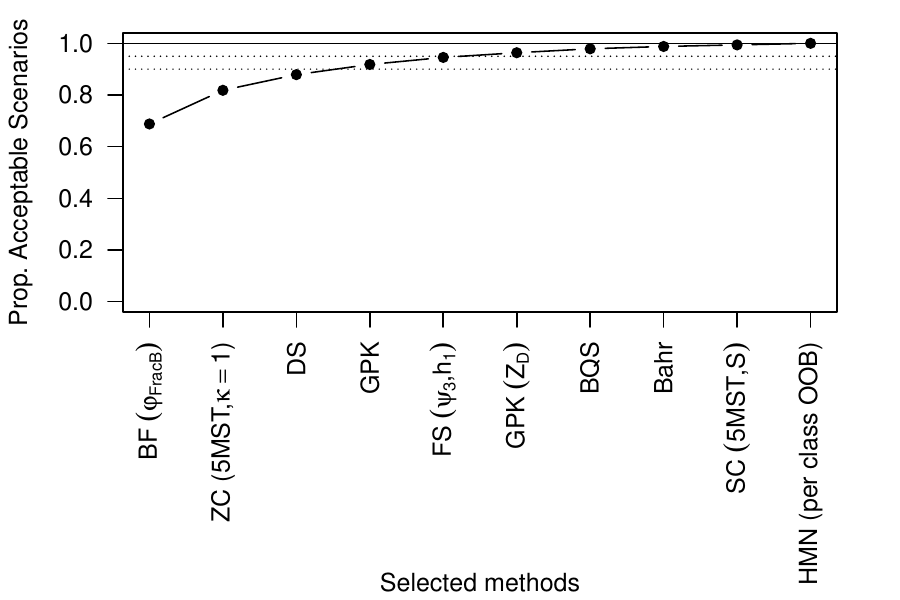}
		\caption{Proportions of scenarios that are covered by the best method combination found by greedy search. Each combination is given by the union of the methods on the $x$-axis starting from the left up to the respective point. Dotted lines indicate a 90\% and 95\% threshold, respectively.}\label{fig:meth.sel.greedy.two}
	\end{figure}
	
	Since calculating the exact solution of the best combination is infeasible run-time-wise, for five methods or more, the solutions of a greedy search are presented instead. 
	For small combinations, an exhaustive search was performed in addition. 
	The results are given in Figure~\ref{fig:meth.sel} in Appendix~\ref{app:comp.greedy.exhaustive}. 
	The combinations from the exhaustive search are only slightly better than those identified by the greedy search.
	The proportion of covered scenarios for the greedy search is shown in Figure~\ref{fig:meth.sel.greedy.two}. 
	On the $x$-axis, the methods are added sequentially to the combination. 
	For the best method, BF~($\varphi_{\text{FracB}}$), on its own, already almost 70\% of the scenarios are covered. 
	The methods that are added in the following only add small increases to the proportion of covered scenarios. 
	With the combination of BF~($\varphi_{\text{FracB}}$), ZC~(5MST,$\kappa=1$), DS and GPK, already 90\% of the scenarios are covered. 
	When adding also FS~($\psi_3$,$h_1$) and GPK~($Z_D$), 95\% are reached. 
	For reaching 100\%, BQS, Bahr, SC~(5MST,S), and the HMN~(per class OOB) are also added. 
	The scenarios that are not covered by the best combination of four methods are mostly scenarios of the skewness and kurtosis alternative. 
	As discussed previously, the methods that are good at detecting this alternative differ strongly from the best methods for the other alternatives. 
	Therefore, these are added later in the greedy search as they do not cover many scenarios overall, but specific scenarios that are not yet covered by the other methods. 
	These scenarios are then mostly covered by the best six method combination through the addition of the FS~($\psi_3$, $h_1$) and GPK~($Z_D$). 
	The scenarios that are then left uncovered are shift alternatives for log normal data with high $p$ and balanced sample sizes, which are covered later by the HMN, and normal vs. $t$ alternatives that need the Bahr or the SH variants. 
	For a list of all uncovered scenarios for the best four or six method combination, see Tables~\ref{tab:scens.not.covered.top4.two} and~\ref{tab:scens.not.covered.top6.two} in Appendix~\ref{app:scens.uncovered}.
	These methods, which are added last, are either rather bad in the overall comparisons (BQS, HMN) or cover mostly scenarios that are already covered by other methods selected earlier (Bahr, SC) and therefore can only contribute a few, specific scenarios that are additionally covered. 
	
	In the exhaustive search for small combinations, the proportions of covered scenarios are slightly increased by replacing BF~($\varphi_{\text{FracB}}$) by BF~($\varphi_{\text{FracA}}$) in the combinations of two or more methods.

	\FloatBarrier

	\subsection{Multi-Sample Case}
	In the following, the results of the evaluation of methods for the multi-sample case are presented. 
	First, the results regarding computational aspects are summarized. 
	Afterwards, the overall method ranking according to the methods' performances in detecting differences between datasets is established. 
	Based on this, decision rules for finding the best method and an overall best combination of methods are proposed. 
	
	\subsubsection{Evaluation of Computational Aspects}

\begin{figure}[!t]
		\centering
		\includegraphics[width=\linewidth]{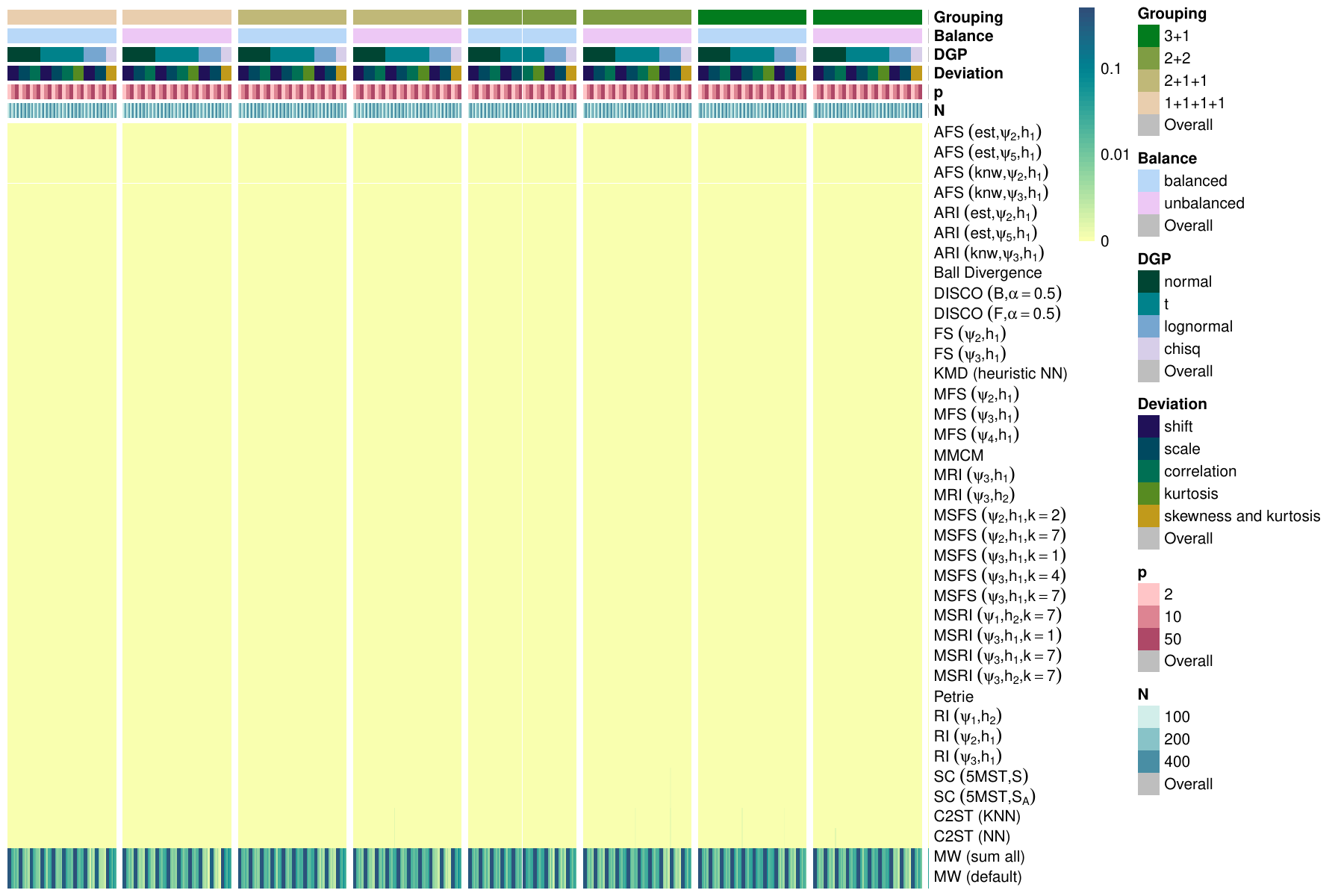}
		\caption{Proportion of iterations per simulation scenario in which computational errors occurred. The total iteration number per scenario is 500. Proportions are shown on a $\log_{10}$-scale with 0 set to $1/1000$. Grey tiles in the heatmap indicate missing values that occur for methods that do not consider a target variable in scenarios where the deviation is in the OGM. Methods are ordered from top to bottom by increasing row means of the proportions and alphabetically in case of ties.}\label{fig:heat.err.multi}
	\end{figure}

	In the following, the computational errors encountered during the simulation process, as well as the runtime and memory consumption in the benchmark setting, are compared between all methods in the multi-sample setting. 
	Figure~\ref{fig:heat.err.multi} shows a heatmap of the proportion of iterations in which each method encountered a computational error during the simulation process. 
	Mostly, these proportions are zero, except for the C2ST, which encountered sporadic errors in three iterations across all scenarios, and for the MW, for which the proportions are high and even close to one in many scenarios. 
	The errors for the MW result from non-convergence of the clustering algorithm that is used during the calculation of the test statistic. 
	Note that for some iterations, there were problems with the compute cluster where only a small proportion of the requested RAM was available leading to failure of all methods. 
	These iterations were removed from all analyses. 
	The only scenario in which this happened in a large number of iterations was for unbalanced sample sizes, $t$-distributed data, the ``2+1+1'' grouping, $N = 200$, $p = 50$, and a shift of $0.5$. 
	In that case, 294 iterations were affected. 
	In two other cases, only two or six iterations were affected, respectively.
	
\begin{figure}[!t]
		\centering
		\includegraphics[width=\linewidth]{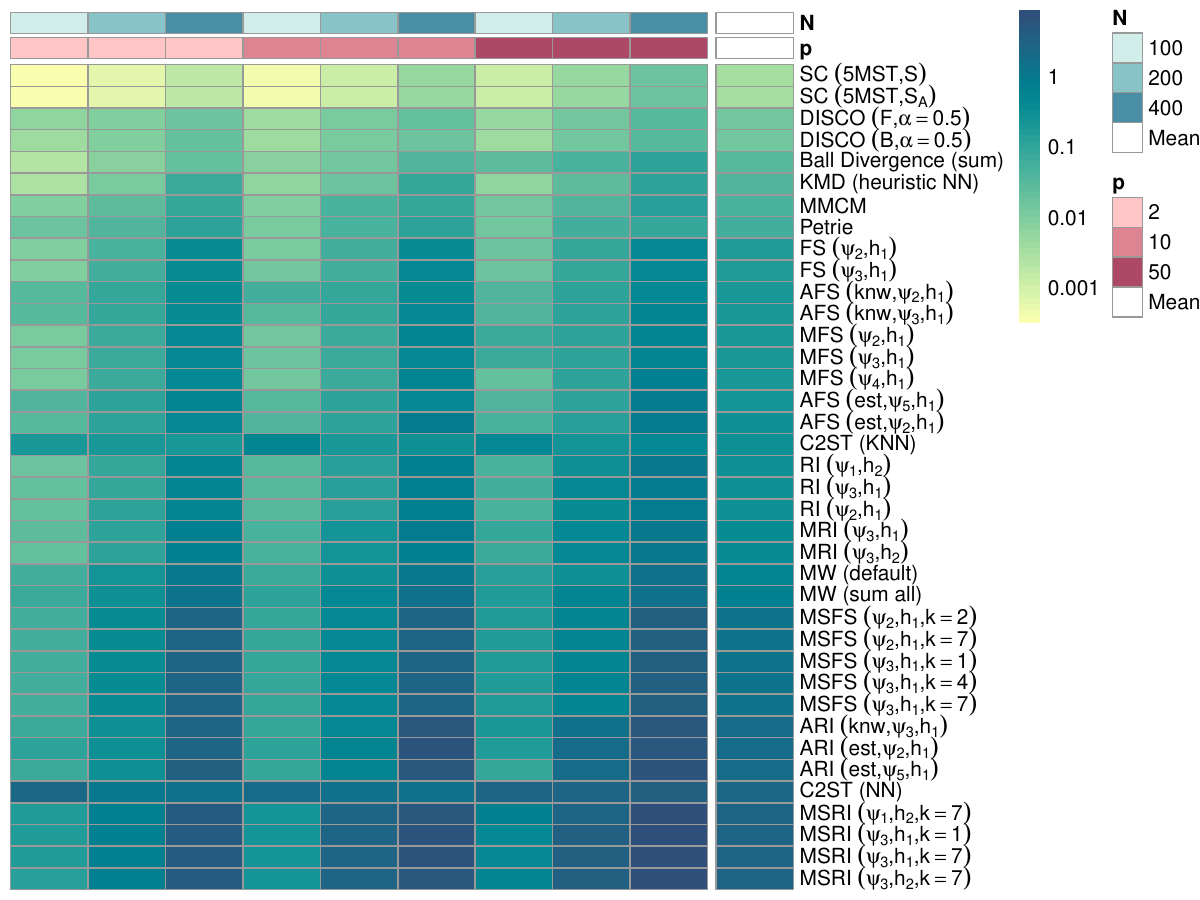}
		\caption{Median runtime in seconds for each multi-sample method in the benchmark scenarios. The scenarios all consider the null situation for normal data with equal sample sizes, and only $N$ and $p$ are varied. Each method is run at least ten times. If necessary, the number of runs is increased to guarantee that the method is run for at least 1\,s. The runtime is shown on a $\log_{10}$-scale. Methods are ordered from top to bottom by increasing row means and alphabetically in case of ties.}\label{fig:heat.time.multi}
	\end{figure}

	Figure~\ref{fig:heat.time.multi} shows a heatmap of the runtimes of the methods in the benchmark setting. 
	The SC has the lowest runtimes for all $N$ and $p$ combinations. 
	The variants of SC do not differ in their runtimes since both statistics are always calculated together in the implementation that was used. 
	The next methods in the ranking by mean runtime are the DISCO variants, KMD, Ball Divergence, MMCM, and Petrie's method. 
	The FS and RI test variants, as well as the C2ST and MW variants, take longer.
	The RI is taking more time than the FS test, and the multi-scale versions of both tests take longer than the other variants since they require the calculation of multiple statistics for different numbers of clusters. 
	Note that for a given $\psi$ and $h$ function, in the multi-scale case, all statistics for different $k$ are calculated simultaneously, so the runtimes do not differ depending on $k$. 
	The C2ST variants take notably longer than the other methods for low $N$. 
	The C2ST using the neural net (NN) classifier takes clearly longer than using the $K$-NN classifier. 

\begin{figure}[!t]
		\centering
		\includegraphics[width=\linewidth]{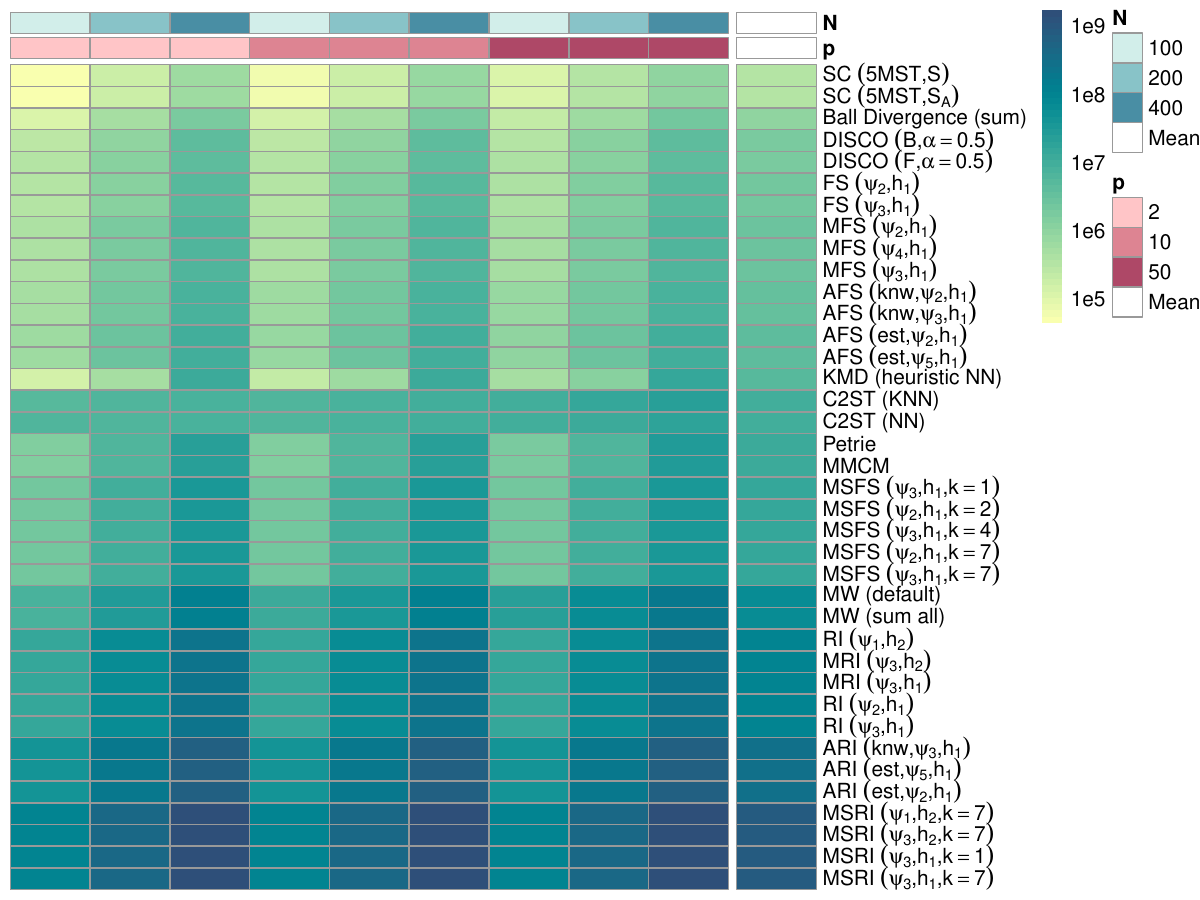}
		\caption{Memory allocation in bytes for each multi-sample method in the benchmark scenarios. The scenarios all consider the null situation for normal data with equal sample sizes, and only $N$ and $p$ are varied. Each method is run once. The memory consumption is shown on a $\log_{10}$-scale. Methods are ordered from top to bottom by increasing row means and alphabetically in case of ties.}\label{fig:heat.mem.multi}
	\end{figure}

	Figure~\ref{fig:heat.mem.multi} shows a heatmap of the memory allocations of the methods in the benchmark setting.
	The ranking is similar to that for the runtime. 
	The DISCO variants need more memory than the KMD and Ball divergence, while being faster. 
	Petrie's method and the MMCM are now in the mid-range (ranked worse than with regard to runtime), as well as the C2ST variants (ranked better than with regard to runtime).

	In summary, all methods except for MW appear computationally stable. 
	SC, DISCO, KMD, and the Ball divergence show low computational resource consumption. 
	The classifier-based tests and especially the multi-scale versions of the FS and RI tests, on the other hand, showed the highest runtimes and memory consumption. 
	
	\subsubsection{Overall Method Ranking}
\begin{figure}[!t]
		\centering
		\includegraphics[width=\linewidth]{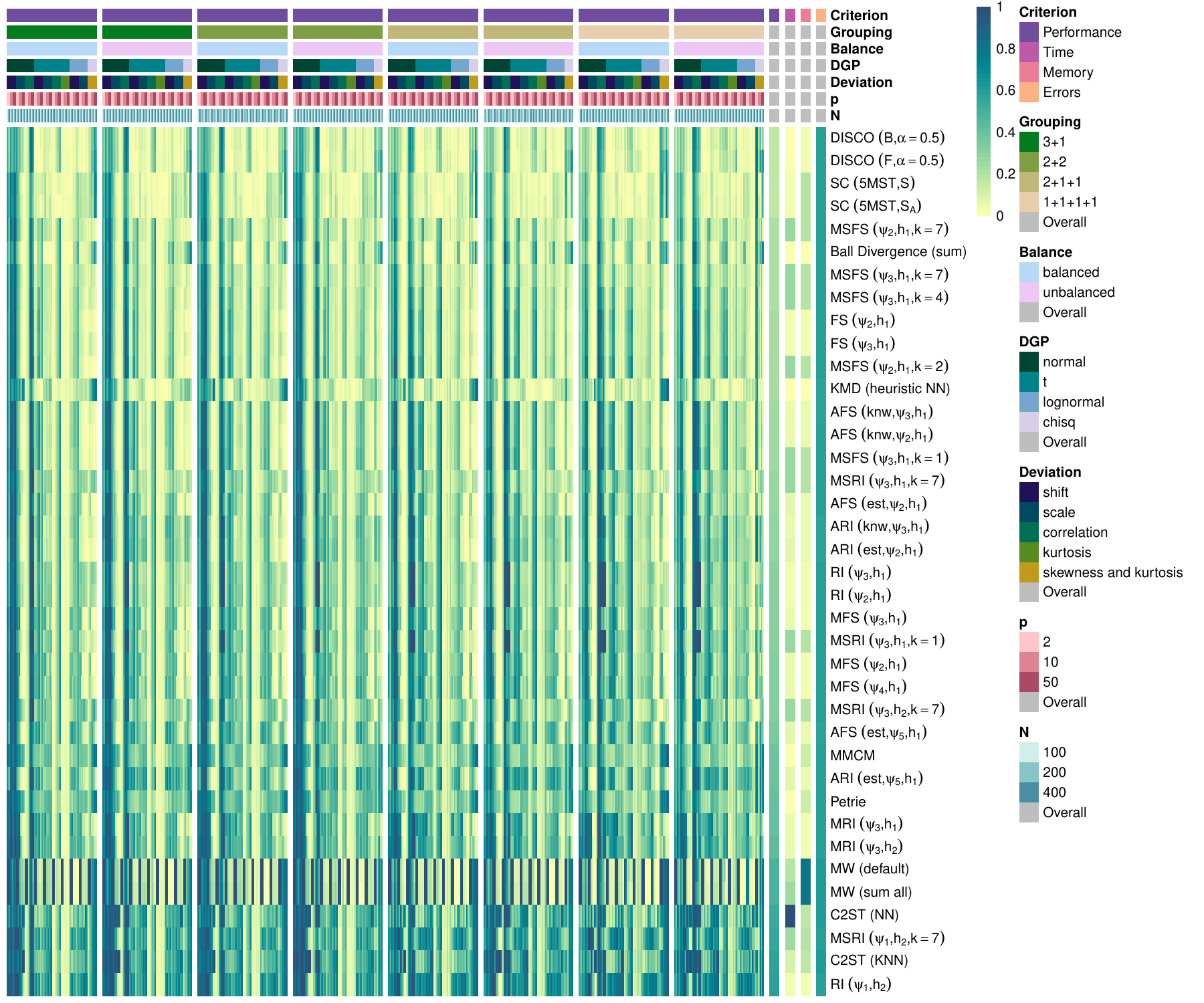}
		\caption{Mean difference to best method per scenario. Methods are ordered from top to bottom by increasing mean difference over all scenarios (``Overall Performance'') and alphabetically in case of ties.}\label{fig:heat.diff.best.multi}
	\end{figure}

	In the following, the methods are ranked based on the mean differences of their PESR values to that of the ideal method.
	Figure~\ref{fig:heat.diff.best.multi} shows these differences in a heatmap. 
	Again, differences between the performances within the same method are clearly visible depending on the deviation, and the dataset dimensions $N$, $p$, and sample size balance.
	Again, the performance of most methods is better for higher $N$, lower $p$, and balanced sample sizes. 
	There are only slight differences between the groupings, with a slight increase in performance for most methods from ``3+1'' to ``2+2'' to ``2+1+1'' to ``1+1+1+1''. 
	Overall, the DISCO and SC variants have the lowest overall differences to the ideal method, followed by the MSFS and FS variants, and the Ball Divergence. 
	The KMD, which is also ranked in the top half, shows a complementary pattern to the FS methods and could therefore be a valuable addition in combinations. 
	The RI variants are mostly ranked in the mid-range to the lower half.
	The MMCM and Petrie's method are also among the worst methods, along with MW and the C2ST variants. 
	The MSFS and MSRI variants, C2ST, and MW had also comparably high runtimes and memory consumption and might therefore be preferred less for practical application.
	For MW, the high error proportion also limits the applicability in practice.

	The performances again are dependent on the deviation; therefore, the deviation-wise rankings are considered as well in the following. 
	For shift alternatives (see Figure~\ref{fig:heat.diff.best.multi.shift} in Appendix~\ref{app:diff.best.dev}), the DISCO variants are best followed by KMD, SC, and Ball divergence. 
	MMCM, Petrie's method, and C2ST~(NN) perform comparably better than overall and are ranked in the mid-range.
	MW is either not working ($p = 2$) or quite good ($p > 2$), and worse for normal data than for other DGPs.

	For scale alternatives (see Figure~\ref{fig:heat.diff.best.multi.scale} in Appendix~\ref{app:diff.best.dev}), the Ball divergence is clearly the best method, followed by some FS and both DISCO variants, and then the SC variants.
	The RI variants and the remaining FS variants are ranked in the mid-range to the bottom half.
	The KMD is ranked in the lower mid-range (often okay but bad for some scenarios), and MMCM, Petrie, and C2ST are among the worst.
	MW is also again among the worst methods, but mainly because of its bad performance for $p = 2$.

	For correlation alternatives (see Figure~\ref{fig:heat.diff.best.multi.cor} in Appendix~\ref{app:diff.best.dev}), the SC variants perform best, followed by the KMD and the MSFS ($k = 7$) variants. 
	The DISCO variants, MMCM, Petrie's method, and the Ball Divergence are also at the top.
	The C2ST and MW are again worse; however, MW only for low $p$.
	For high $p$, MW is better than many methods from the mid-range.

	For kurtosis for the $t$-distributed data (see Figure~\ref{fig:heat.diff.best.multi.kurtosis} in Appendix~\ref{app:diff.best.dev}), there are almost no differences between the methods except for MW, which is much worse than the other methods for $p = 2$. 
	The SC variants and the Ball Divergence are best, followed by DISCO~(B,$\alpha=0.5$) and KMD, and then the FS and RI variants. 
	DISCO~(F,$\alpha=0.5$) is ranked in the middle, and MMCM, Petrie's method, and the C2ST variants are at the bottom. 
		
	For the skewness and kurtosis alternatives for $\chi^2$-distributed data (see Figure~\ref{fig:heat.diff.best.multi.skew.kurtosis} in Appendix~\ref{app:diff.best.dev}), the FS variants perform best, followed by the RI variants.
	DISCO, SC, Ball Divergence, KMD, Petrie's method, MMCM, C2ST, and MW make up the bottom half of the ranking.
	The good performance of the best RI variants is, in part, complementary to that of the best FS variants.

\begin{figure}[!t]
		\centering
		\includegraphics[width=\linewidth]{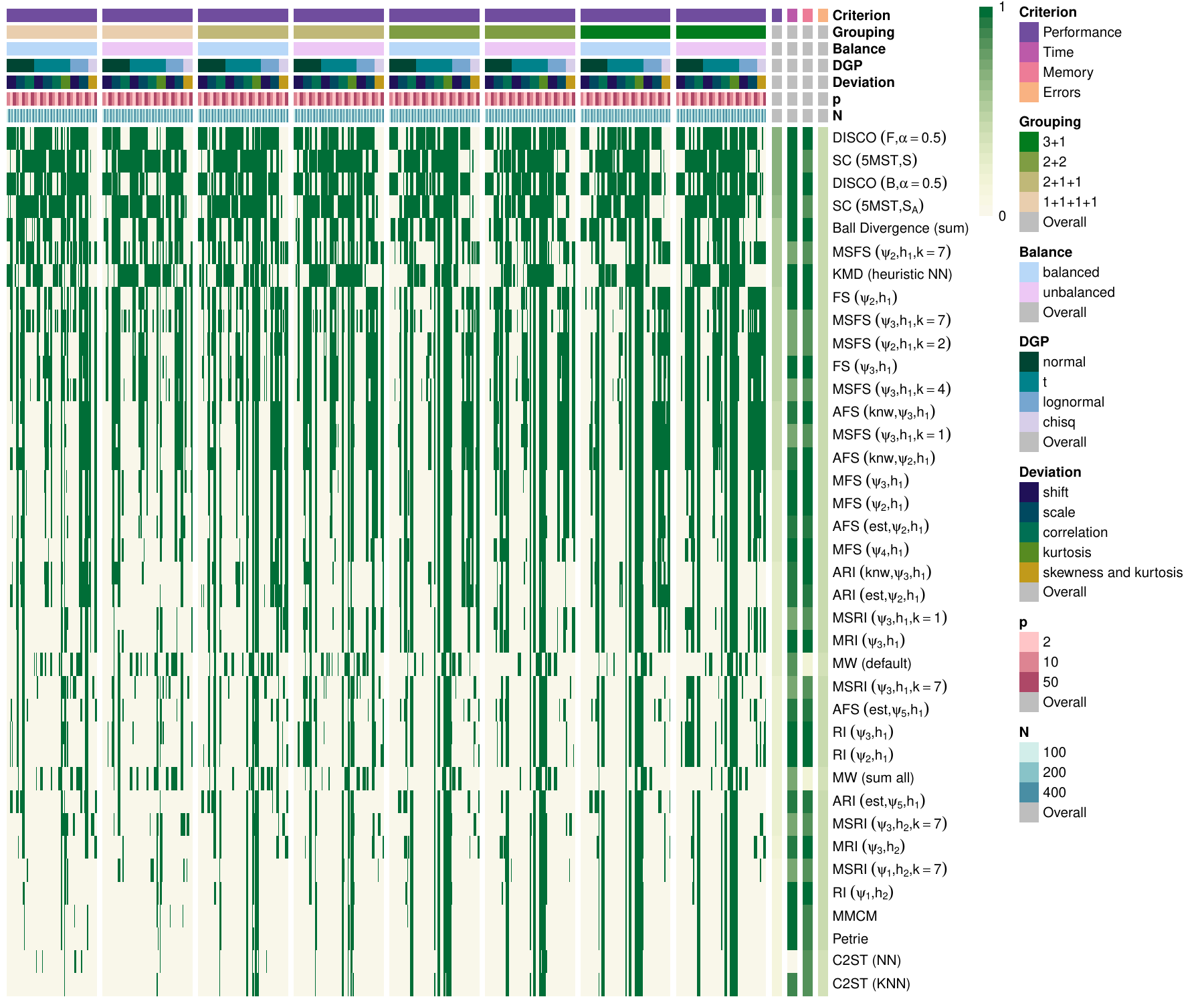}
		\caption{Acceptable methods per scenario. Light beige corresponds to non-acceptable, green corresponds to acceptable. Methods are ordered from top to bottom by increasing proportion of scenarios for which they are acceptable and alphabetically in case of ties.}\label{fig:heat.accept.multi}
	\end{figure}

	Figure~\ref{fig:heat.accept.multi} shows the scenarios for which each of the methods is considered acceptable.
	The DISCO~(F,$\alpha=0.5$) covers the highest number of scenarios, followed by SC~(5MST,S), DISCO~(B,$\alpha=0.5$), and the Ball Divergence. 
	The scenarios covered by DISCO and SC are, in part, complementary to each other.
	The KMD is ranked 7\textsuperscript{th}, surrounded by FS variants. 
	After that, the RI variants and MW follow. 
	The MMCM, Petrie's method, and C2ST variants cover the fewest scenarios.
	Again, the ranking according to the number of acceptable scenarios is similar to that based on the mean differences to the ideal method, with larger differences in the mid-range or worse performing methods (see Figure~\ref{fig:comp.rank.multi} in Appendix~\ref{app:comp.rank}). 
	Of the better-performing methods, the KMD is ranked better with regard to the number of acceptable scenarios, with a rank of seven instead of twelve, almost switching places with the MSFS~($\psi_3$, $h_1$,$k=4$) that is ranked twelfth instead of eighth.

	\subsubsection{Decision Rules for Finding the Best Method}
	
	In the following, decision rules are presented for finding the methods that cover the most scenarios based on the dimensions of the datasets ($N$, $p$, sample size balance).
	
\begin{figure}[!t]
		\centering
		\includegraphics[width=\linewidth]{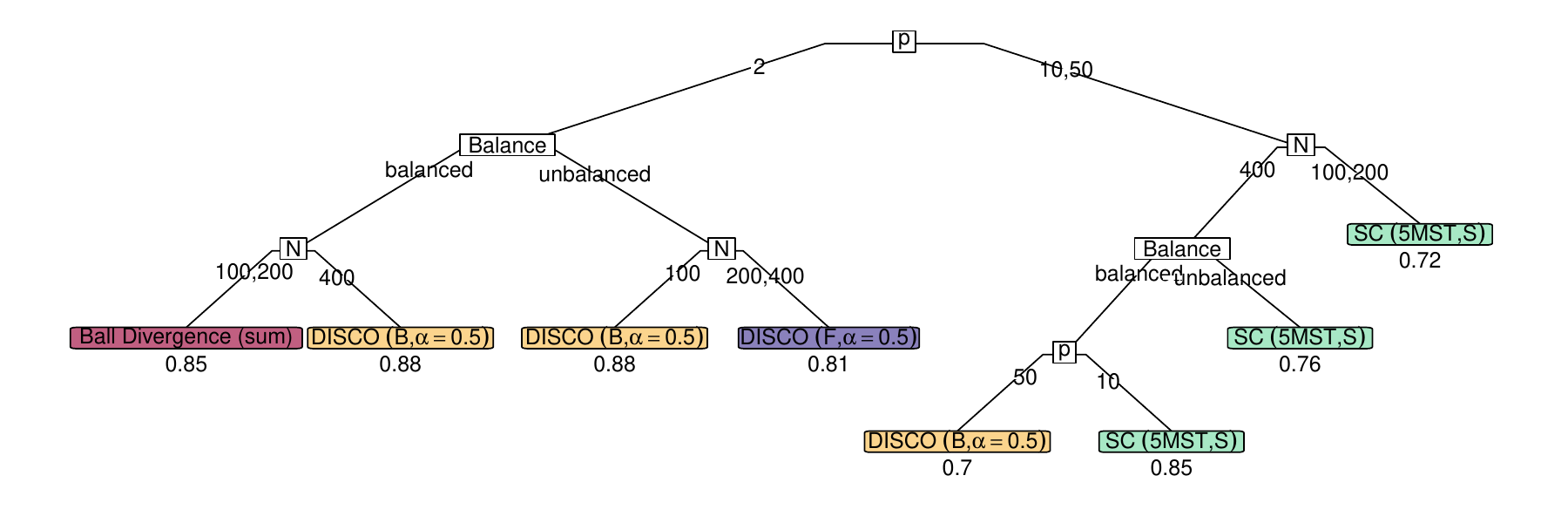}
		\caption{Decision rules for choosing a method that covers most scenarios for two numeric datasets given only the information about the dataset dimensions. Numbers underneath the leaf nodes give the proportion of scenarios in the respective leaf node that are covered by the chosen method.} \label{fig:tree.multi}
	\end{figure}

	Figure~\ref{fig:tree.multi} presents the overall decision rules given no other information than the dataset dimensions. 
	For $p = 10, 50$, the use of the SC~(5MST,S) is recommended, except for $p = 50$, $N = 400$, and balanced sample sizes, where DISCO~(B,$\alpha = 0.5$) is recommended instead. 
	For $p = 2$, the situation is less clear. 
	For balanced sample sizes and $N = 100, 200$, the Ball divergence is recommended, and for $N = 400$, DISCO~(B,$\alpha = 0.5$) is recommended. 
	For unbalanced sample sizes and $N = 100$, DISCO~(B,$\alpha = 0.5$) is recommended as well, while for larger $N$, DISCO~(F,$\alpha = 0.5$) is recommended instead.
	The selected methods cover 70\,\% to 88\,\% of scenarios within the respective leaf node.
	A better coverage can be reached when additional information about the deviation is used (or about the DGP, see Appendix~\ref{app:dec.trees}). 
	The recommendations by deviation are discussed in the following.

	For shift alternatives, DISCO~(F,$\alpha = 0.5$) is always selected as the best method regardless of $N$, $p$, and the balance.
	No tree is plotted for this case.
	
	
	
\begin{figure}[!t]
		\centering
		\includegraphics[width=\linewidth]{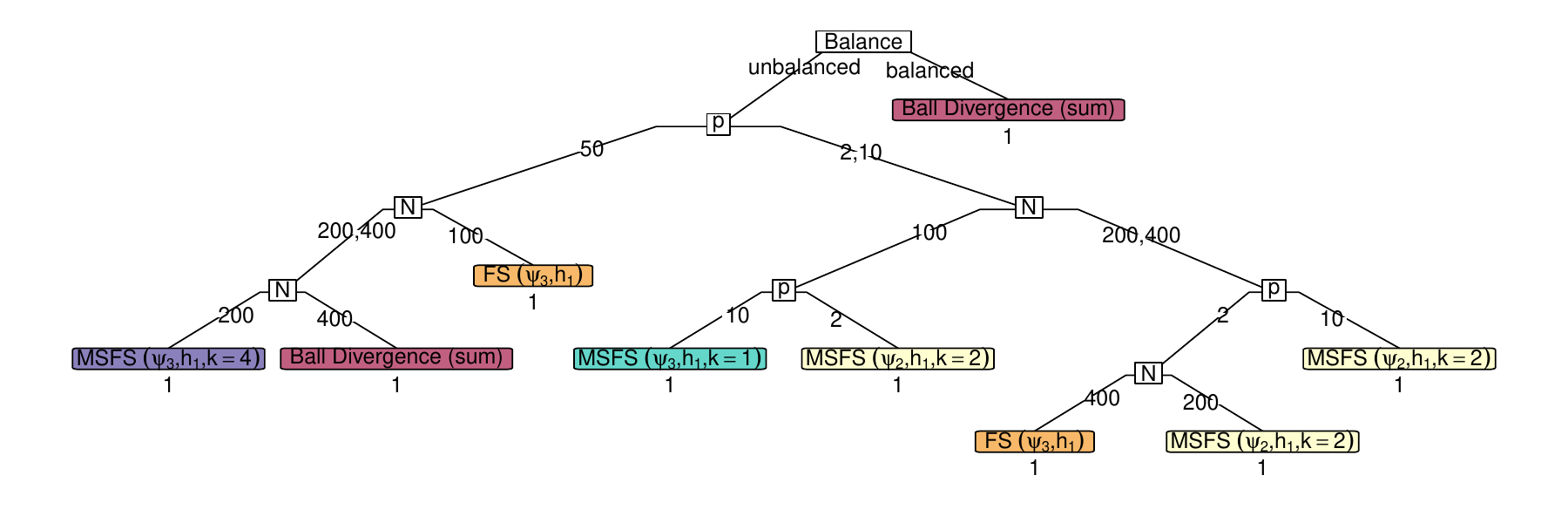}
		\caption{Decision rules for choosing a method that covers most scenarios for multiple numeric datasets and scale alternatives given only the information about the dataset dimensions. }\label{fig:tree.scale.multi}
	\end{figure}

	For scale alternatives (Figure~\ref{fig:tree.scale.multi}), for balanced sample sizes, the Ball divergence is recommended regardless of $N$ and $p$. 
	For unbalanced sample sizes, the rules are less clear, and Ball divergence is only selected for the highest $N$ and $p$ combination.
	Depending on $N$ and $p$ the use of different FS variants, often MSFS with low $k$, is recommended.
	Mostly MSFS~($\psi_2$,$h_1$,$k=2$) is suggested for $p = 2,10$.
	In each leaf node, now a coverage of 100\,\% of the scenarios is reached.
	
\begin{figure}[!b]
		\centering
		\includegraphics[width=\linewidth]{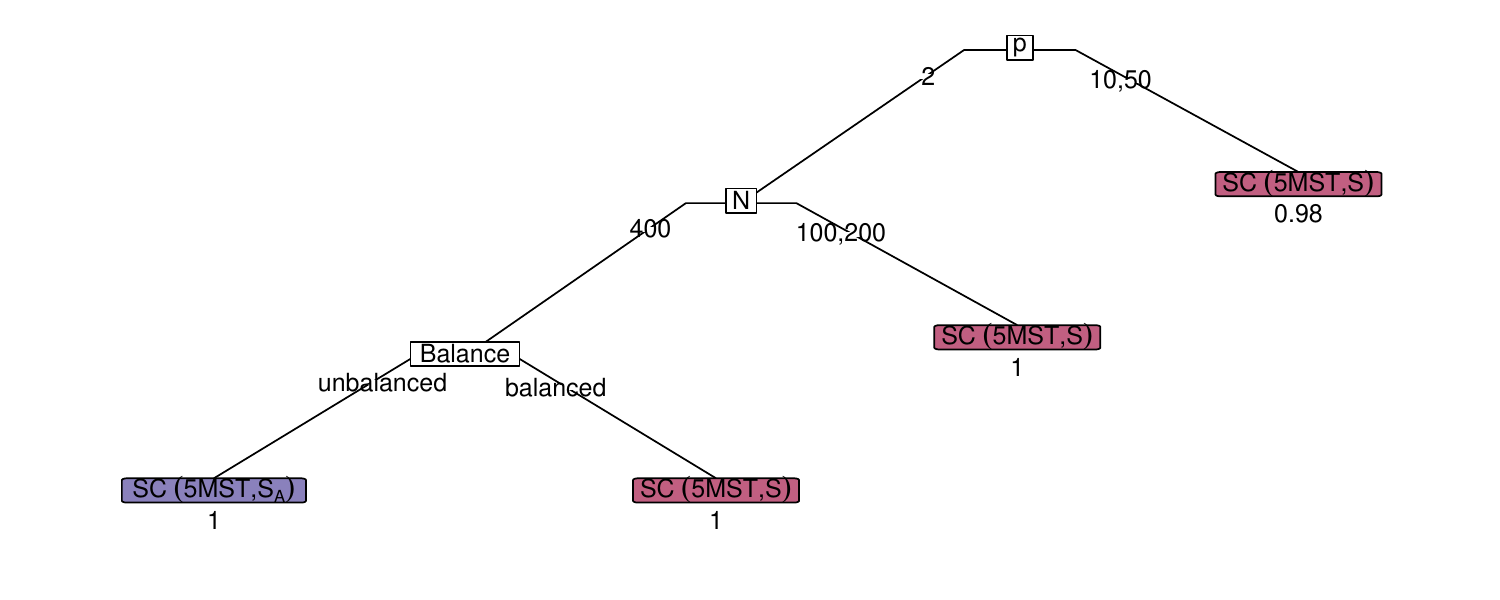}
		\caption{Decision rules for choosing a method that covers most scenarios for multiple numeric datasets and correlation alternatives given only the information about the dataset dimensions. Numbers underneath the leaf nodes give the proportion of scenarios in the respective leaf node that are covered by the chosen method. }\label{fig:tree.cor.multi}
	\end{figure}

	For correlation alternatives (Figure~\ref{fig:tree.cor.multi}), the SC is recommended. 
	Mostly, SC~(5MST,S) is recommended, only for $p = 2$, $N = 400$, and unbalanced sample sizes S$_\text{A}$ performs best.

\begin{figure}[!t]
		\centering
		\includegraphics[width=\linewidth]{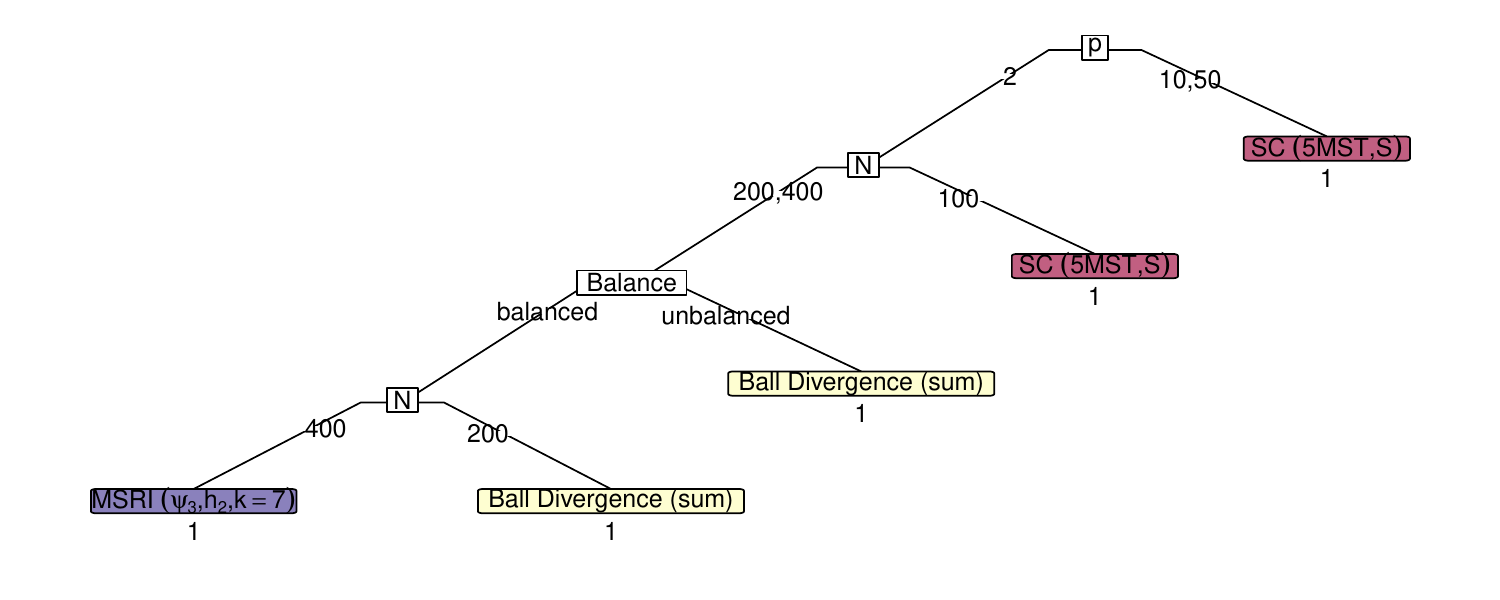}
		\caption{Decision rules for choosing a method that covers most scenarios for multiple numeric datasets and kurtosis alternatives given only the information about the dataset dimensions. }\label{fig:tree.kurtosis.multi}
	\end{figure}

	For the kurtosis alternative in $t$-distributed data (Figure~\ref{fig:tree.kurtosis.multi}), and $p = 10, 50$ or $p = 2, N = 100$, the SC~(5MST,S) is recommended as well. 
	Otherwise, the Ball Divergence, or for $N = 400$, and balanced sample sizes the MSRI~($\psi_3$,$h_2$,$k=7$) are better.

\begin{figure}[!t]
		\centering
		\includegraphics[width=\linewidth]{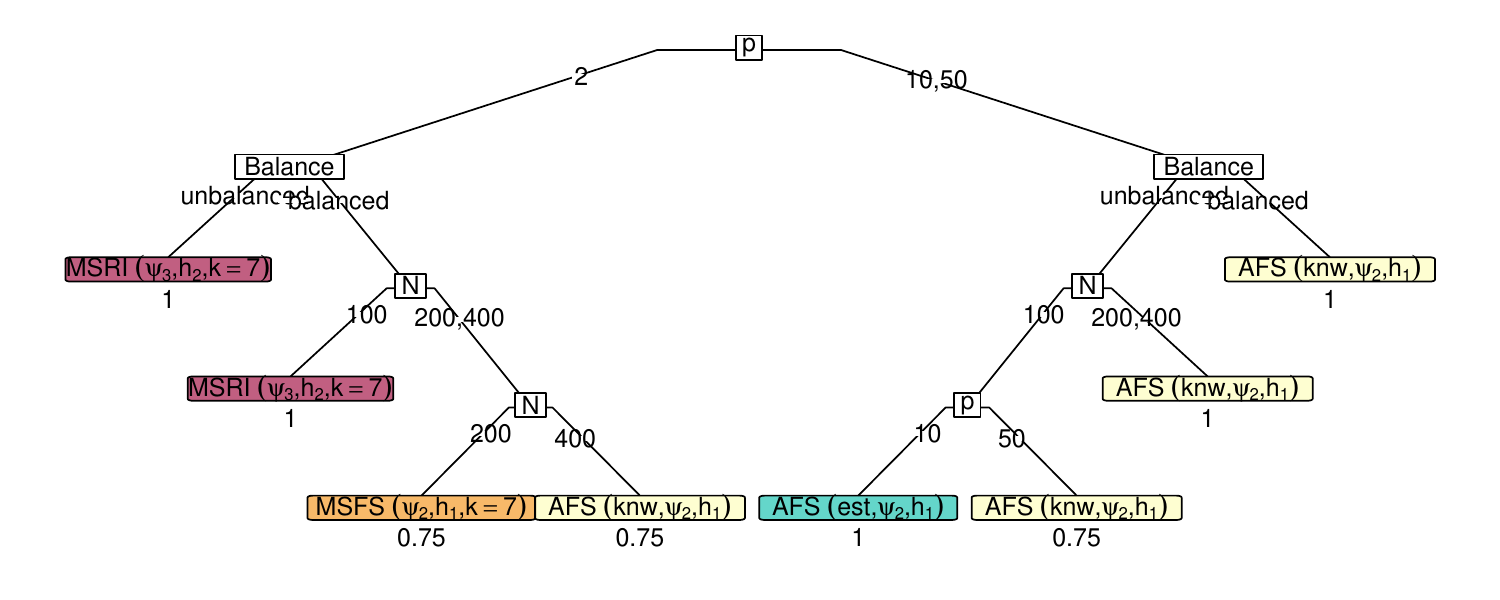}
		\caption{Decision rules for choosing a method that covers most scenarios for multiple numeric datasets and shift alternatives given only the information about the dataset dimensions. Numbers underneath the leaf nodes give the proportion of scenarios in the respective leaf node that are covered by the chosen method. }\label{fig:tree.skew.kurtosis.multi}
	\end{figure}

	For the skewness and kurtosis alternative on $\chi^2$-distributed data (Figure~\ref{fig:tree.skew.kurtosis.multi}), the recommendations are again less clear and no longer always cover 100\,\% in leaf nodes as for other alternatives.
	Always, some FS variant is chosen, but which one depends on $N$, $p$, and the sample size balance. 
	For $p = 10,50$, mostly AFS~(knw,$\psi_2$,$h_1$) is recommended, except for unbalanced sample sizes, $N=100$, and $p=10$. 
	For $p = 2$, the MSRI~($\psi_3$,$h_2$,$k=7$) is suggested in all unbalanced sample size settings, and balanced sample size settings with $N = 100$.
	For balanced sample sizes and $N = 200$, MSFS~($\psi_2$,$h_1$,$k=7$), is chosen, while for $N=400$, AFS~(knw,$\psi_2$,$h_1$) again recommended.
	
	\subsubsection{Best Method Combination}	
	
\begin{figure}[!tb]
		\centering
		\includegraphics[width=\linewidth]{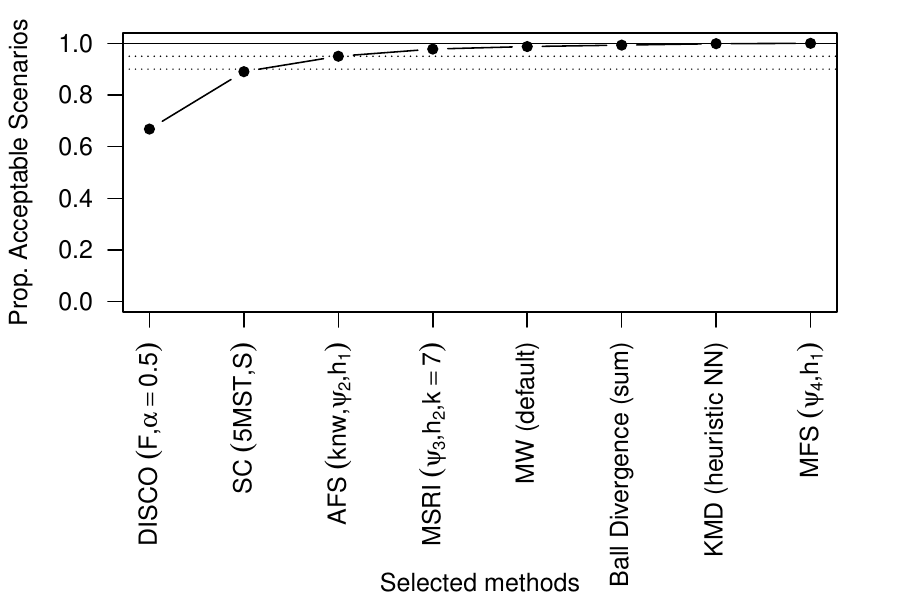}
		\caption{Proportions of scenarios that are covered by the best method combination found by greedy search. Each combination is given by the union of the methods on the $x$-axis starting from the left up to the respective point. Dotted lines indicate a 90\% and 95\% threshold, respectively.}\label{fig:meth.sel.greedy.multi}
	\end{figure}

	A greedy search for the method combinations that cover most scenarios is performed. 
	An exhaustive search for small combinations where it is still feasible, gives almost identical results  (see Appendix~\ref{app:comp.greedy.exhaustive}). 
	The resulting proportions of scenarios that are covered by each combination identified by the greedy search are shown in Figure~\ref{fig:meth.sel.greedy.multi}.
	The DISCO~(F,$\alpha=0.5$) alone covers about 67\,\% of the scenarios. 
	Adding SC~(5MST,S) is already sufficient to reach almost 90\,\%.
	Together with the AFS~(knw,$\psi_2$,$h_1$), already 95\,\% coverage is reached. 
	For covering 100\,\% of the scenarios, the following five methods are additionally needed according to the greedy search: MSRI~($\psi_3$,$h_2$,$k=7$), MW (default), Ball Divergence, KMD, MFS~($\psi_4$,$h_1$).
	These five methods that are added last in the combination only add very few scenarios that were not covered by the preceding methods.
	The scenarios that are not covered by the best combination of two or three methods are given in Tables~\ref{tab:scens.not.covered.top2.multi} and~\ref{tab:scens.not.covered.top3.multi} in Appendix~\ref{app:scens.uncovered}.

	\section{Discussion and Conclusion}\label{sec:summary}
	Methods for quantifying dataset similarity can be applied in all situations where it is of interest to distinguish between datasets or their underlying distributions. 
	Examples include two- or $k$-sample testing, and deciding which datasets are similar enough in meta-learning or for synthetic data generation and simulation. 
	Here, methods for quantifying the similarity or distance of two or more multivariate numeric datasets are compared with regard to their practical utility and empirical performance. 
	An extensive simulation study was conducted and evaluated, including the most promising dataset similarity methods from a preceding theoretical method comparison \parencite{stolte_methods_2024}. 
	The simulation set-up includes various deviations, like shift, scale, or higher moments, for multiple distribution types, including normal, $t$, log-normal, and $\chi^2$-distributed datasets.
	The methods were compared regarding their runtime, memory requirements, computational errors, and especially with respect to their ability to detect the differences between datasets. 
	No classical power comparison could be conducted since not all methods are statistical tests, but a quantity, the proportion of extreme simulation repetitions (PESR), which was motivated by power, was used in the comparison. 
	
	With regard to the methods' performances, clear differences between the deviations were observed both in the two-sample and in the multi-sample case. 
	Shift and scale alternatives could be well detected by many methods, while the detection of differences in higher moments was more difficult. 
	In the presence of a target variable, methods that could take such a target into account were mostly unable to detect differences that only affect the generation of this target. 
	The methods that were best in detecting the differences between datasets depend on the deviation of interest, and in part also on the data-generating mechanism, i.e.\ on the distribution of the variables in the datasets. 
	The overall best method for two samples was the BF test \parencite{baringhaus_rigid_2010} using $\varphi_{\text{FracB}}$ or $\varphi_{\text{FracA}}$ as the distance function. 
	It was shown to be numerically stable, have low resource consumption, the lowest mean difference in PESR to the ideal method, and the highest number of scenarios covered where it is acceptable, i.e.\ not much worse than the best method for the respective scenario. 
	However, depending on the deviation, other methods might be better, which becomes especially clear since the BF test was rarely the best according to the derived decision rules per deviation.
	Thus, the BF test can be seen as an overall good compromise that is often not far from the best, but if more information on the DGP or deviation is given, this information can be used to find a better-fitting method. 
	Alternatively, the BF in combination with the ZC~(5MST,$\kappa=1$) test \parencite{zhang_graph-based_2022}, the DS test \parencite{deb_multivariate_2021}, and the GPK test \parencite{song_generalized_2021} covers 90\,\% of all scenarios considered in the simulation study, i.e.\ in each of these 90\% of the scenarios at least one of these methods is almost as good as the best method in the respective scenario. 
	When using the FS~($\psi_3$,$h_1$) test \parencite{paul_clustering-based_2022} and the GPK~($Z_D$) component in addition, 95\,\% of scenarios can be covered already. 
	These combinations might be applied in a multiple testing framework to simultaneously check for various differences between the distributions.
	
	For the multi-sample case, the DISCO~(F,$\alpha=0.5$) \parencite{rizzo_disco_2010} can be recommended overall as it showed no computational errors, comparably low runtime and memory consumption, and was ranked best according to PESR and number of scenarios that are covered as acceptable.
	However, as for the best method in the two-sample case, it is rarely the best method according to the decision rules, but other methods might be better if more knowledge about the DGP or deviations is given. 
	The best method combinations are the DISCO~(F,$\alpha=0.5$), together with the SC~(5MST,S) test \parencite{song_new_2022} already covering 90\,\% of the scenarios and in combination with the AFS~(knw,$\psi_2$,$h_1$) test \parencite{paul_clustering-based_2022}, covering 95\,\%. 
	
	The results of the current study might depend on the chosen set of scenarios and especially on the considered deviations, since it could be observed that the methods performed quite differently regarding different deviations. 
	The scenarios here combine the ones used in previous studies such that they can be considered to cover a range of settings that are regarded as relevant by different groups of researchers in the field and allow for a method comparison in a very controlled setting that is not focused on one particular application. 
	Interesting additional scenarios might include more applied data situations, either based on real-world datasets or using real-world data directly, as these would provide valuable additional insights for the practical use of the methods for certain applications. 
	The current simulation results using well-understood parametric scenarios and thereby granting full control of the data-generating mechanism can be used to guide such real-world data based evaluations tailored to specific applications.
	
	One aspect that would, in particular, be worthwhile to examine in greater detail is comparing the similarity of datasets that include a target variable. 
	Based on the results here, the best choice would be to disregard the target variable and use the best of the methods that do not take the target variable into account. 
	However, good methods for detecting changes in the relationship of the covariates and the target variable are needed, and it was not analysed how treating a target variable like the covariates would affect the performance of the methods that do not take a target variable into account. 
	
	Since many of the methods that were considered here are proposed as two- or $k$-sample tests, real power comparisons of these tests and, where applicable, evaluations of the asymptotic properties of the tests under different conditions, would be another aspect worth investigating, especially for the methods that performed well in the current comparison. 
	
	\subsection*{Acknowledgements}
	This work has been supported (in part) by the Research Training Group ``Biostatistical Methods for High-Dimensional Data in Toxicology'' (RTG 2624, Project P1) funded by the Deutsche Forschungsgemeinschaft (DFG, German Research Foundation -- Project Number 427806116).\\
	The authors gratefully acknowledge the computing time provided on the Linux HPC cluster at TU Dortmund University (LiDO3), partially funded in the course of the Large-Scale Equipment Initiative by the Deutsche Forschungsgemeinschaft (DFG, German Research Foundation) as project 271512359.

	\subsection*{Competing Interests}
	The authors have no competing interests to declare.

	\subsection*{Data and Code Availability}
	The full code of the simulation study can be found on Zenodo \parencite{code_zenodo_num} along with all simulation results.
	
	\subsection*{Author Contributions}
	Conceptualization: Marieke Stolte, Jörg Rahnenführer, Andrea Bommert; Formal analysis: Marieke Stolte; Methodology: Marieke Stolte; Project administration: Jörg Rahnenführer, Andrea Bommert; Software: Marieke Stolte; Supervision: Jörg Rahnenführer, Andrea Bommert; Visualization: Marieke Stolte; Writing -- original draft: Marieke Stolte; Writing -- review \& editing: Jörg Rahnenführer, Andrea Bommert

	\begin{sloppypar}
		\begin{footnotesize}
			\printbibliography[title = References]
		\end{footnotesize}
	\end{sloppypar}
	
	\clearpage
	\appendix
	\FloatBarrier
	
	\section{Scenario Parameter Settings}\label{app:scen.tabs}

	\begin{table}[!h]
		\centering
		\begin{tabular}{lp{6.5cm}p{7cm}}
			\toprule
			Distribution & Alternative & Parameters\\
			\midrule
			Normal & Null: $F_j = N_p(0, I_p),$ $ j = 1, 2$ & -- \\
			& Shift: $F_j = N_p(\delta_j / \sqrt{p}\cdot\mathbf{1}, I_p),$ $ j = 1, 2$ & $\delta_1 = 0$, $\delta_2 = 0.1, 0.25, 0.5, 0.75, $ $ 1, 1.5$ \\
			& Scale: $F_j = N_p(0, \sigma_j^{1/p}\cdot I_p),$ $ j = 1, 2$ & $\sigma_1 = 1$, $\sigma_2 = 1/10, 1/3, 1/2, 2/3, $ $ 4/5, 5/4, 3/2, 2, 3, 10$ \\
			& Correlation: $F_j = N_p(0, \Sigma_j),$ $ j = 1, 2$ & $\Sigma_1 = I_p$, $\Sigma_2 = \rho \cdot \mathbf{1}_{p, p} $ $+ (1-\rho)\cdot I_p,$ $\rho = 0.05, 0.1, 0.2, 0.3, 0.4, 0.6, 0.8$ \\
			& $t$: $F_1 = N_p(0, I), F_2 = t_{p, \nu}(0, (\nu-2)/\nu \cdot I_p)$ & $\nu = 30, 20, 10, 5, 3$ \\
			Log-Normal & Null: $F_j = LN_p(\mu, \sigma\cdot I_p),$ $ j = 1, 2$ & $\mu$ and $\sigma$ chosen such that $\E\left(X^{(j)}_i\right) = \mathbf{1}_p$ and $\Cov\left(X^{(j)}_i\right) = I_p$ \\
			& Shift: $X^{(j)}_i = X^{\prime(j)}_i + \delta_j/\sqrt{p} \mathbf{1}_p, $ $X^{\prime(j)}_i\sim LN_p(\mu, \sigma\cdot I_p),$ $ j = 1, 2$ & $\mu$ and $\sigma$ such that $\E\left(X^{\prime(j)}_i\right) = \mathbf{1}_p$, $\Cov\left(X^{\prime(j)}_i\right) = I_p,$ $\delta_1 = 0$, $\delta_2 = 0.1, 0.25, 0.5, 0.75, 1, 1.5$ \\
			& Scale: $X^{(j)}_i = \sigma_j^{1/p} \cdot X^{\prime(j)}_i, $ $X^{\prime(j)}_i\sim LN_p(\mu, \sigma\cdot I_p),$ $ j = 1, 2$ & $\mu$ and $\sigma$ such that $\E\left(X^{\prime(j)}_i\right) = \mathbf{1}_p$, $\Cov\left(X^{\prime(j)}_i\right) = I_p,$, $\sigma_1 = 1$ $\sigma_2= 1/10, 1/3, 1/2,2/3,4/5, 5/4,  $ $3/2, 2, 3, 10$ \\
			$t$ & Null: $F_j = t_{p, 3}(0, \sigma\cdot I_p),$ $ j = 1, 2$ & $\sigma = 1/3$ chosen such that $\Cov\left(X^{(j)}_i\right) = I_p$ \\
			& Shift: $F_j = t_{p, 3}(\delta_j / \sqrt{p}\cdot\mathbf{1}, 1/3\cdot I_p),$ $ j = 1, 2$ & $\delta_1 = 0$, $\delta_2 = 0.1, 0.25, 0.5, 0.75, $ $ 1, 1.5$ \\
			& Scale: $F_j = t_{p, 3}(0, 1/3\sigma_j^{1/p}\cdot I_p),$ $ j = 1, 2$ & $\sigma_1 = 1$, $\sigma_2 = 1/10, 1/3, 1/2, 2/3, $ $ 4/5, 5/4, 3/2, 2, 3, 10$ \\
			& Correlation: $F_j = t_{p, 3}(0, \Sigma_j),$ $ j = 1, 2$ & $\Sigma_1 = 1/3 I_p$, $\Sigma_2 = 1/3 (\rho \cdot \mathbf{1}_{p, p} $ $+ (1-\rho)\cdot I_p),$ $\rho = 0.05, 0.1, 0.2, 0.3, 0.4, 0.6, 0.8$ \\
			& Kurtosis: $F_j = t_{p, \nu_j}(0, (\nu_j-2)/\nu \cdot I_p),$ $ j = 1, 2$ & $\nu_1 = 3$, $\nu_2 = 3.05, 3.1, 3.2, 3.3, 3.4$ \\
			$\chi^2$ & Null: $F_j = \chi^2_{p, 1},$ $ j = 1, 2$ & Each variable is standardized \\
			& Skewness and kurtosis: $F_j = \chi^2_{p, \nu_j},$ $ j = 1, 2$ & $\nu_1 = 1$, $\nu_2 = 1.1, 1.5, 2, 3, 4, 5$, each variable is standardized \\
			\bottomrule
		\end{tabular}
		\caption{Scenarios for numerical data generation for $k = 2$. $X_i^{(j)}\stackrel{\text{iid}}{\sim}F_j$, $i = 1, \dots, n_j,$ $j = 1, 2$.}
		\label{tab:scen.num.no.y}
	\end{table}
	
	\begin{table}[!h]
		\centering
		\begin{tabular}{lp{8cm}p{5.5cm}}
			\toprule
			Distribution & Alternative & Parameters\\
			\midrule
			Each & Null: $\Prob\left(Y^{(j)}_i = 1\right) = \exp(\eta_i^{(j)}) / (1+\exp(\eta_i^{(j)}))$, $\eta_i^{(j)} = -1/2 + X^{(j)}_i\beta$  & $\beta = 1/2\cdot(\mathbf{1}_{p/2}^T, -\mathbf{1}_{p/2}^T)^T$ \\
			& Wrong sign: $\Prob\left(Y^{(j)}_i = 1\right) = \exp(\eta_i^{(j)}) / (1+\exp(\eta_i^{(j)}))$, $\eta_i^{(j)} = -1/2 + X^{(j)}_i\beta^{(j)}$ & $\beta^{(1)} = 1/2\cdot(\mathbf{1}_{p/2}^T, -\mathbf{1}_{p/2}^T)^T$, $\beta^{(2)} = 1/2\cdot(-\mathbf{1}_{p/2}^T, \mathbf{1}_{p/2}^T)^T$ \\
			& Wrong size: $\Prob\left(Y^{(j)}_i = 1\right) = \exp(\eta_i^{(j)}) / (1+\exp(\eta_i^{(j)}))$, $\eta_i^{(j)} = -1/2 + X^{(j)}_i\beta^{(j)}$ & $\beta^{(1)} = 1/2\cdot(\mathbf{1}_{p/2}^T, -\mathbf{1}_{p/2}^T)^T$, $\beta^{(2)} = 1/4\cdot(\mathbf{1}_{p/2}^T, -\mathbf{1}_{p/2}^T)^T$ \\
			\bottomrule
		\end{tabular}
		\caption{Additional scenarios for numerical data generation for $k = 2$ with target variable. $X_i^{(j)}\stackrel{\text{iid}}{\sim}F_j$, $i = 1, \dots, n_j,$ $j = 1, 2$.}
		\label{tab:scen.num.y}
	\end{table}
	\FloatBarrier
	
	\begin{longtable}{lp{6.5cm}p{7cm}}
		\toprule
		Distribution & Alternative & Parameters\\
		\midrule
		\endhead
		\bottomrule
		\caption{Scenarios for numerical data generation for $k = 4$. $X_i^{(j)}\stackrel{\text{iid}}{\sim}F_j$, $i = 1, \dots, n_j,$ $j = 1, \dots, 4$.}
		\endfoot
		\bottomrule
		\caption{Scenarios for numerical data generation for $k = 4$. $X_i^{(j)}\stackrel{\text{iid}}{\sim}F_j$, $i = 1, \dots, n_j,$ $j = 1, \dots, 4$.}\label{tab:scen.num.multi}\\
		\endlastfoot
		Normal & Null: $F_j = N_p(0, I_p),$ $ j = 1, \dots, k$ & -- \\
		& Shift: $F_j = N_p(\delta_j / \sqrt{p}\cdot\mathbf{1}, I_p),$ $j = 1, \dots, 4$ & $\delta_1 = \delta_2 = \delta_3 = 0$, $\delta_4 = 0.1, 0.25, 0.5, 0.75, $ $ 1, 1.5$ \\
		&  & $\delta_1 = \delta_2 = 0$, $\delta_3 = \delta_4 = 0.1, 0.25, 0.5, 0.75, $ $ 1, 1.5$ \\
		& & $\delta_1 = \delta_2 = 0$, $\delta_3 = \delta$,  $\delta_4 = 2\delta$, $\delta = 0.1, 0.25, 0.5, 0.75, $ $ 1, 1.5$ \\
		& & $\delta_j = (j-1)\cdot \delta$, $\delta = 0.1, 0.25, 0.5, 0.75, $ $ 1, 1.5$ \\
		& Scale: $F_j = N_p(0, \sigma_j^{1/p}\cdot I_p),$ $j = 1, \dots, 4$ & $\sigma_1 = \sigma_2 = \sigma_3 = 1$, $\sigma_4 = 1/10, 1/3, 1/2, 2/3, $ $ 4/5, 5/4, 3/2, 2, 3, 10$ \\
		& & $\sigma_1 = \sigma_2 = 1$, $\sigma_3 = \sigma_4 = 1/10, 1/3, 1/2, 2/3, $ $ 4/5, 5/4, 3/2, 2, 3, 10$ \\
		& & $\sigma_1 = \sigma_2 = 1$, $\sigma_3 = \sigma$, $\sigma_4 = \sigma^2$, $\sigma = 1/10, 1/3, 1/2, 2/3, $ $ 4/5, 5/4, 3/2, 2, 3, 10$ \\
		& & $\sigma_j = \sigma^{(j-1)}$, $\sigma = 1/10, 1/3, 1/2, 2/3, $ $ 4/5, 5/4, 3/2, 2, 3, 10$ \\
		& Correlation: $F_j = N_p(0, \Sigma_j),$ $j = 1, \dots, 4$ & $\Sigma_1 = \Sigma_2 = \Sigma_3 = I_p$, $\Sigma_4 = \rho \cdot \mathbf{1}_{p, p} $$+ (1-\rho)\cdot I_p,$ $\rho = 0.05, 0.1, 0.2, 0.3$ \\
		& & $\Sigma_1 = \Sigma_2 = I_p$, $\Sigma_4 = \Sigma_3 = \rho \cdot \mathbf{1}_{p, p} $$+ (1-\rho)\cdot I_p,$ $\rho = 0.05, 0.1, 0.2, 0.3$ \\
		& & $\Sigma_1 = \Sigma_2 =I_p$, $\Sigma_3 = \rho \cdot \mathbf{1}_{p, p} $$+ (1-\rho)\cdot I_p,$, $\Sigma_4 = 2\rho \cdot \mathbf{1}_{p, p} $$+ (1-2\rho)\cdot I_p,$ $\rho = 0.05, 0.1, 0.2, 0.3$ \\
		& & $\Sigma_1 = I_p$, $\Sigma_2 = \rho \cdot \mathbf{1}_{p, p} $$+ (1-\rho)\cdot I_p,$ $\Sigma_3 = 2\rho \cdot \mathbf{1}_{p, p} $$+ (1-2\rho)\cdot I_p,$ $\Sigma_4 = 3\rho \cdot \mathbf{1}_{p, p} $$+ (1-3\rho)\cdot I_p,$ $\rho = 0.05, 0.1, 0.2, 0.3$ \\
		Log-Normal & Null: $F_j = LN_p(\mu, \sigma\cdot I_p),$ $j = 1, \dots, 4$ & $\mu$ and $\sigma$ chosen such that $\E\left(X^{(j)}_i\right) = \mathbf{1}_p$ and $\Cov\left(X^{(j)}_i\right) = I_p$ \\
		& Shift: $X^{(j)}_i = X^{\prime(j)}_i + \delta_j/\sqrt{p} \mathbf{1}_p, $ $X^{\prime(j)}_i\sim LN_p(\mu, \sigma\cdot I_p),$ $\mu$ and $\sigma$ such that $\E\left(X^{\prime(j)}_i\right) = \mathbf{1}_p$, $\Cov\left(X^{\prime(j)}_i\right) = I_p,$, $j = 1, \dots, 4$ &$\delta_1 =\delta_2 =\delta_3 = 0$, $\delta_4 = 0.1, 0.25, 0.5, 0.75, 1, 1.5$ \\
		& & $\delta_1 =\delta_2 = 0$, $\delta_3 = \delta_4 = 0.1, 0.25, 0.5, 0.75, 1, 1.5$ \\
		&  & $\delta_1 = \delta_2 = 0$, $\delta_3 = \delta$, $\delta_4 = 2\delta$, $\delta = 0.1, 0.25, 0.5, 0.75, 1, 1.5$ \\
		&  & $\delta_j = (j-1)\delta$, $\delta = 0.1, 0.25, 0.5, 0.75, 1, 1.5$ \\
		& Scale: $X^{(j)}_i = \sigma_j^{1/p} \cdot X^{\prime(j)}_i, $ $X^{\prime(j)}_i\sim LN_p(\mu, \sigma\cdot I_p),$ $\mu$ and $\sigma$ such that $\E\left(X^{\prime(j)}_i\right) = \mathbf{1}_p$, $\Cov\left(X^{\prime(j)}_i\right) = I_p,$, $j = 1, \dots, 4$ & $\sigma_1 = \sigma_2= \sigma_3= 1,$ $\sigma_4= 1/10, 1/3, 1/2,2/3,4/5, 5/4,  $ $3/2, 2, 3, 10$ \\
		&  & $\sigma_1 = \sigma_2= 1,$ $\sigma_3= \sigma_4= 1/10, 1/3, 1/2,2/3,4/5, 5/4,  $ $3/2, 2, 3, 10$ \\
		&  & $\Cov\left(X^{\prime(j)}_i\right) = I_p,$, $\sigma_1 = \sigma_2= 1,$ $\sigma_3 = s$, $\sigma_4= s^2$, $s = 1/10, 1/3, 1/2,2/3,4/5, 5/4,  $ $3/2, 2, 3, 10$ \\
		& & $\sigma_j = s^{j-1}$ $s= 1/10, 1/3, 1/2,2/3,4/5, 5/4,  $ $3/2, 2, 3, 10$ \\
		$t$ & Null: $F_j = t_{p, 3}(0, \sigma\cdot I_p),$ $j = 1, \dots, 4$ & $\sigma = 1/3$ chosen such that $\Cov\left(X^{(j)}_i\right) = I_p$ \\
		& Shift: $F_j = t_{p, 3}(\delta_j / \sqrt{p}\cdot\mathbf{1}, 1/3\cdot I_p),$ $j = 1, \dots, 4$ & $\delta_1 = \delta_2 = \delta_3 = 0$, $\delta_4 = 0.1, 0.25, 0.5, 0.75, $ $ 1, 1.5$ \\
		& & $\delta_1 = \delta_2 = 0$, $\delta_3 = \delta_4 = 0.1, 0.25, 0.5, 0.75, $ $ 1, 1.5$ \\
		& & $\delta_1 = \delta_2 = 0$, $\delta_3 = \delta$, $\delta_4 = 2\delta$, $\delta = 0.1, 0.25, 0.5, 0.75, $ $ 1, 1.5$ \\
		& & $\delta_j = (j-1) \delta$, $\delta = 0.1, 0.25, 0.5, 0.75, $ $ 1, 1.5$ \\
		& Scale: $F_j = t_{p, 3}(0, 1/3\sigma_j^{1/p}\cdot I_p),$ $j = 1, \dots, 4$ & $\sigma_1 = \sigma_2 = \sigma_3 = 1$, $\sigma_4 = 1/10, 1/3, 1/2, 2/3, $ $ 4/5, 5/4, 3/2, 2, 3, 10$ \\
		& & $\sigma_1 = \sigma_2 = 1$, $\sigma_3 = \sigma_4 = 1/10, 1/3, 1/2, 2/3, $ $ 4/5, 5/4, 3/2, 2, 3, 10$ \\
		& & $\sigma_1 = \sigma_2 = 1$, $\sigma_3 = s$, $\sigma_4 = s^2,$ $s = 1/10, 1/3, 1/2, 2/3, $ $ 4/5, 5/4, 3/2, 2, 3, 10$ \\
		& & $\sigma_j = s^{j-1}$, $s = 1/10, 1/3, 1/2, 2/3, $ $ 4/5, 5/4, 3/2, 2, 3, 10$ \\
		& Correlation: $F_j = t_{p, 3}(0, \Sigma_j),$ $j = 1, \dots, 4$ & $\Sigma_1 = \Sigma_2 = \Sigma_3 = 1/3 I_p$, $\Sigma_4 = 1/3 (\rho \cdot \mathbf{1}_{p, p} $$+ (1-\rho)\cdot I_p),$ $\rho = 0.05, 0.1, 0.2, 0.3$ \\
		& & $\Sigma_1 = \Sigma_2 = 1/3 I_p$, $\Sigma_3 = \Sigma_4 = 1/3 (\rho \cdot \mathbf{1}_{p, p} $$+ (1-\rho)\cdot I_p),$ $\rho = 0.05, 0.1, 0.2, 0.3$ \\
		& & $\Sigma_1 = \Sigma_2 = 1/3 I_p$, $\Sigma_3 = 1/3 (\rho \cdot \mathbf{1}_{p, p} $$+ (1-\rho)\cdot I_p),$ $\Sigma_3 = 1/3 (2\rho \cdot \mathbf{1}_{p, p} $$+ (1-2\rho)\cdot I_p),$ $\rho = 0.05, 0.1, 0.2, 0.3$ \\
		& & $\Sigma_j = 1/3 ((j-1)\rho \cdot \mathbf{1}_{p, p} $$+ (1-(j-1)\rho)\cdot I_p),$ $\rho = 0.05, 0.1, 0.2, 0.3$ \\
		& Kurtosis: $F_j = t_{p, \nu_j}(0, (\nu_j-2)/\nu \cdot I_p),$ $j = 1, \dots, 4$ & $\nu_1 = \nu_2 = \nu_3 = 3$, $\nu_4 = 3.05, 3.1, 3.2, 3.3, 3.4$ \\
		& & $\nu_1 = \nu_2 = 3$, $\nu_3 = \nu_4 = 3.05, 3.1, 3.2, 3.3, 3.4$ \\
		& & $\nu_1 = \nu_2 = 3$, $\nu_3 = 3 + d$, $\nu_4 = 3 + 2d$, $d = 0.05, 0.1, 0.2, 0.3, 0.4$ \\
		& & $\nu_j = 3 + (j-1)d$, $d = 0.05, 0.1, 0.2, 0.3, 0.4$ \\
		$\chi^2$ & Null: $F_j = \chi^2_{p, 1},$ $j = 1, \dots, 4$ & Each variable is standardized \\
		& Skewness and kurtosis: $F_j = \chi^2_{p, \nu_j},$ $j = 1, \dots, 4$ & $\nu_1 = \nu_2 = \nu_3 = 1$, $\nu_4 = 1.1, 1.5, 2, 3, 4, 5$, each variable is standardized \\
		& & $\nu_1 = \nu_2 = 1$, $\nu_3 = \nu_4 = 1.1, 1.5, 2, 3, 4, 5$, each variable is standardized \\
		& & $\nu_1 = \nu_2 = 1$, $\nu_3 = 1 + d$, $\nu_4 = 1 + 2d$, $d = 0.1, 0.5, 1, 2, 3, 4$, each variable is standardized \\
		& & $\nu_j = 1 + (j-1) d$, $d = 0.1, 0.5, 1, 2, 3, 4$,, each variable is standardized \\
	\end{longtable}
	\FloatBarrier
	
	\section{Complete Method list}\label{app:meth.tabs}

	\begin{landscape}
		
		\begin{longtable}{p{4cm}p{4cm}p{1.4cm}p{5cm}p{2.8cm}p{0.4cm}p{0.9cm}p{0.8cm}p{1.4cm}}
			\toprule
			Method &  (Sub)class & No. fulfilled & Implementation & Inclusion & $y$? & Num.? & Cat.? & $K > 2$?\\
			\midrule
			\endhead
			\bottomrule
			\endfoot
			\bottomrule
			\caption{Methods from theoretical comparison chosen for the empirical comparison. Methods applicable to two or more categorical datasets without target variables are highlighted. These are compared in the current study. No. fulfilled: Number of fulfilled criteria in theoretical comparison. $y$?: Can the method deal with a target variable in the dataset? Num.?: Is the method as implemented applicable to numeric data? Cat.?: Is the method as implemented applicable to categorical data? $K > 2$?: Is the method as implemented applicable to more than two datasets at a time? \faTimes$^*$: Method is, in theory, applicable, but implementation is not. \faCheck$^*$: Implementation is applicable, although this case is not described in the literature.}\label{tab:methods}\\
			\endlastfoot
			KMD \textcite{huang_kernel_2022} & Kernel-based & 15 & R package KMD \textcite{KMD} & Implemented \& $\ge 11$ criteria & \faTimes & \faCheck & \faTimes$^*$ & \faCheck\\
			\textcite{mukherjee_distribution-free_2022} & Graph-based & 13 & R package multicross \textcite{multicross} & Implemented \& $\ge 11$ criteria & \faTimes & \faCheck & \faCheck & \faCheck \\
			\textcite{biswas_distribution-free_2014} & Graph-based & 12 & Own implementation & $\ge 11$ criteria & \faTimes & \faCheck & \faTimes & \faCheck\\
			\textcite{friedman_multivariate_1979} & Graph-based & 13 & R package gTests \textcite{gTests} & Implemented \& $\ge 11$ criteria & \faTimes & \faCheck & \faCheck & \faTimes \\
			Cross-match test \textcite{rosenbaum_exact_2005} & Graph-based & 13 & R package crossmatch \textcite{crossmatch} & Implemented \& $\ge 11$ criteria & \faTimes & \faCheck & \faTimes & \faTimes \\
			Cramér test \textcite{baringhaus_new_2004} & Inter-point distances & 11 & R package cramer \textcite{cramer} & Implemented \& $\ge 11$ criteria & \faTimes & \faCheck & \faTimes & \faTimes \\
			Energy statistic \textcite{szekely_energy_2017} & Inter-point distances & 13 & R package energy \textcite{energy} & Implemented \& $\ge 11$ criteria & \faTimes & \faCheck & \faTimes & \faCheck \\
			\textcite{deb_multivariate_2021} & Inter-point distances / Rank-based & 12 & Implementation based on R code for paper (\url{https://github.com/NabarunD/MultiDistFree}) & Implemented \& $\ge 11$ criteria & \faTimes & \faCheck & \faTimes & \faTimes \\
			\textcite{ntoutsi_general_2008} & Comparison of density functions & 11 & Own implementation & $\ge 11$ criteria & \faCheck & \faCheck & \faTimes & \faTimes \\
			\textcite{ganti_framework_1999} & Comparison of density functions & 11 & Own implementation & $\ge 11$ criteria & \faCheck & \faCheck & \faTimes$^*$ & \faTimes \\
			\textcite{hediger_use_2021} & Binary classification & 11 & R package hypoRF \textcite{hypoRF} & Implemented \& $\ge 11$ criteria & \faTimes & \faCheck & \faCheck & \faTimes \\
			\textcite{petrie_graph-theoretic_2016} & Graph-based & 13 & R package multicross \textcite{multicross} & Implemented \& $\ge 11$ criteria & \faTimes & \faCheck & \faCheck$^*$ & \faCheck \\
			OTDD \textcite{alvarez-melis_geometric_2020} & Distance/ similarity measure for datasets & 11 & own implementation based on python implementation (\url{https://github.com/microsoft/otdd}) & $\ge 11$ criteria & \faCheck & \faCheck & \faCheck & \faTimes \\
			Jeffreys divergence & Divergence & 11 & Own implementation &  $\ge 11$ criteria & \faTimes & \faCheck & \faTimes & \faTimes \\
			\textcite{baringhaus_rigid_2010} & Inter-point distances & 11 & R package cramer \textcite{cramer} & Implemented \& $\ge 11$ criteria & \faTimes & \faCheck & \faTimes & \faTimes \\
			\textcite{bahr_ein_1996} & Inter-point distances & 11 & R package cramer \textcite{cramer} & Implemented \& $\ge 11$ criteria & \faTimes & \faCheck & \faTimes & \faTimes \\
			\textcite{biswas_nonparametric_2014} & Inter-point distances & 11 & Own implementation &  $\ge 11$ criteria & \faTimes & \faCheck & \faTimes & \faTimes \\
			\textcite{schilling_multivariate_1986} and \textcite{henze_multivariate_1988} & Graph-based & 11 & Own implementation &  $\ge 11$ criteria & \faTimes & \faCheck & \faTimes & \faTimes \\
			\textcite{yu_two-sample_2007} & Binary classification & 11 & Own implementation using R package Ecume \textcite{Ecume} & (Almost) implemented \& $\ge 11$ criteria & \faTimes & \faCheck & \faCheck & \faTimes \\
			Wasserstein distance & Probability metric & 9 & R package Ecume \textcite{Ecume} & Implemented & \faTimes & \faCheck & \faTimes & \faTimes \\   
			\textcite{chen_new_2017} & Graph-based & 11 & R package gTests \textcite{gTests} & Implemented \& $\ge 11$ & \faTimes & \faCheck & \faCheck & \faTimes \\
			\textcite{chen_weighted_2018} & Graph-based & 12 & R package gTests \textcite{gTests} & Implemented \& $\ge 11$ & \faTimes & \faCheck & \faCheck & \faTimes\\
			Ball divergence \textcite{pan_ball_2018} & Testing & 9 & R package ball \textcite{ball} & Implemented & \faTimes & \faCheck & \faTimes & \faCheck \\
			\textcite{song_new_2022} & Graph-based & 11 & R package gTestsMulti \textcite{gTestsMulti} & Implemented \& $\ge 11$ & \faTimes & \faCheck & \faTimes & \faCheck\\
			DISCO \textcite{rizzo_disco_2010} & Inter-point distances & 10 & R package energy \textcite{energy} & Implemented & \faTimes & \faCheck & \faTimes & \faCheck\\
			\textcite{li_measuring_2022} & Comparison of cha\-rac\-teris\-tic func\-tions & 9 & Own implementation & Best in (sub)class & \faTimes & \faCheck & \faTimes & \faTimes \\
			Maximum Mean Discrepancy (MMD) \textcite{gretton_kernel_2006} & Kernel-based (MMD) & 9 & R packages kernlab \textcite{kernlab} and Ecume \textcite{Ecume} & Implemented & \faTimes & \faCheck & \faTimes$^*$ & \faTimes\\
			\textcite{mukhopadhyay_nonparametric_2020} & Graph-based & 9 & R package LPKsample \textcite{LPKsample} & Implemented & \faTimes & \faCheck & \faTimes$^*$ & \faCheck \\
			\textcite{chen_ensemble_2013} & Graph-based & 9 & R packages gTests \textcite{gTests}, gCat \textcite{gCat} & Implemented & \faTimes & \faCheck & \faCheck & \faTimes \\
			Block MMD \textcite{zaremba_b-test_2013} & Kernel-based (MMD) & 8 & R implementation based on matlab code for paper (\url{https://github.com/wojzaremba/btest}) & Implemented & \faTimes & \faCheck & \faTimes$^*$ & \faTimes \\
			\textcite{song_generalized_2021} & Kernel-based (MMD) & 8 & R package kerTests \textcite{kerTests} & Implemented & \faTimes & \faCheck & \faTimes$^*$ & \faTimes \\
			Constrained Minimum Distance \textcite{tatti_distances_2007} & Comparison based on summary statistics & 8 & Own implementation & Best in (sub)class & \faTimes & \faTimes & \faCheck & \faTimes \\
			\textcite{biau_asymptotic_2005} & Comparison of CDFs & 8 & Own implementation & Best in (sub)class & \faTimes & \faCheck & \faTimes & \faCheck \\
			Classifier Two-Sample Test \textcite{lopez-paz_revisiting_2017} & Binary classification & 7 & R package Ecume \textcite{Ecume} & Implemented & \faTimes & \faCheck & \faCheck & \faCheck \\
			&  &  &  &  &  & &  \\
			DiProPerm test \textcite{wei_direction-projection-permutation_2016} & Binary classification & 5 & Own implementation & Implemented & \faTimes & \faCheck & \faTimes & \faTimes\\
		\end{longtable}
	\end{landscape}
	\FloatBarrier
	
	\section{Method Parameter Settings}\label{app:meth.pars}
	All methods are used with default parameters based on recommendations from the literature, if available. 
	If no sensible default is available, different options are compared.
	These are described in the following. 
	
	For the KMD \textcite{huang_kernel_2022}, there is no clear recommendation for setting the graph type. Possible options are an MST or an undirected $K$-nearest neighbour graph. 
	\textcite{huang_kernel_2022} recommend using the $K$-NN graph for its flexibility, but the choice of $K$ is unclear. 
	Based on the simulation results in the original article, the default is $K = 0.1 N$ for testing and $K = 1$ for estimation. 
	For increasing power, it is beneficial to choose large $K$s. For consistency of the tests, $K = o(N / \log(N))$ together with a continuous distribution of inter-point distances is sufficient, thus $K$ cannot be chosen too large compared to $N$. 
	On the other hand, in the context of estimating the KMD, choosing $K$ is a bias-variance trade-off with small $K$s decreasing the bias and larger $K$s decreasing the variance (see discussion in Appendix D.3 of \textcite{huang_kernel_2022}). 
	Therefore, here, $K = 1, 5, 0.1N$ are used. 
	Additionally, the MST is used as the similarity graph. 	
	
	For the Cramér statistic of \textcite{baringhaus_rigid_2010}, three different kernel functions are used, of which the first, 
	\[\varphi_{\text{log}}(x) = \log(1 + x),\]
	is recommended preferably for location alternatives, and the other two choices 
	\[\varphi_{\text{FracA}}(x) = 1 - \frac{1}{1+x}\] and
	\[\varphi_{\text{FracB}}(x) = 1 - \frac{1}{(1+x)^2}\]
	are recommended preferably for dispersion alternatives.
	All three variants are used here since both location and dispersion alternatives are considered. 

	The power $\alpha\in(0, 2]$ to which the inter-point distances are taken in calculating the DISCO statistics \textcite{rizzo_disco_2010} is chosen as one by default. 
	All values $\alpha < 2$ lead to a consistent test. 
	Here, $\alpha \in \{0.5, 1, 1.5\}$ are used to cover the whole range for both the DISCO $F$ statistic as well as the DISCO between-sample ($B$) statistic since, e.g., lower powers are recommended for heavy-tailed data which is also included here.	
	
	The classifier for the classifier two-sample test (C2ST) of \textcite{lopez-paz_revisiting_2017} is chosen as a $K$-nearest neighbour classifier or as a multilayer perceptron network optimized by stochastic gradient descent. 
	A $K$-nearest neighbour classifier as well as a one-layer neural network were compared in the simulations in the original publication. The neural network performed superior there. 
	The $K$-NN classifier is the default in the R implementation of the test. 
	Note that \textcite{lopez-paz_revisiting_2017} did not tune the hyperparameters of the classifiers, which might affect the performance. 
	The implementation that is used here tunes the hyperparameters.
	
	For the random forest-based test of \textcite{hediger_use_2021}, both methods of estimating the prediction accuracy, namely taking the overall OOB accuracy or averaging the per-class OOB accuracies, are compared.
	
	All three suggested options are used for the classifier for the DiProPerm statistic \textcite{wei_direction-projection-permutation_2016}, including DWD, MD, and SVM.
	Each of the classifiers is used in combination with each of the suggested univariate statistics, i.e.\ the mean difference (MD), $t$ statistic, and the area under the receiver operator curve (AUC).  
	
	For the edge count tests (FR, CF, CCS, ZC), a similarity graph has to be chosen. 
	Typical choices are the $K$-minimum spanning tree (MST) and the $K$-nearest neighbour (NN) graph. 
	The choice of $K$ is difficult, and there are no guidelines on how to choose this parameter. 
	Many of the tests were originally proposed for $K = 1$, therefore, this value is used. 
	Previous simulation studies show better empirical performance for higher values of $K$ \textcite{zhu_limiting_2024}. 
	Often, $K = 5$ is used, so this is included as a second choice here.
	
	For the max-type test of \textcite{zhang_graph-based_2022}, all recommended values for the parameter $\kappa$ are compared, i.e.\ $\kappa \in\{1,$$ 1.14,$$1.31\}$.

	Similarly, for the multi-sample edge count test of \textcite{song_new_2022}, a similarity graph has to be chosen. 
	The $K$-MST is suggested. 
	Analogously to the other edge-count tests, $K = 1, 5$ is considered. 
	Both test statistics $S$ and $S_A$ are considered as no recommendations are made which one should be used when. 
	
	For the nearest-neighbour test of \textcite{schilling_multivariate_1986} and \textcite{henze_multivariate_1988}, the choice of the number of nearest neighbours $k$ to consider is unclear. 
	The common choices of $K = 1, 5$ are used here.
	
	For the statistic of \textcite{mukhopadhyay_nonparametric_2020}, there are different options for the clustering algorithm and for combining the results of the multiple components. 
	The number of components is kept at the default of 4 since higher moments are not investigated here. 
	For the combination criterion, both options of combining the $p$ values using Fisher's method and computing an overall kernel based on the significant components are used. 
	As the statistic is always zero if no component is significant, adding all components instead of only the significant ones is tried additionally. 
	As the clustering algorithm \texttt{mclust} has a default number of maximal iterations of $2147483647$, only $K$-means clustering is used for the clustering. 
	If the \texttt{mclust} algorithm does not converge according to the tolerance criterion within a few iterations, its runtime becomes infeasibly large. 
	As this happened for some scenarios in the tests of the simulation, and the maximum number of iterations cannot be changed from outside the \texttt{LPKsample::GLP} function, the \texttt{mclust} algorithm is not used within the simulations.
	
	For Jeffreys' divergence, different methods for estimating the density ratio are compared. 
	Implemented options include the options of using the Kullback-Leibler importance estimation procedure (KLIEP) or the (relative) unrestricted least-squares importance fitting (rULSIF/ULSIF). 
	For the KLIEP approach, the idea is to minimize the KL divergence of $f_1(x)$ and $r(x)f_2(x)$, for two densities $f_1, f_2$ and their ratio $r(x) = f_1(x) / f_(x)$. 
	LSIF approach, the ratio $r$ is fitted using a least-squares approach for minimizing the squared error of the ratio. 
	In the unconstrained fitting, the solutions are not constrained to be non-negative. 
	For an overview of density ratio estimation, see \textcite{sugiyama_density_2012}.
	
	For the $k$-sample Ball divergence \textcite{pan_ball_2018}, all three strategies, sum, summax, and max, are compared to combine the pairwise comparison results.
	These correspond to summing up the pairwise Ball divergences, or taking the maximum of sums of the Ball divergences from each sample to all other samples, or summing up the largest $k-1$ pairwise Ball divergences, respectively.
	The multi-sample case for the Ball divergence is implemented in the \texttt{Ball} package \textcite{ball}, but not described in the original publication. 
	Since it is available, it is included in the simulations, although it was not included in the theoretical comparison before.
	
	For the GPK test \textcite{song_generalized_2021}, all proposed test statistics $GPK$, $Z_{W, 0.8}$, $Z_{W, 1.2}$, and $Z_D$ are used as well.
	
	The partition for the \textcite{biau_asymptotic_2005} statistic is chosen as a rectangle partition, with the number of elements in the rectangle partition set to 
	\[m_n = n^{\varepsilon}, \varepsilon\in\{0.5, 0.8, 0.9\}\]
	which fulfils the requirements on $m_n$ described in that article.
	There are no recommendations in the article on how to choose the partition and the authors themselves state that the choice is difficult.
	The rectangle partition is chosen for simplicity here. 
	
	For the FS and RI test \textcite{paul_clustering-based_2022}, all proposed variants of the original, the modified, the multi-scale and the aggregated test statistics are use in combination with all proposed $\psi$ and $h$ functions as there are no recommendations which test and which functions should be used. 
	The aggregated and multi-scale versions are only sensible in the multi-sample case. 
	For the aggregated tests AFS and ARI, either the original or modified tests can be used. 
	Both variants are done, where the original variant is denoted as ``knw'' as the number of clusters is assumed to be known while the modified variant is denoted as ``est'' as the number of clusters is estimated. 
	For the multi-scale tests, the maximum number of clusters is kept as the default at twice the number of datasets, i.e.\ eight here for four datasets. 
	Later, we will give the parameter $k$ which is the index of the test statistic, i.e.\ the number of clusters used minus one since always at least two clusters are used. 
	
	For the engineer metric, only the typical choice of $q = 2$ is considered corresponding to the simple Euclidean distance of the mean vectors of the datasets. 
	
	For the tree-based methods of \textcite{ganti_framework_1999} and \textcite{ntoutsi_general_2008}, using trees with default parameters and tuning the parameters of the trees are compared. 
	For \textcite{ntoutsi_general_2008}, all three proposed methods for calculating the dataset similarity are used. 
	For \textcite{ganti_framework_1999}, both $f_s$ and $f_a$ are used as difference functions.
	
	For the optimal transport dataset distance (OTDD) \textcite{alvarez-melis_geometric_2020}, the exact version is used along with different speed-ups. 
	One version where both the inner and outer optimal transport problems are approximated using the Sinkhorn divergence is compared as well as the data augmentation version, the Gaussian approximation for the inner OT problem, the Gaussian approximation assuming equal covariances for the inner OT problem, and the naive upper bound for the inner OT problem. 
	All of these are approximations of the problem that can potentially speed up the computation, but some come with restrictive assumptions. 
	Therefore, it is of interest how much computation speed can be gained and how much performance of the dataset distance in recognizing differences between datasets is lost. 
	
	Note that for $k = 2$, the statistics of \textcite{schilling_multivariate_1986} and \textcite{henze_multivariate_1988} and of \textcite{huang_kernel_2022} are equivalent when, for the latter, the $K$-NN graph is used together with the discrete kernel \textcite{huang_kernel_2022}. 
	Due to different implementations, both are used nonetheless.
	Note that the method of \textcite{petrie_graph-theoretic_2016} and \textcite{mukherjee_distribution-free_2022} are equal to the one of \textcite{rosenbaum_exact_2005} for $K = 2$. 
	Since the implementations of the \textcite{petrie_graph-theoretic_2016} and \textcite{mukherjee_distribution-free_2022} methods use the implementation of the \textcite{rosenbaum_exact_2005} method, only the \textcite{rosenbaum_exact_2005} test is used for $K = 2$.

	
	\section{Additional Figures}\label{app:add.figs}
	
	\subsection{Pre-Selection of Methods}\label{app:meth.sel}
	
	\subsubsection[k = 2]{$k = 2$}
	\paragraph{Shift}\mbox{} 

	\begin{figure}[!h]
		\centering
		\includegraphics[width=\linewidth]{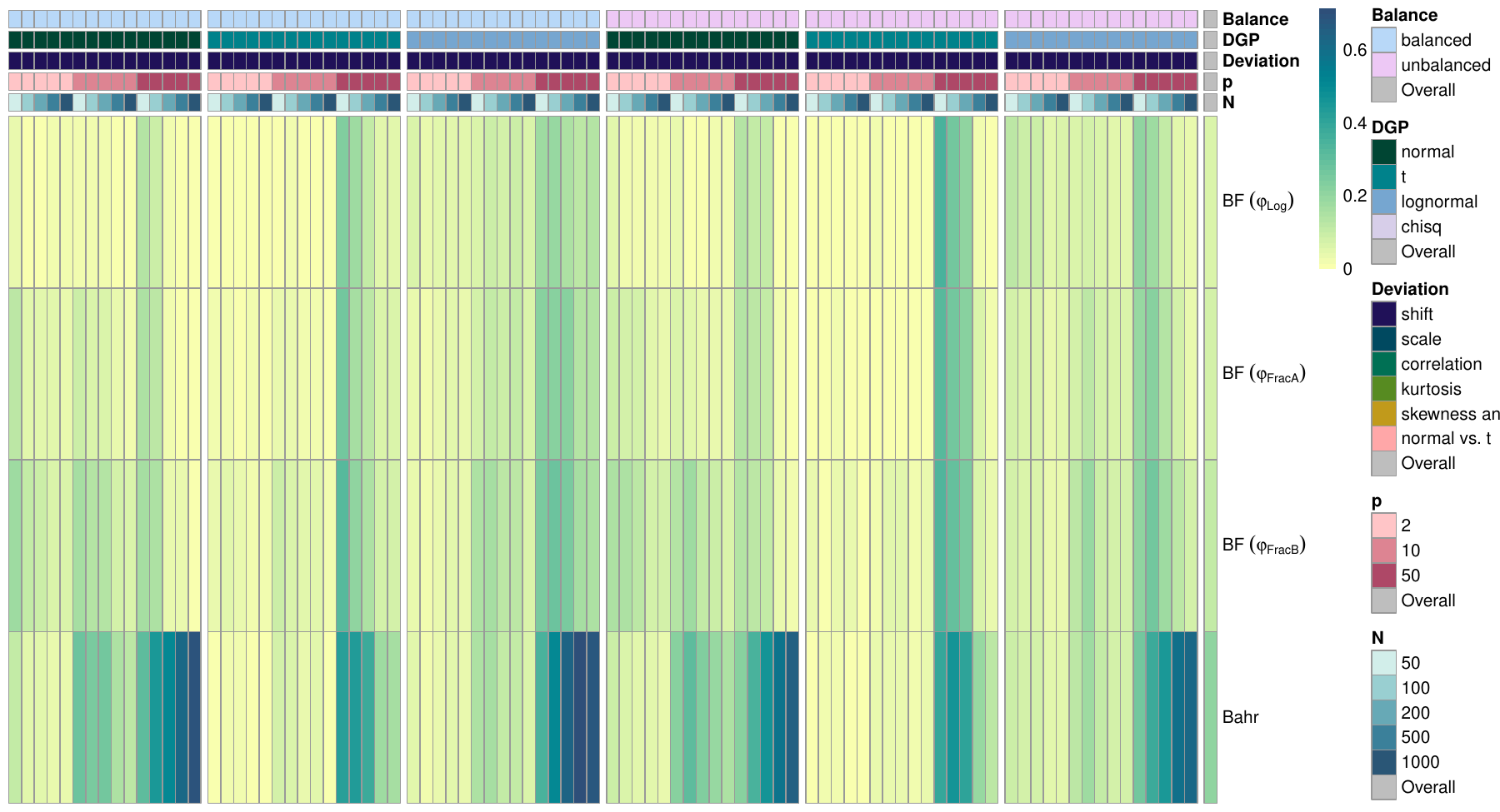}
		\caption{Mean PESR difference to best method per scenario for shift alternatives and $k = 2$ datasets for all variants of the BF method \textcite{baringhaus_rigid_2010}. Selected variant: BF~($\varphi_{\text{Log}}$).}
	\end{figure}
	
	\begin{figure}[!h]
		\centering
		\includegraphics[width=\linewidth]{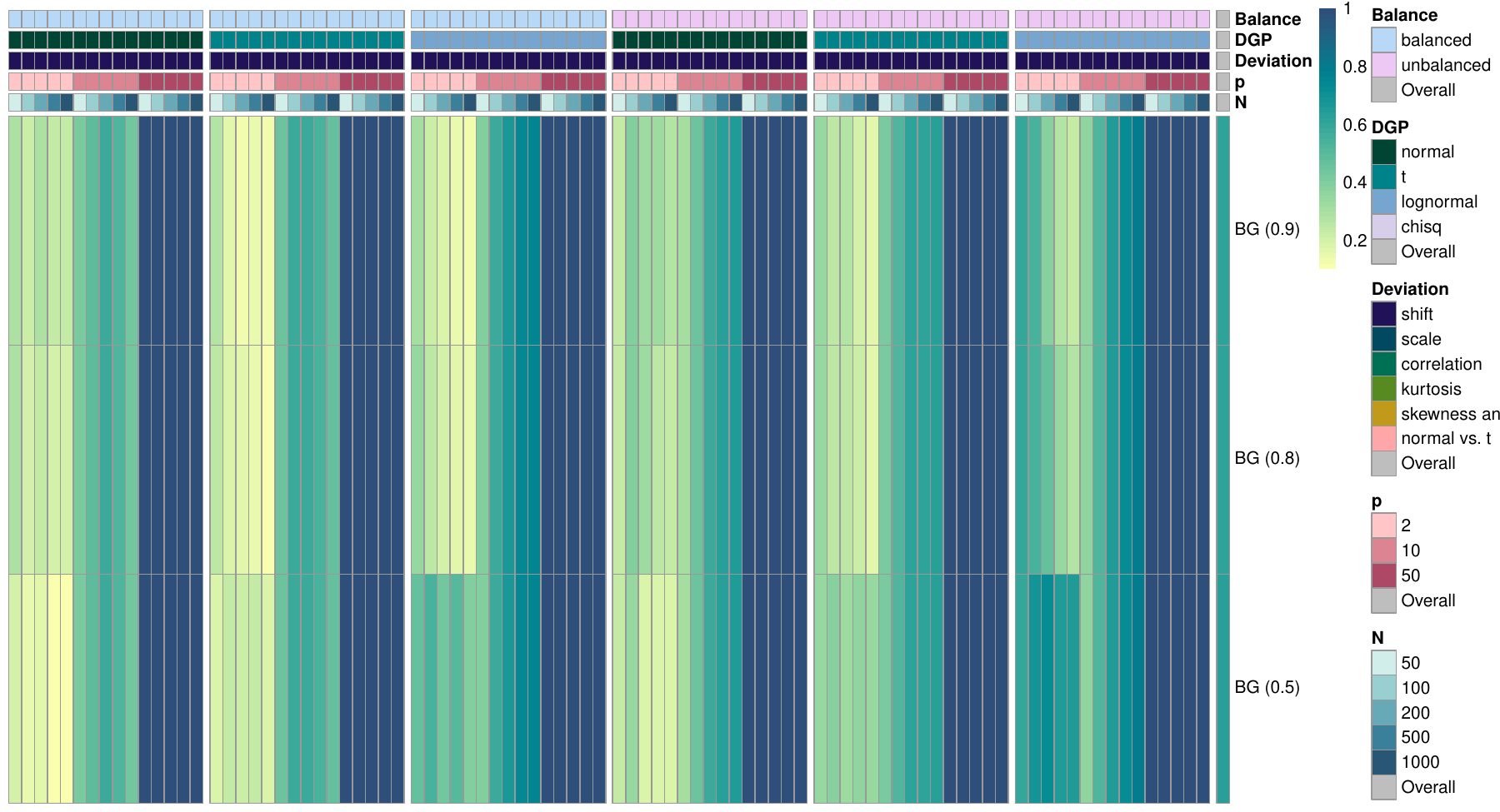}
		\caption{Mean PESR difference to best method per scenario for shift alternatives and $k = 2$ datasets for all variants of the BG method \textcite{biau_asymptotic_2005}. Selected variant: BG~(0.8) (almost identical performance to BG~(0.9) but feasible for larger $p$).}
	\end{figure}
	
	\begin{figure}[!h]
		\centering
		\includegraphics[width=\linewidth]{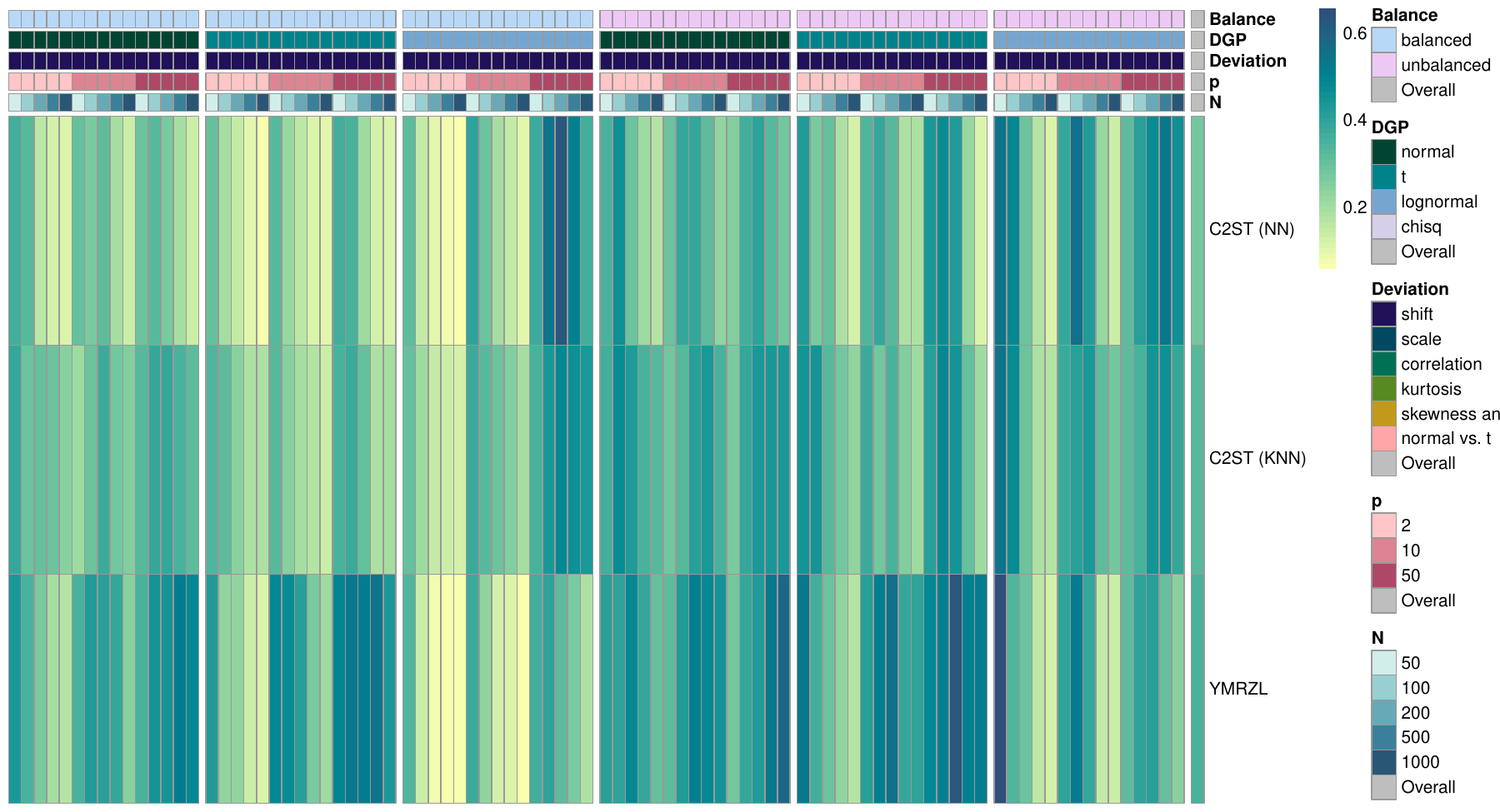}
		\caption{Mean PESR difference to best method per scenario for shift alternatives and $k = 2$ datasets for all variants of the YMRZL and C2ST method \textcite{yu_two-sample_2007, lopez-paz_revisiting_2017}. Selected variant: C2ST~(NN).}
	\end{figure}
	
	\begin{figure}[!h]
		\centering
		\includegraphics[width=\linewidth]{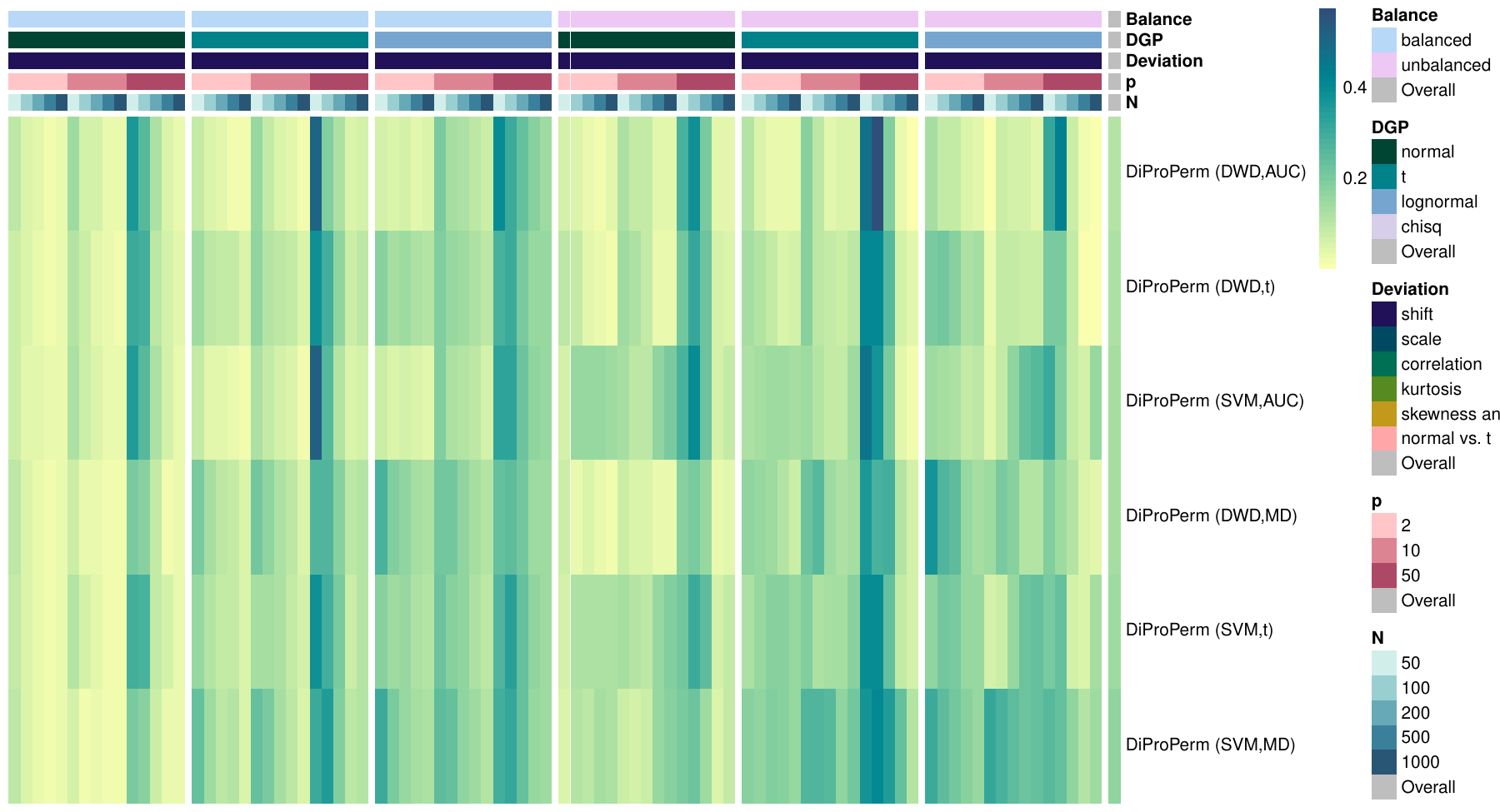}
		\caption{Mean PESR difference to best method per scenario for shift alternatives and $k = 2$ datasets for all variants of the DiProPerm method \textcite{wei_direction-projection-permutation_2016}. Selected variant: DiProPerm~(DWD,AUC).}
	\end{figure}
	
	\begin{figure}[!h]
		\centering
		\includegraphics[width=\linewidth]{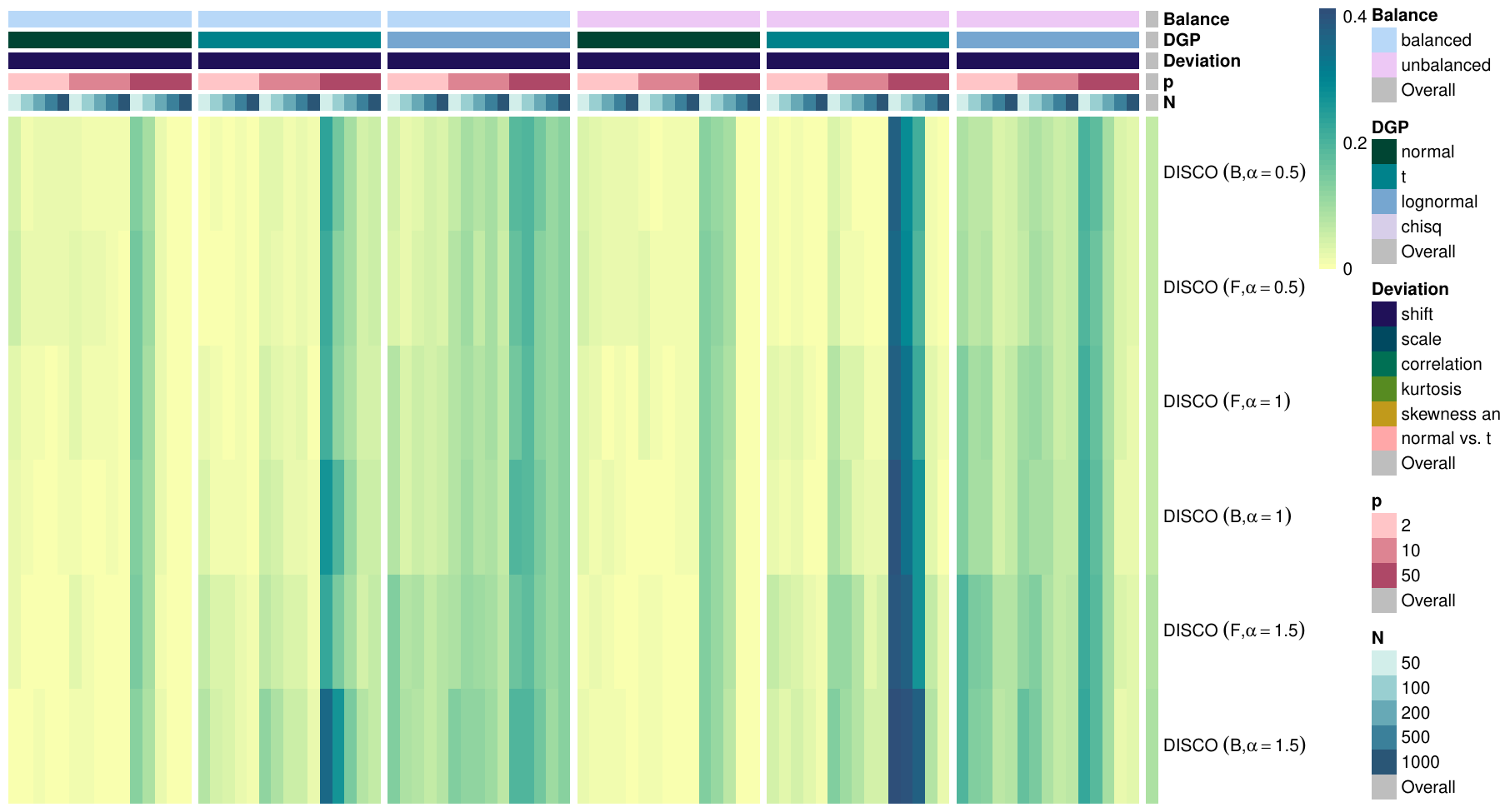}
		\caption{Mean PESR difference to best method per scenario for shift alternatives and $k = 2$ datasets for all variants of the DISCO method \textcite{rizzo_disco_2010}. Selected variant: DISCO~(B,$\alpha=0.5$).}
	\end{figure}
	
	\begin{figure}[!h]
		\centering
		\includegraphics[width=\linewidth]{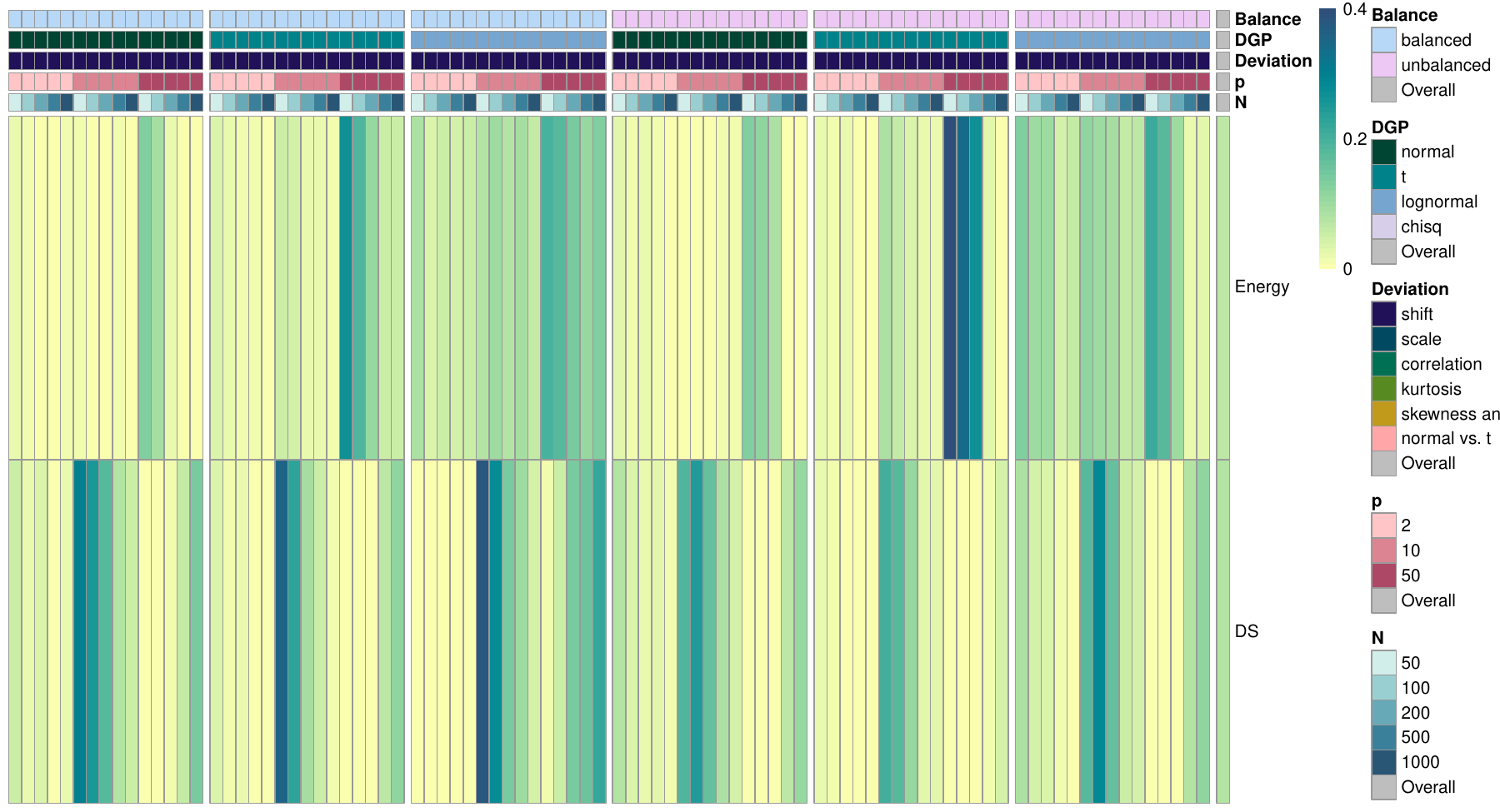}
		\caption{Mean PESR difference to best method per scenario for shift alternatives and $k = 2$ datasets for all variants of the Energy method \textcite{szekely_testing_2004}. Selected variants: Energy and DS (complement each other).}
	\end{figure}
	
	\begin{figure}[!h]
		\centering
		\includegraphics[width=\linewidth]{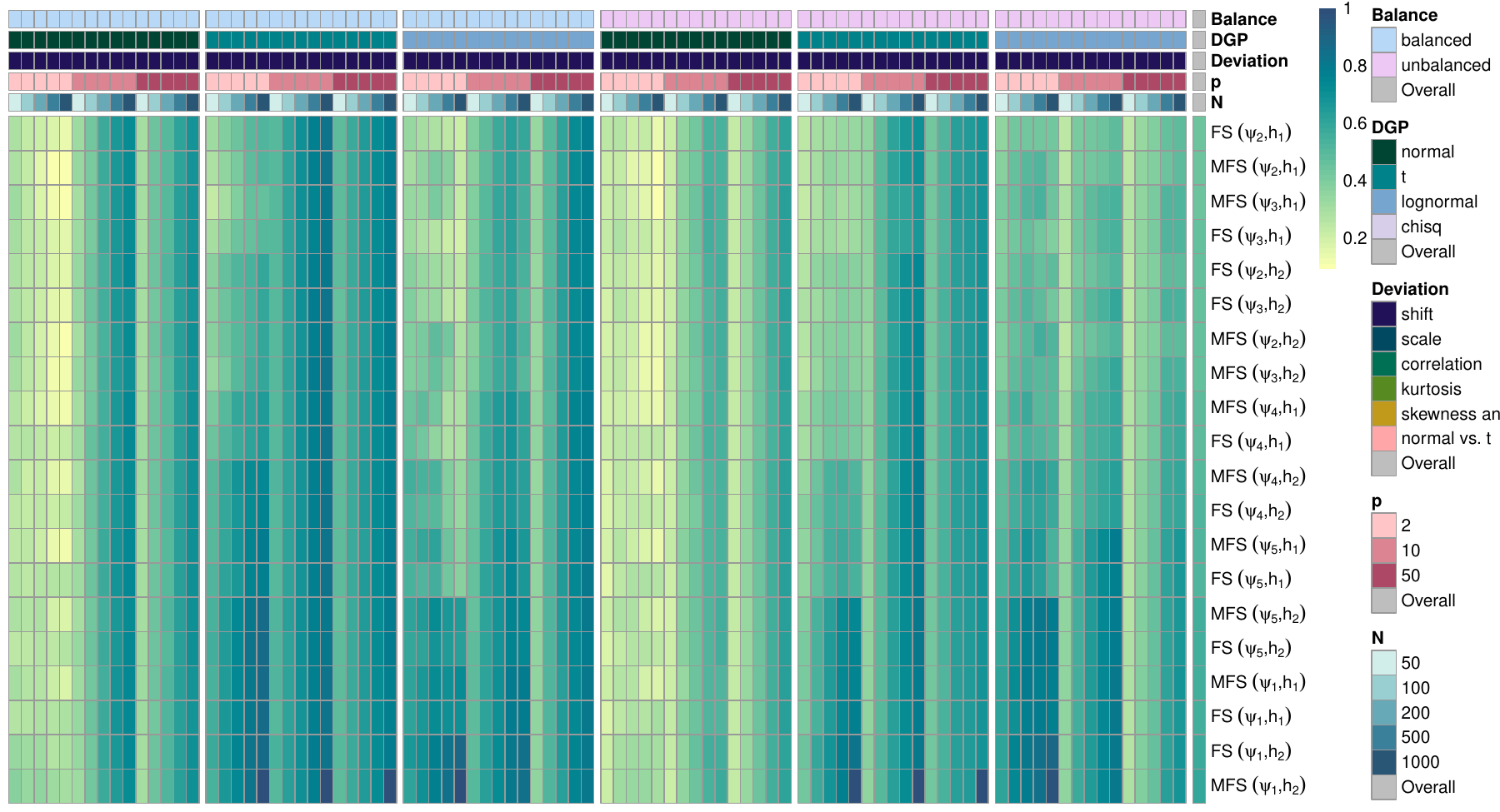}
		\caption{Mean PESR difference to best method per scenario for shift alternatives and $k = 2$ datasets for all variants of the FS test \textcite{paul_clustering-based_2022}. Selected variant: FS~($\psi_2, h_1$).}
	\end{figure}
	
	\begin{figure}[!h]
		\centering
		\includegraphics[width=\linewidth]{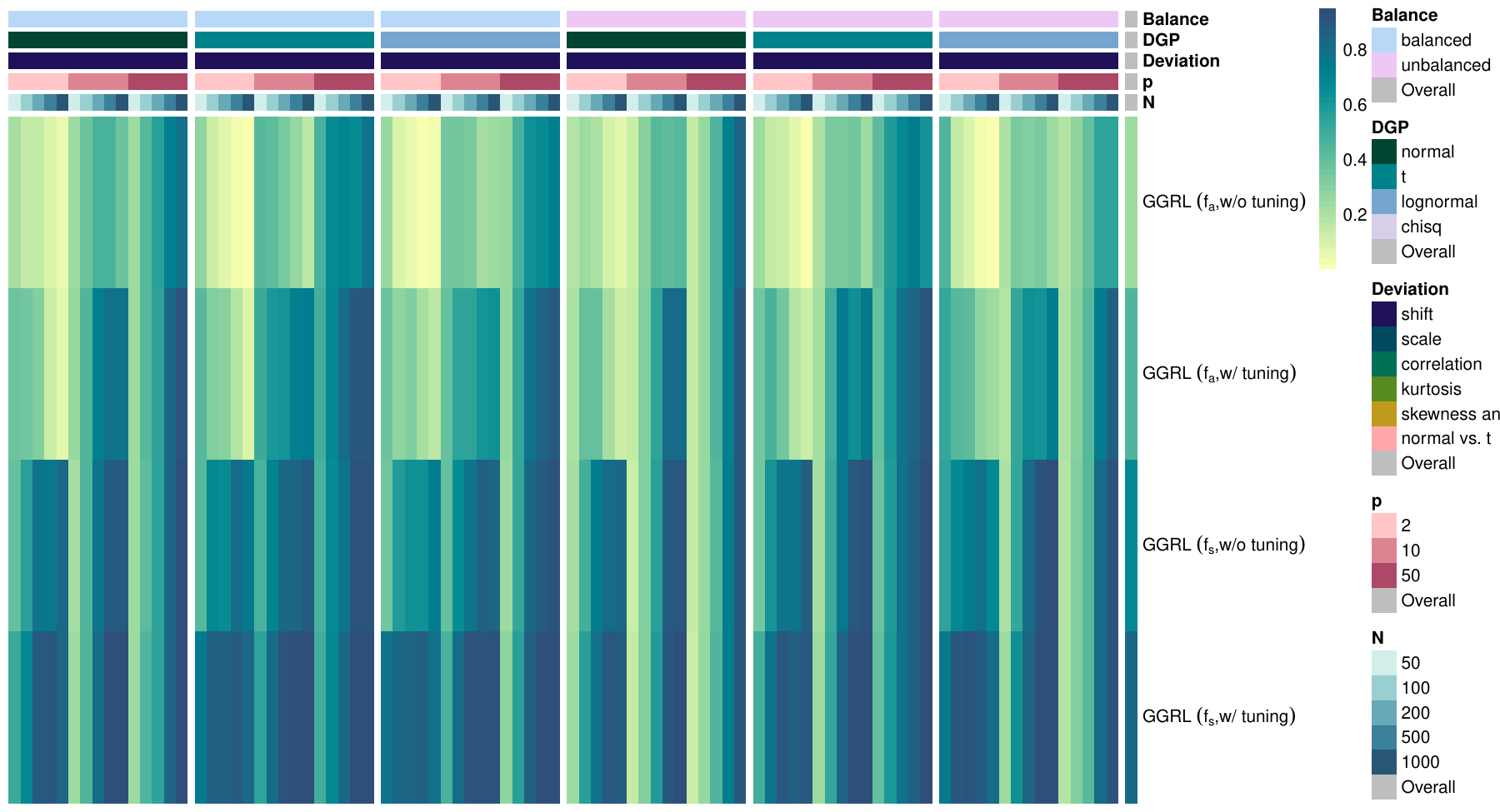}
		\caption{Mean PESR difference to best method per scenario for shift alternatives and $k = 2$ datasets for all variants of the GGRL method \textcite{ganti_framework_1999}. Selected variant: GGRL~($f_a$,w/o tuning).}
	\end{figure}
	
	\begin{figure}[!h]
		\centering
		\includegraphics[width=\linewidth]{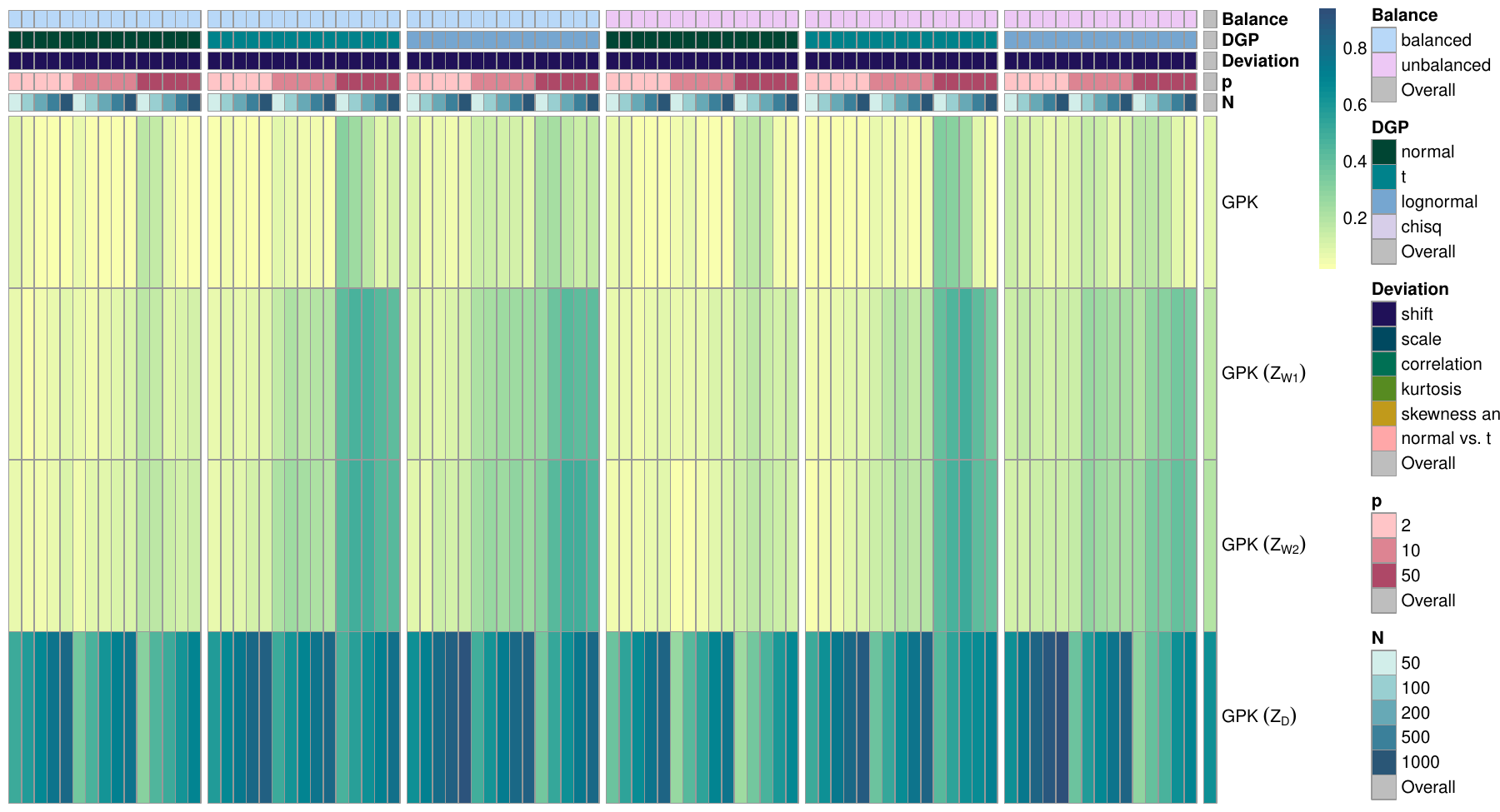}
		\caption{Mean PESR difference to best method per scenario for shift alternatives and $k = 2$ datasets for all variants of the GPK test \textcite{song_generalized_2021}. Selected variant: GPK.}
	\end{figure}
	
	\begin{figure}[!h]
		\centering
		\includegraphics[width=\linewidth]{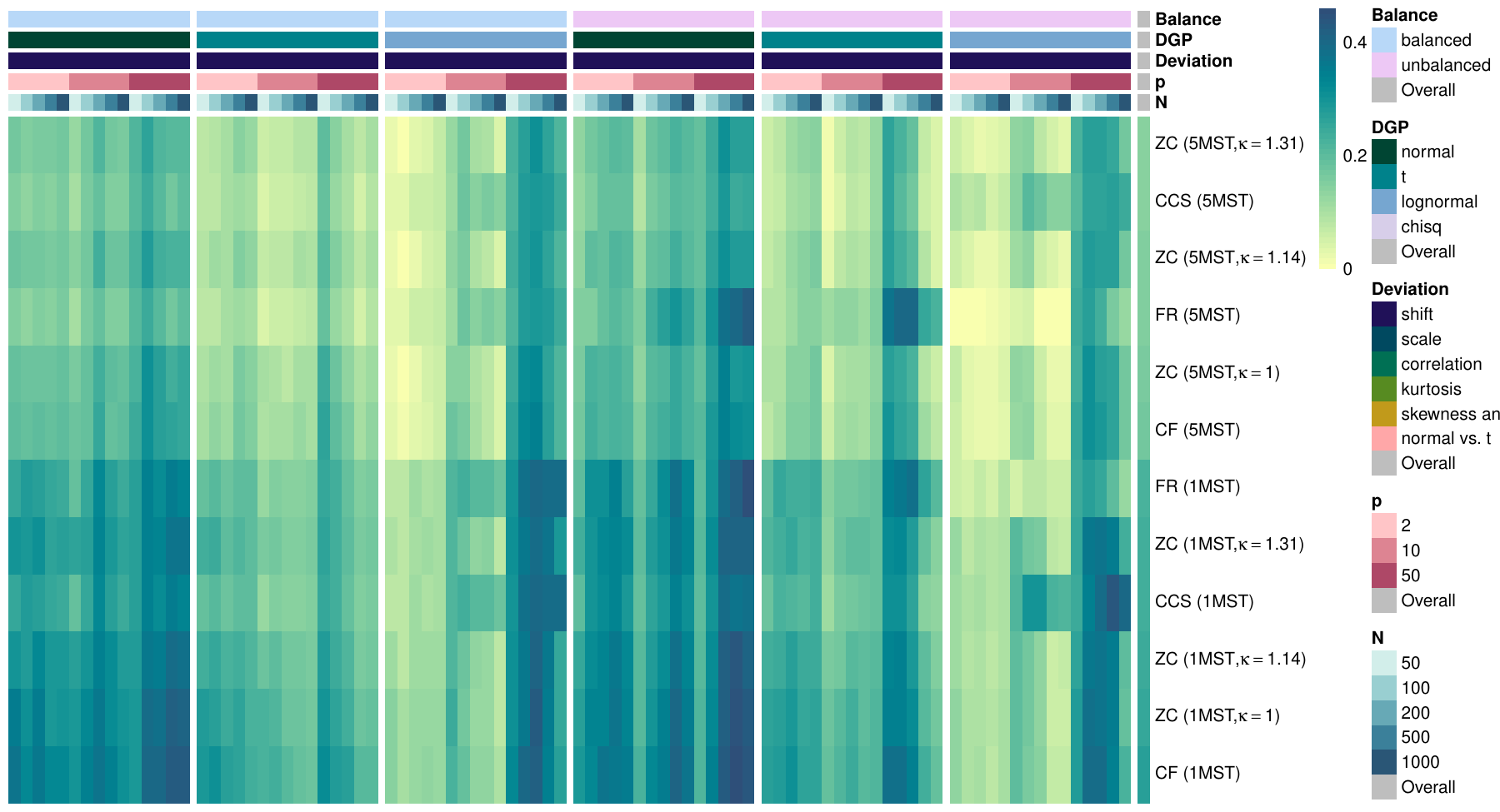}
		\caption{Mean PESR difference to best method per scenario for shift alternatives and $k = 2$ datasets for all variants of the graph-based tests FR, CF, CCS, ZC \textcite{friedman_multivariate_1979, chen_new_2017, chen_weighted_2018, zhang_graph-based_2022}. Selected variant: ZC~(5MST,$\kappa=1.31$).}
	\end{figure}
	
	\begin{figure}[!h]
		\centering
		\includegraphics[width=\linewidth]{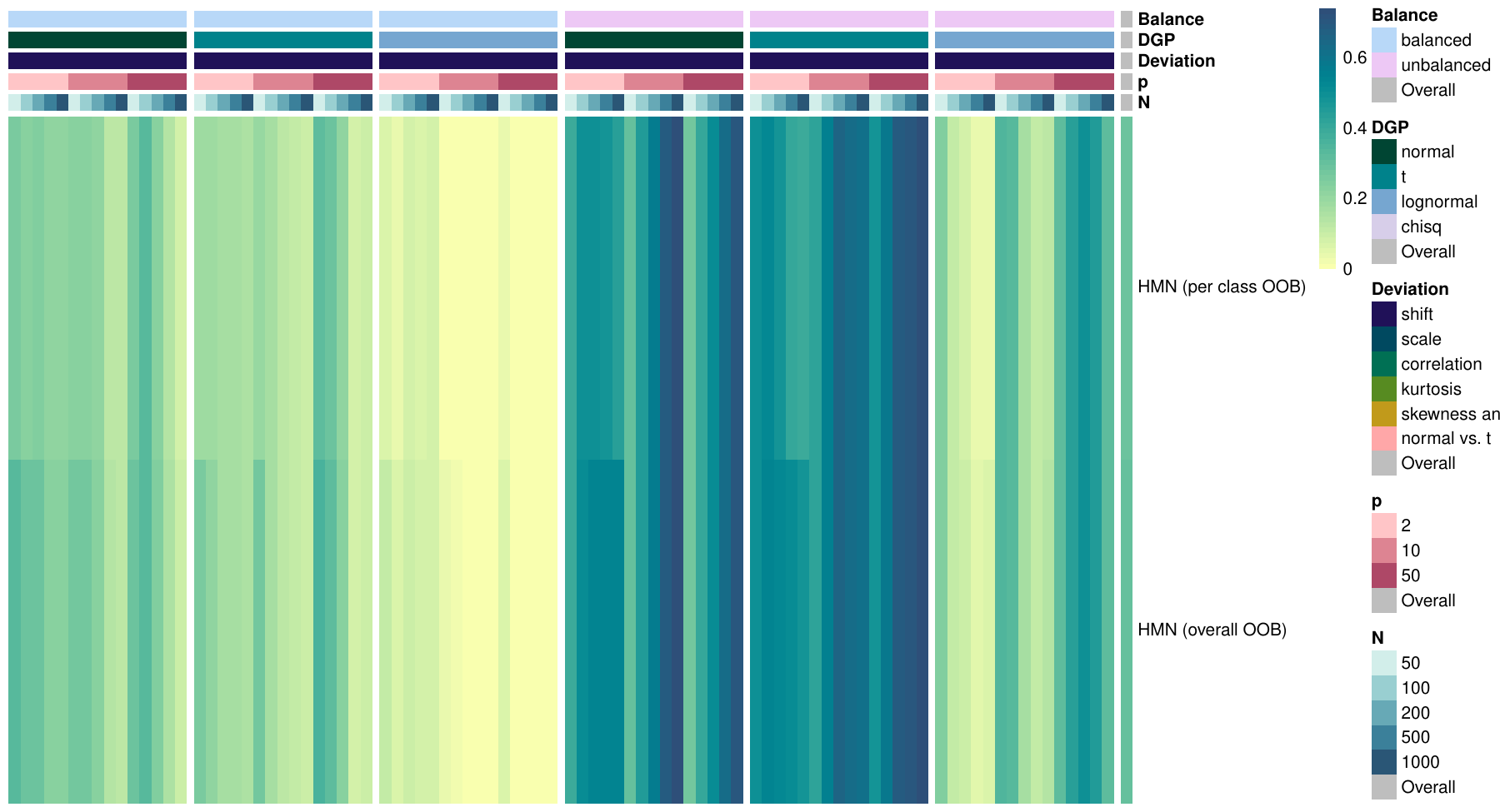}
		\caption{Mean PESR difference to best method per scenario for shift alternatives and $k = 2$ datasets for all variants of the HMN method \textcite{hediger_use_2021}. Selected variant: HMN~(per class OOB).}
	\end{figure}
	
	\begin{figure}[!h]
		\centering
		\includegraphics[width=\linewidth]{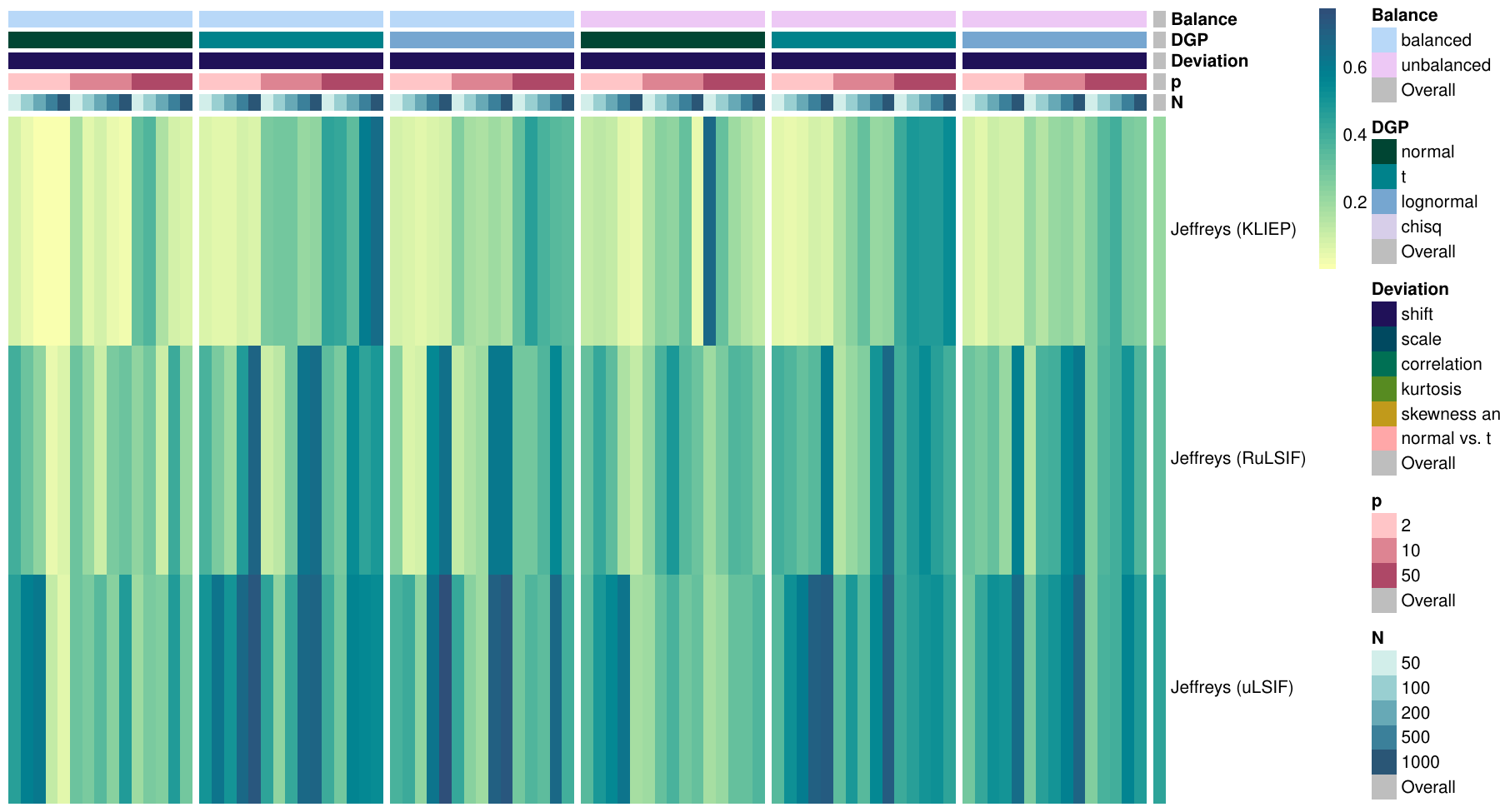}
		\caption{Mean PESR difference to best method per scenario for shift alternatives and $k = 2$ datasets for all variants of the Jeffreys divergence \textcite{jeffreys_invariant_1997}. Selected variant: Jeffreys~(KLIEP).}
	\end{figure}
	
	\begin{figure}[!h]
		\centering
		\includegraphics[width=\linewidth]{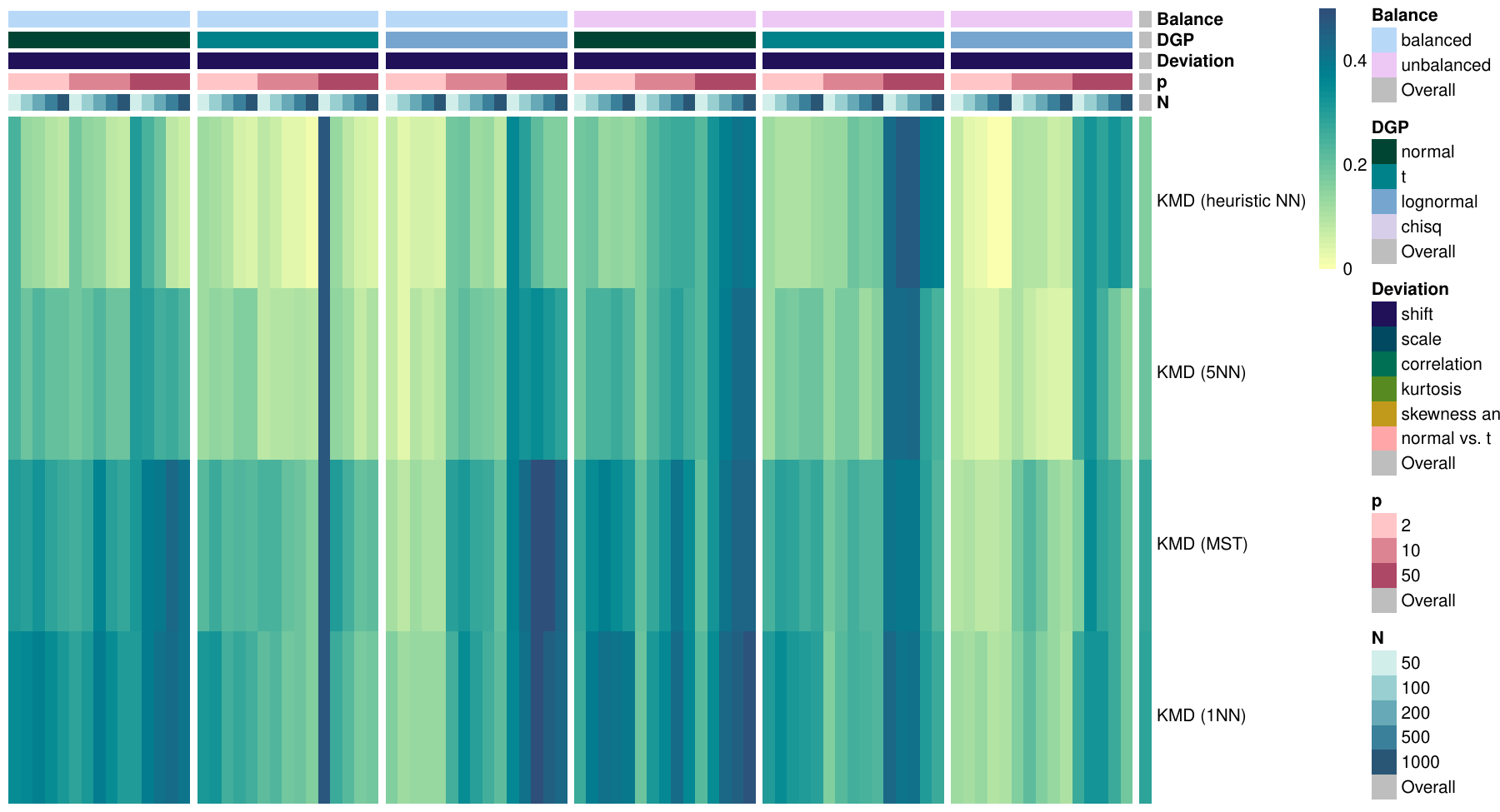}
		\caption{Mean PESR difference to best method per scenario for shift alternatives and $k = 2$ datasets for all variants of the KMD \textcite{huang_kernel_2022}. Selected variant: KMD~(heuristic NN).}
	\end{figure}
	
	\begin{figure}[!h]
		\centering
		\includegraphics[width=\linewidth]{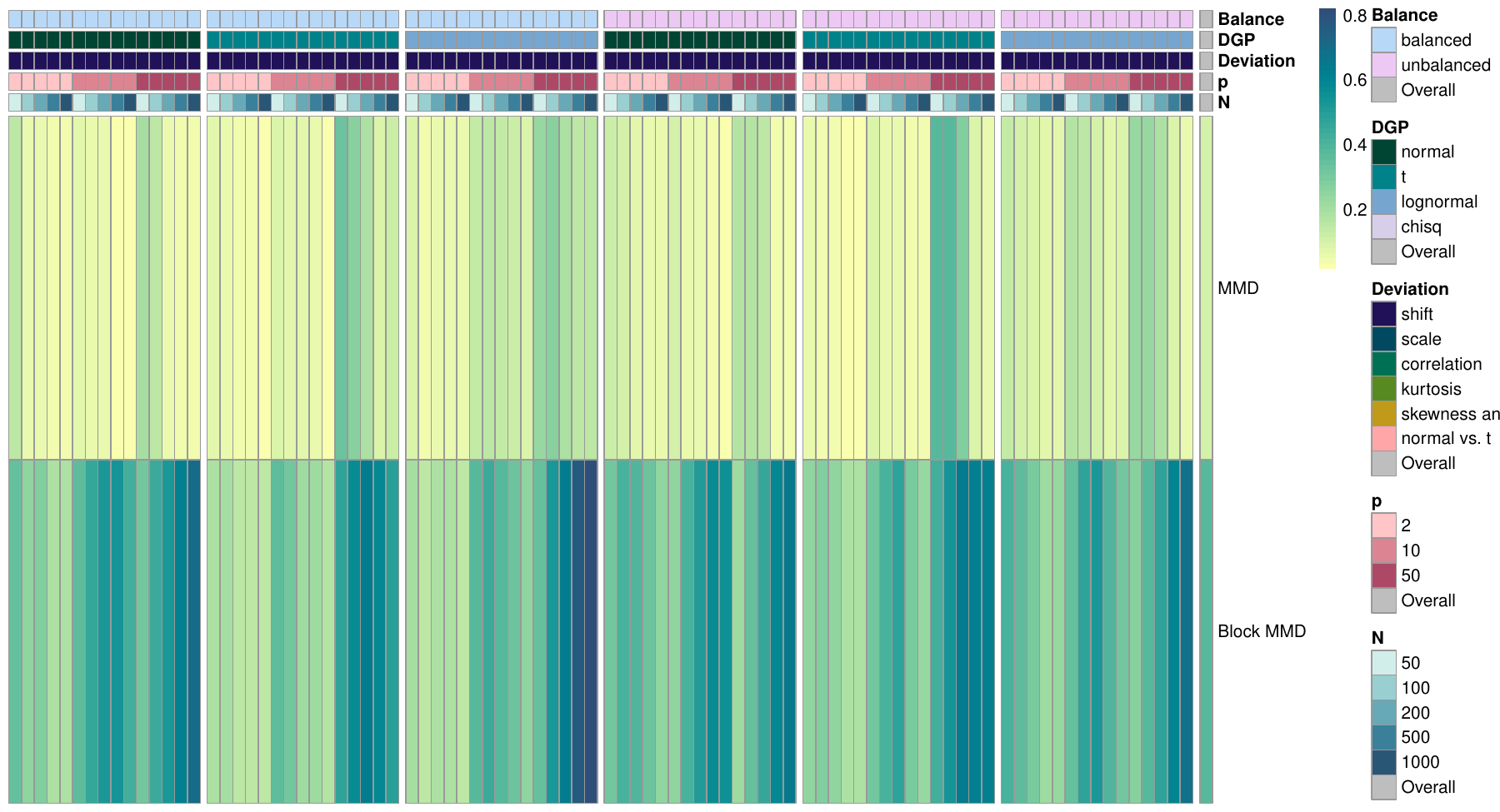}
		\caption{Mean PESR difference to best method per scenario for shift alternatives and $k = 2$ datasets for all variants of the MMD \textcite{gretton_kernel_2006, zaremba_b-test_2013}. Selected variant: MMD.}
	\end{figure}
	
	\begin{figure}[!h]
		\centering
		\includegraphics[width=\linewidth]{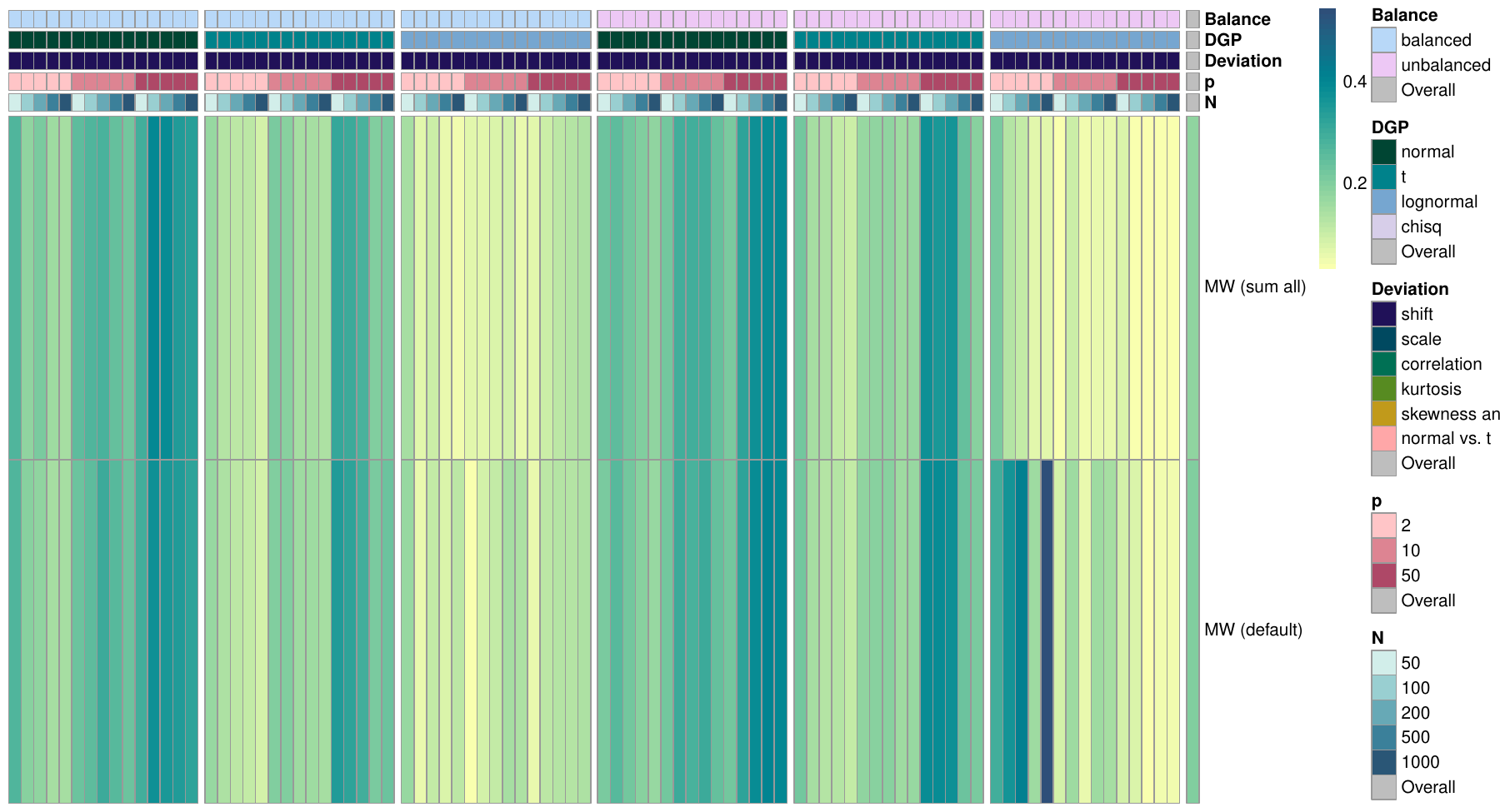}
		\caption{Mean PESR difference to best method per scenario for shift alternatives and $k = 2$ datasets for all variants of the MW method \textcite{mukhopadhyay_nonparametric_2020}. Selected variant: MW~(sum all).}
	\end{figure}
	
	\begin{figure}[!h]
		\centering
		\includegraphics[width=\linewidth]{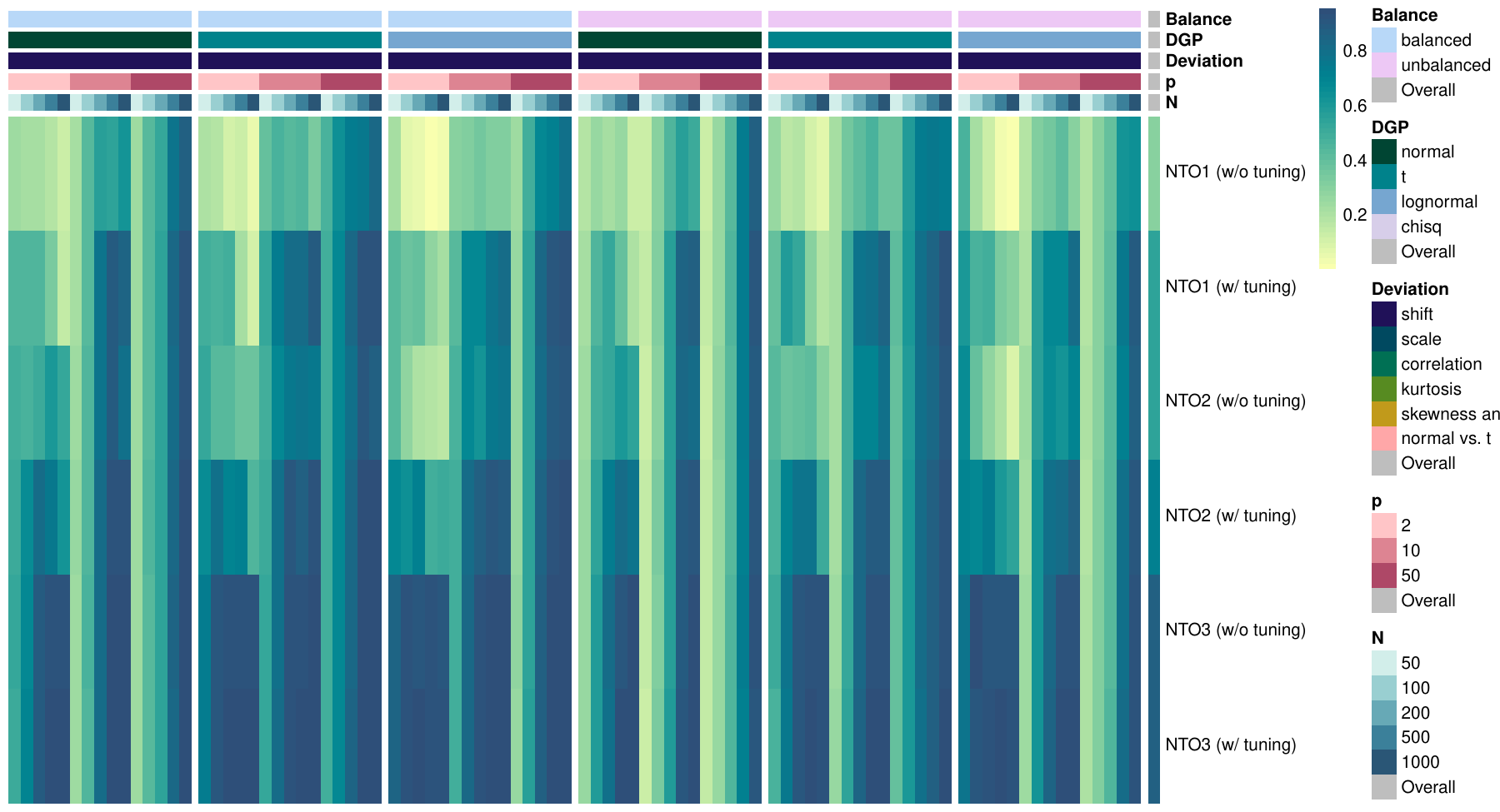}
		\caption{Mean PESR difference to best method per scenario for shift alternatives and $k = 2$ datasets for all variants of the NKT method \textcite{ntoutsi_general_2008}. Selected variant: NTO1~(w/o tuning).}
	\end{figure}
	
	\begin{figure}[!h]
		\centering
		\includegraphics[width=\linewidth]{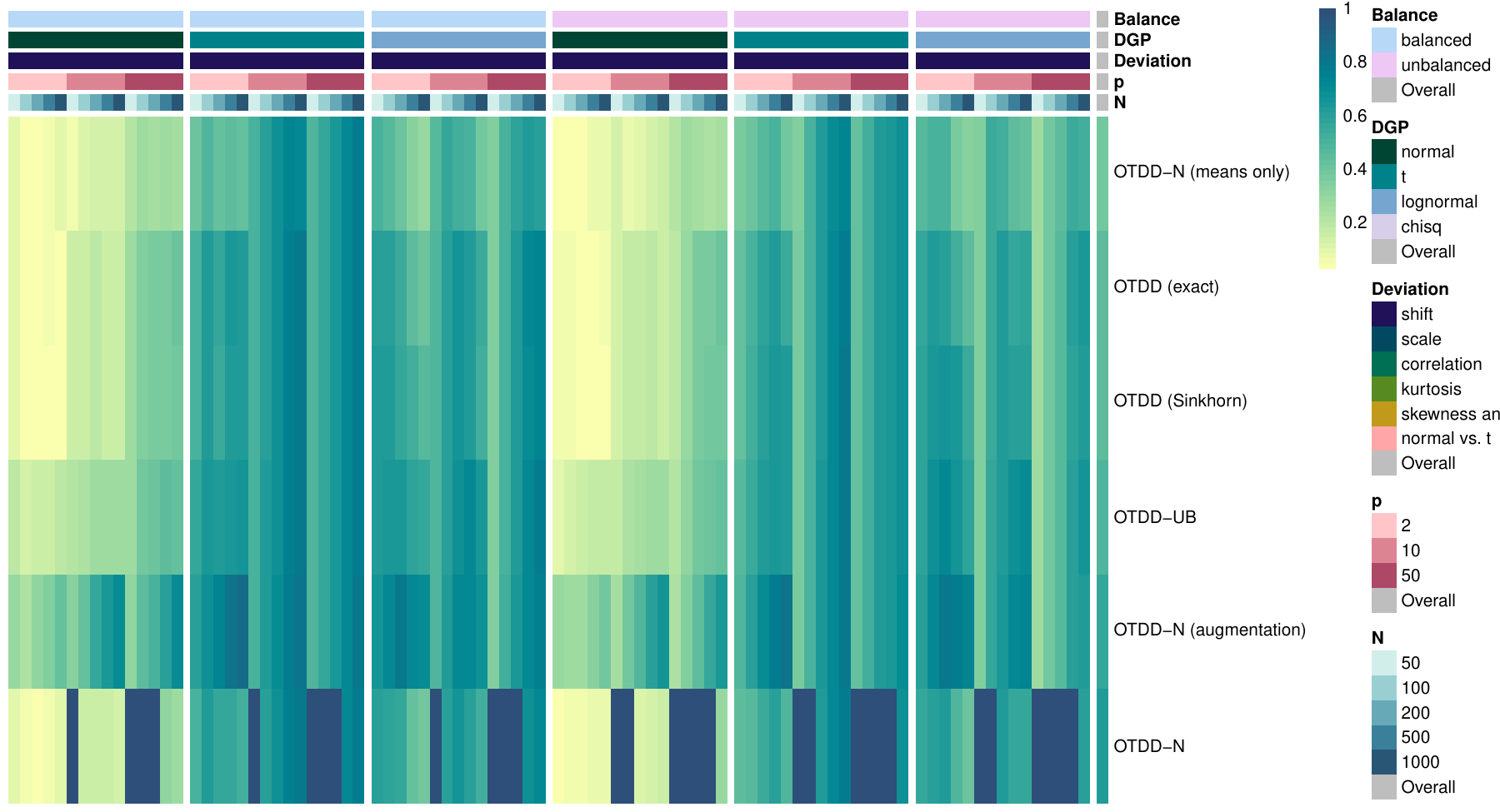}
		\caption{Mean PESR difference to best method per scenario for shift alternatives and $k = 2$ datasets for all variants of the OTDD \textcite{alvarez-melis_geometric_2020}. Selected variant: OTDD-N~(means only).}
	\end{figure}
	
	\begin{figure}[!h]
		\centering
		\includegraphics[width=\linewidth]{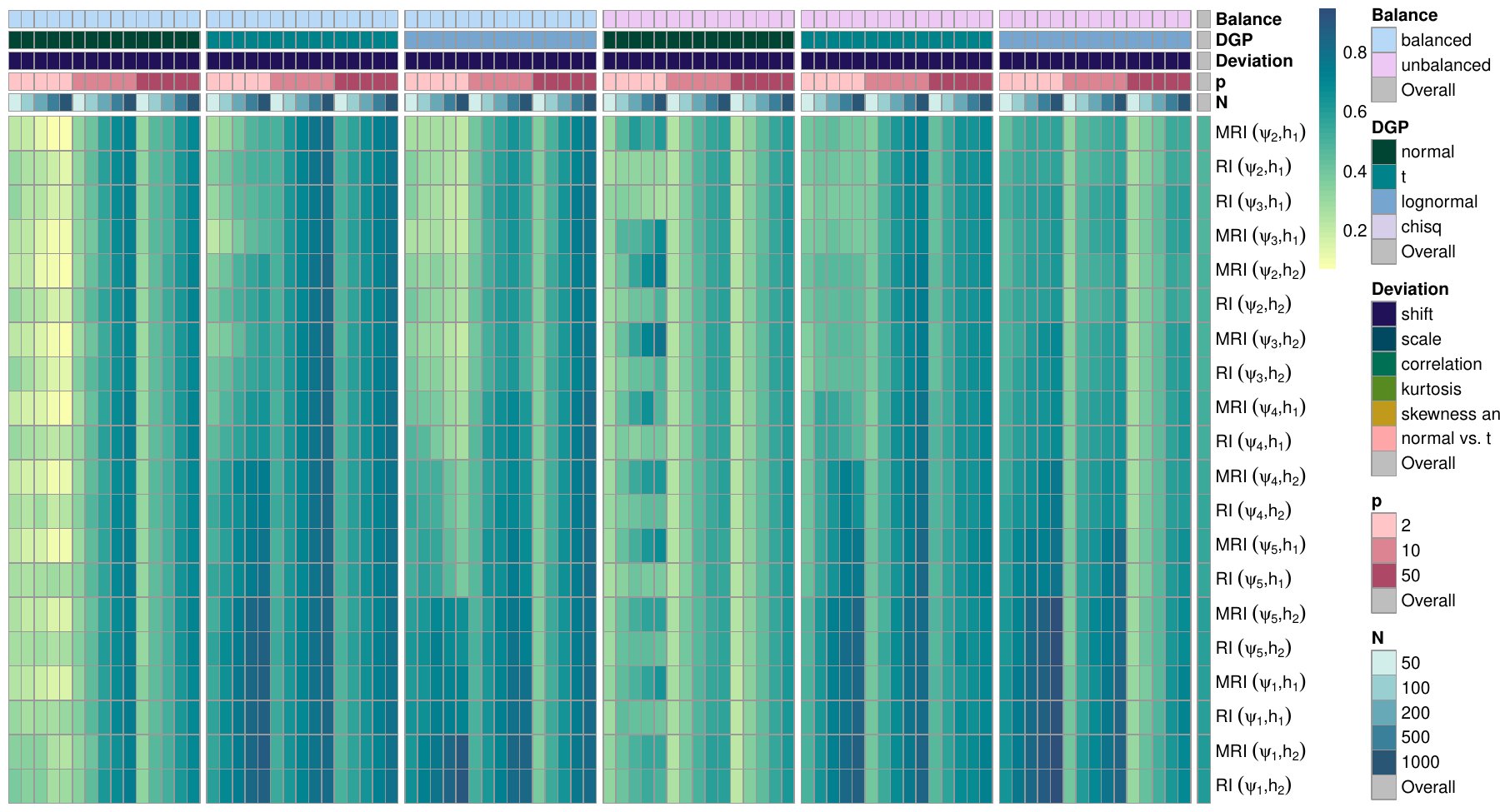}
		\caption{Mean PESR difference to best method per scenario for shift alternatives and $k = 2$ datasets for all variants of the RI test \textcite{paul_clustering-based_2022}. Selected variant: MRI~($\psi_2, h_1$).}
	\end{figure}
	
	\begin{figure}[!h]
		\centering
		\includegraphics[width=\linewidth]{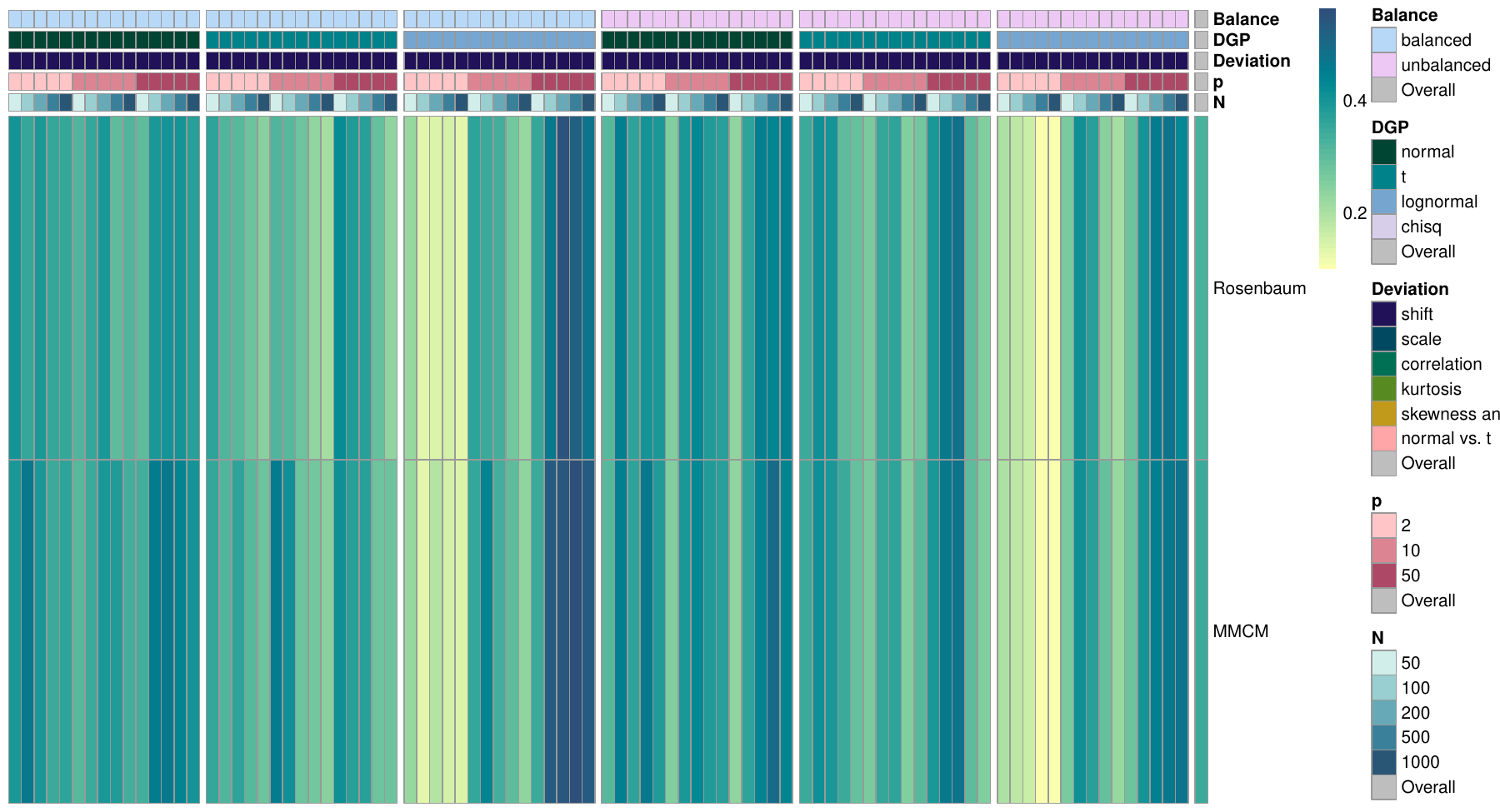}
		\caption{Mean PESR difference to best method per scenario for shift alternatives and $k = 2$ datasets for all variants of the Rosenbaum test \textcite{rosenbaum_exact_2005, mukherjee_distribution-free_2022}. Selected variant: Rosenbaum.}
	\end{figure}
	
	\begin{figure}[!h]
		\centering
		\includegraphics[width=\linewidth]{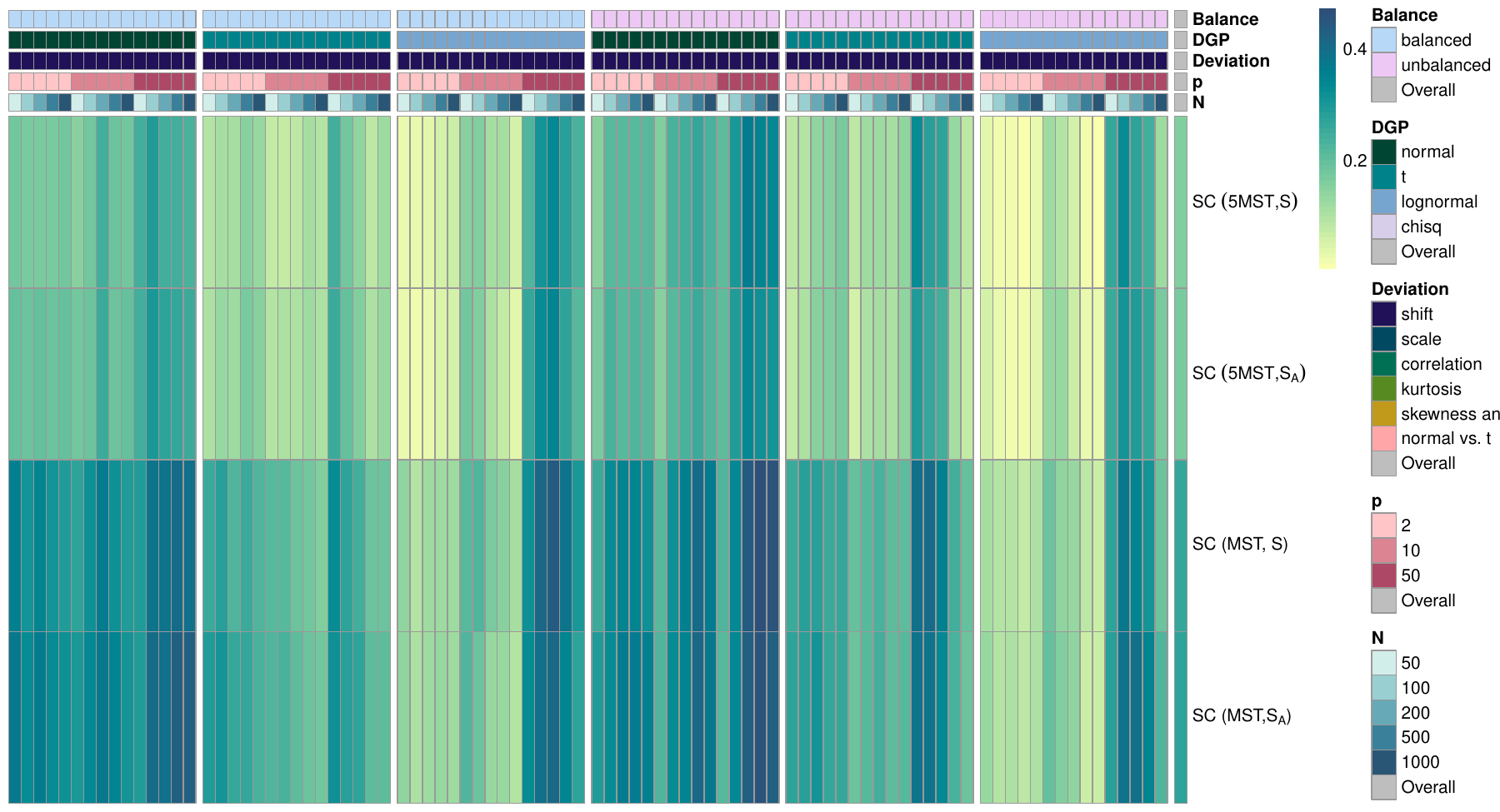}
		\caption{Mean PESR difference to best method per scenario for shift alternatives and $k = 2$ datasets for all variants of the SC test \textcite{song_new_2022}. Selected variant: SC(5MST,S).}
	\end{figure}
	
	\begin{figure}[!h]
		\centering
		\includegraphics[width=\linewidth]{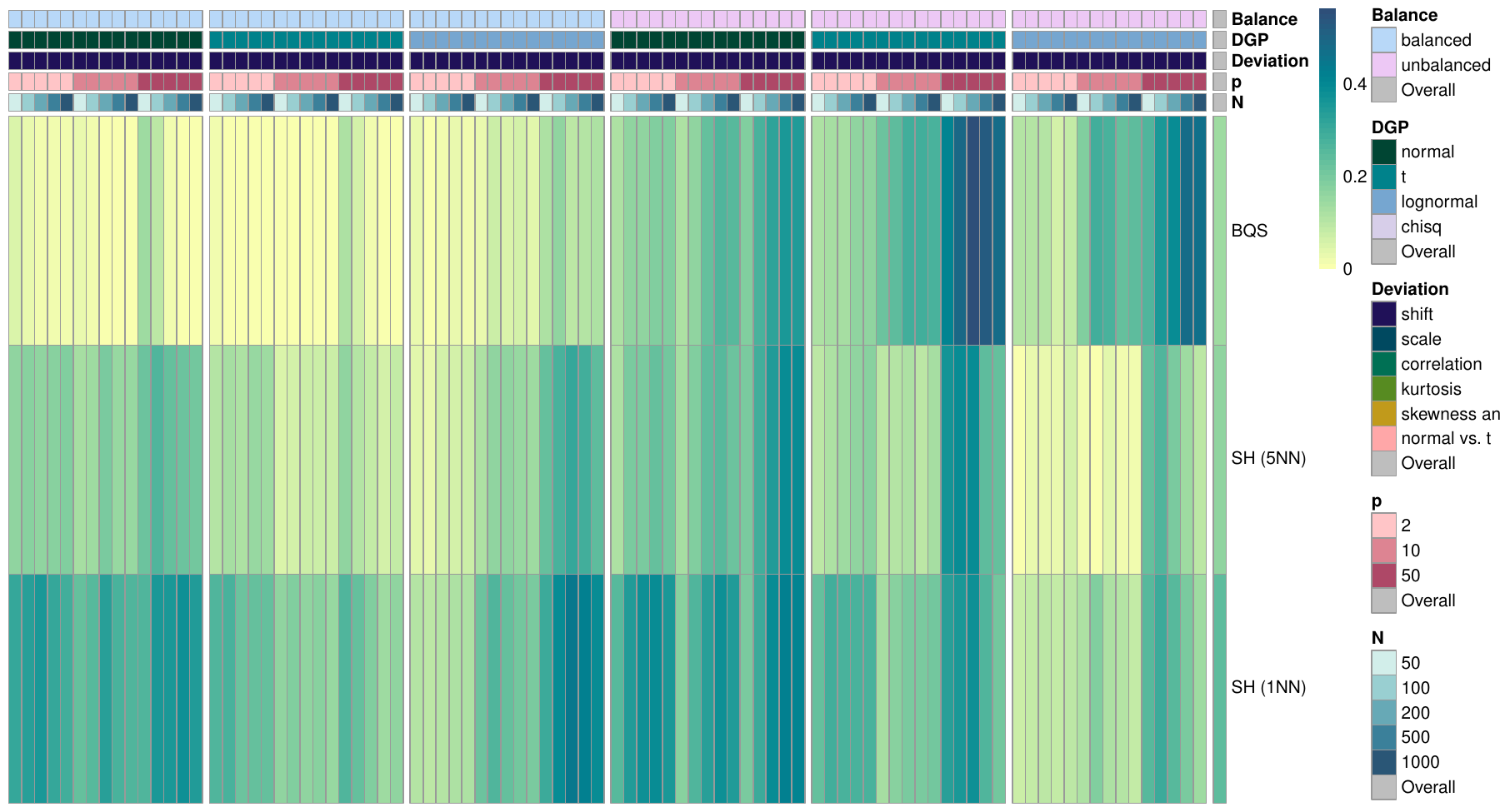}
		\caption{Mean PESR difference to best method per scenario for shift alternatives and $k = 2$ datasets for all variants of the SH test \textcite{schilling_multivariate_1986, henze_multivariate_1988, barakat_multivariate_1996}. Selected variants: BQS, SH(5NN) (complement each other).}
	\end{figure}
	\clearpage
	
	\paragraph{Scale}\mbox{} 
	
	\begin{figure}[!h]
		\centering
		\includegraphics[width=\linewidth]{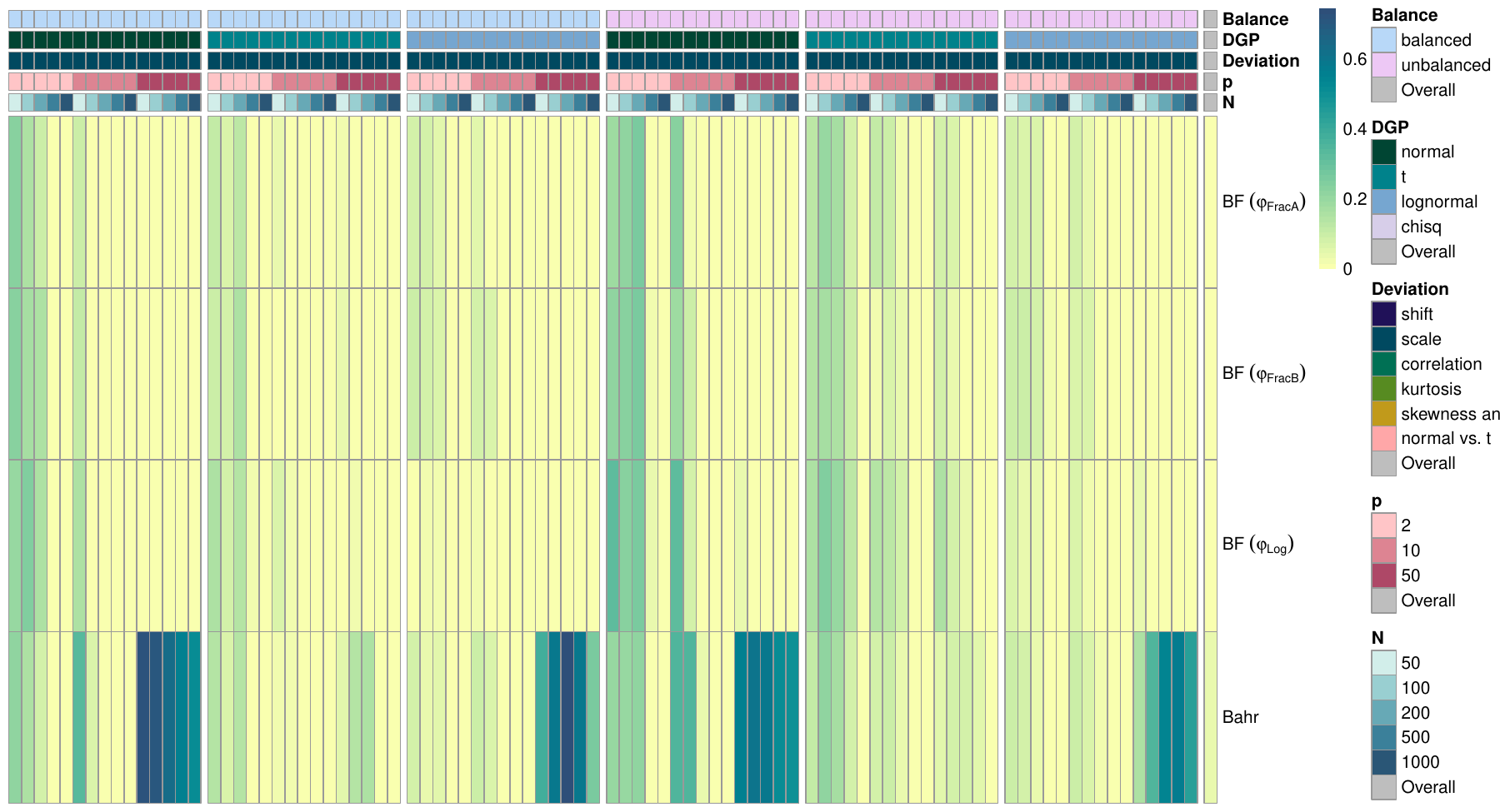}
		\caption{Mean PESR difference to best method per scenario for scale alternatives and $k = 2$ datasets for all variants of the BF method \textcite{baringhaus_rigid_2010}. Selected variant: BF~($\varphi_{\text{FracA}}$).}
	\end{figure}
	
	\begin{figure}[!h]
		\centering
		\includegraphics[width=\linewidth]{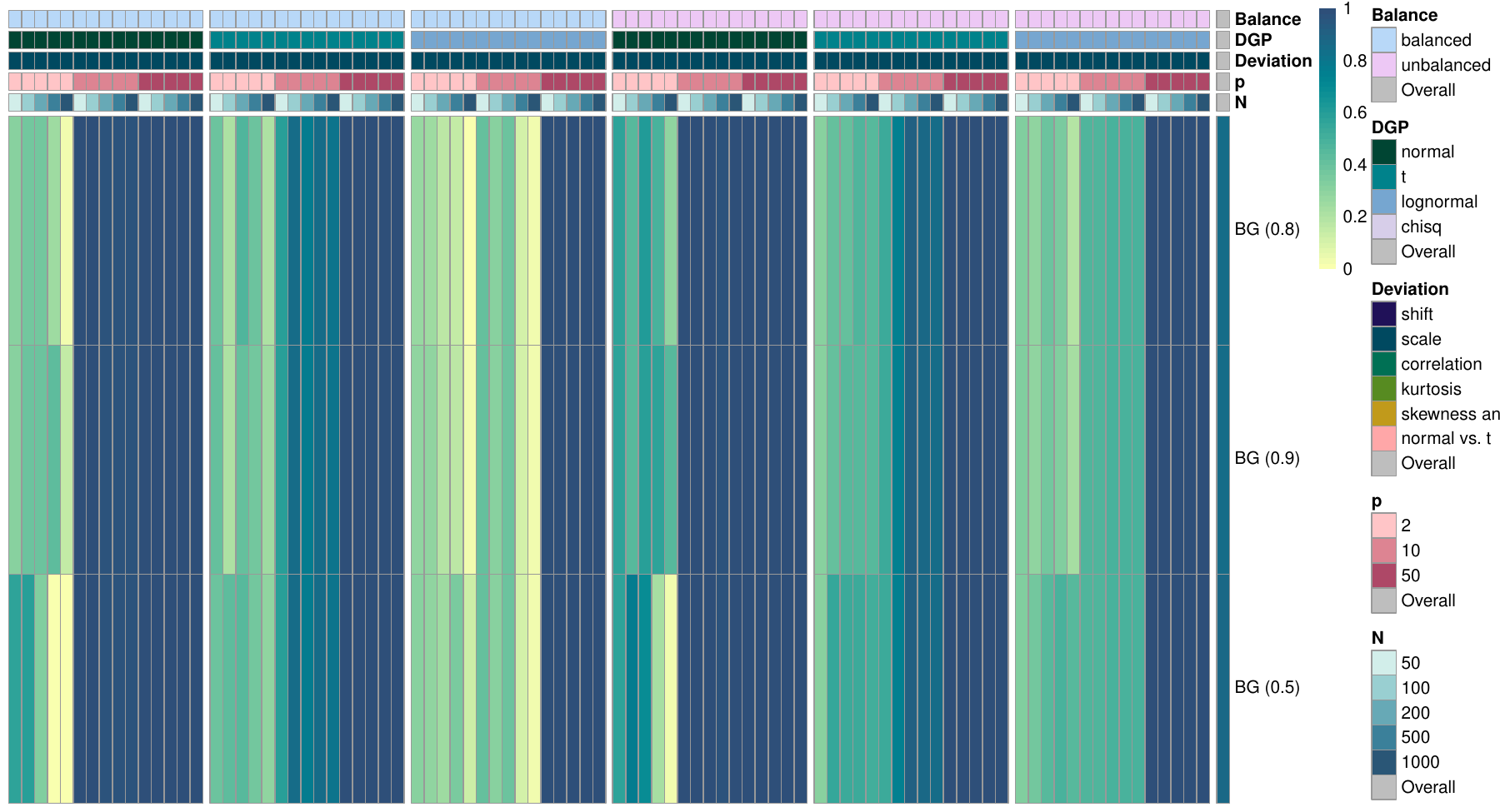}
		\caption{Mean PESR difference to best method per scenario for scale alternatives and $k = 2$ datasets for all variants of the BG method \textcite{biau_asymptotic_2005}. Selected variant: BG~(0.8).}
	\end{figure}
	
	\begin{figure}[!h]
		\centering
		\includegraphics[width=\linewidth]{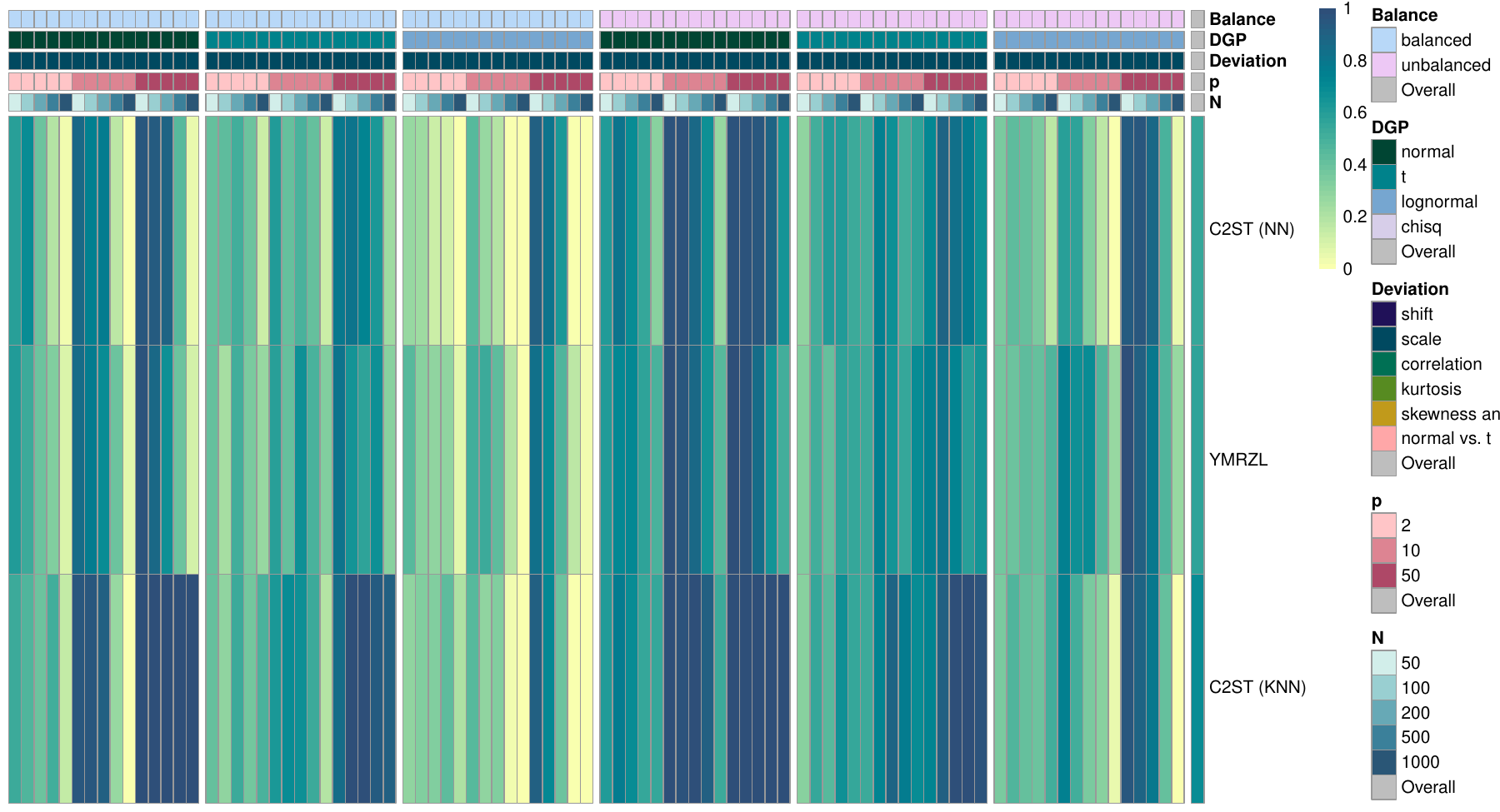}
		\caption{Mean PESR difference to best method per scenario for scale alternatives and $k = 2$ datasets for all variants of the YMRZL and C2ST method \textcite{yu_two-sample_2007, lopez-paz_revisiting_2017}. Selected variant: C2ST~(NN).}
	\end{figure}
	
	\begin{figure}[!h]
		\centering
		\includegraphics[width=\linewidth]{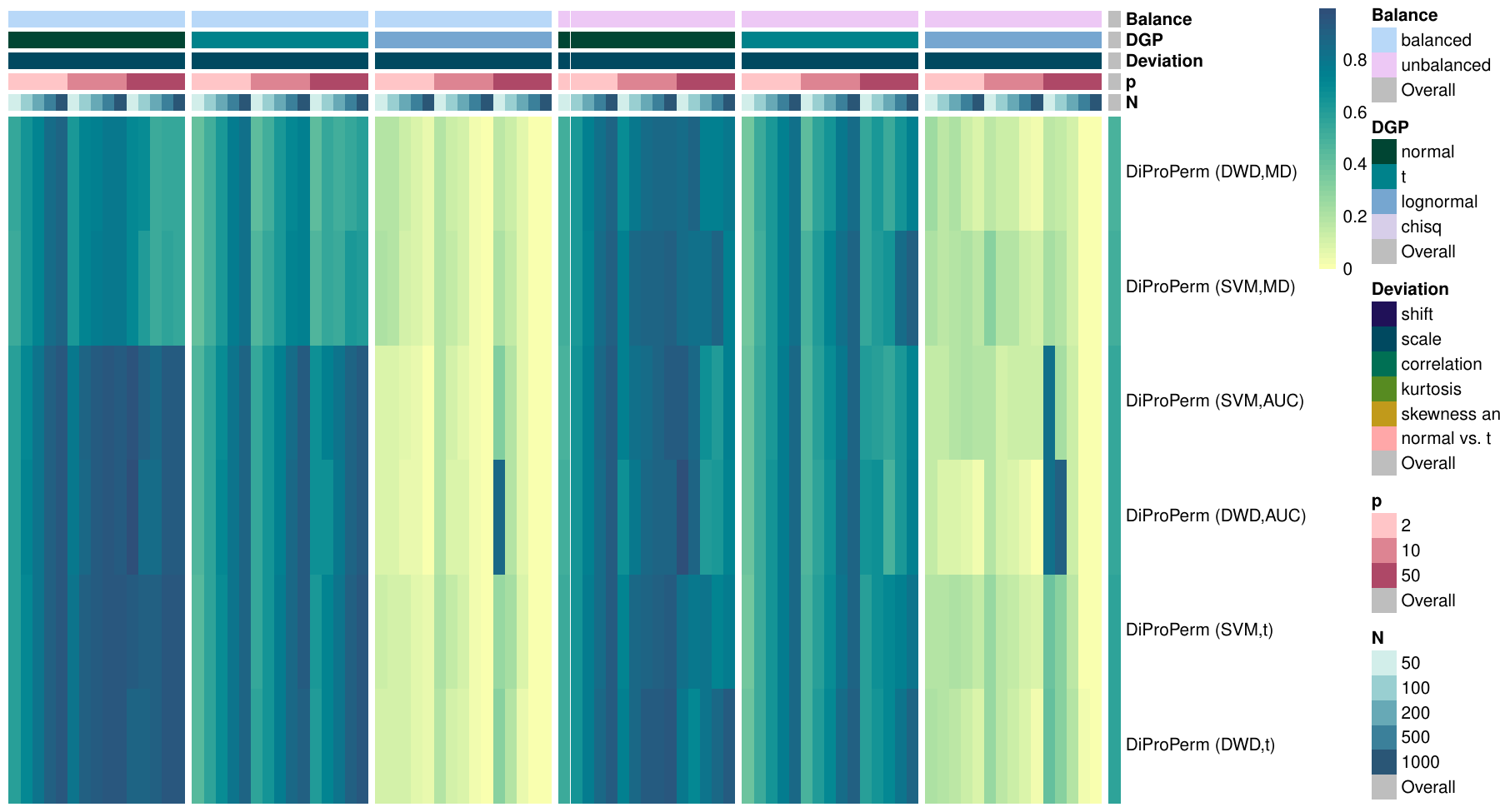}
		\caption{Mean PESR difference to best method per scenario for scale alternatives and $k = 2$ datasets for all variants of the DiProPerm method \textcite{wei_direction-projection-permutation_2016}. Selected variant: DiProPerm~(DWD,MD).}
	\end{figure}
	
	\begin{figure}[!h]
		\centering
		\includegraphics[width=\linewidth]{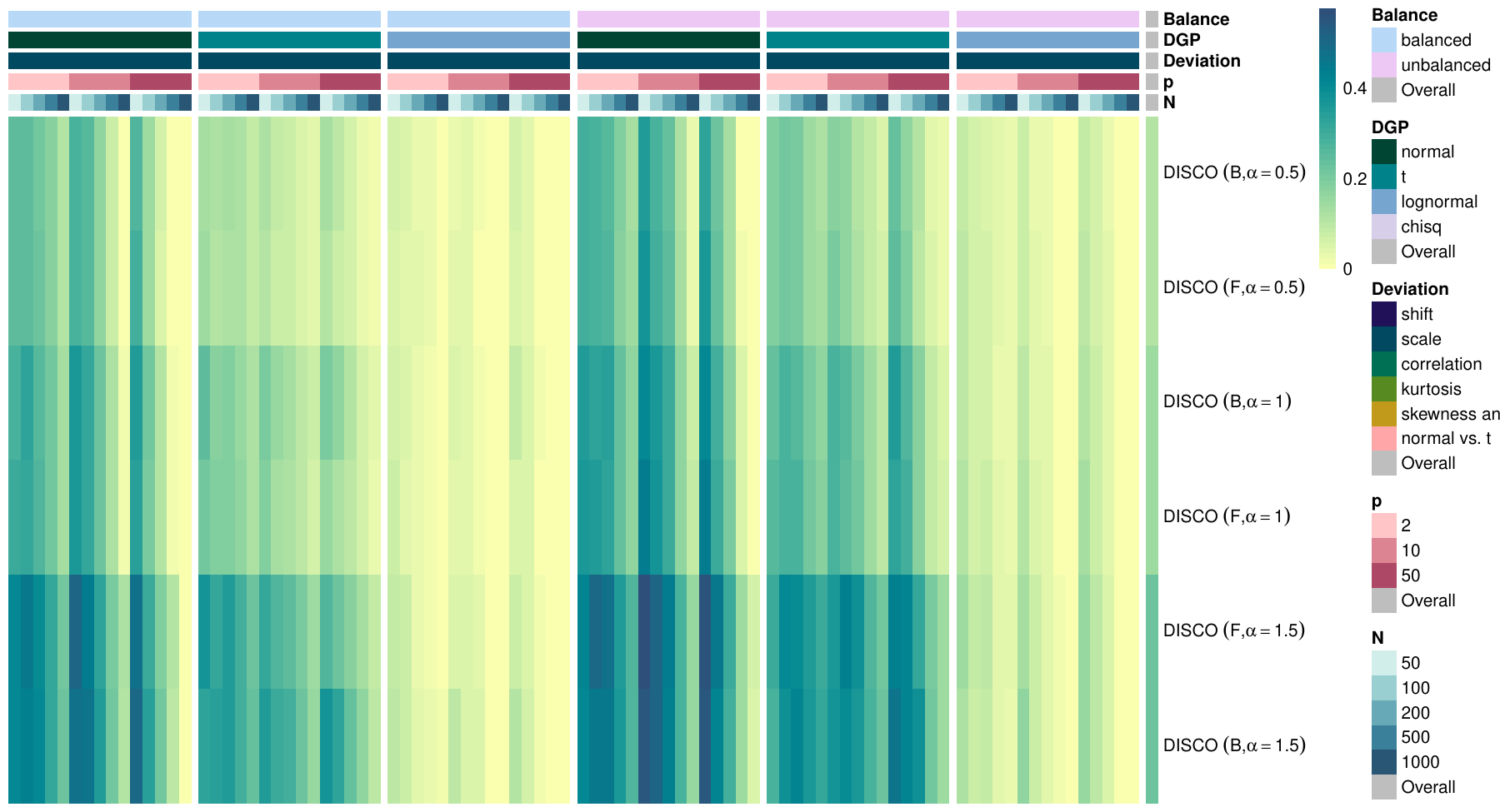}
		\caption{Mean PESR difference to best method per scenario for scale alternatives and $k = 2$ datasets for all variants of the DISCO method \textcite{rizzo_disco_2010}. Selected variant: DISCO~(B,$\alpha=0.5$).}
	\end{figure}
	
	\begin{figure}[!h]
		\centering
		\includegraphics[width=\linewidth]{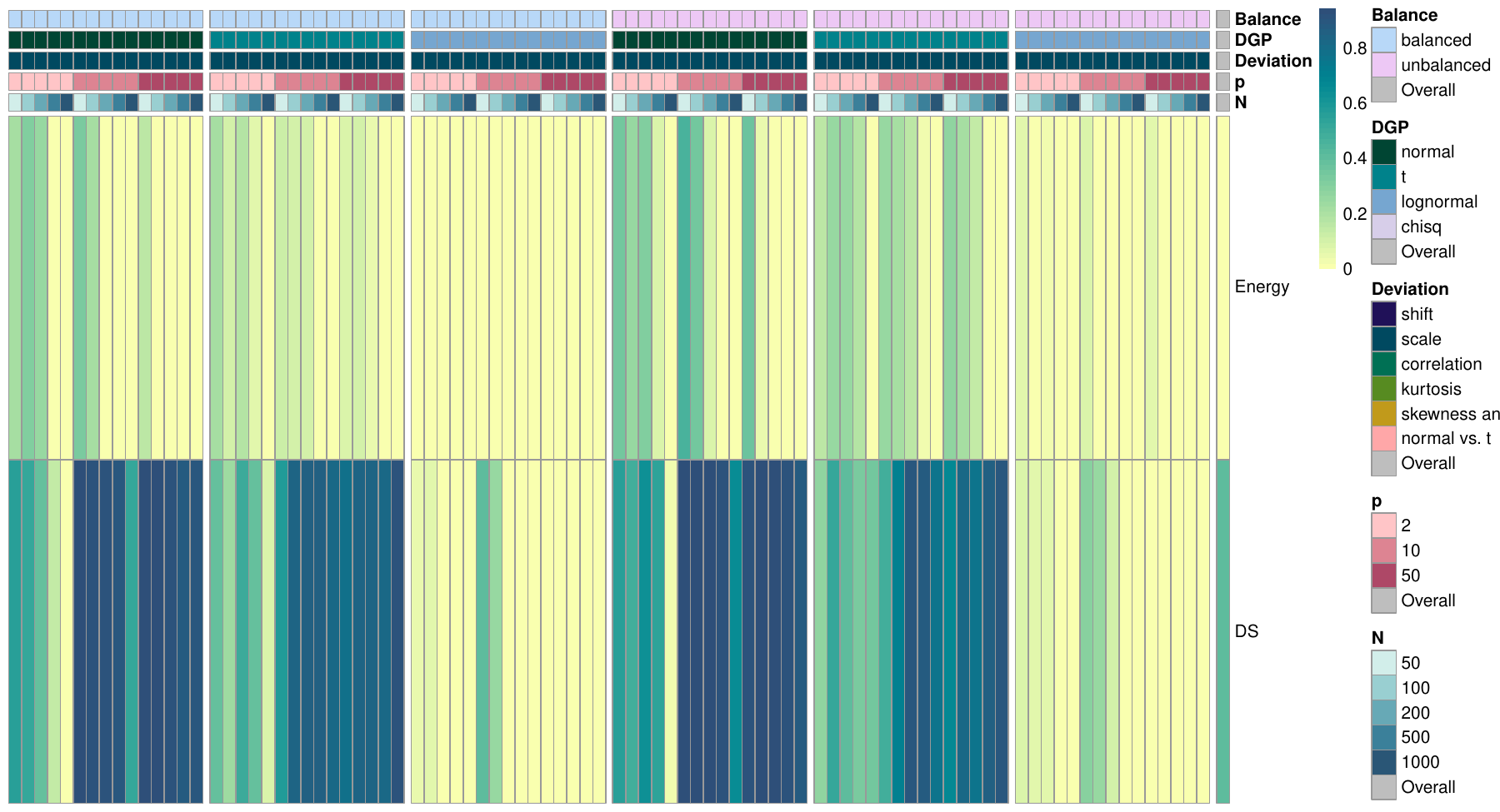}
		\caption{Mean PESR difference to best method per scenario for scale alternatives and $k = 2$ datasets for all variants of the Energy method \textcite{szekely_testing_2004}. Selected variant: Energy.}
	\end{figure}
	
	\begin{figure}[!h]
		\centering
		\includegraphics[width=\linewidth]{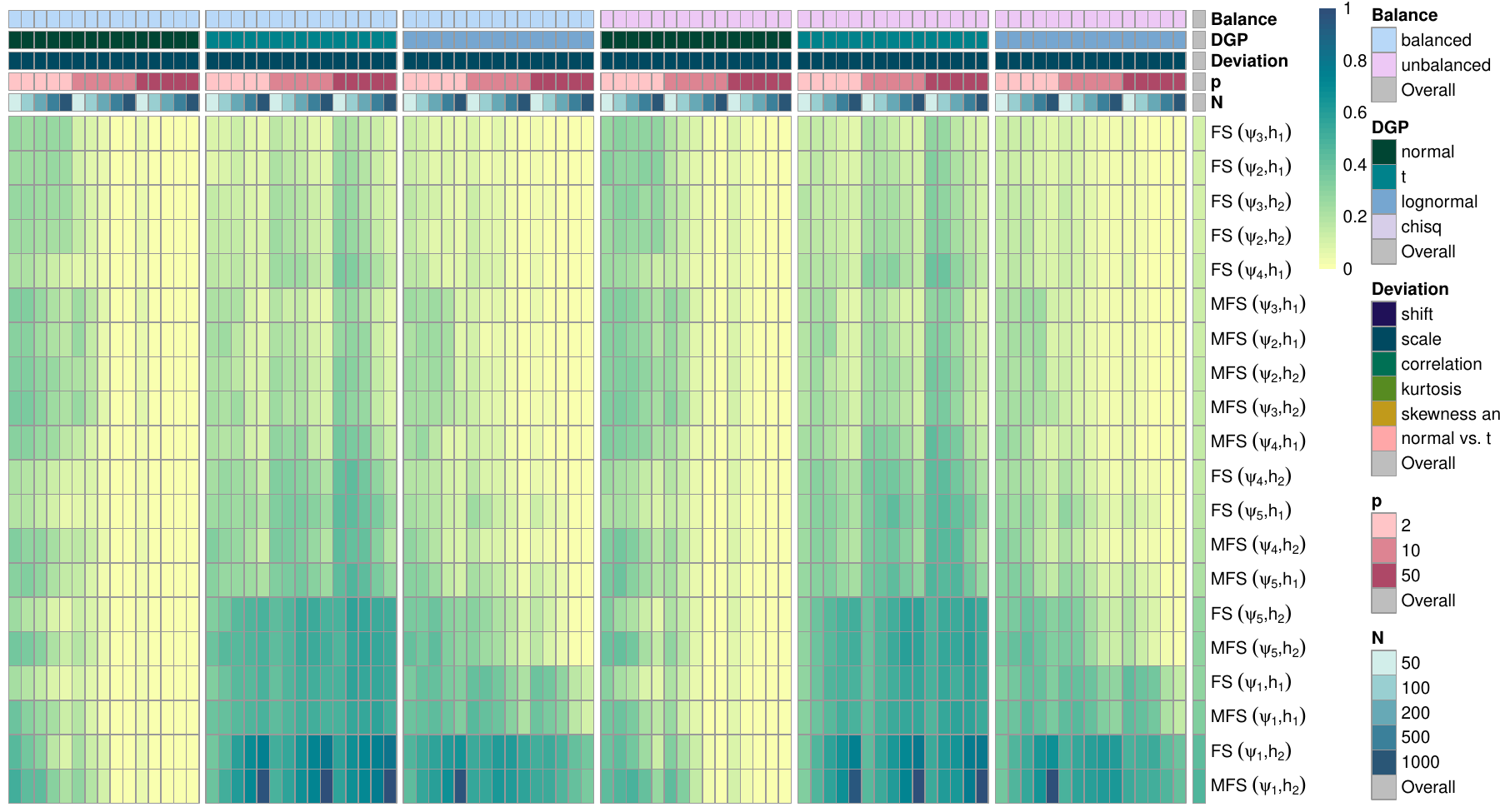}
		\caption{Mean PESR difference to best method per scenario for scale alternatives and $k = 2$ datasets for all variants of the FS test \textcite{paul_clustering-based_2022}. Selected variant: FS~($\psi_3, h_1$).}
	\end{figure}
	
	\begin{figure}[!h]
		\centering
		\includegraphics[width=\linewidth]{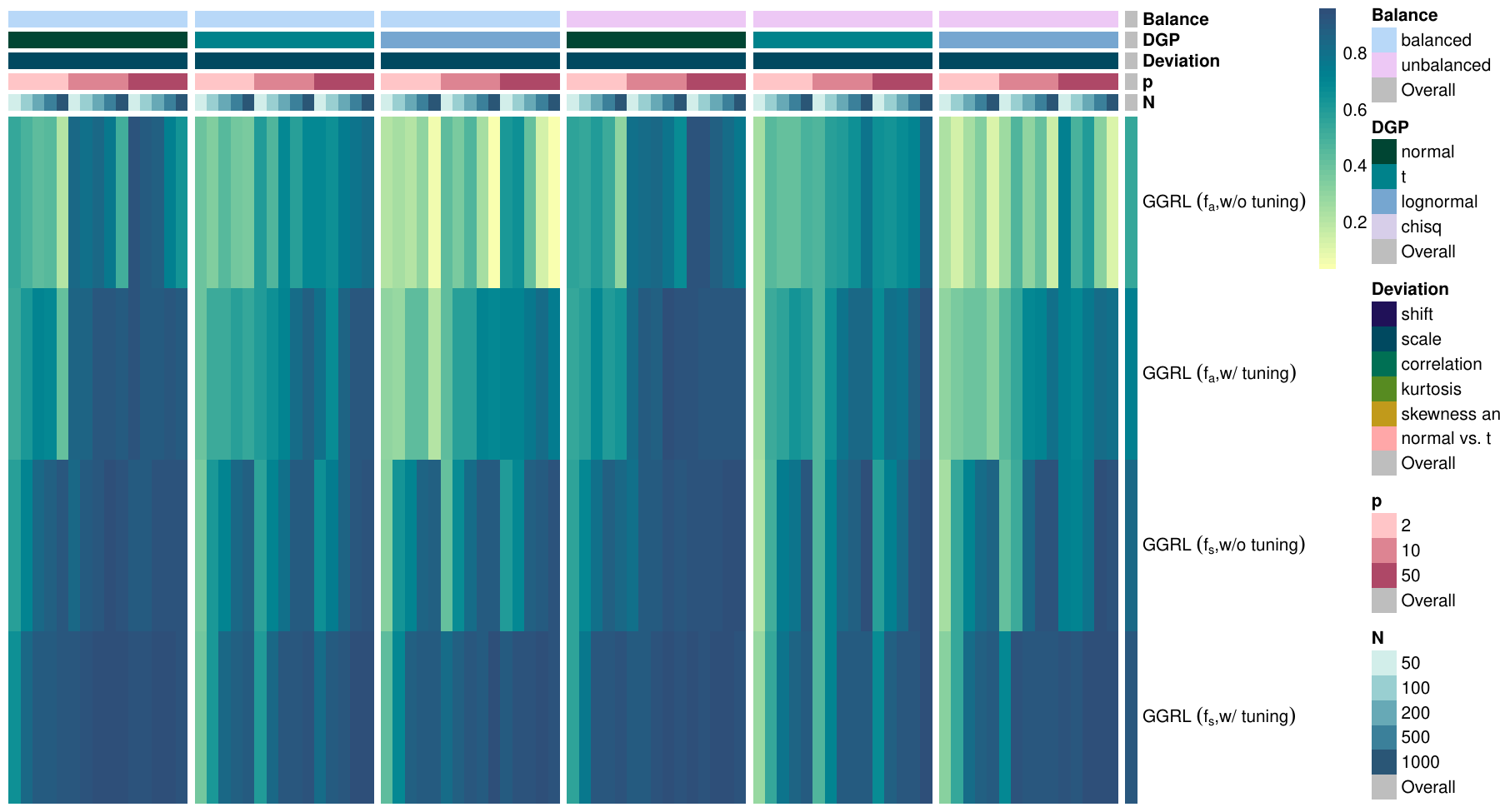}
		\caption{Mean PESR difference to best method per scenario for scale alternatives and $k = 2$ datasets for all variants of the GGRL method \textcite{ganti_framework_1999}. Selected variant: GGRL~($f_a$,w/o~tuning).}
	\end{figure}
	
	\begin{figure}[!h]
		\centering
		\includegraphics[width=\linewidth]{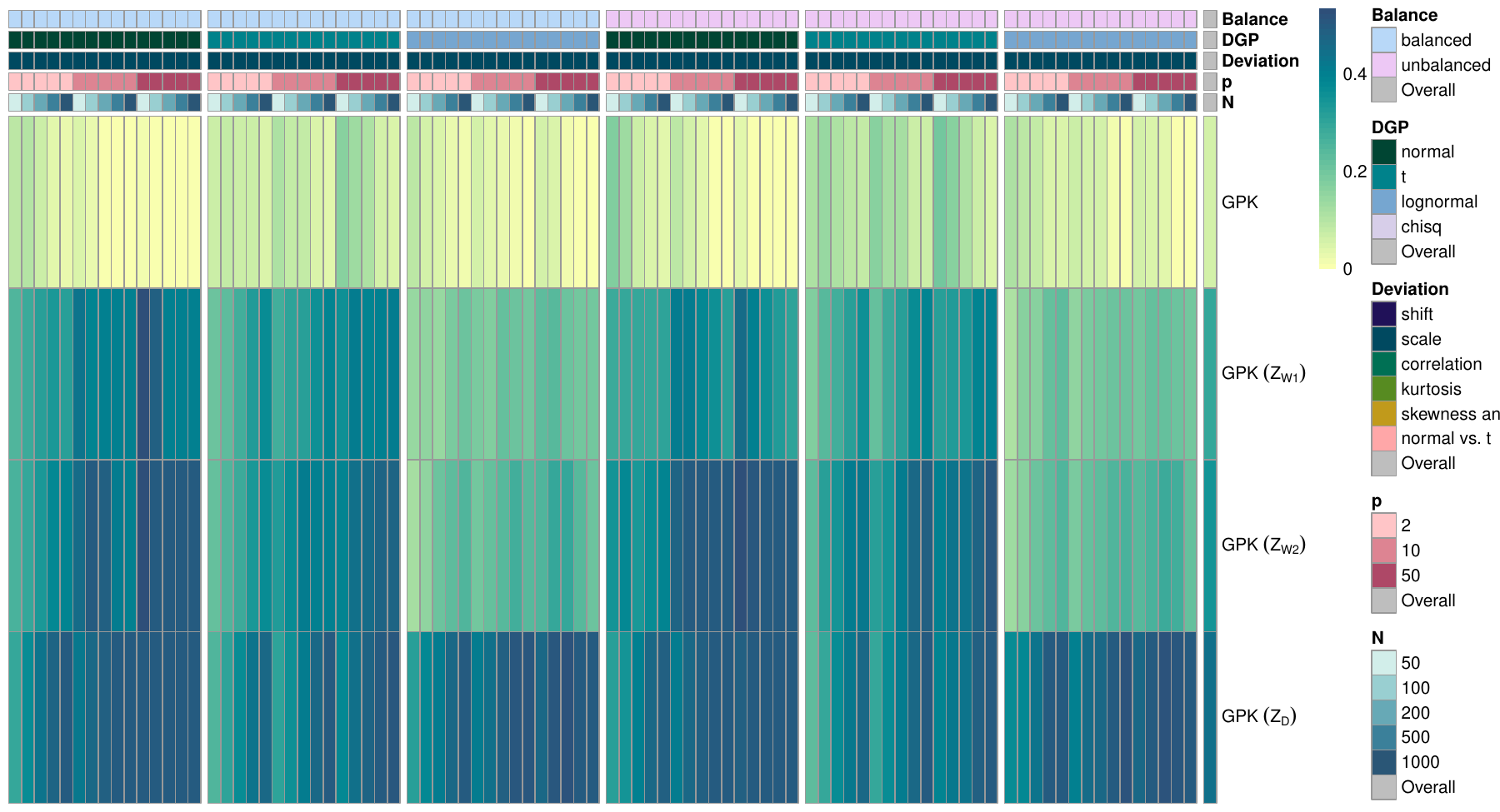}
		\caption{Mean PESR difference to best method per scenario for scale alternatives and $k = 2$ datasets for all variants of the GPK test \textcite{song_generalized_2021}. Selected variant: GPK.}
	\end{figure}
	
	\begin{figure}[!h]
		\centering
		\includegraphics[width=\linewidth]{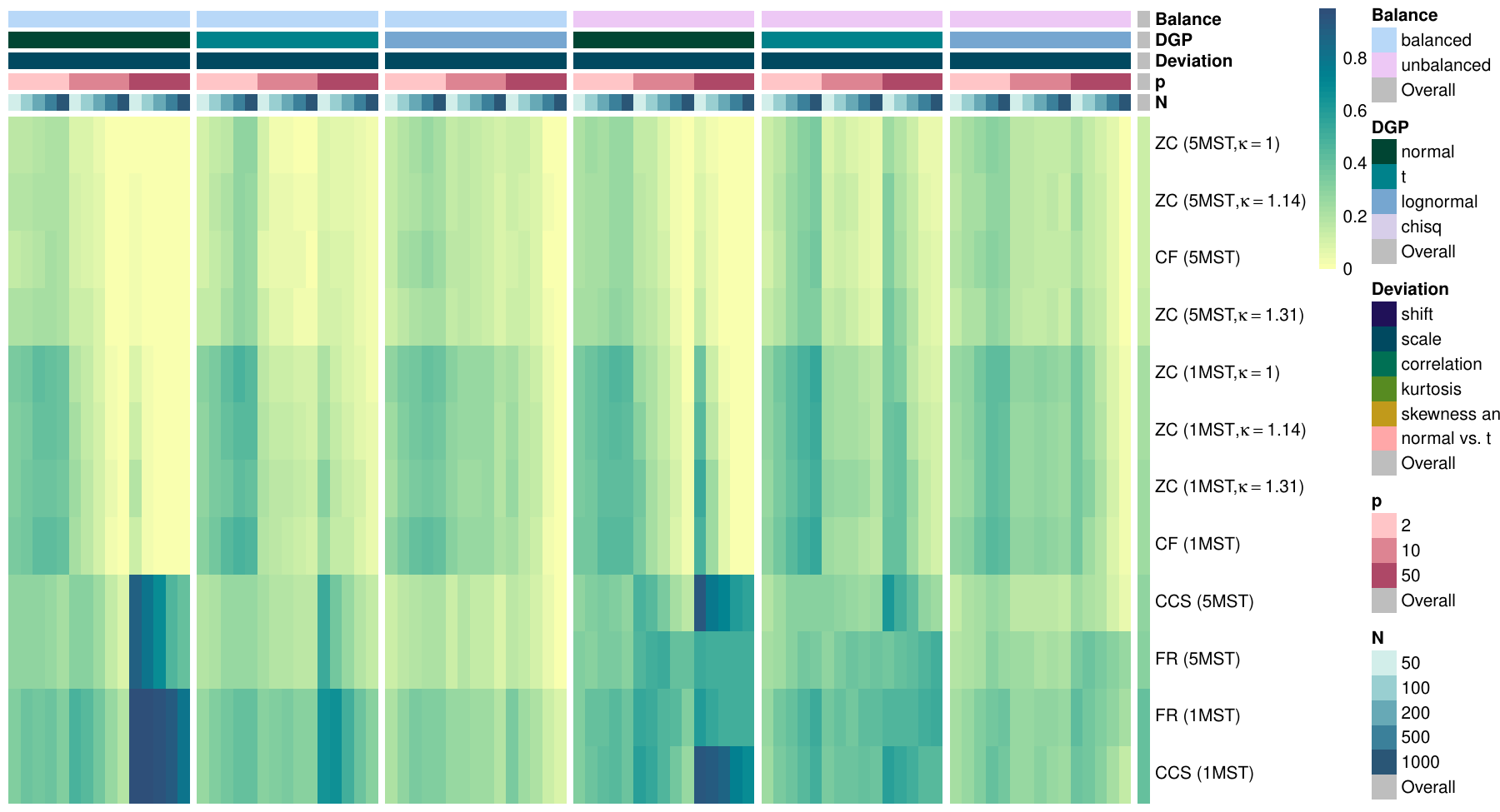}
		\caption{Mean PESR difference to best method per scenario for scale alternatives and $k = 2$ datasets for all variants of the graph-based tests FR, CF, CCS, ZC \textcite{friedman_multivariate_1979, chen_new_2017, chen_weighted_2018, zhang_graph-based_2022}. Selected variant: ZC~(5MST,$\kappa=1$).}
	\end{figure}
	
	\begin{figure}[!h]
		\centering
		\includegraphics[width=\linewidth]{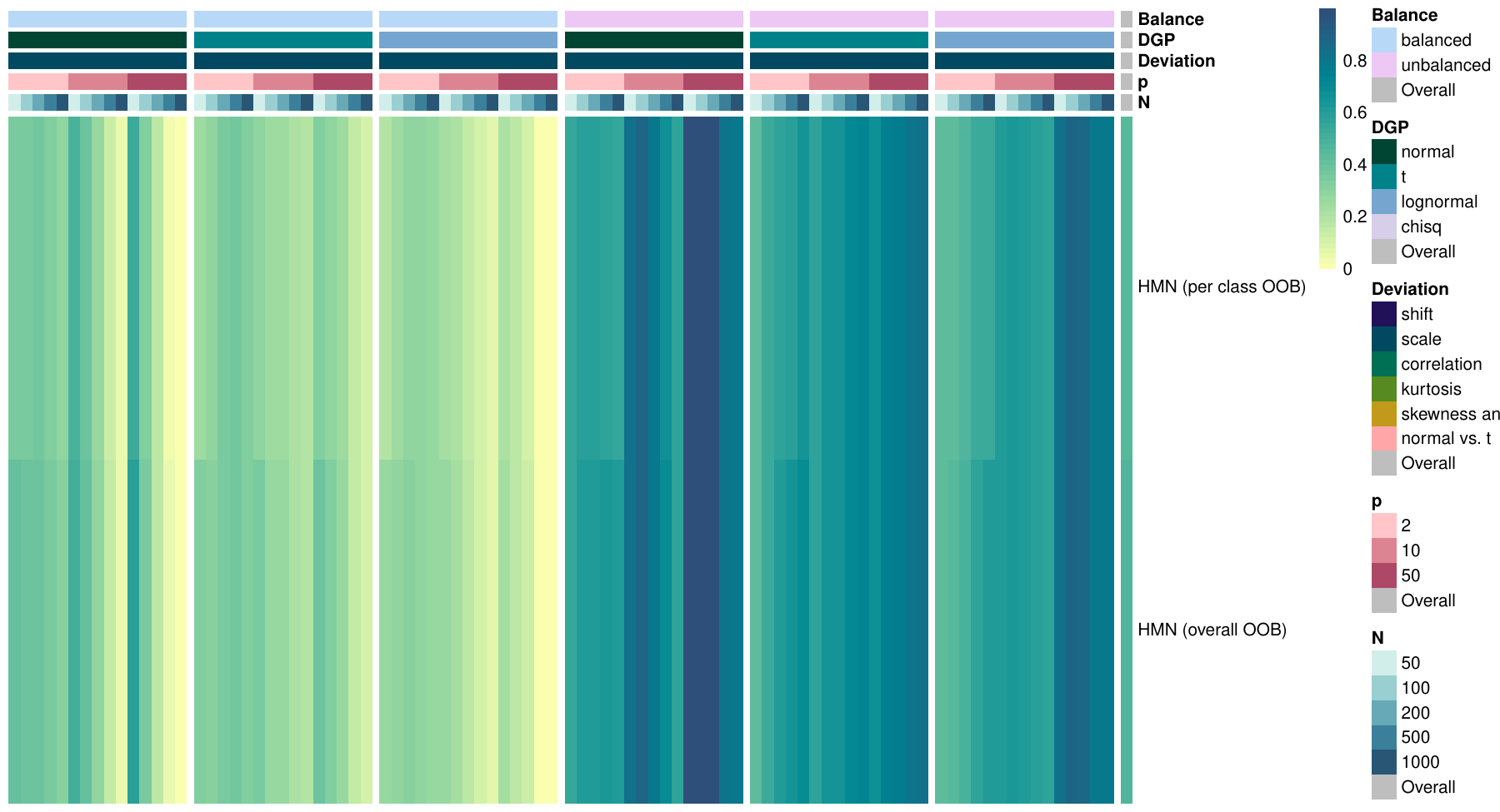}
		\caption{Mean PESR difference to best method per scenario for scale alternatives and $k = 2$ datasets for all variants of the HMN method \textcite{hediger_use_2021}. Selected variant: HMN~(per class OOB).}
	\end{figure}
	
	\begin{figure}[!h]
		\centering
		\includegraphics[width=\linewidth]{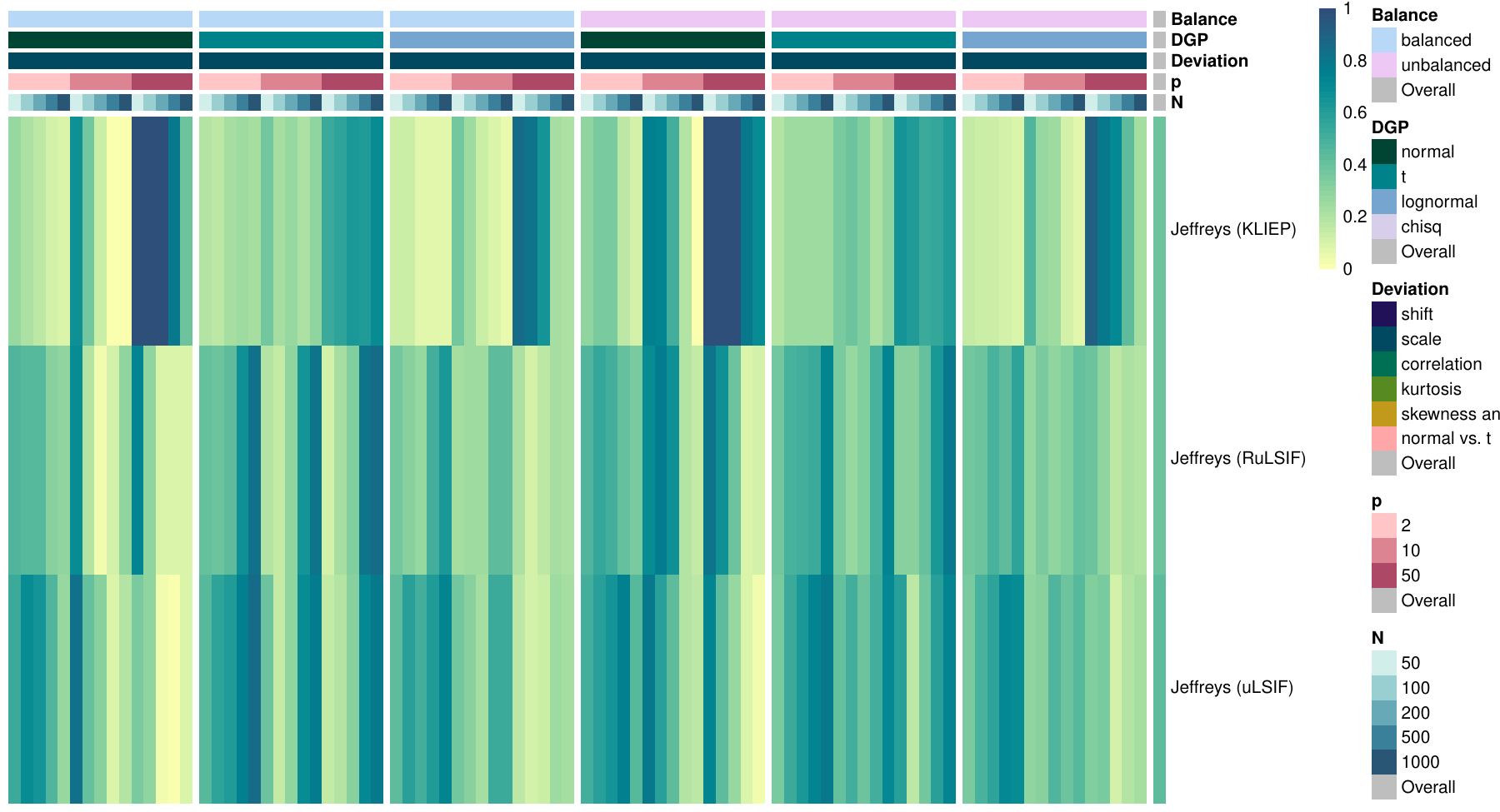}
		\caption{Mean PESR difference to best method per scenario for scale alternatives and $k = 2$ datasets for all variants of the Jeffreys divergence \textcite{jeffreys_invariant_1997}. Selected variant: Jeffreys~(KLIEP).}
	\end{figure}
	
	\begin{figure}[!h]
		\centering
		\includegraphics[width=\linewidth]{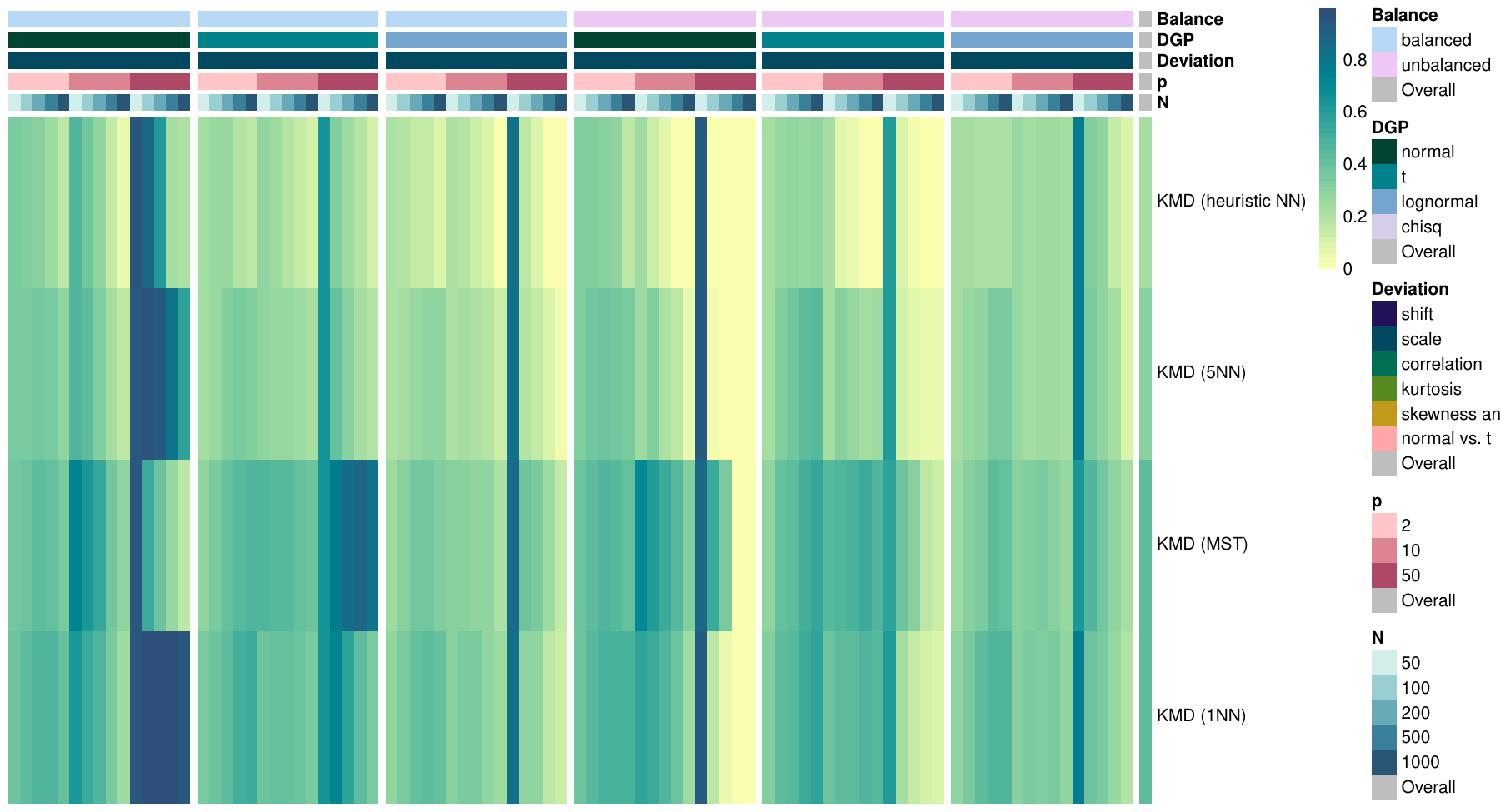}
		\caption{Mean PESR difference to best method per scenario for scale alternatives and $k = 2$ datasets for all variants of the KMD \textcite{huang_kernel_2022}. Selected variant: KMD~(heuristic NN).}
	\end{figure}
	
	\begin{figure}[!h]
		\centering
		\includegraphics[width=\linewidth]{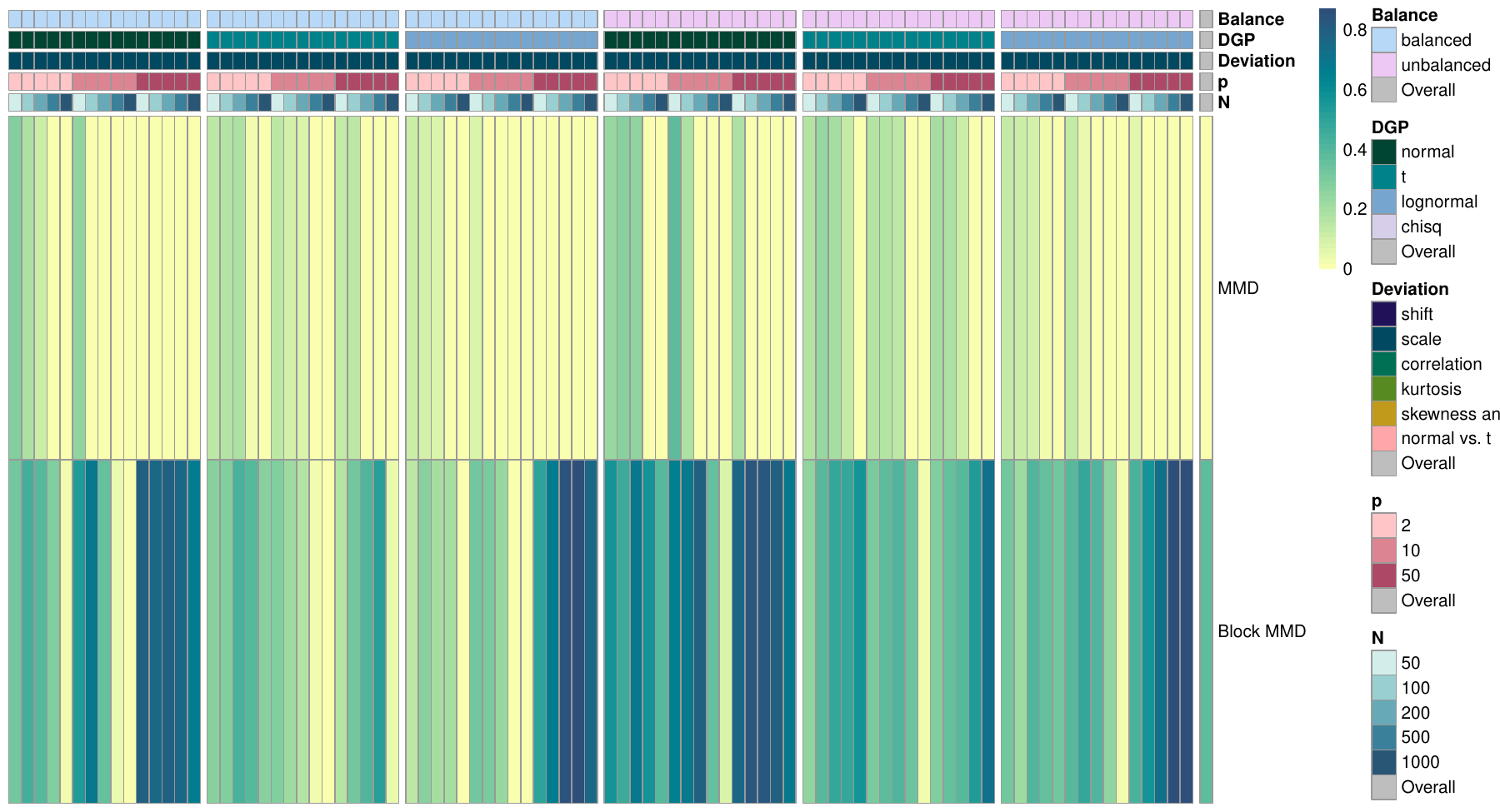}
		\caption{Mean PESR difference to best method per scenario for scale alternatives and $k = 2$ datasets for all variants of the MMD \textcite{gretton_kernel_2006, zaremba_b-test_2013}. Selected variant: MMD.}
	\end{figure}
	
	\begin{figure}[!h]
		\centering
		\includegraphics[width=\linewidth]{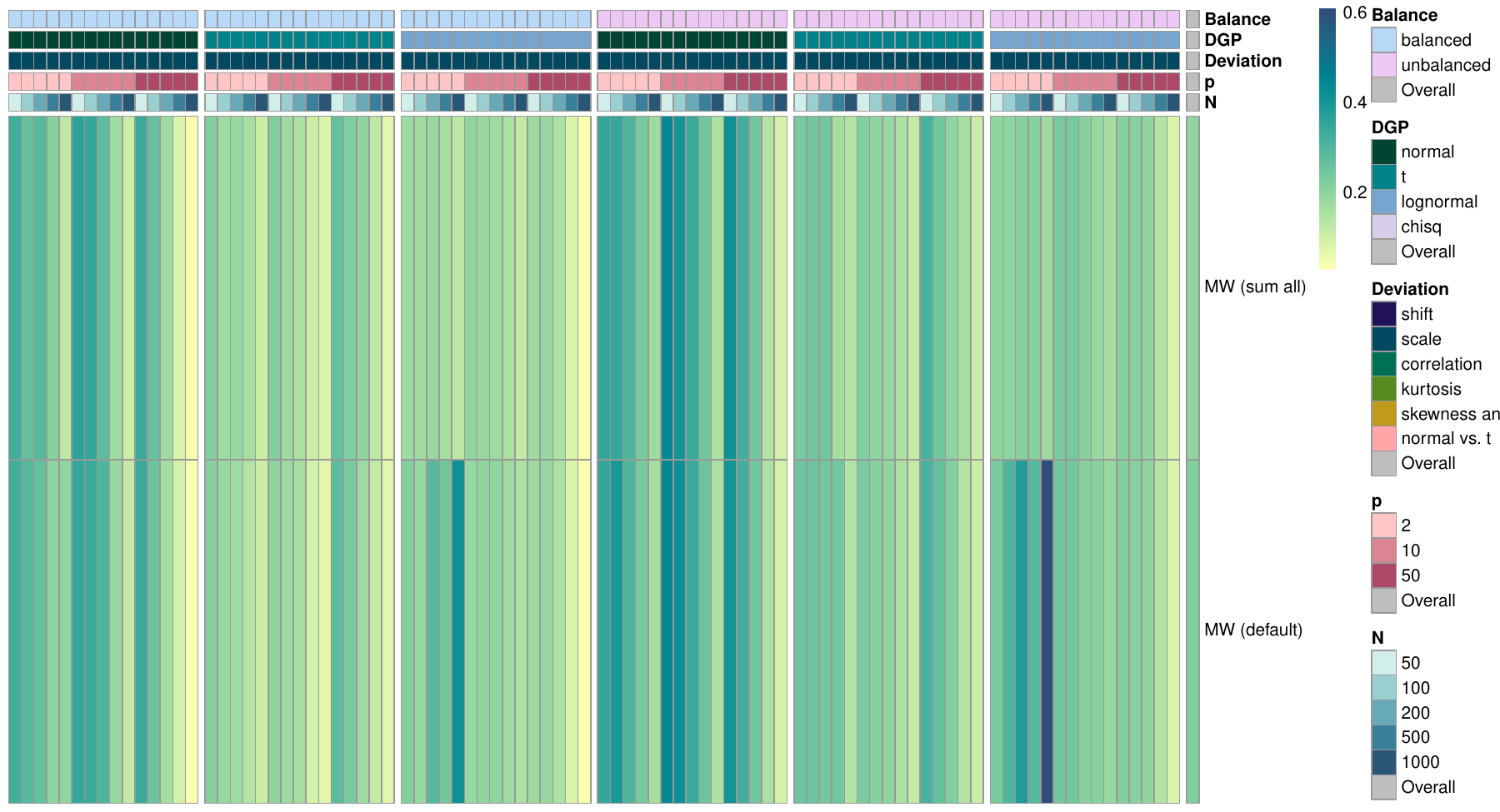}
		\caption{Mean PESR difference to best method per scenario for scale alternatives and $k = 2$ datasets for all variants of the MW method \textcite{mukhopadhyay_nonparametric_2020}. Selected variant: MW~(sum all).}
	\end{figure}
	
	\begin{figure}[!h]
		\centering
		\includegraphics[width=\linewidth]{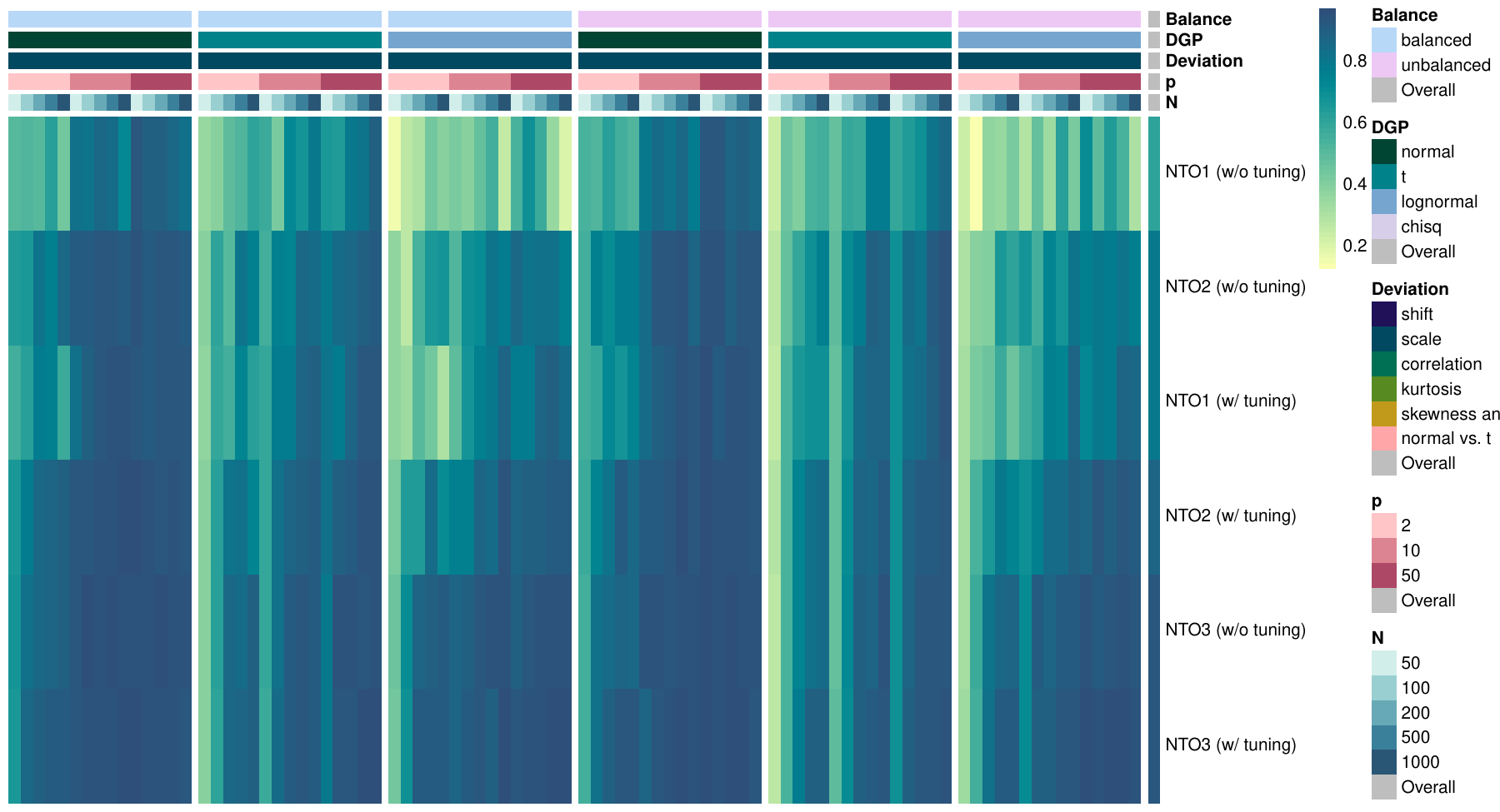}
		\caption{Mean PESR difference to best method per scenario for scale alternatives and $k = 2$ datasets for all variants of the NKT method \textcite{ntoutsi_general_2008}. Selected variant: NTO1~(w/o~tuning).}
	\end{figure}
	
	\begin{figure}[!h]
		\centering
		\includegraphics[width=\linewidth]{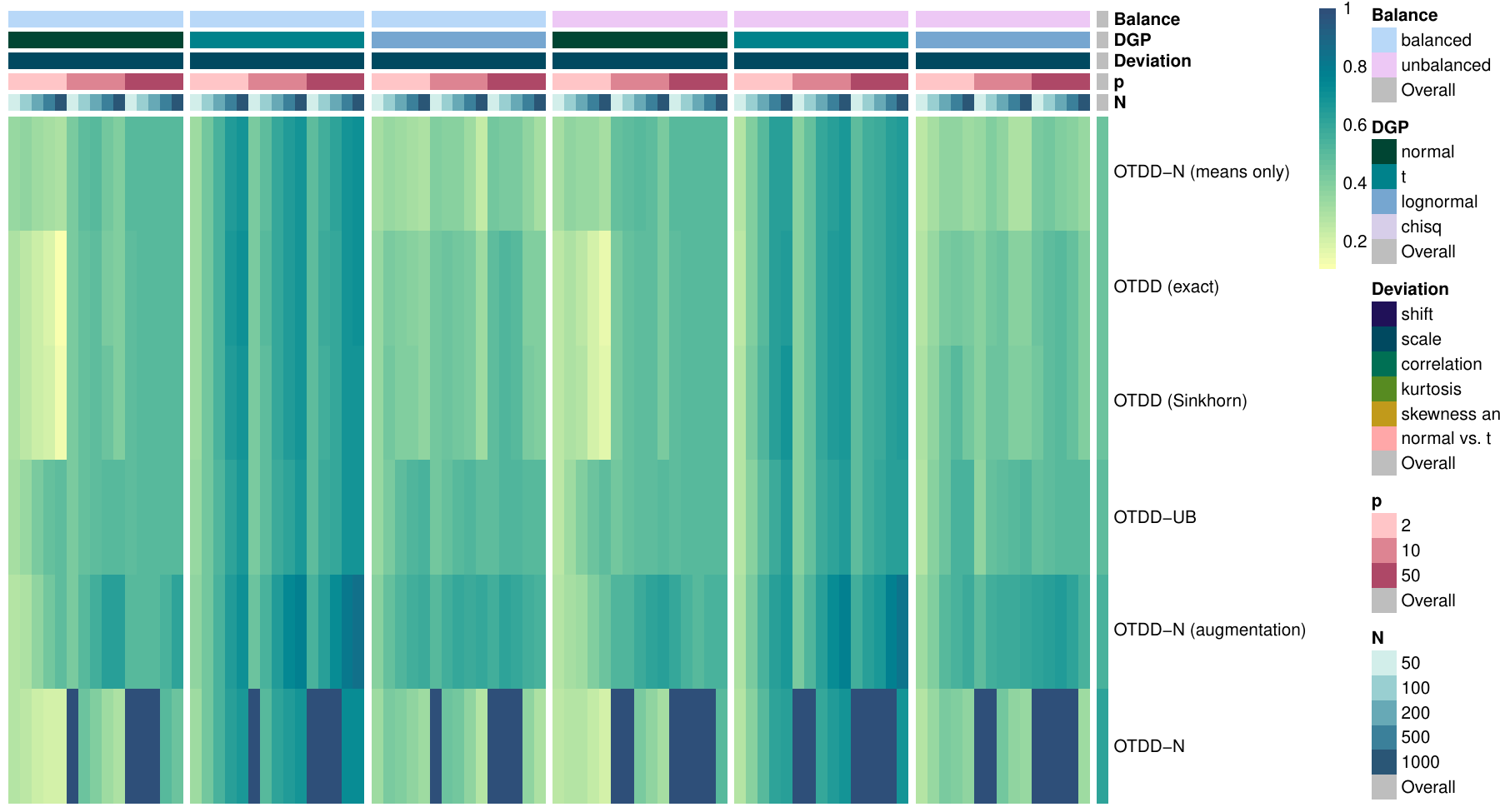}
		\caption{Mean PESR difference to best method per scenario for scale alternatives and $k = 2$ datasets for all variants of the OTDD \textcite{alvarez-melis_geometric_2020}. Selected variant: OTDD-N (means only).}
	\end{figure}
	
	\begin{figure}[!h]
		\centering
		\includegraphics[width=\linewidth]{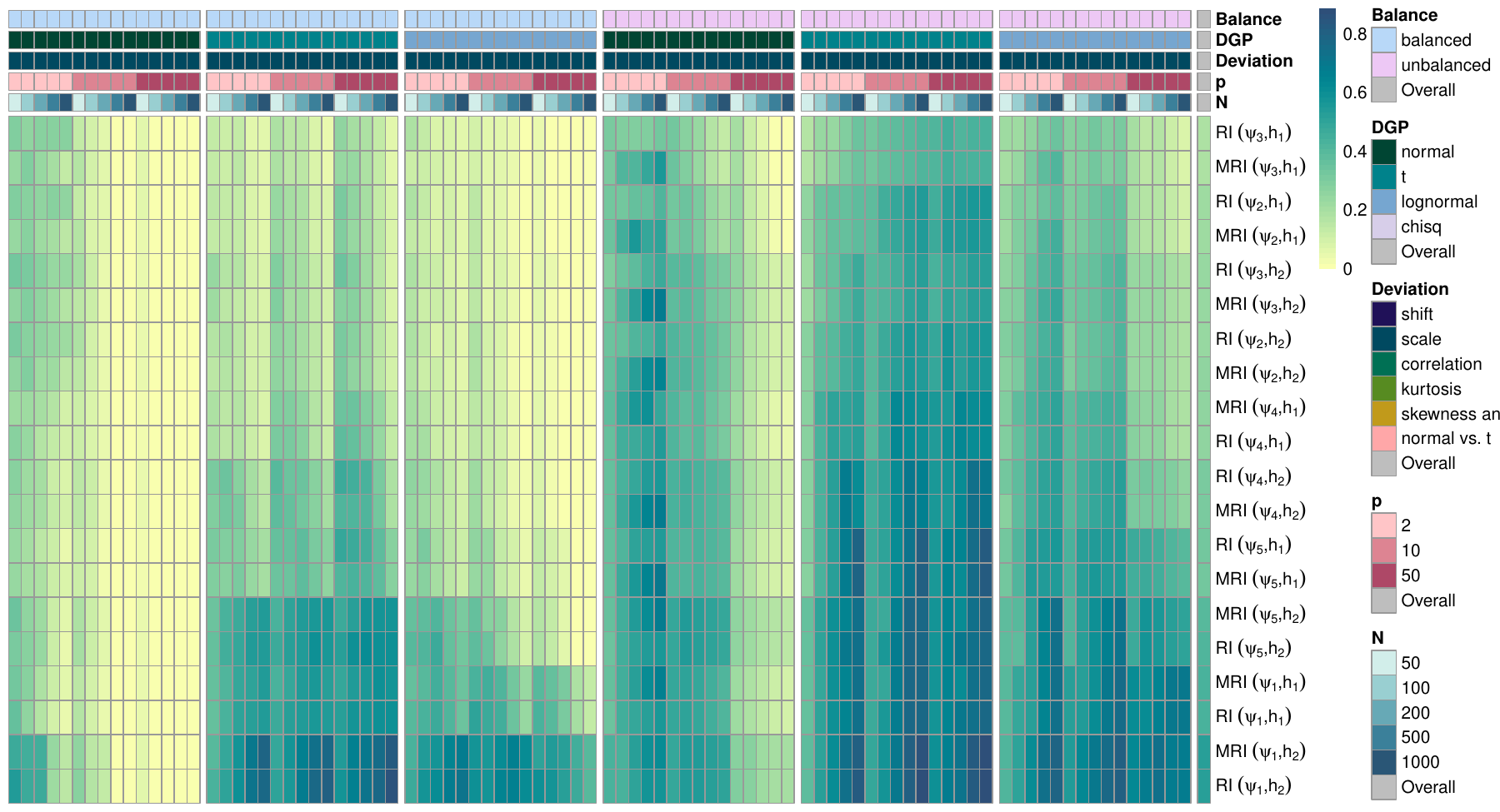}
		\caption{Mean PESR difference to best method per scenario for scale alternatives and $k = 2$ datasets for all variants of the RI test \textcite{paul_clustering-based_2022}. Selected variant: RI~($\psi_3, h_1$).}
	\end{figure}
	
	\begin{figure}[!h]
		\centering
		\includegraphics[width=\linewidth]{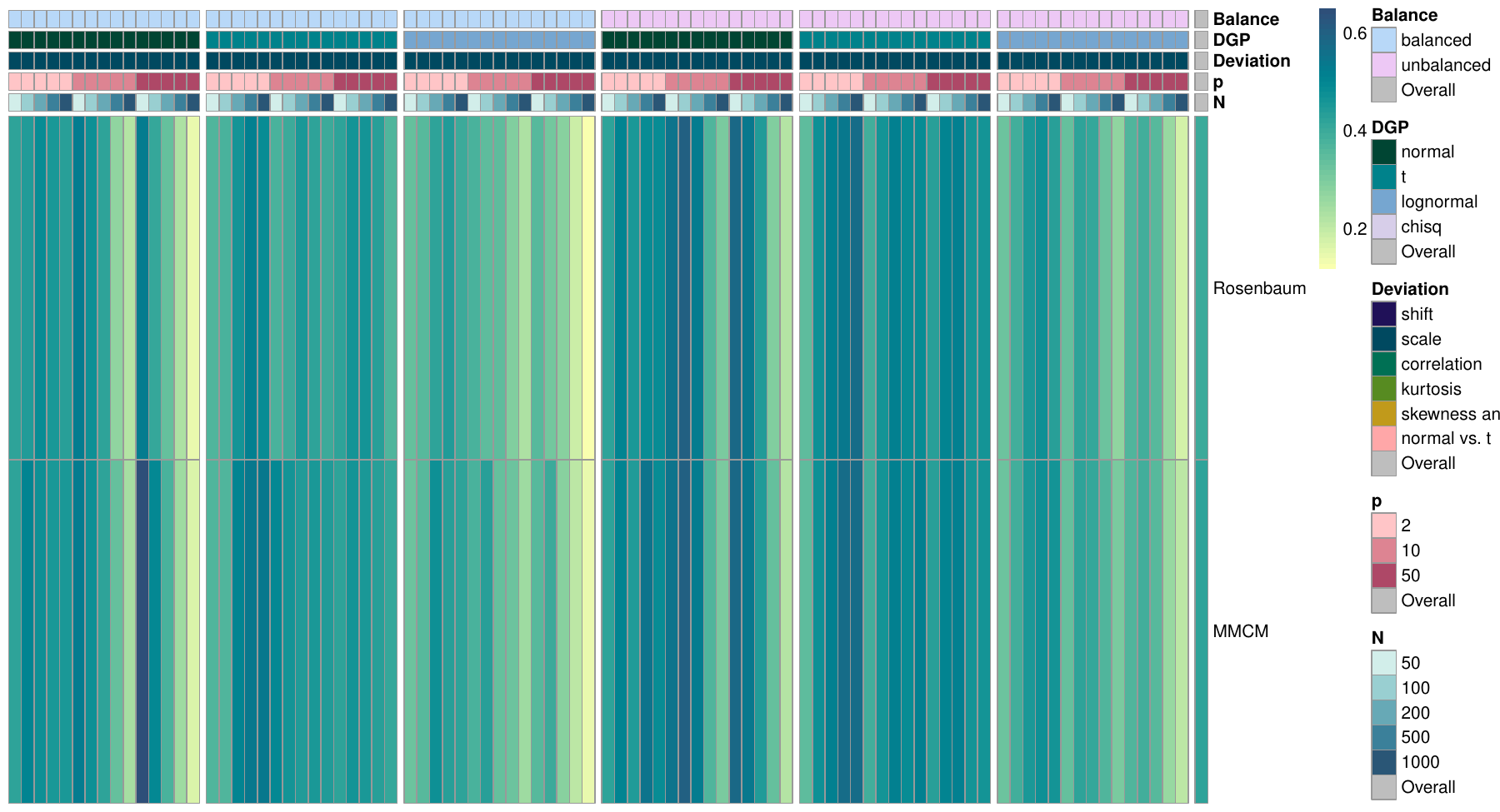}
		\caption{Mean PESR difference to best method per scenario for scale alternatives and $k = 2$ datasets for all variants of the Rosenbaum test \textcite{rosenbaum_exact_2005, mukherjee_distribution-free_2022}. Selected variant: Rosenbaum.}
	\end{figure}
	
	\begin{figure}[!h]
		\centering
		\includegraphics[width=\linewidth]{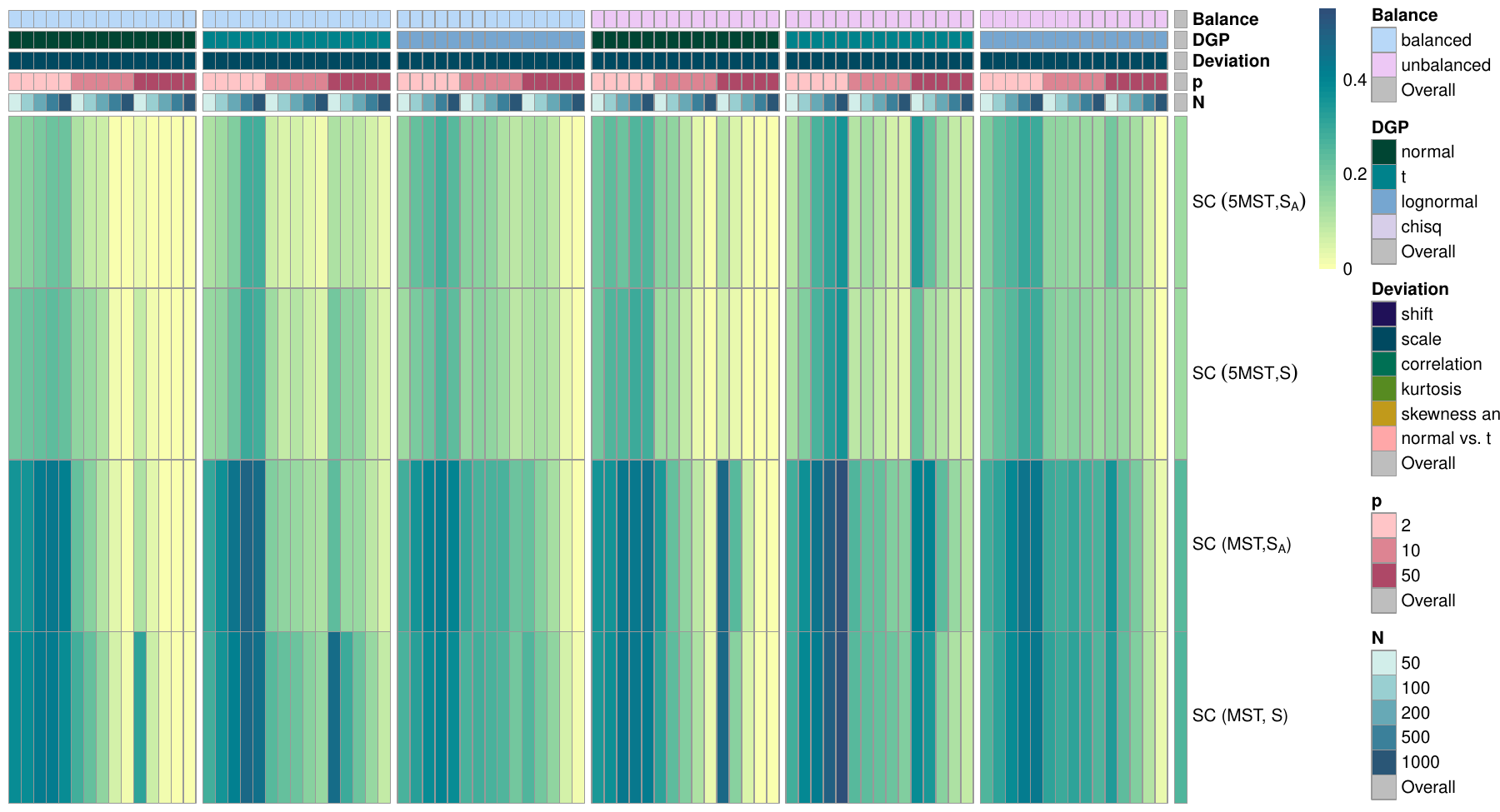}
		\caption{Mean PESR difference to best method per scenario for scale alternatives and $k = 2$ datasets for all variants of the SC test \textcite{song_new_2022}. Selected variant: SC(5MST,S$_{\text{A}}$).}
	\end{figure}
	
	\begin{figure}[!h]
		\centering
		\includegraphics[width=\linewidth]{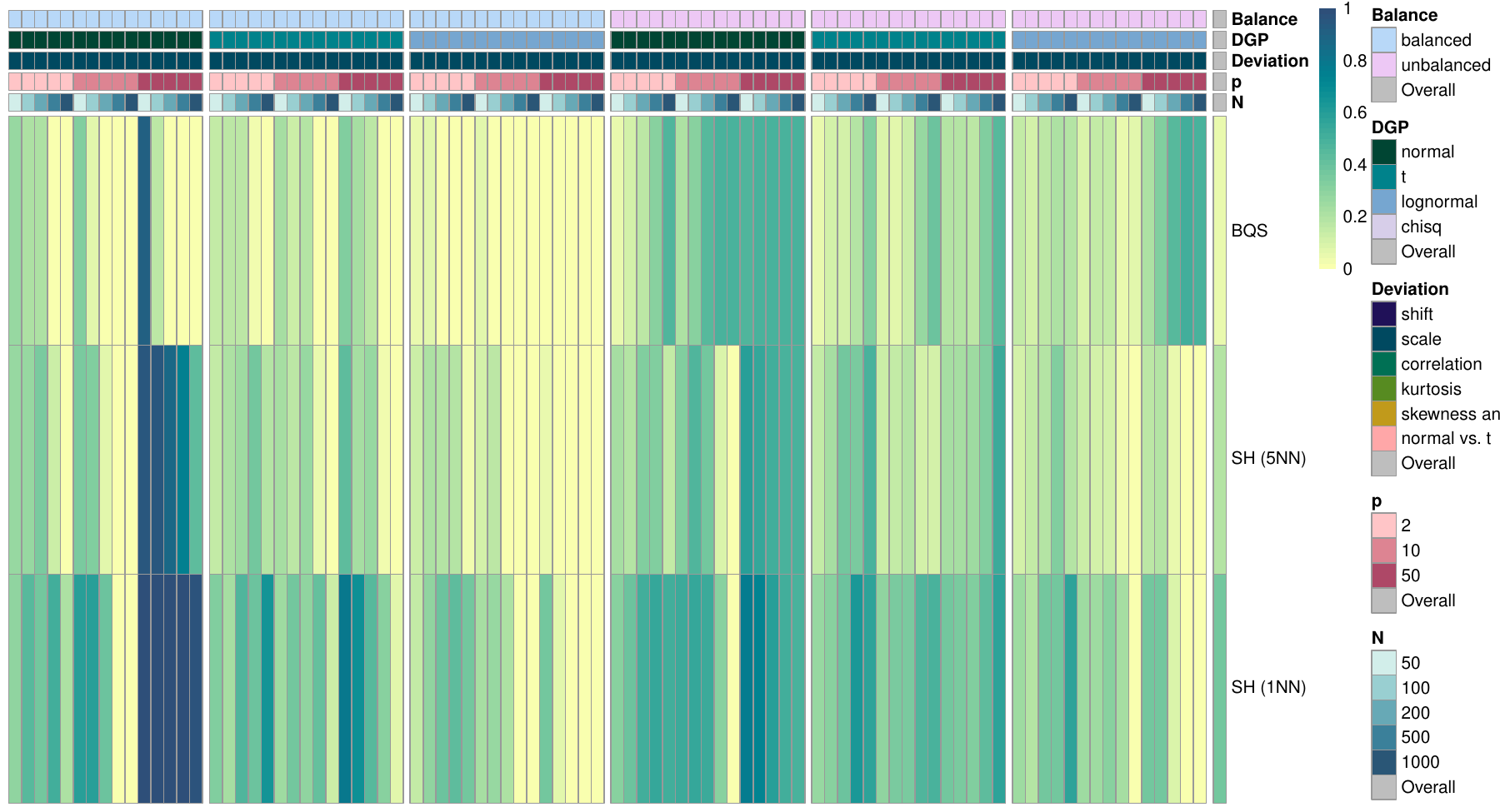}
		\caption{Mean PESR difference to best method per scenario for scale alternatives and $k = 2$ datasets for all variants of the SH test \textcite{schilling_multivariate_1986, henze_multivariate_1988, barakat_multivariate_1996}. Selected variants: BQS, SH(5NN) (complement each other).}
	\end{figure}
	\clearpage
	
	\paragraph{Correlation}\mbox{} 
	
	\begin{figure}[!h]
		\centering
		\includegraphics[width=\linewidth]{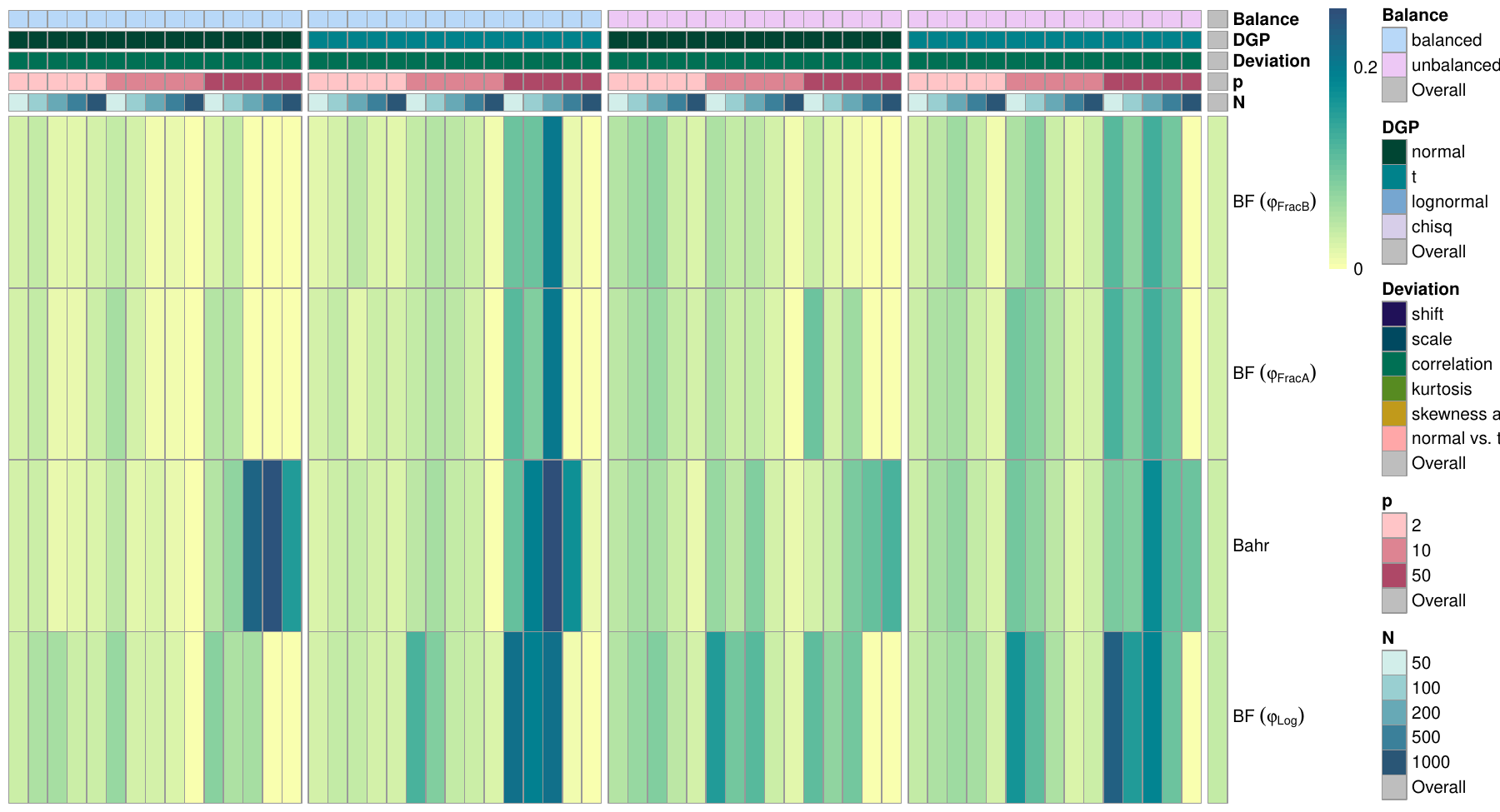}
		\caption{Mean PESR difference to best method per scenario for correlation alternatives and $k = 2$ datasets for all variants of the BF method \textcite{baringhaus_rigid_2010}. Selected variant: BF~($\varphi_{\text{FracB}}$).}
	\end{figure}
	
	\begin{figure}[!h]
		\centering
		\includegraphics[width=\linewidth]{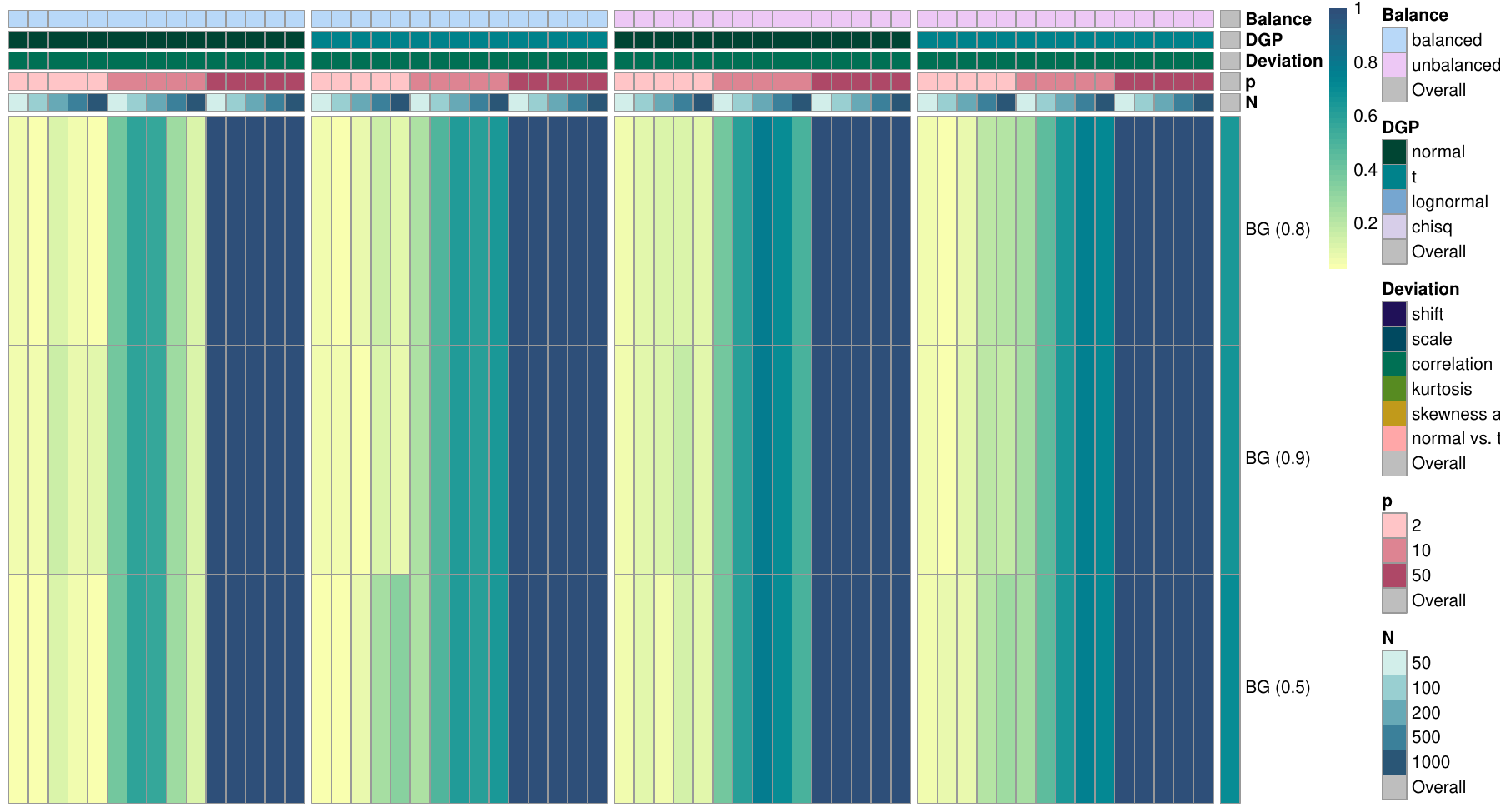}
		\caption{Mean PESR difference to best method per scenario for correlation alternatives and $k = 2$ datasets for all variants of the BG method \textcite{biau_asymptotic_2005}. Selected variant: BG~(0.8).}
	\end{figure}
	
	\begin{figure}[!h]
		\centering
		\includegraphics[width=\linewidth]{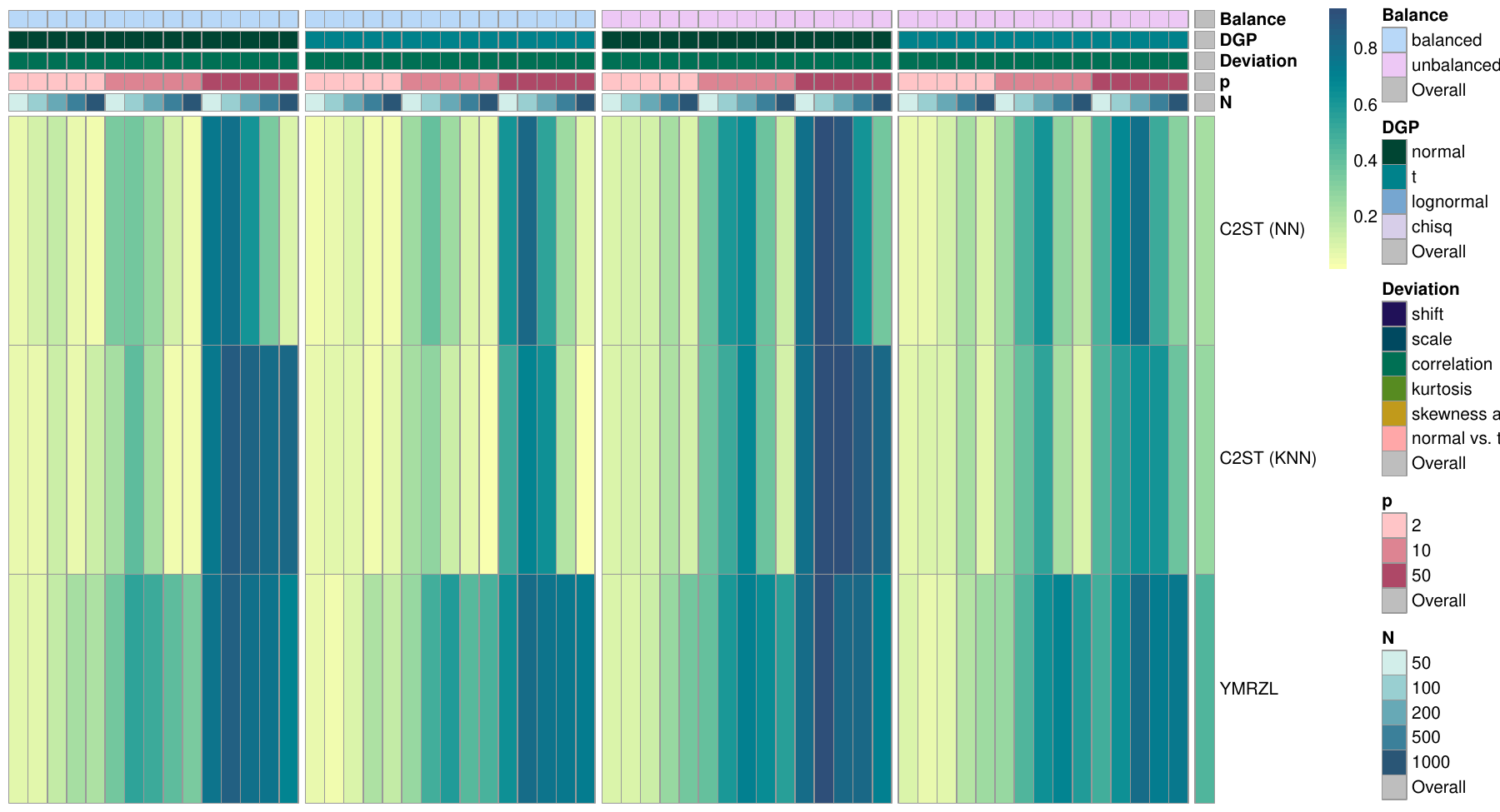}
		\caption{Mean PESR difference to best method per scenario for correlation alternatives and $k = 2$ datasets for all variants of the YMRZL and C2ST method \textcite{yu_two-sample_2007, lopez-paz_revisiting_2017}. Selected variant: C2ST~(NN).}
	\end{figure}
	
	\begin{figure}[!h]
		\centering
		\includegraphics[width=\linewidth]{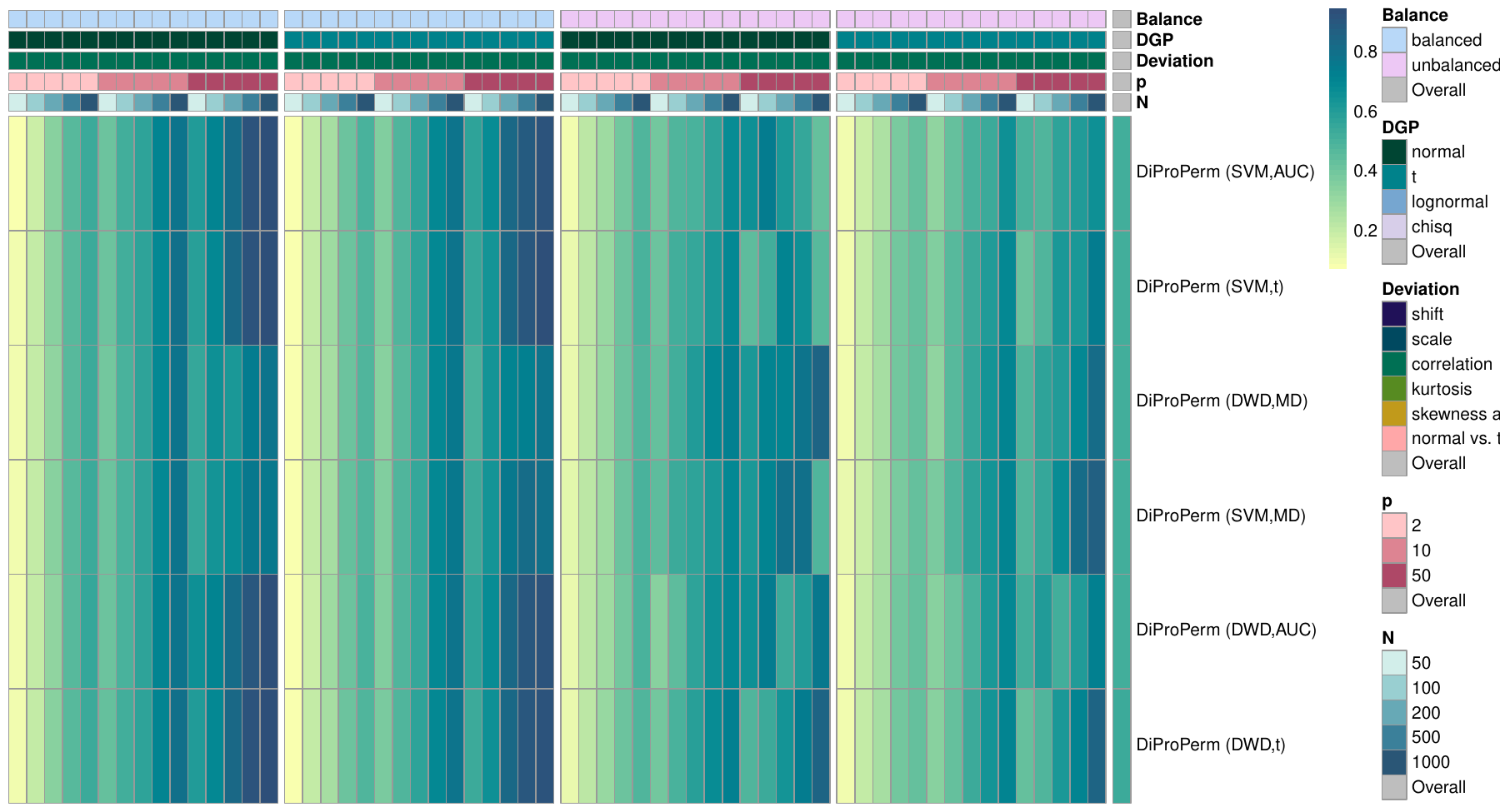}
		\caption{Mean PESR difference to best method per scenario for correlation alternatives and $k = 2$ datasets for all variants of the DiProPerm method \textcite{wei_direction-projection-permutation_2016}. Selected variant: DiProPerm~(SVM,AUC).}
	\end{figure}
	
	\begin{figure}[!h]
		\centering
		\includegraphics[width=\linewidth]{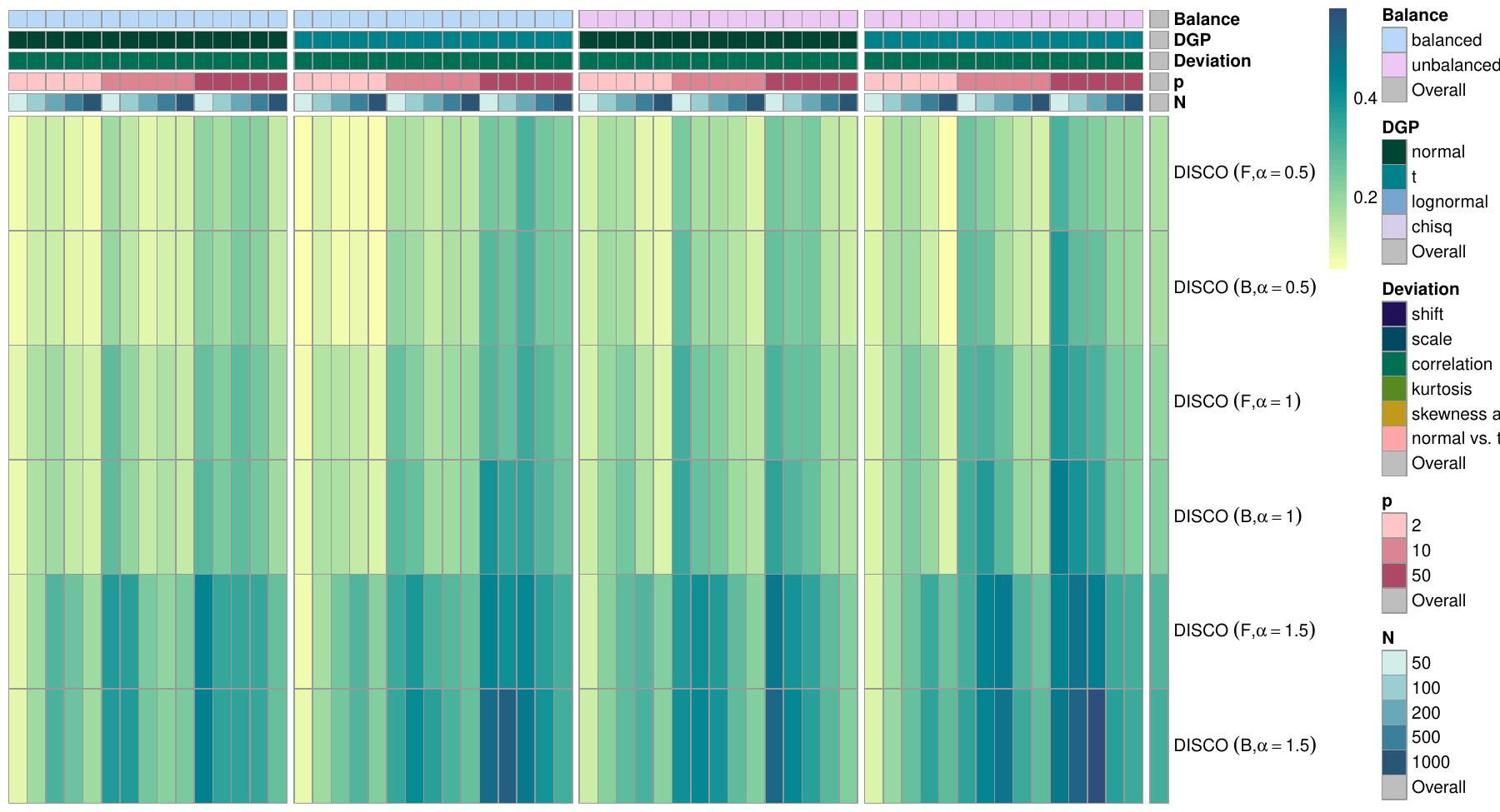}
		\caption{Mean PESR difference to best method per scenario for correlation alternatives and $k = 2$ datasets for all variants of the DISCO method \textcite{rizzo_disco_2010}. Selected variant: DISCO~(F,$\alpha=0.5$).}
	\end{figure}
	
	\begin{figure}[!h]
		\centering
		\includegraphics[width=\linewidth]{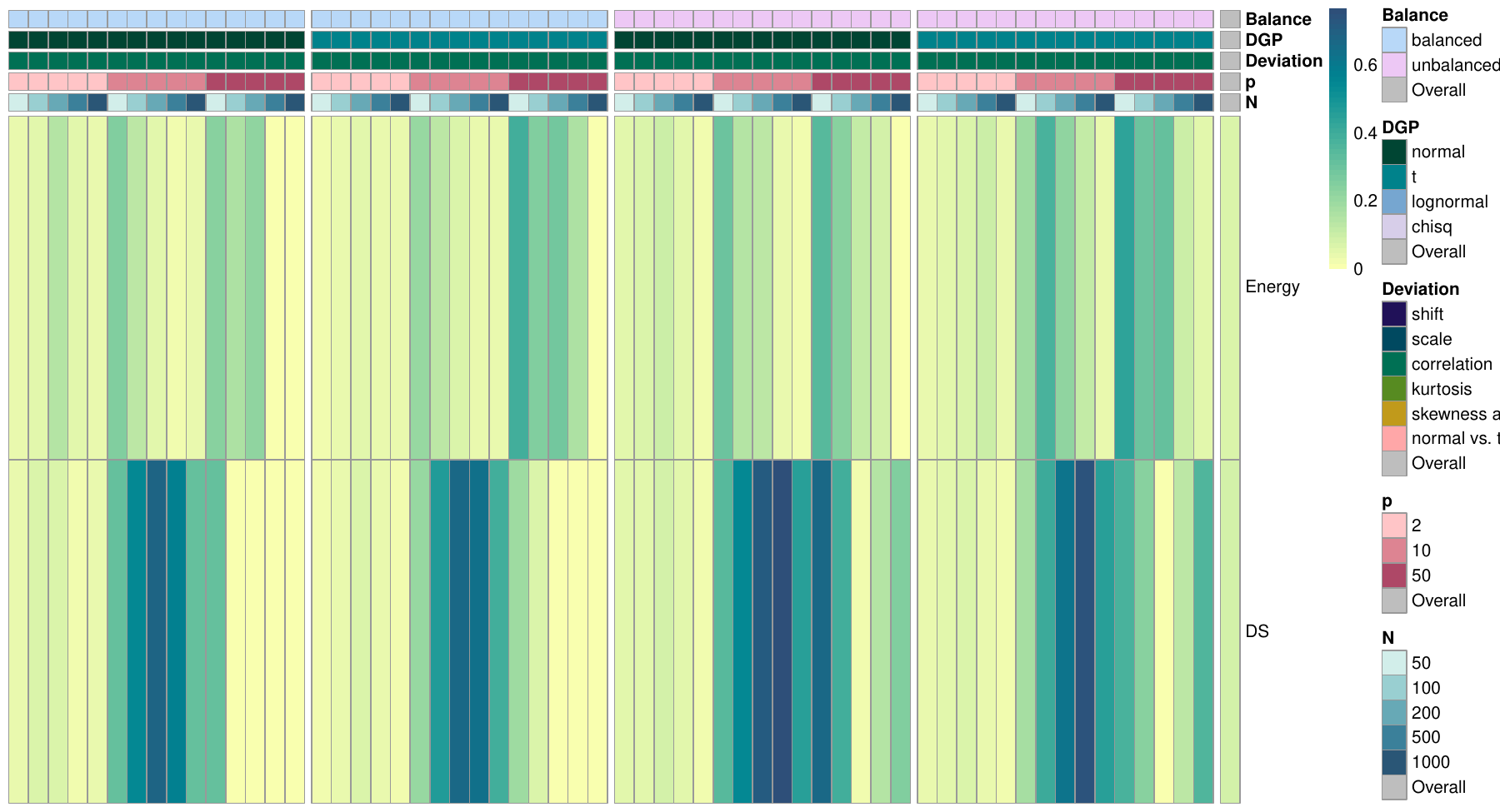}
		\caption{Mean PESR difference to best method per scenario for correlation alternatives and $k = 2$ datasets for all variants of the Energy method \textcite{szekely_testing_2004}. Selected variant: Energy.}
	\end{figure}
	
	\begin{figure}[!h]
		\centering
		\includegraphics[width=\linewidth]{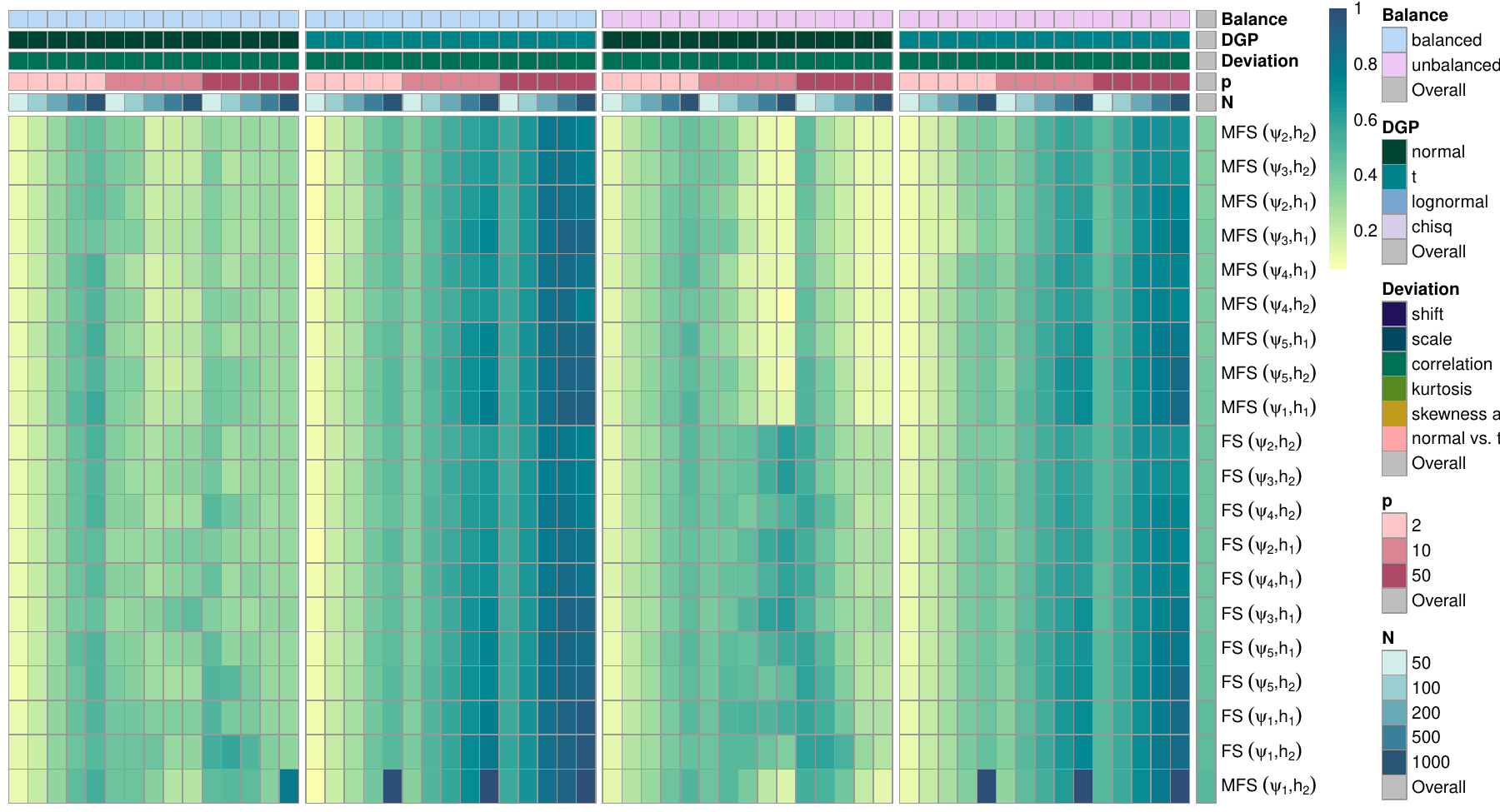}
		\caption{Mean PESR difference to best method per scenario for correlation alternatives and $k = 2$ datasets for all variants of the FS test \textcite{paul_clustering-based_2022}. Selected variant: MFS~($\psi_2, h_2$).}
	\end{figure}
	
	\begin{figure}[!h]
		\centering
		\includegraphics[width=\linewidth]{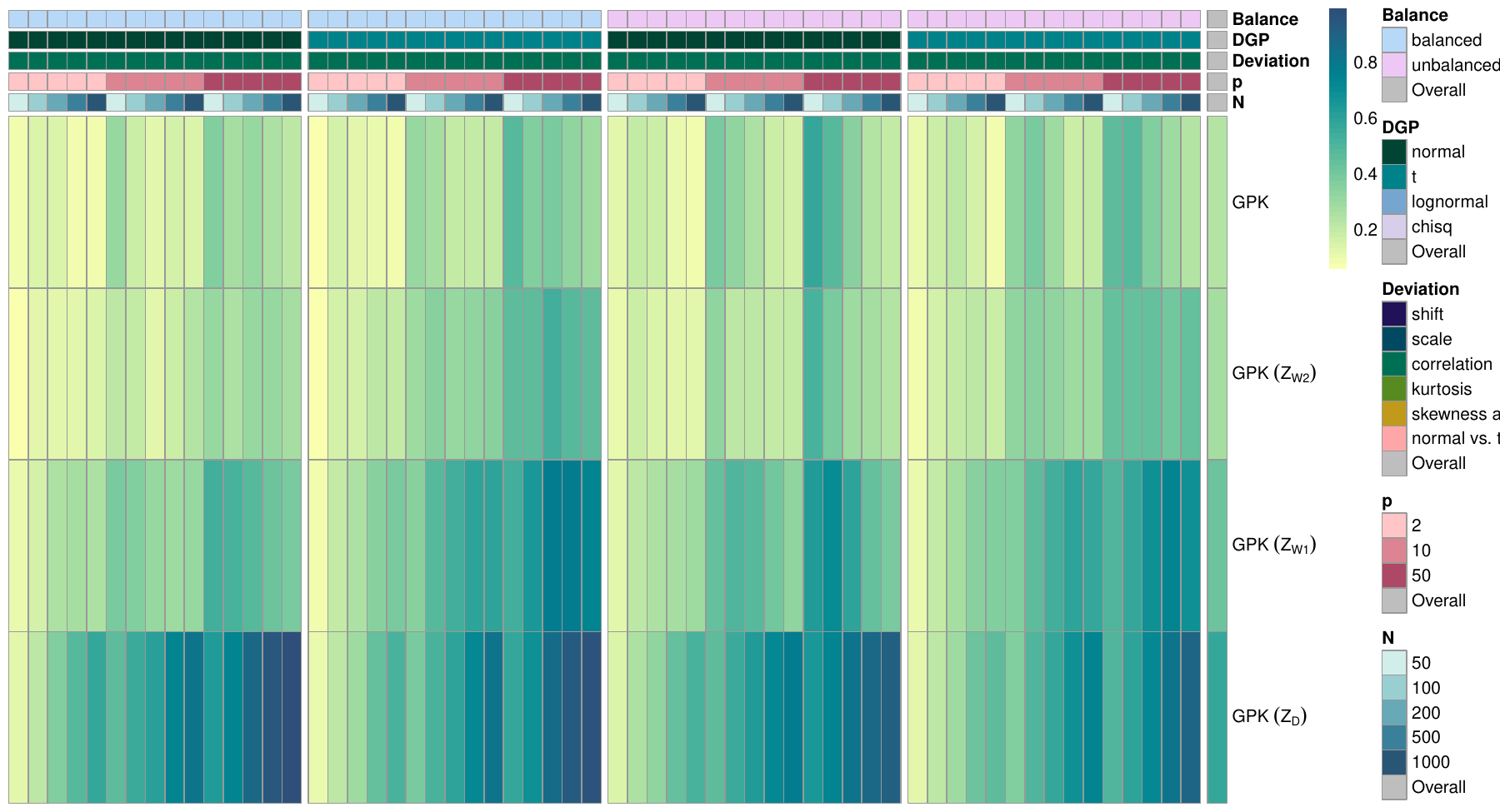}
		\caption{Mean PESR difference to best method per scenario for correlation alternatives and $k = 2$ datasets for all variants of the GPK test \textcite{song_generalized_2021}. Selected variant: GPK.}
	\end{figure}
	
	\begin{figure}[!h]
		\centering
		\includegraphics[width=\linewidth]{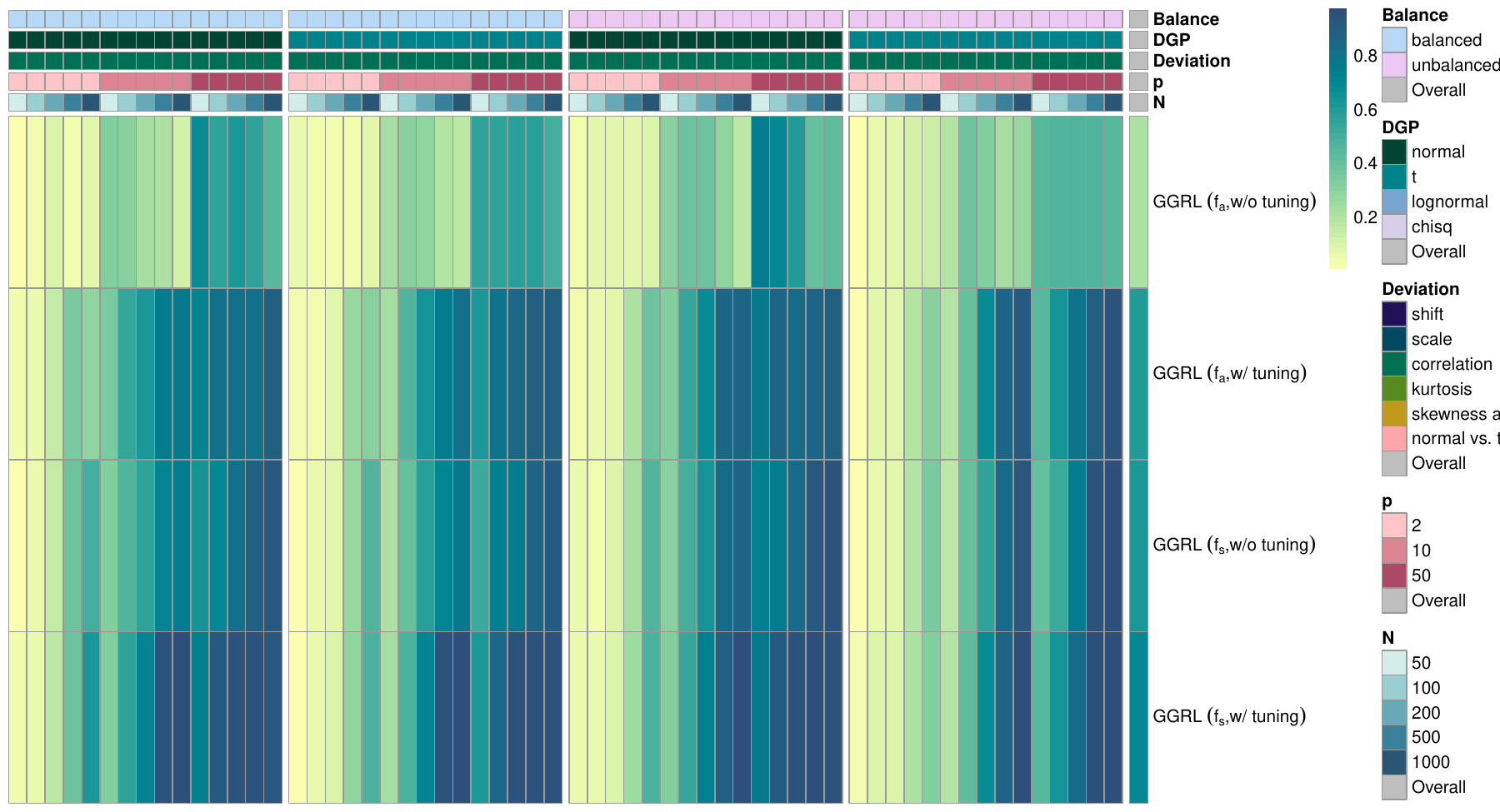}
		\caption{Mean PESR difference to best method per scenario for correlation alternatives and $k = 2$ datasets for all variants of the GGRL method \textcite{ganti_framework_1999}. Selected variant: GGRL~($f_a$,w/o~tuning).}
	\end{figure}
	
	\begin{figure}[!h]
		\centering
		\includegraphics[width=\linewidth]{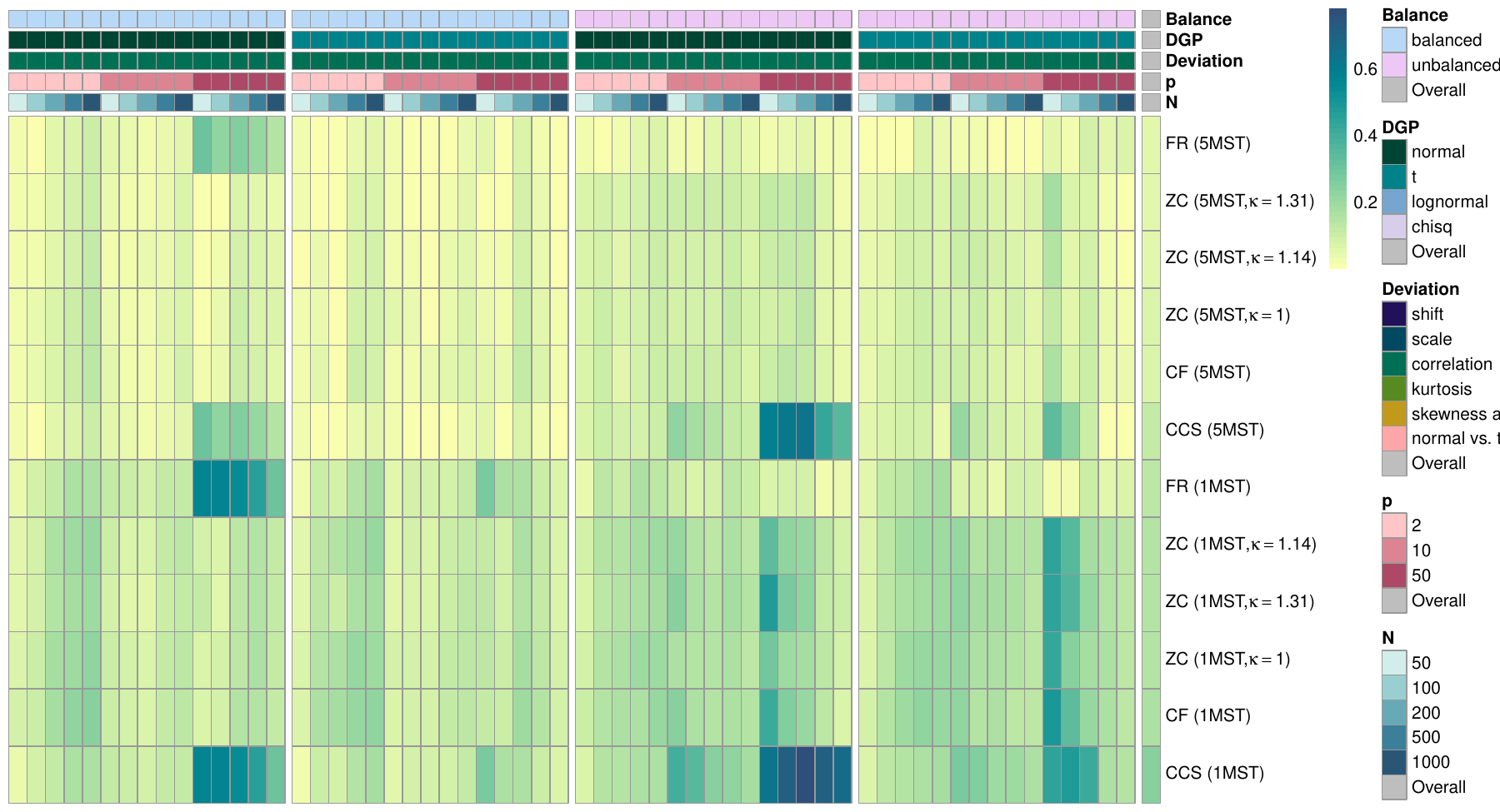}
		\caption{Mean PESR difference to best method per scenario for correlation alternatives and $k = 2$ datasets for all variants of the graph-based tests FR, CF, CCS, ZC \textcite{friedman_multivariate_1979, chen_new_2017, chen_weighted_2018, zhang_graph-based_2022}. Selected variant: FR~(5MST).}
	\end{figure}
	
	\begin{figure}[!h]
		\centering
		\includegraphics[width=\linewidth]{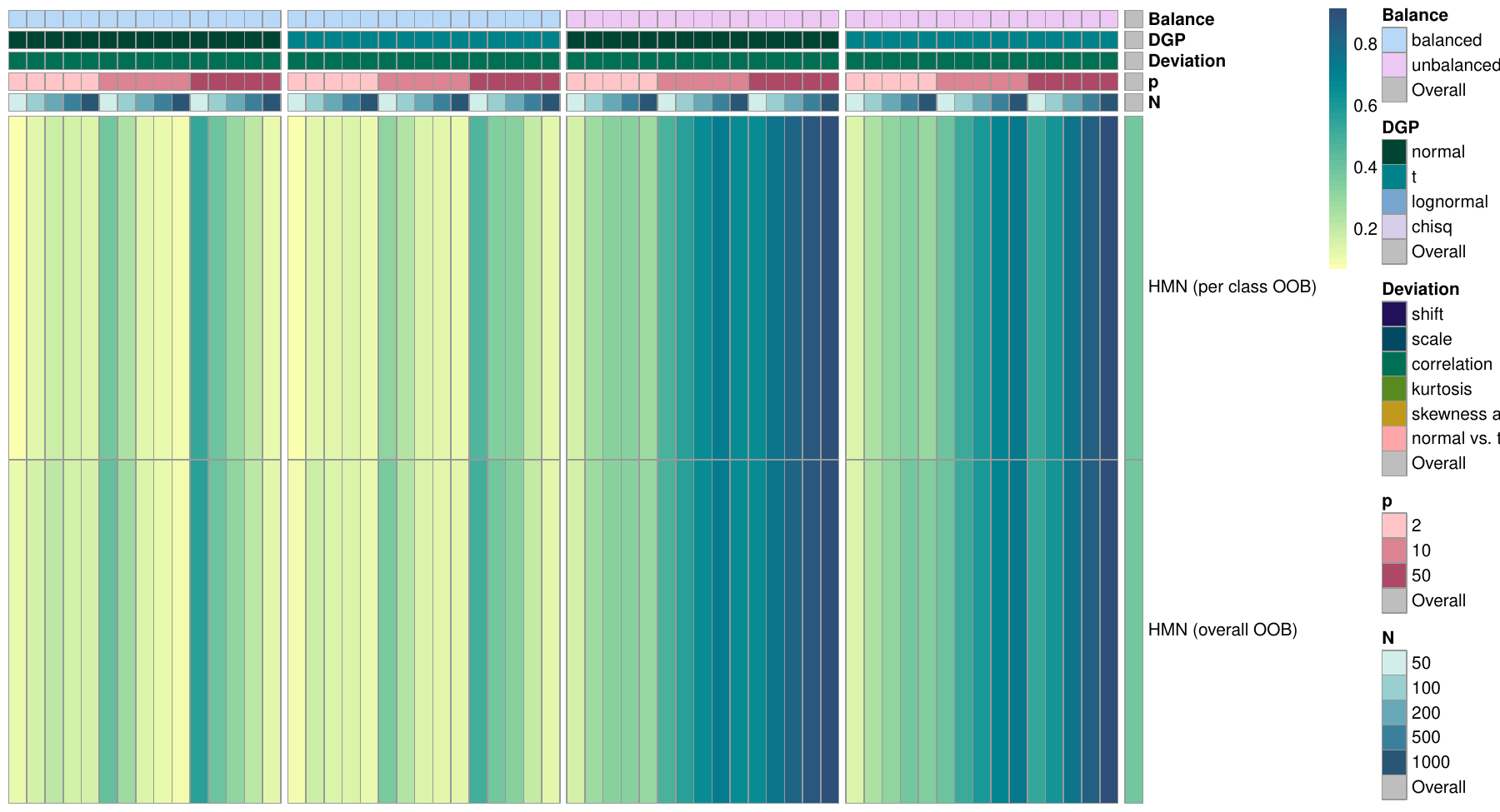}
		\caption{Mean PESR difference to best method per scenario for correlation alternatives and $k = 2$ datasets for all variants of the HMN method \textcite{hediger_use_2021}. Selected variant: HMN~(per class OOB).}
	\end{figure}
	
	\begin{figure}[!h]
		\centering
		\includegraphics[width=\linewidth]{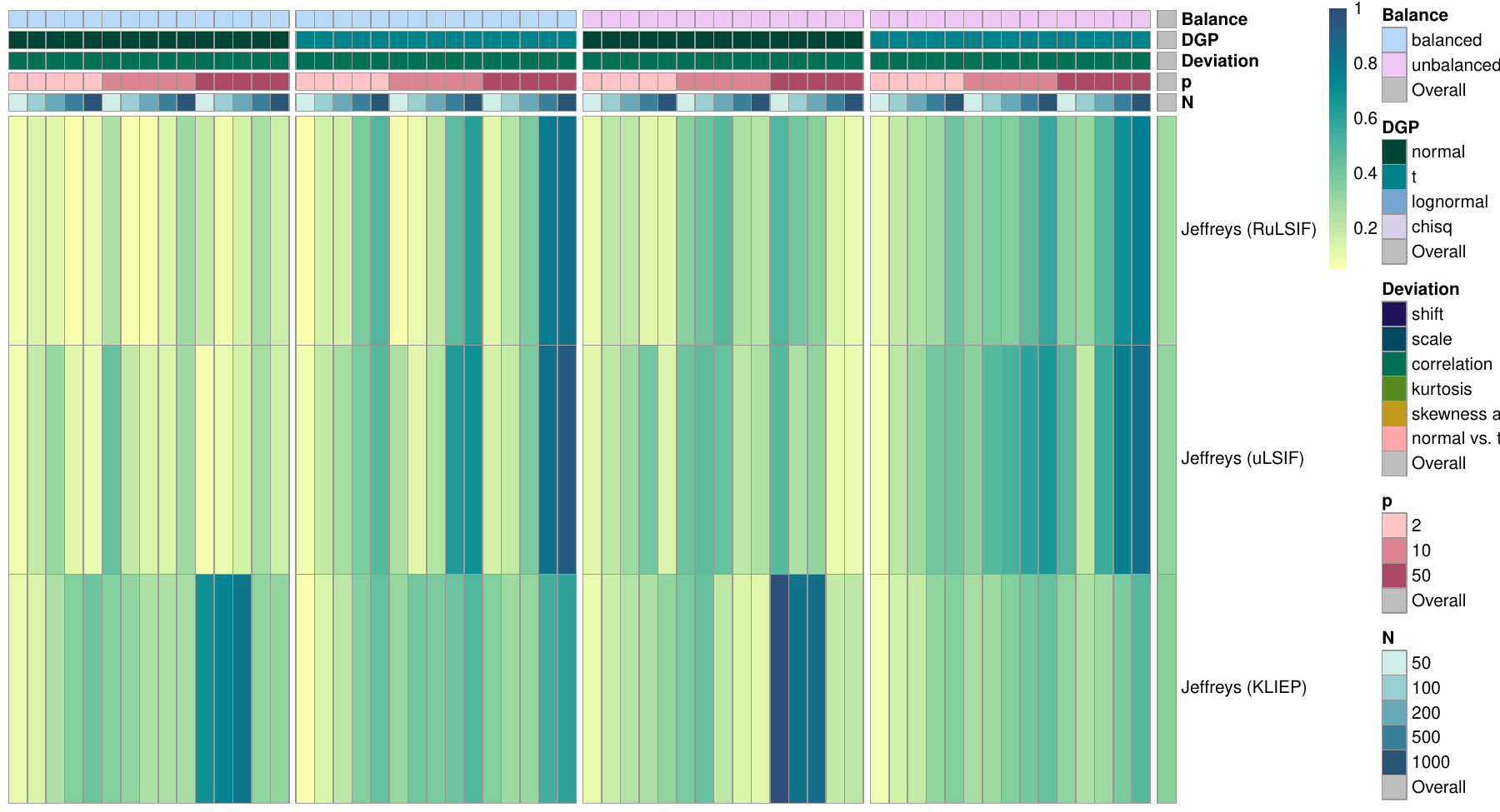}
		\caption{Mean PESR difference to best method per scenario for correlation alternatives and $k = 2$ datasets for all variants of the Jeffreys divergence \textcite{jeffreys_invariant_1997}. Selected variant: Jeffreys~(RuLSIF).}
	\end{figure}
	
	\begin{figure}[!h]
		\centering
		\includegraphics[width=\linewidth]{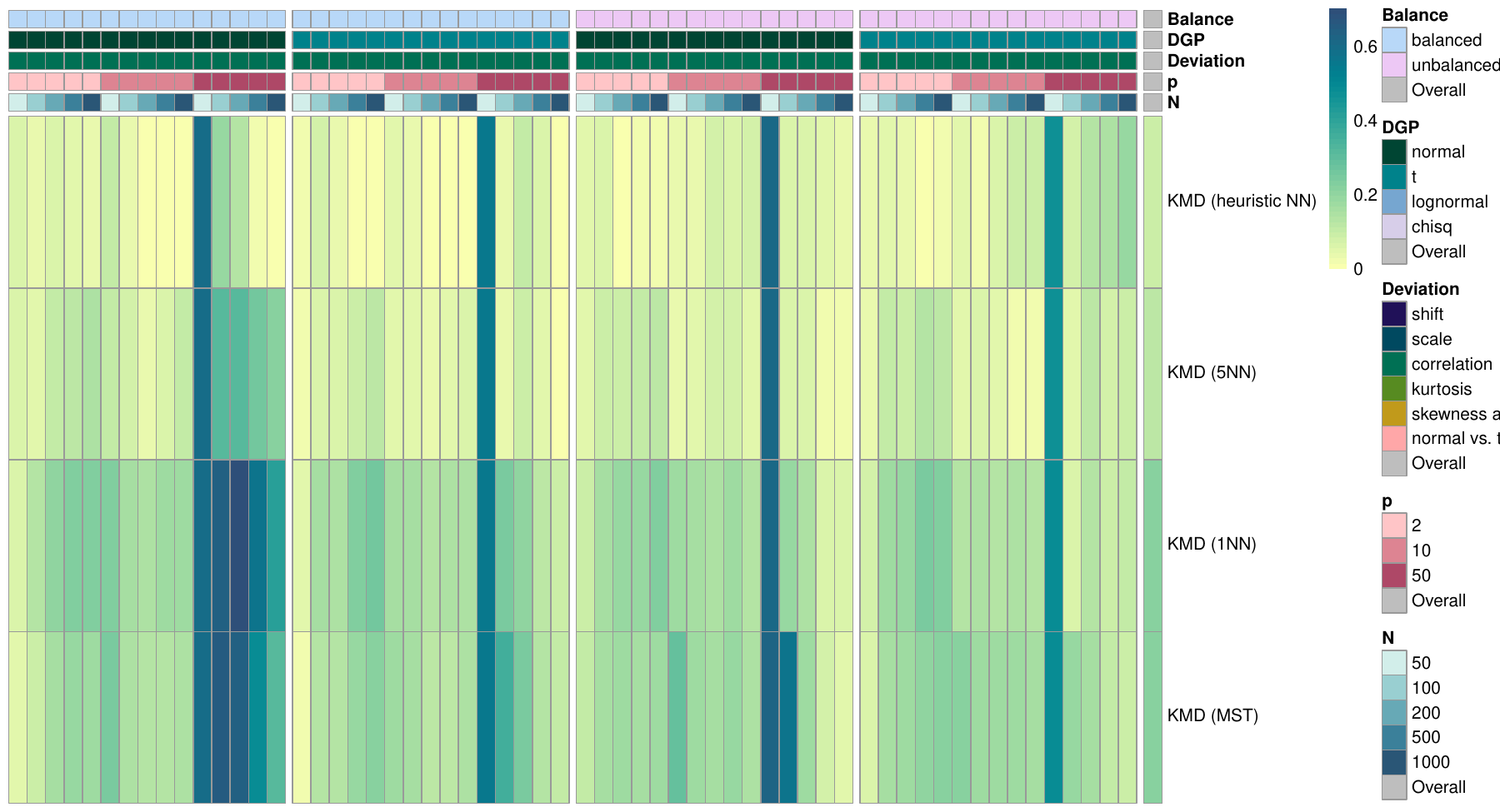}
		\caption{Mean PESR difference to best method per scenario for correlation alternatives and $k = 2$ datasets for all variants of the KMD \textcite{huang_kernel_2022}. Selected variant: KMD~(heuristic NN).}
	\end{figure}
	
	\begin{figure}[!h]
		\centering
		\includegraphics[width=\linewidth]{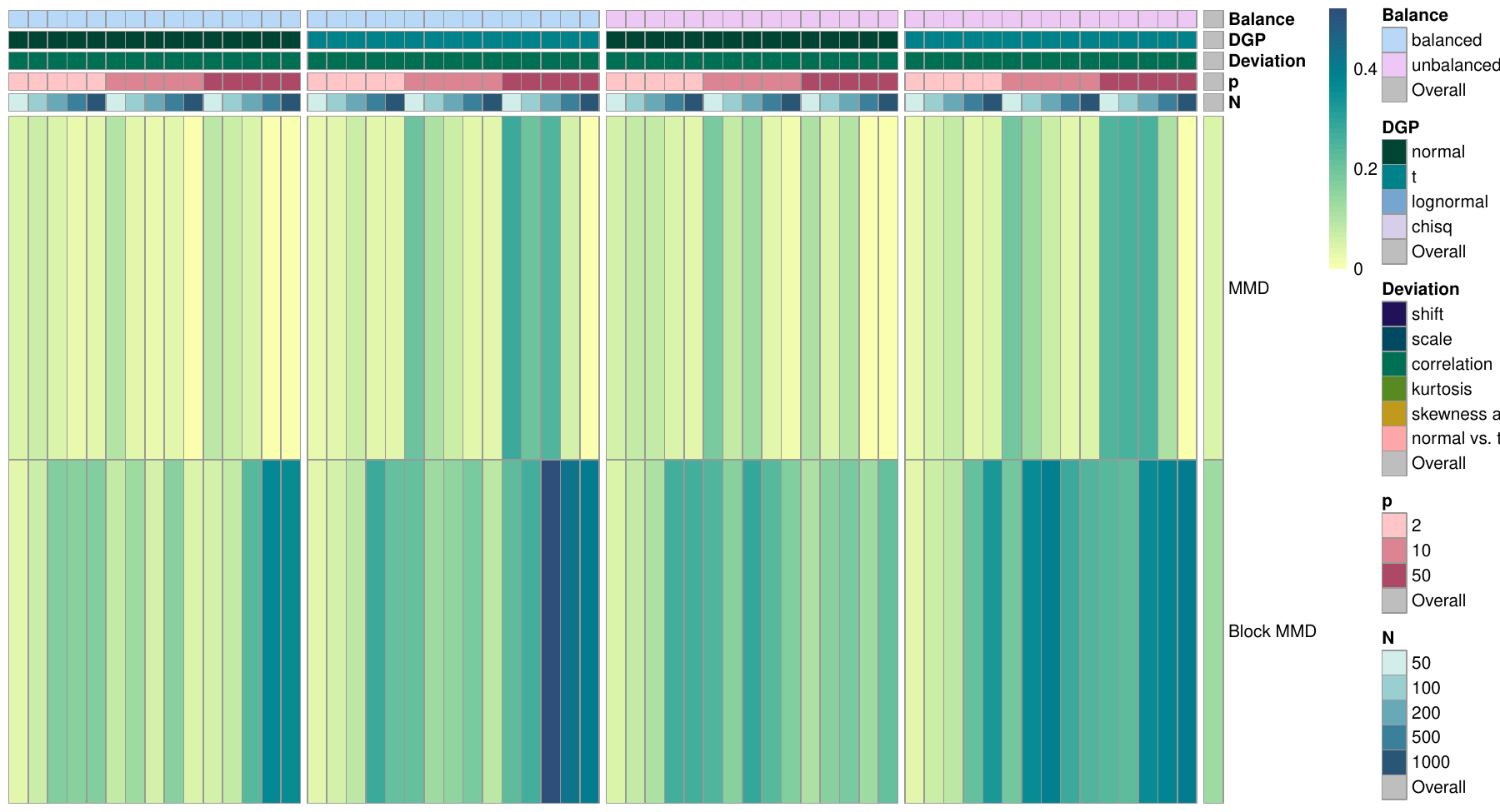}
		\caption{Mean PESR difference to best method per scenario for correlation alternatives and $k = 2$ datasets for all variants of the MMD \textcite{gretton_kernel_2006, zaremba_b-test_2013}. Selected variant: MMD.}
	\end{figure}
	
	\begin{figure}[!h]
		\centering
		\includegraphics[width=\linewidth]{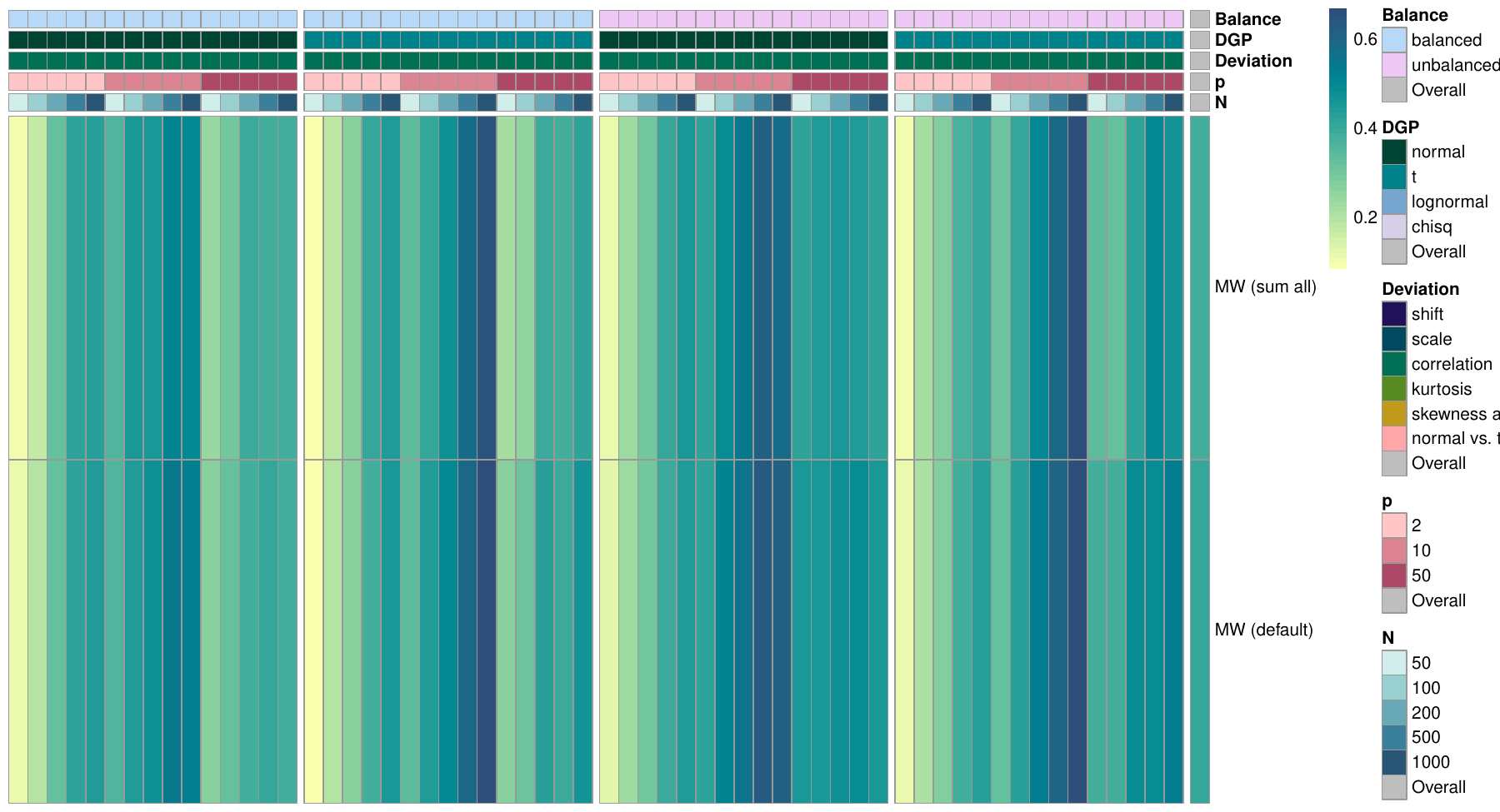}
		\caption{Mean PESR difference to best method per scenario for correlation alternatives and $k = 2$ datasets for all variants of the MW method \textcite{mukhopadhyay_nonparametric_2020}. Selected variant: MW~(sum all).}
	\end{figure}
	
	\begin{figure}[!h]
		\centering
		\includegraphics[width=\linewidth]{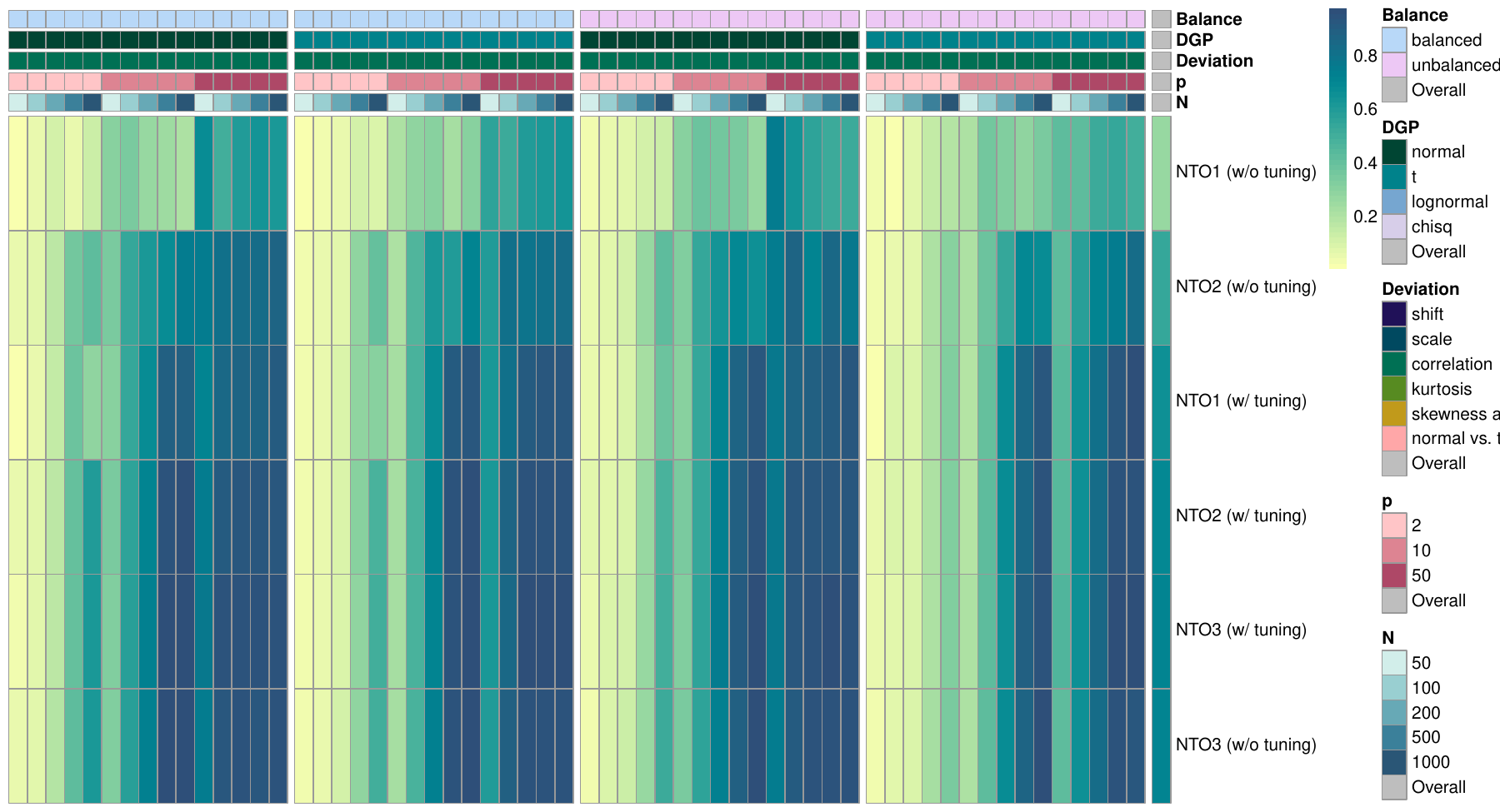}
		\caption{Mean PESR difference to best method per scenario for correlation alternatives and $k = 2$ datasets for all variants of the NKT method \textcite{ntoutsi_general_2008}. Selected variant: NTO1~(w/o~tuning).}
	\end{figure}
	
	\begin{figure}[!h]
		\centering
		\includegraphics[width=\linewidth]{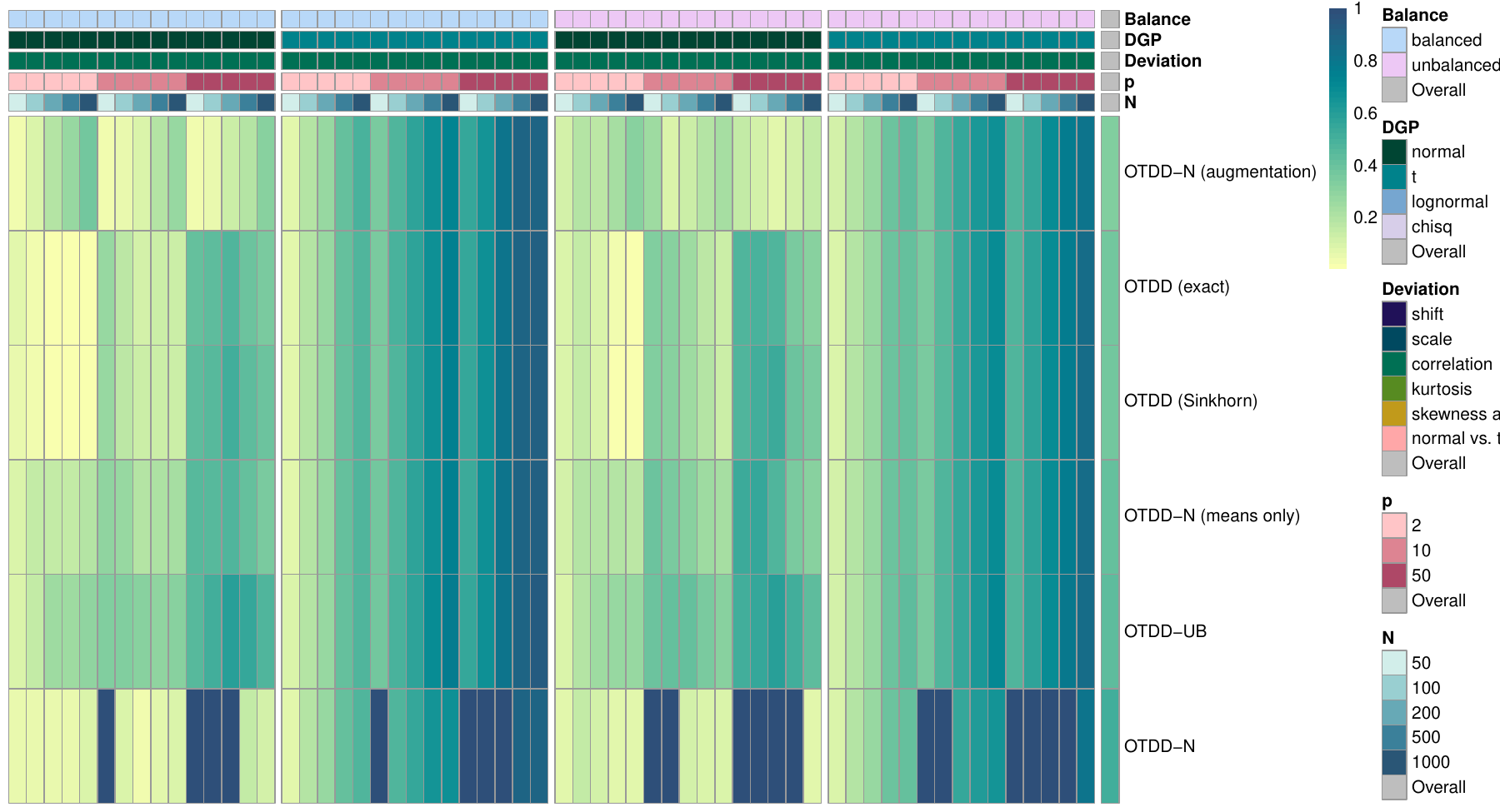}
		\caption{Mean PESR difference to best method per scenario for correlation alternatives and $k = 2$ datasets for all variants of the OTDD \textcite{alvarez-melis_geometric_2020}. Selected variant: OTDD~(augmentation).}
	\end{figure}
	
	\begin{figure}[!h]
		\centering
		\includegraphics[width=\linewidth]{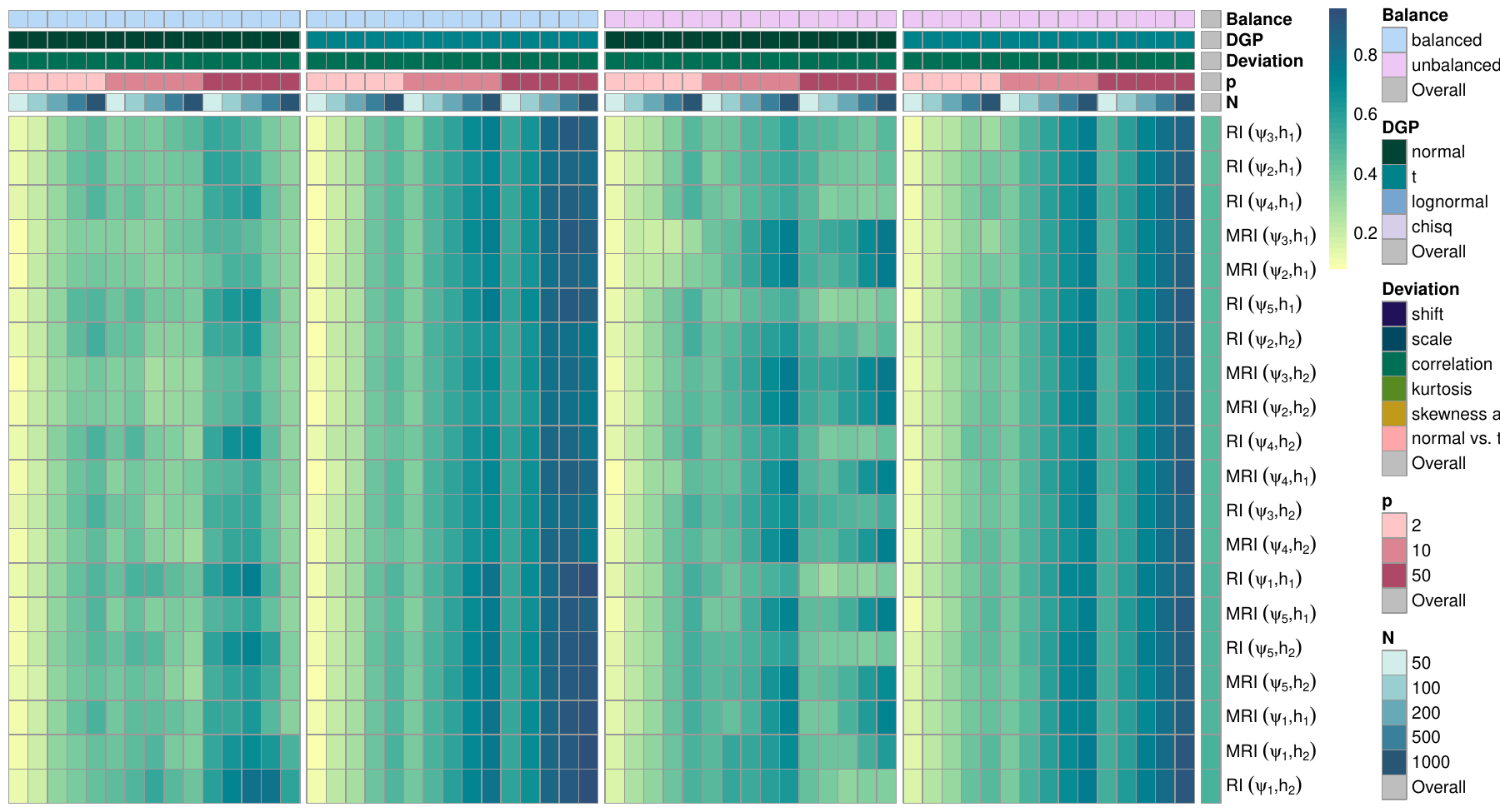}
		\caption{Mean PESR difference to best method per scenario for correlation alternatives and $k = 2$ datasets for all variants of the RI test \textcite{paul_clustering-based_2022}. Selected variant: RI~($\psi_3, h_1$).}
	\end{figure}
	
	\begin{figure}[!h]
		\centering
		\includegraphics[width=\linewidth]{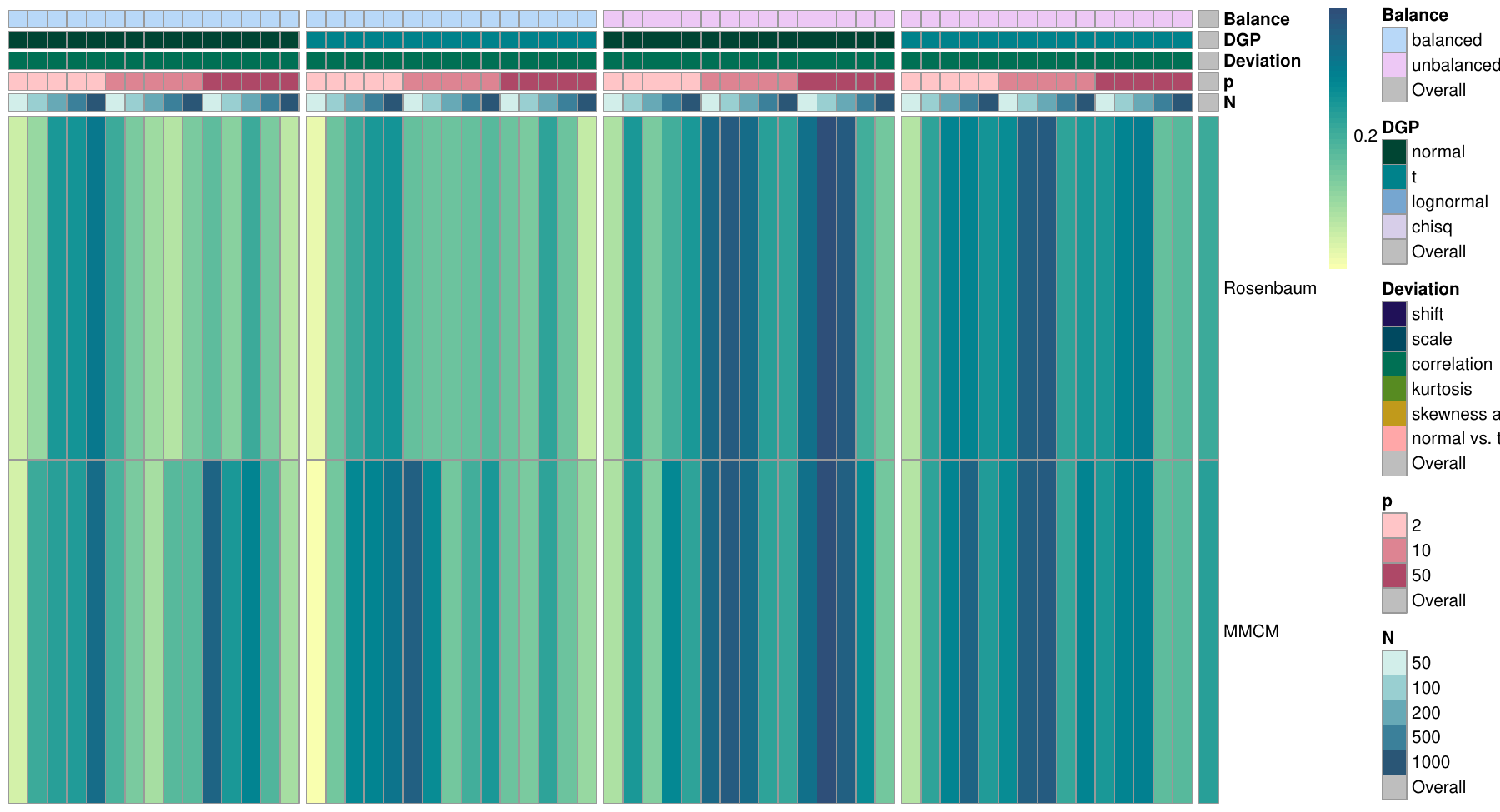}
		\caption{Mean PESR difference to best method per scenario for correlation alternatives and $k = 2$ datasets for all variants of the Rosenbaum test \textcite{rosenbaum_exact_2005, mukherjee_distribution-free_2022}. Selected variant: Rosenbaum.}
	\end{figure}
	
	\begin{figure}[!h]
		\centering
		\includegraphics[width=\linewidth]{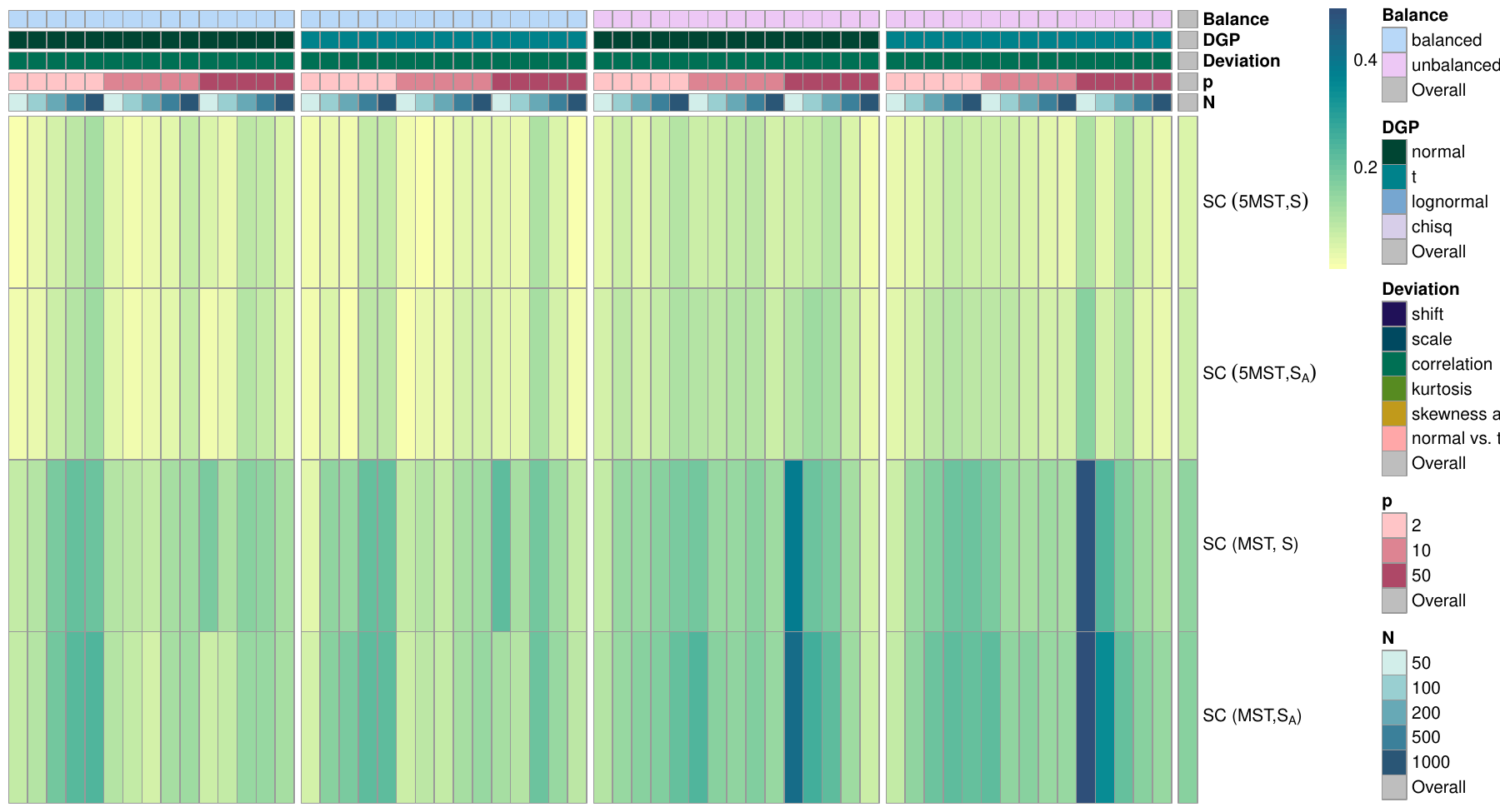}
		\caption{Mean PESR difference to best method per scenario for correlation alternatives and $k = 2$ datasets for all variants of the SC test \textcite{song_new_2022}. Selected variant: SC(5MST,S).}
	\end{figure}
	
	\begin{figure}[!h]
		\centering
		\includegraphics[width=\linewidth]{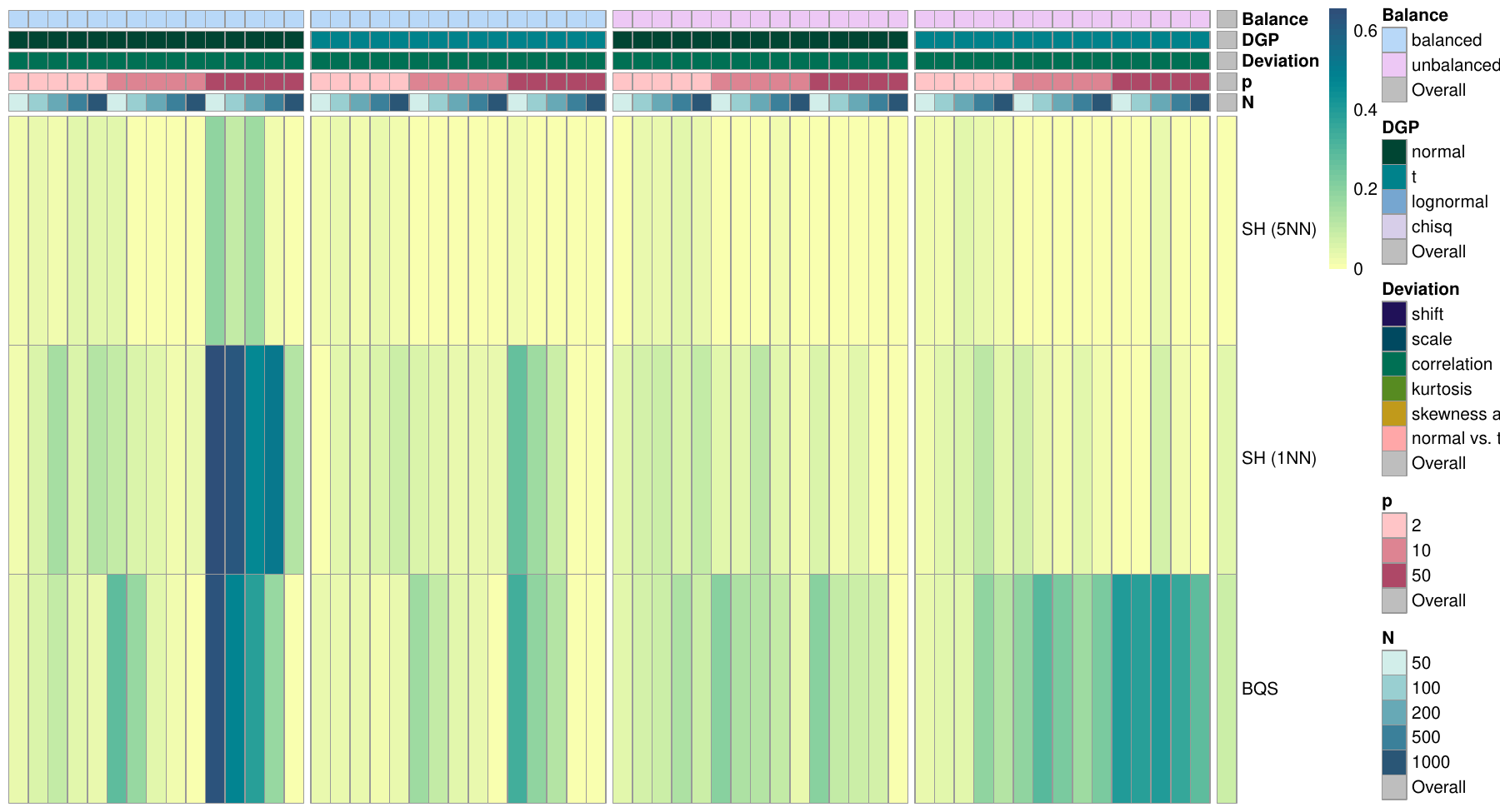}
		\caption{Mean PESR difference to best method per scenario for correlation alternatives and $k = 2$ datasets for all variants of the SH test \textcite{schilling_multivariate_1986, henze_multivariate_1988, barakat_multivariate_1996}. Selected variants: SH(5NN).}
	\end{figure}
	\clearpage
	
	\paragraph[Normal vs. t]{Normal vs.\ $t$}\mbox{} 
	
	\begin{figure}[!h]
		\centering
		\includegraphics[width=\linewidth]{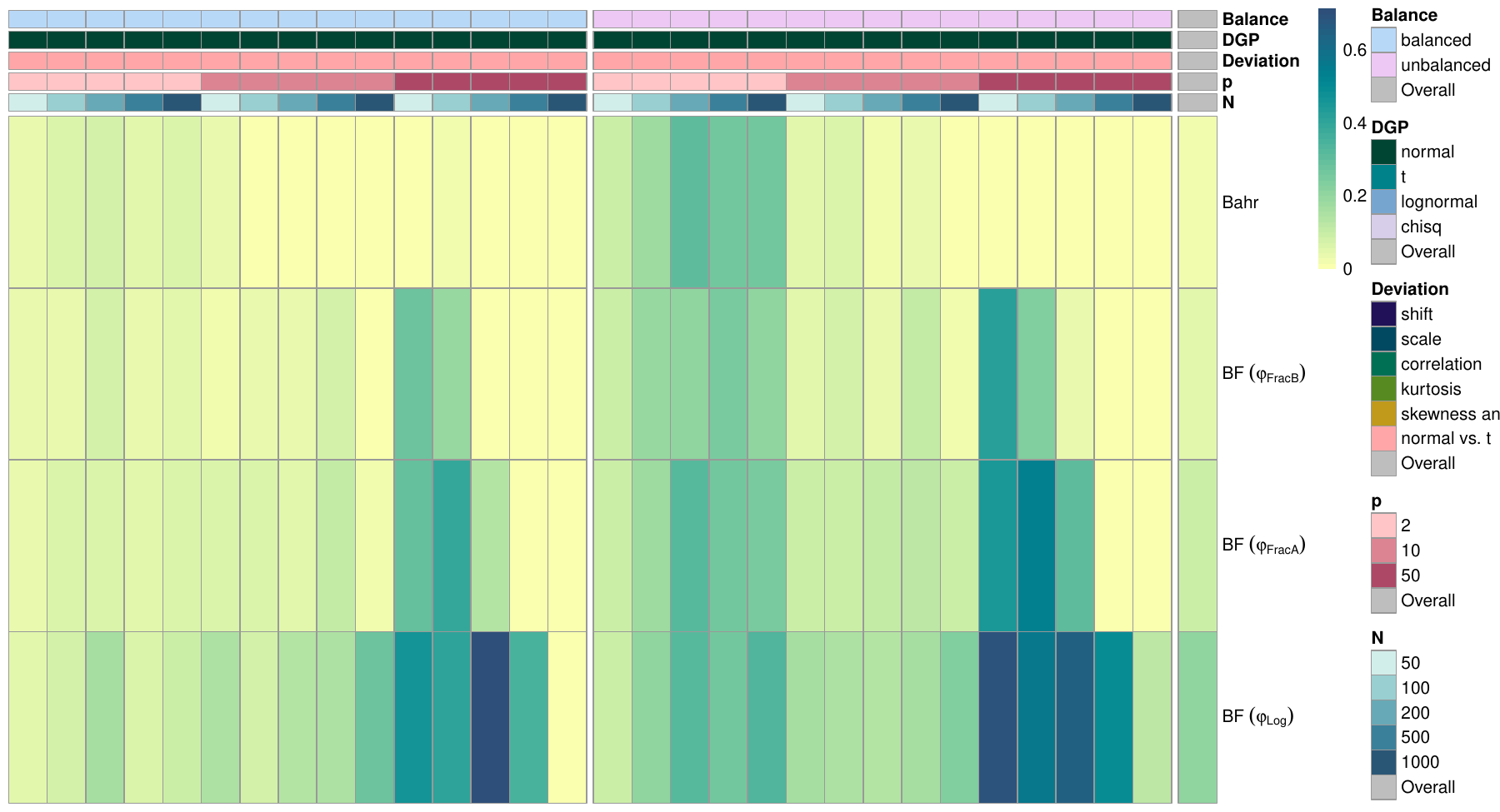}
		\caption{Mean PESR difference to best method per scenario for the normal vs.\ $t$ alternative and $k = 2$ datasets for all variants of the BF method \textcite{baringhaus_rigid_2010}. Selected variant: Bahr.}
	\end{figure}
	
	\begin{figure}[!h]
		\centering
		\includegraphics[width=\linewidth]{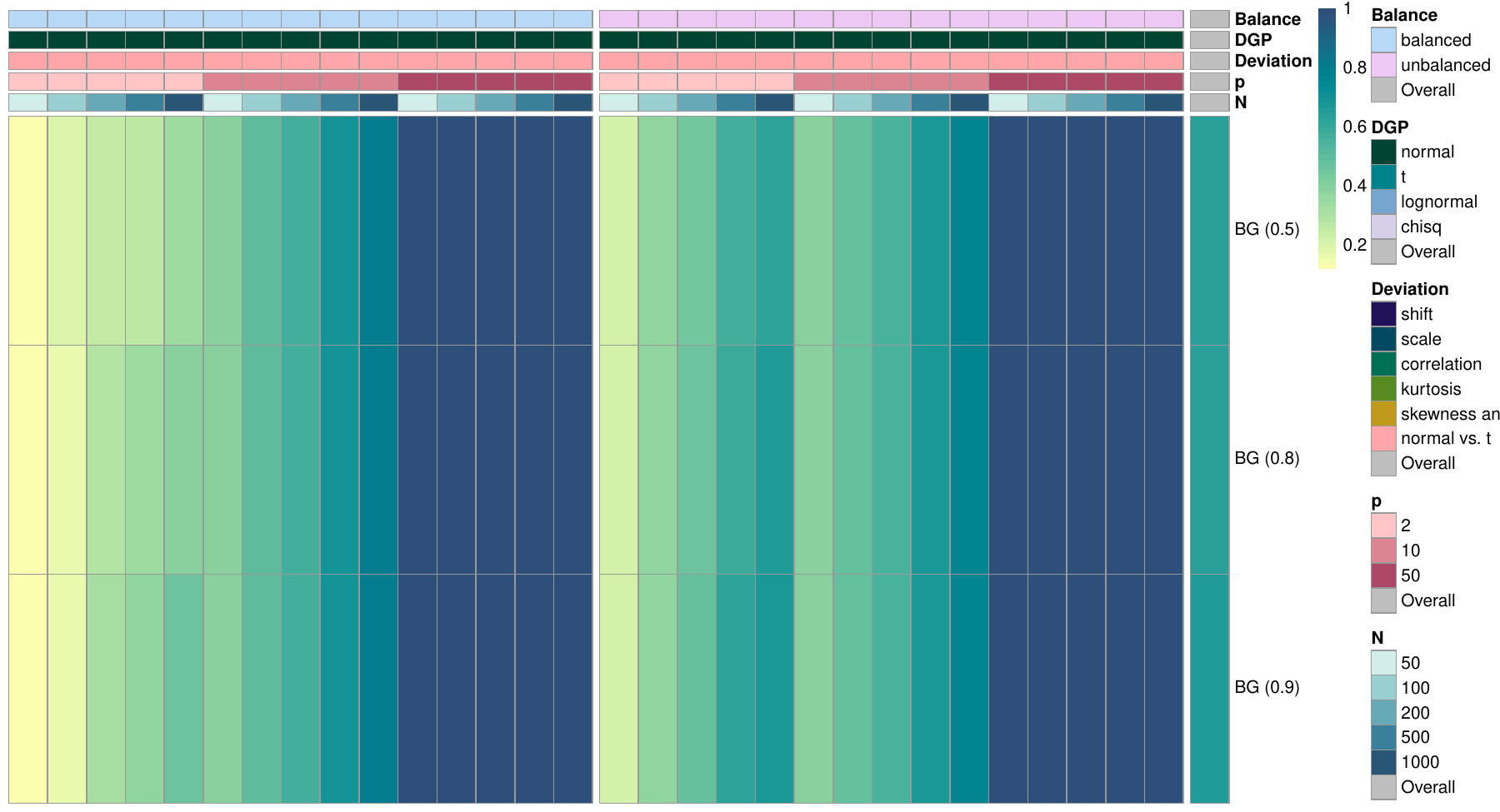}
		\caption{Mean PESR difference to best method per scenario for the normal vs.\ $t$ alternative and $k = 2$ datasets for all variants of the BG method \textcite{biau_asymptotic_2005}. Selected variant: BG~(0.5).}
	\end{figure}
	
	\begin{figure}[!h]
		\centering
		\includegraphics[width=\linewidth]{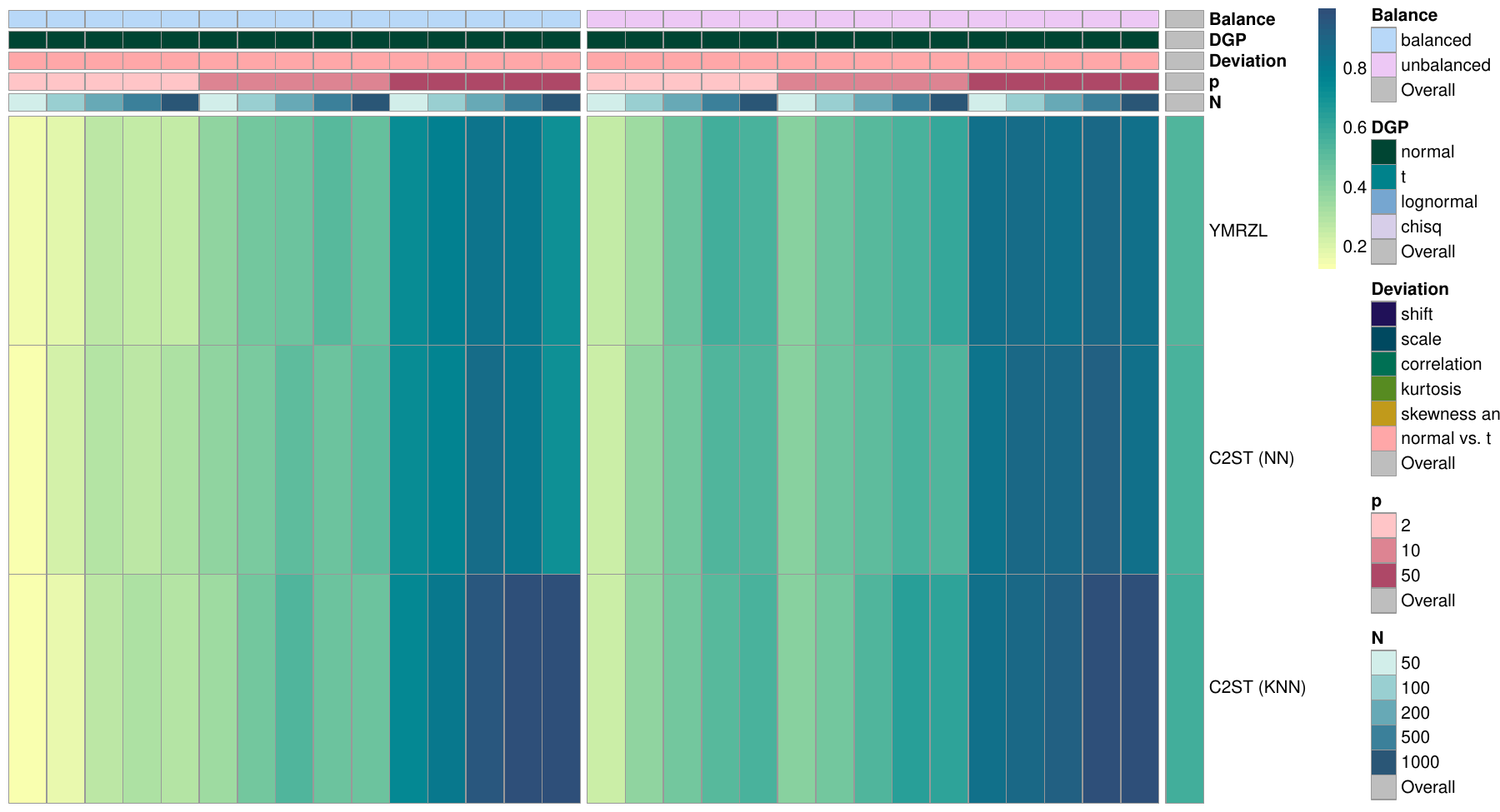}
		\caption{Mean PESR difference to best method per scenario for the normal vs.\ $t$ alternative and $k = 2$ datasets for all variants of the YMRZL and C2ST method \textcite{yu_two-sample_2007, lopez-paz_revisiting_2017}. Selected variant: YMRZL.}
	\end{figure}
	
	\begin{figure}[!h]
		\centering
		\includegraphics[width=\linewidth]{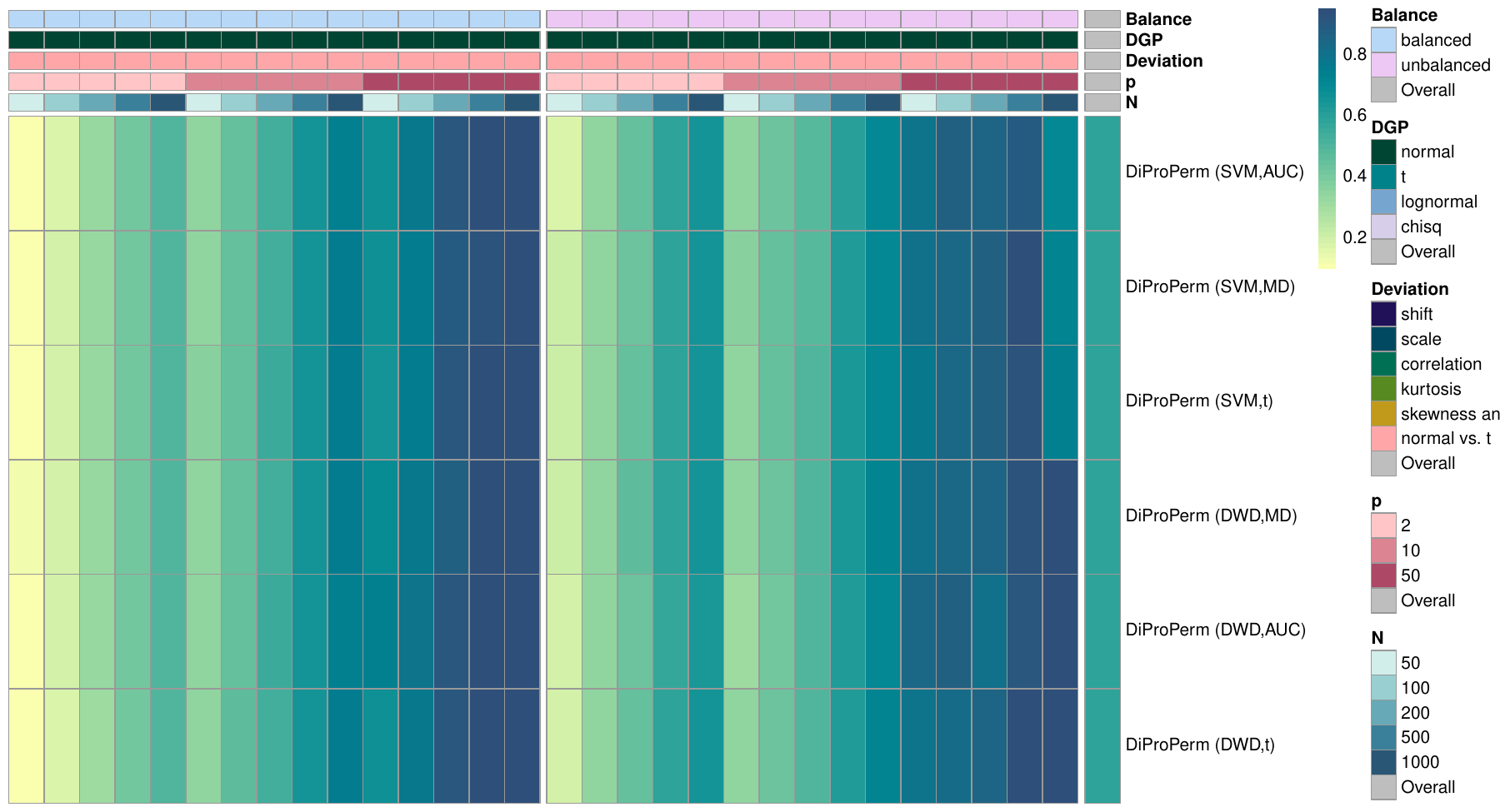}
		\caption{Mean PESR difference to best method per scenario for the normal vs.\ $t$ alternative and $k = 2$ datasets for all variants of the DiProPerm method \textcite{wei_direction-projection-permutation_2016}. Selected variant: DiProPerm~(SVM,AUC).}
	\end{figure}
	
	\begin{figure}[!h]
		\centering
		\includegraphics[width=\linewidth]{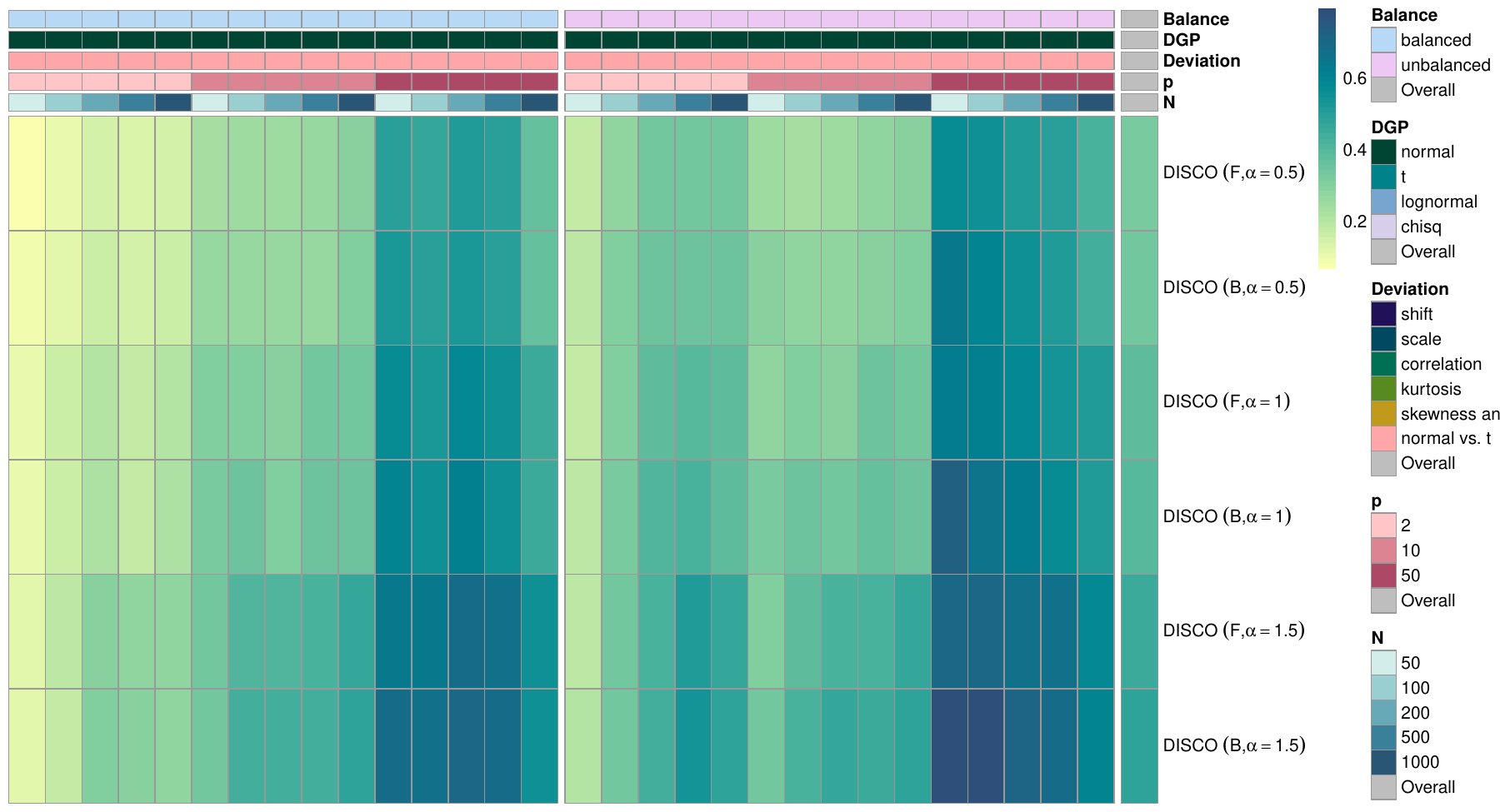}
		\caption{Mean PESR difference to best method per scenario for the normal vs.\ $t$ alternative and $k = 2$ datasets for all variants of the DISCO method \textcite{rizzo_disco_2010}. Selected variant: DISCO~(F,$\alpha=0.5$).}
	\end{figure}
	
	\begin{figure}[!h]
		\centering
		\includegraphics[width=\linewidth]{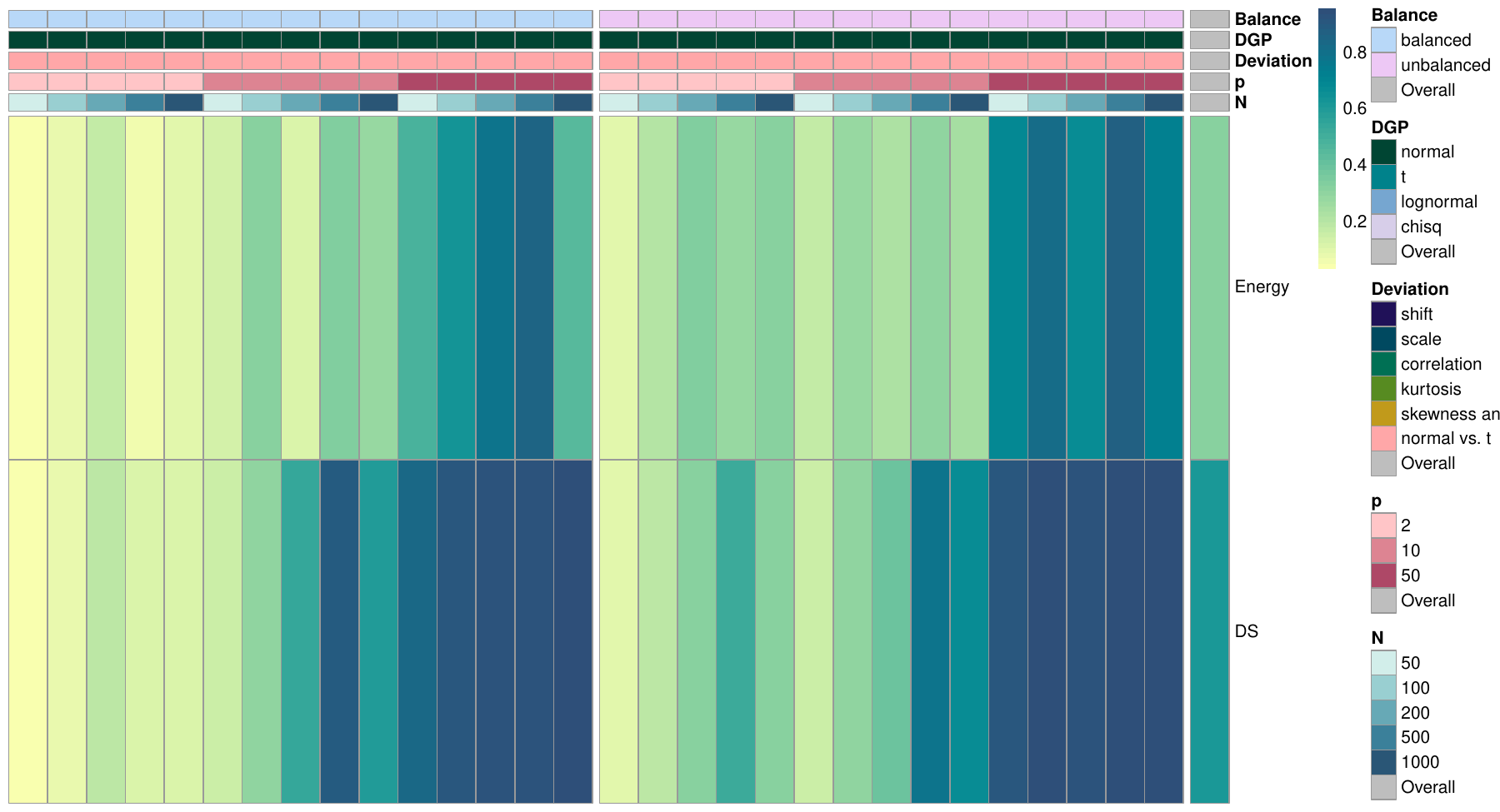}
		\caption{Mean PESR difference to best method per scenario for the normal vs.\ $t$ alternative and $k = 2$ datasets for all variants of the Energy method \textcite{szekely_testing_2004}. Selected variant: Energy.}
	\end{figure}
	
	\begin{figure}[!h]
		\centering
		\includegraphics[width=\linewidth]{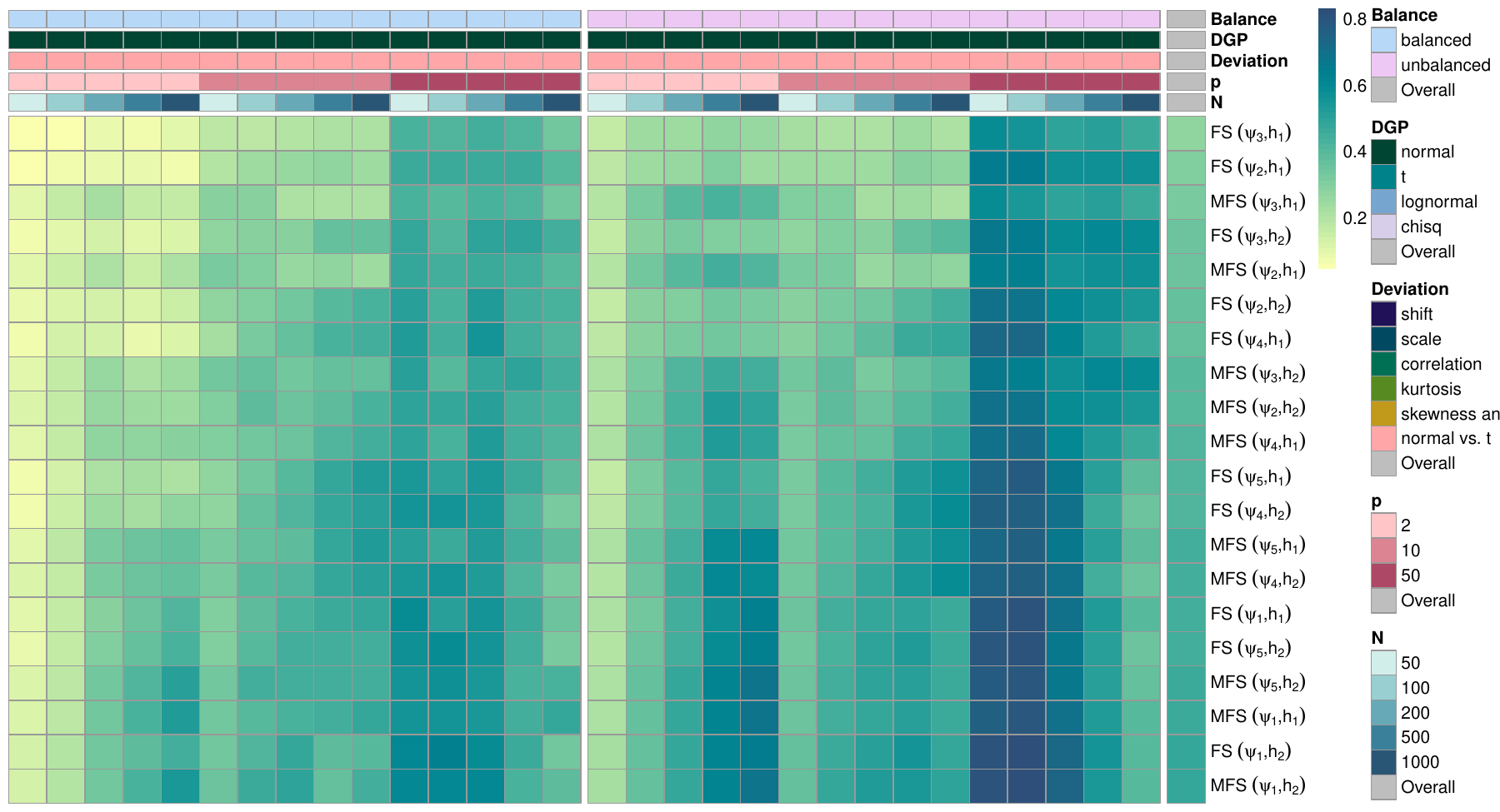}
		\caption{Mean PESR difference to best method per scenario for the normal vs.\ $t$ alternative and $k = 2$ datasets for all variants of the FS test \textcite{paul_clustering-based_2022}. Selected variant: FS~($\psi_3, h_1$).}
	\end{figure}
	
	\begin{figure}[!h]
		\centering
		\includegraphics[width=\linewidth]{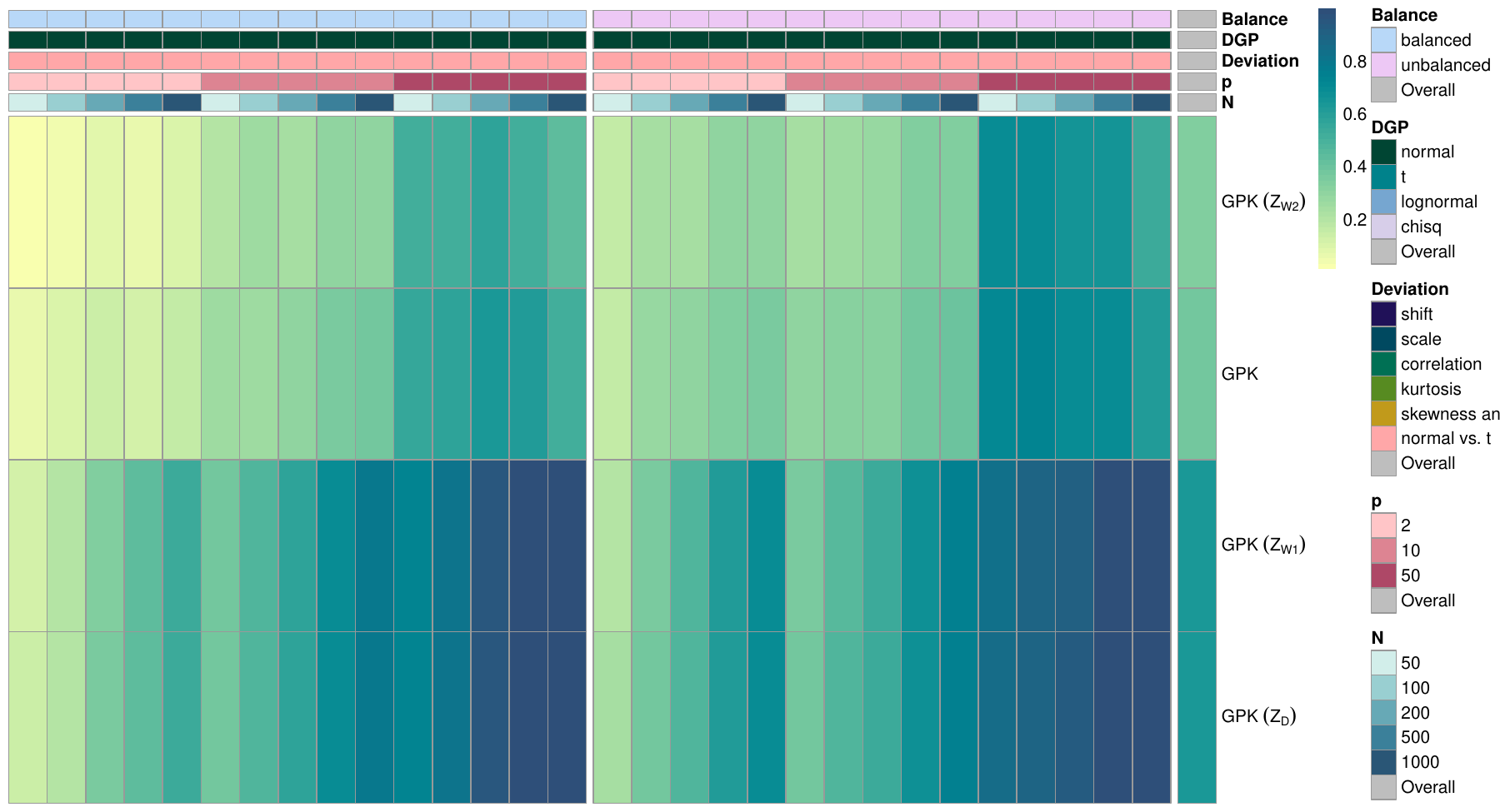}
		\caption{Mean PESR difference to best method per scenario for the normal vs.\ $t$ alternative and $k = 2$ datasets for all variants of the GPK test \textcite{song_generalized_2021}. Selected variant: GPK~(Z$_{\text{W}2}$).}
	\end{figure}
	
	\begin{figure}[!h]
		\centering
		\includegraphics[width=\linewidth]{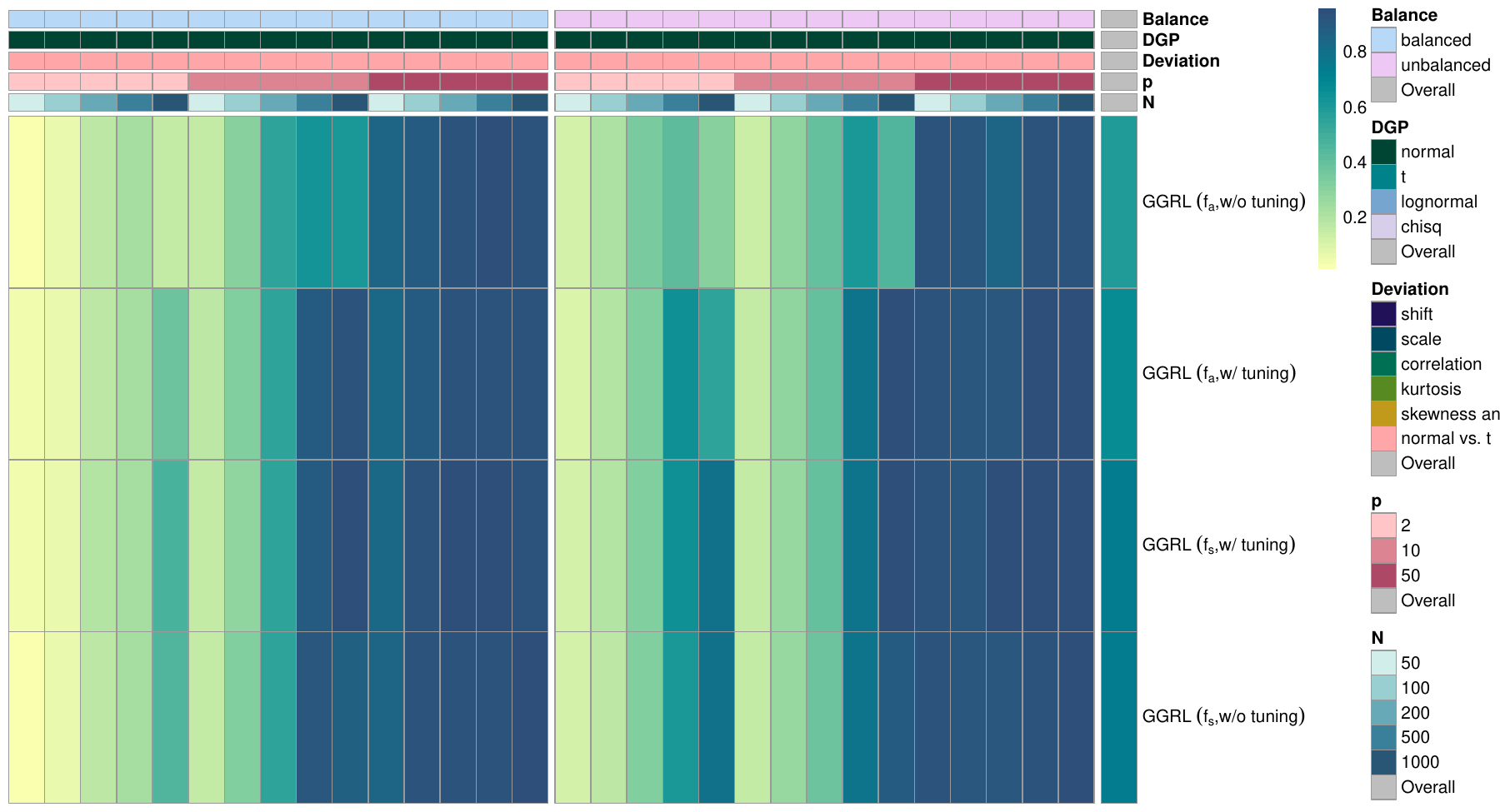}
		\caption{Mean PESR difference to best method per scenario for the normal vs.\ $t$ alternative and $k = 2$ datasets for all variants of the GGRL method \textcite{ganti_framework_1999}. Selected variant: GGRL~($f_a$,w/o~tuning).}
	\end{figure}
	
	\begin{figure}[!h]
		\centering
		\includegraphics[width=\linewidth]{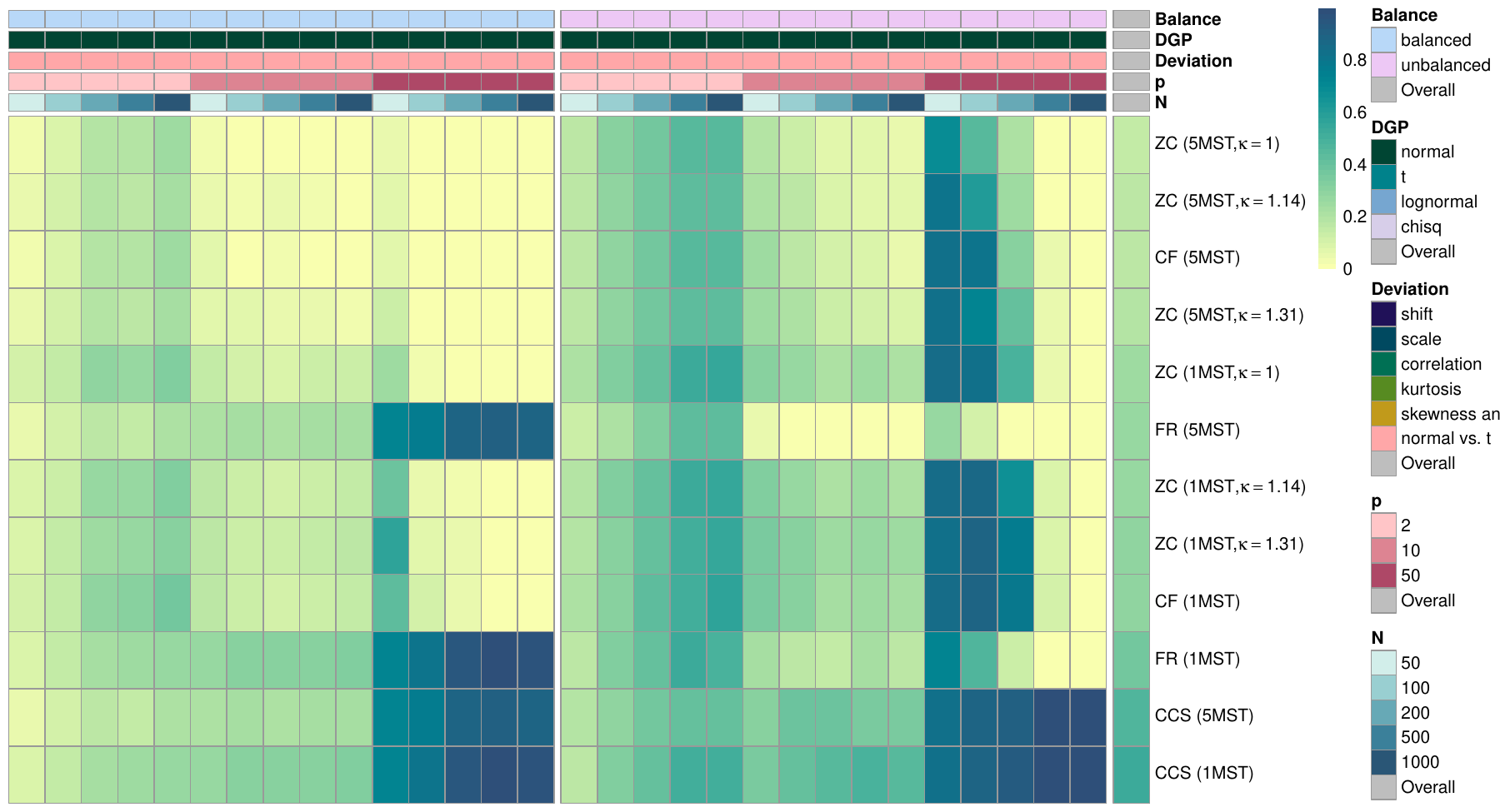}
		\caption{Mean PESR difference to best method per scenario for the normal vs.\ $t$ alternative and $k = 2$ datasets for all variants of the graph-based tests FR, CF, CCS, ZC \textcite{friedman_multivariate_1979, chen_new_2017, chen_weighted_2018, zhang_graph-based_2022}. Selected variant: ZC~(5MST,$\kappa=1$).}
	\end{figure}
	
	\begin{figure}[!h]
		\centering
		\includegraphics[width=\linewidth]{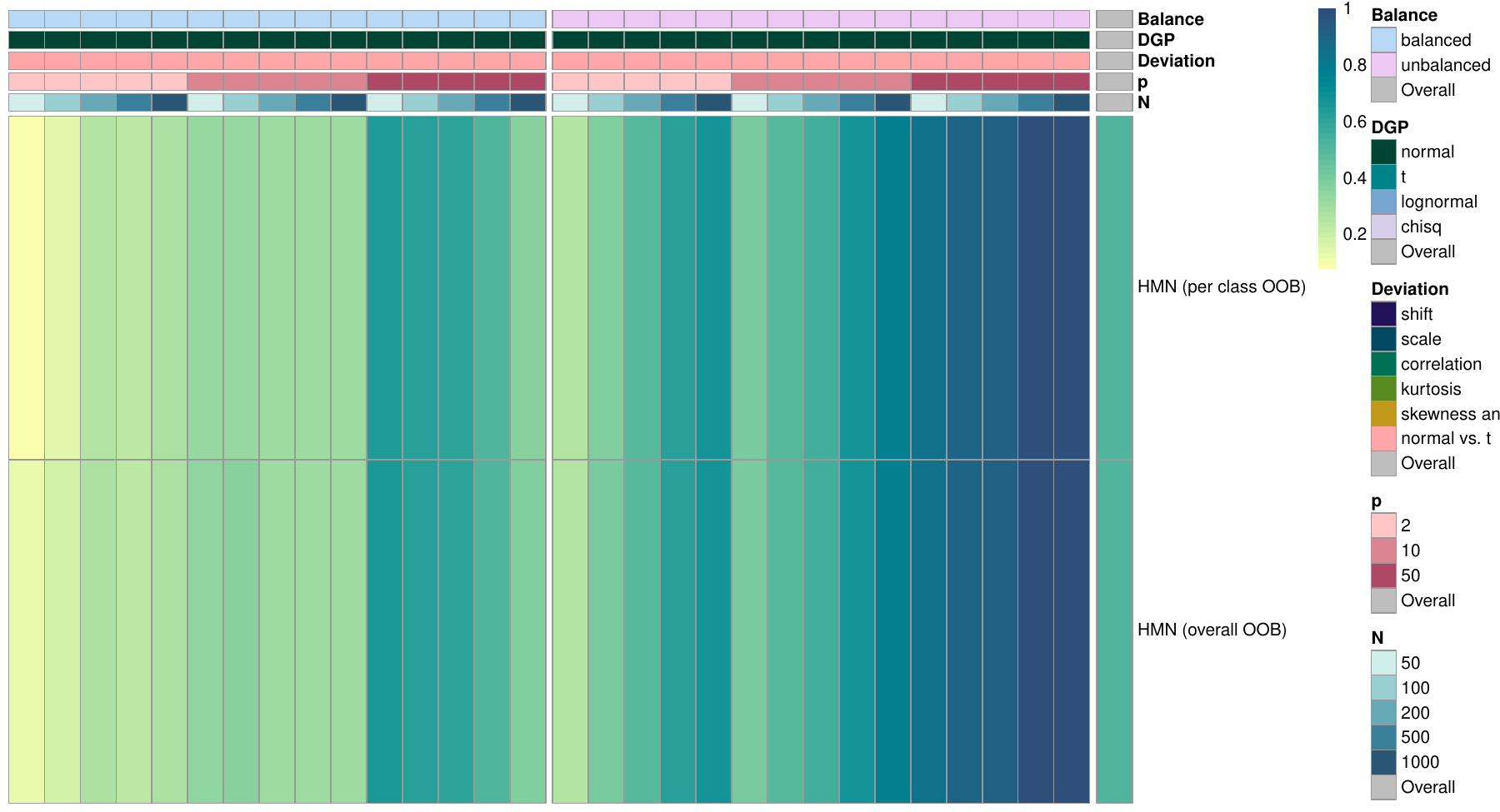}
		\caption{Mean PESR difference to best method per scenario for the normal vs.\ $t$ alternative and $k = 2$ datasets for all variants of the HMN method \textcite{hediger_use_2021}. Selected variant: HMN~(per class OOB).}
	\end{figure}
	
	\begin{figure}[!h]
		\centering
		\includegraphics[width=\linewidth]{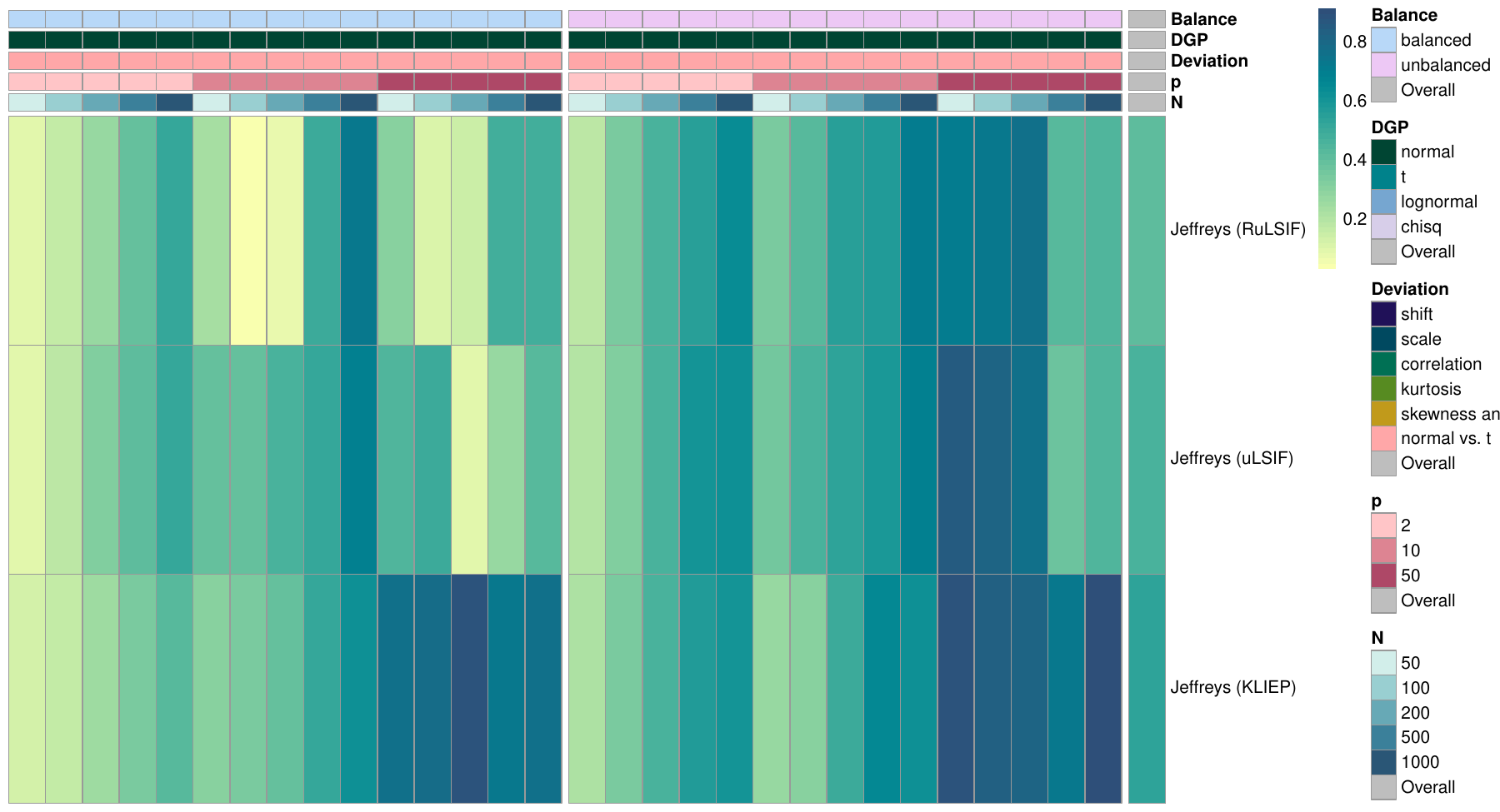}
		\caption{Mean PESR difference to best method per scenario for the normal vs.\ $t$ alternative and $k = 2$ datasets for all variants of the Jeffreys divergence \textcite{jeffreys_invariant_1997}. Selected variant: Jeffreys~(RuLSIF).}
	\end{figure}
	
	\begin{figure}[!h]
		\centering
		\includegraphics[width=\linewidth]{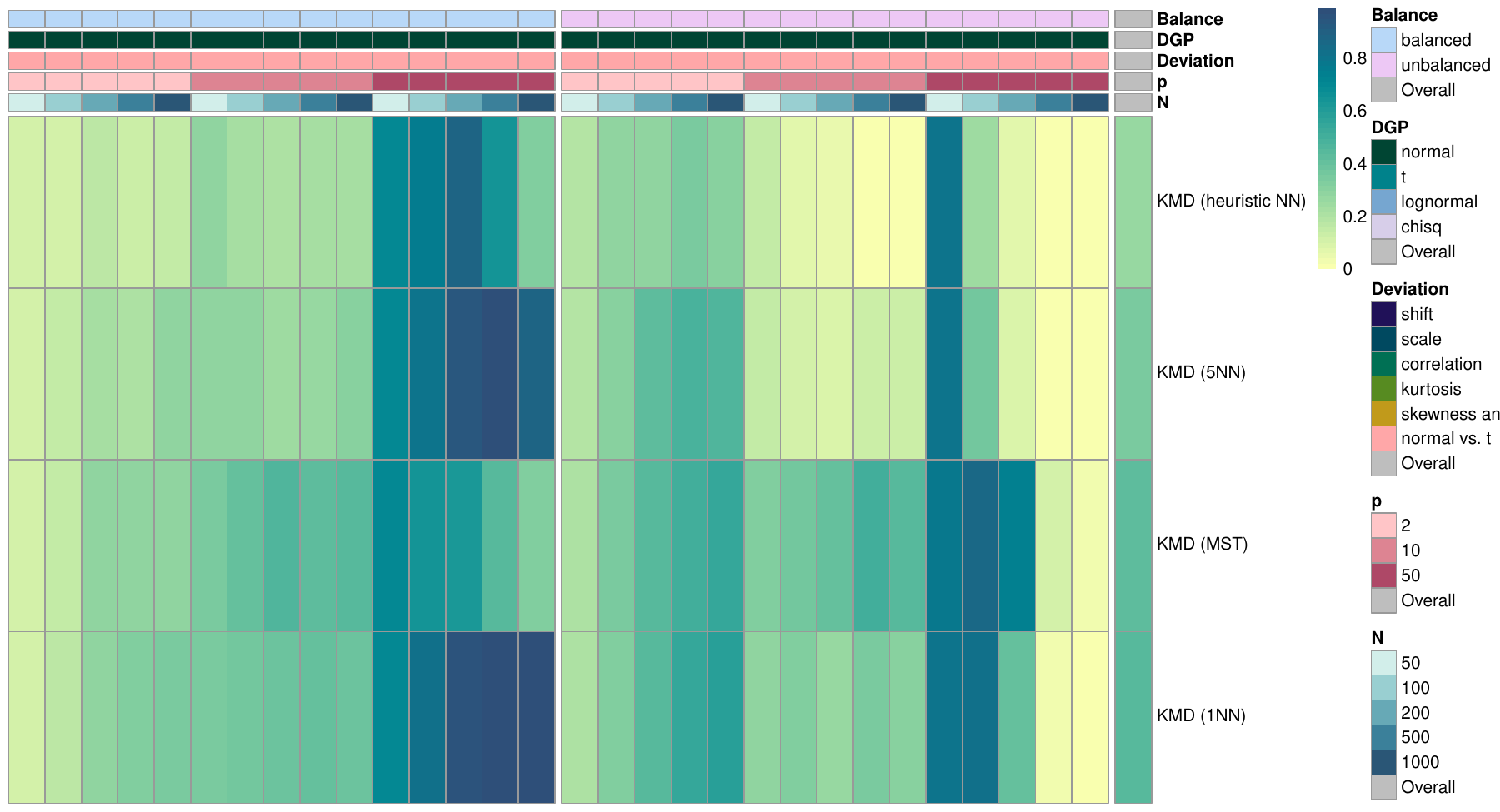}
		\caption{Mean PESR difference to best method per scenario for the normal vs.\ $t$ alternative and $k = 2$ datasets for all variants of the KMD \textcite{huang_kernel_2022}. Selected variant: KMD~(heuristic NN).}
	\end{figure}
	
	\begin{figure}[!h]
		\centering
		\includegraphics[width=\linewidth]{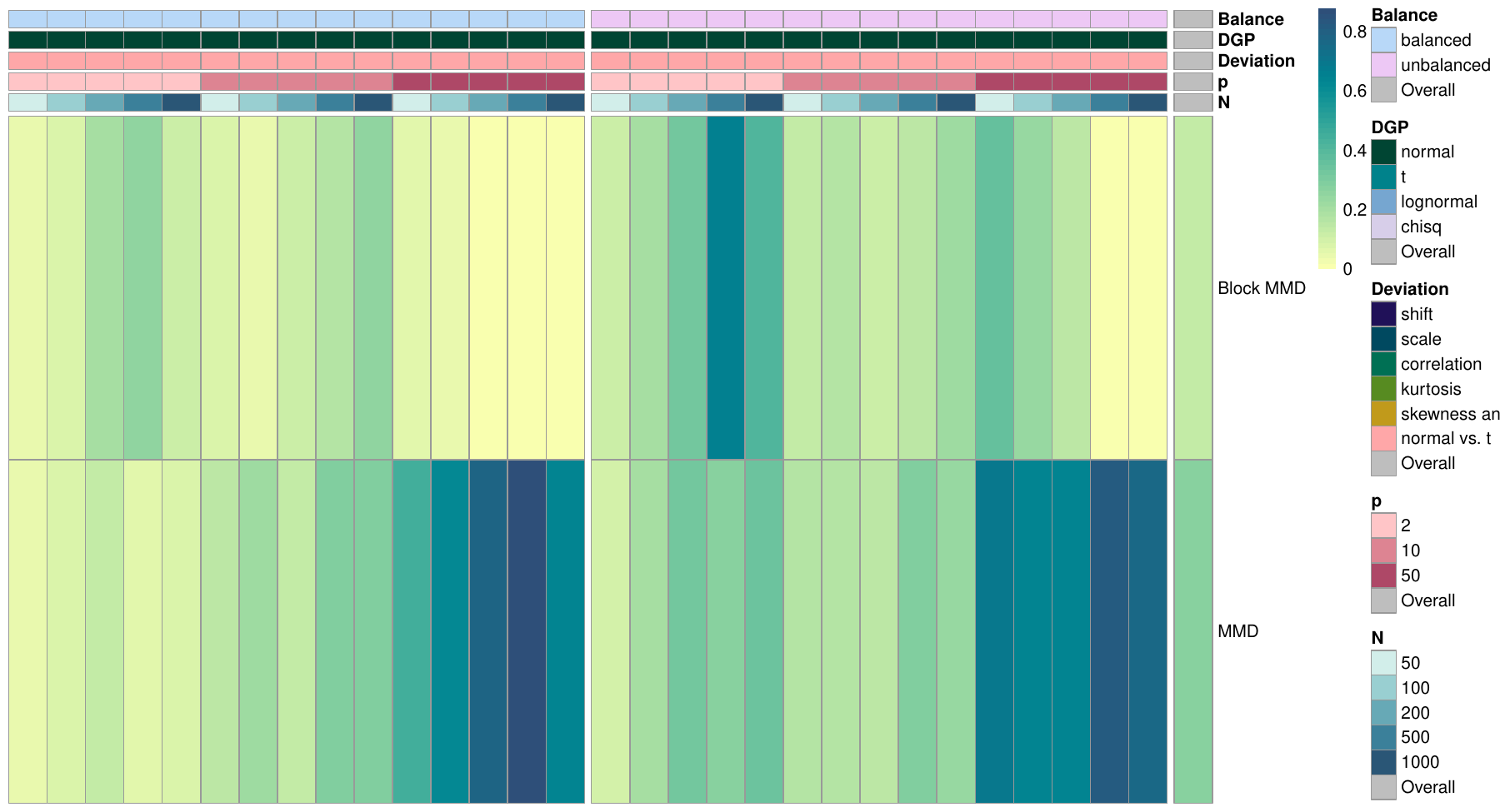}
		\caption{Mean PESR difference to best method per scenario for the normal vs.\ $t$ alternative and $k = 2$ datasets for all variants of the MMD \textcite{gretton_kernel_2006, zaremba_b-test_2013}. Selected variant: Block MMD.}
	\end{figure}
	
	\begin{figure}[!h]
		\centering
		\includegraphics[width=\linewidth]{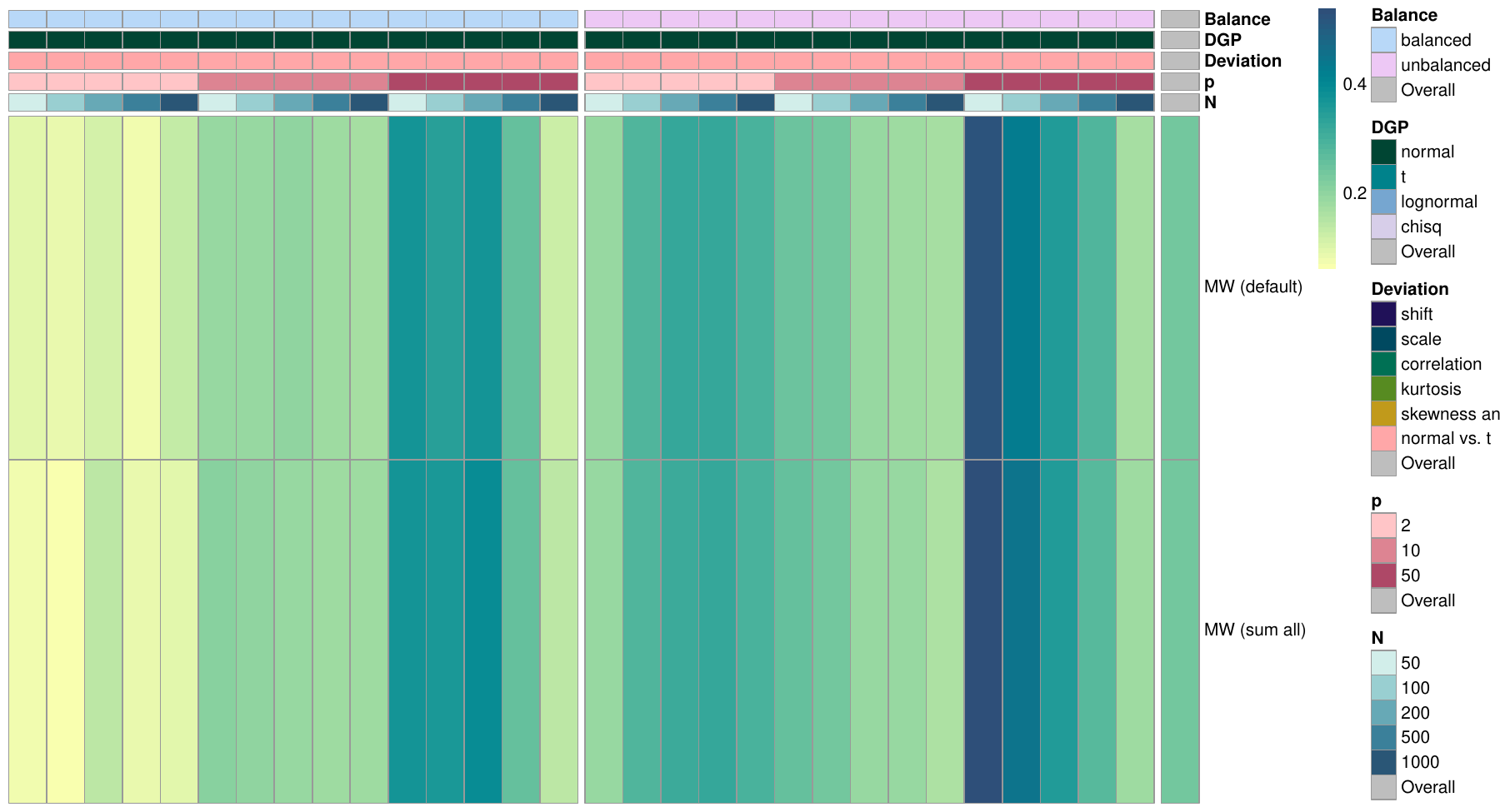}
		\caption{Mean PESR difference to best method per scenario for the normal vs.\ $t$ alternative and $k = 2$ datasets for all variants of the MW method \textcite{mukhopadhyay_nonparametric_2020}. Selected variant: MW~(default).}
	\end{figure}
	
	\begin{figure}[!h]
		\centering
		\includegraphics[width=\linewidth]{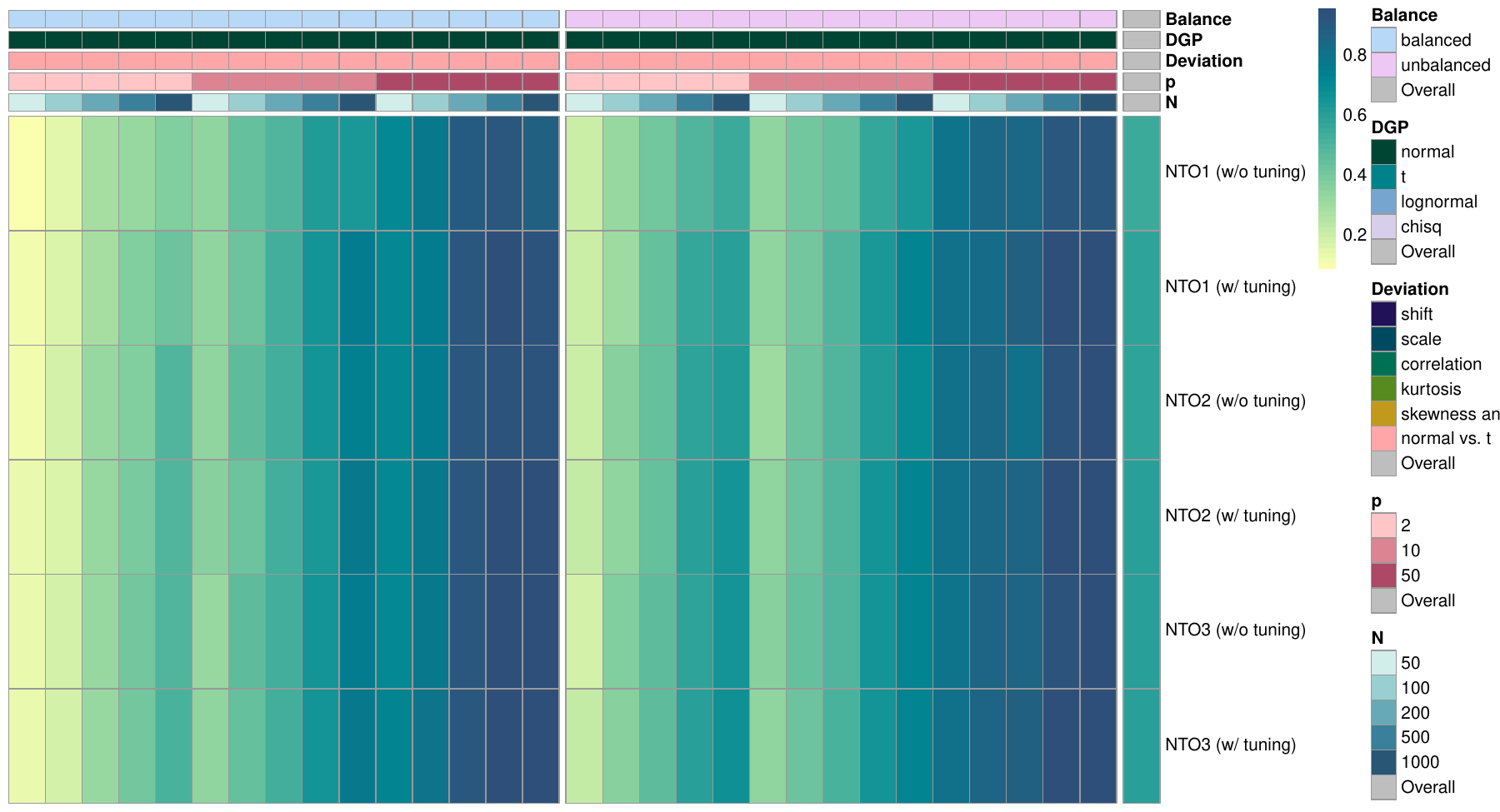}
		\caption{Mean PESR difference to best method per scenario for the normal vs.\ $t$ alternative and $k = 2$ datasets for all variants of the NKT method \textcite{ntoutsi_general_2008}. Selected variant: NTO1~(w/o~tuning).}
	\end{figure}
	
	\begin{figure}[!h]
		\centering
		\includegraphics[width=\linewidth]{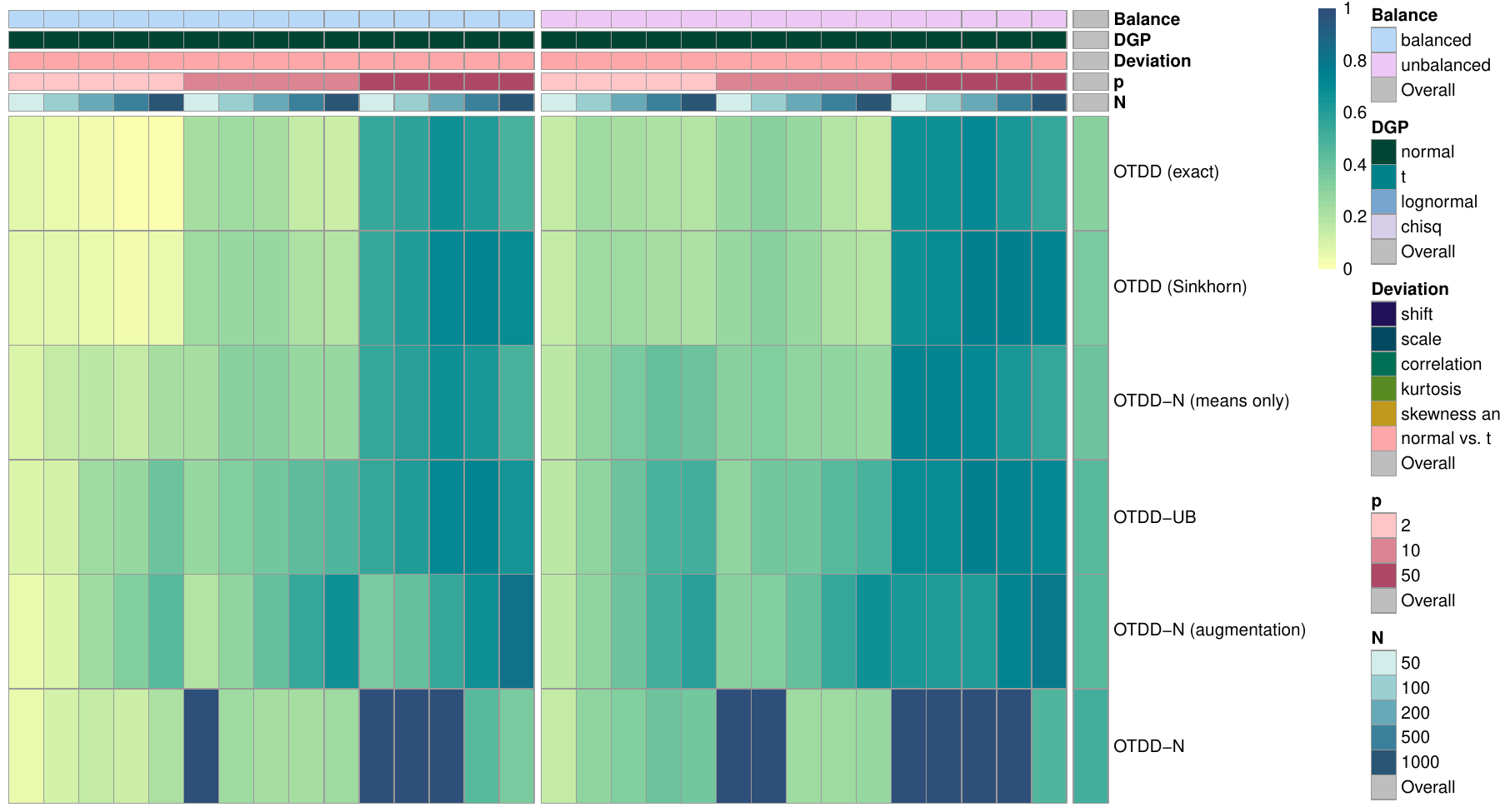}
		\caption{Mean PESR difference to best method per scenario for the normal vs.\ $t$ alternative and $k = 2$ datasets for all variants of the OTDD \textcite{alvarez-melis_geometric_2020}. Selected variant: OTDD~(exact).}
	\end{figure}
	
	\begin{figure}[!h]
		\centering
		\includegraphics[width=\linewidth]{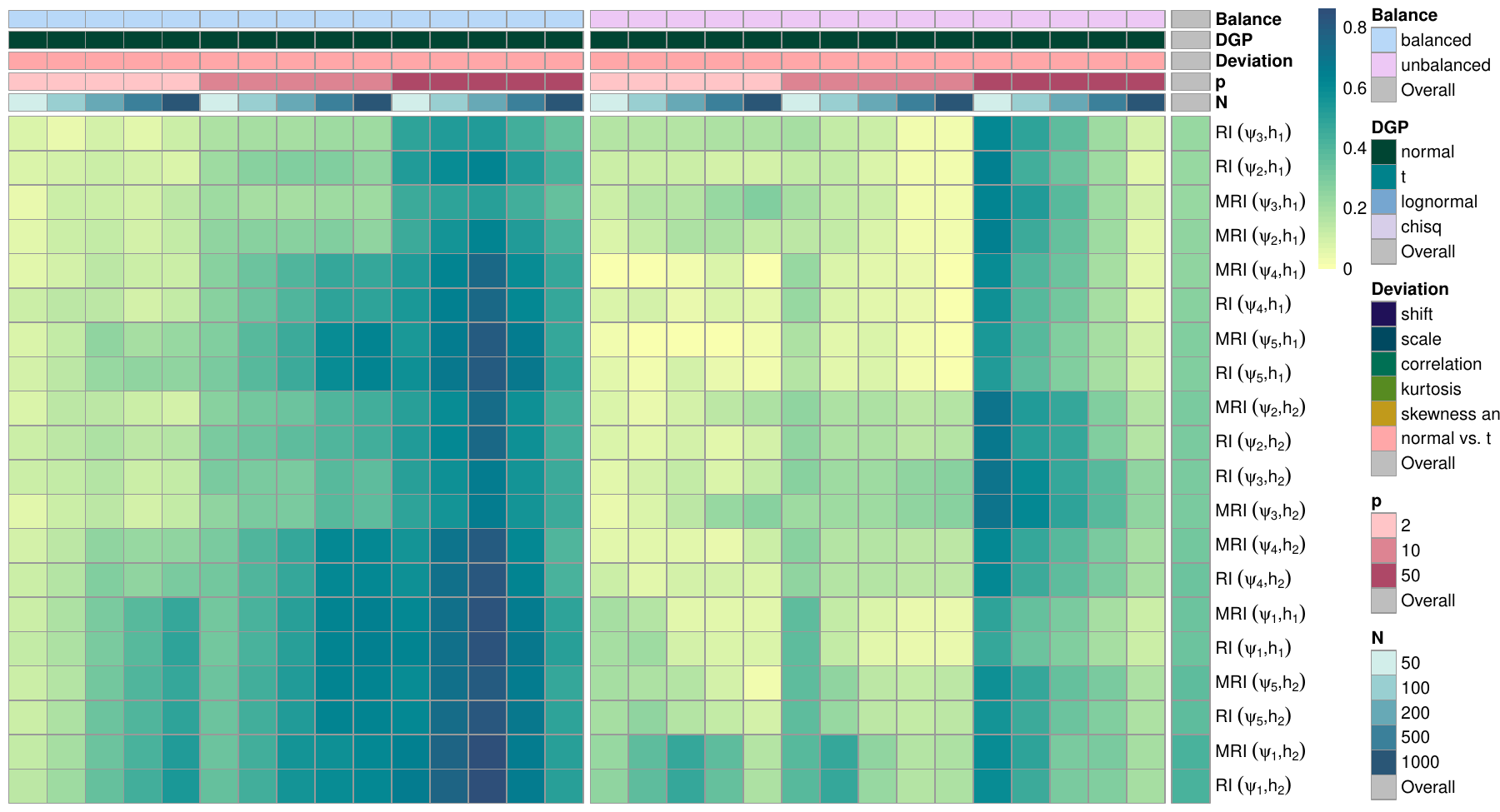}
		\caption{Mean PESR difference to best method per scenario for the normal vs.\ $t$ alternative and $k = 2$ datasets for all variants of the RI test \textcite{paul_clustering-based_2022}. Selected variant: RI~($\psi_3, h_1$).}
	\end{figure}
	
	\begin{figure}[!h]
		\centering
		\includegraphics[width=\linewidth]{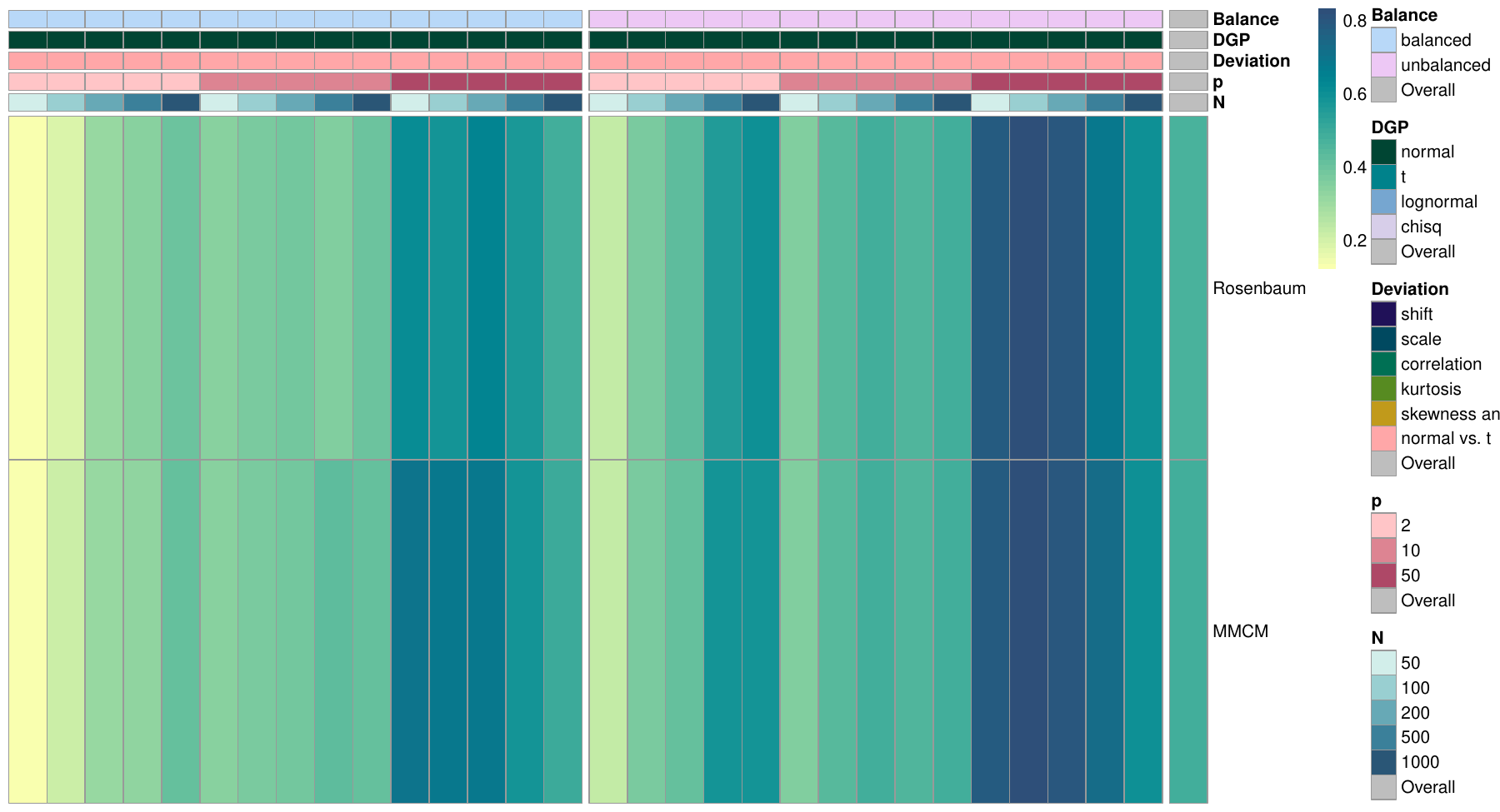}
		\caption{Mean PESR difference to best method per scenario for the normal vs.\ $t$ alternative and $k = 2$ datasets for all variants of the Rosenbaum test \textcite{rosenbaum_exact_2005, mukherjee_distribution-free_2022}. Selected variant: Rosenbaum.}
	\end{figure}
	
	\begin{figure}[!h]
		\centering
		\includegraphics[width=\linewidth]{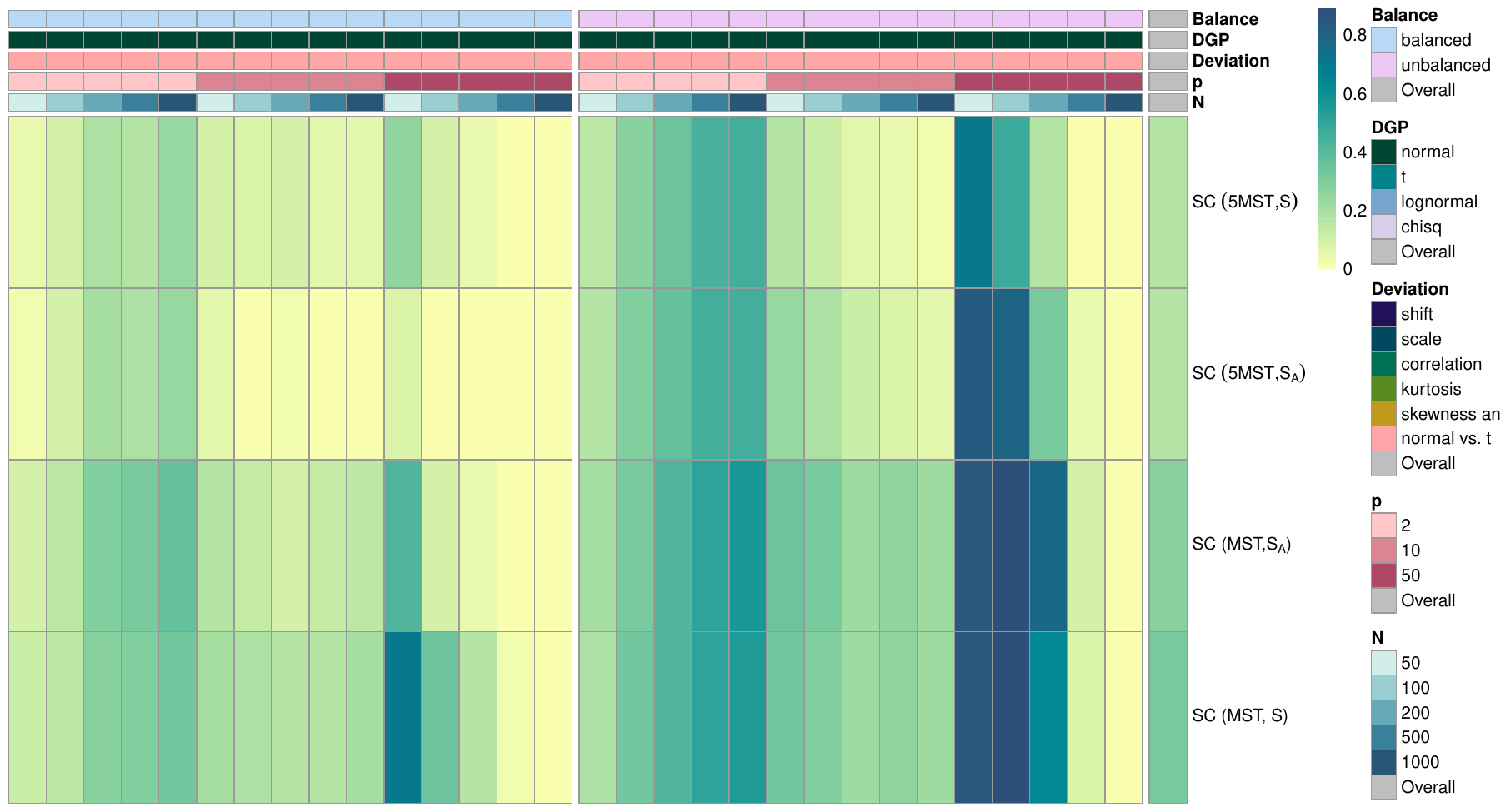}
		\caption{Mean PESR difference to best method per scenario for the normal vs.\ $t$ alternative and $k = 2$ datasets for all variants of the SC test \textcite{song_new_2022}. Selected variant: SC~(5MST,S).}
	\end{figure}
	
	\begin{figure}[!h]
		\centering
		\includegraphics[width=\linewidth]{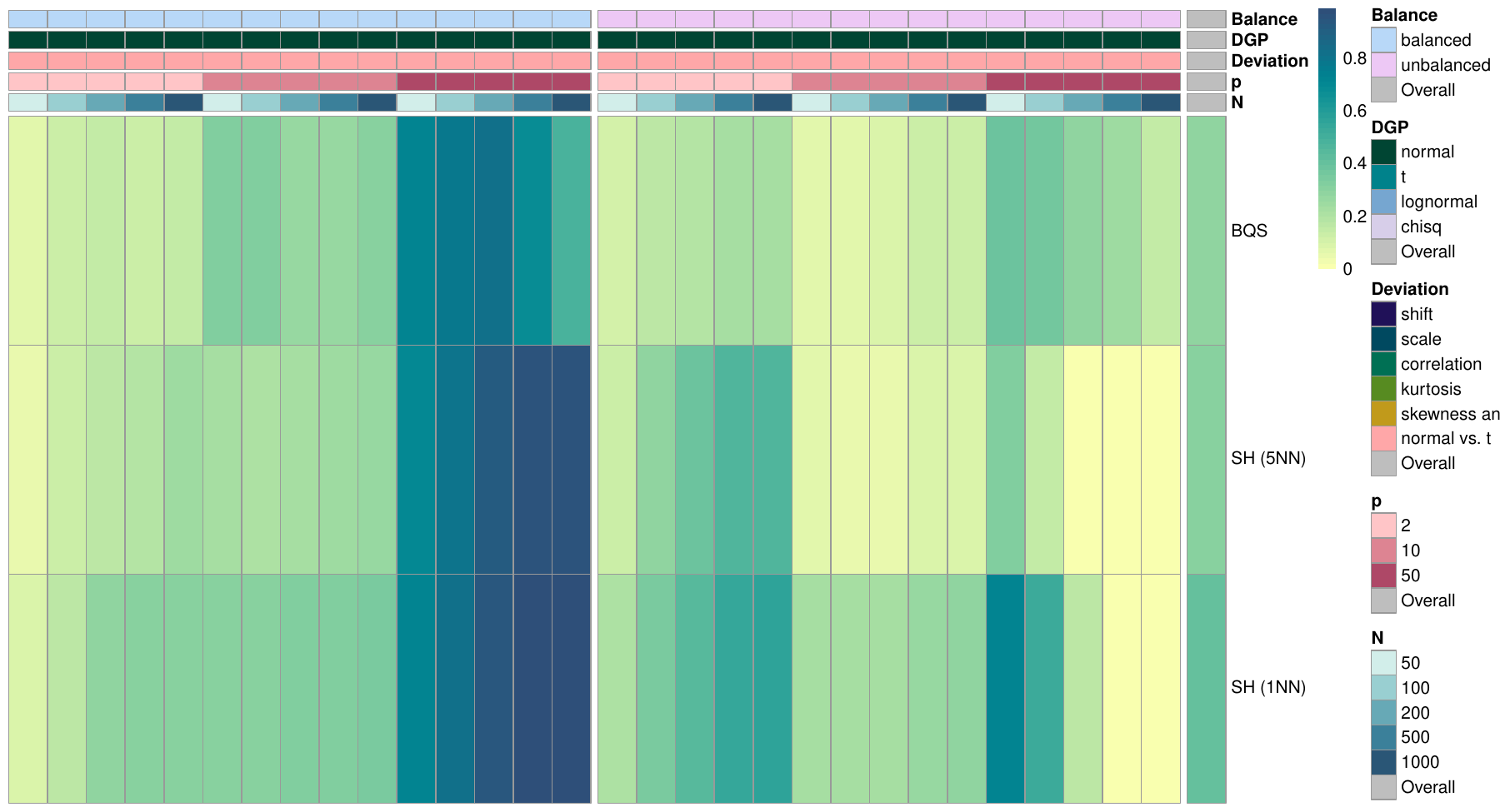}
		\caption{Mean PESR difference to best method per scenario for the normal vs.\ $t$ alternative and $k = 2$ datasets for all variants of the SH test \textcite{schilling_multivariate_1986, henze_multivariate_1988, barakat_multivariate_1996}. Selected variants: BQS.}
	\end{figure}
	\clearpage
	
	\paragraph{Kurtosis}\mbox{} 
	
	\begin{figure}[!h]
		\centering
		\includegraphics[width=\linewidth]{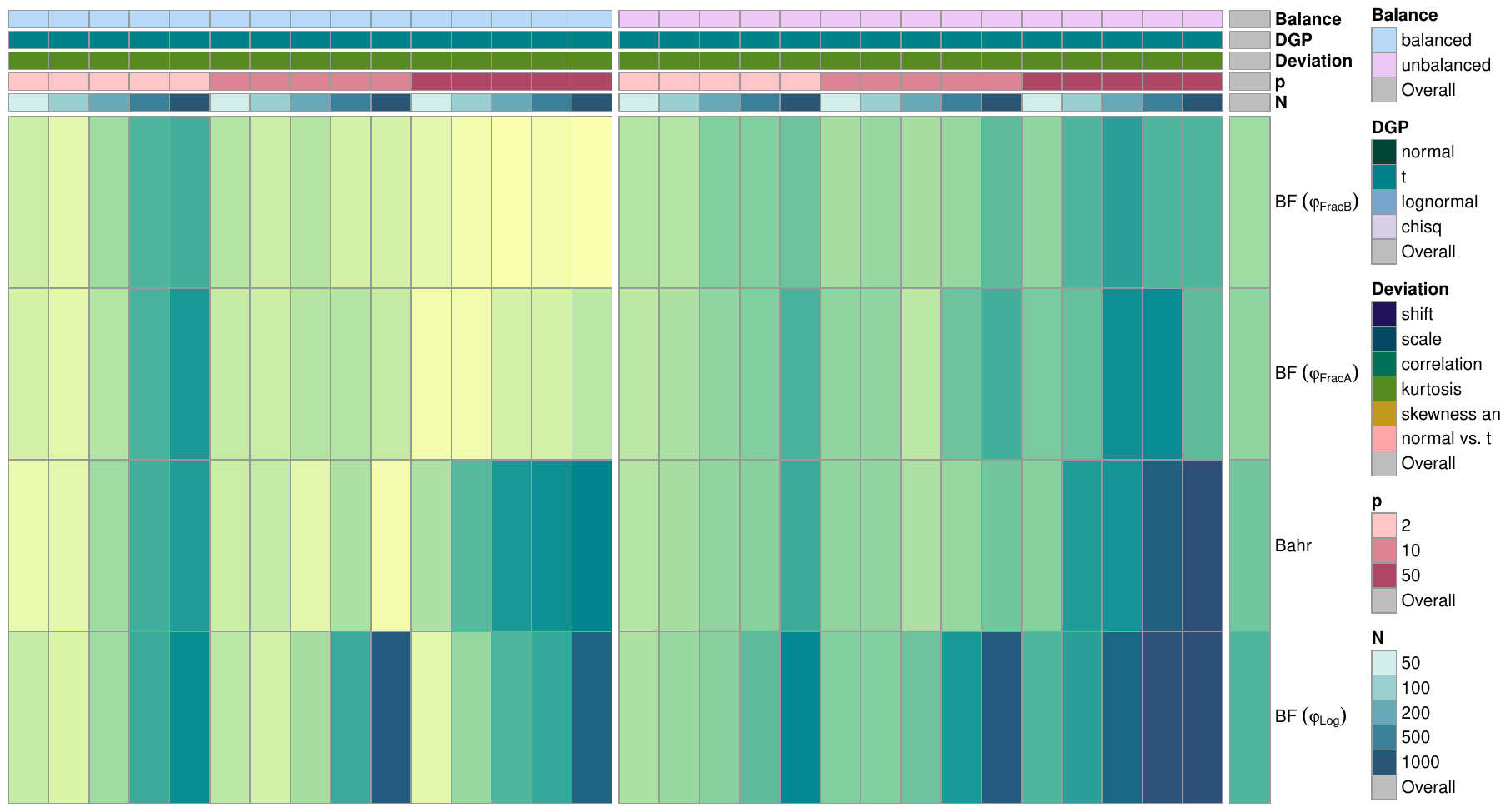}
		\caption{Mean PESR difference to best method per scenario for the kurtosis alternative and $k = 2$ datasets for all variants of the BF method \textcite{baringhaus_rigid_2010}. Selected variant: BF~($\varphi_{\text{FracB}}$).}
	\end{figure}
	
	\begin{figure}[!h]
		\centering
		\includegraphics[width=\linewidth]{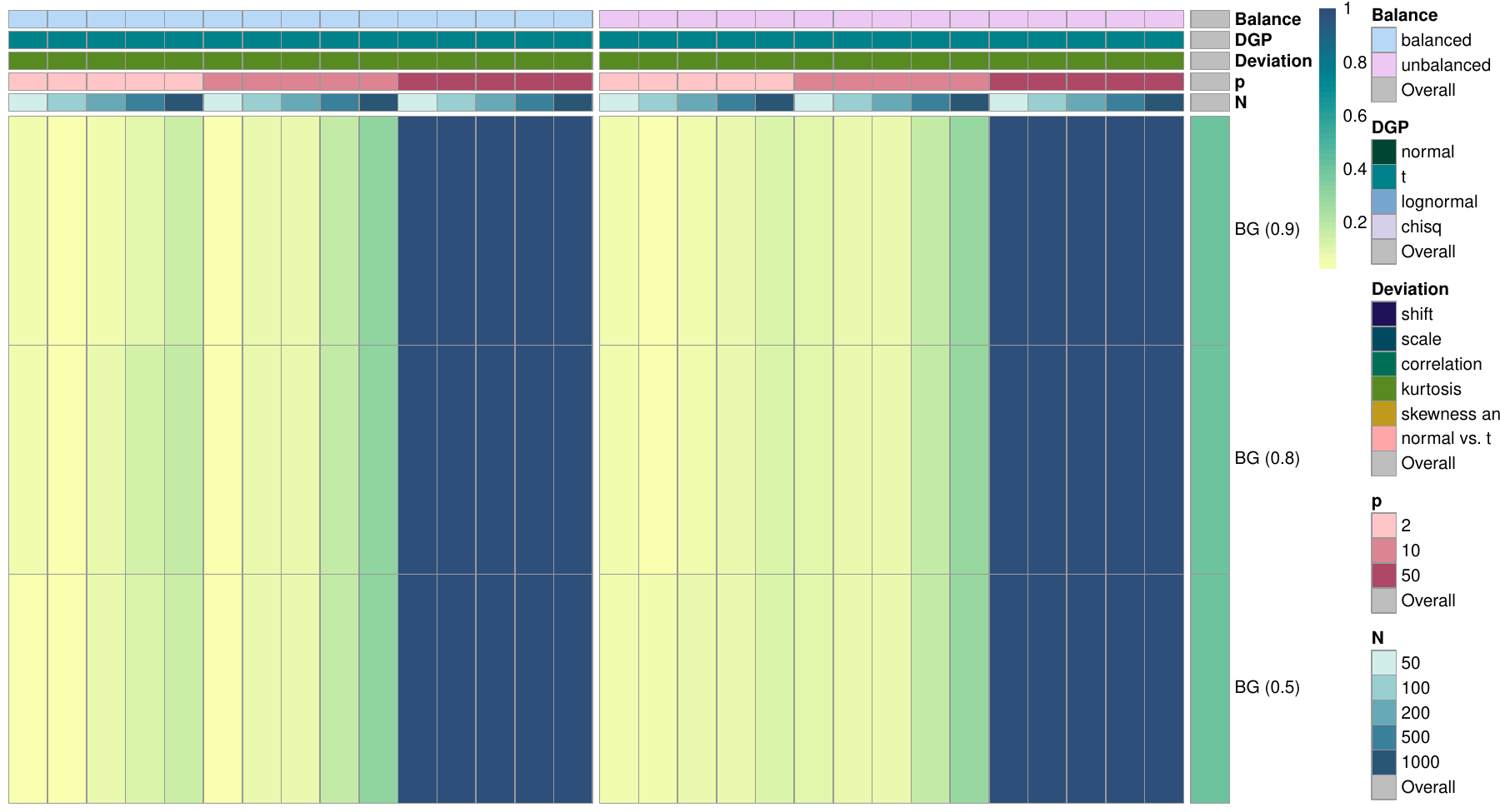}
		\caption{Mean PESR difference to best method per scenario for the kurtosis alternative and $k = 2$ datasets for all variants of the BG method \textcite{biau_asymptotic_2005}. Selected variant: BG~(0.9).}
	\end{figure}
	
	\begin{figure}[!h]
		\centering
		\includegraphics[width=\linewidth]{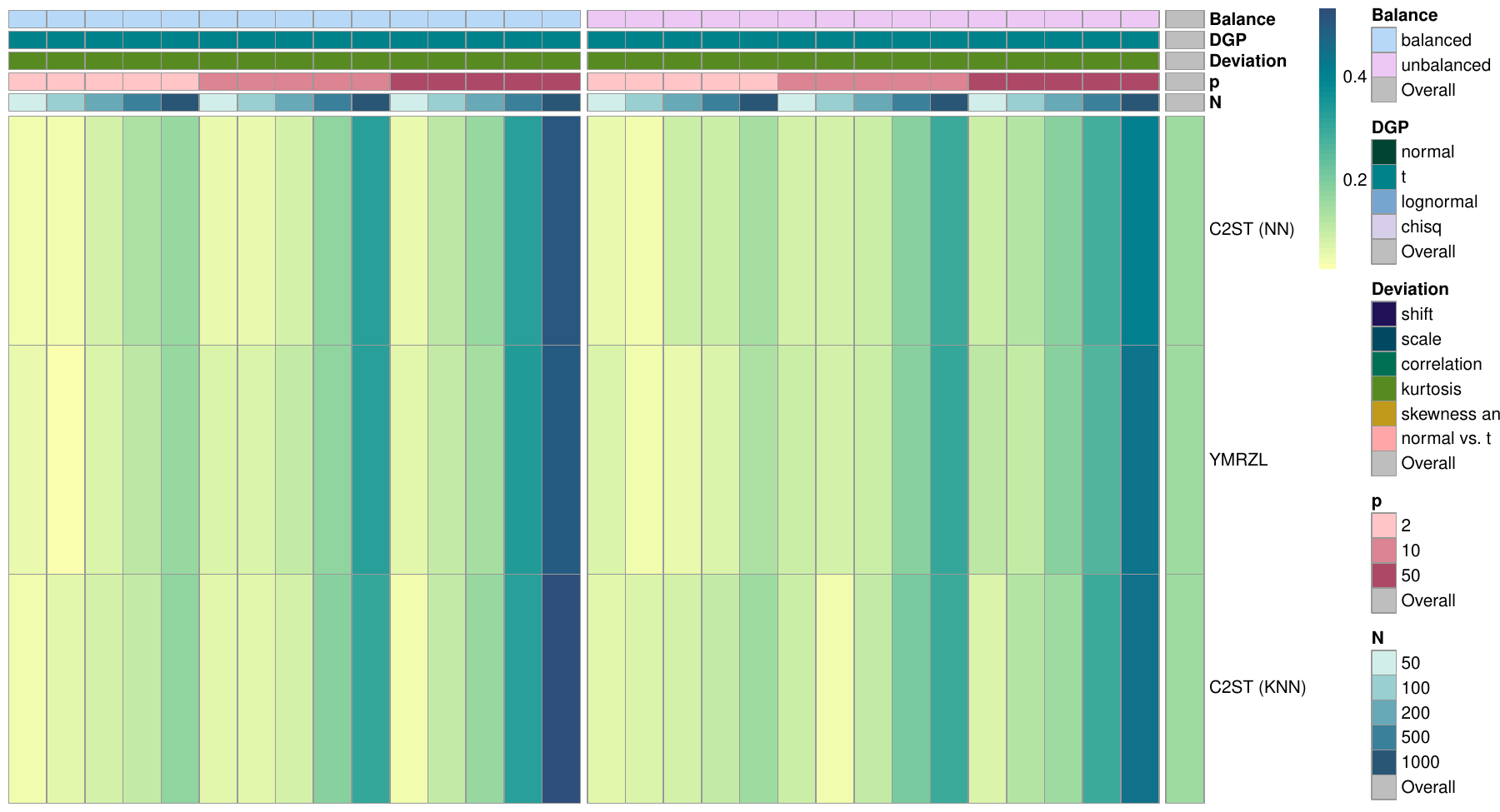}
		\caption{Mean PESR difference to best method per scenario for the kurtosis alternative and $k = 2$ datasets for all variants of the YMRZL and C2ST method \textcite{yu_two-sample_2007, lopez-paz_revisiting_2017}. Selected variant: C2ST~(NN).}
	\end{figure}
	
	\begin{figure}[!h]
		\centering
		\includegraphics[width=\linewidth]{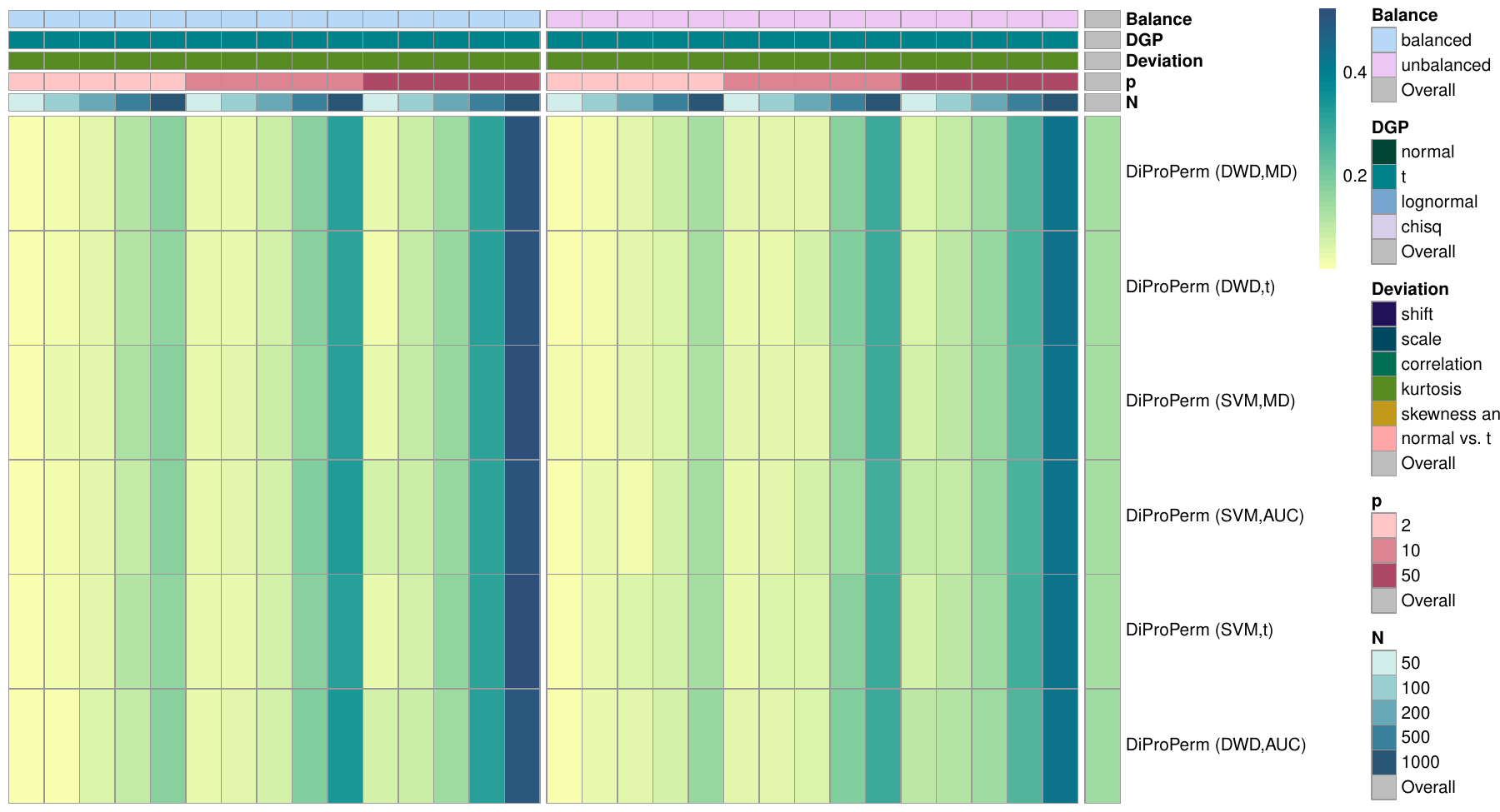}
		\caption{Mean PESR difference to best method per scenario for the kurtosis alternative and $k = 2$ datasets for all variants of the DiProPerm method \textcite{wei_direction-projection-permutation_2016}. Selected variant: DiProPerm~(DWD,MD).}
	\end{figure}
	
	\begin{figure}[!h]
		\centering
		\includegraphics[width=\linewidth]{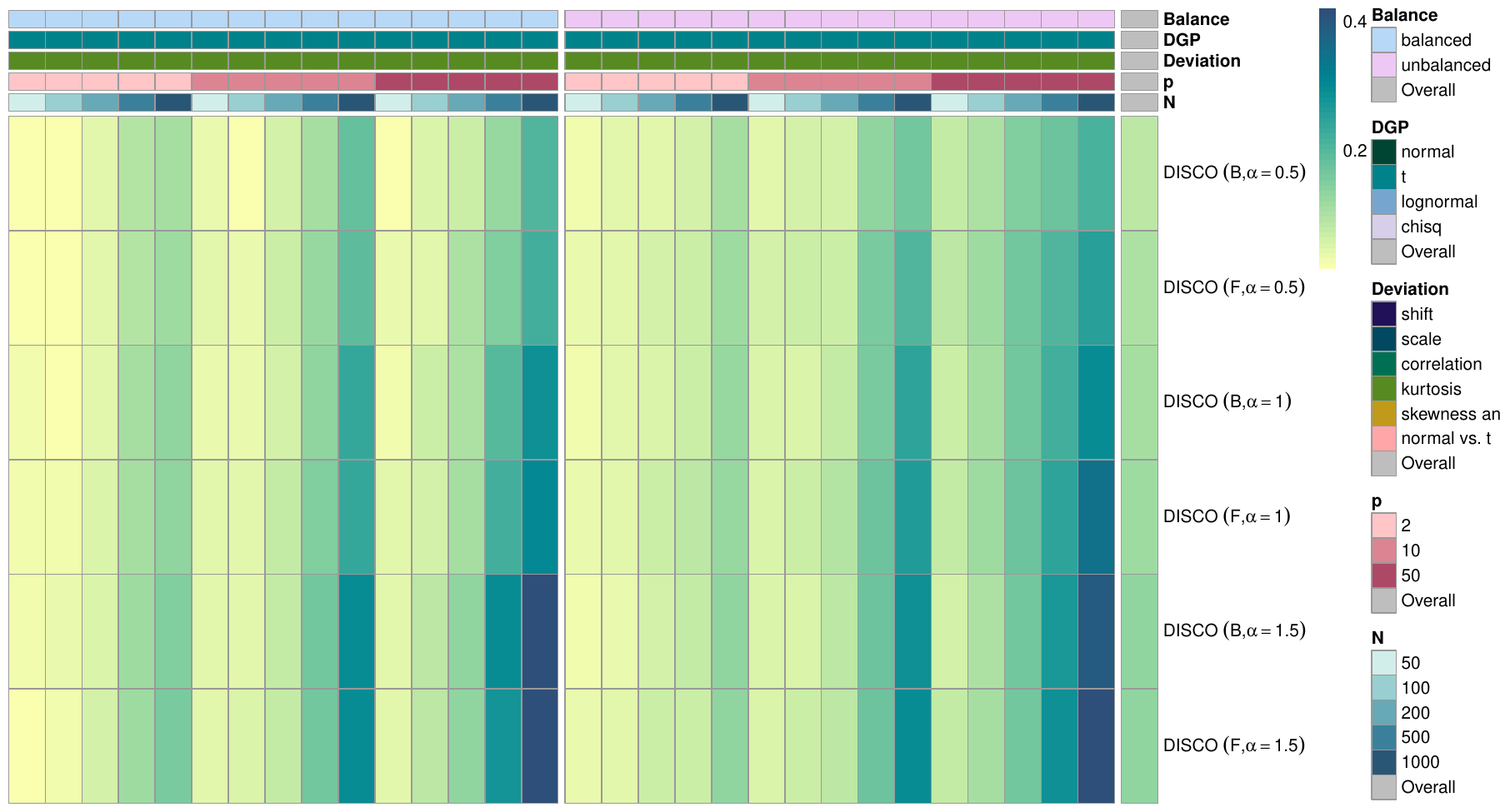}
		\caption{Mean PESR difference to best method per scenario for the kurtosis alternative and $k = 2$ datasets for all variants of the DISCO method \textcite{rizzo_disco_2010}. Selected variant: DISCO~(B,$\alpha=0.5$).}
	\end{figure}
	
	\begin{figure}[!h]
		\centering
		\includegraphics[width=\linewidth]{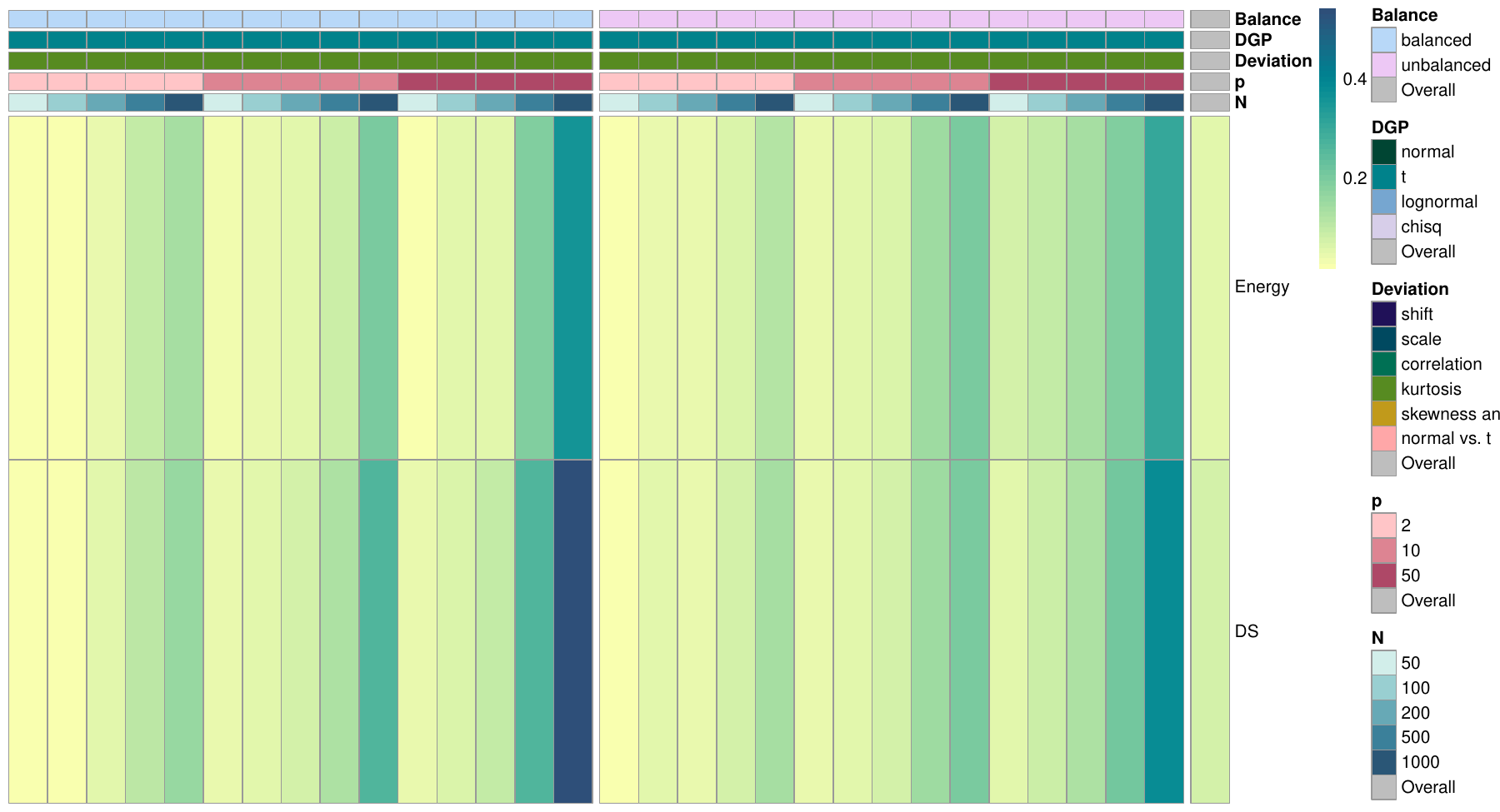}
		\caption{Mean PESR difference to best method per scenario for the kurtosis alternative and $k = 2$ datasets for all variants of the Energy method \textcite{szekely_testing_2004}. Selected variant: Energy.}
	\end{figure}
	
	\begin{figure}[!h]
		\centering
		\includegraphics[width=\linewidth]{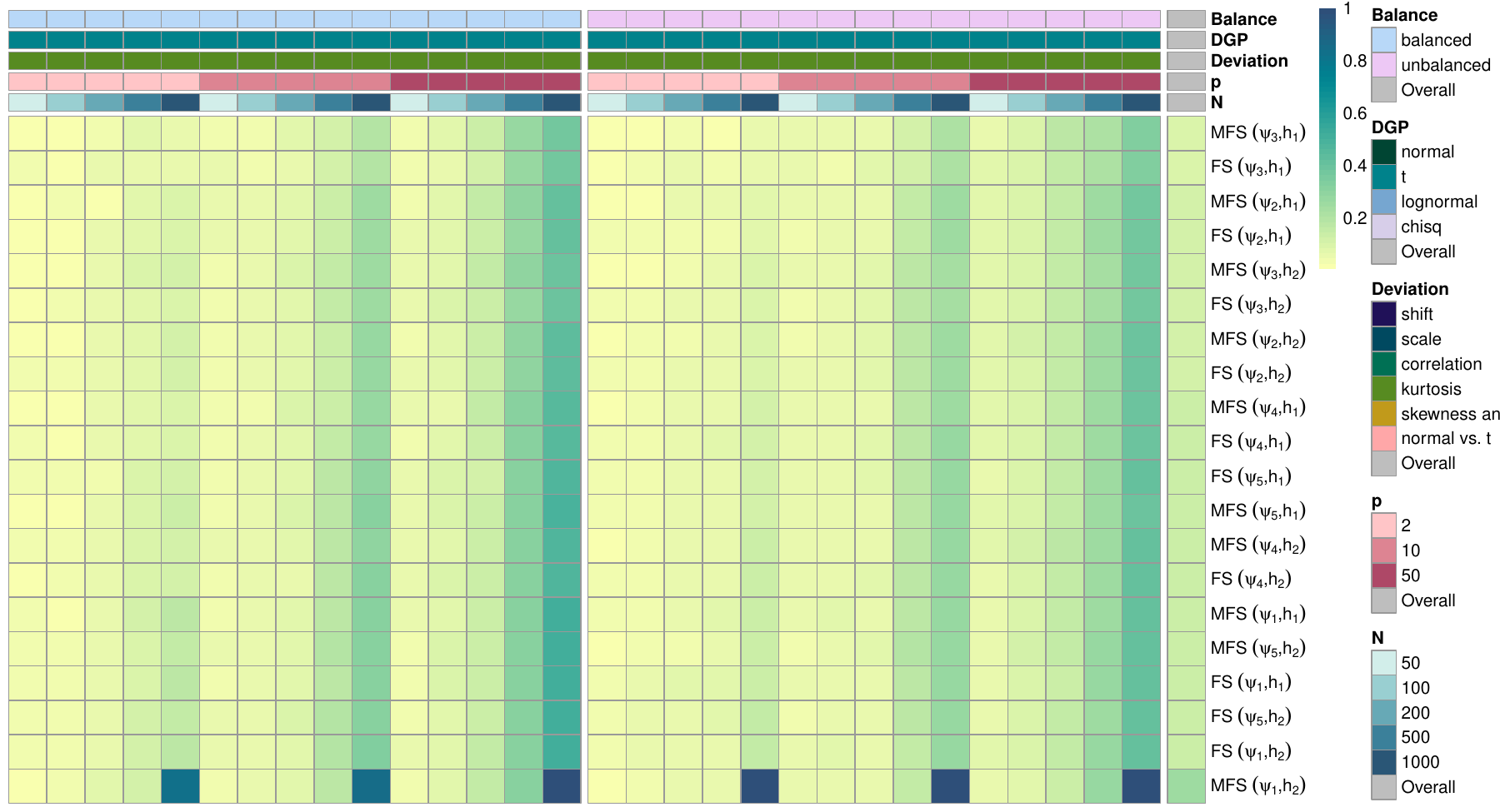}
		\caption{Mean PESR difference to best method per scenario for the kurtosis alternative and $k = 2$ datasets for all variants of the FS test \textcite{paul_clustering-based_2022}. Selected variant: MFS~($\psi_3, h_1$).}
	\end{figure}
	
	\begin{figure}[!h]
		\centering
		\includegraphics[width=\linewidth]{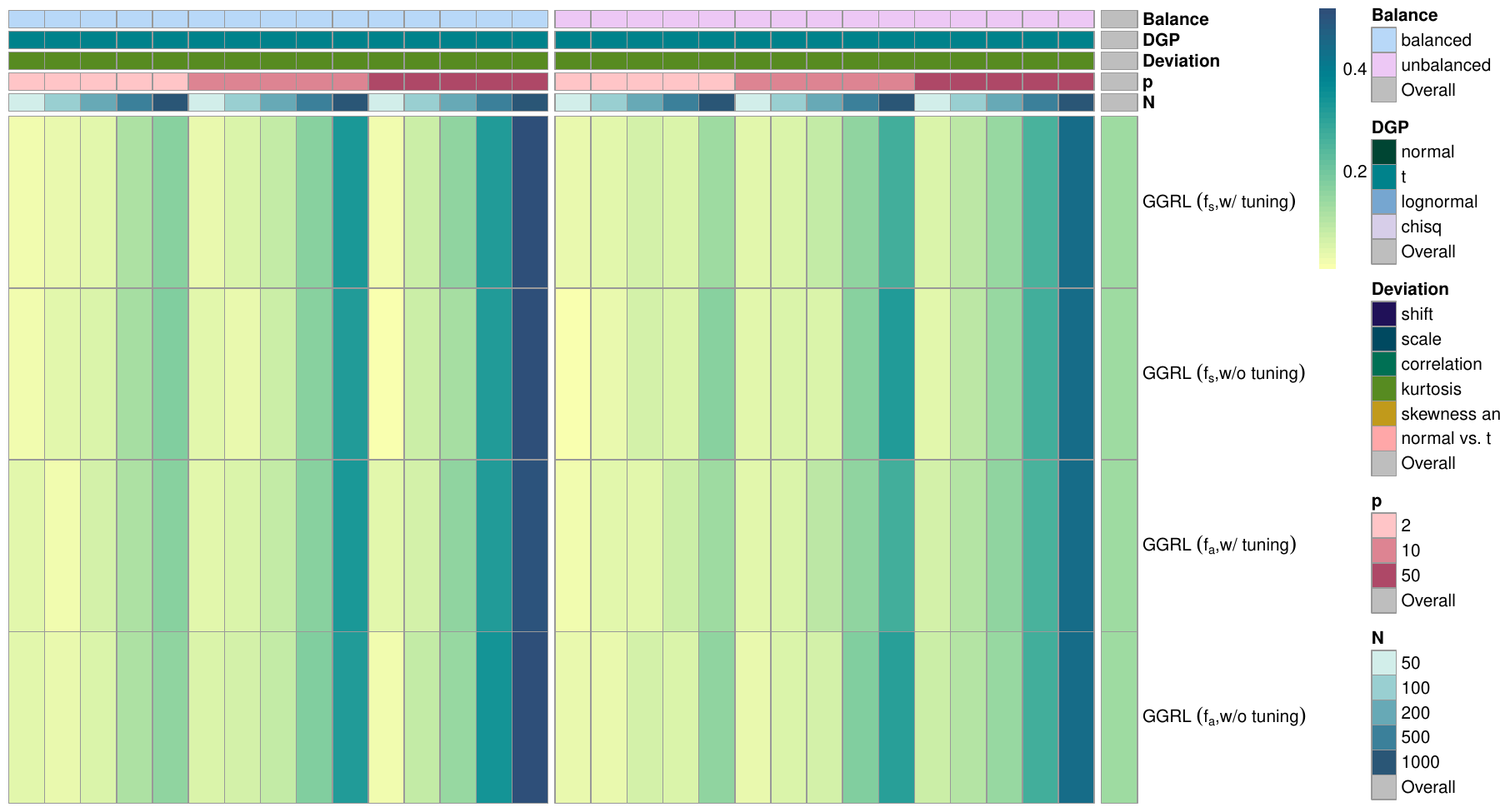}
		\caption{Mean PESR difference to best method per scenario for the kurtosis alternative and $k = 2$ datasets for all variants of the GGRL method \textcite{ganti_framework_1999}. Selected variant: GGRL~($f_a$,w/o~tuning).}
	\end{figure}
	
	\begin{figure}[!h]
		\centering
		\includegraphics[width=\linewidth]{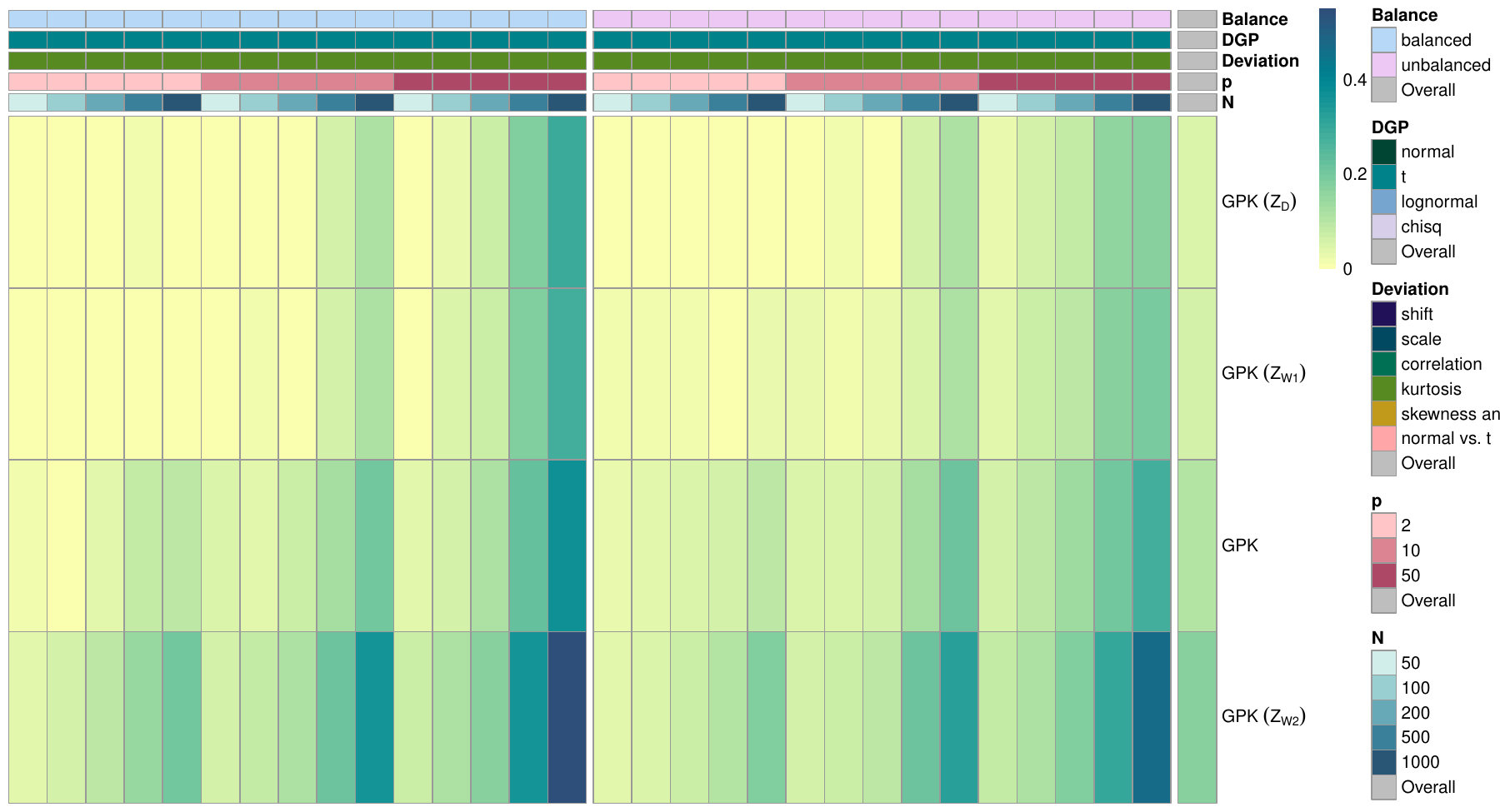}
		\caption{Mean PESR difference to best method per scenario for the kurtosis alternative and $k = 2$ datasets for all variants of the GPK test \textcite{song_generalized_2021}. Selected variant: GPK~(Z$_{\text{D}}$).}
	\end{figure}
	
	\begin{figure}[!h]
		\centering
		\includegraphics[width=\linewidth]{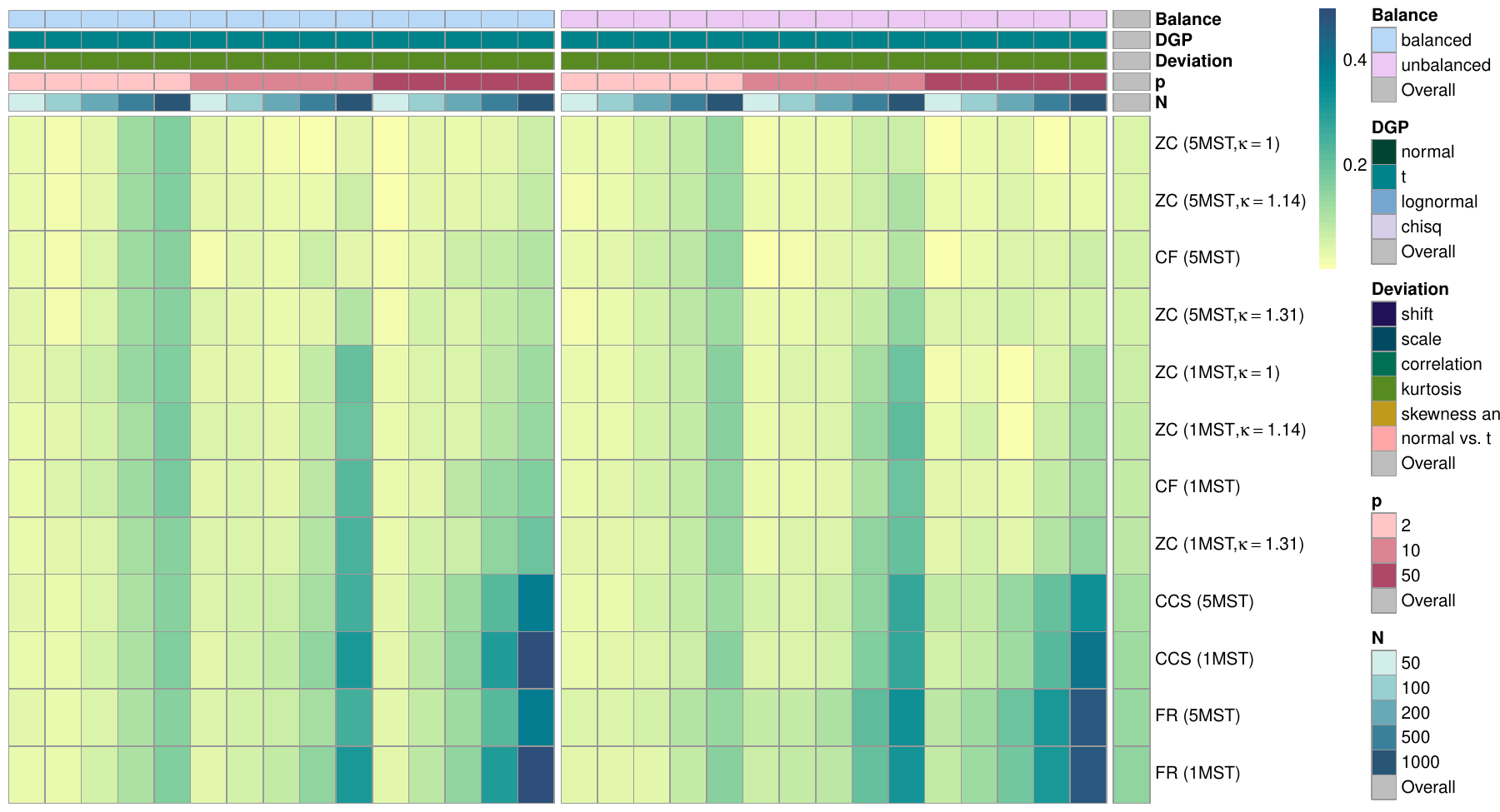}
		\caption{Mean PESR difference to best method per scenario for the kurtosis alternative and $k = 2$ datasets for all variants of the graph-based tests FR, CF, CCS, ZC \textcite{friedman_multivariate_1979, chen_new_2017, chen_weighted_2018, zhang_graph-based_2022}. Selected variant: ZC~(5MST,$\kappa=1$).}
	\end{figure}
	
	\begin{figure}[!h]
		\centering
		\includegraphics[width=\linewidth]{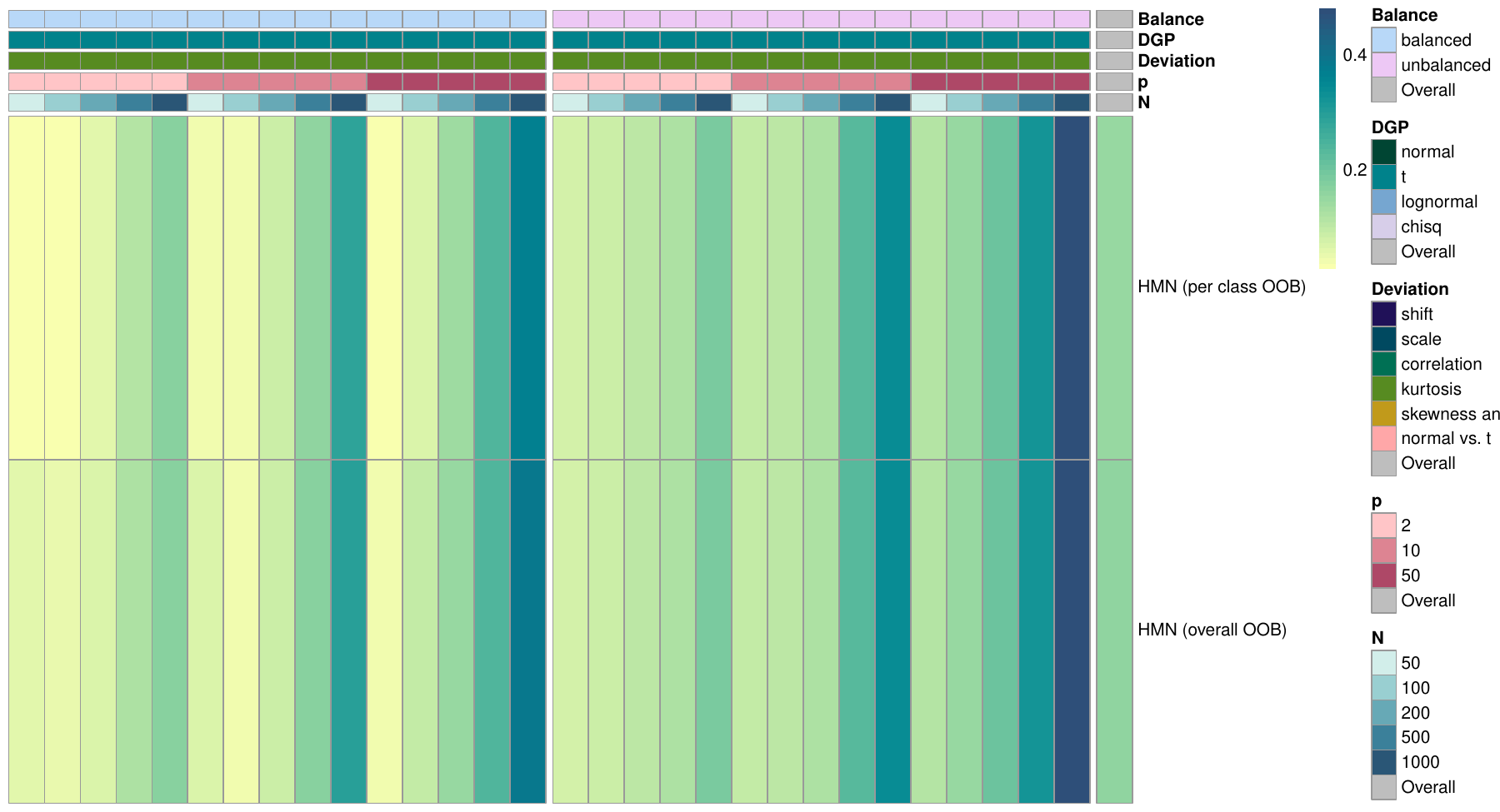}
		\caption{Mean PESR difference to best method per scenario for the kurtosis alternative and $k = 2$ datasets for all variants of the HMN method \textcite{hediger_use_2021}. Selected variant: HMN~(per class OOB).}
	\end{figure}
	
	\begin{figure}[!h]
		\centering
		\includegraphics[width=\linewidth]{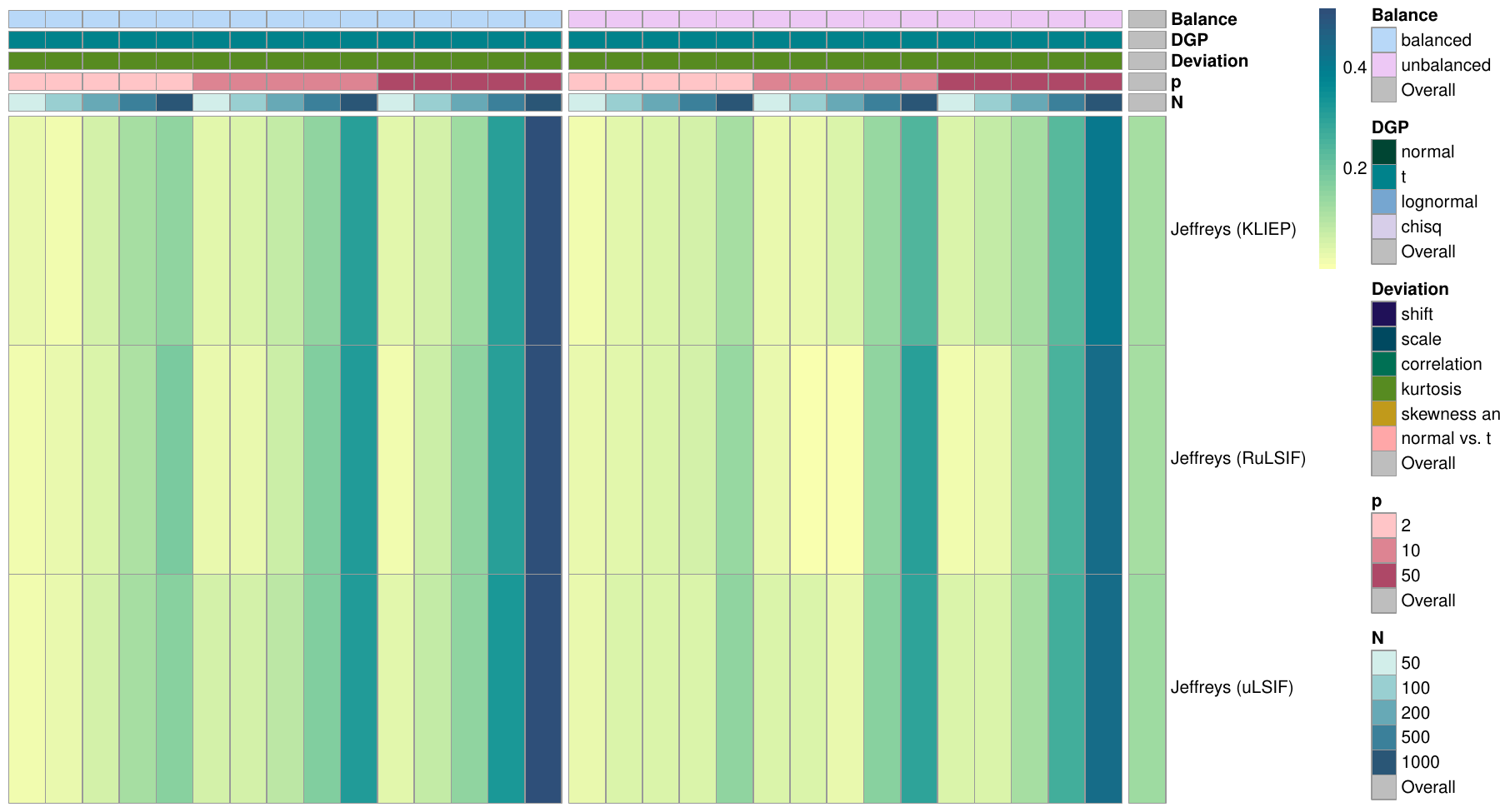}
		\caption{Mean PESR difference to best method per scenario for the kurtosis alternative and $k = 2$ datasets for all variants of the Jeffreys divergence \textcite{jeffreys_invariant_1997}. Selected variant: Jeffreys~(KLIEP).}
	\end{figure}
	
	\begin{figure}[!h]
		\centering
		\includegraphics[width=\linewidth]{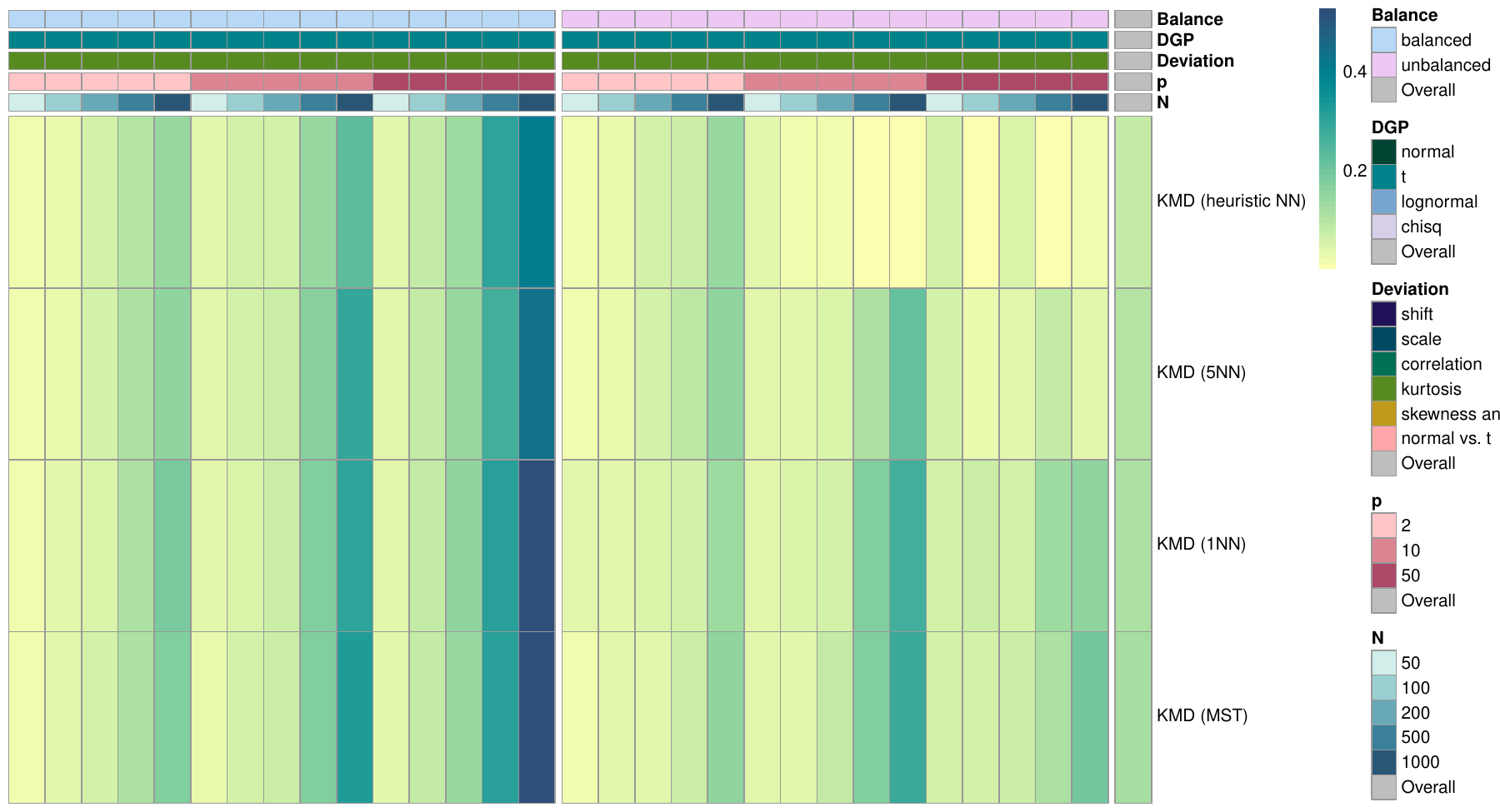}
		\caption{Mean PESR difference to best method per scenario for the kurtosis alternative and $k = 2$ datasets for all variants of the KMD \textcite{huang_kernel_2022}. Selected variant: KMD~(heuristic NN).}
	\end{figure}
	
	\begin{figure}[!h]
		\centering
		\includegraphics[width=\linewidth]{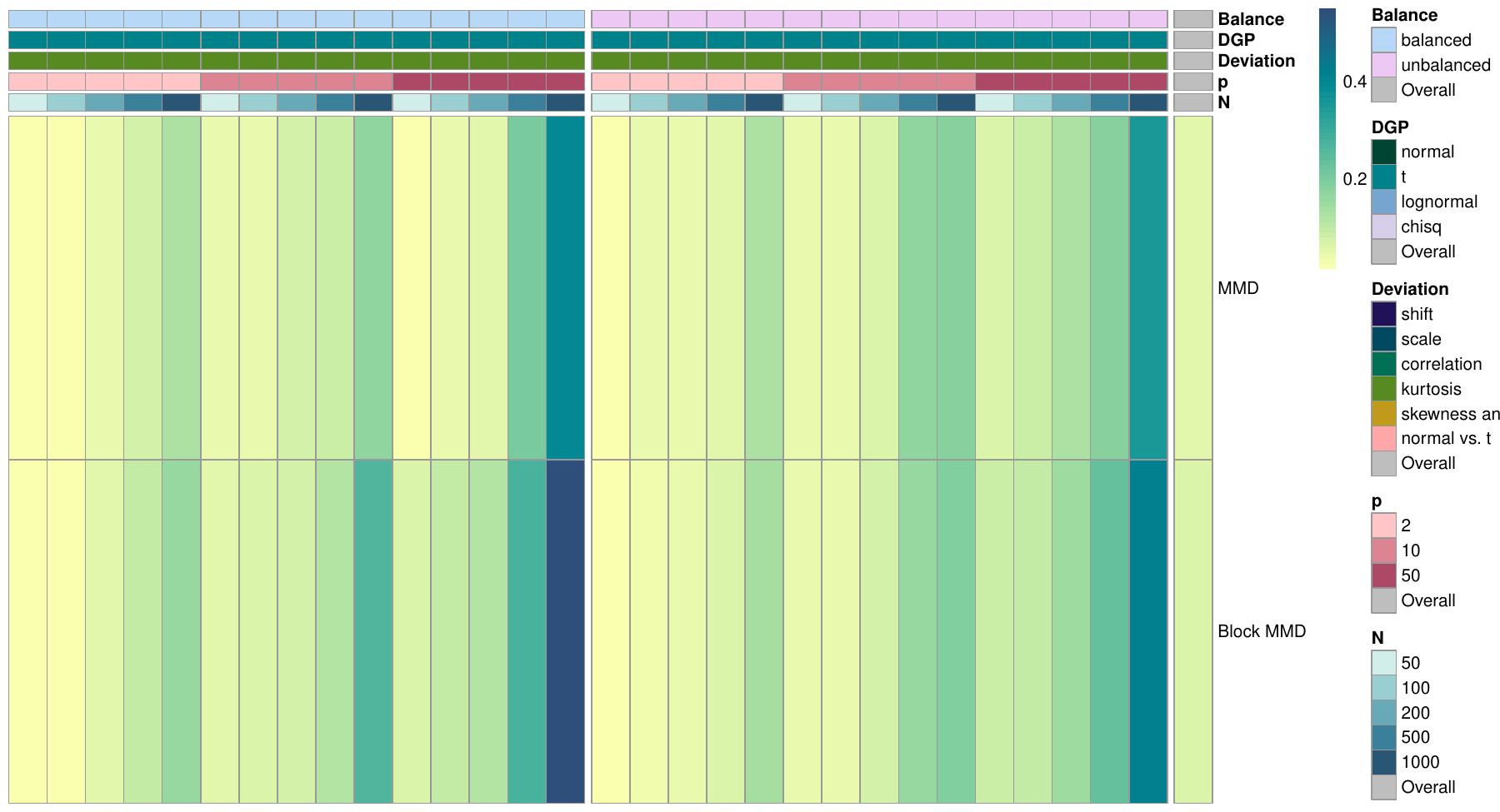}
		\caption{Mean PESR difference to best method per scenario for the kurtosis alternative and $k = 2$ datasets for all variants of the MMD \textcite{gretton_kernel_2006, zaremba_b-test_2013}. Selected variant: MMD.}
	\end{figure}
	
	\begin{figure}[!h]
		\centering
		\includegraphics[width=\linewidth]{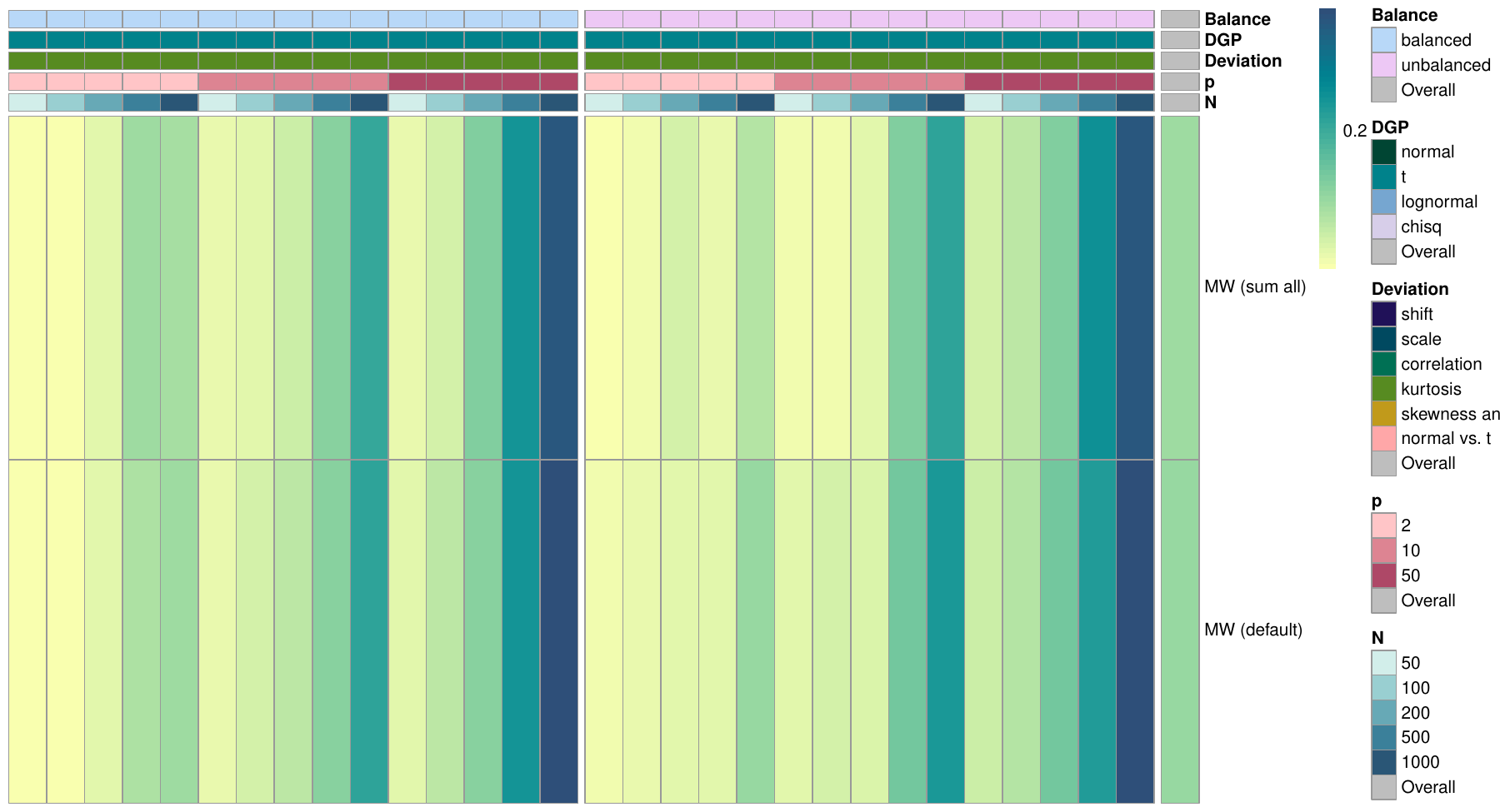}
		\caption{Mean PESR difference to best method per scenario for the kurtosis alternative and $k = 2$ datasets for all variants of the MW method \textcite{mukhopadhyay_nonparametric_2020}. Selected variant: MW~(sum all).}
	\end{figure}
	
	\begin{figure}[!h]
		\centering
		\includegraphics[width=\linewidth]{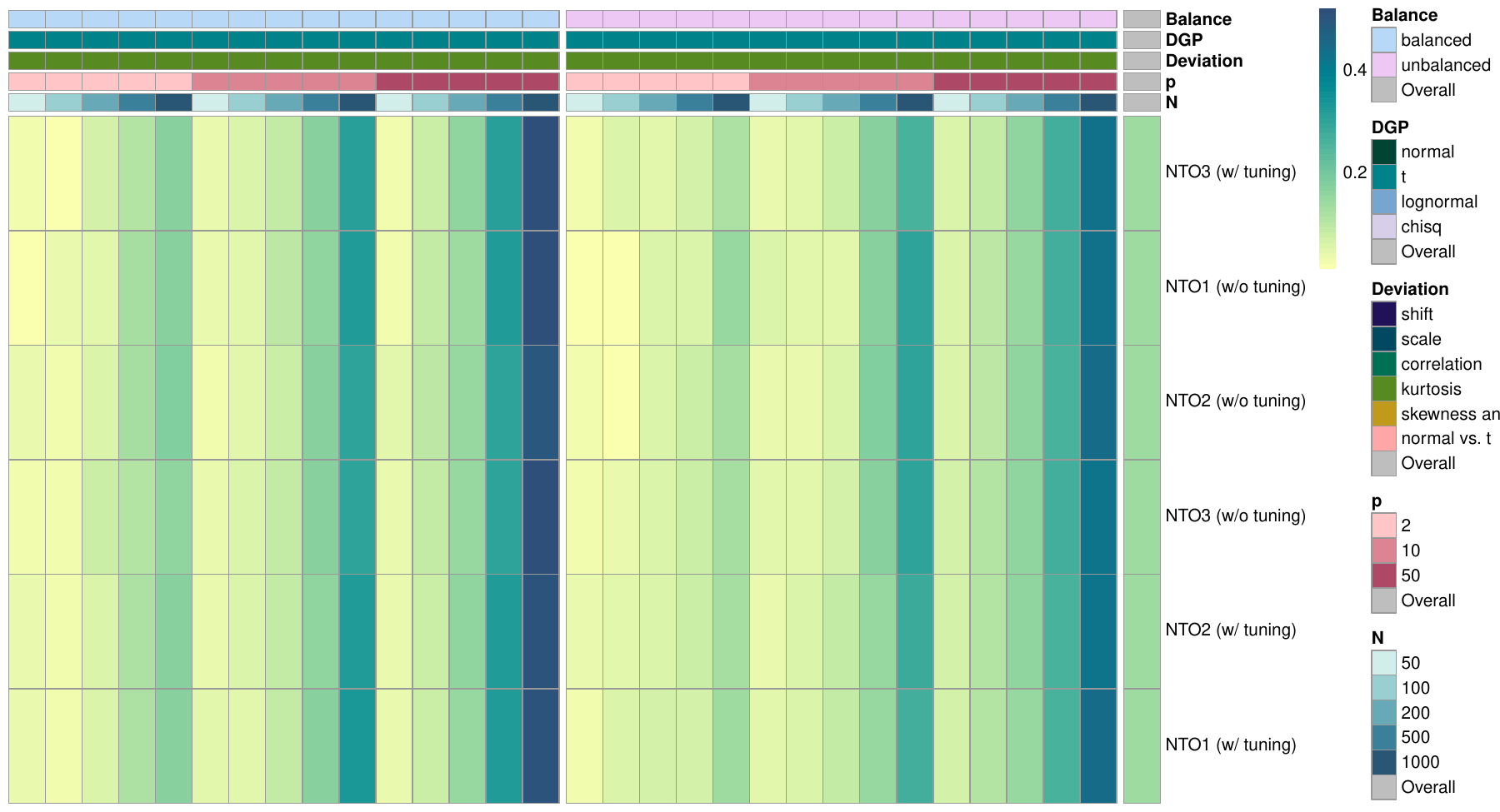}
		\caption{Mean PESR difference to best method per scenario for the kurtosis alternative and $k = 2$ datasets for all variants of the NKT method \textcite{ntoutsi_general_2008}. Selected variant: NTO3~(tuning).}
	\end{figure}
	
	\begin{figure}[!h]
		\centering
		\includegraphics[width=\linewidth]{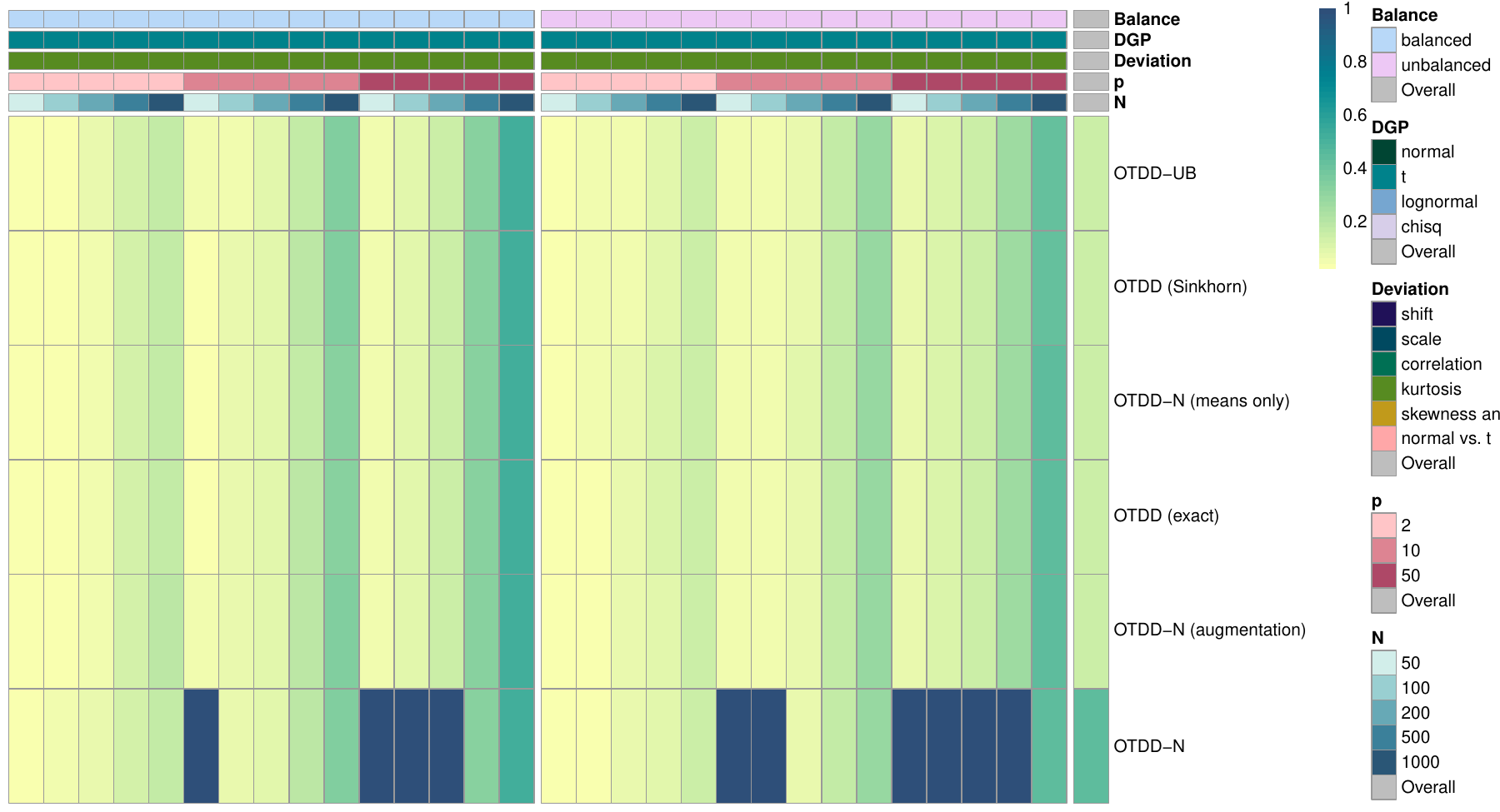}
		\caption{Mean PESR difference to best method per scenario for the kurtosis alternative and $k = 2$ datasets for all variants of the OTDD \textcite{alvarez-melis_geometric_2020}. Selected variant: OTDD-UB.}
	\end{figure}
	
	\begin{figure}[!h]
		\centering
		\includegraphics[width=\linewidth]{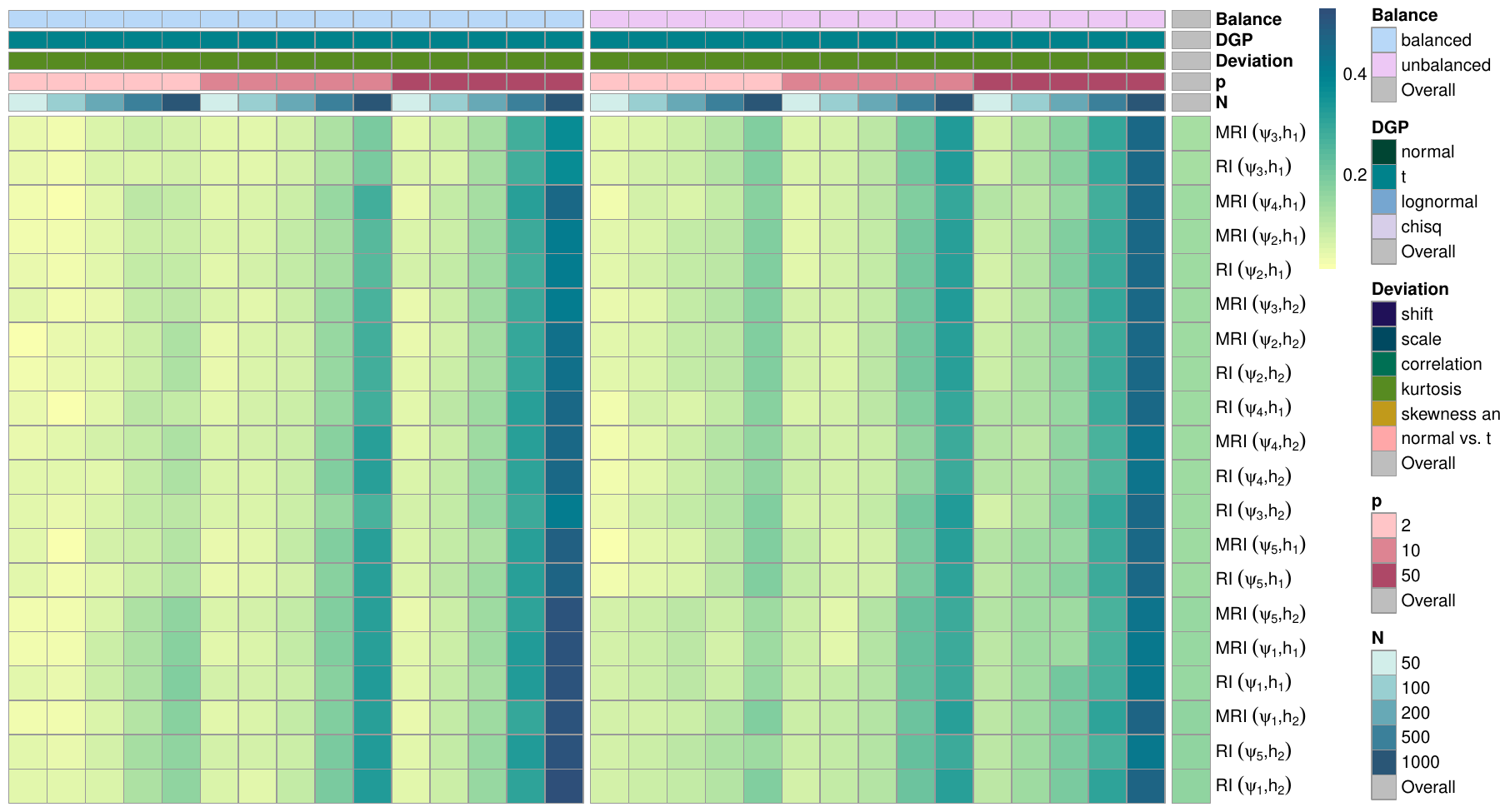}
		\caption{Mean PESR difference to best method per scenario for the kurtosis alternative and $k = 2$ datasets for all variants of the RI test \textcite{paul_clustering-based_2022}. Selected variant: MRI~($\psi_3, h_1$).}
	\end{figure}
	
	\begin{figure}[!h]
		\centering
		\includegraphics[width=\linewidth]{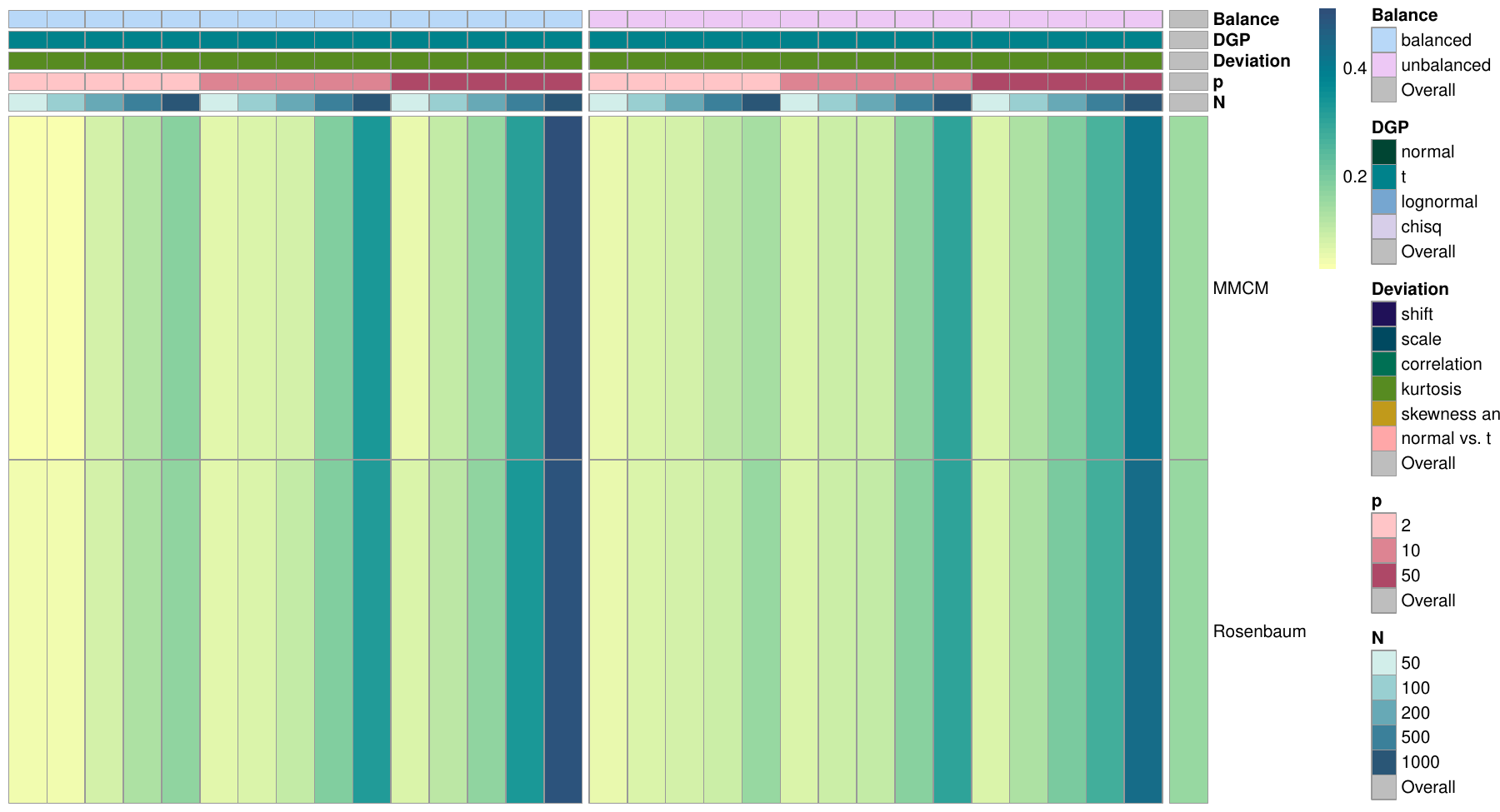}
		\caption{Mean PESR difference to best method per scenario for the kurtosis alternative and $k = 2$ datasets for all variants of the Rosenbaum test \textcite{rosenbaum_exact_2005, mukherjee_distribution-free_2022}. Selected variant: Rosenbaum (almost identical performance).}
	\end{figure}
	
	\begin{figure}[!h]
		\centering
		\includegraphics[width=\linewidth]{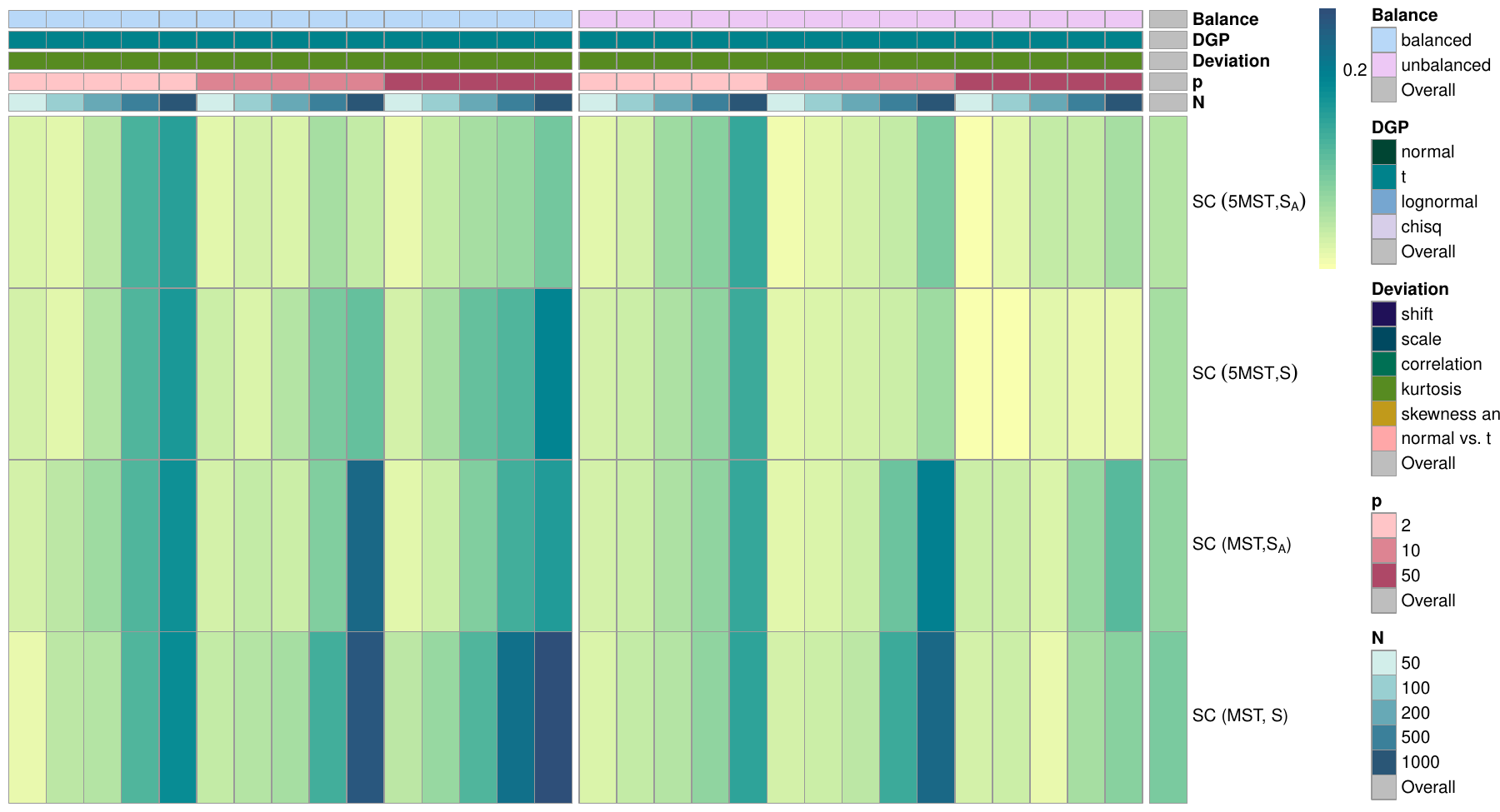}
		\caption{Mean PESR difference to best method per scenario for the kurtosis alternative and $k = 2$ datasets for all variants of the SC test \textcite{song_new_2022}. Selected variant: SC~(5MST,S$_{\text{A}}$).}
	\end{figure}
	
	\begin{figure}[!h]
		\centering
		\includegraphics[width=\linewidth]{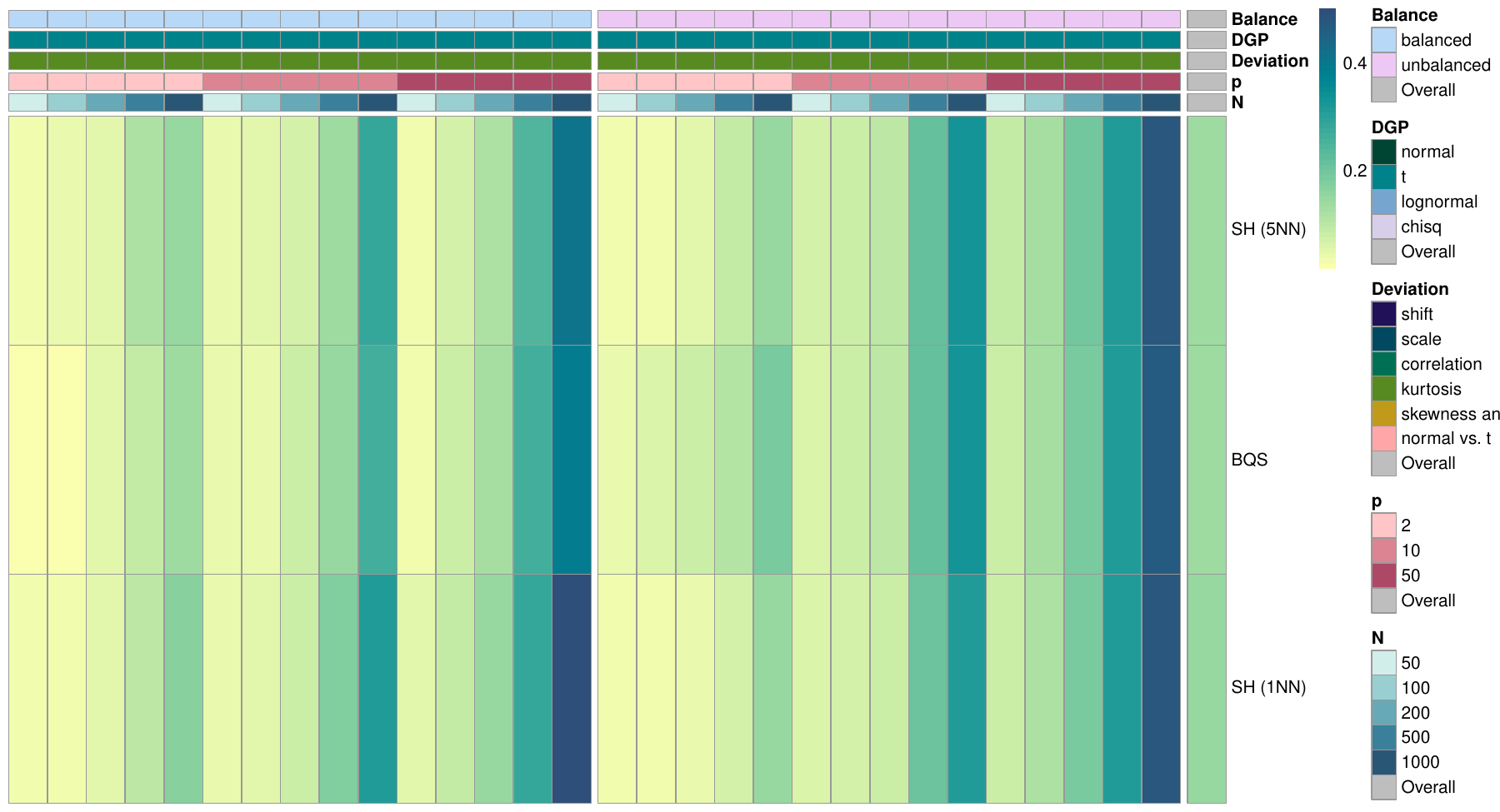}
		\caption{Mean PESR difference to best method per scenario for the kurtosis alternative and $k = 2$ datasets for all variants of the SH test \textcite{schilling_multivariate_1986, henze_multivariate_1988, barakat_multivariate_1996}. Selected variants: SH~(5NN).}
	\end{figure}
	\clearpage
	
	\paragraph{Skewness and Kurtosis}\mbox{} 
	
	\begin{figure}[!h]
		\centering
		\includegraphics[width=\linewidth]{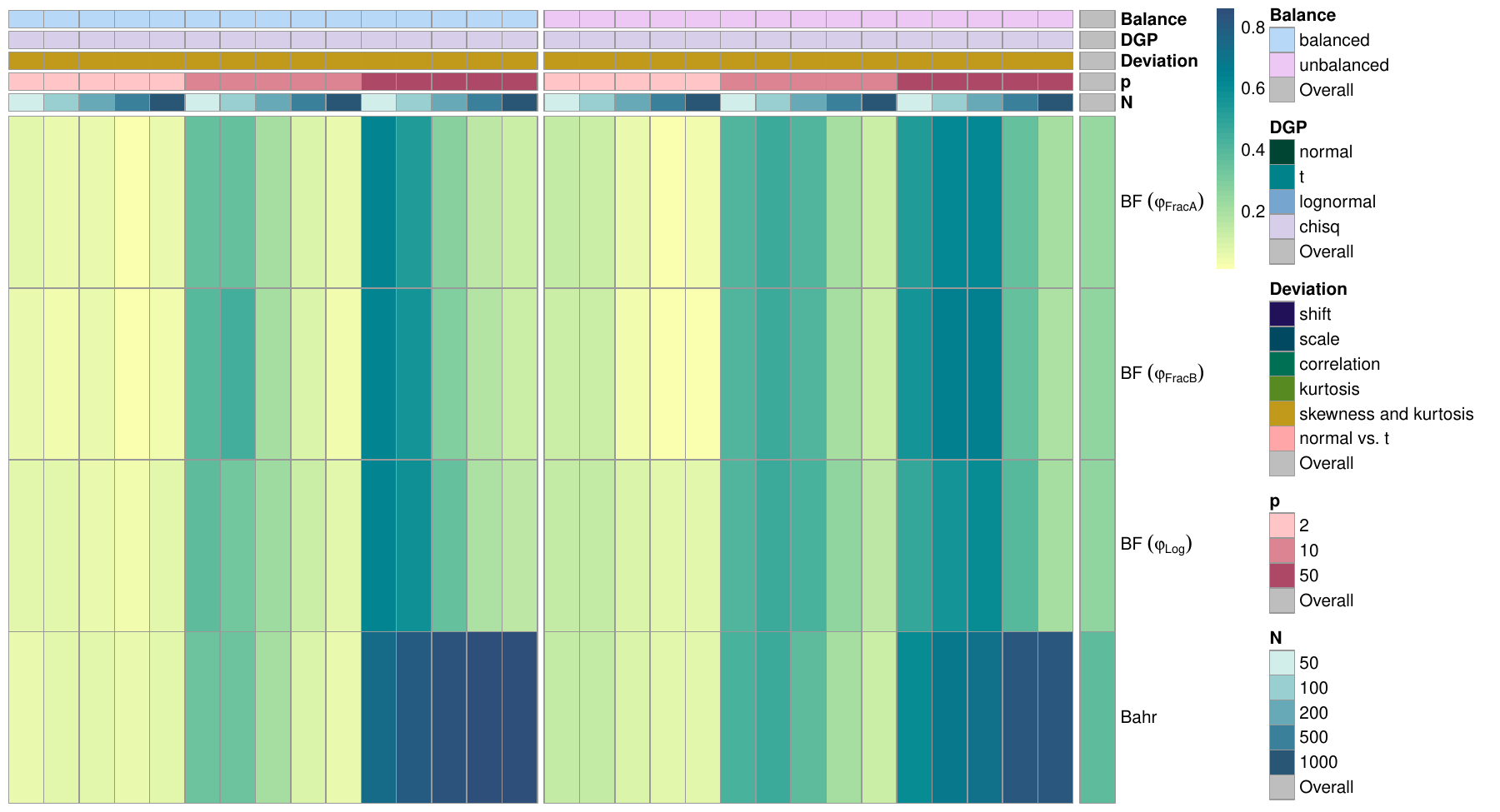}
		\caption{Mean PESR difference to best method per scenario for the skewness and kurtosis alternative and $k = 2$ datasets for all variants of the BF method \textcite{baringhaus_rigid_2010}. Selected variant: BF~($\varphi_{\text{FracA}}$).}
	\end{figure}
	
	\begin{figure}[!h]
		\centering
		\includegraphics[width=\linewidth]{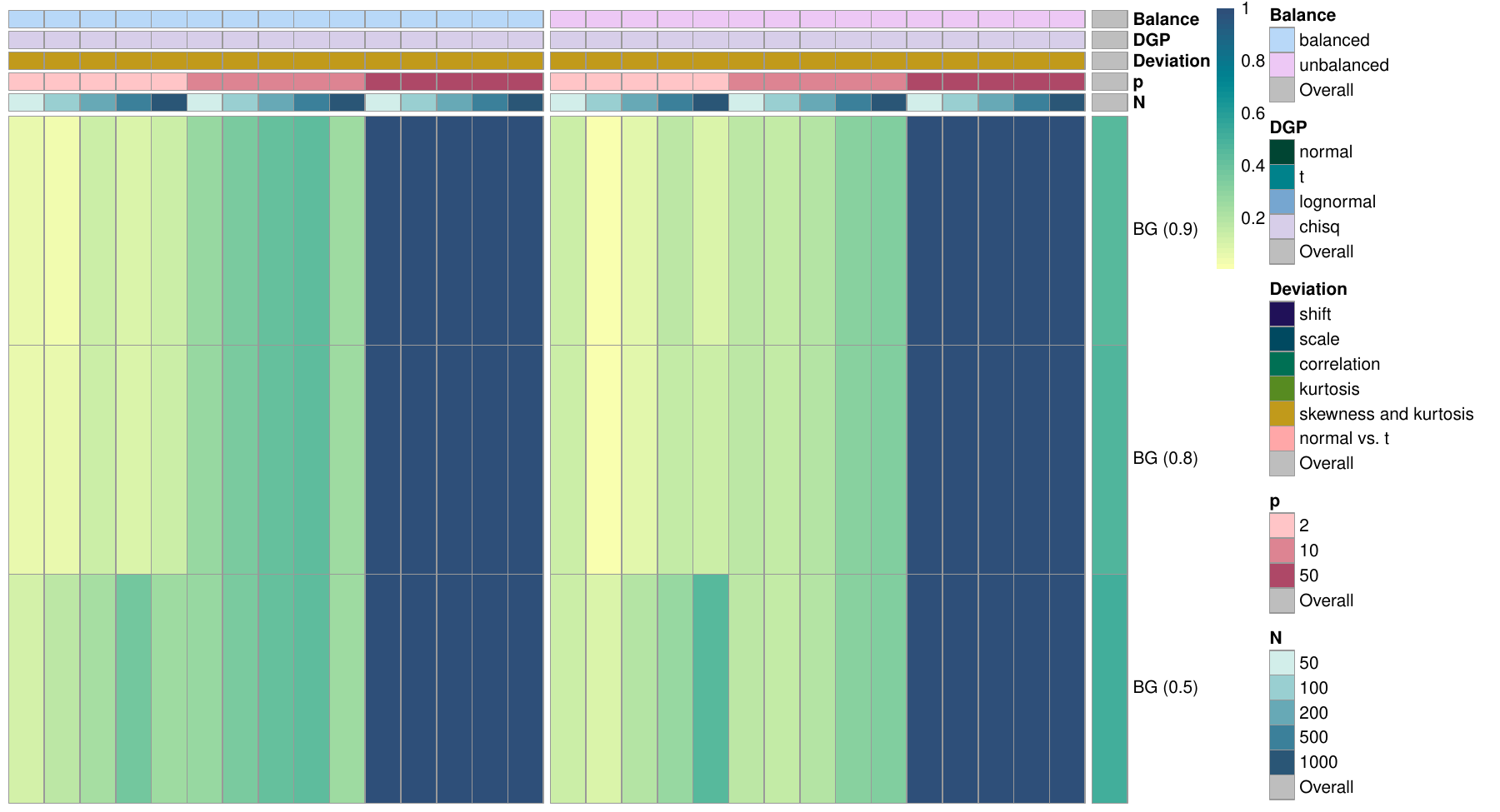}
		\caption{Mean PESR difference to best method per scenario for the skewness and kurtosis alternative and $k = 2$ datasets for all variants of the BG method \textcite{biau_asymptotic_2005}. Selected variant: BG~(0.9).}
	\end{figure}
	
	\begin{figure}[!h]
		\centering
		\includegraphics[width=\linewidth]{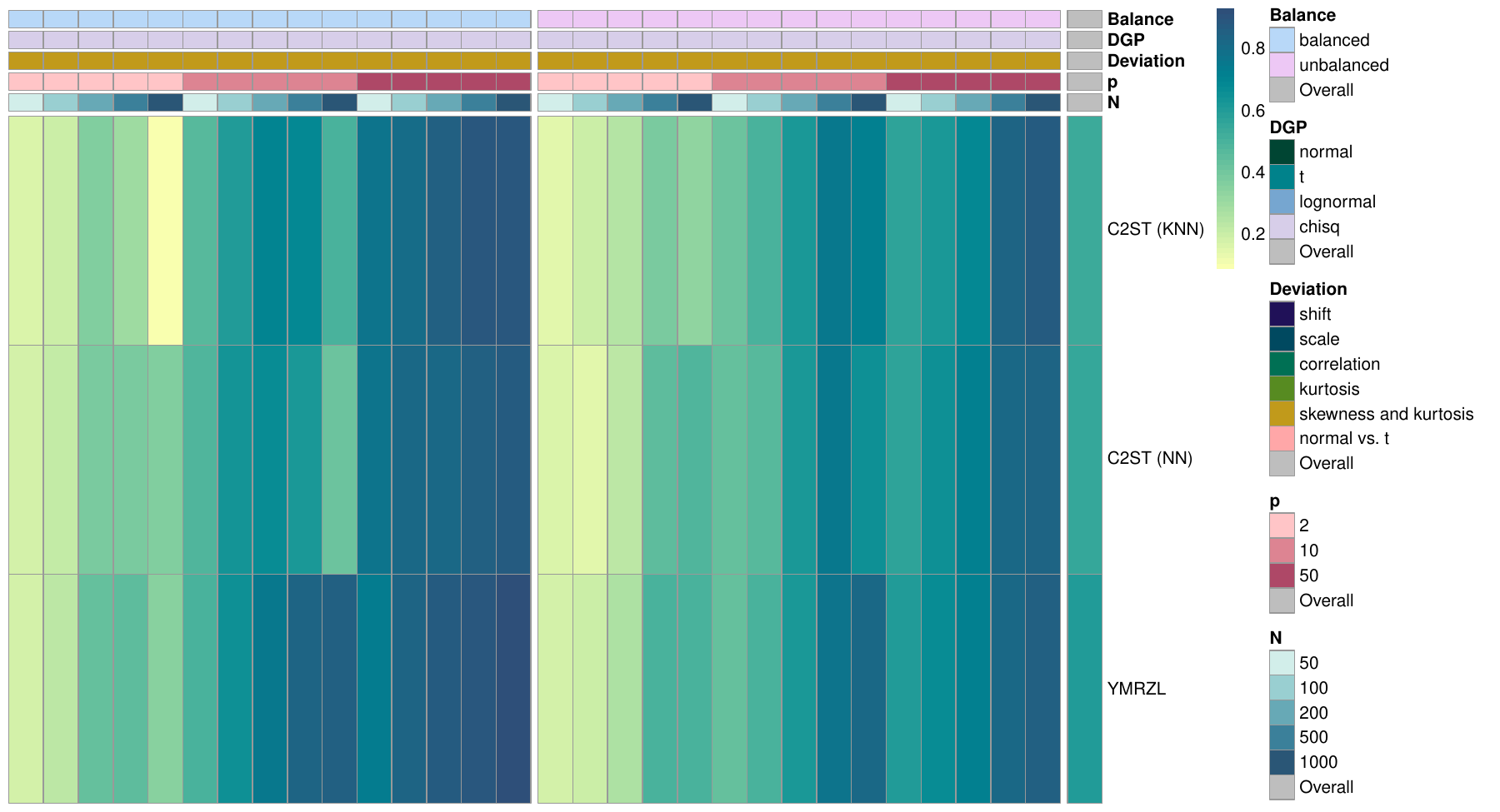}
		\caption{Mean PESR difference to best method per scenario for the skewness and kurtosis alternative and $k = 2$ datasets for all variants of the YMRZL and C2ST method \textcite{yu_two-sample_2007, lopez-paz_revisiting_2017}. Selected variant: C2ST~(KNN).}
	\end{figure}
	
	\begin{figure}[!h]
		\centering
		\includegraphics[width=\linewidth]{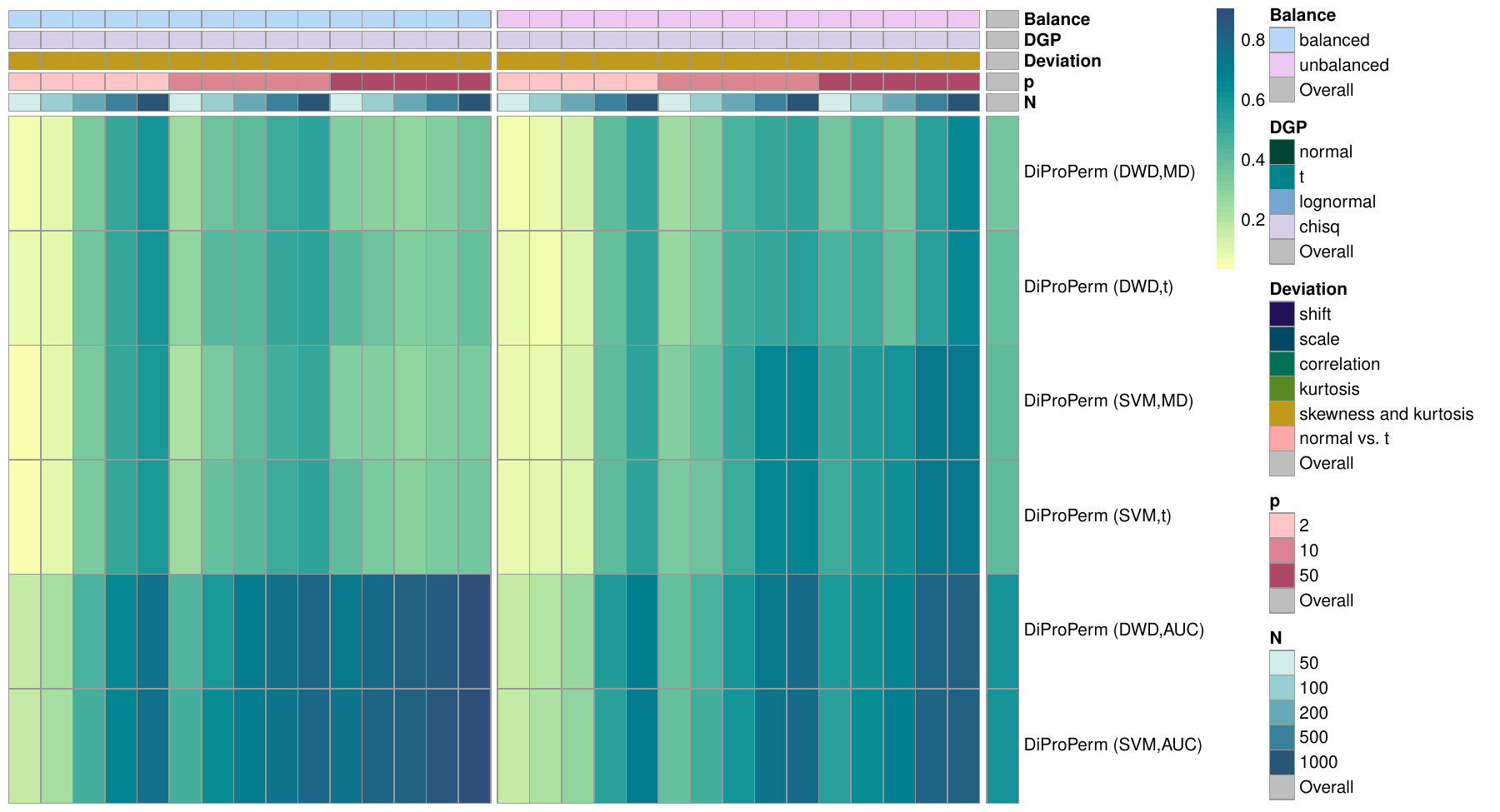}
		\caption{Mean PESR difference to best method per scenario for the kurtosis alternative and $k = 2$ datasets for all variants of the DiProPerm method \textcite{wei_direction-projection-permutation_2016}. Selected variant: DiProPerm~(DWD,MD).}
	\end{figure}
	
	\begin{figure}[!h]
		\centering
		\includegraphics[width=\linewidth]{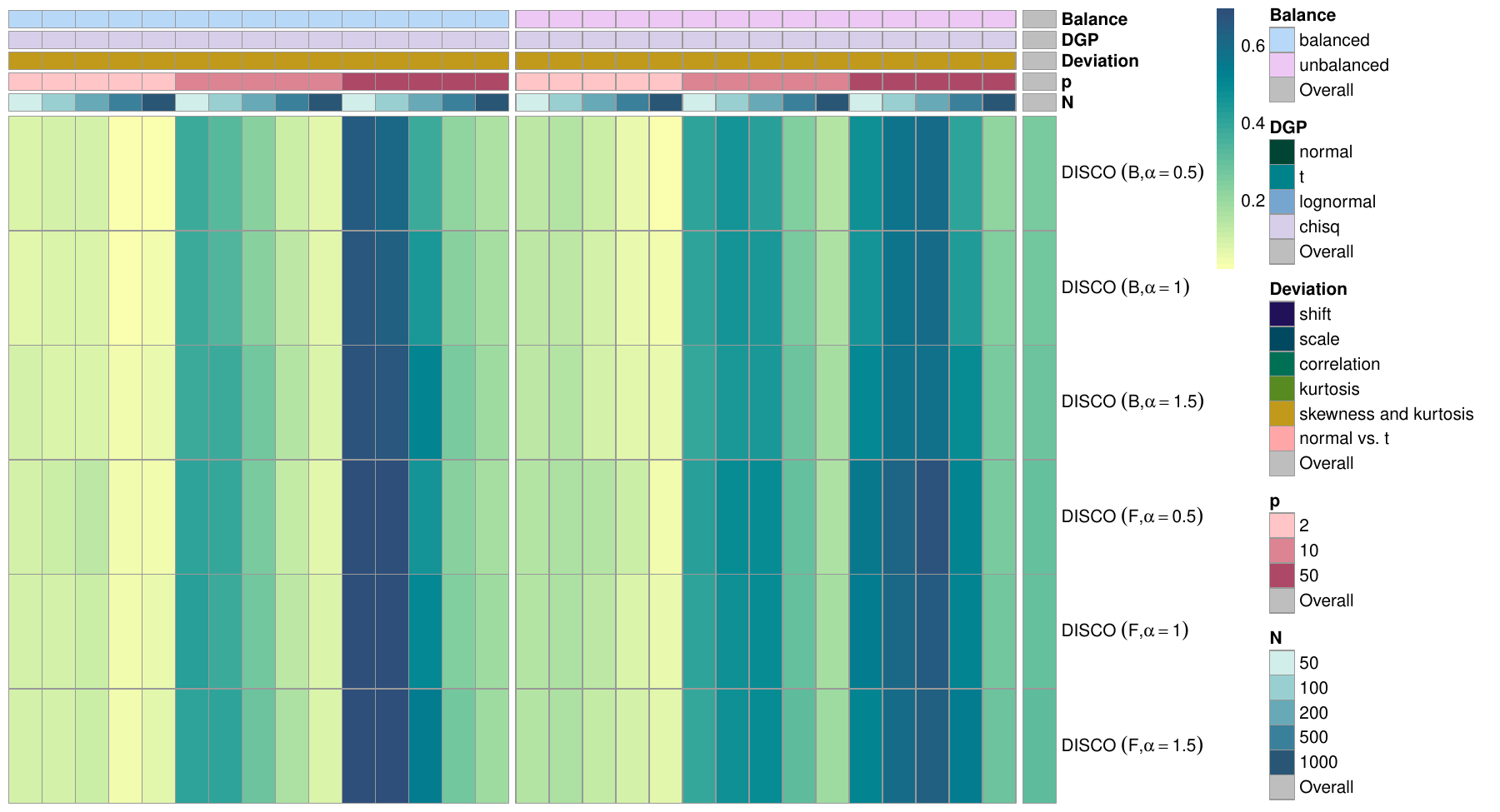}
		\caption{Mean PESR difference to best method per scenario for the skewness and kurtosis alternative and $k = 2$ datasets for all variants of the DISCO method \textcite{rizzo_disco_2010}. Selected variant: DISCO~(B,$\alpha=0.5$).}
	\end{figure}
	
	\begin{figure}[!h]
		\centering
		\includegraphics[width=\linewidth]{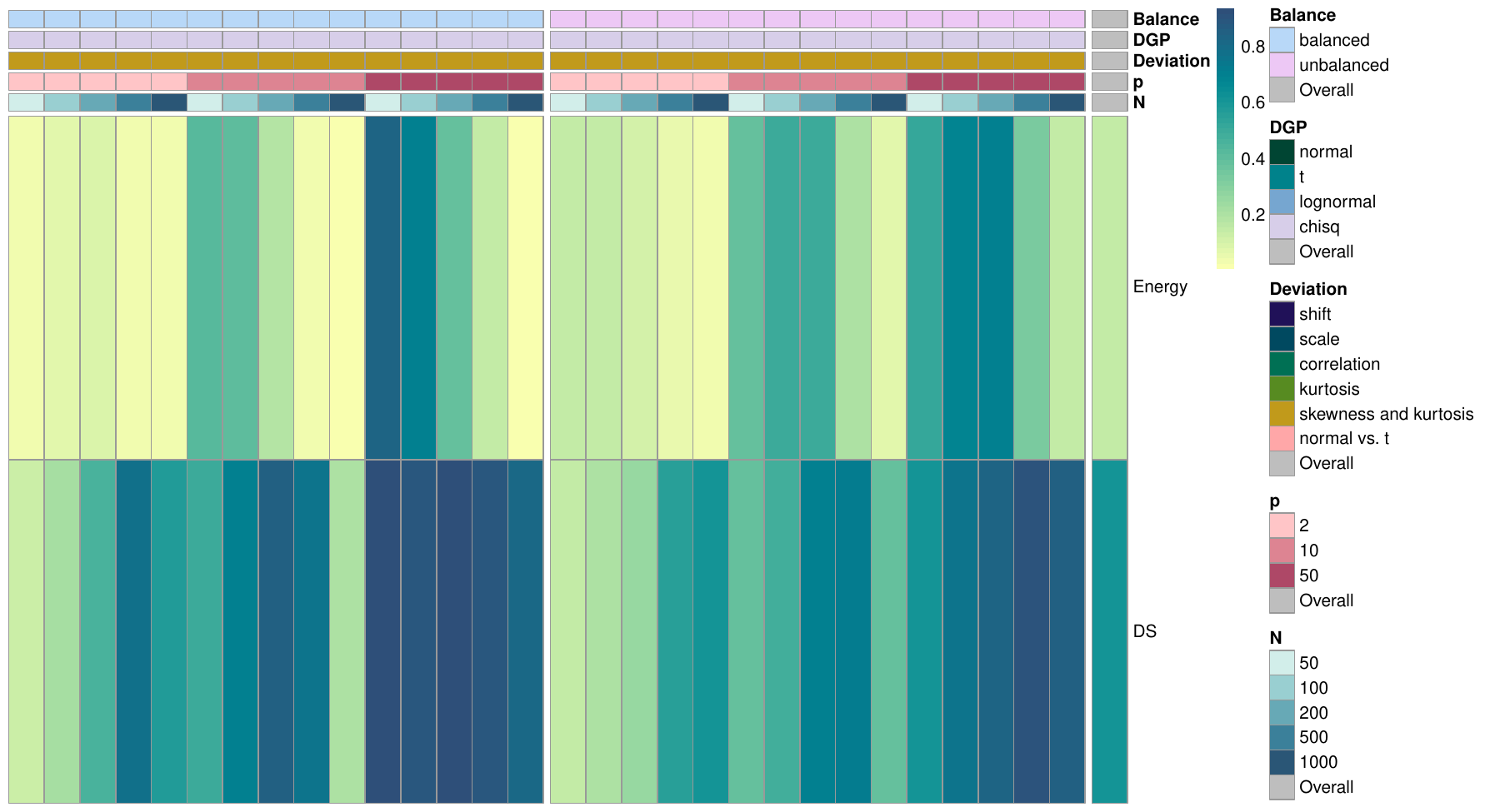}
		\caption{Mean PESR difference to best method per scenario for the skewness and kurtosis alternative and $k = 2$ datasets for all variants of the Energy method \textcite{szekely_testing_2004}. Selected variant: Energy.}
	\end{figure}
	
	\begin{figure}[!h]
		\centering
		\includegraphics[width=\linewidth]{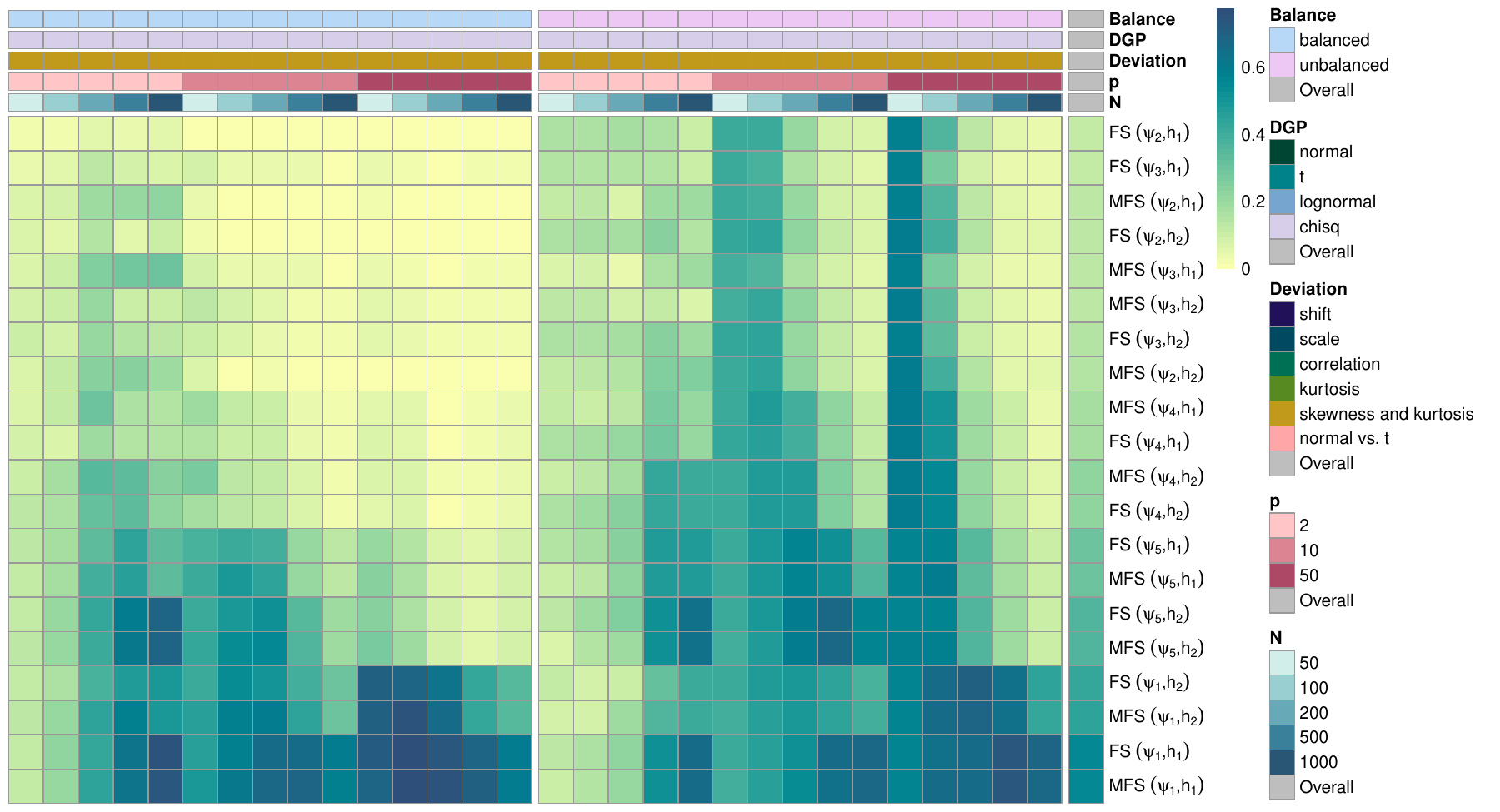}
		\caption{Mean PESR difference to best method per scenario for the skewness and kurtosis alternative and $k = 2$ datasets for all variants of the FS test \textcite{paul_clustering-based_2022}. Selected variant: FS~($\psi_2, h_1$).}
	\end{figure}
	
	\begin{figure}[!h]
		\centering
		\includegraphics[width=\linewidth]{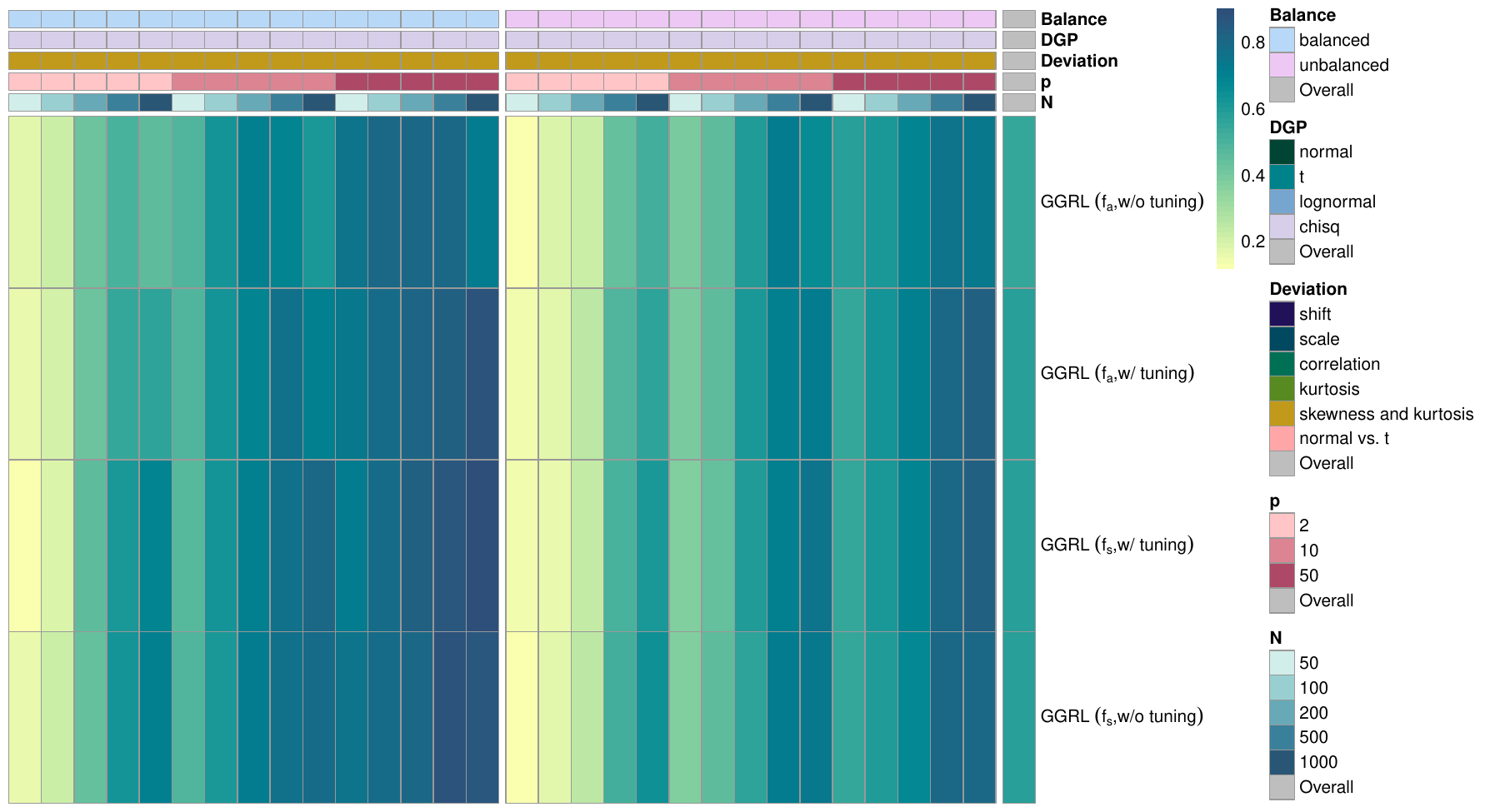}
		\caption{Mean PESR difference to best method per scenario for the skewness and kurtosis alternative and $k = 2$ datasets for all variants of the GGRL method \textcite{ganti_framework_1999}. Selected variant: GGRL~($f_a$,w/o~tuning).}
	\end{figure}
	
	\begin{figure}[!h]
		\centering
		\includegraphics[width=\linewidth]{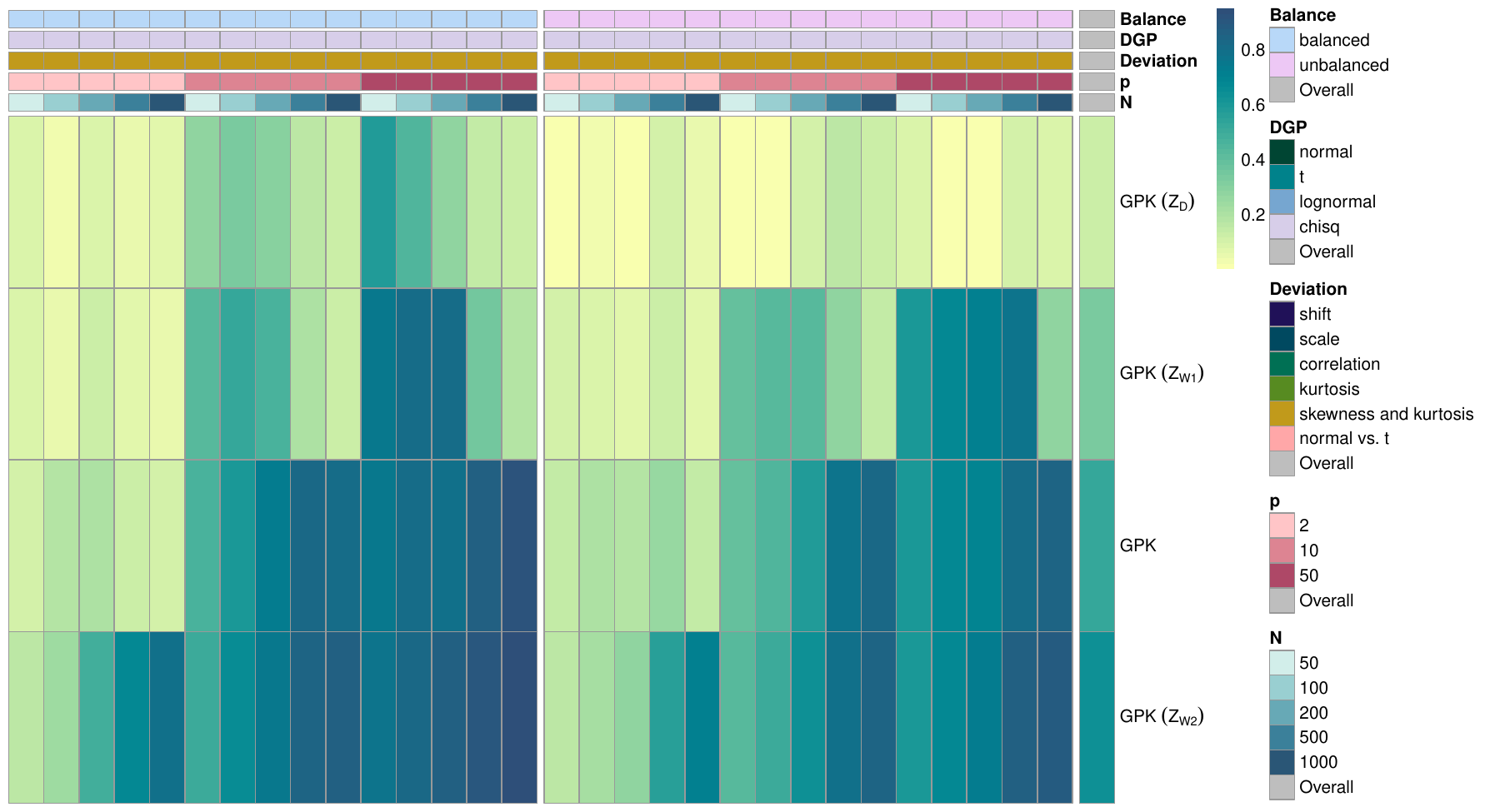}
		\caption{Mean PESR difference to best method per scenario for the skewness and kurtosis alternative and $k = 2$ datasets for all variants of the GPK test \textcite{song_generalized_2021}. Selected variant: GPK~(Z$_{\text{D}}$).}
	\end{figure}
	
	\begin{figure}[!h]
		\centering
		\includegraphics[width=\linewidth]{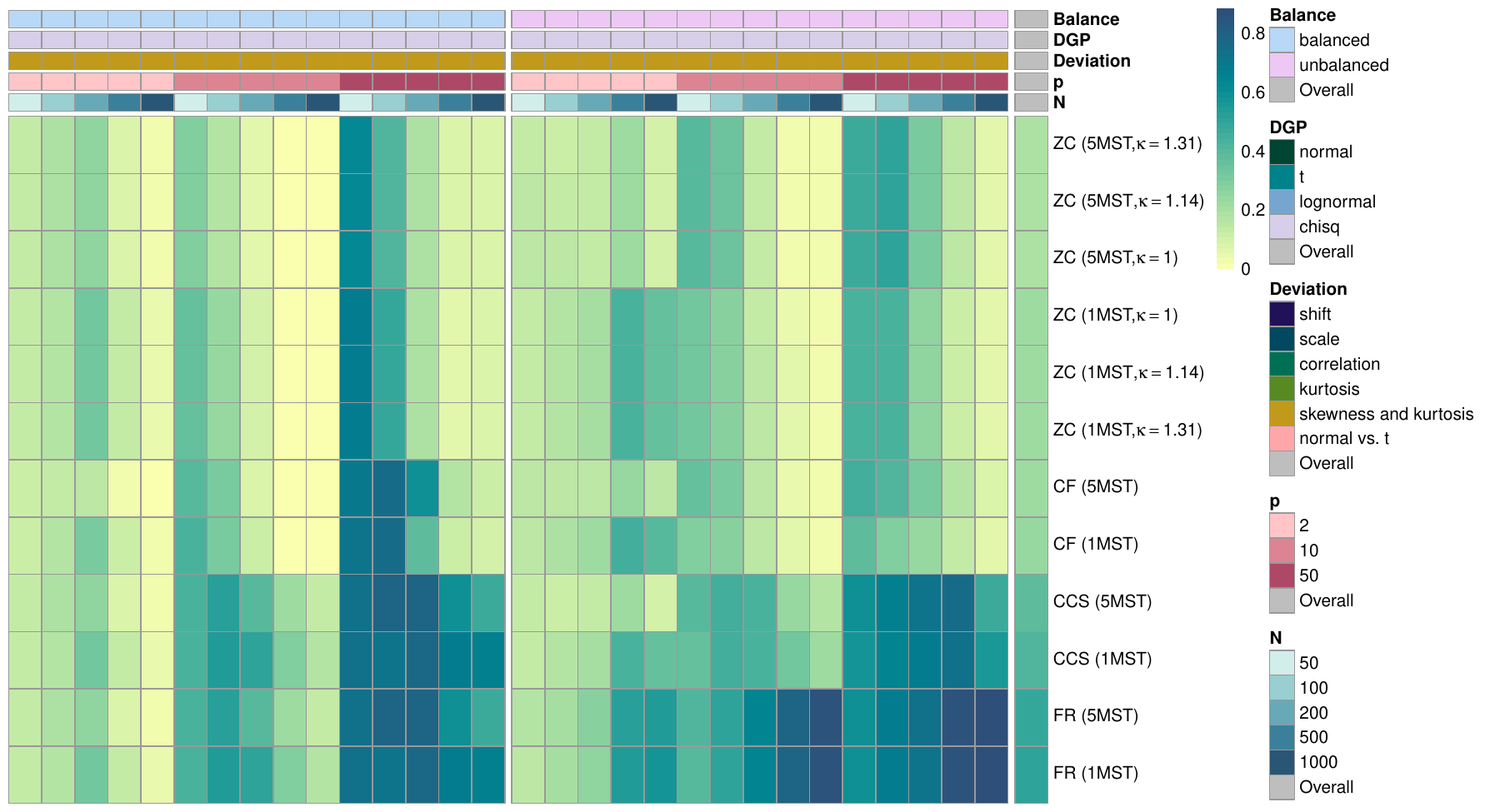}
		\caption{Mean PESR difference to best method per scenario for the skewness and kurtosis alternative and $k = 2$ datasets for all variants of the graph-based tests FR, CF, CCS, ZC \textcite{friedman_multivariate_1979, chen_new_2017, chen_weighted_2018, zhang_graph-based_2022}. Selected variant: ZC~(5MST,$\kappa=1.31$).}
	\end{figure}
	
	\begin{figure}[!h]
		\centering
		\includegraphics[width=\linewidth]{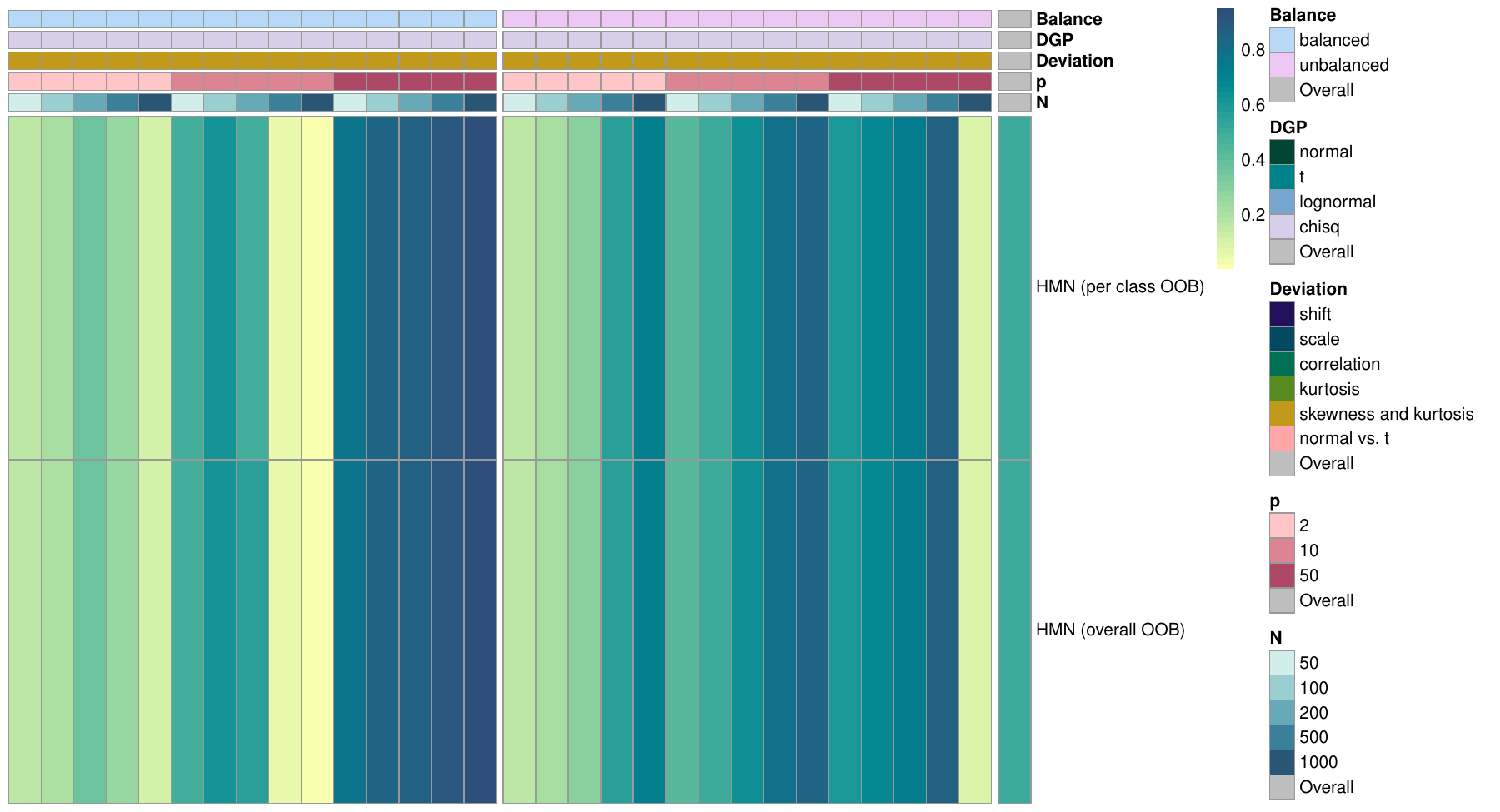}
		\caption{Mean PESR difference to best method per scenario for the skewness and kurtosis alternative and $k = 2$ datasets for all variants of the HMN method \textcite{hediger_use_2021}. Selected variant: HMN~(per class OOB).}
	\end{figure}
	
	\begin{figure}[!h]
		\centering
		\includegraphics[width=\linewidth]{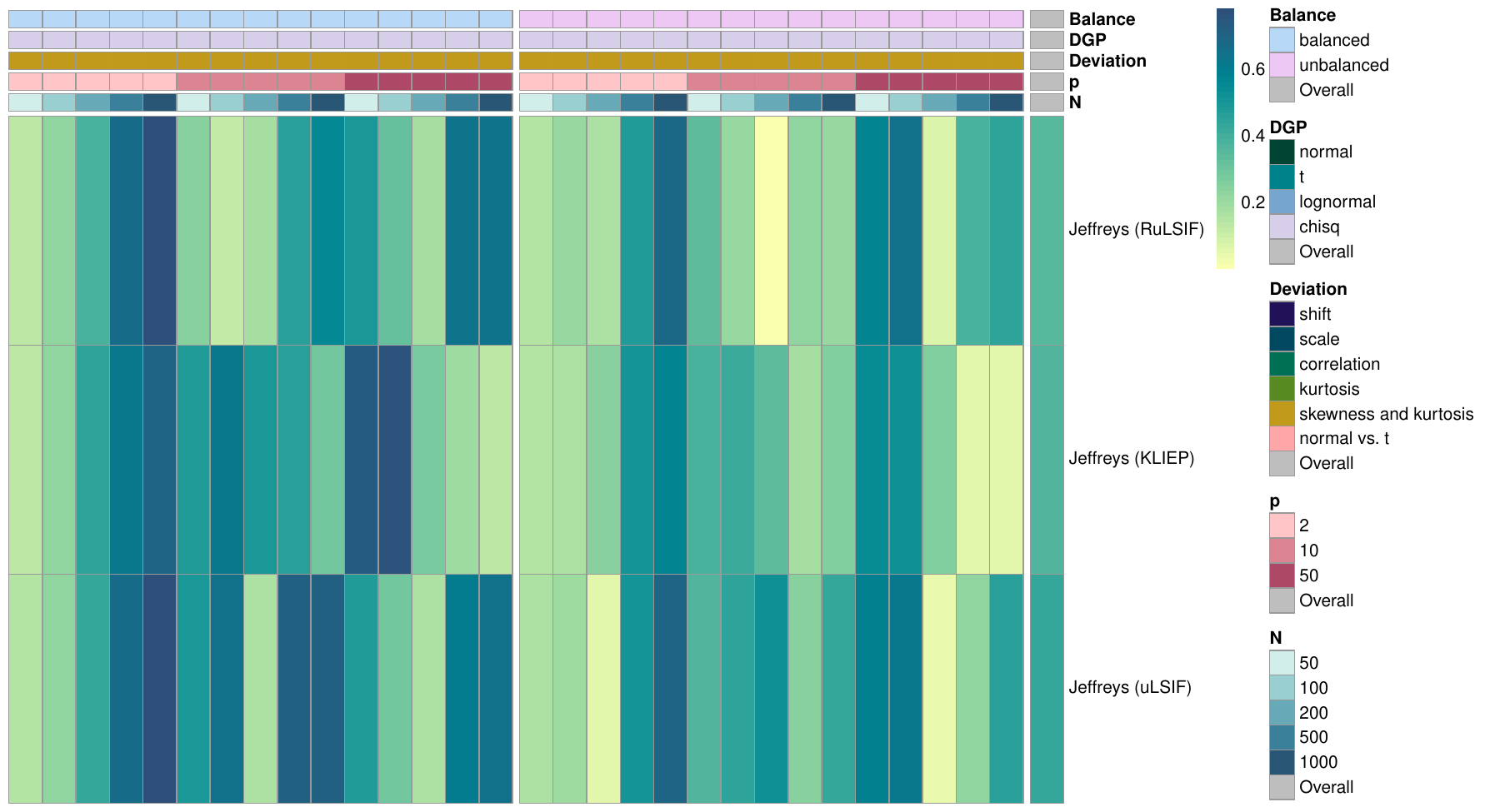}
		\caption{Mean PESR difference to best method per scenario for the skewness and kurtosis alternative and $k = 2$ datasets for all variants of the Jeffreys divergence \textcite{jeffreys_invariant_1997}. Selected variant: Jeffreys~(RuLSIF).}
	\end{figure}
	
	\begin{figure}[!h]
		\centering
		\includegraphics[width=\linewidth]{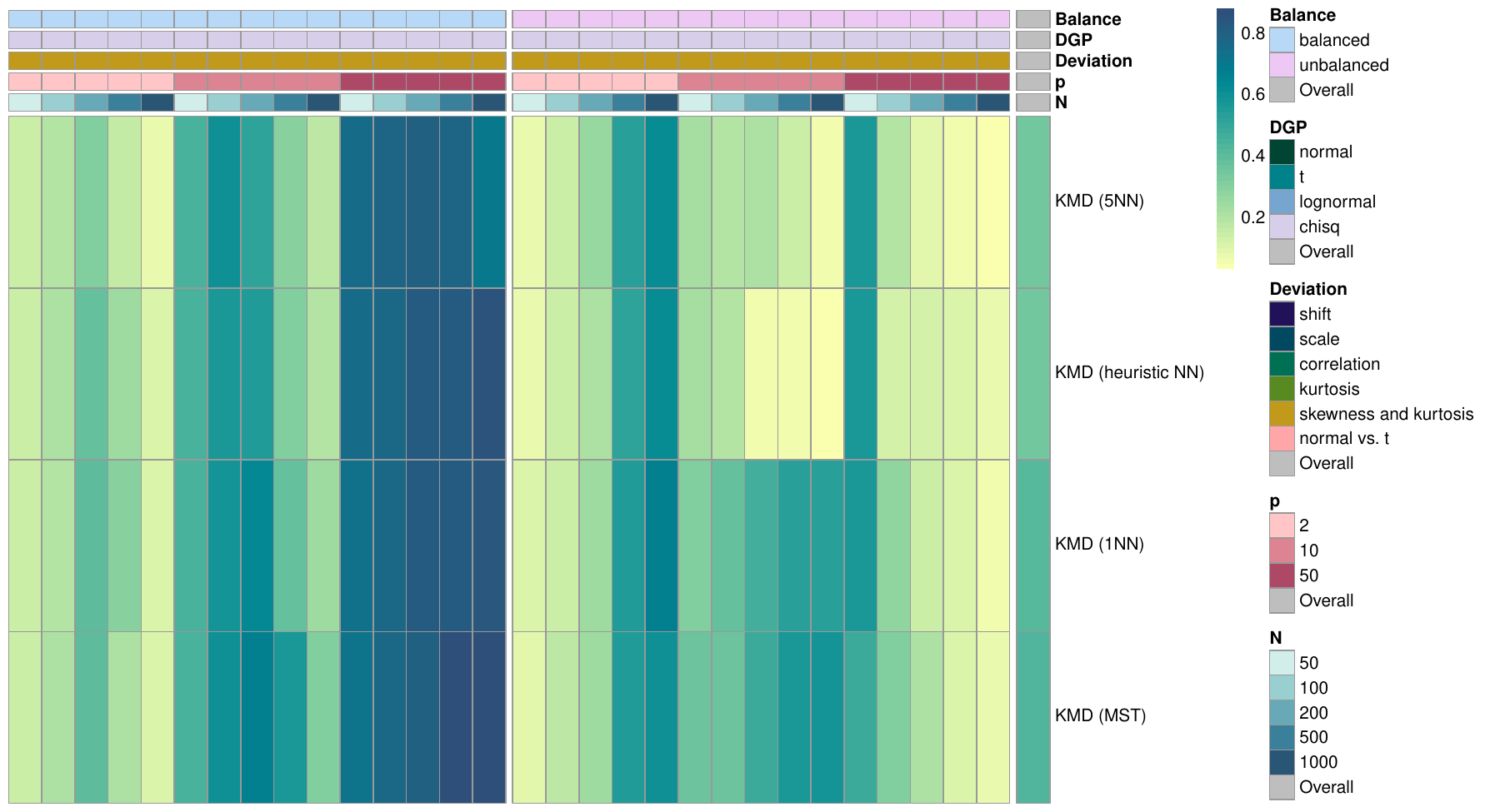}
		\caption{Mean PESR difference to best method per scenario for the skewness and kurtosis alternative and $k = 2$ datasets for all variants of the KMD \textcite{huang_kernel_2022}. Selected variant: KMD~(5NN).}
	\end{figure}
	
	\begin{figure}[!h]
		\centering
		\includegraphics[width=\linewidth]{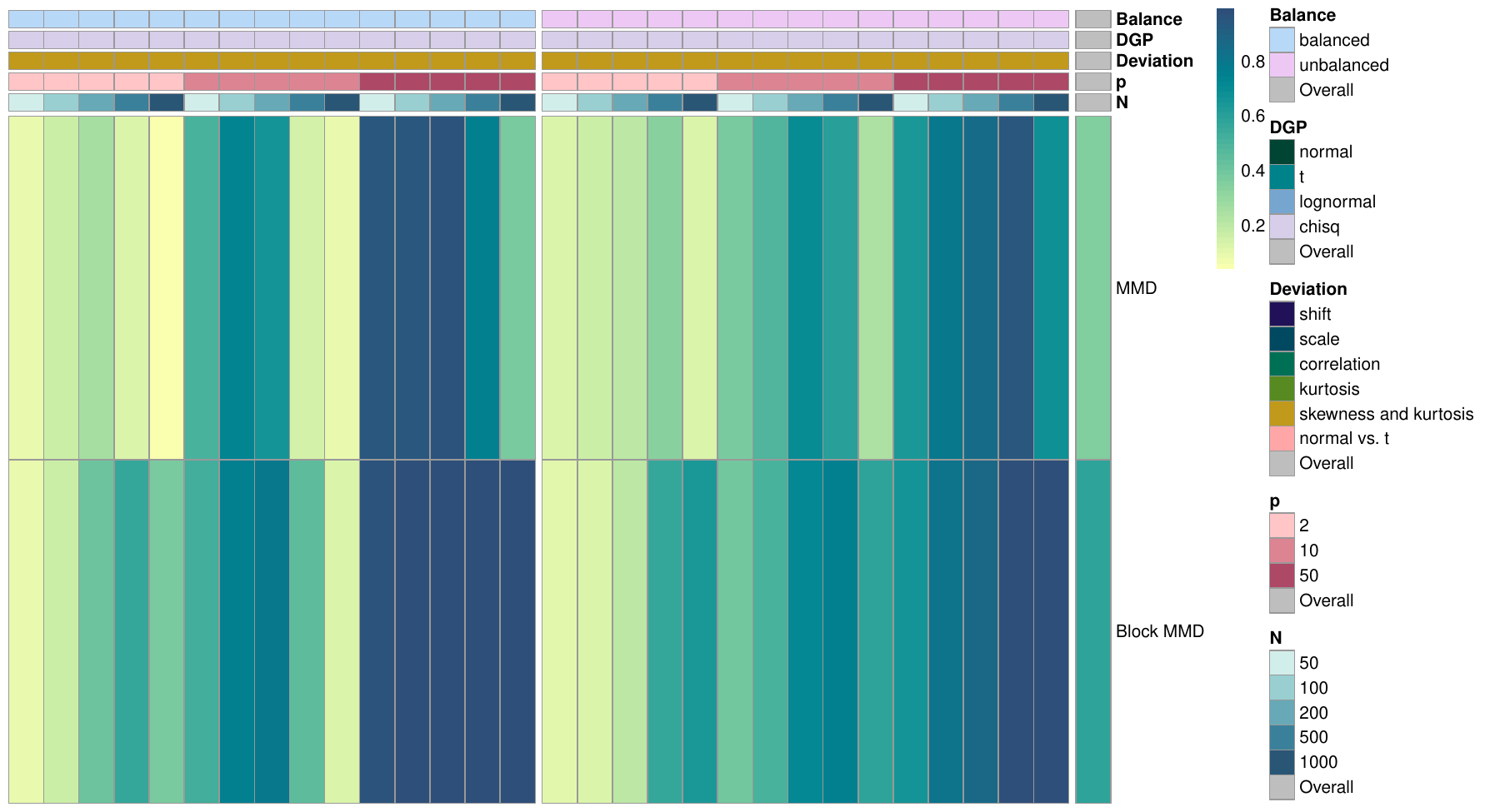}
		\caption{Mean PESR difference to best method per scenario for the skewness and kurtosis alternative and $k = 2$ datasets for all variants of the MMD \textcite{gretton_kernel_2006, zaremba_b-test_2013}. Selected variant: MMD.}
	\end{figure}
	
	\begin{figure}[!h]
		\centering
		\includegraphics[width=\linewidth]{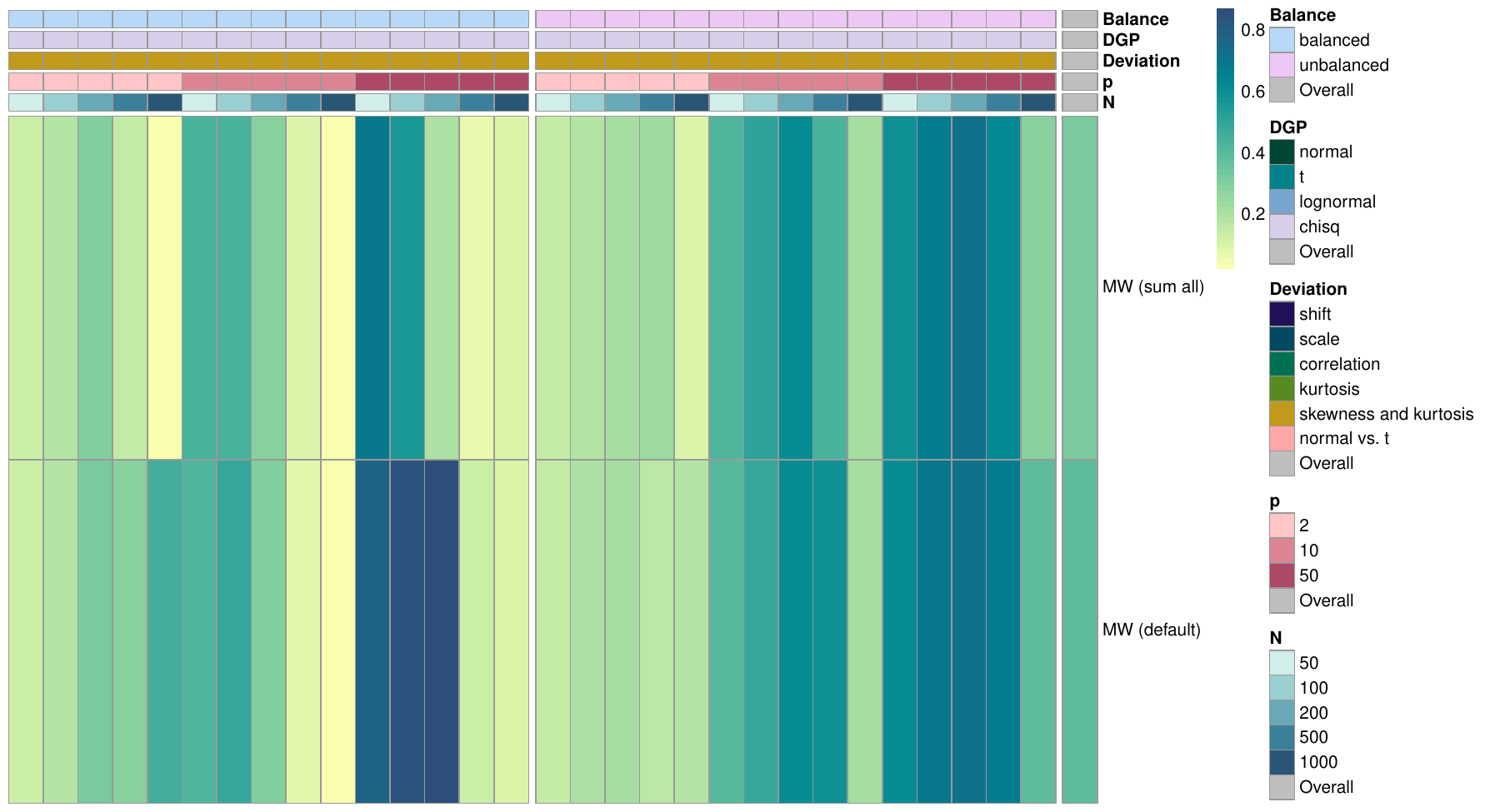}
		\caption{Mean PESR difference to best method per scenario for the skewness and kurtosis alternative and $k = 2$ datasets for all variants of the MW method \textcite{mukhopadhyay_nonparametric_2020}. Selected variant: MW~(sum all).}
	\end{figure}
	
	\begin{figure}[!h]
		\centering
		\includegraphics[width=\linewidth]{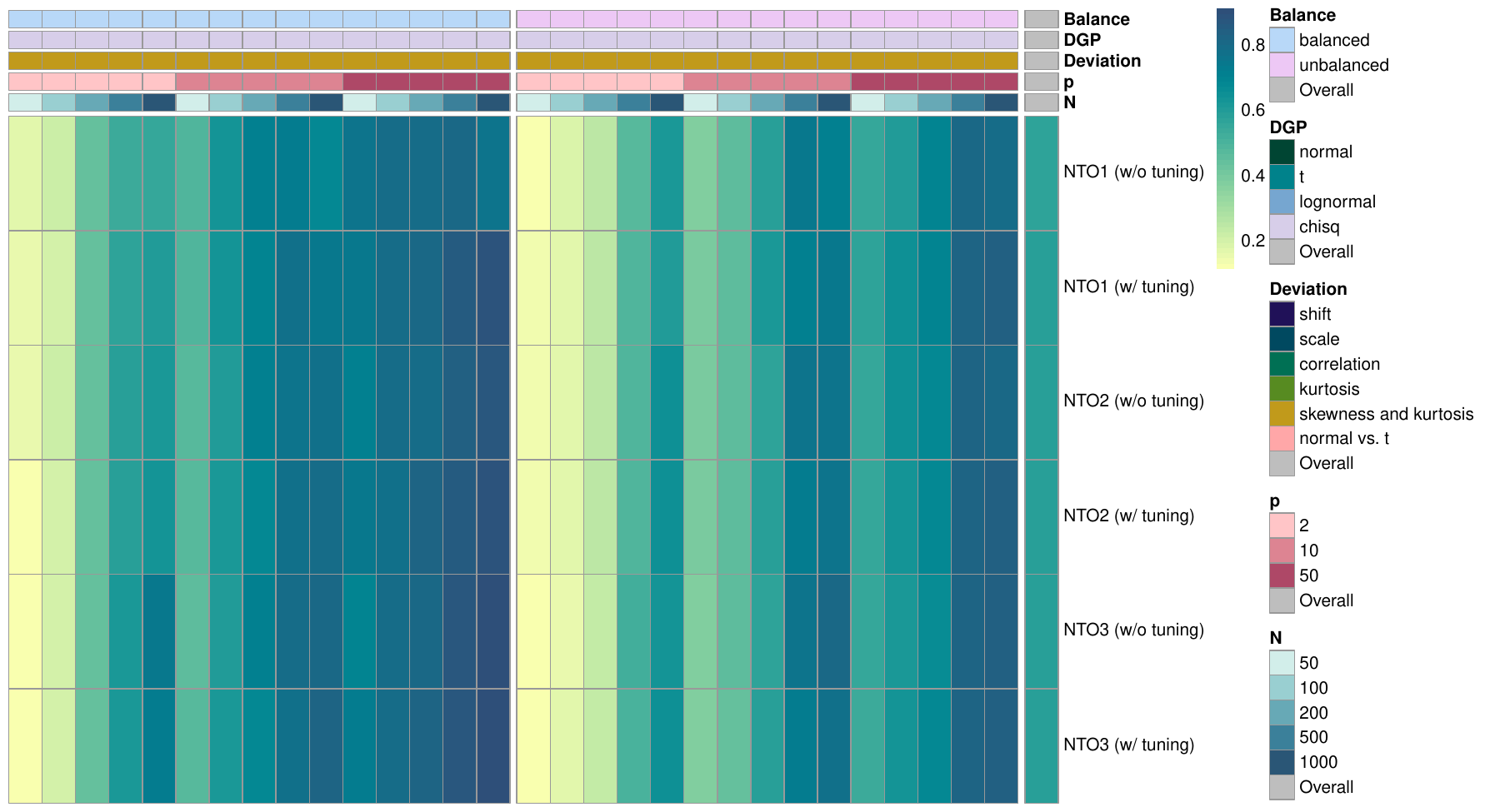}
		\caption{Mean PESR difference to best method per scenario for the skewness and kurtosis alternative and $k = 2$ datasets for all variants of the NKT method \textcite{ntoutsi_general_2008}. Selected variant: NTO1~(w/o~tuning).}
	\end{figure}
	
	\begin{figure}[!h]
		\centering
		\includegraphics[width=\linewidth]{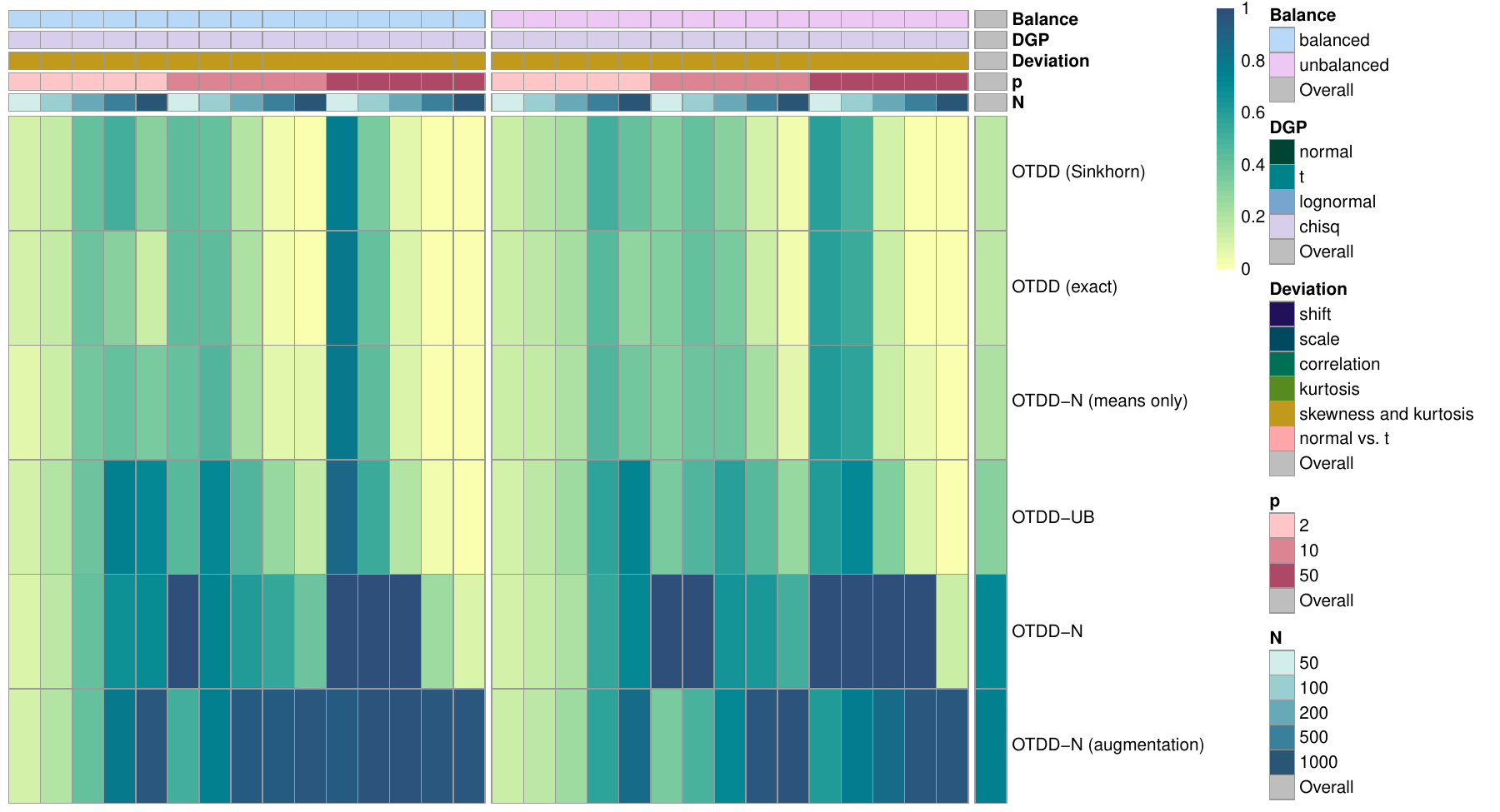}
		\caption{Mean PESR difference to best method per scenario for the skewness and kurtosis alternative and $k = 2$ datasets for all variants of the OTDD \textcite{alvarez-melis_geometric_2020}. Selected variant: OTDD~(Sinkhorn).}
	\end{figure}
	
	\begin{figure}[!h]
		\centering
		\includegraphics[width=\linewidth]{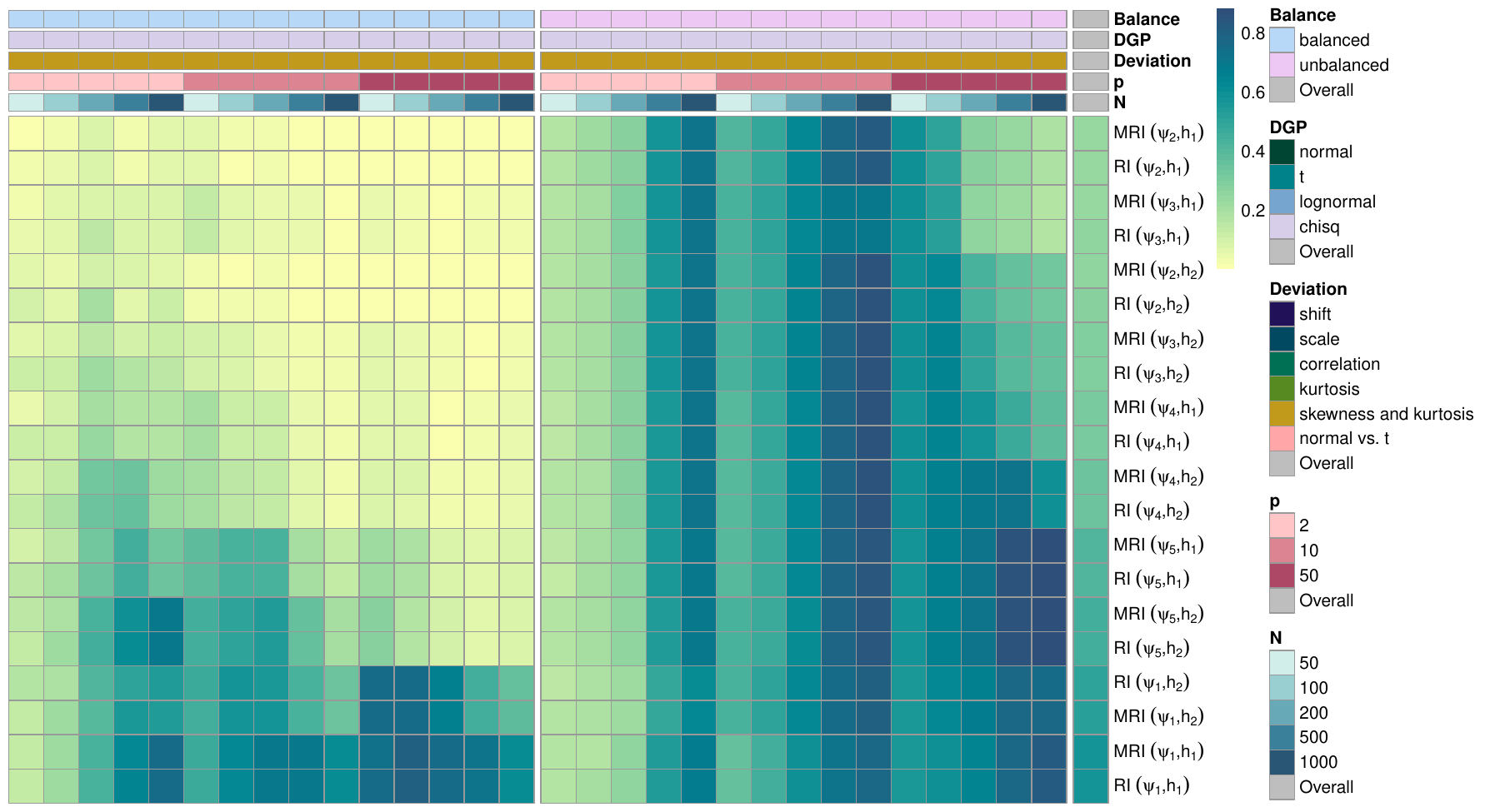}
		\caption{Mean PESR difference to best method per scenario for the skewness and kurtosis alternative and $k = 2$ datasets for all variants of the RI test \textcite{paul_clustering-based_2022}. Selected variant: MRI~($\psi_2, h_1$).}
	\end{figure}
	
	\begin{figure}[!h]
		\centering
		\includegraphics[width=\linewidth]{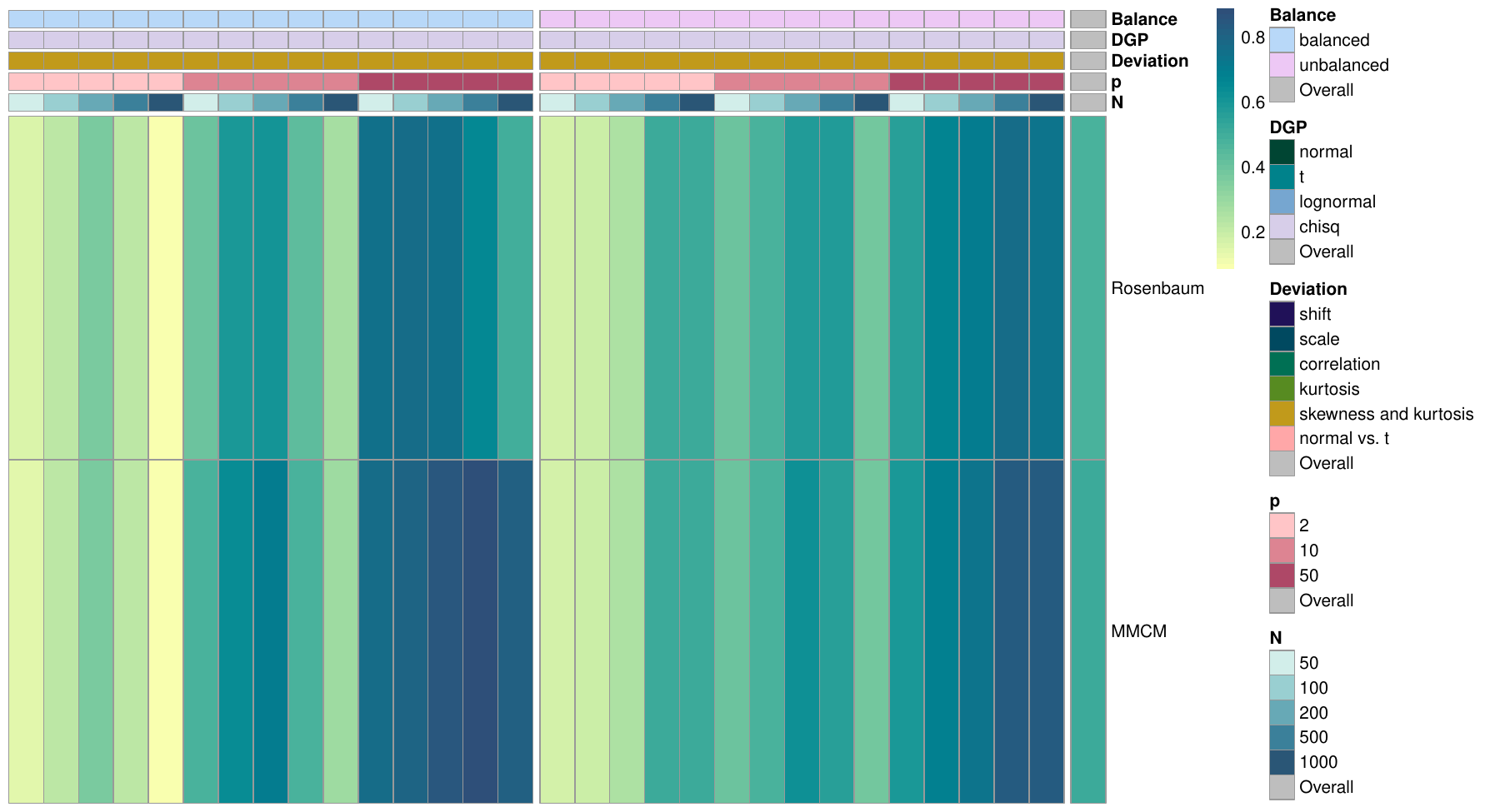}
		\caption{Mean PESR difference to best method per scenario for the skewness and kurtosis alternative and $k = 2$ datasets for all variants of the Rosenbaum test \textcite{rosenbaum_exact_2005, mukherjee_distribution-free_2022}. Selected variant: Rosenbaum.}
	\end{figure}
	
	\begin{figure}[!h]
		\centering
		\includegraphics[width=\linewidth]{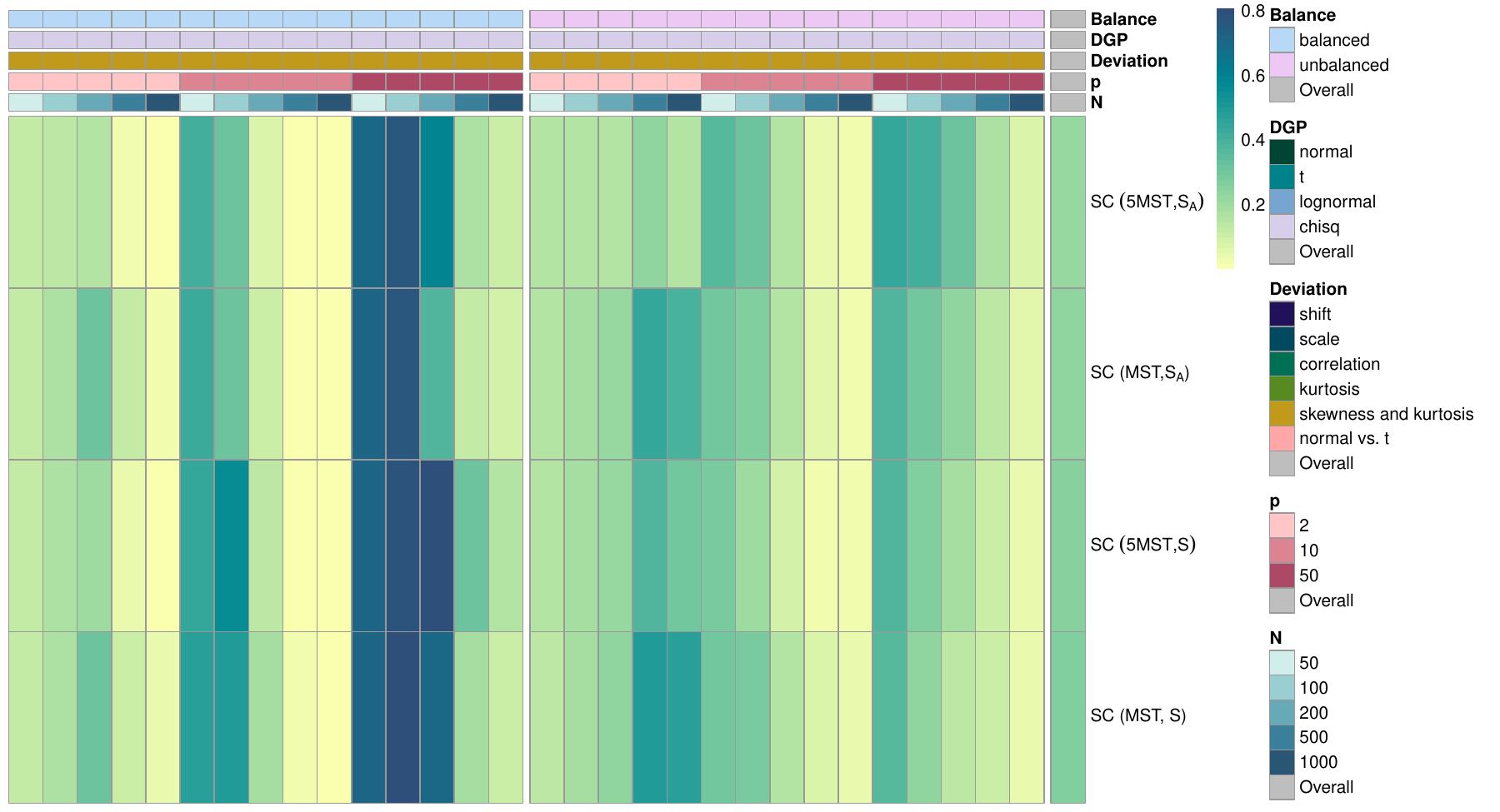}
		\caption{Mean PESR difference to best method per scenario for the kurtosis alternative and $k = 2$ datasets for all variants of the SC test \textcite{song_new_2022}. Selected variant: SC(5MST,S$_{\text{A}}$).}
	\end{figure}
	
	\begin{figure}[!h]
		\centering
		\includegraphics[width=\linewidth]{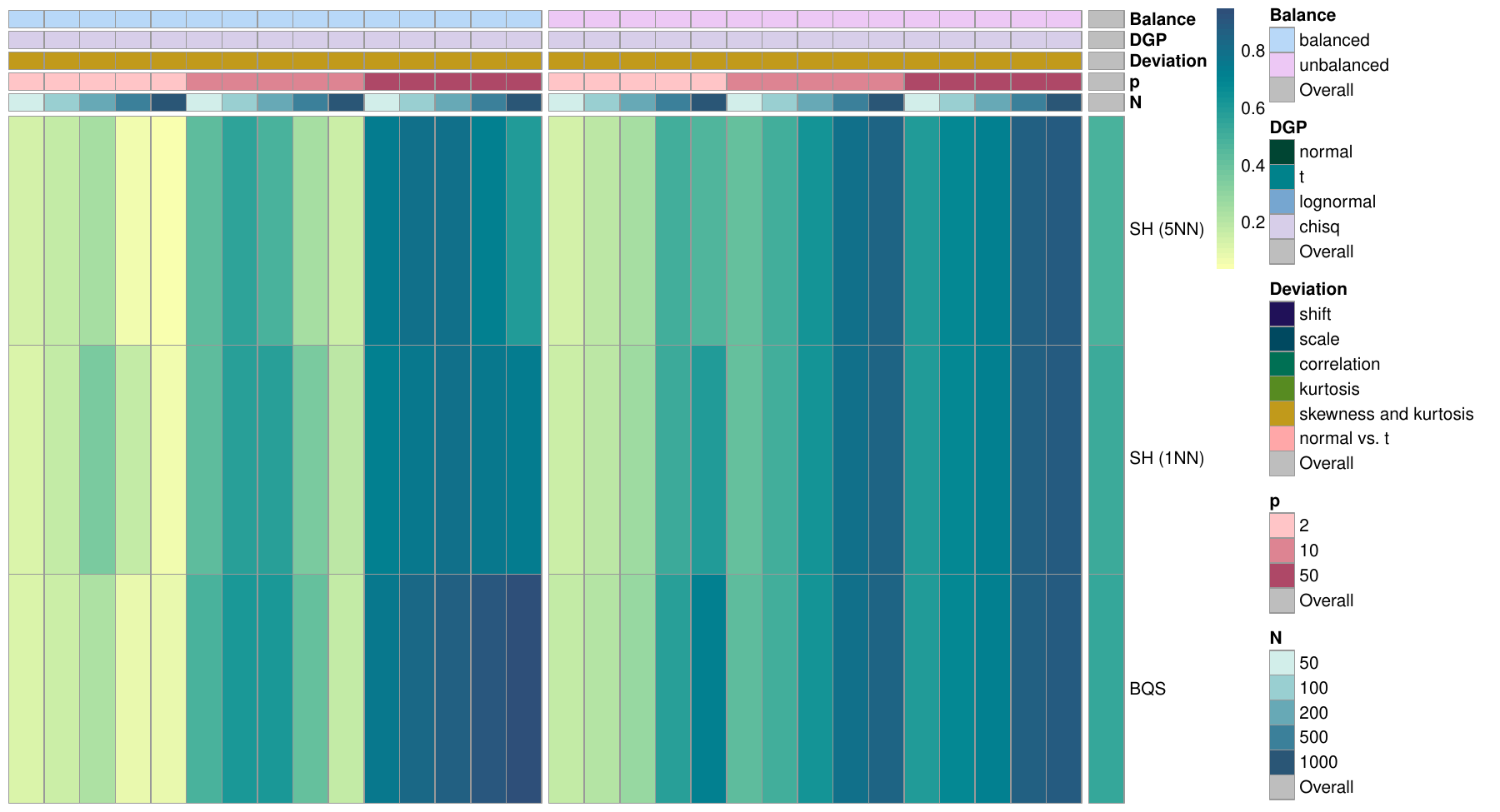}
		\caption{Mean PESR difference to best method per scenario for the skewness and kurtosis alternative and $k = 2$ datasets for all variants of the SH test \textcite{schilling_multivariate_1986, henze_multivariate_1988, barakat_multivariate_1996}. Selected variants: SH~(5NN).}
	\end{figure}
	\clearpage
	
	\paragraph{OGM}\mbox{} 
	
	\begin{figure}[!h]
		\centering
		\includegraphics[width=\linewidth]{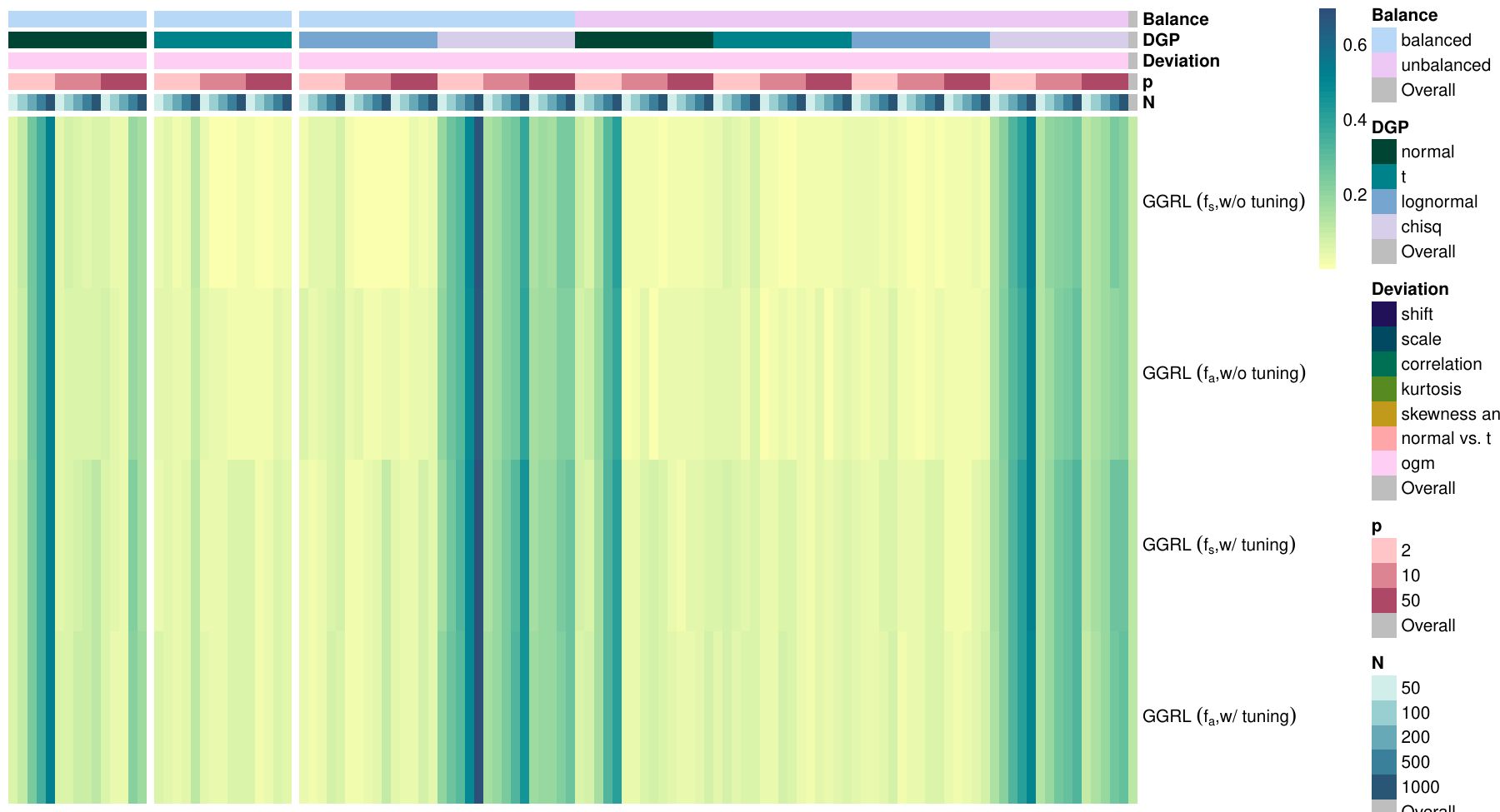}
		\caption{Mean PESR difference to best method per scenario for OGM alternatives and $k = 2$ datasets for all variants of the GGRL method \textcite{ganti_framework_1999}. Selected variant: GGRL~($f_s$,w/o~tuning).}
	\end{figure}
	
	\begin{figure}[!h]
		\centering
		\includegraphics[width=\linewidth]{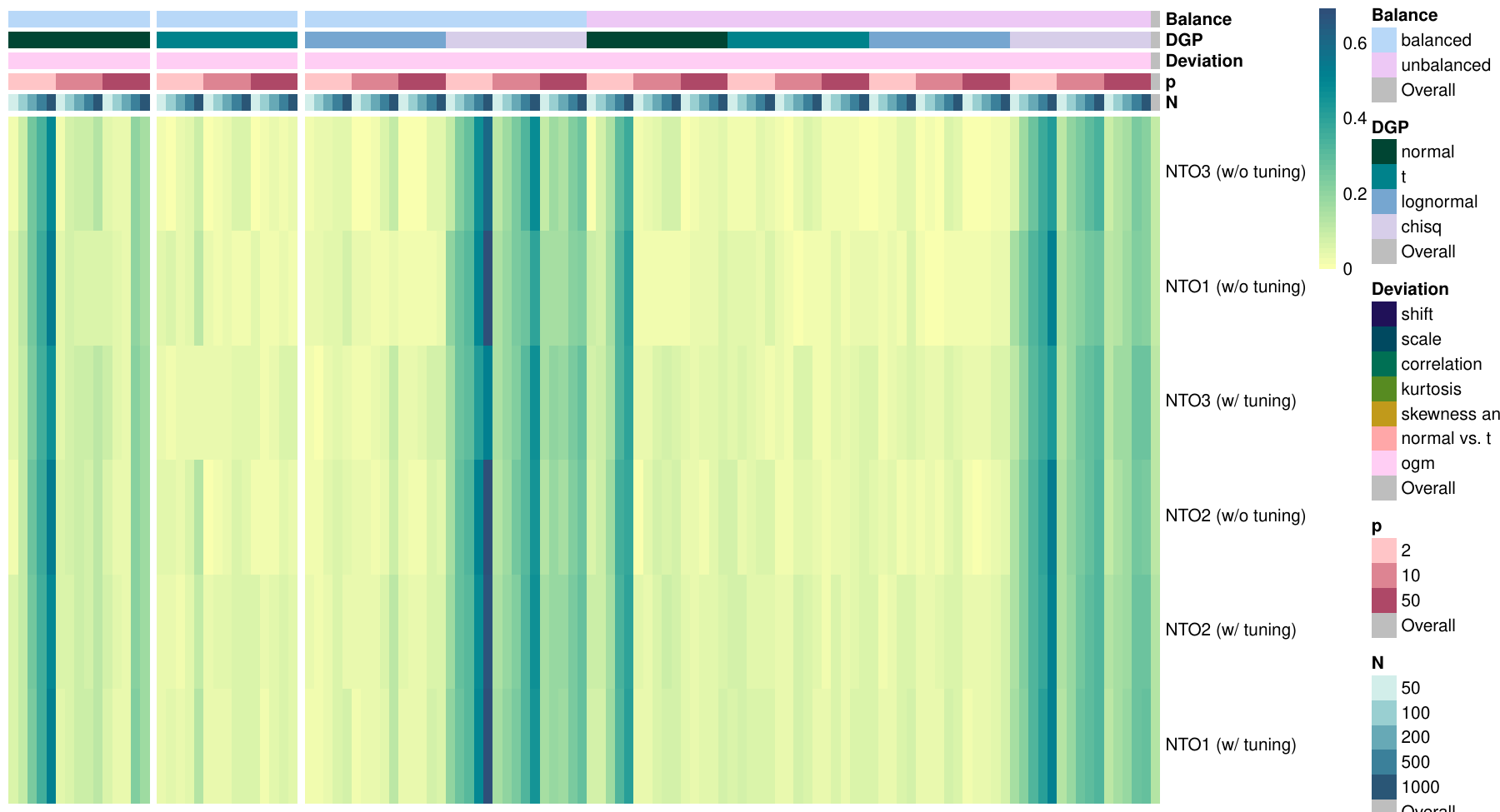}
		\caption{Mean PESR difference to best method per scenario for OGM alternatives and $k = 2$ datasets for all variants of the NKT method \textcite{ntoutsi_general_2008}. Selected variant: NTO3~(w/o~tuning).}
	\end{figure}
	
	\begin{figure}[!h]
		\centering
		\includegraphics[width=\linewidth]{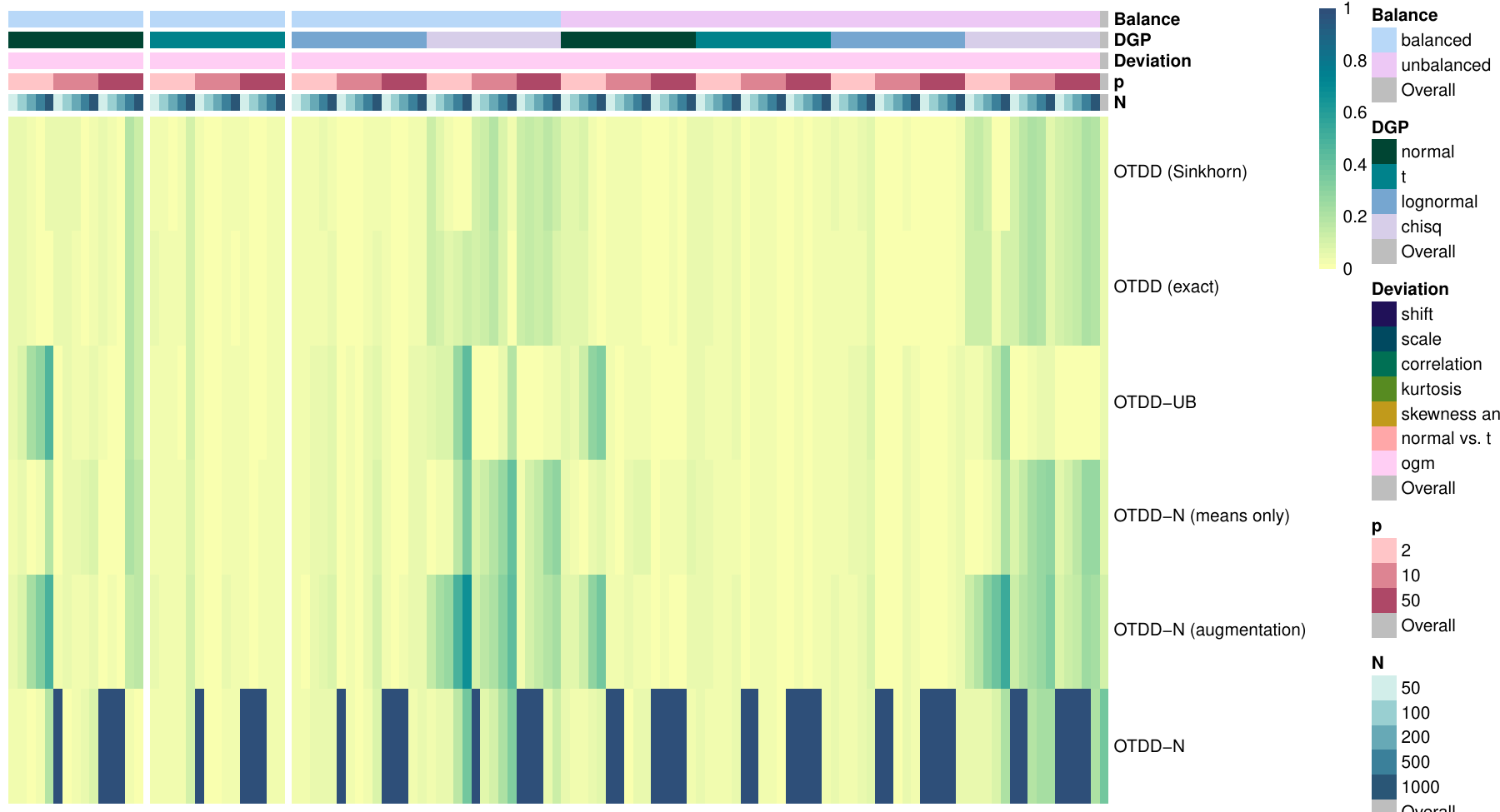}
		\caption{Mean PESR difference to best method per scenario for OGM alternatives and $k = 2$ datasets for all variants of the OTDD \textcite{alvarez-melis_geometric_2020}. Selected variant: OTDD~(Sinkhorn).}
	\end{figure}
	\clearpage
	
	\subsubsection[k = 4]{$k = 4$}
	
	\paragraph{Shift}\mbox{} 
	
	\begin{figure}[!h]
		\centering
		\includegraphics[width=\linewidth]{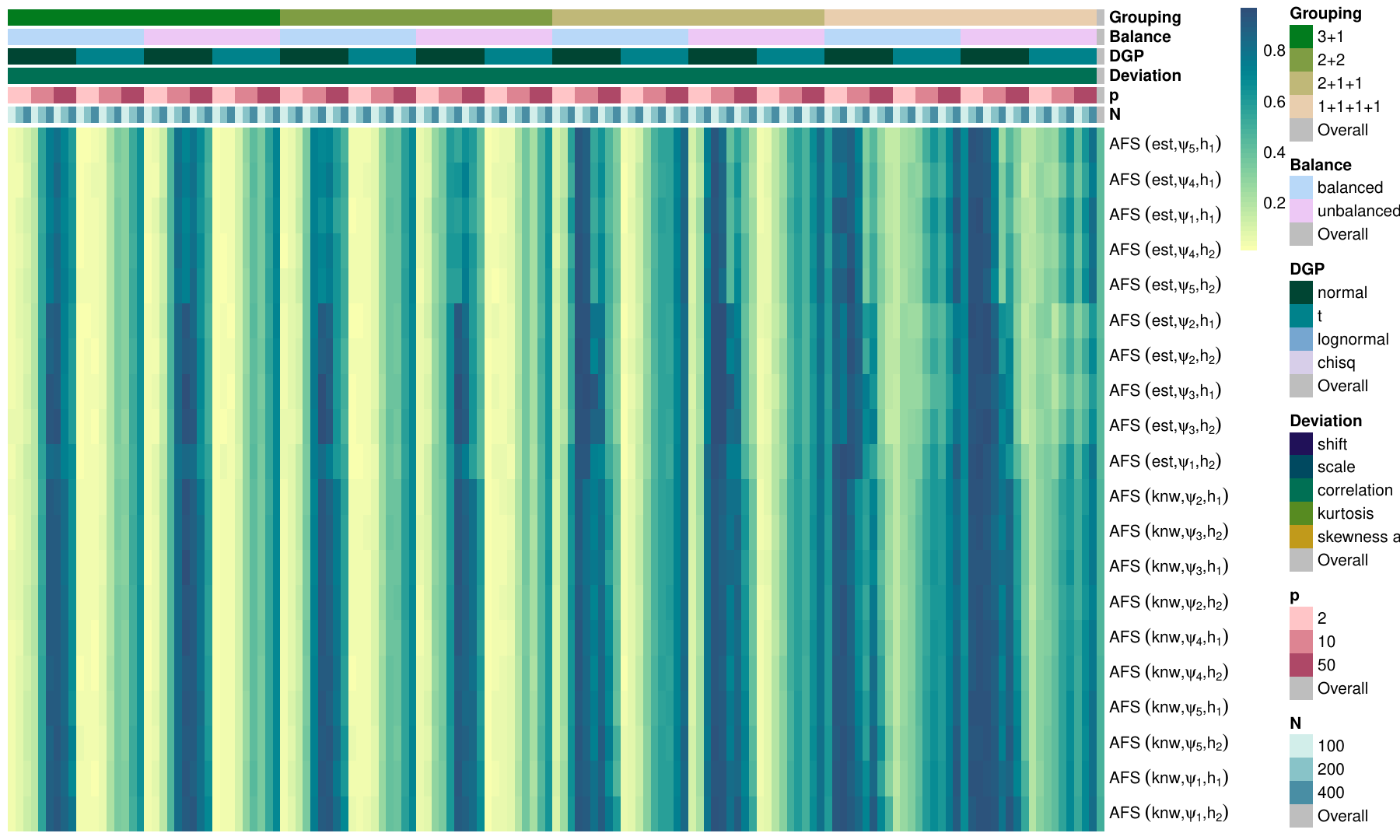}
		\caption{Mean PESR difference to best method per scenario for the shift alternative and $k = 4$ datasets for all variants of the aggregated FS test \textcite{paul_clustering-based_2022}. Selected variant: AFS~(est,$\psi_5,h_1$).}
	\end{figure}
	
	\begin{figure}[!h]
		\centering
		\includegraphics[width=\linewidth]{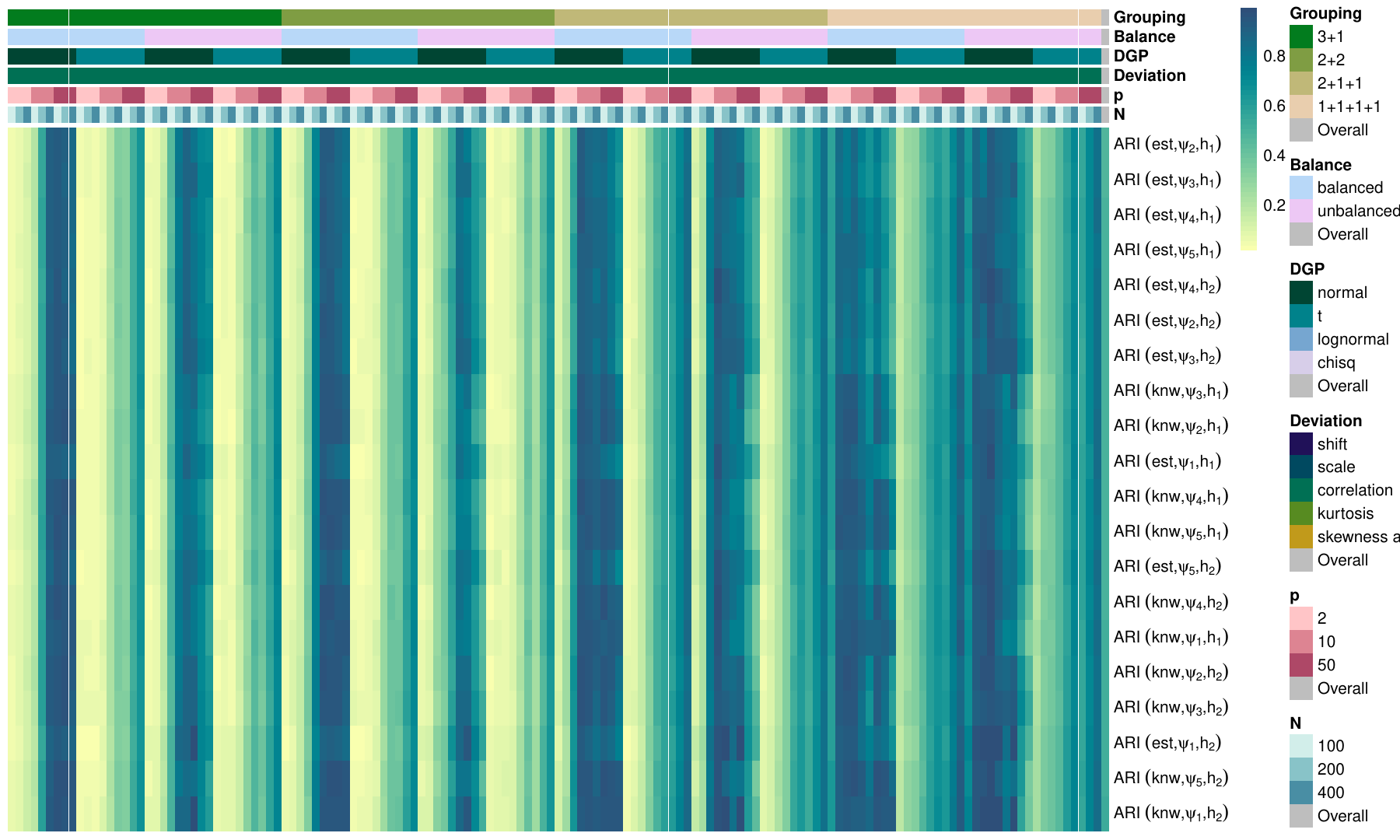}
		\caption{Mean PESR difference to best method per scenario for the shift alternative and $k = 4$ datasets for all variants of the aggregated RI test \textcite{paul_clustering-based_2022}. Selected variant: ARI~(est,$\psi_2,h_1$).}
	\end{figure}
	
	\begin{figure}[!h]
		\centering
		\includegraphics[width=\linewidth]{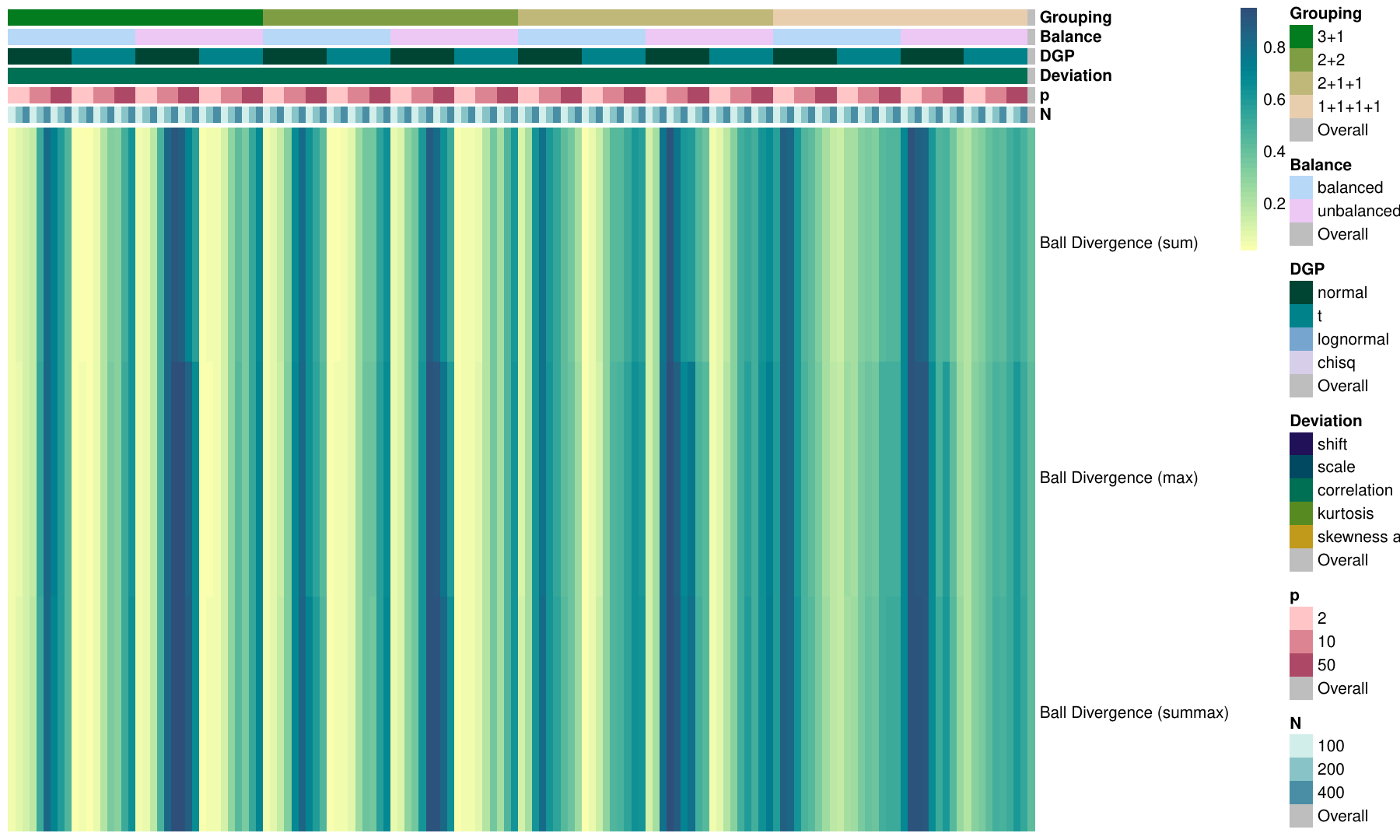}
		\caption{Mean PESR difference to best method per scenario for the shift alternative and $k = 4$ datasets for all variants of the Ball divergence \textcite{pan_ball_2018}. Selected variant: Ball divergence~(sum).}
	\end{figure}
	
	\begin{figure}[!h]
		\centering
		\includegraphics[width=\linewidth]{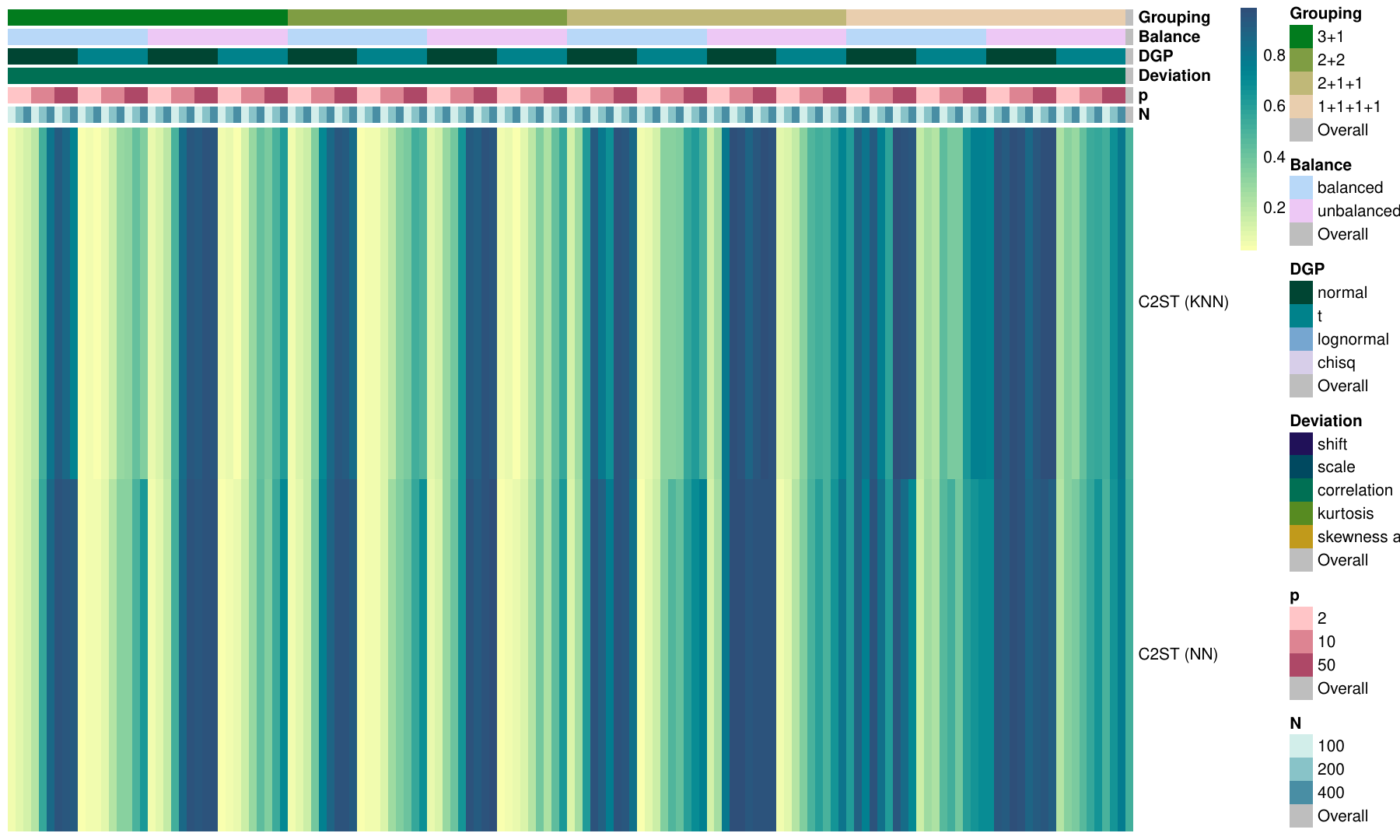}
		\caption{Mean PESR difference to best method per scenario for the shift alternative and $k = 4$ datasets for all variants of the C2ST \textcite{alvarez-melis_geometric_2020}. Selected variant: C2ST~(KNN).}
	\end{figure}
	
	\begin{figure}[!h]
		\centering
		\includegraphics[width=\linewidth]{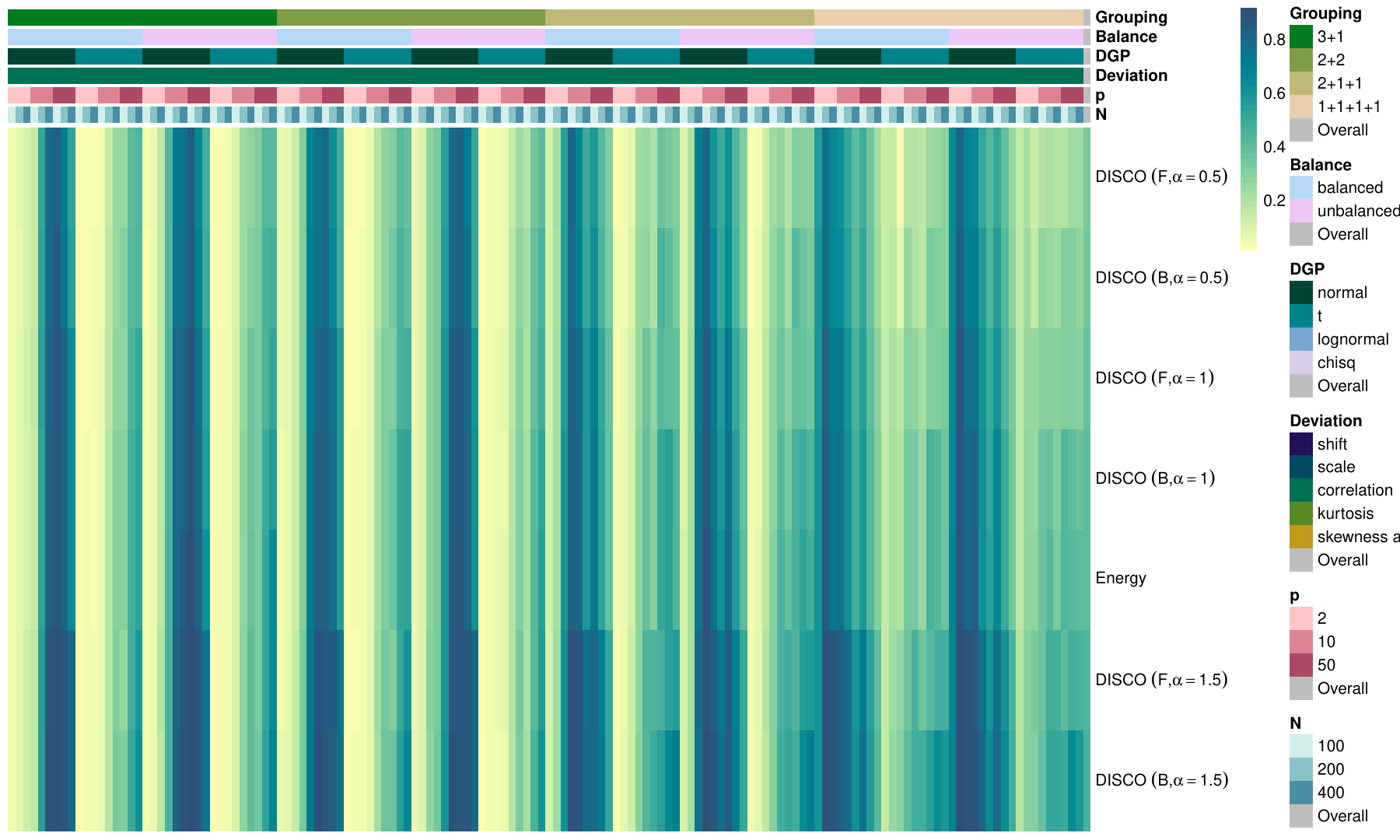}
		\caption{Mean PESR difference to best method per scenario for the shift alternative and $k = 4$ datasets for all variants of the Energy distance / DISCO \textcite{szekely_testing_2004, rizzo_disco_2010}. Selected variant: DISCO~(F,$\alpha=0.5$).}
	\end{figure}
	
	\begin{figure}[!h]
		\centering
		\includegraphics[width=\linewidth]{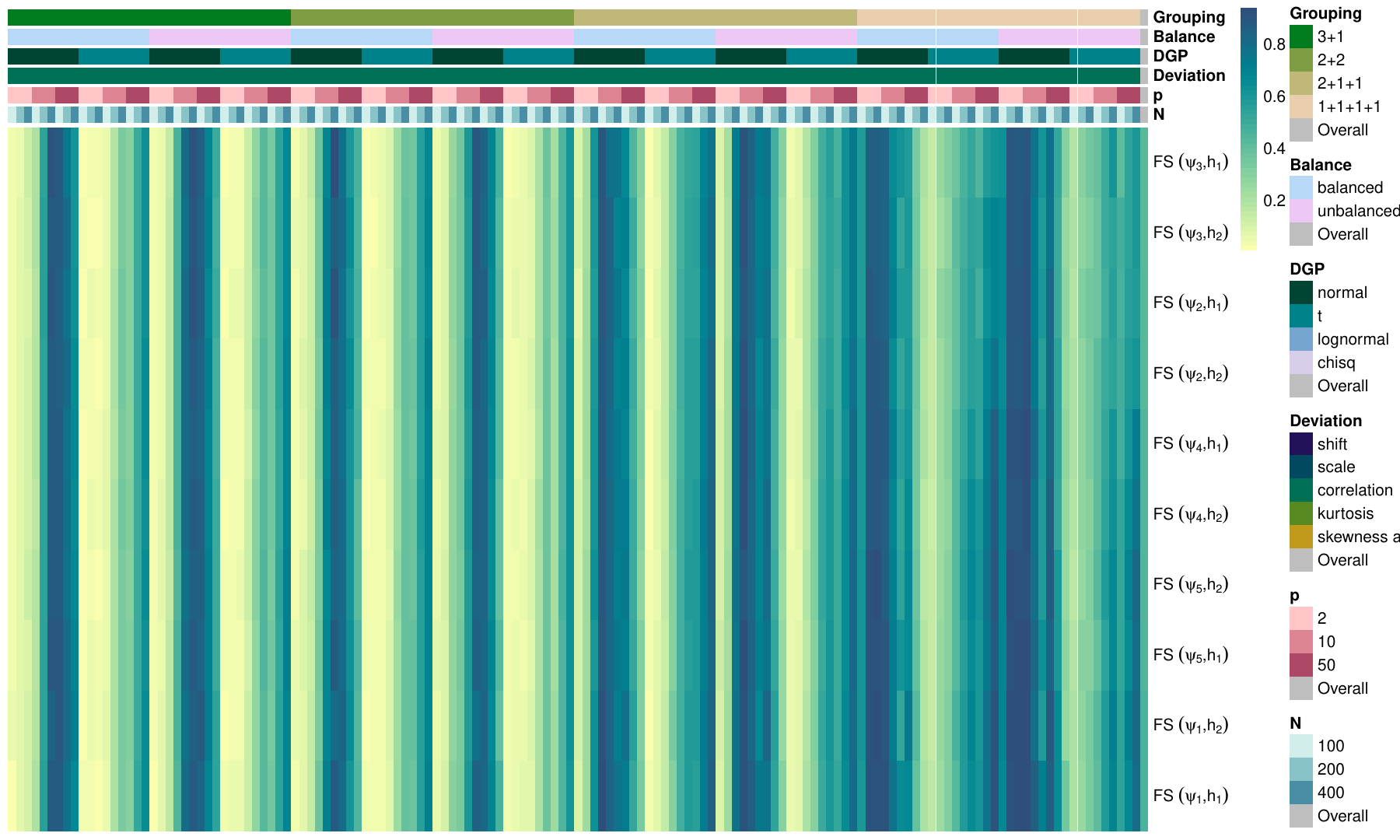}
		\caption{Mean PESR difference to best method per scenario for the shift alternative and $k = 4$ datasets for all variants of the FS test \textcite{paul_clustering-based_2022}. Selected variant: FS~($\psi_3$,$h_1$).}
	\end{figure}
	
	\begin{figure}[!h]
		\centering
		\includegraphics[width=\linewidth]{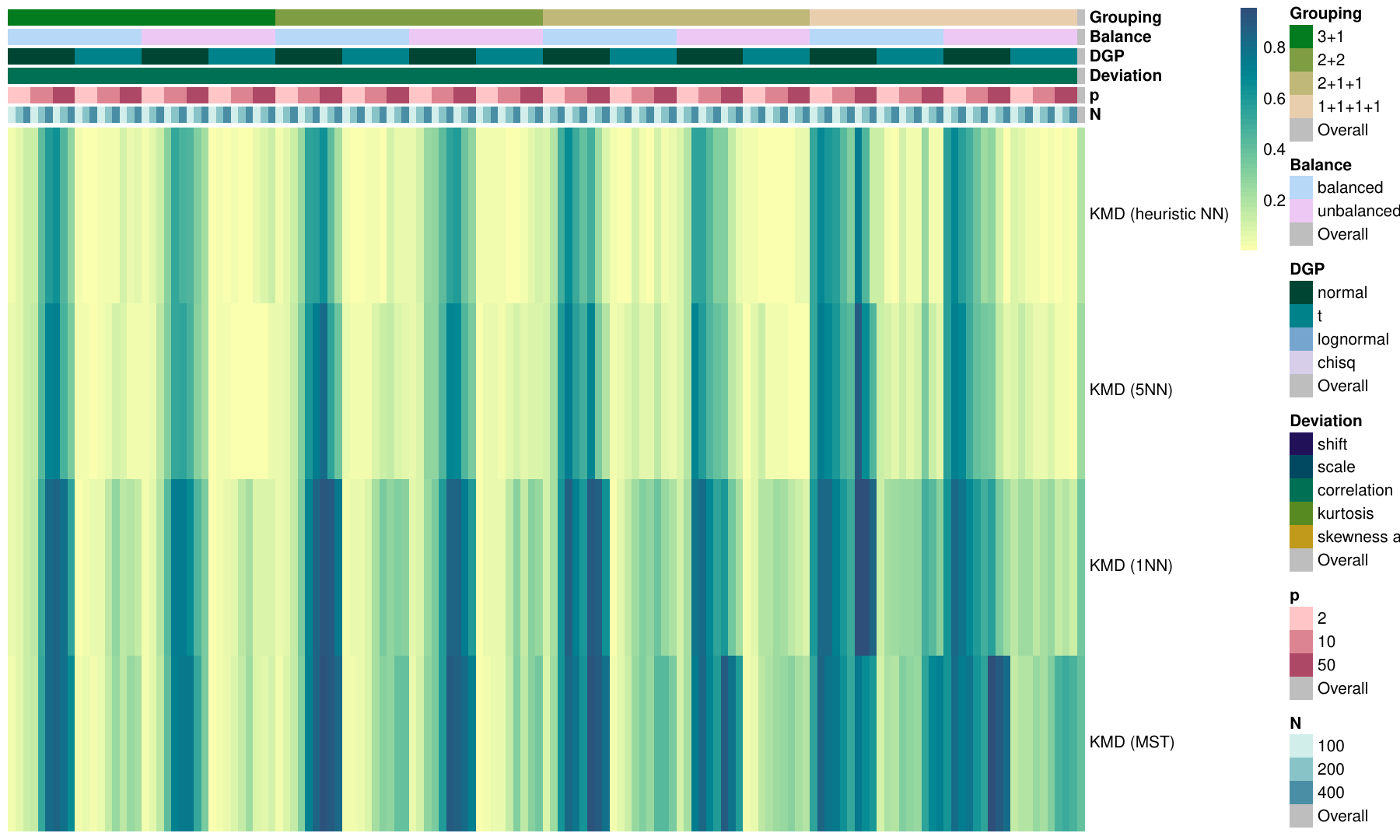}
		\caption{Mean PESR difference to best method per scenario for the shift alternative and $k = 4$ datasets for all variants of the modified KMD \textcite{huang_kernel_2022}. Selected variant: KMD~(heuristic NN).}
	\end{figure}
	
	\begin{figure}[!h]
		\centering
		\includegraphics[width=\linewidth]{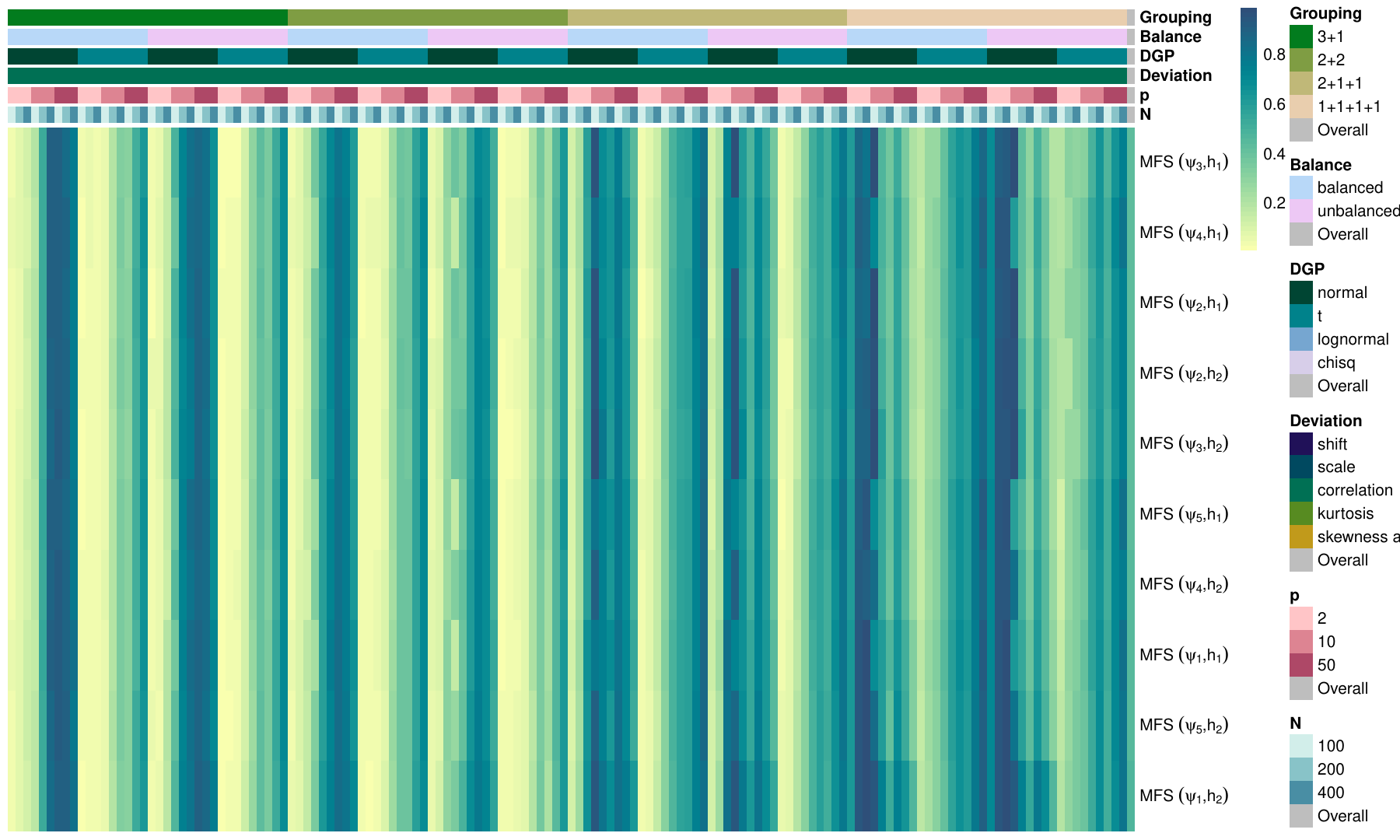}
		\caption{Mean PESR difference to best method per scenario for the shift alternative and $k = 4$ datasets for all variants of the modified FS test \textcite{paul_clustering-based_2022}. Selected variant: MFS~($\psi_3$,$h_1$).}
	\end{figure}
	
	\begin{figure}[!h]
		\centering
		\includegraphics[width=\linewidth]{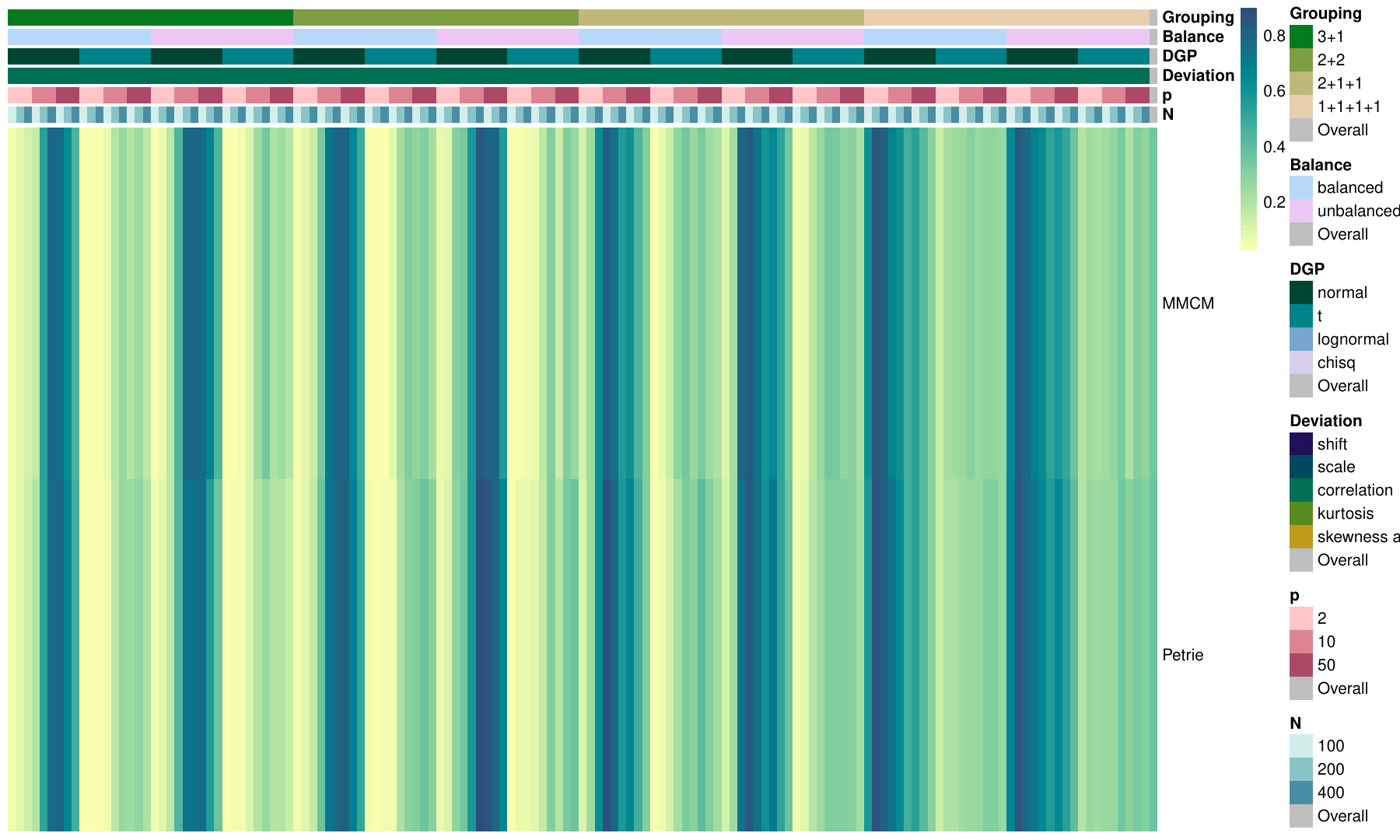}
		\caption{Mean PESR difference to best method per scenario for the shift alternative and $k = 4$ datasets for the MMCM and Petrie's test \textcite{mukherjee_distribution-free_2022, petrie_graph-theoretic_2016}. Selected variant: MMCM.}
	\end{figure}
	
	\begin{figure}[!h]
		\centering
		\includegraphics[width=\linewidth]{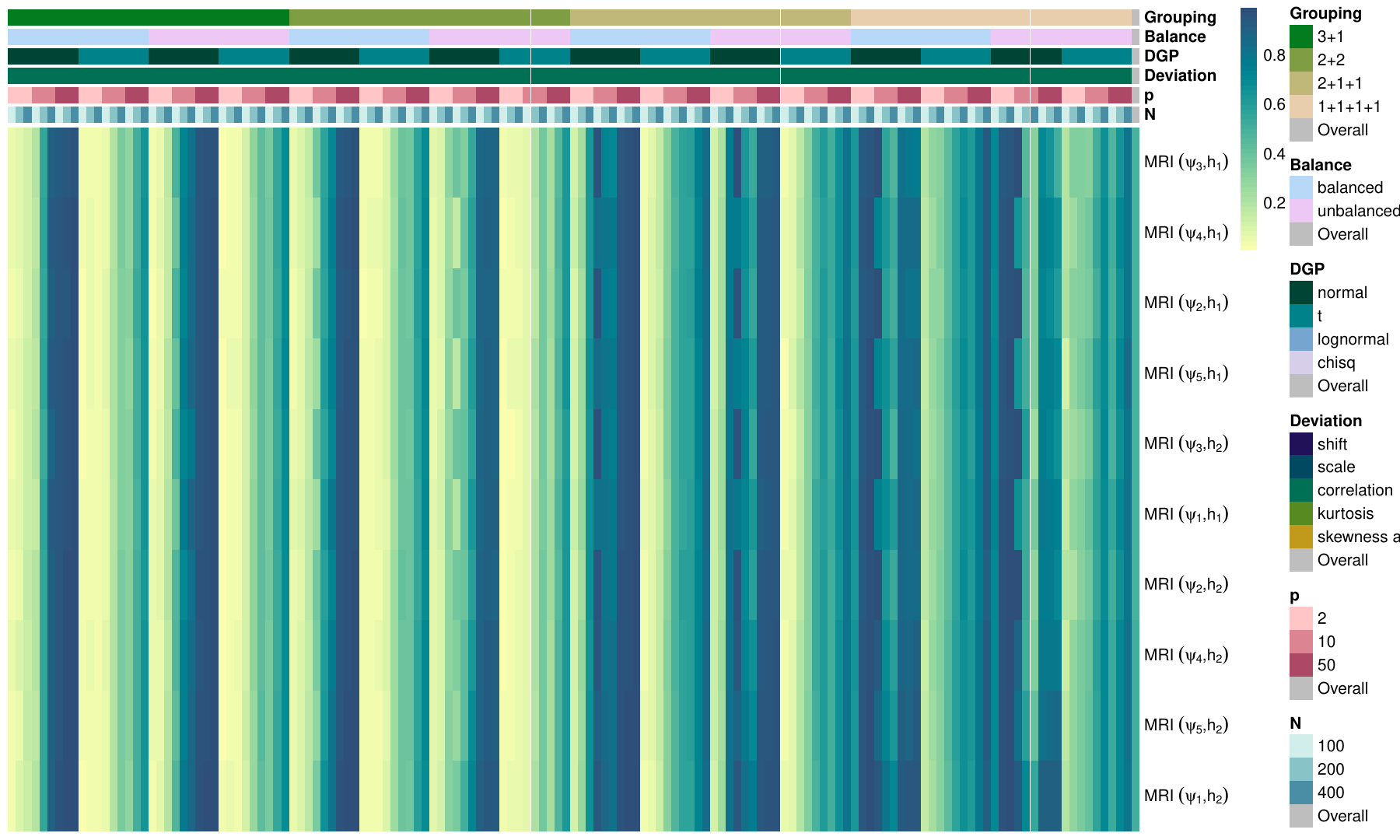}
		\caption{Mean PESR difference to best method per scenario for the shift alternative and $k = 4$ datasets for all variants of the modified RI test \textcite{paul_clustering-based_2022}. Selected variant: MRI~($\psi_3$,$h_1$).}
	\end{figure}
	
	\begin{figure}[!h]
		\centering
		\includegraphics[width=\linewidth]{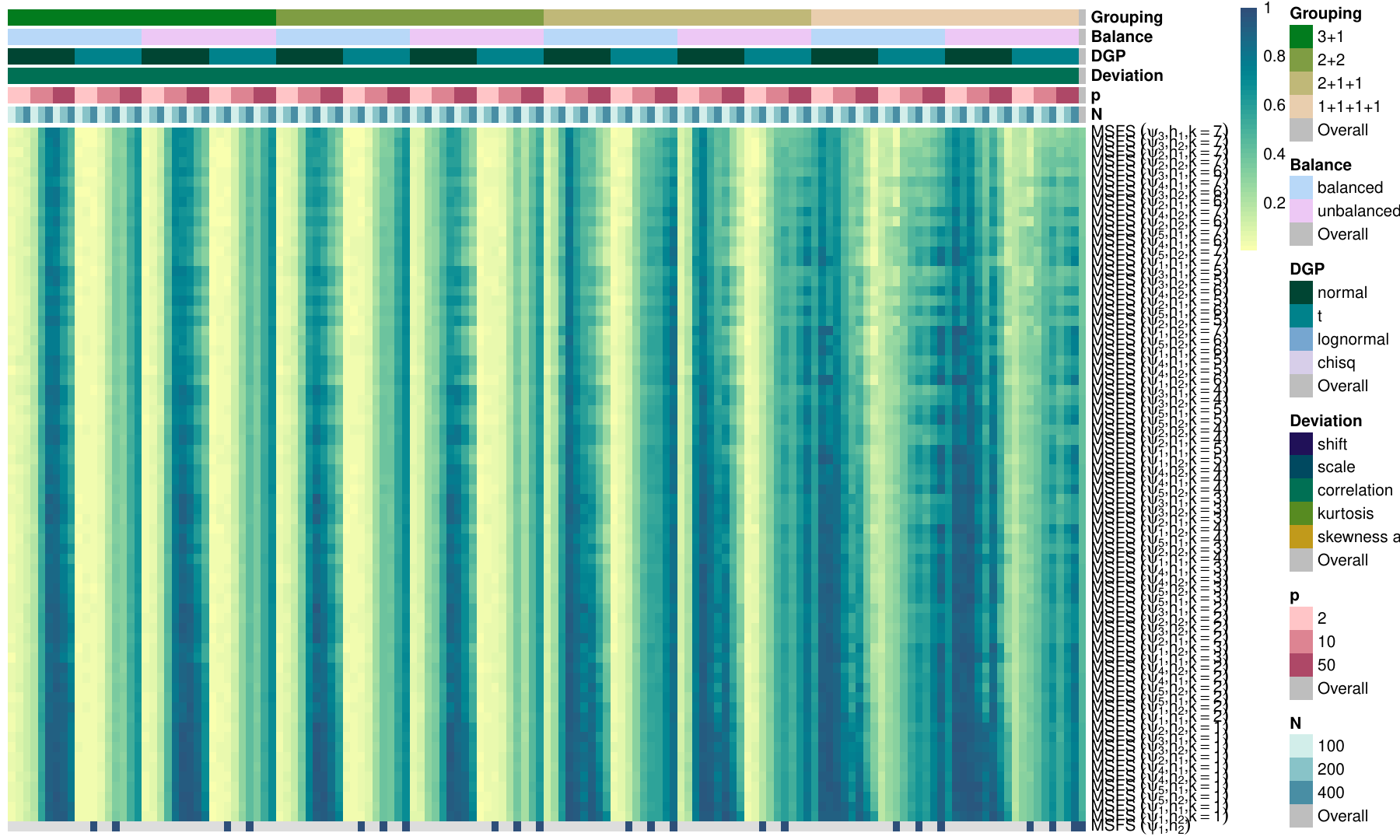}
		\caption{Mean PESR difference to best method per scenario for the shift alternative and $k = 4$ datasets for all variants of the multiscale FS test \textcite{paul_clustering-based_2022}. Selected variant: MSFS~($\psi_3$,$h_1$,$k=7$).}
	\end{figure}
	
	\begin{figure}[!h]
		\centering
		\includegraphics[width=\linewidth]{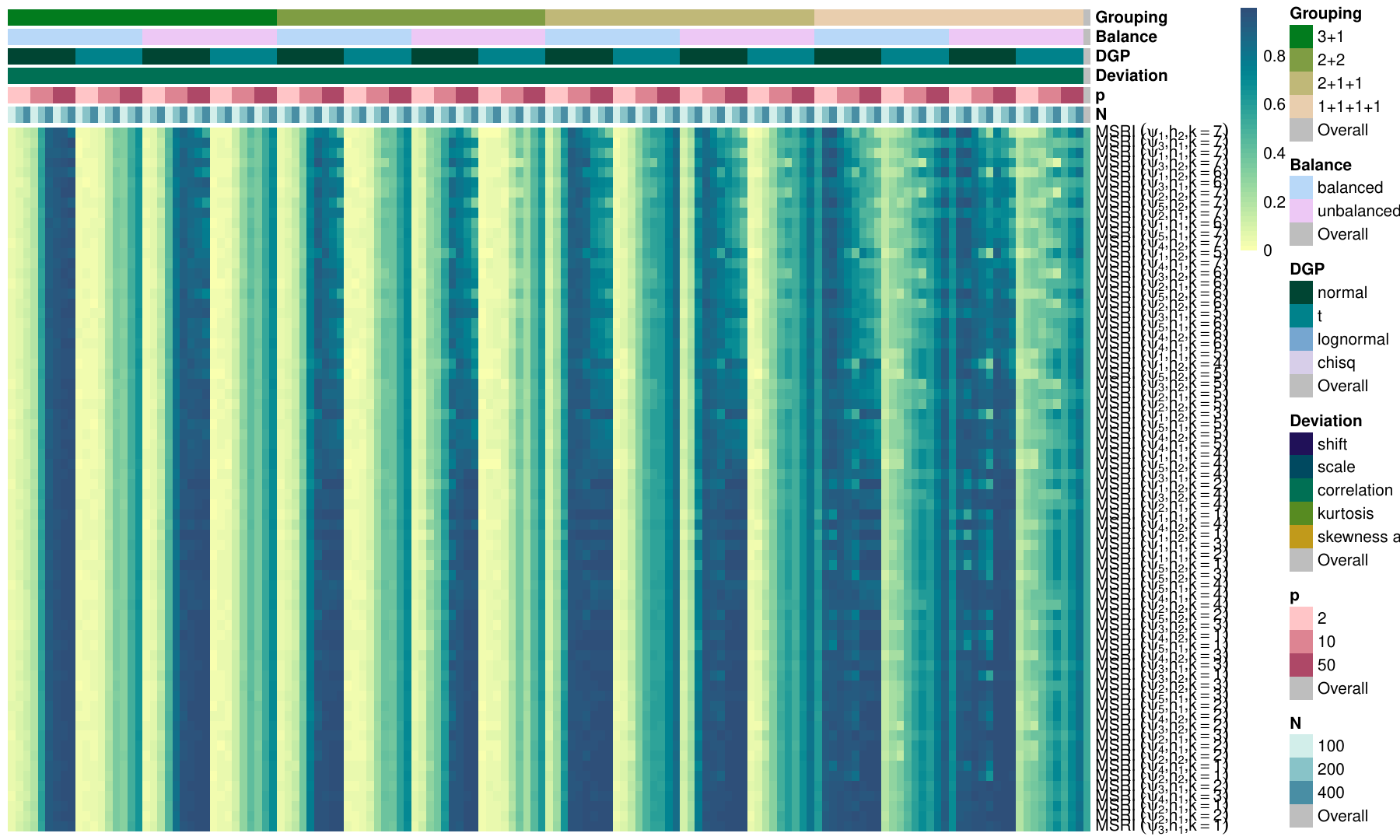}
		\caption{Mean PESR difference to best method per scenario for the shift alternative and $k = 4$ datasets for all variants of the multiscale RI test \textcite{paul_clustering-based_2022}. Selected variant: MSRI~($\psi_1$,$h_2$,$k=7$).}
	\end{figure}
	
	\begin{figure}[!h]
		\centering
		\includegraphics[width=\linewidth]{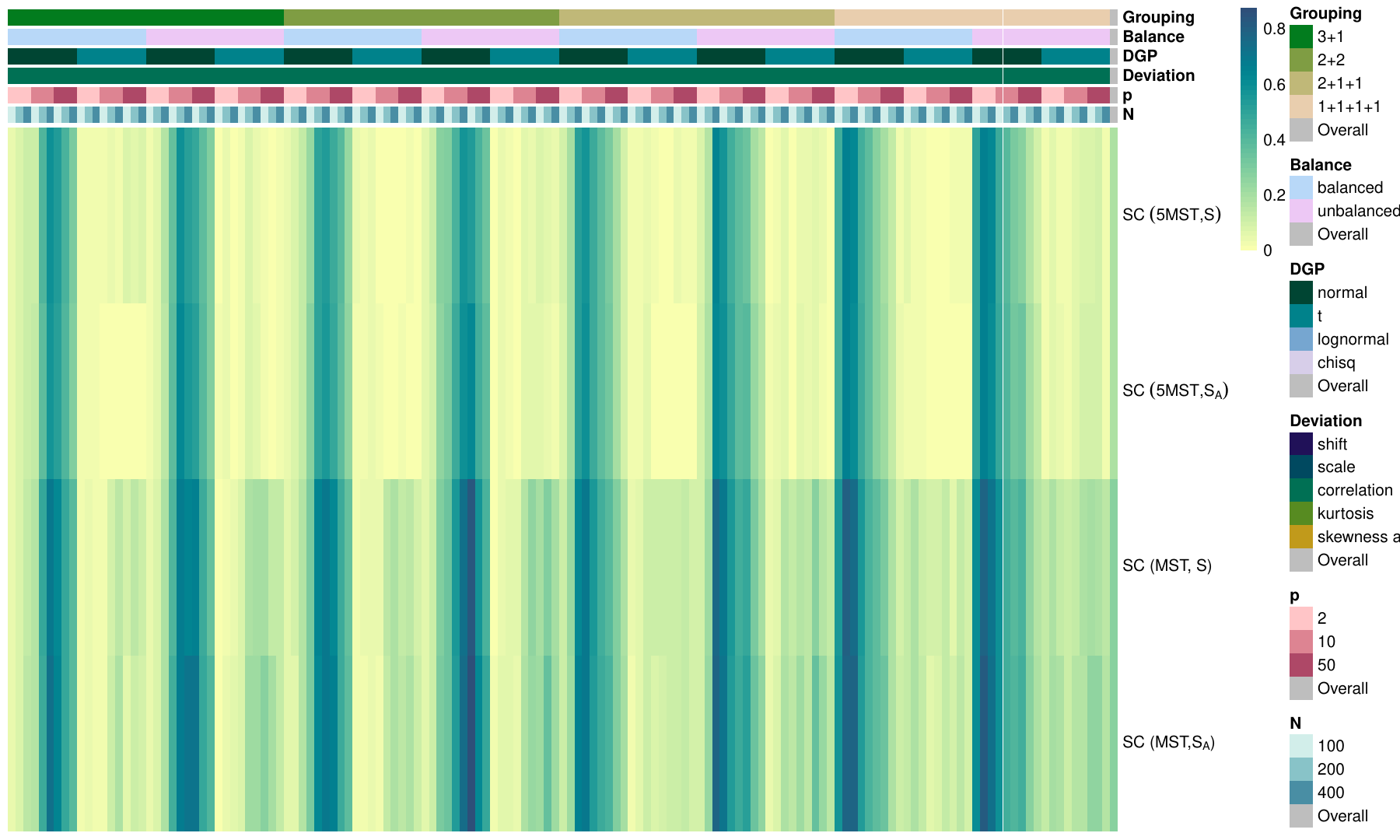}
		\caption{Mean PESR difference to best method per scenario for the shift alternative and $k = 4$ datasets for all variants of the SC test \textcite{song_generalized_2021}. Selected variant: SC~(5MST,$S$).}
	\end{figure}
	\clearpage
	
	\paragraph{Scale}	\mbox{} 
	
	\begin{figure}[!h]
		\centering
		\includegraphics[width=\linewidth]{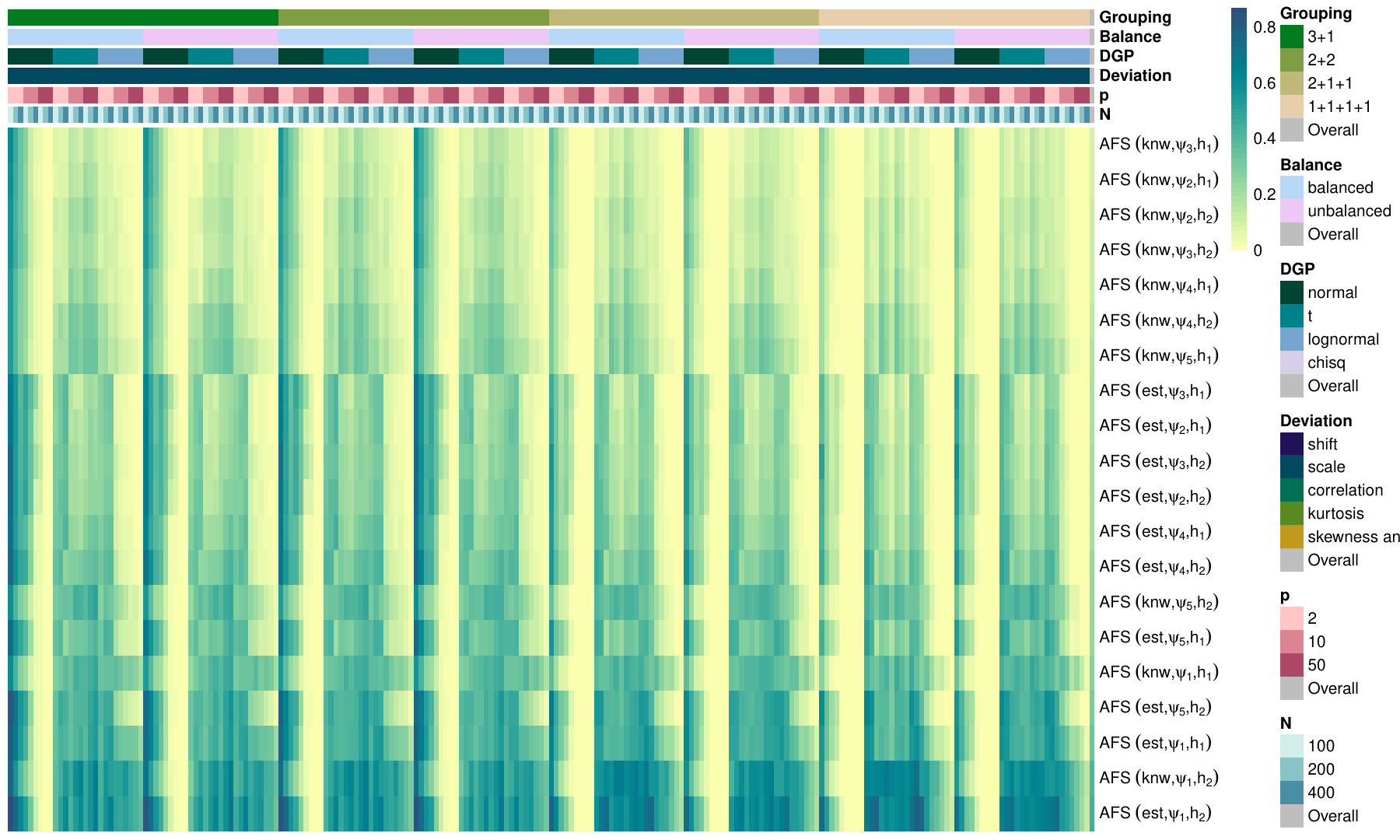}
		\caption{Mean PESR difference to best method per scenario for the scale alternative and $k = 4$ datasets for all variants of the aggregated FS test \textcite{paul_clustering-based_2022}. Selected variant: AFS~(knw,$\psi_3,h_1$).}
	\end{figure}
	
	\begin{figure}[!h]
		\centering
		\includegraphics[width=\linewidth]{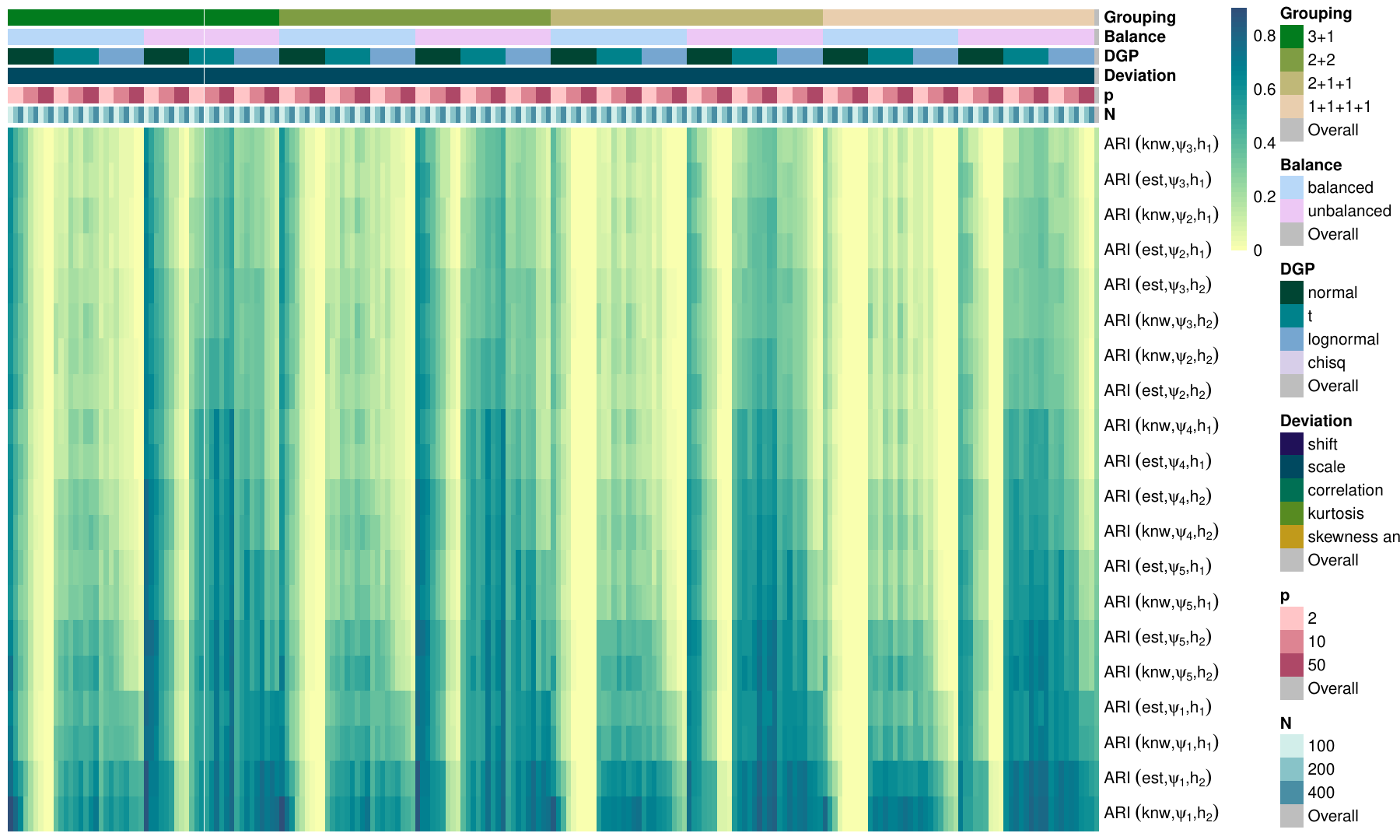}
		\caption{Mean PESR difference to best method per scenario for the scale alternative and $k = 4$ datasets for all variants of the aggregated RI test \textcite{paul_clustering-based_2022}. Selected variant: ARI~(knw,$\psi_3,h_1$).}
	\end{figure}
	
	\begin{figure}[!h]
		\centering
		\includegraphics[width=\linewidth]{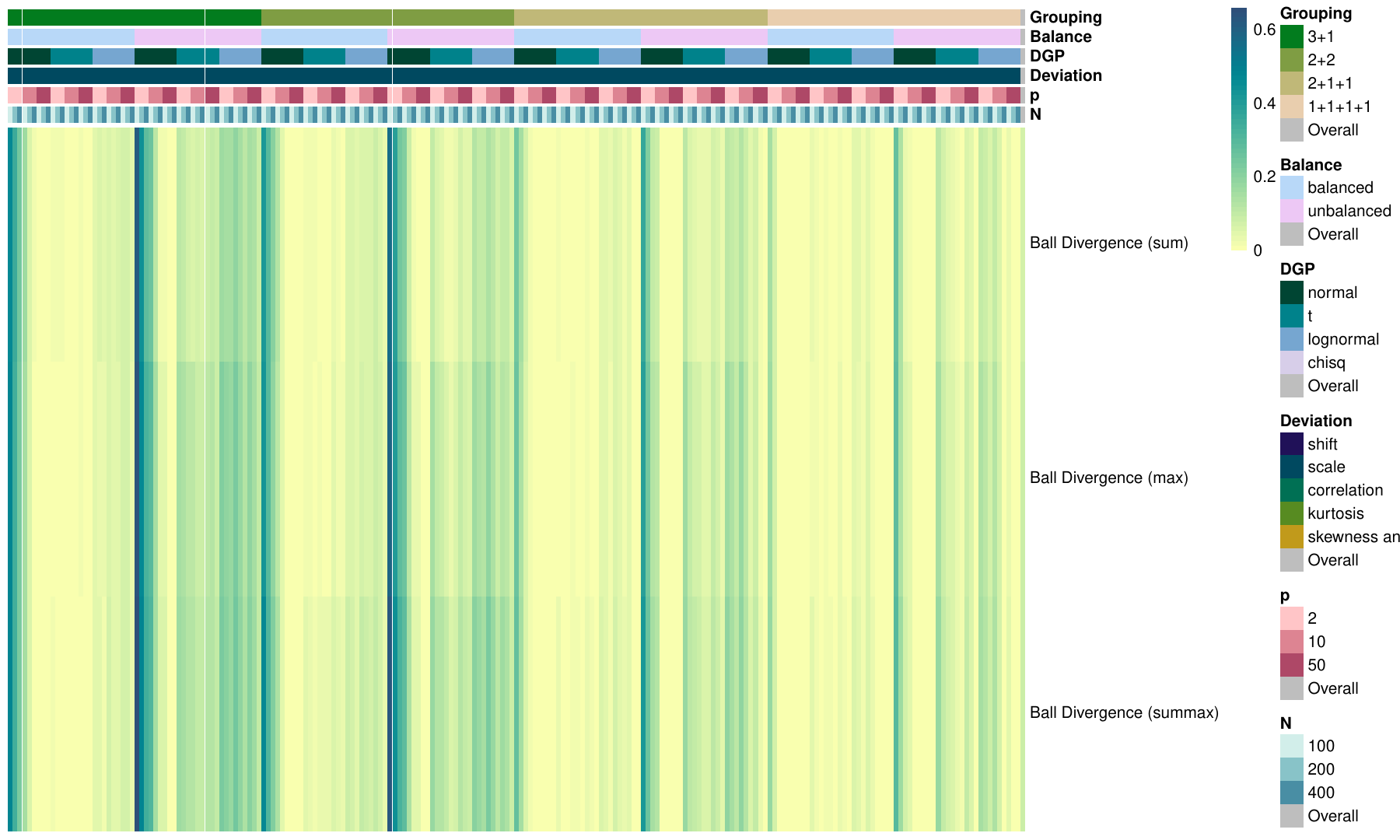}
		\caption{Mean PESR difference to best method per scenario for the scale alternative and $k = 4$ datasets for all variants of the Ball divergence \textcite{pan_ball_2018}. Selected variant: Ball divergence~(sum).}
	\end{figure}
	
	\begin{figure}[!h]
		\centering
		\includegraphics[width=\linewidth]{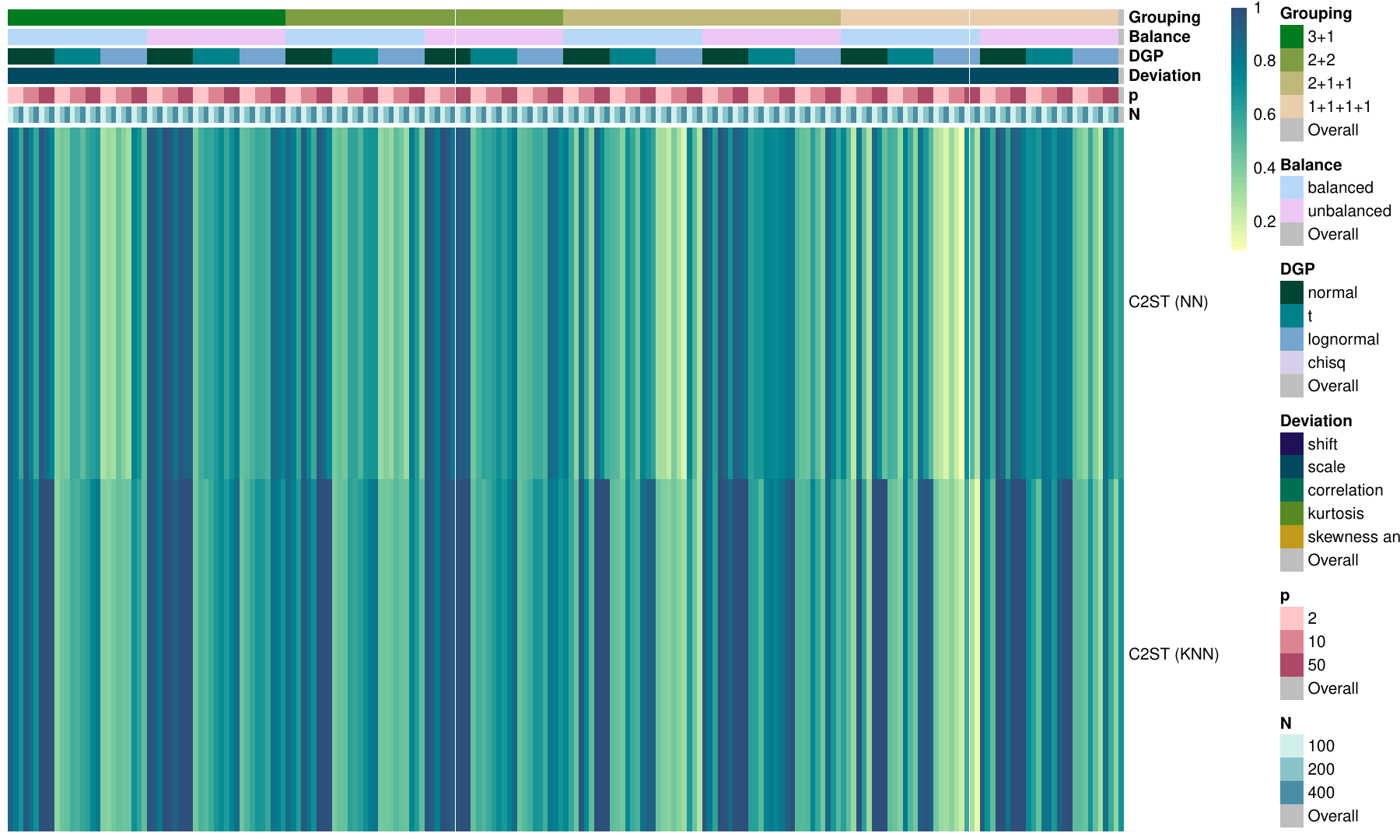}
		\caption{Mean PESR difference to best method per scenario for the scale alternative and $k = 4$ datasets for all variants of the C2ST \textcite{alvarez-melis_geometric_2020}. Selected variant: C2ST~(NN).}
	\end{figure}
	
	\begin{figure}[!h]
		\centering
		\includegraphics[width=\linewidth]{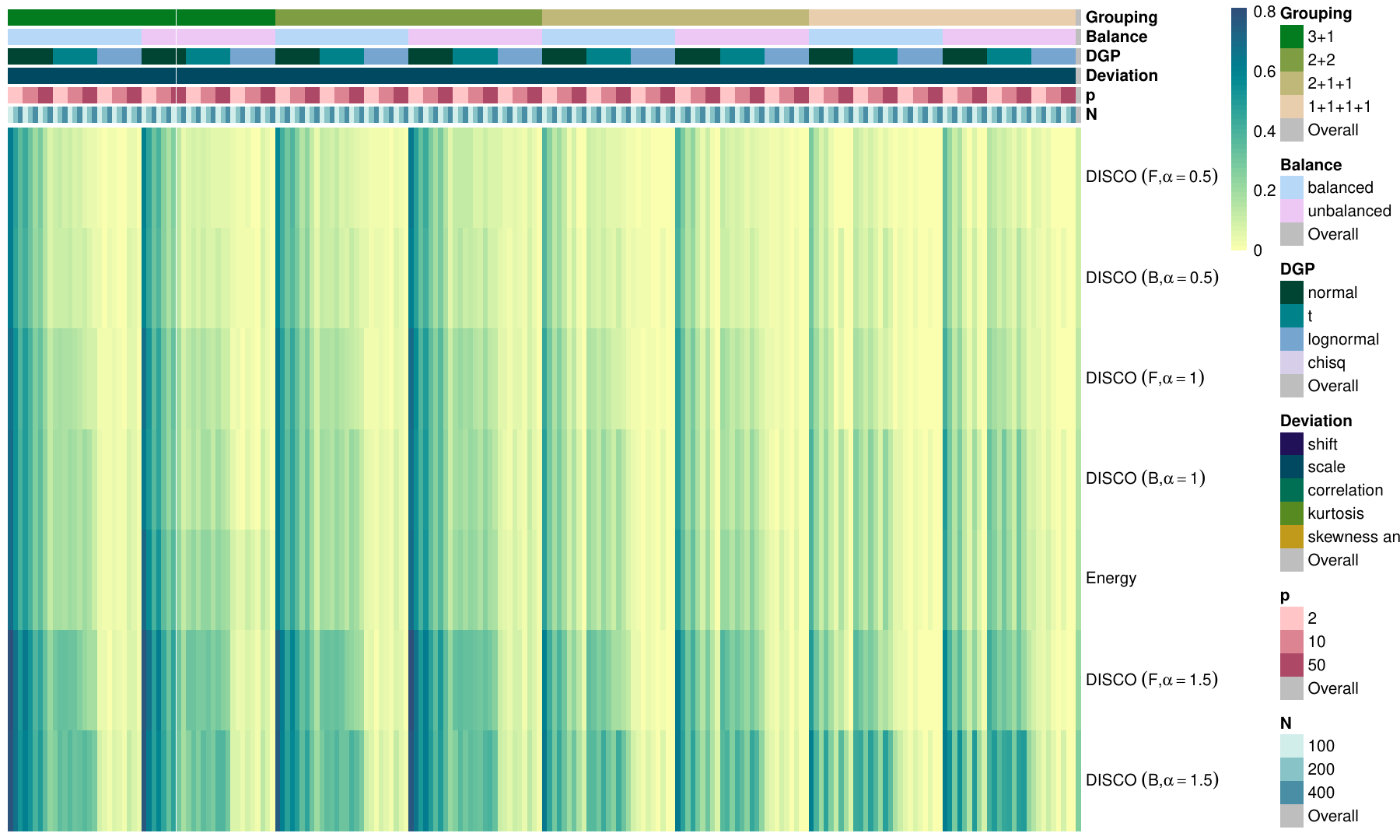}
		\caption{Mean PESR difference to best method per scenario for the scale alternative and $k = 4$ datasets for all variants of the Energy distance / DISCO \textcite{szekely_testing_2004, rizzo_disco_2010}. Selected variant: DISCO~(F,$\alpha=0.5$).}
	\end{figure}
	
	\begin{figure}[!h]
		\centering
		\includegraphics[width=\linewidth]{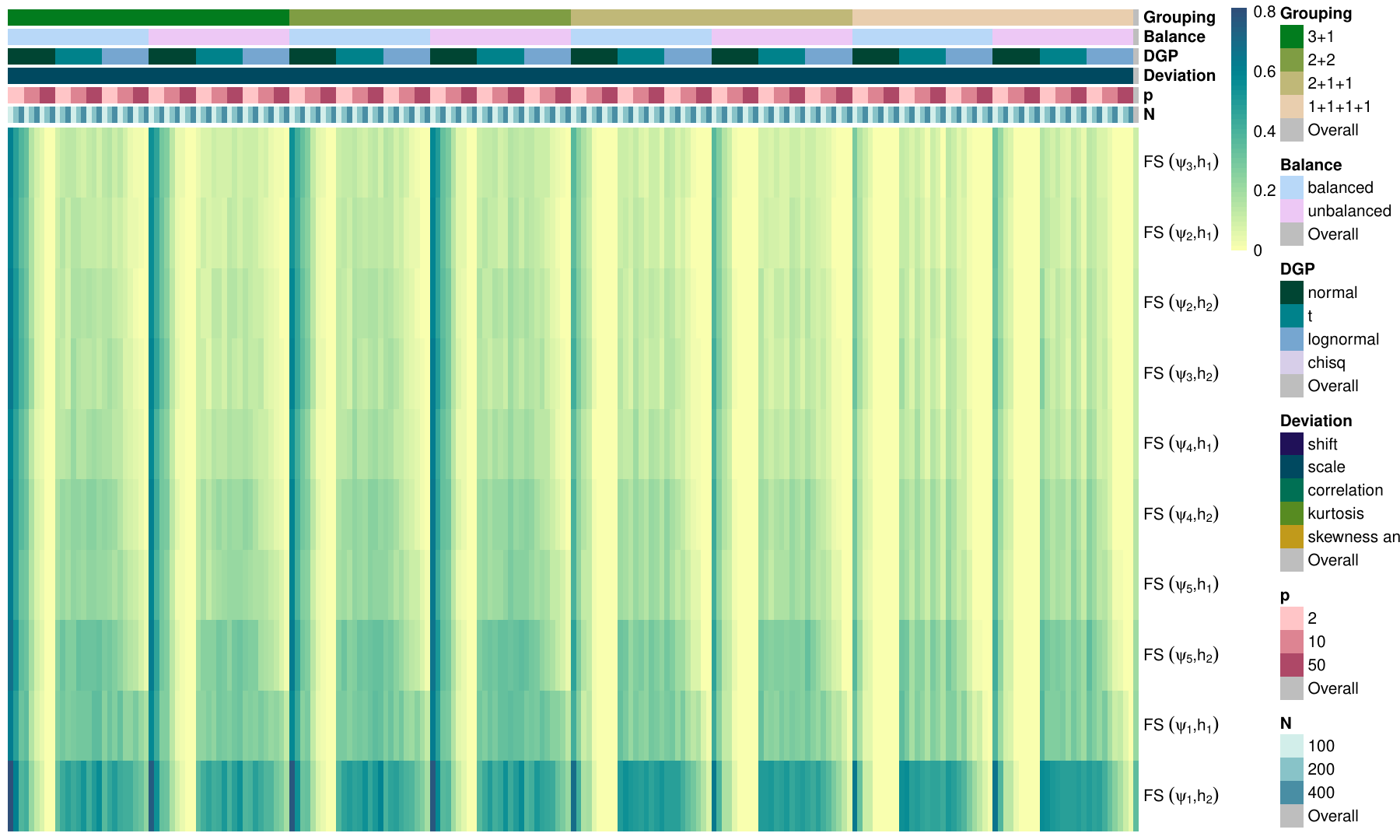}
		\caption{Mean PESR difference to best method per scenario for the scale alternative and $k = 4$ datasets for all variants of the FS test \textcite{paul_clustering-based_2022}. Selected variant: FS~($\psi_3$,$h_1$).}
	\end{figure}
	
	\begin{figure}[!h]
		\centering
		\includegraphics[width=\linewidth]{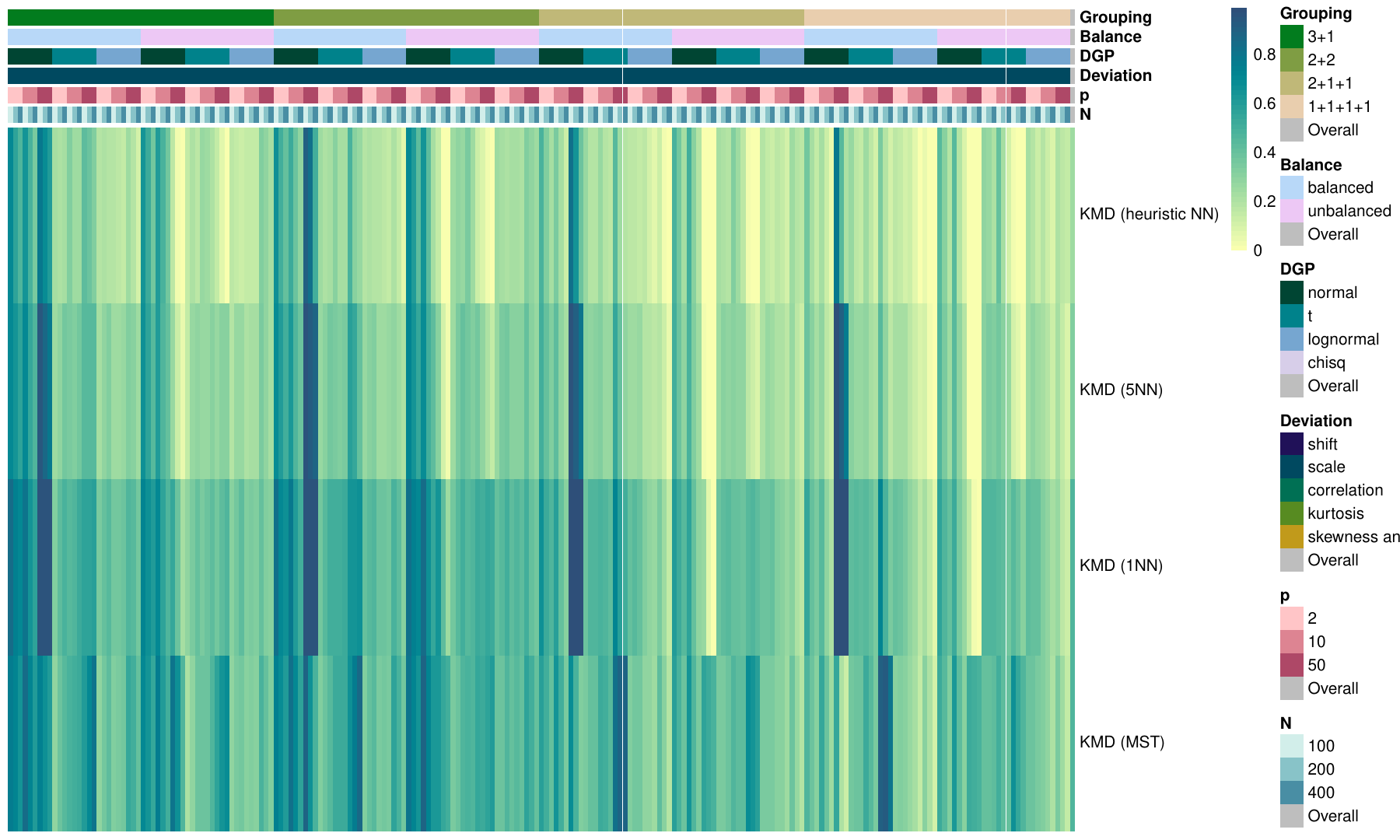}
		\caption{Mean PESR difference to best method per scenario for the scale alternative and $k = 4$ datasets for all variants of the modified KMD \textcite{huang_kernel_2022}. Selected variant: KMD~(heuristic NN).}
	\end{figure}
	
	\begin{figure}[!h]
		\centering
		\includegraphics[width=\linewidth]{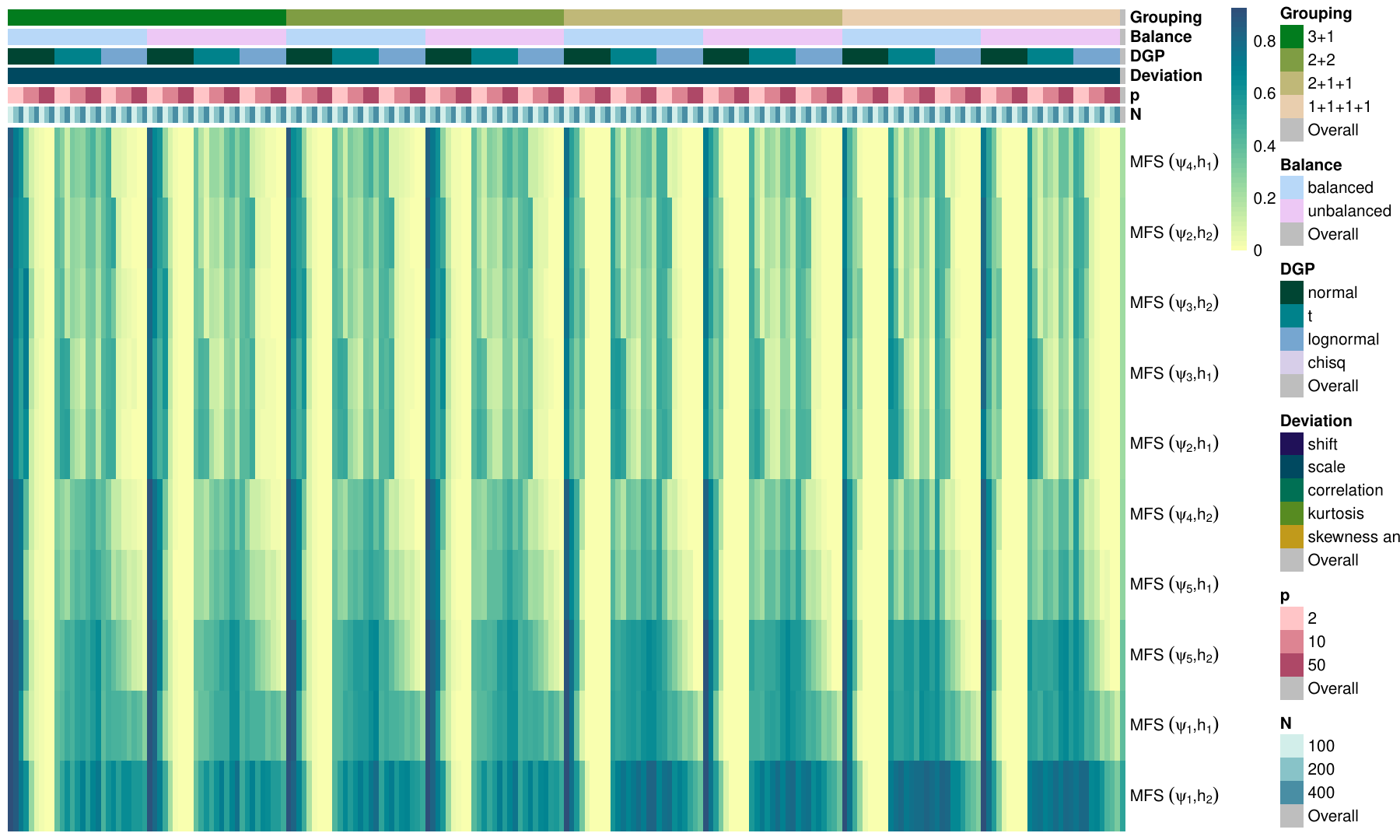}
		\caption{Mean PESR difference to best method per scenario for the scale alternative and $k = 4$ datasets for all variants of the modified FS test \textcite{paul_clustering-based_2022}. Selected variant: MFS~($\psi_4$,$h_1$).}
	\end{figure}
	
	\begin{figure}[!h]
		\centering
		\includegraphics[width=\linewidth]{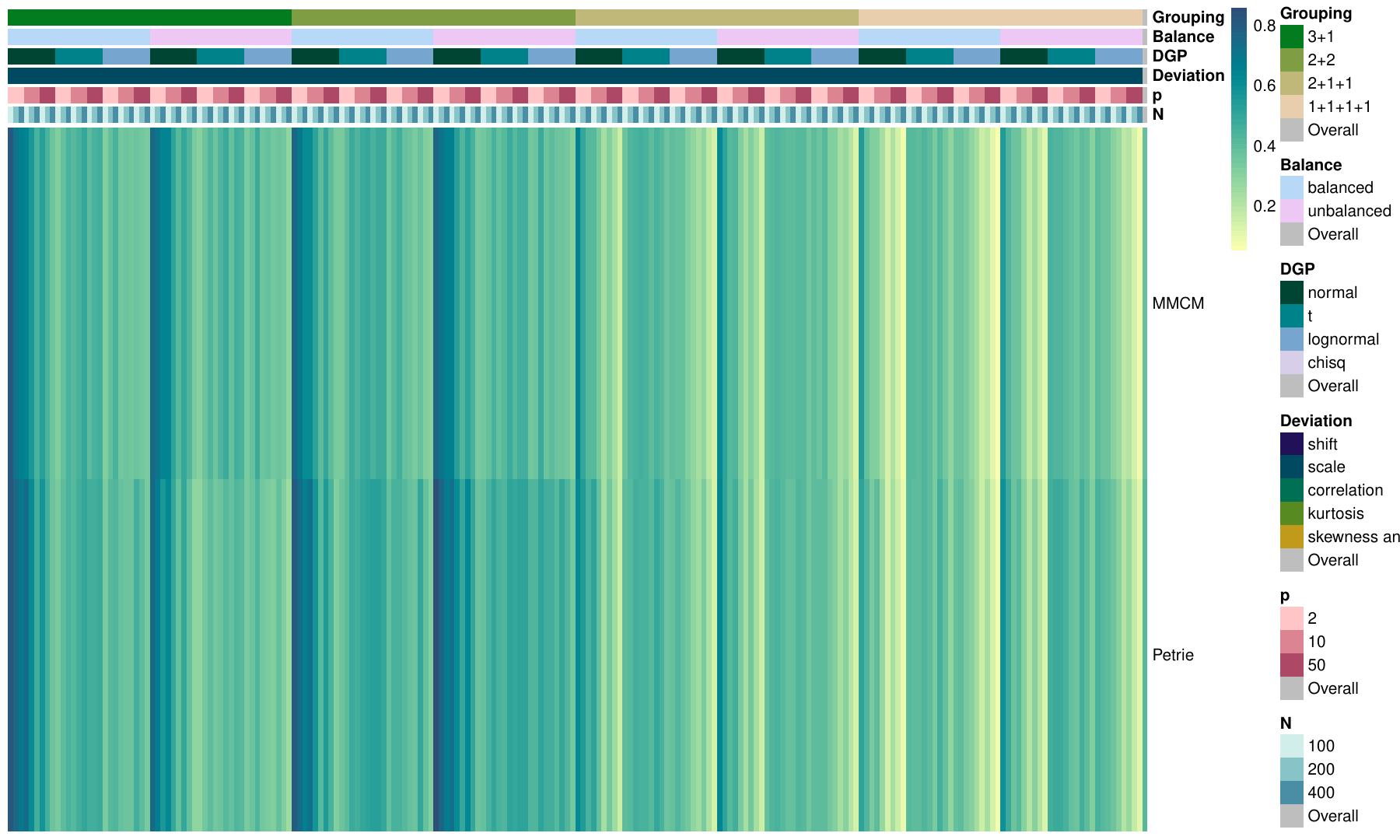}
		\caption{Mean PESR difference to best method per scenario for the scale alternative and $k = 4$ datasets for the MMCM and Petrie's test \textcite{mukherjee_distribution-free_2022, petrie_graph-theoretic_2016}. Selected variant: MMCM.}
	\end{figure}
	
	\begin{figure}[!h]
		\centering
		\includegraphics[width=\linewidth]{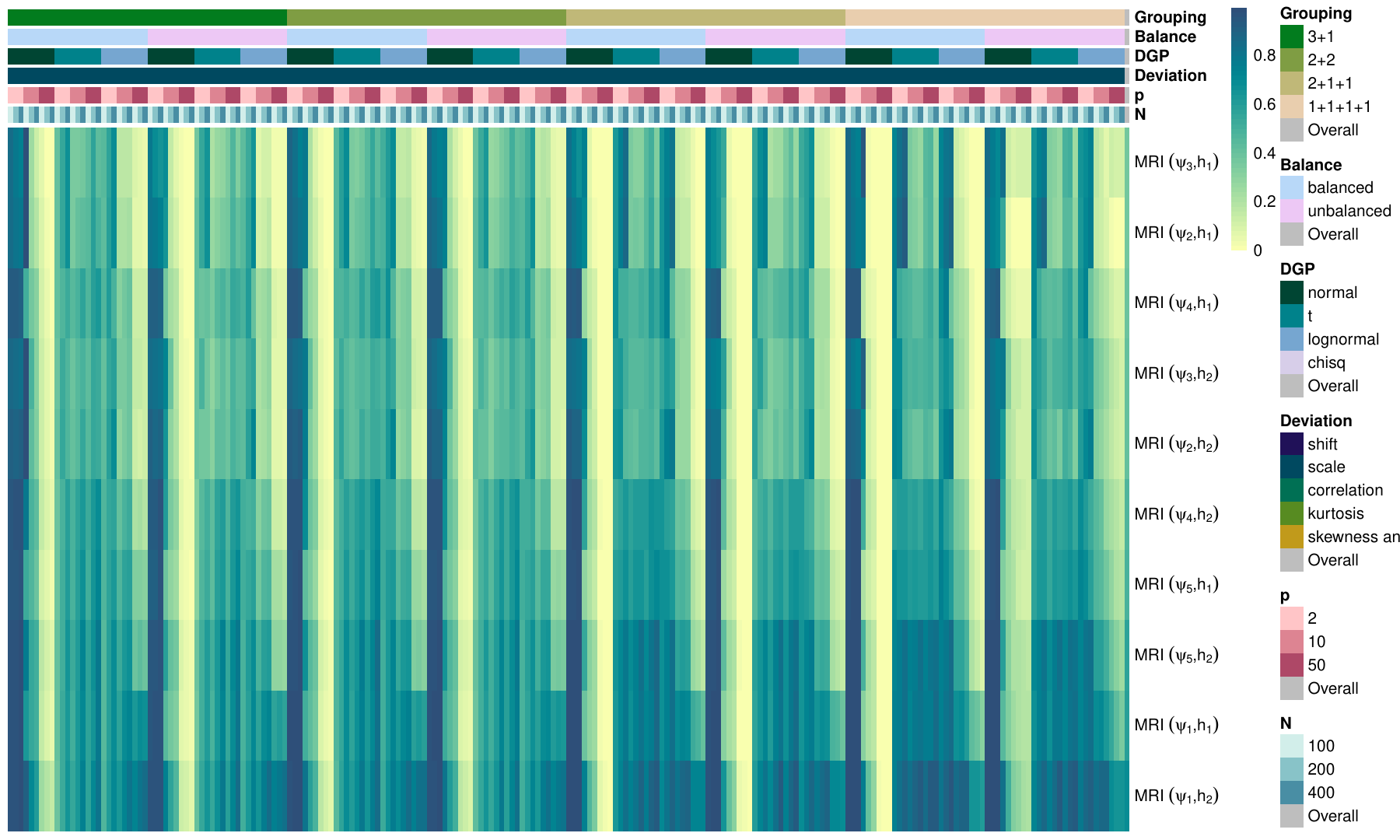}
		\caption{Mean PESR difference to best method per scenario for the scale alternative and $k = 4$ datasets for all variants of the modified RI test \textcite{paul_clustering-based_2022}. Selected variant: MRI~($\psi_3$,$h_1$).}
	\end{figure}
	
	\begin{figure}[!h]
		\centering
		\includegraphics[width=\linewidth]{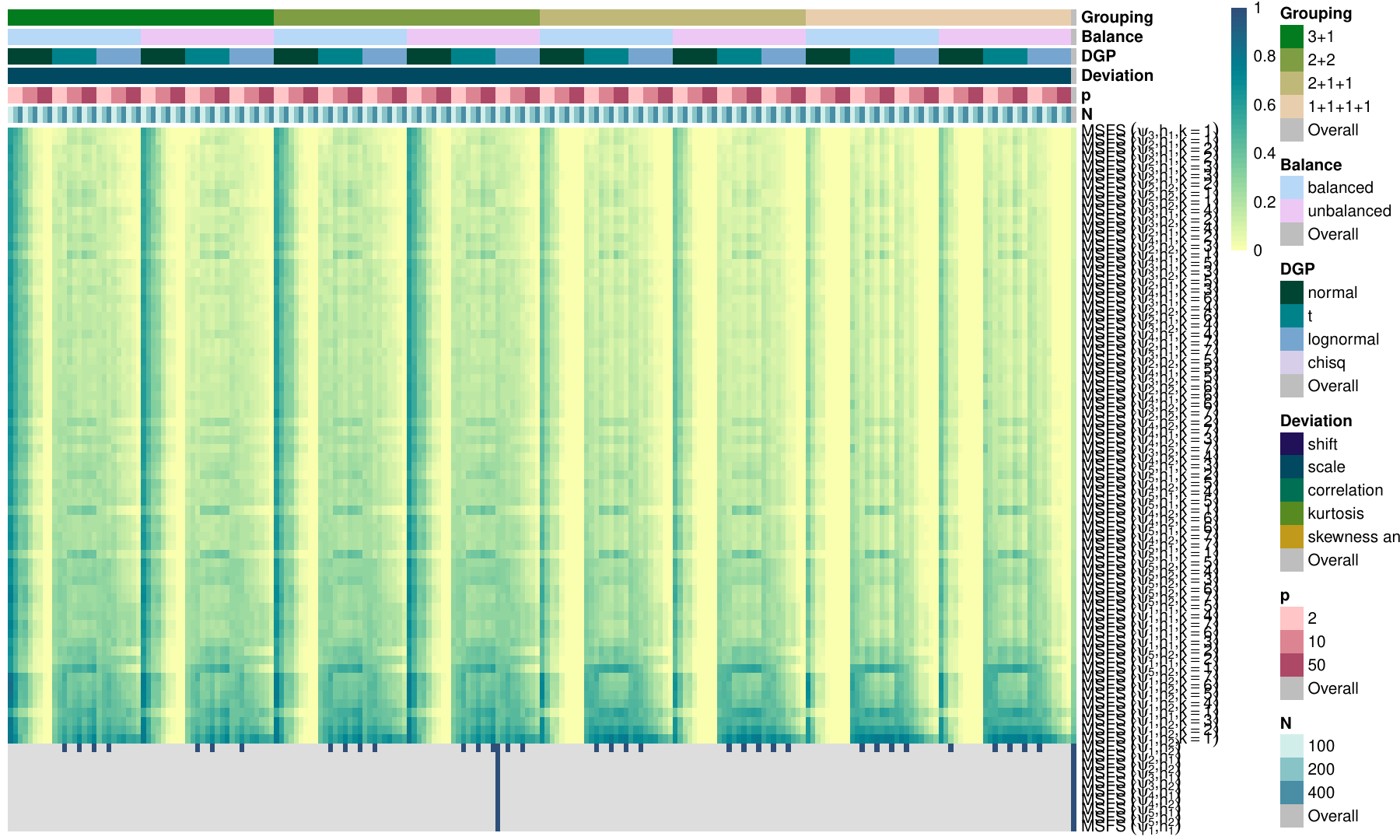}
		\caption{Mean PESR difference to best method per scenario for the scale alternative and $k = 4$ datasets for all variants of the multiscale FS test \textcite{paul_clustering-based_2022}. Selected variant: MSFS~($\psi_3$,$h_1$,$k=1$).}
	\end{figure}
	
	\begin{figure}[!h]
		\centering
		\includegraphics[width=\linewidth]{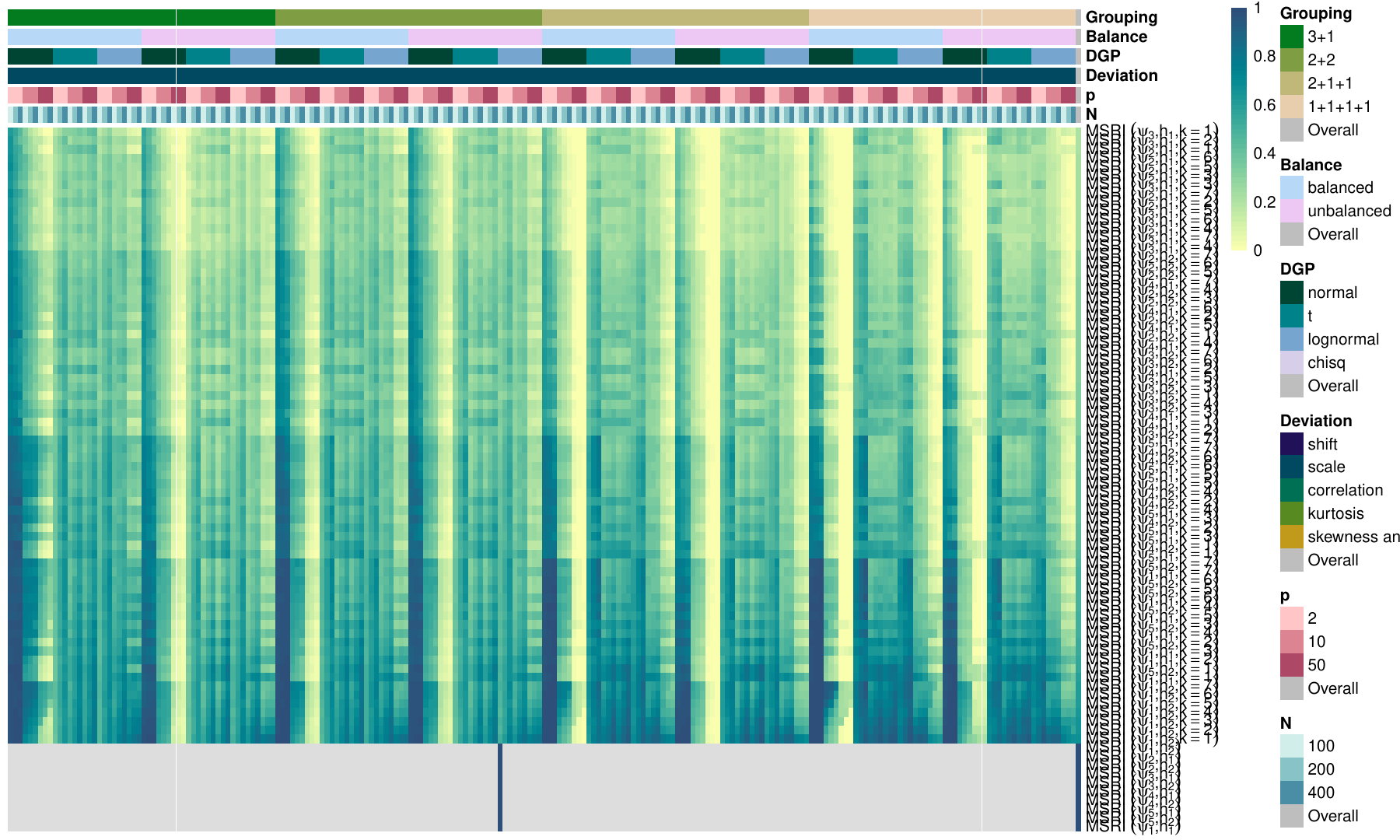}
		\caption{Mean PESR difference to best method per scenario for the scale alternative and $k = 4$ datasets for all variants of the multiscale RI test \textcite{paul_clustering-based_2022}. Selected variant: MSRI~($\psi_3$,$h_1$,$k=1$).}
	\end{figure}
	
	\begin{figure}[!h]
		\centering
		\includegraphics[width=\linewidth]{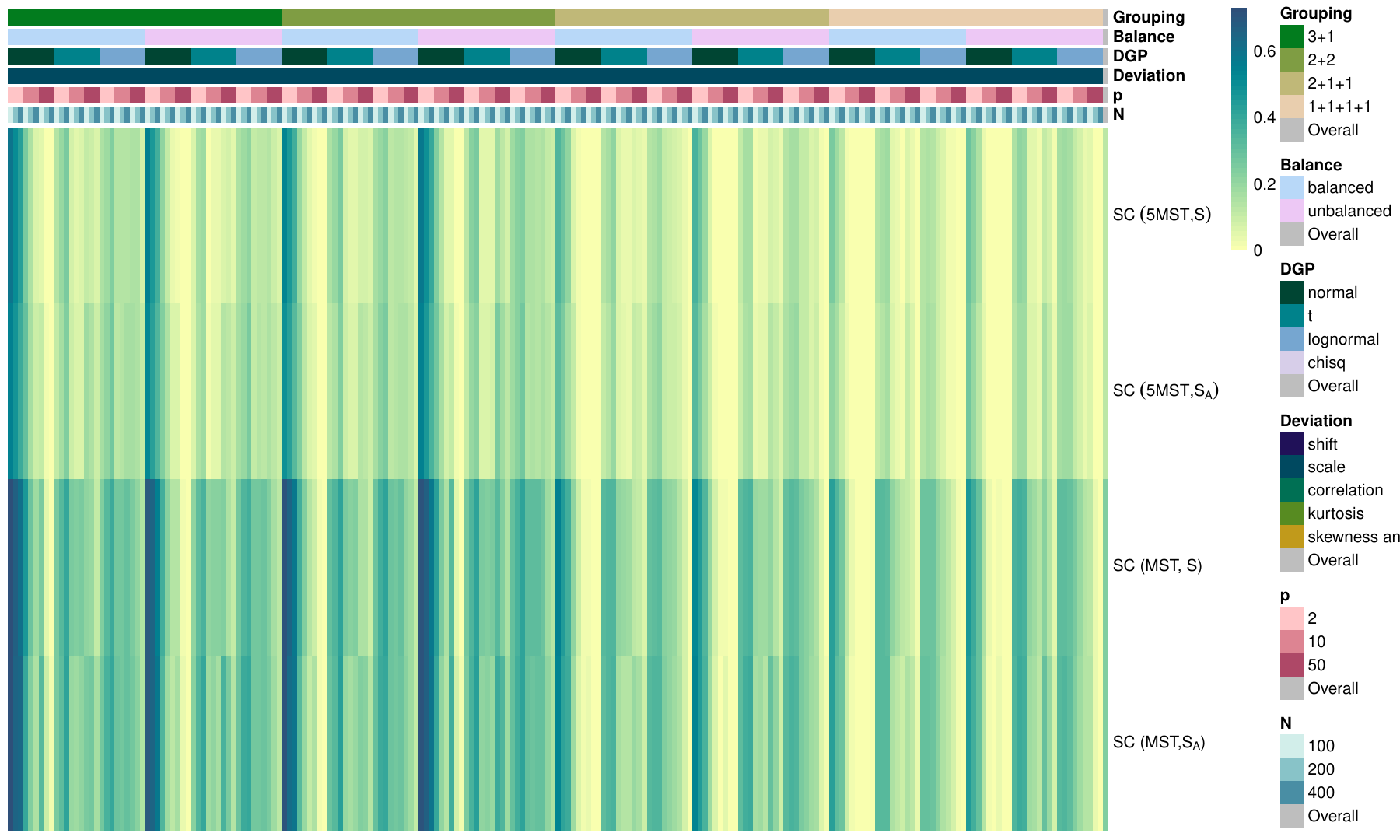}
		\caption{Mean PESR difference to best method per scenario for the scale alternative and $k = 4$ datasets for all variants of the SC test \textcite{song_generalized_2021}. Selected variant: SC~(5MST,$S$).}
	\end{figure}
	\clearpage
	
	\paragraph{Correlation}	\mbox{} 
	
	\begin{figure}[!h]
		\centering
		\includegraphics[width=\linewidth]{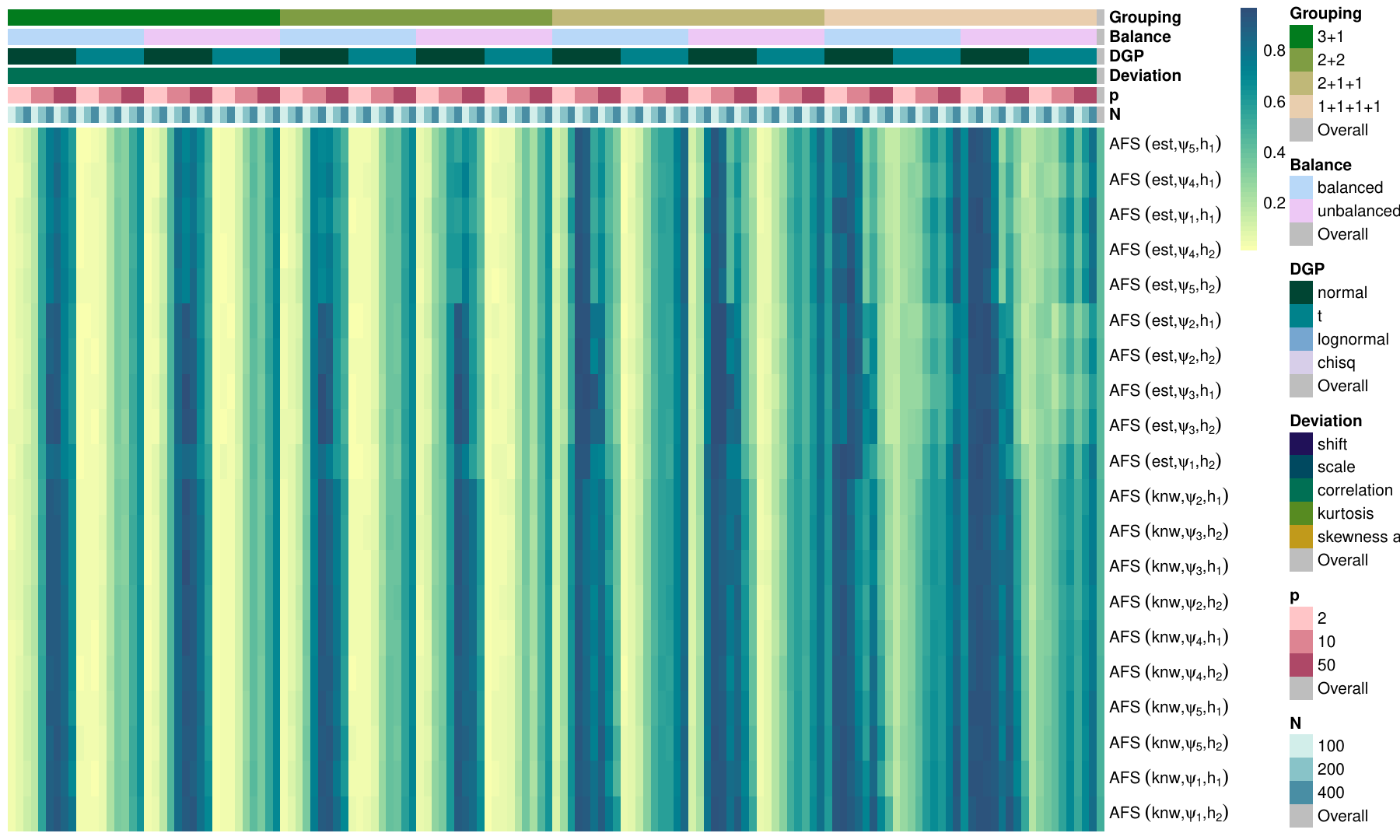}
		\caption{Mean PESR difference to best method per scenario for the correlation alternative and $k = 4$ datasets for all variants of the aggregated FS test \textcite{paul_clustering-based_2022}. Selected variant: AFS~(est,$\psi_5,h_1$).}
	\end{figure}
	
	\begin{figure}[!h]
		\centering
		\includegraphics[width=\linewidth]{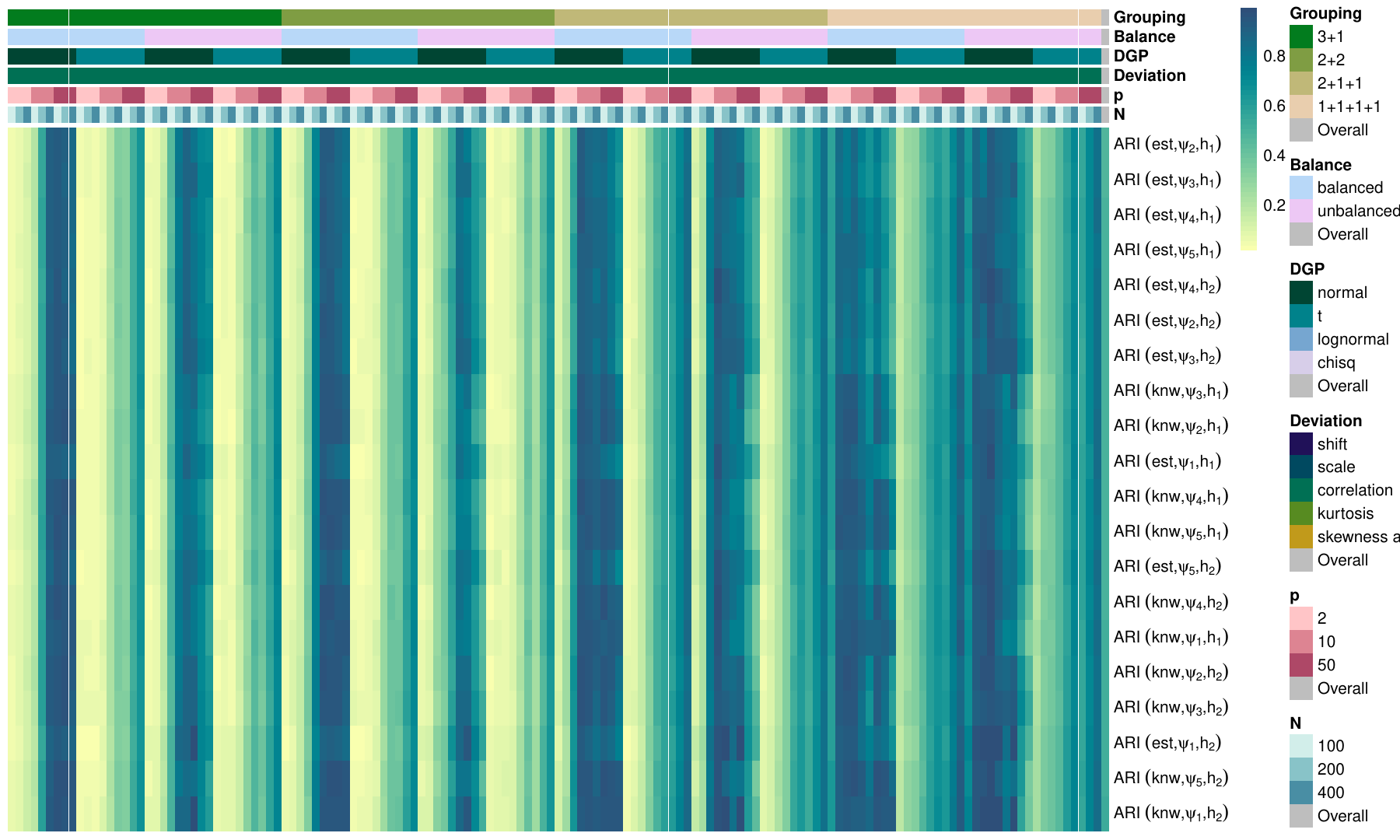}
		\caption{Mean PESR difference to best method per scenario for the correlation alternative and $k = 4$ datasets for all variants of the aggregated RI test \textcite{paul_clustering-based_2022}. Selected variant: ARI~(est,$\psi_2,h_1$).}
	\end{figure}
	
	\begin{figure}[!h]
		\centering
		\includegraphics[width=\linewidth]{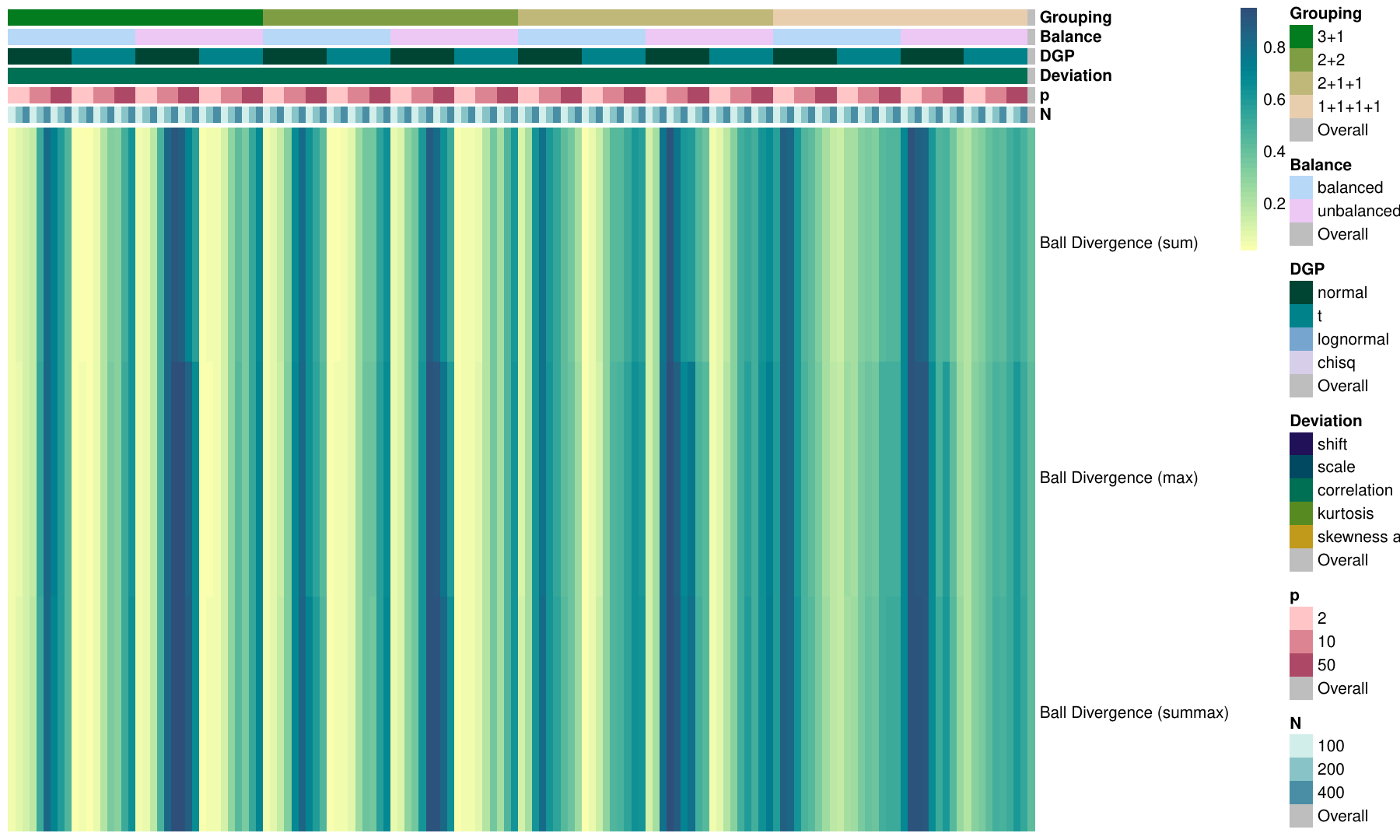}
		\caption{Mean PESR difference to best method per scenario for the correlation alternative and $k = 4$ datasets for all variants of the Ball divergence \textcite{pan_ball_2018}. Selected variant: Ball divergence~(sum).}
	\end{figure}
	
	\begin{figure}[!h]
		\centering
		\includegraphics[width=\linewidth]{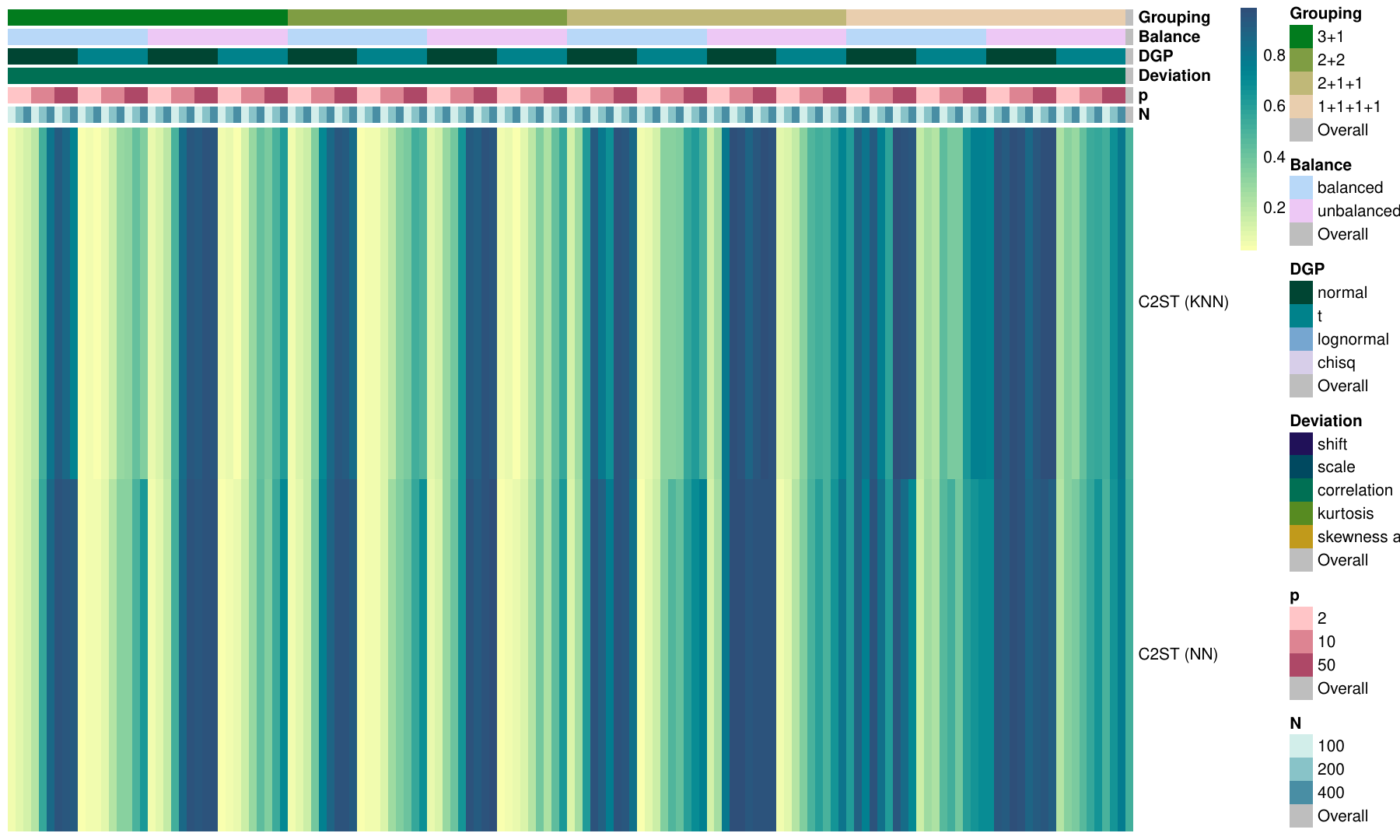}
		\caption{Mean PESR difference to best method per scenario for the correlation alternative and $k = 4$ datasets for all variants of the C2ST \textcite{alvarez-melis_geometric_2020}. Selected variant: C2ST~(KNN).}
	\end{figure}
	
	\begin{figure}[!h]
		\centering
		\includegraphics[width=\linewidth]{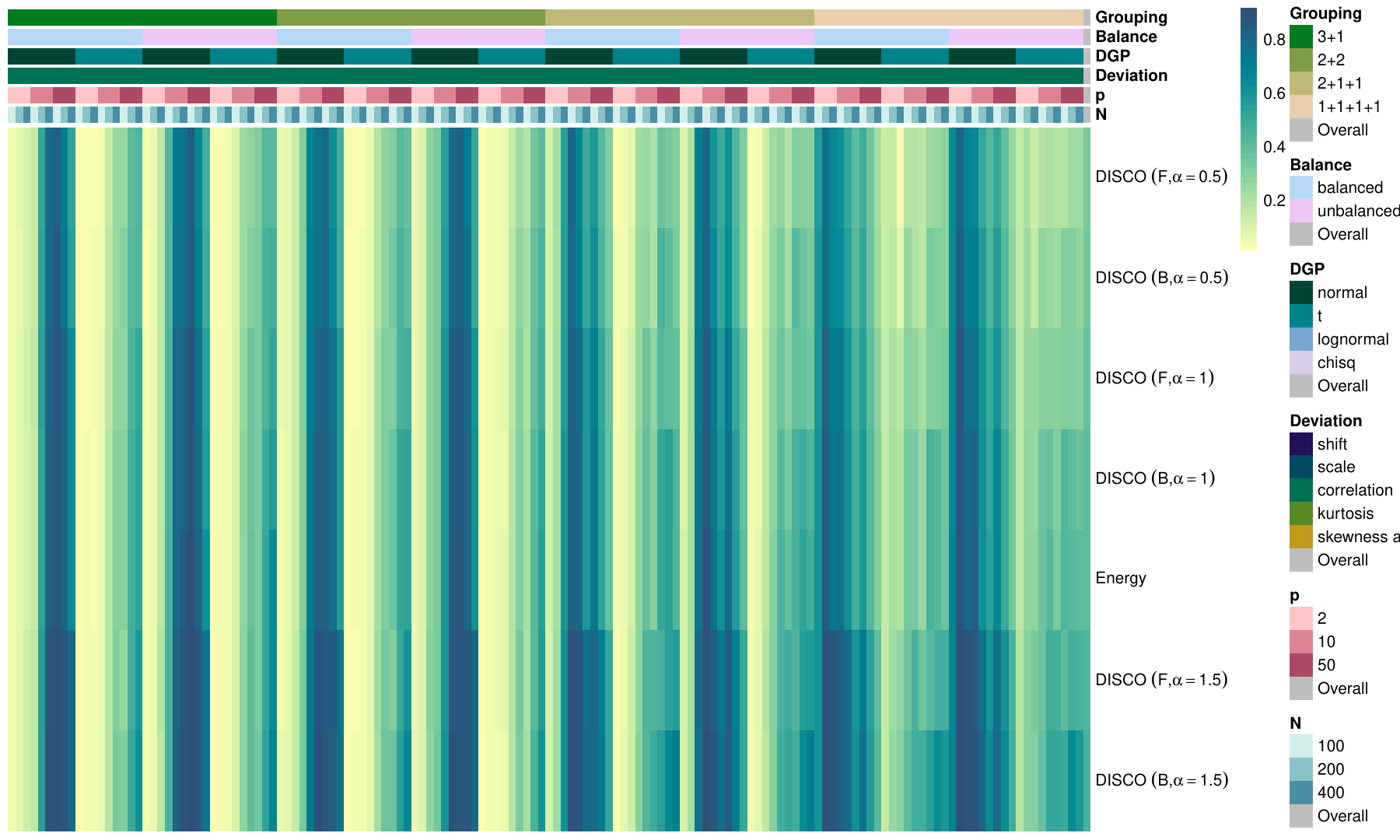}
		\caption{Mean PESR difference to best method per scenario for the correlation alternative and $k = 4$ datasets for all variants of the Energy distance / DISCO \textcite{szekely_testing_2004, rizzo_disco_2010}. Selected variant: DISCO~(F,$\alpha=0.5$).}
	\end{figure}
	
	\begin{figure}[!h]
		\centering
		\includegraphics[width=\linewidth]{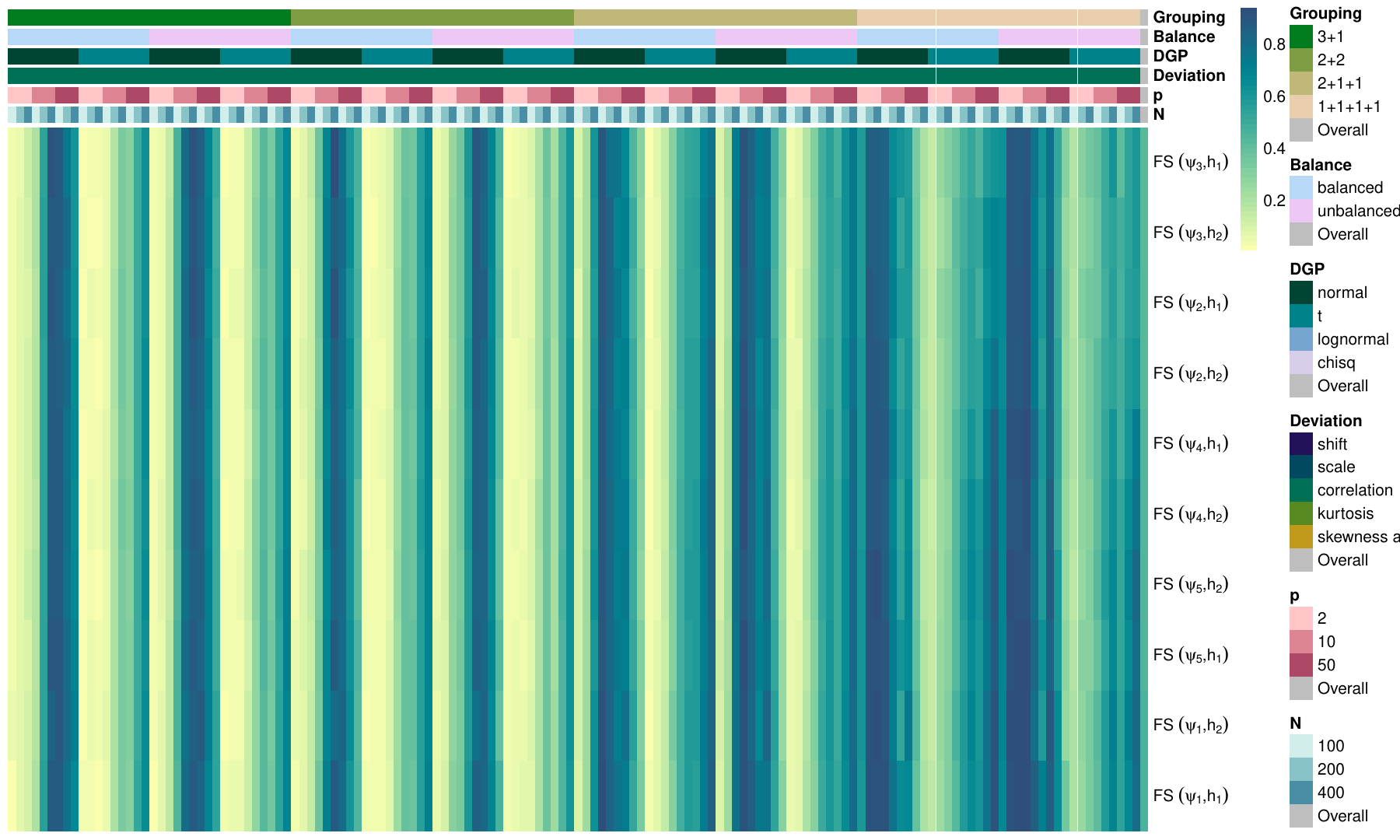}
		\caption{Mean PESR difference to best method per scenario for the correlation alternative and $k = 4$ datasets for all variants of the FS test \textcite{paul_clustering-based_2022}. Selected variant: FS~($\psi_3$,$h_1$).}
	\end{figure}
	
	\begin{figure}[!h]
		\centering
		\includegraphics[width=\linewidth]{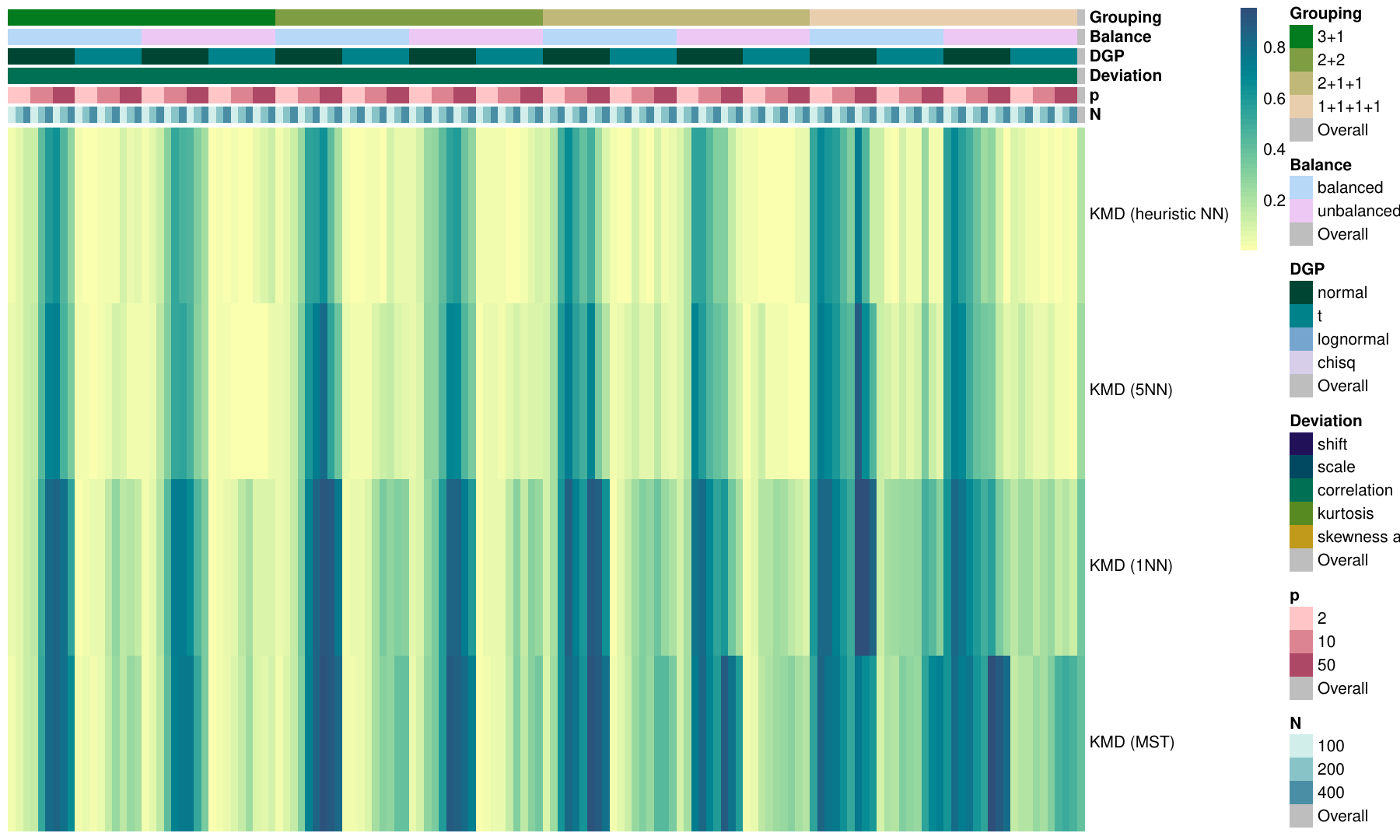}
		\caption{Mean PESR difference to best method per scenario for the correlation alternative and $k = 4$ datasets for all variants of the modified KMD \textcite{huang_kernel_2022}. Selected variant: KMD~(heuristic NN).}
	\end{figure}
	
	\begin{figure}[!h]
		\centering
		\includegraphics[width=\linewidth]{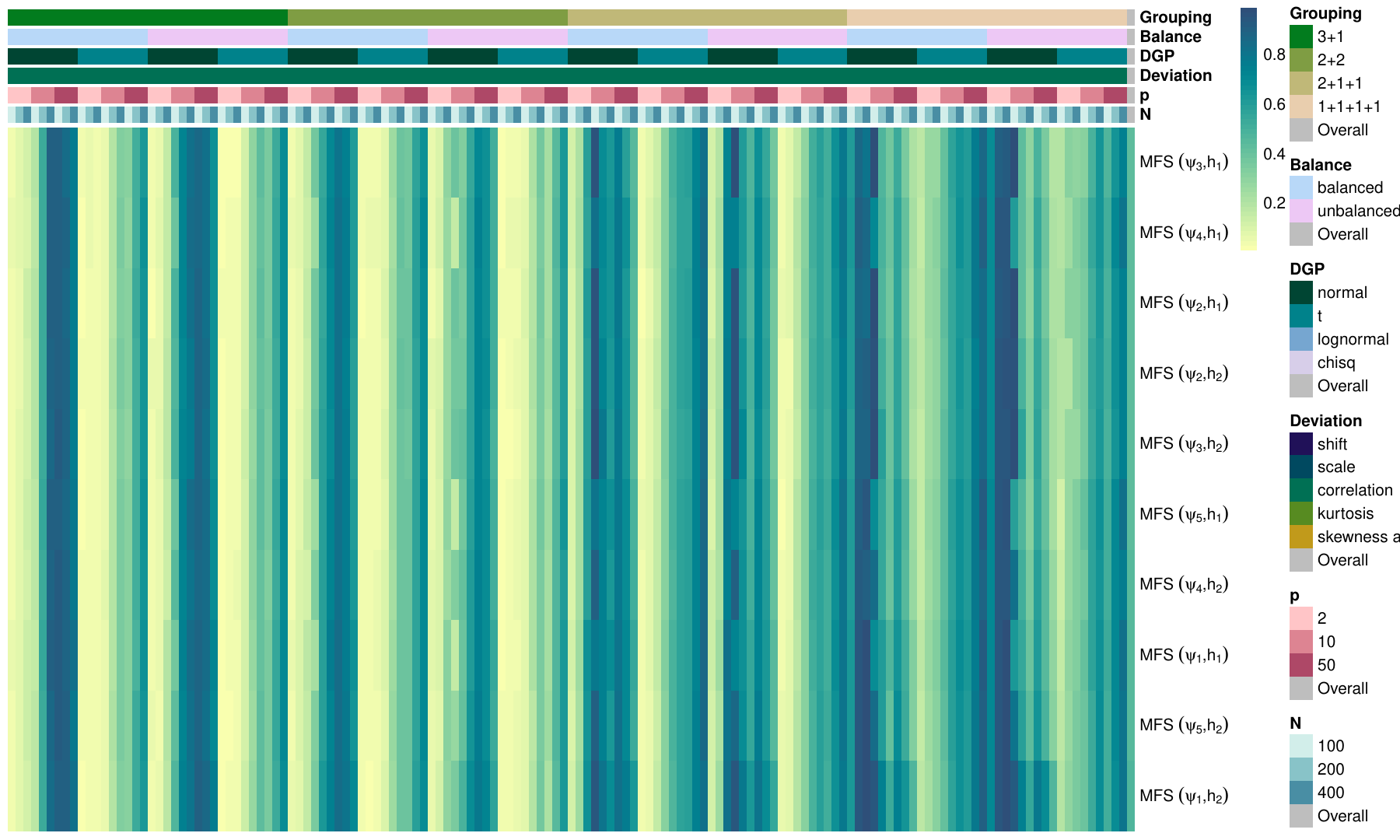}
		\caption{Mean PESR difference to best method per scenario for the correlation alternative and $k = 4$ datasets for all variants of the modified FS test \textcite{paul_clustering-based_2022}. Selected variant: MFS~($\psi_3$,$h_1$).}
	\end{figure}
	
	\begin{figure}[!h]
		\centering
		\includegraphics[width=\linewidth]{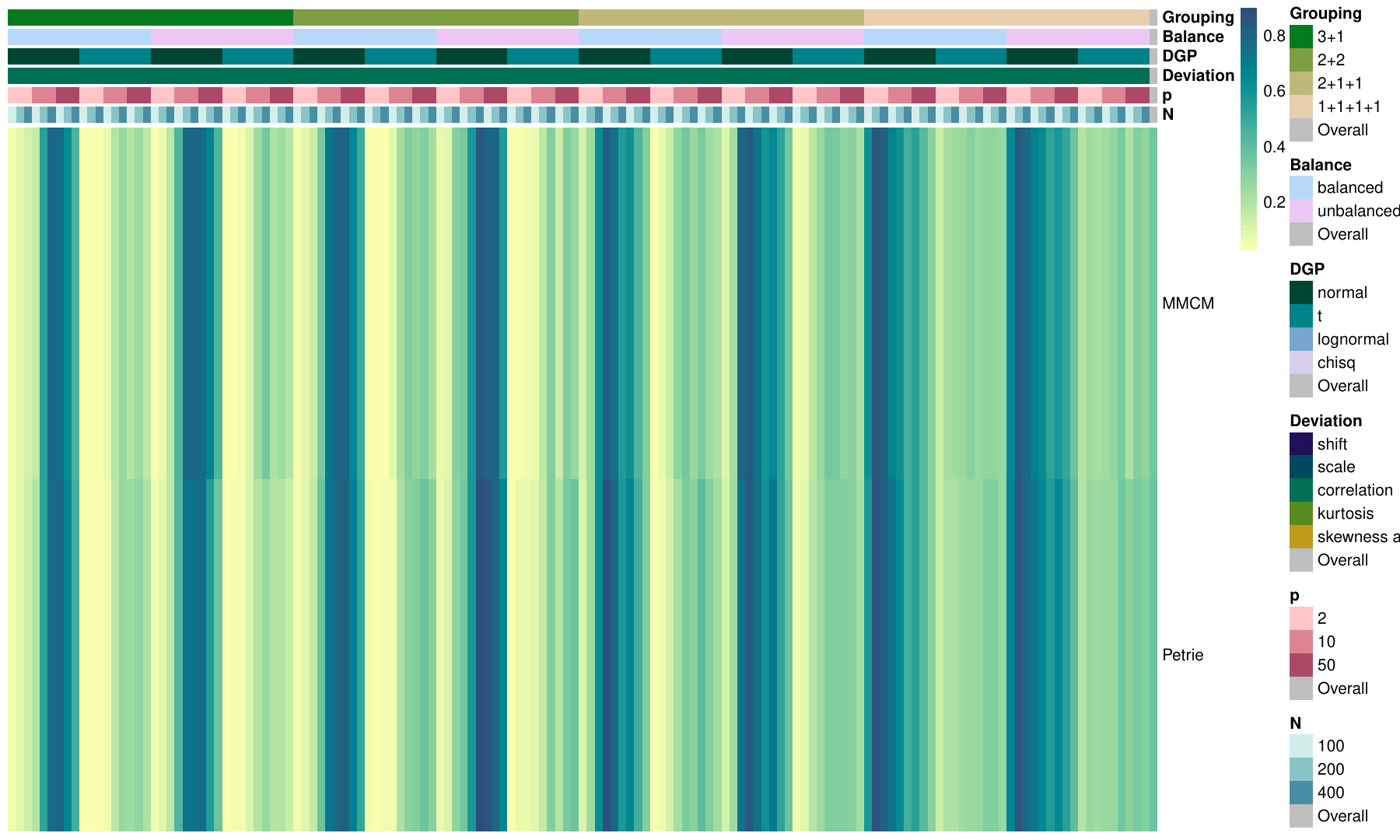}
		\caption{Mean PESR difference to best method per scenario for the correlation alternative and $k = 4$ datasets for the MMCM and Petrie's test \textcite{mukherjee_distribution-free_2022, petrie_graph-theoretic_2016}. Selected variant: MMCM.}
	\end{figure}
	
	\begin{figure}[!h]
		\centering
		\includegraphics[width=\linewidth]{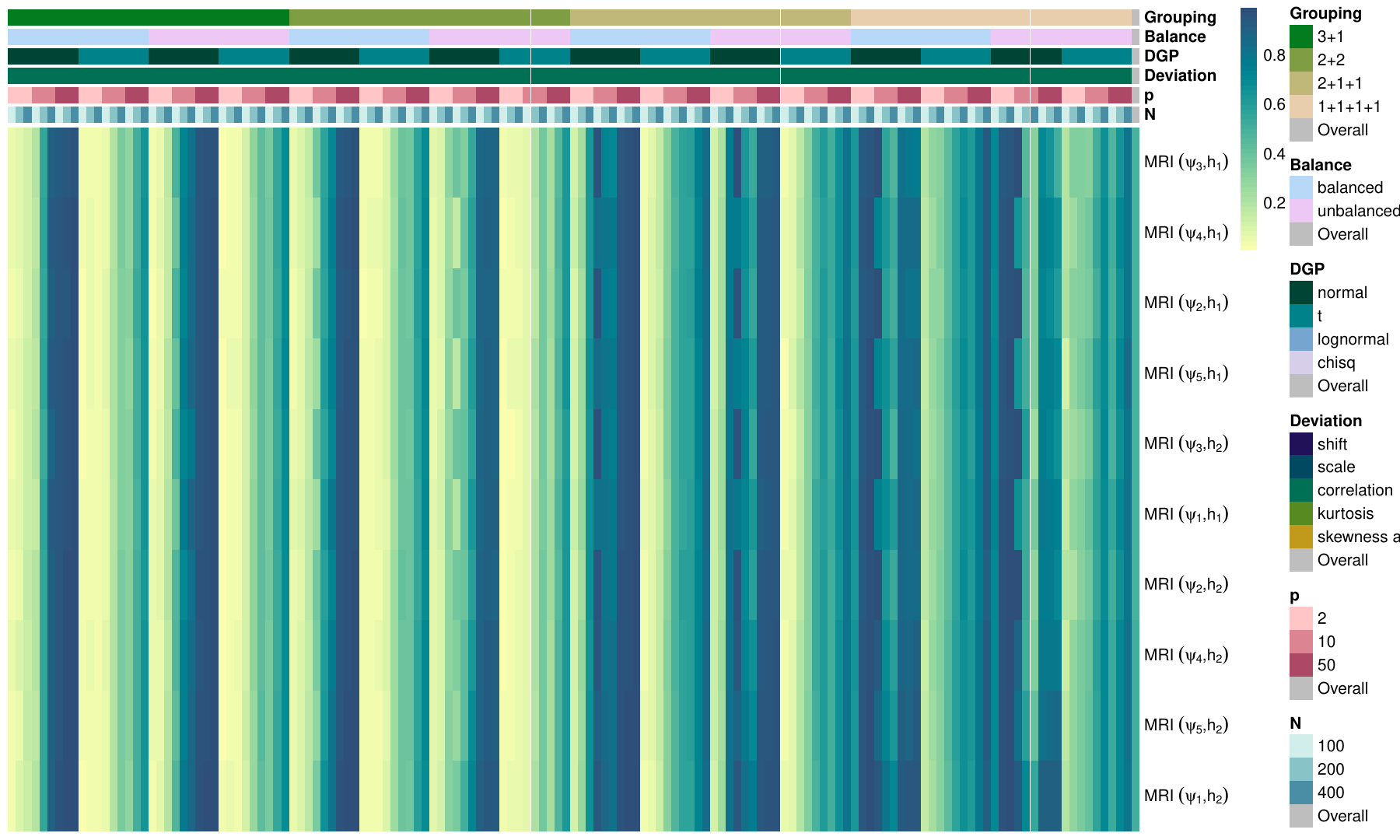}
		\caption{Mean PESR difference to best method per scenario for the correlation alternative and $k = 4$ datasets for all variants of the modified RI test \textcite{paul_clustering-based_2022}. Selected variant: MRI~($\psi_3$,$h_1$).}
	\end{figure}
	
	\begin{figure}[!h]
		\centering
		\includegraphics[width=\linewidth]{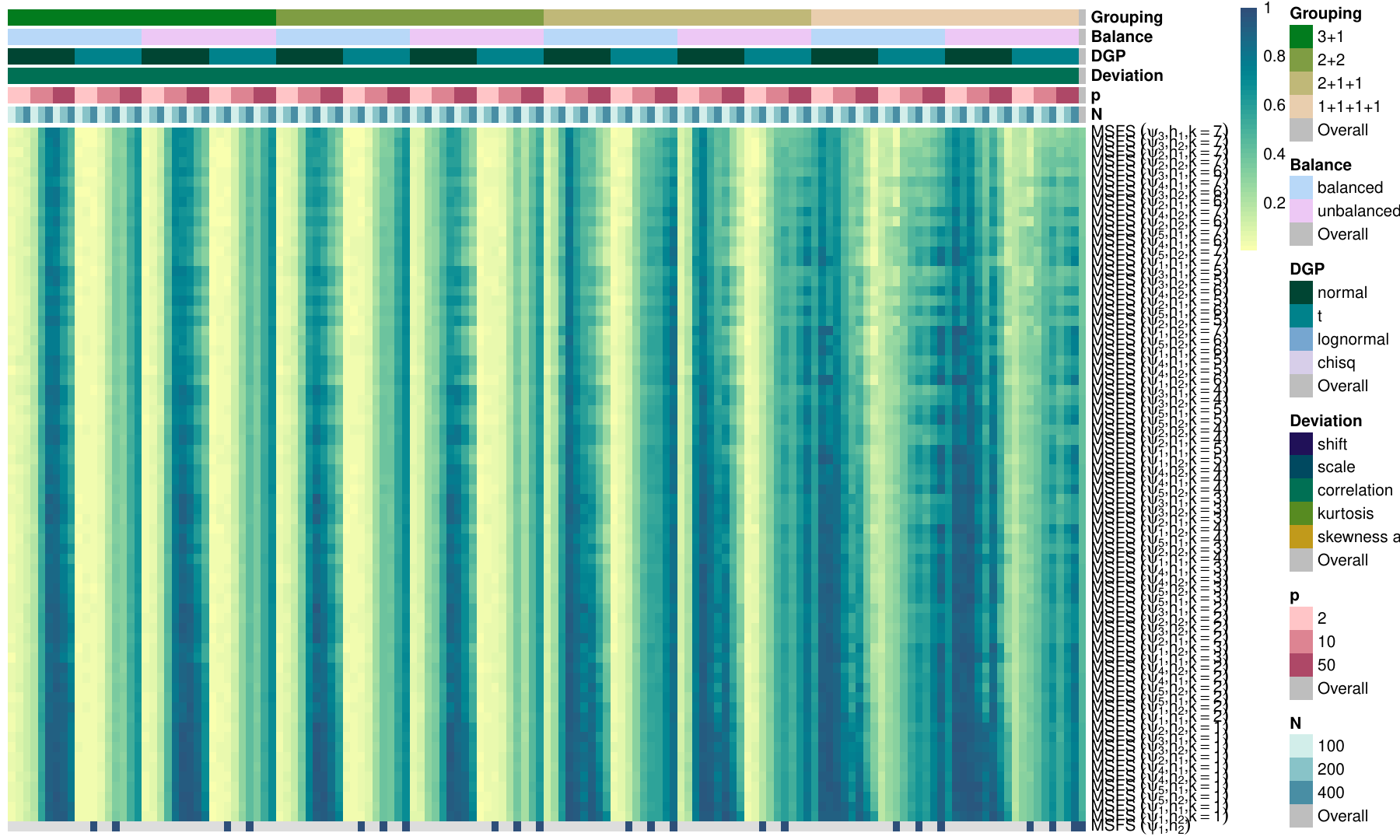}
		\caption{Mean PESR difference to best method per scenario for the correlation alternative and $k = 4$ datasets for all variants of the multiscale FS test \textcite{paul_clustering-based_2022}. Selected variant: MSFS~($\psi_3$,$h_1$,$k=7$).}
	\end{figure}
	
	\begin{figure}[!h]
		\centering
		\includegraphics[width=\linewidth]{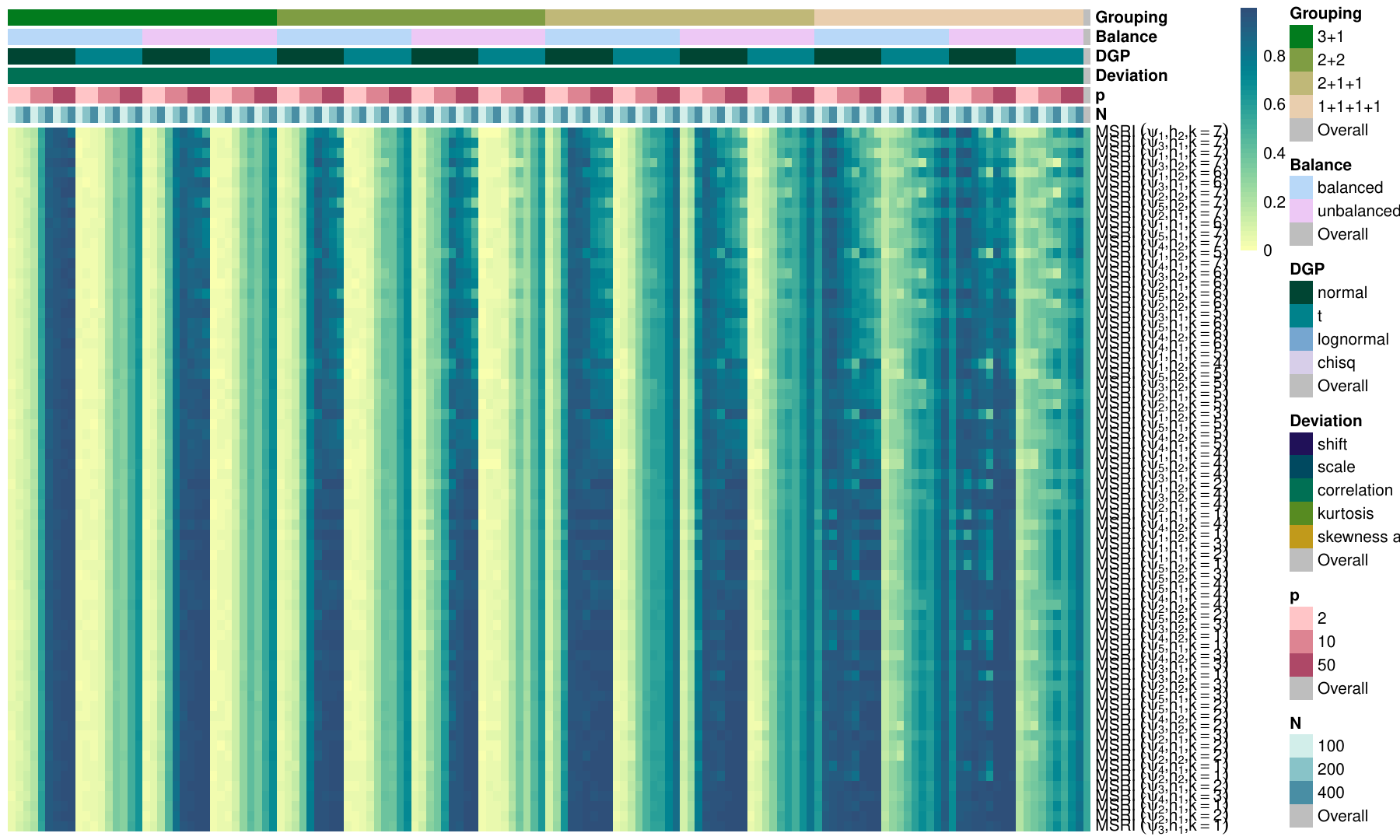}
		\caption{Mean PESR difference to best method per scenario for the correlation alternative and $k = 4$ datasets for all variants of the multiscale RI test \textcite{paul_clustering-based_2022}. Selected variant: MSRI~($\psi_1$,$h_2$,$k=7$).}
	\end{figure}
	
	\begin{figure}[!h]
		\centering
		\includegraphics[width=\linewidth]{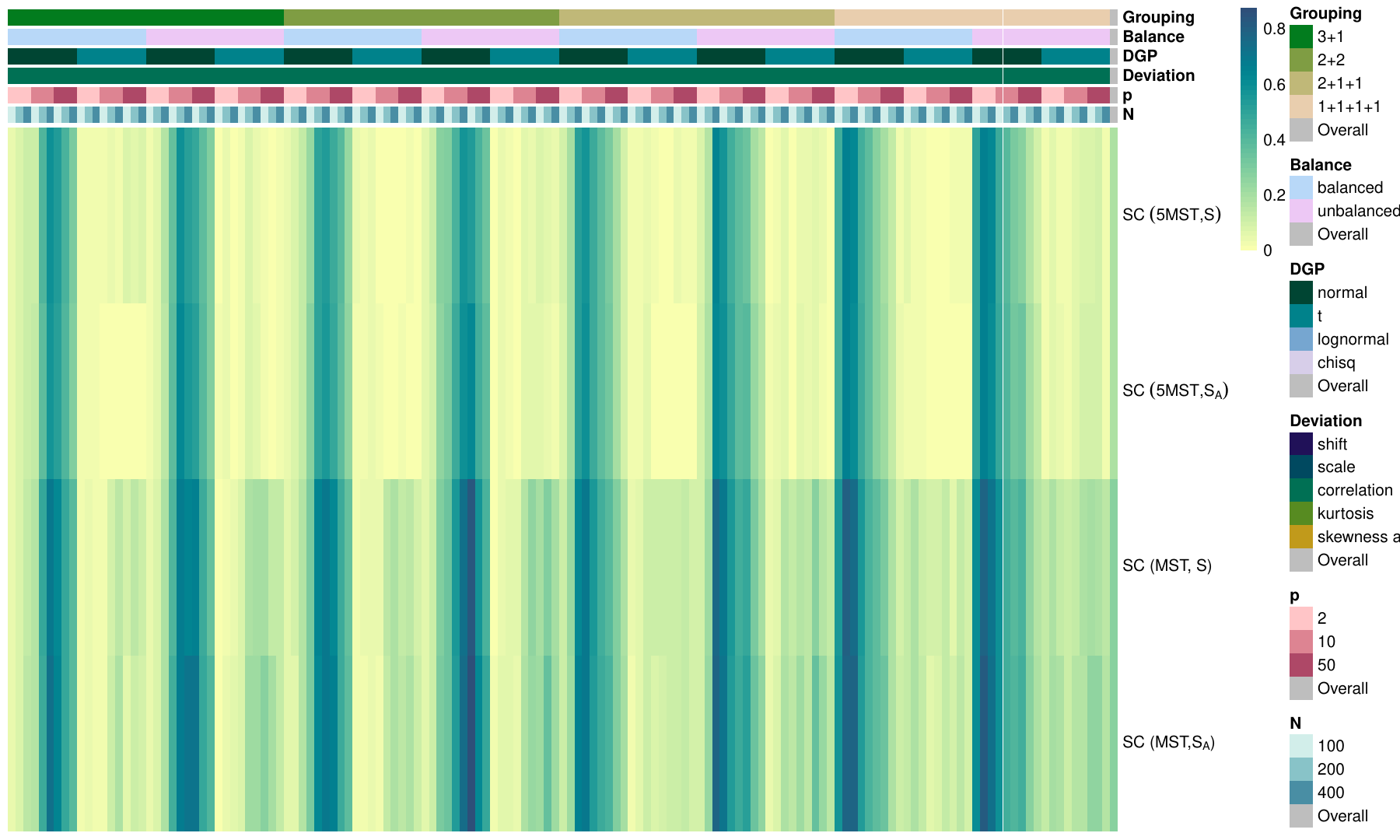}
		\caption{Mean PESR difference to best method per scenario for the correlation alternative and $k = 4$ datasets for all variants of the SC test \textcite{song_generalized_2021}. Selected variant: SC~(5MST,$S$).}
	\end{figure}
	\clearpage

	\paragraph{Kurtosis}\mbox{} 	
	
	\begin{figure}[!h]
		\centering
		\includegraphics[width=\linewidth]{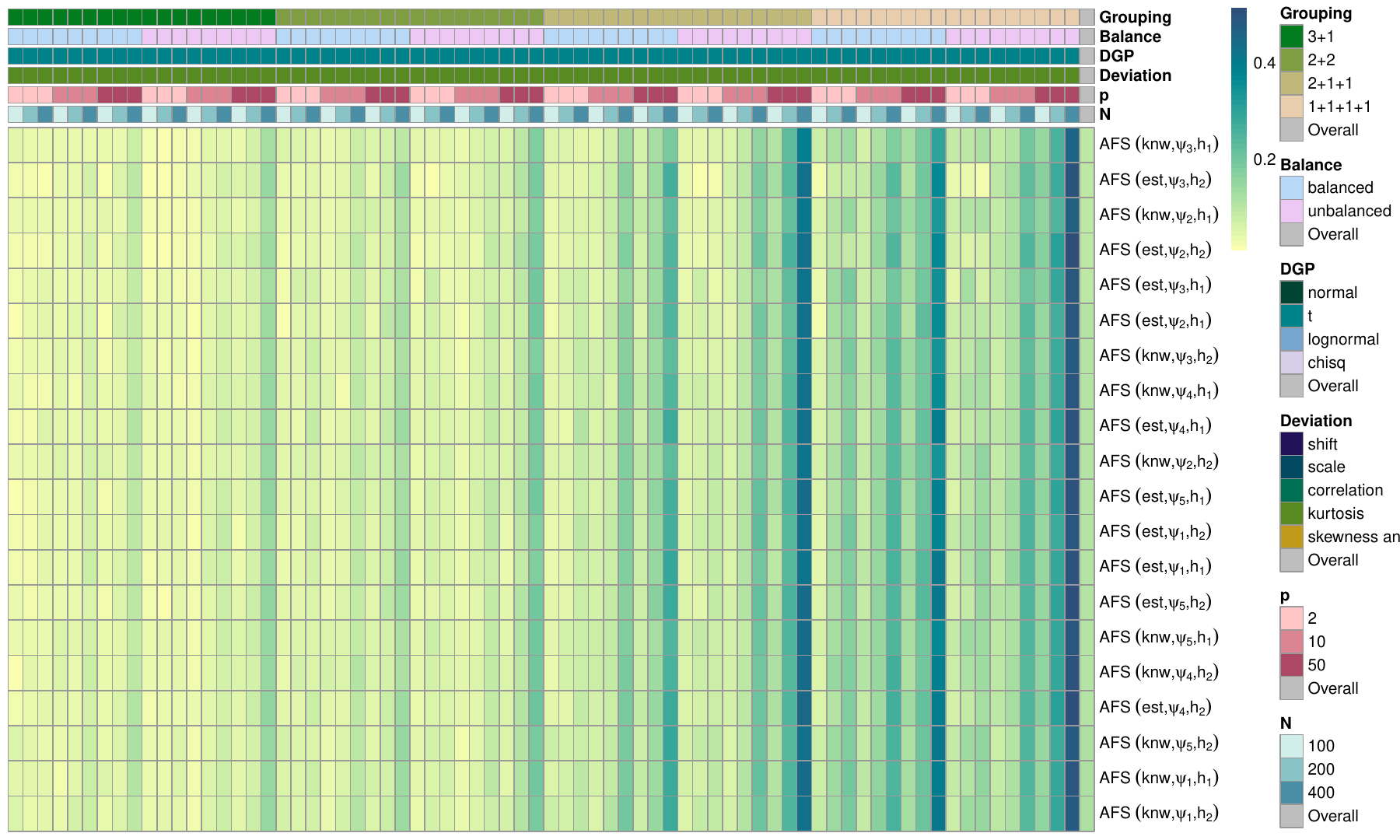}
		\caption{Mean PESR difference to best method per scenario for the kurtosis alternative and $k = 4$ datasets for all variants of the aggregated FS test \textcite{paul_clustering-based_2022}. Selected variant: AFS~(knw,$\psi_3,h_1$).}
	\end{figure}
	
	\begin{figure}[!h]
		\centering
		\includegraphics[width=\linewidth]{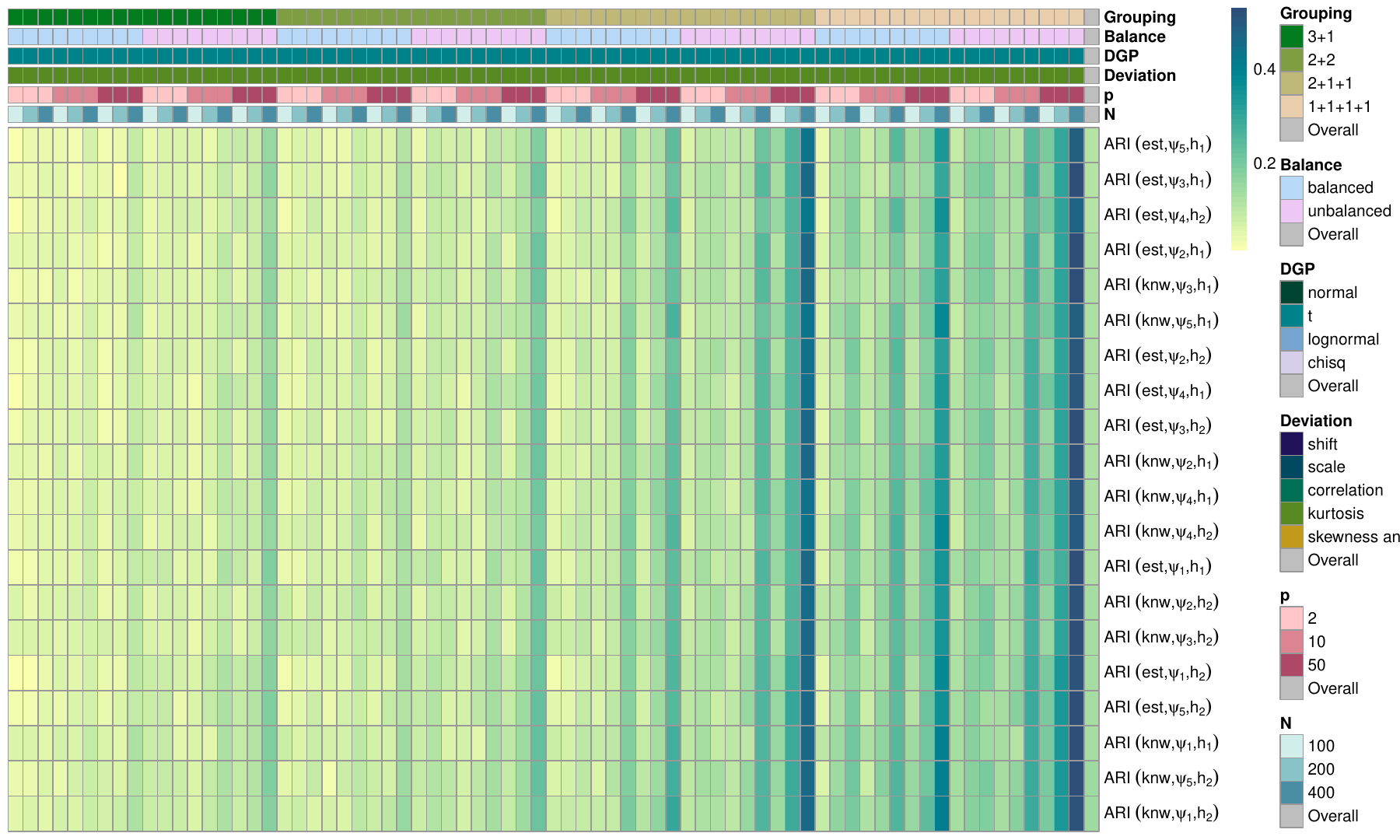}
		\caption{Mean PESR difference to best method per scenario for the kurtosis alternative and $k = 4$ datasets for all variants of the aggregated RI test \textcite{paul_clustering-based_2022}. Selected variant: ARI~(est,$\psi_5,h_1$).}
	\end{figure}
	
	\begin{figure}[!h]
		\centering
		\includegraphics[width=\linewidth]{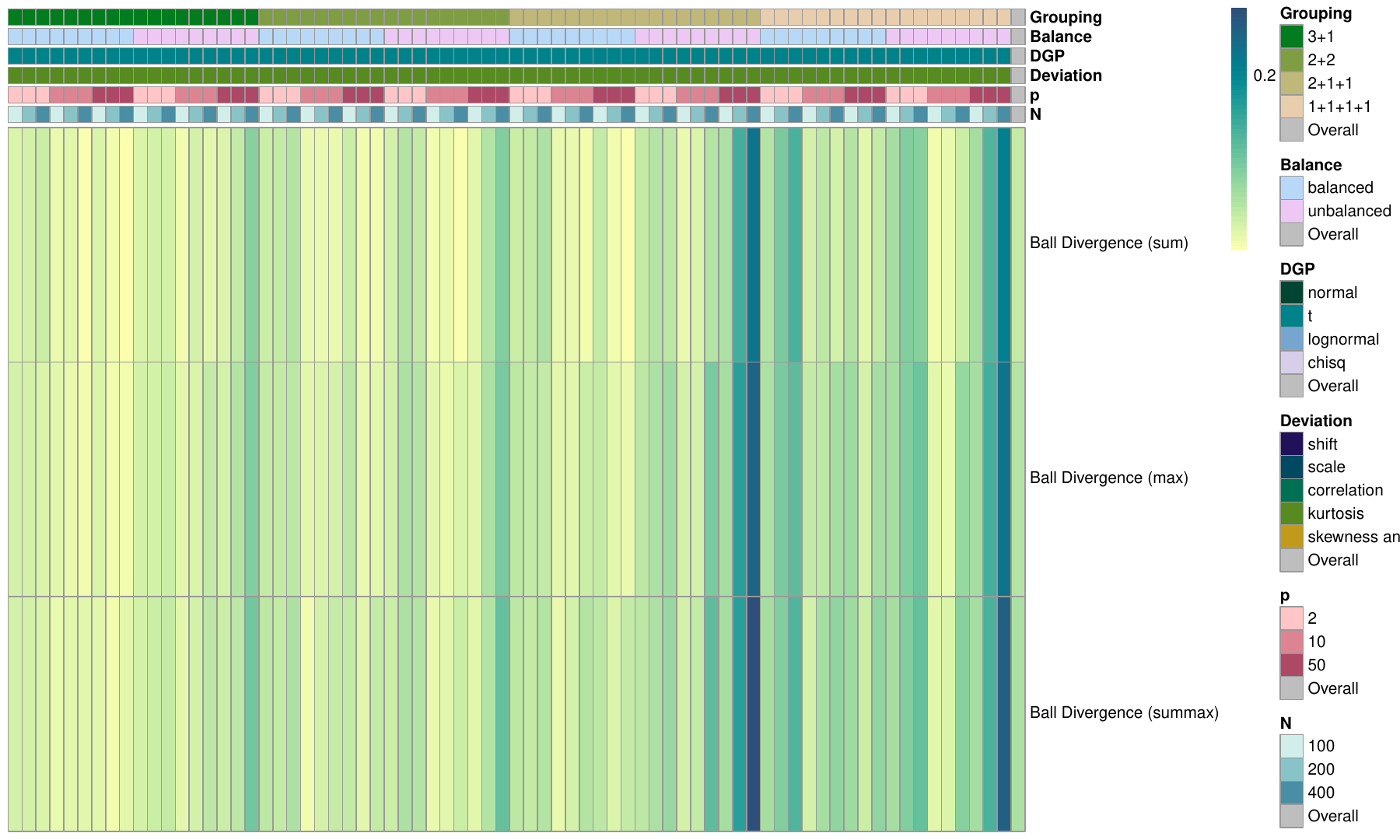}
		\caption{Mean PESR difference to best method per scenario for the kurtosis alternative and $k = 4$ datasets for all variants of the Ball divergence \textcite{pan_ball_2018}. Selected variant: Ball divergence~(sum).}
	\end{figure}
	
	\begin{figure}[!h]
		\centering
		\includegraphics[width=\linewidth]{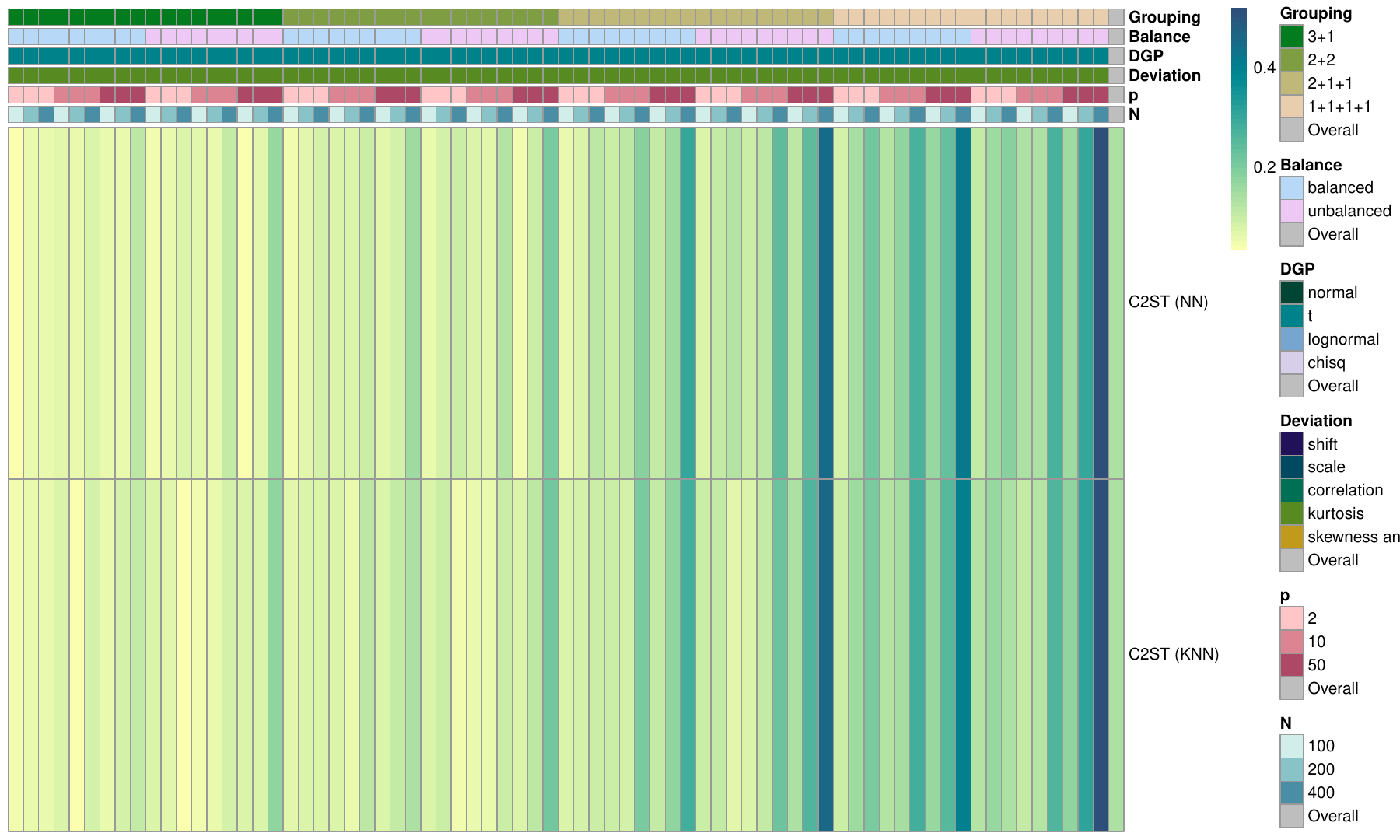}
		\caption{Mean PESR difference to best method per scenario for the kurtosis alternative and $k = 4$ datasets for all variants of the C2ST \textcite{alvarez-melis_geometric_2020}. Selected variant: C2ST~(NN).}
	\end{figure}
	
	\begin{figure}[!h]
		\centering
		\includegraphics[width=\linewidth]{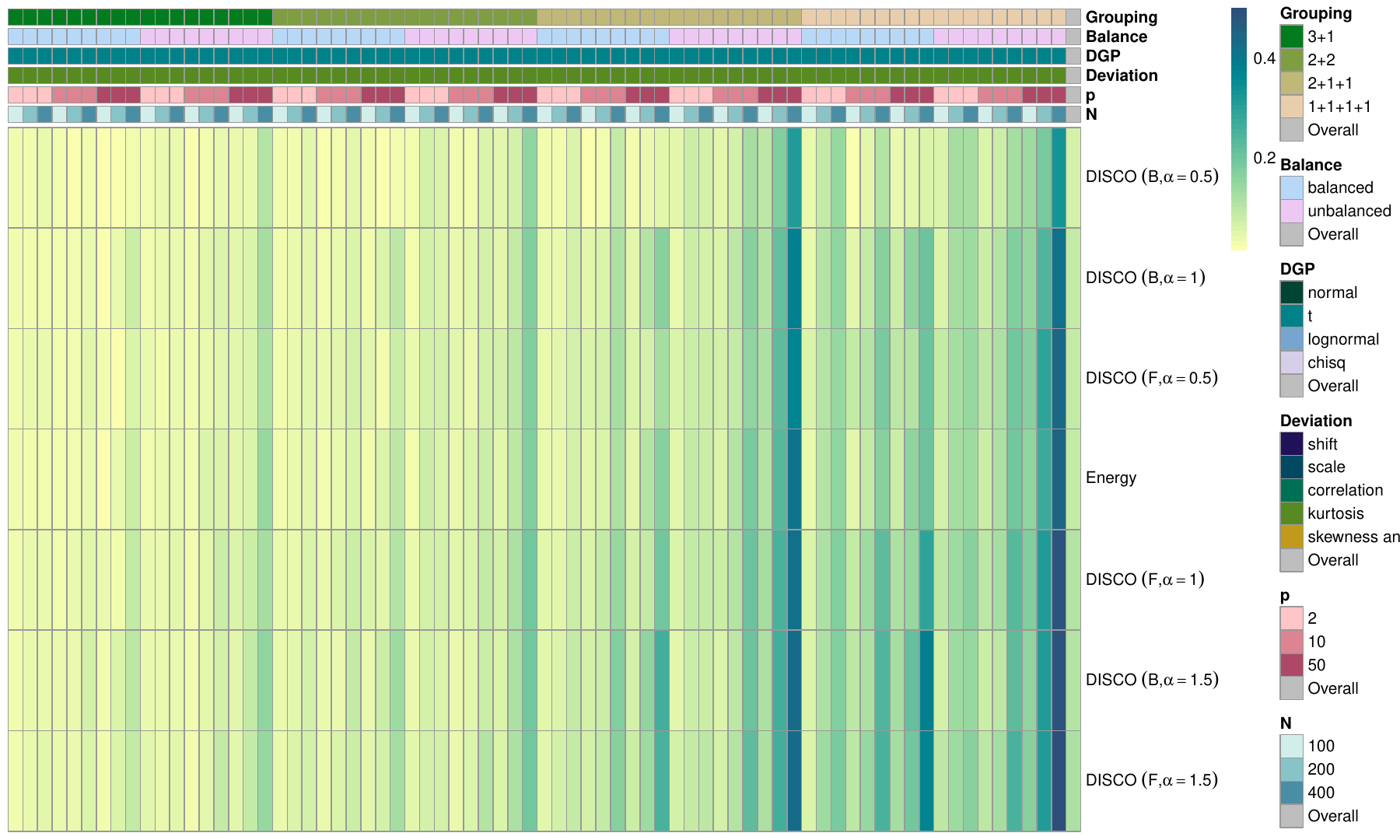}
		\caption{Mean PESR difference to best method per scenario for the kurtosis alternative and $k = 4$ datasets for all variants of the Energy distance / DISCO \textcite{szekely_testing_2004, rizzo_disco_2010}. Selected variant: DISCO~(B,$\alpha=0.5$).}
	\end{figure}
	
	\begin{figure}[!h]
		\centering
		\includegraphics[width=\linewidth]{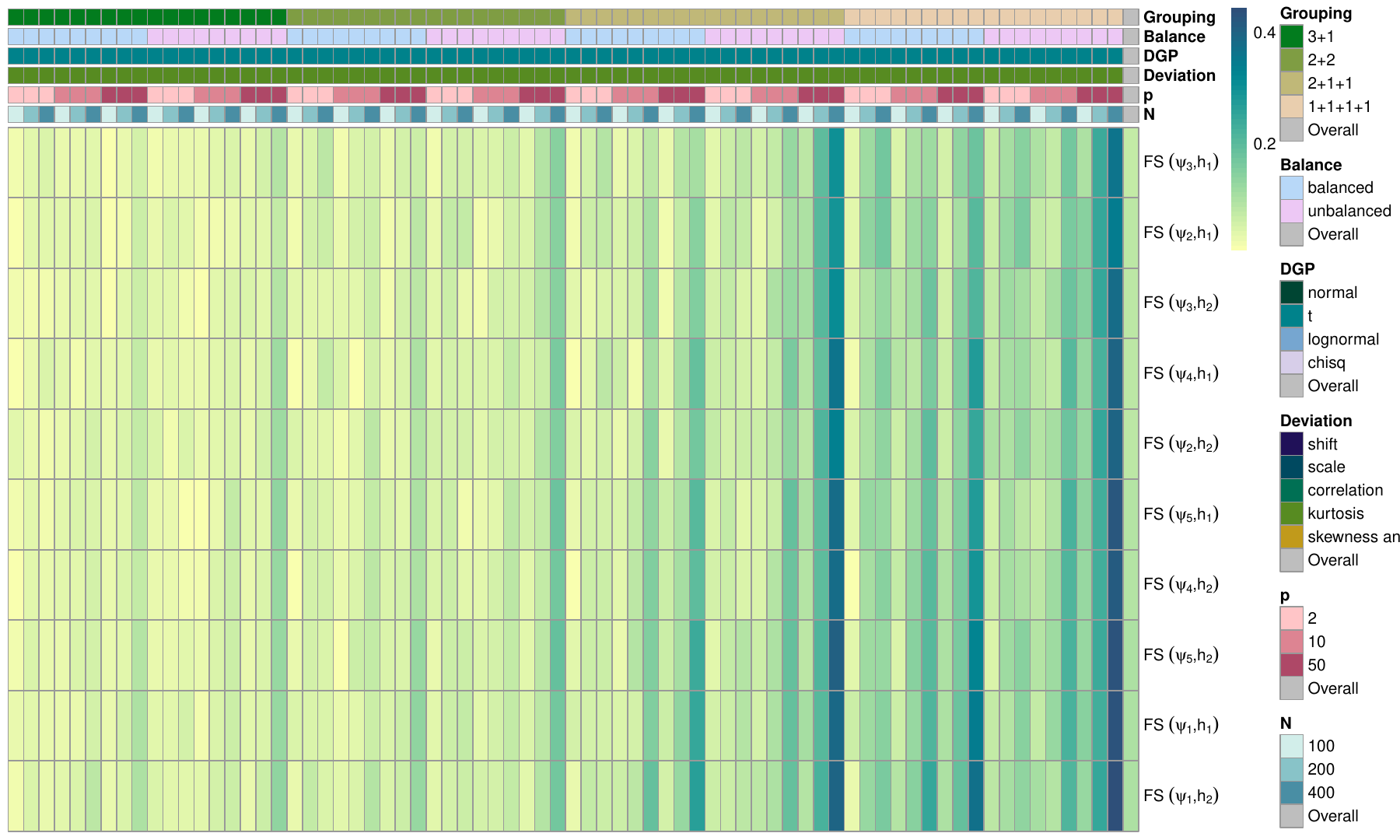}
		\caption{Mean PESR difference to best method per scenario for the kurtosis alternative and $k = 4$ datasets for all variants of the FS test \textcite{paul_clustering-based_2022}. Selected variant: FS~($\psi_3$,$h_1$).}
	\end{figure}
	
	\begin{figure}[!h]
		\centering
		\includegraphics[width=\linewidth]{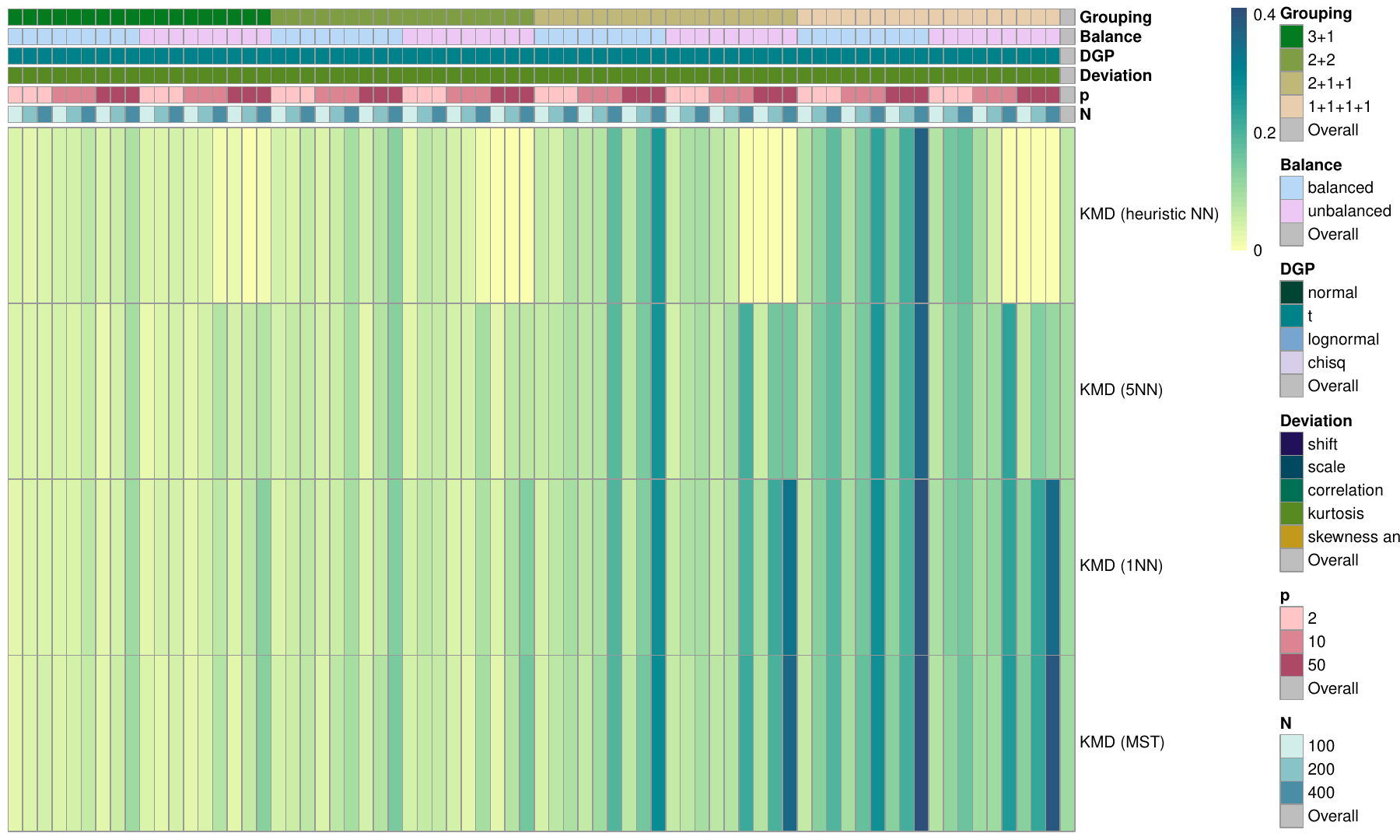}
		\caption{Mean PESR difference to best method per scenario for the kurtosis alternative and $k = 4$ datasets for all variants of the modified KMD \textcite{huang_kernel_2022}. Selected variant: KMD~(heuristic NN).}
	\end{figure}
	
	\begin{figure}[!h]
		\centering
		\includegraphics[width=\linewidth]{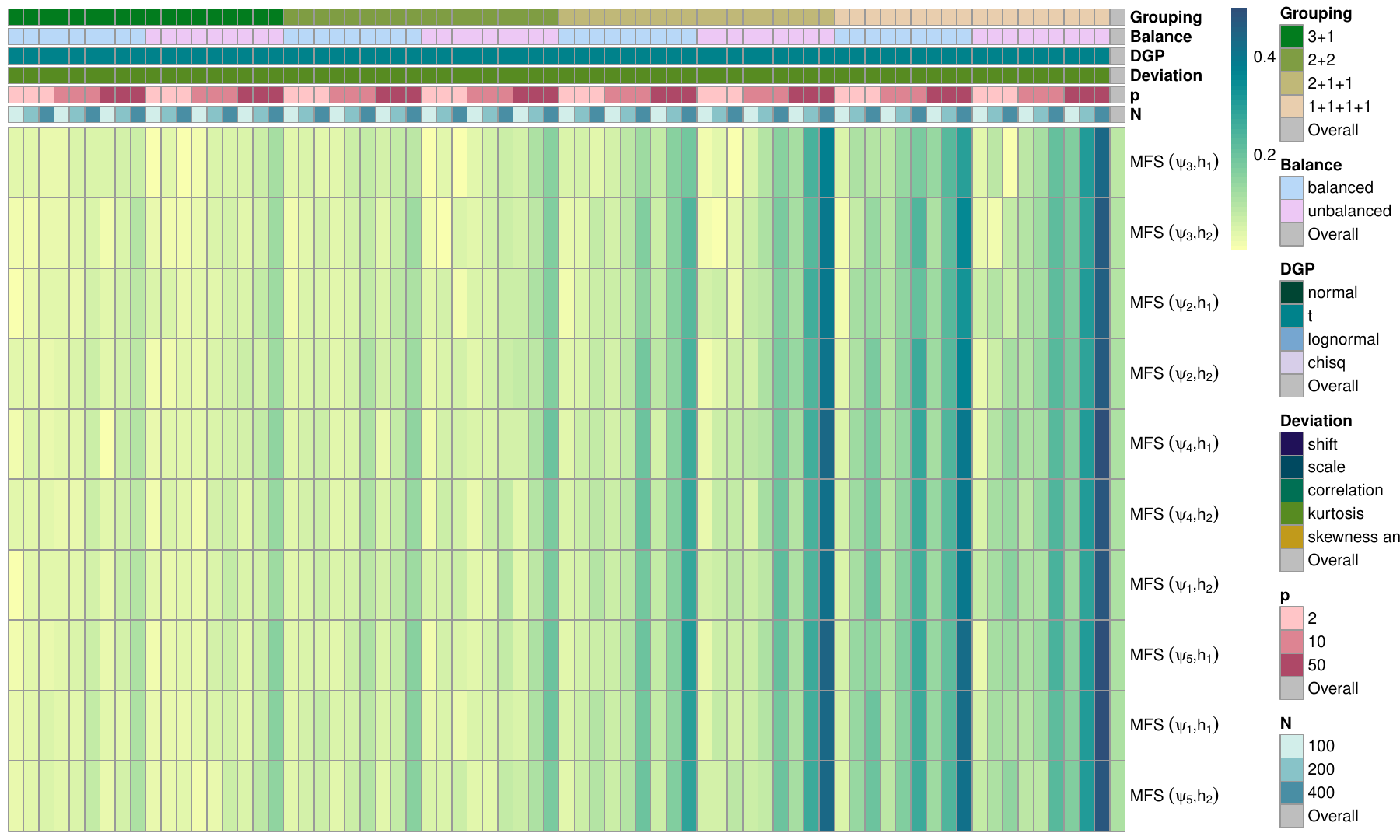}
		\caption{Mean PESR difference to best method per scenario for the kurtosis alternative and $k = 4$ datasets for all variants of the modified FS test \textcite{paul_clustering-based_2022}. Selected variant: MFS~($\psi_3$,$h_1$).}
	\end{figure}
	
	\begin{figure}[!h]
		\centering
		\includegraphics[width=\linewidth]{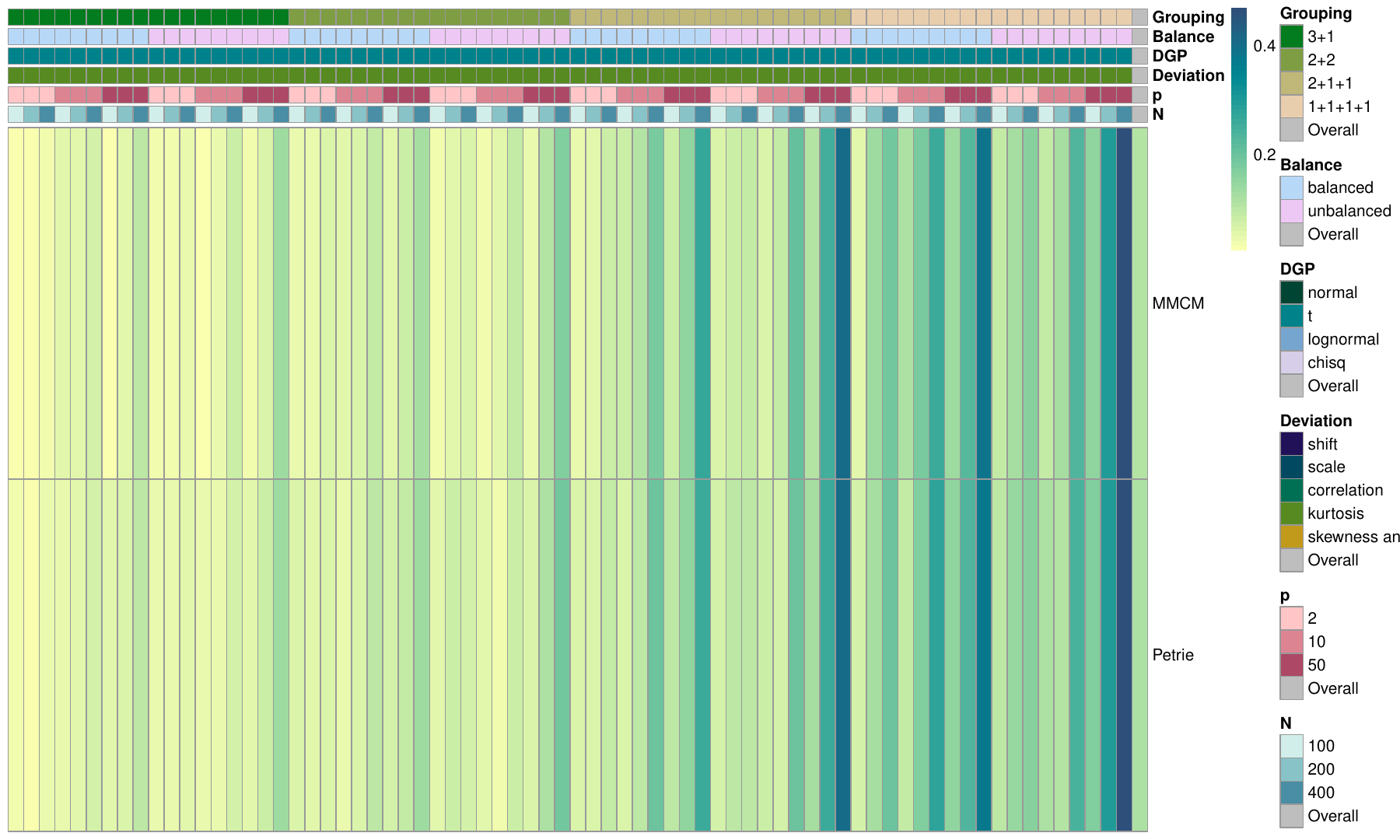}
		\caption{Mean PESR difference to best method per scenario for the kurtosis alternative and $k = 4$ datasets for the MMCM and Petrie's test \textcite{mukherjee_distribution-free_2022, petrie_graph-theoretic_2016}. Selected variant: MMCM.}
	\end{figure}
	
	\begin{figure}[!h]
		\centering
		\includegraphics[width=\linewidth]{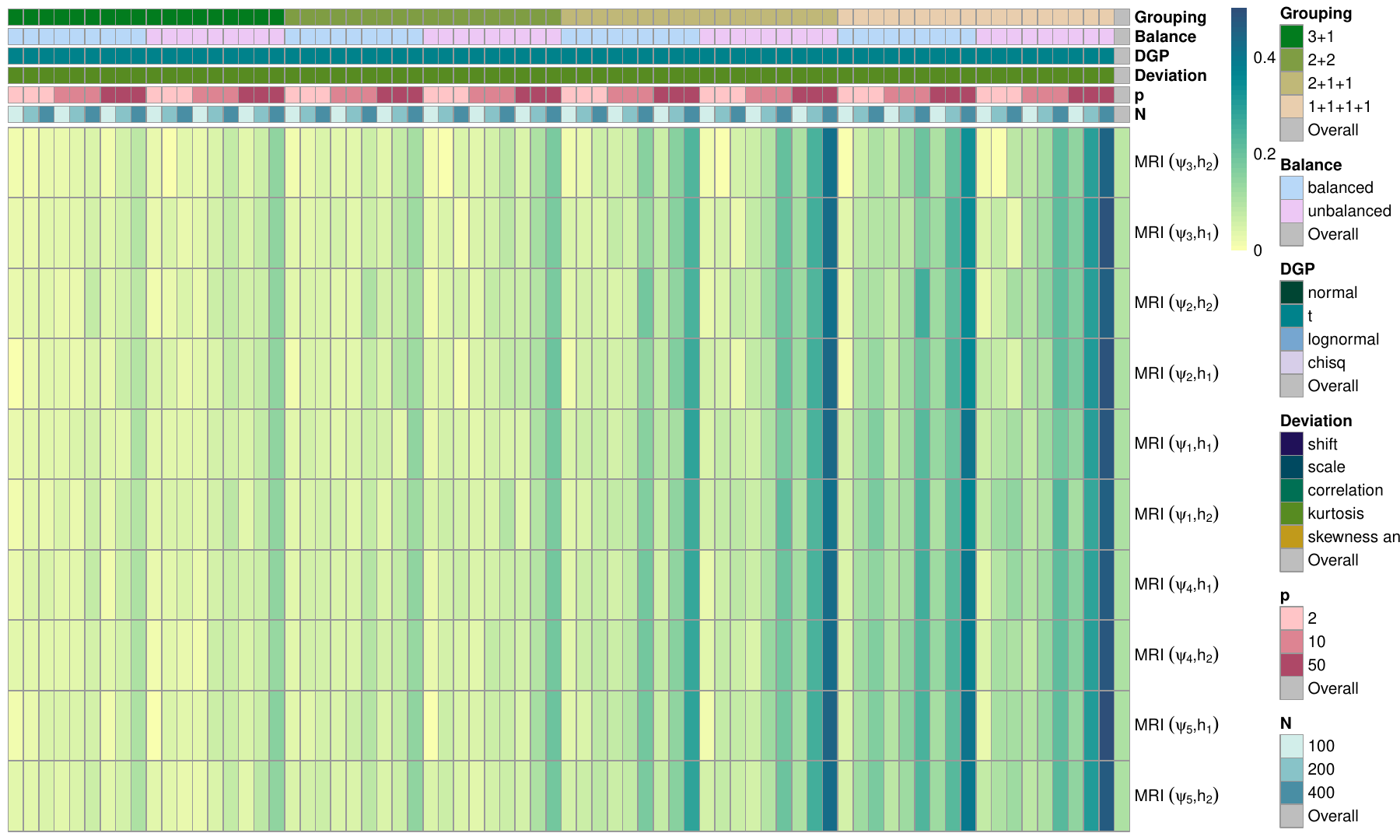}
		\caption{Mean PESR difference to best method per scenario for the kurtosis alternative and $k = 4$ datasets for all variants of the modified RI test \textcite{paul_clustering-based_2022}. Selected variant: MRI~($\psi_3$,$h_2$).}
	\end{figure}
	
	\begin{figure}[!h]
		\centering
		\includegraphics[width=\linewidth]{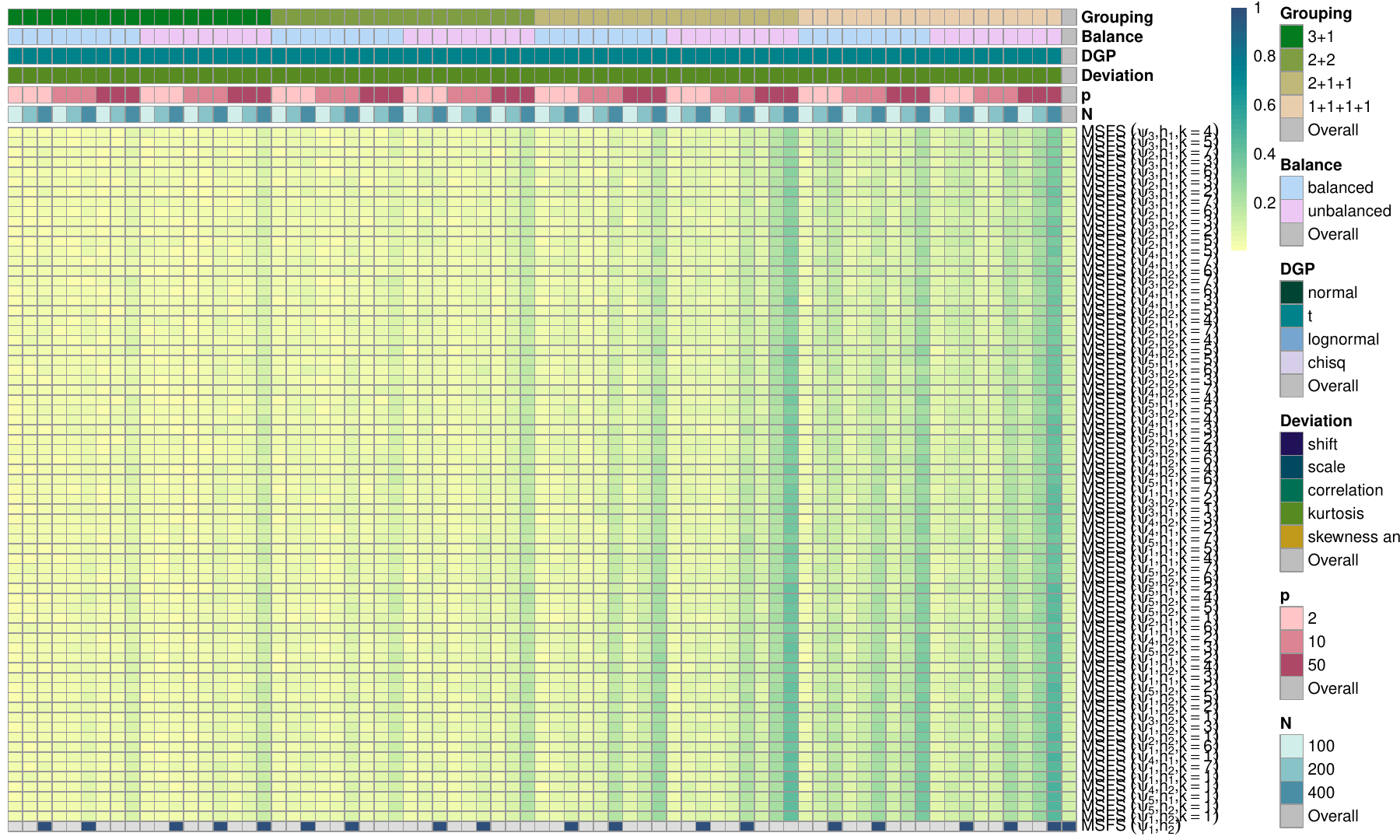}
		\caption{Mean PESR difference to best method per scenario for the kurtosis alternative and $k = 4$ datasets for all variants of the multiscale FS test \textcite{paul_clustering-based_2022}. Selected variant: MSFS~($\psi_3$,$h_1$,$k=4$).}
	\end{figure}
	
	\begin{figure}[!h]
		\centering
		\includegraphics[width=\linewidth]{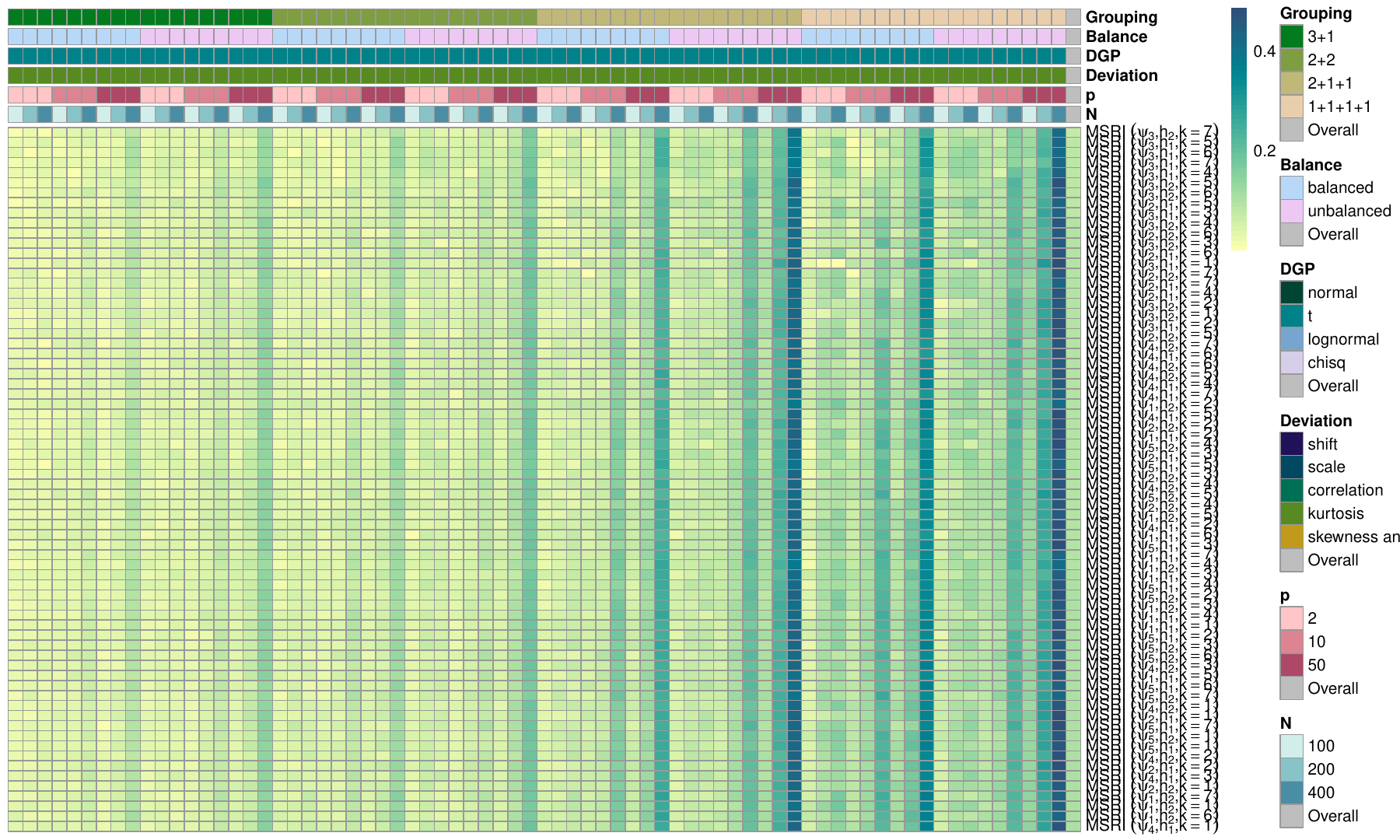}
		\caption{Mean PESR difference to best method per scenario for the kurtosis alternative and $k = 4$ datasets for all variants of the multiscale RI test \textcite{paul_clustering-based_2022}. Selected variant: MSRI~($\psi_3$,$h_2$,$k=7$).}
	\end{figure}
	
	\begin{figure}[!h]
		\centering
		\includegraphics[width=\linewidth]{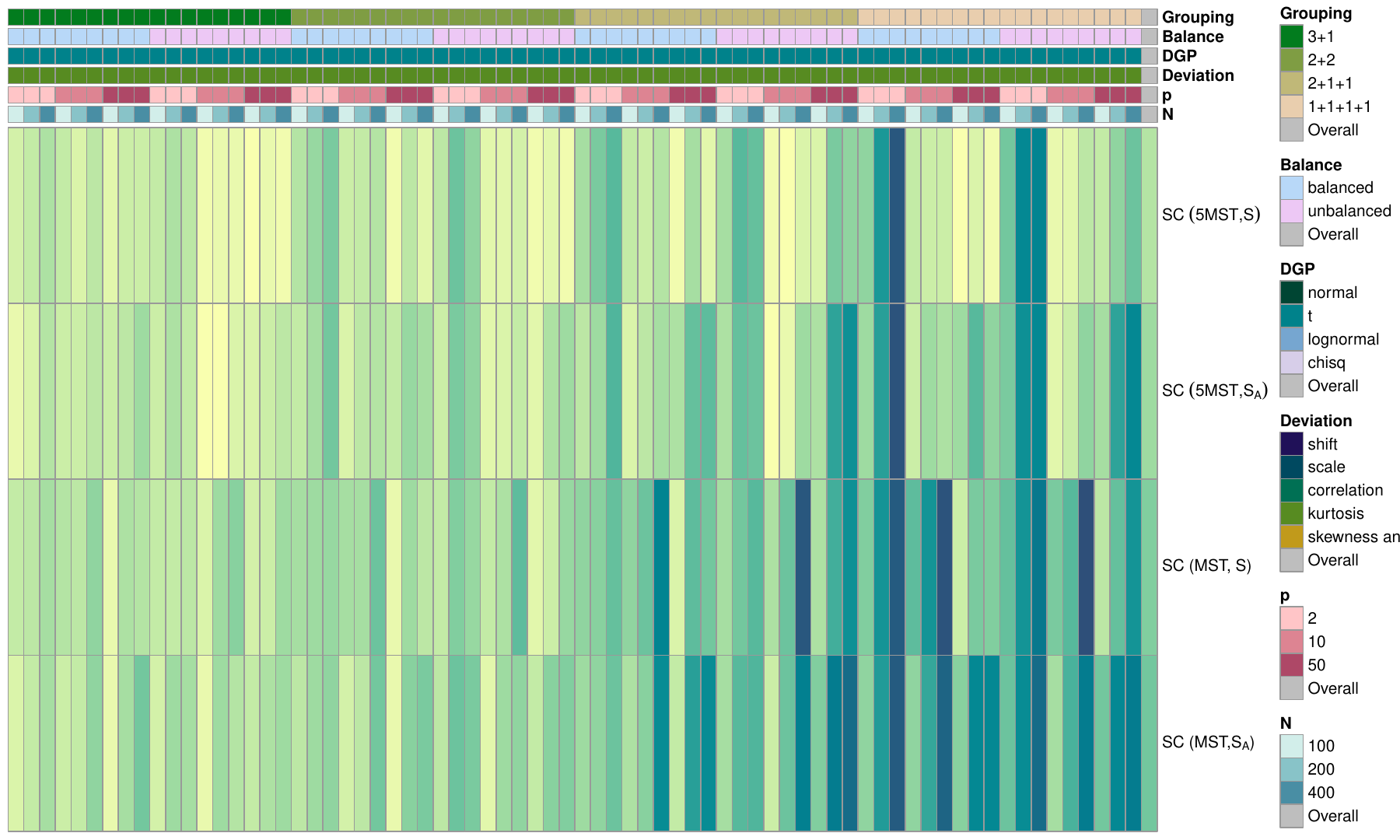}
		\caption{Mean PESR difference to best method per scenario for the kurtosis alternative and $k = 4$ datasets for all variants of the SC test \textcite{song_generalized_2021}. Selected variant: SC~(5MST,$S$).}
	\end{figure}
	\clearpage
	
	\paragraph{Skewness and Kurtosis}	\mbox{} 
	
	\begin{figure}[!h]
		\centering
		\includegraphics[width=\linewidth]{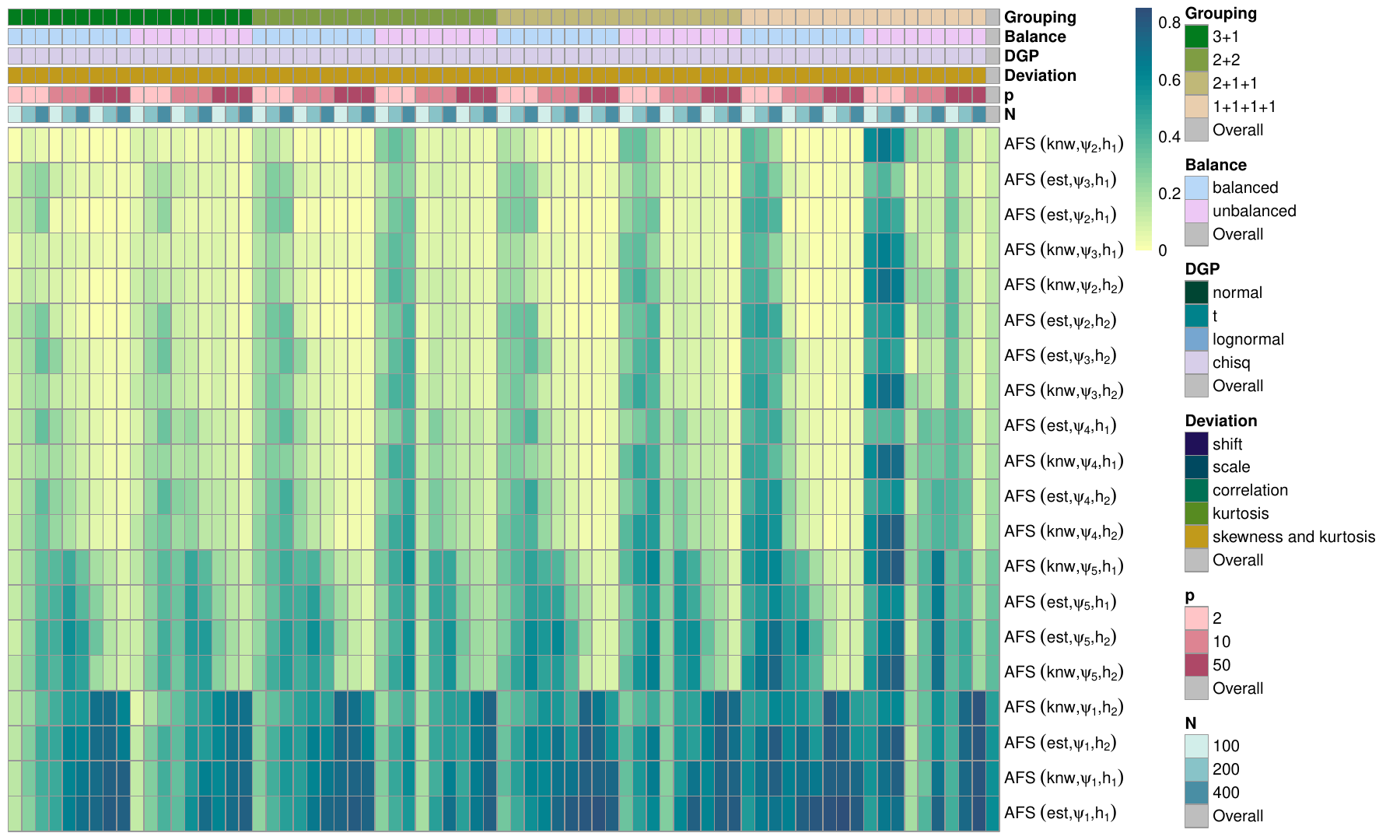}
		\caption{Mean PESR difference to best method per scenario for the skewness and kurtosis alternative and $k = 4$ datasets for all variants of the aggregated FS test \textcite{paul_clustering-based_2022}. Selected variant: AFS~(knw,$\psi_2,h_1$).}
	\end{figure}
	
	\begin{figure}[!h]
		\centering
		\includegraphics[width=\linewidth]{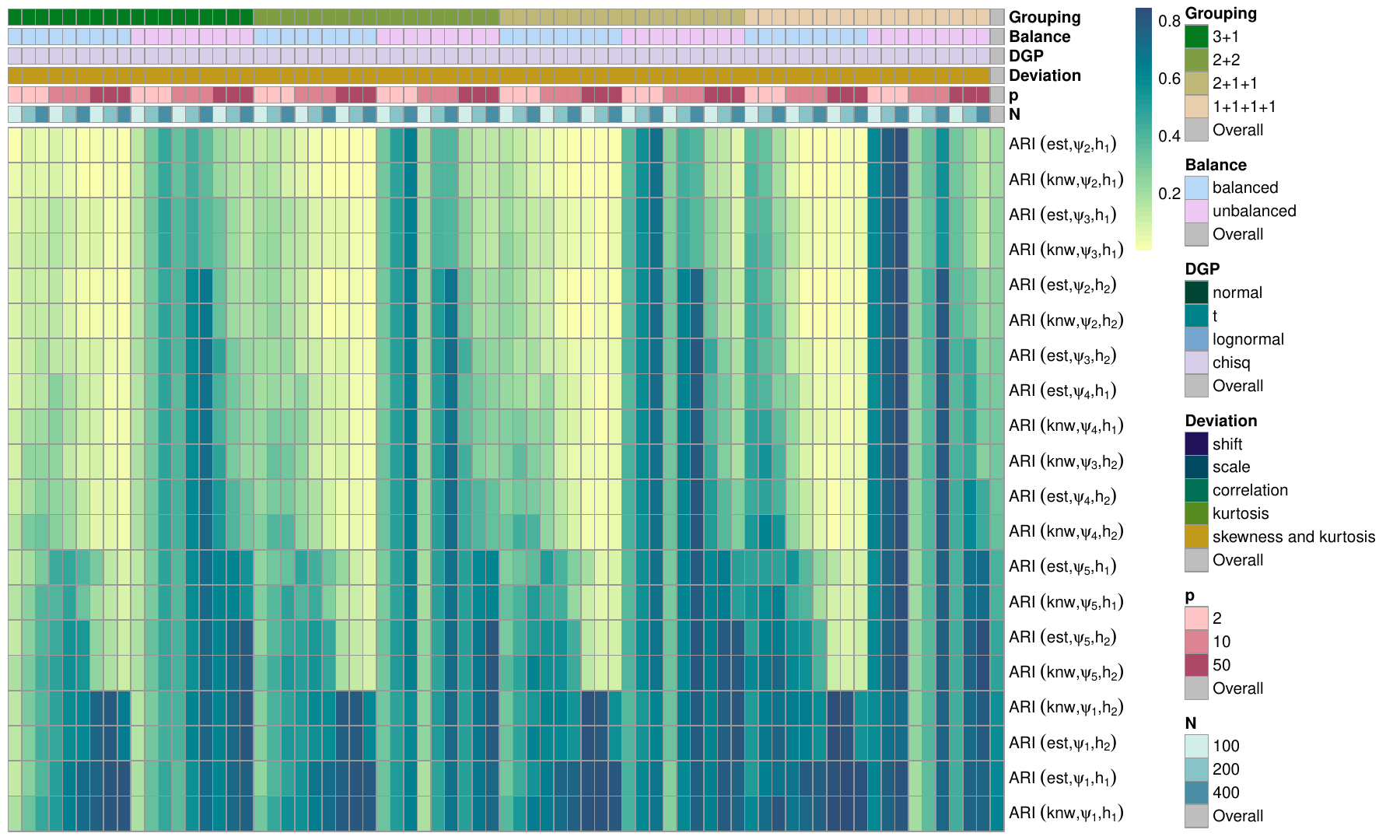}
		\caption{Mean PESR difference to best method per scenario for the skewness and kurtosis alternative and $k = 4$ datasets for all variants of the aggregated RI test \textcite{paul_clustering-based_2022}. Selected variant: ARI~(est,$\psi_2,h_1$).}
	\end{figure}
	
	\begin{figure}[!h]
		\centering
		\includegraphics[width=\linewidth]{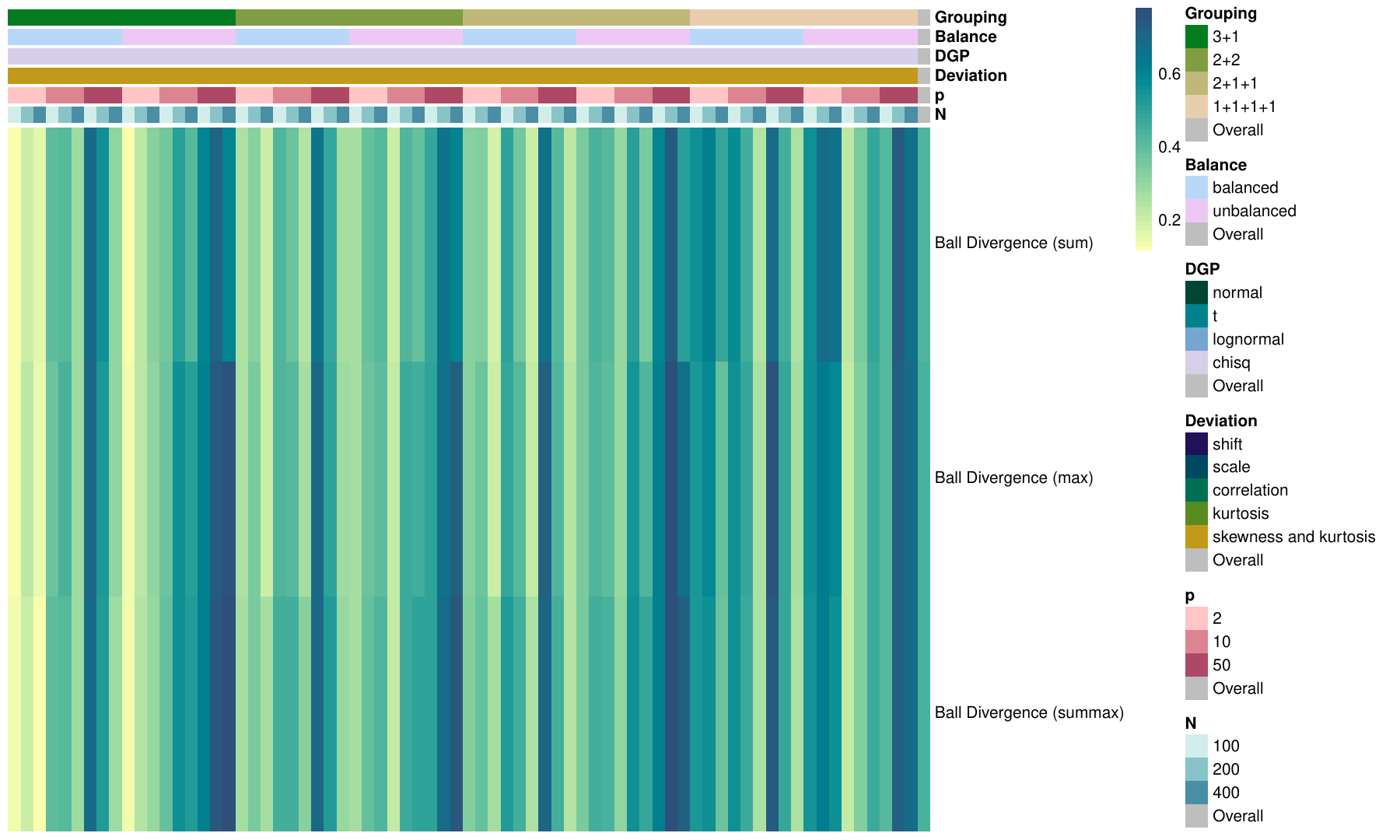}
		\caption{Mean PESR difference to best method per scenario for the skewness and kurtosis alternative and $k = 4$ datasets for all variants of the Ball divergence \textcite{pan_ball_2018}. Selected variant: Ball divergence~(sum).}
	\end{figure}
	
	\begin{figure}[!h]
		\centering
		\includegraphics[width=\linewidth]{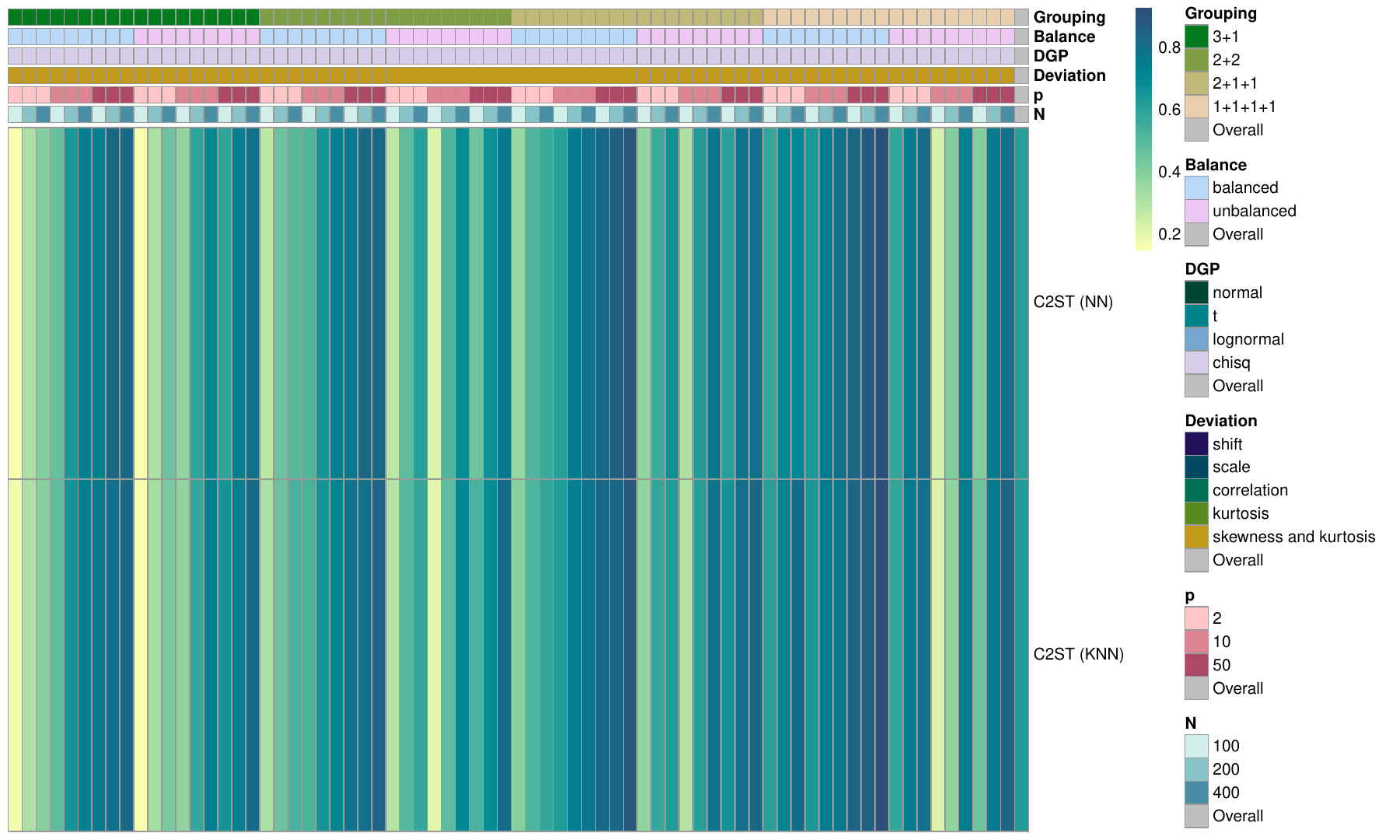}
		\caption{Mean PESR difference to best method per scenario for the skewness and kurtosis alternative and $k = 4$ datasets for all variants of the C2ST \textcite{alvarez-melis_geometric_2020}. Selected variant: C2ST~(NN).}
	\end{figure}
	
	\begin{figure}[!h]
		\centering
		\includegraphics[width=\linewidth]{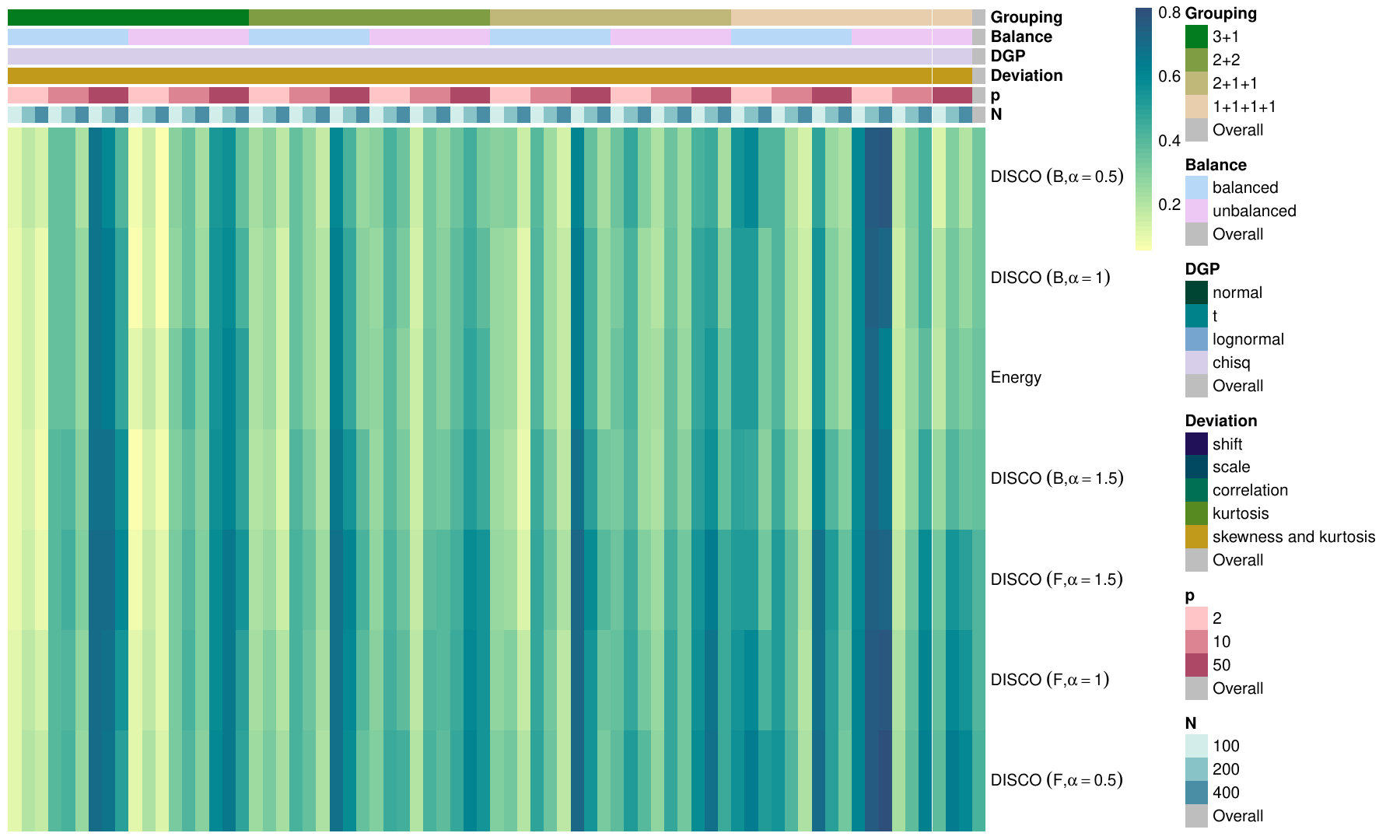}
		\caption{Mean PESR difference to best method per scenario for the skewness and kurtosis alternative and $k = 4$ datasets for all variants of the Energy distance / DISCO \textcite{szekely_testing_2004, rizzo_disco_2010}. Selected variant: DISCO~(B,$\alpha=0.5$).}
	\end{figure}
	
	\begin{figure}[!h]
		\centering
		\includegraphics[width=\linewidth]{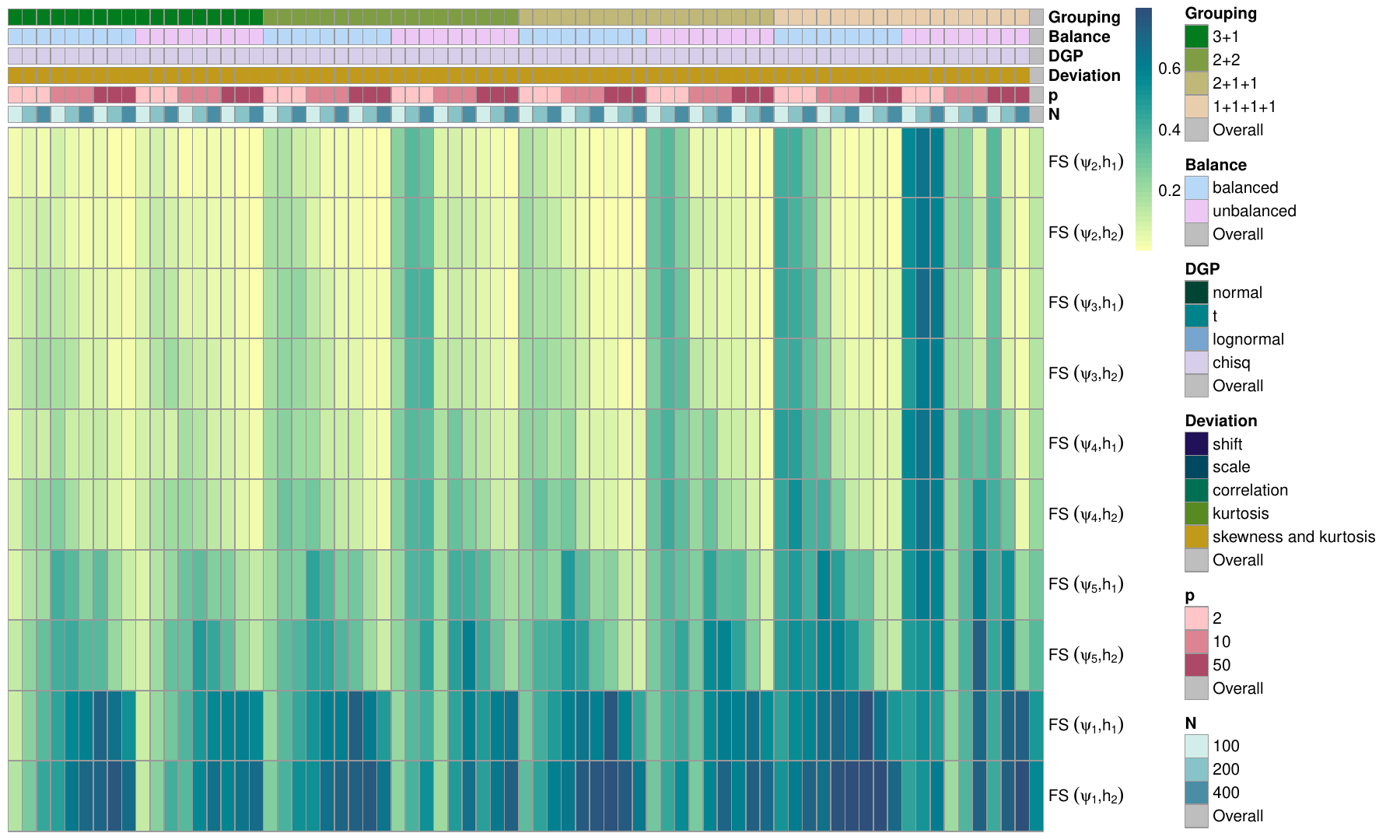}
		\caption{Mean PESR difference to best method per scenario for the skewness and kurtosis alternative and $k = 4$ datasets for all variants of the FS test \textcite{paul_clustering-based_2022}. Selected variant: FS~($\psi_2$,$h_1$).}
	\end{figure}
	
	\begin{figure}[!h]
		\centering
		\includegraphics[width=\linewidth]{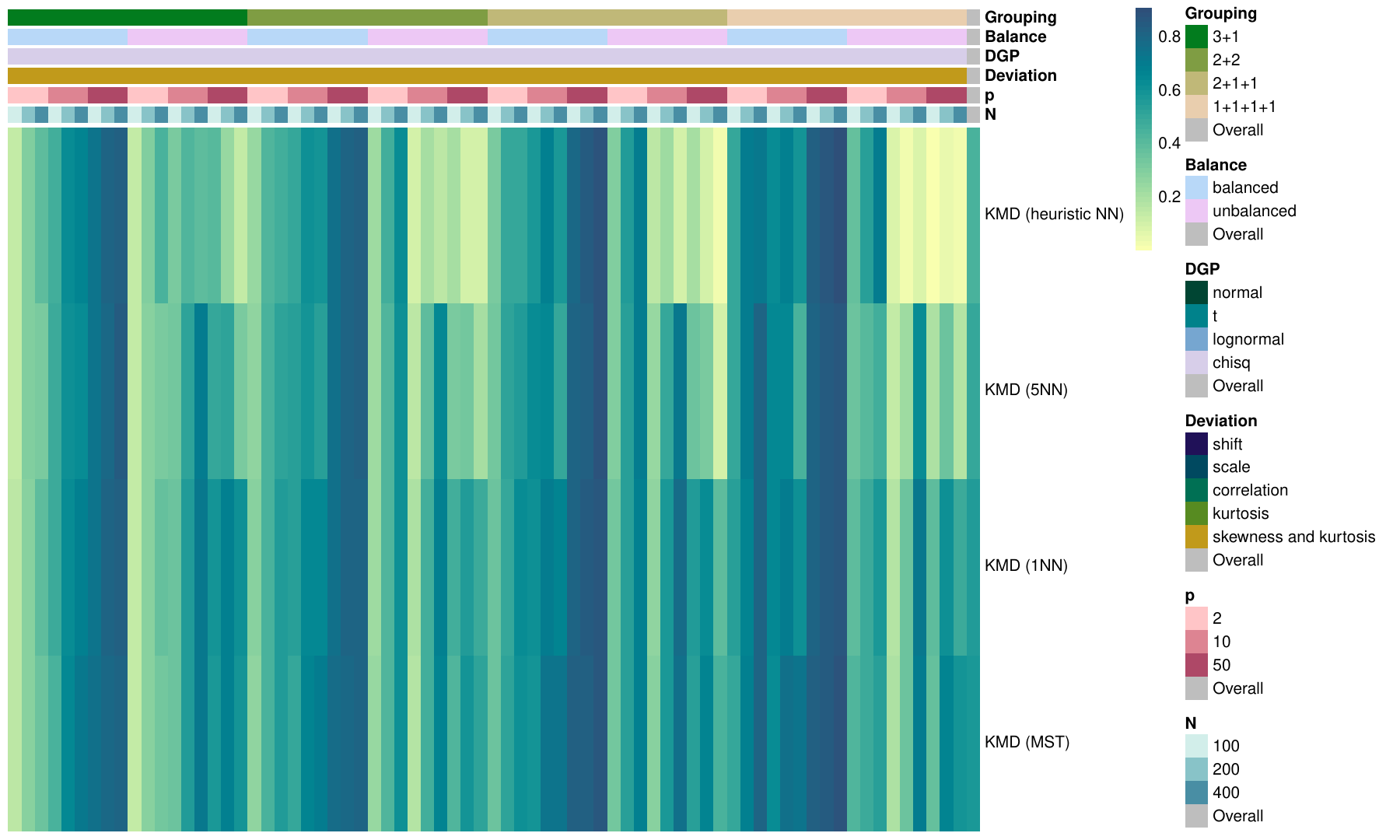}
		\caption{Mean PESR difference to best method per scenario for the skewness and kurtosis alternative and $k = 4$ datasets for all variants of the modified KMD \textcite{huang_kernel_2022}. Selected variant: KMD~(heuristic NN).}
	\end{figure}
	
	\begin{figure}[!h]
		\centering
		\includegraphics[width=\linewidth]{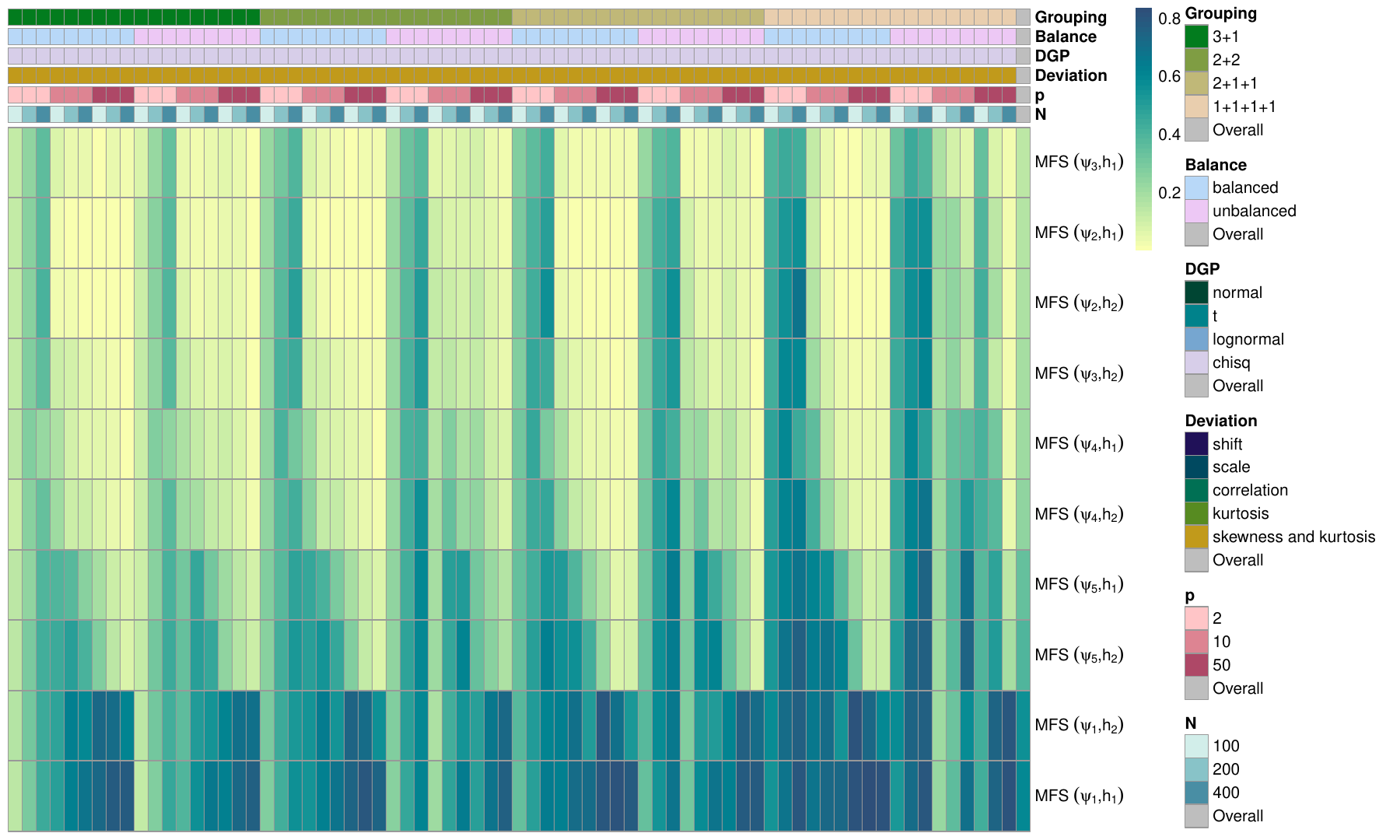}
		\caption{Mean PESR difference to best method per scenario for the skewness and kurtosis alternative and $k = 4$ datasets for all variants of the modified FS test \textcite{paul_clustering-based_2022}. Selected variant: MFS~($\psi_3$,$h_1$).}
	\end{figure}
	
	\begin{figure}[!h]
		\centering
		\includegraphics[width=\linewidth]{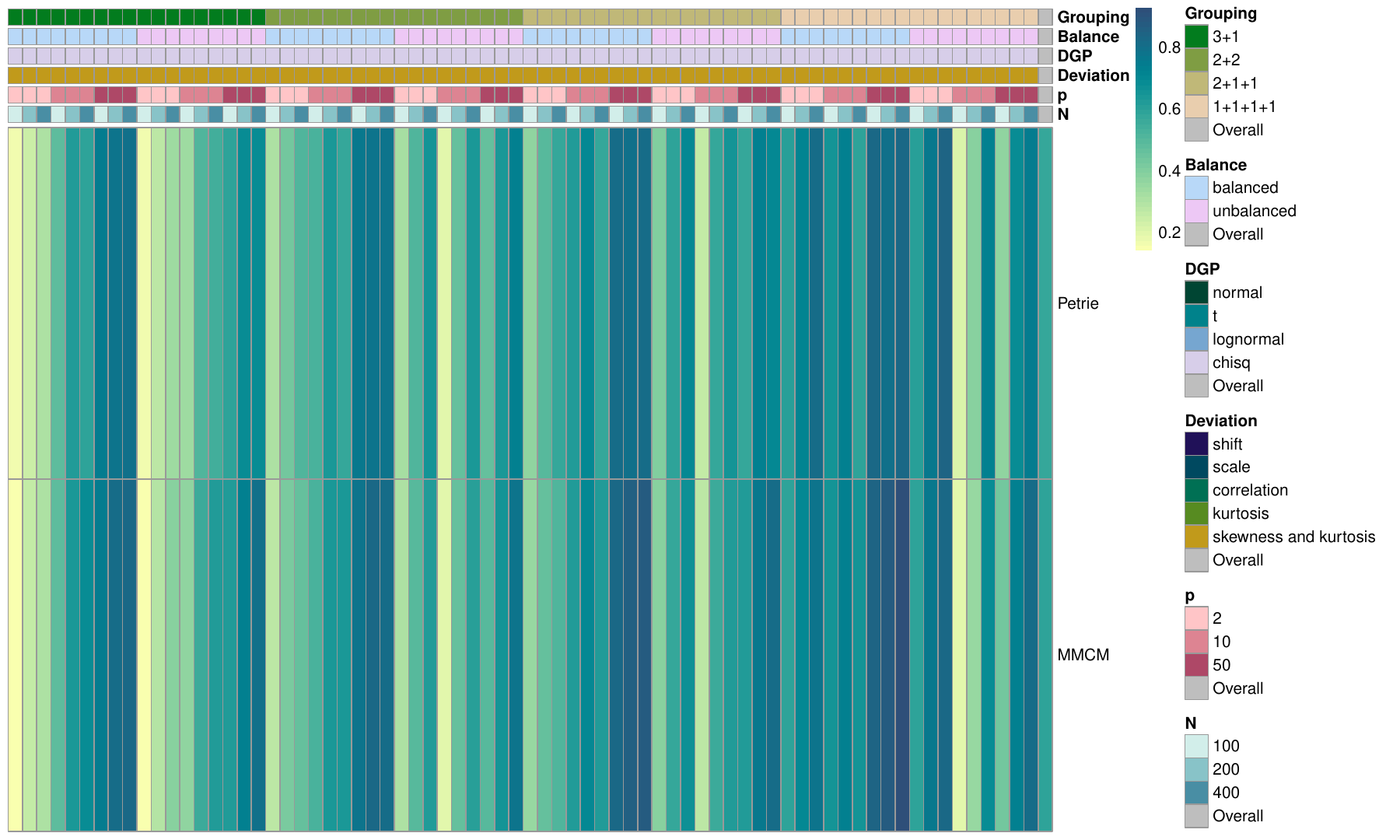}
		\caption{Mean PESR difference to best method per scenario for the skewness and kurtosis alternative and $k = 4$ datasets for the MMCM and Petrie's test \textcite{mukherjee_distribution-free_2022, petrie_graph-theoretic_2016}. Selected variant: Petrie.}
	\end{figure}
	
	\begin{figure}[!h]
		\centering
		\includegraphics[width=\linewidth]{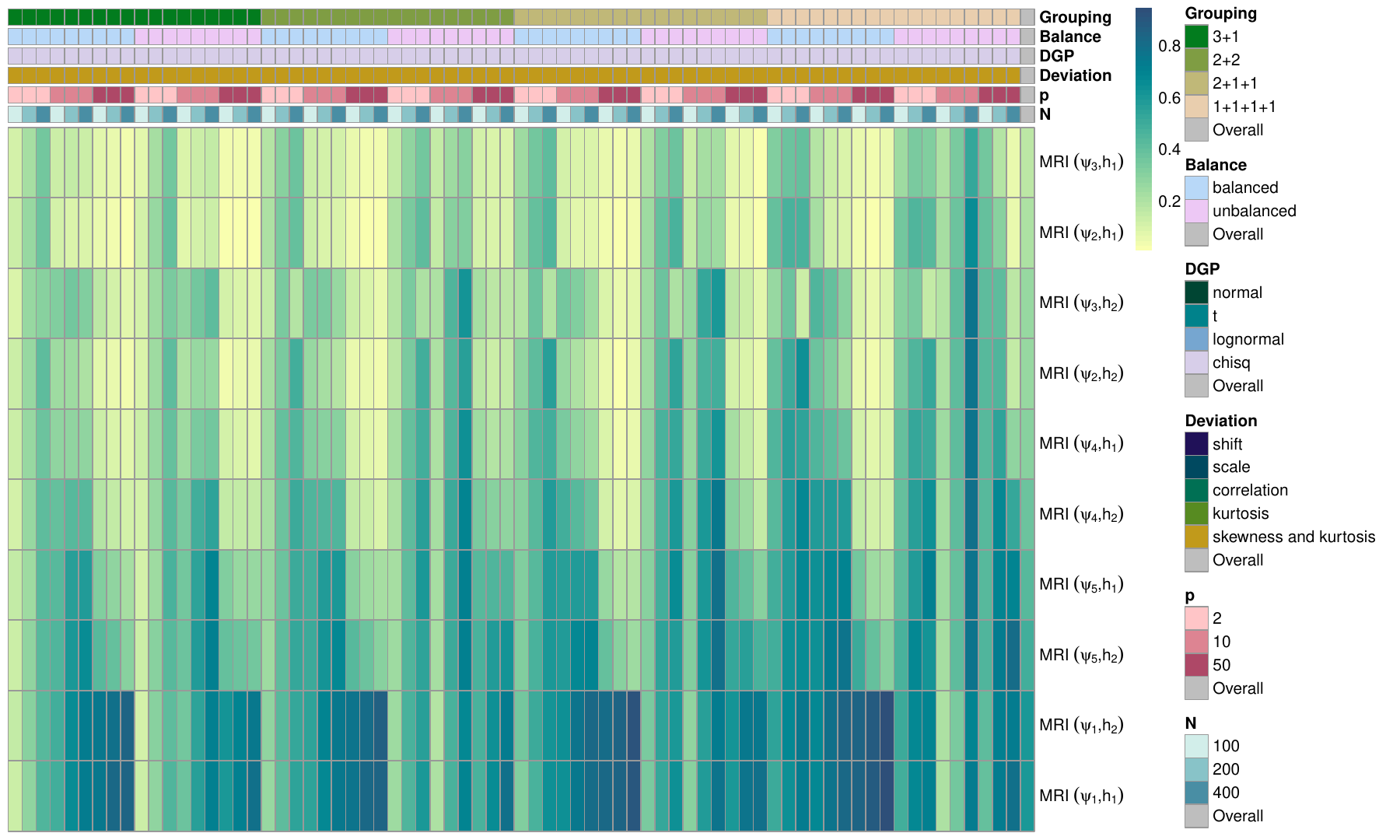}
		\caption{Mean PESR difference to best method per scenario for the skewness and kurtosis alternative and $k = 4$ datasets for all variants of the modified RI test \textcite{paul_clustering-based_2022}. Selected variant: MRI~($\psi_3$,$h_1$).}
	\end{figure}
	
	\begin{figure}[!h]
		\centering
		\includegraphics[width=\linewidth]{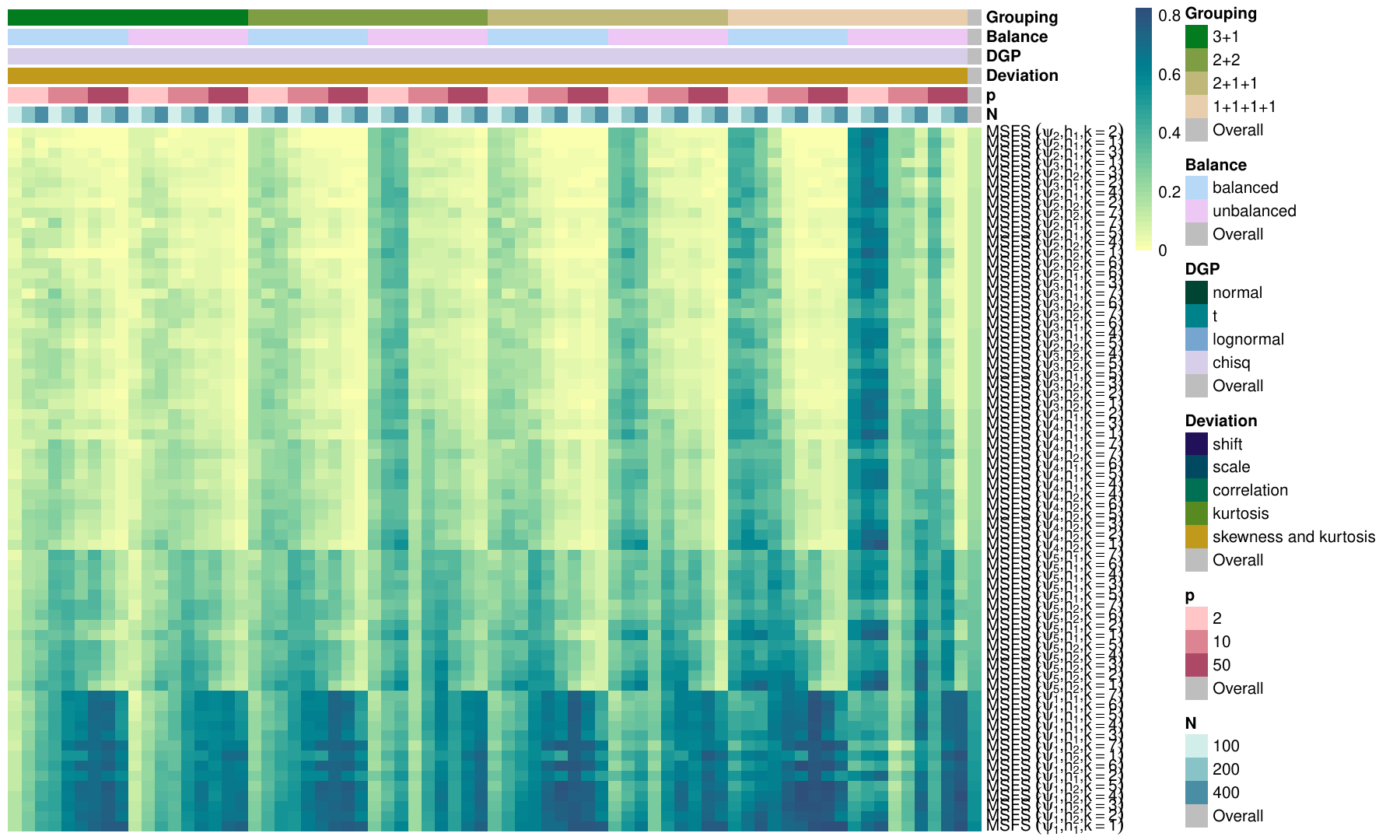}
		\caption{Mean PESR difference to best method per scenario for the skewness and kurtosis alternative and $k = 4$ datasets for all variants of the multiscale FS test \textcite{paul_clustering-based_2022}. Selected variant: MSFS~($\psi_2$,$h_1$,$k=2$).}
	\end{figure}
	
	\begin{figure}[!h]
		\centering
		\includegraphics[width=\linewidth]{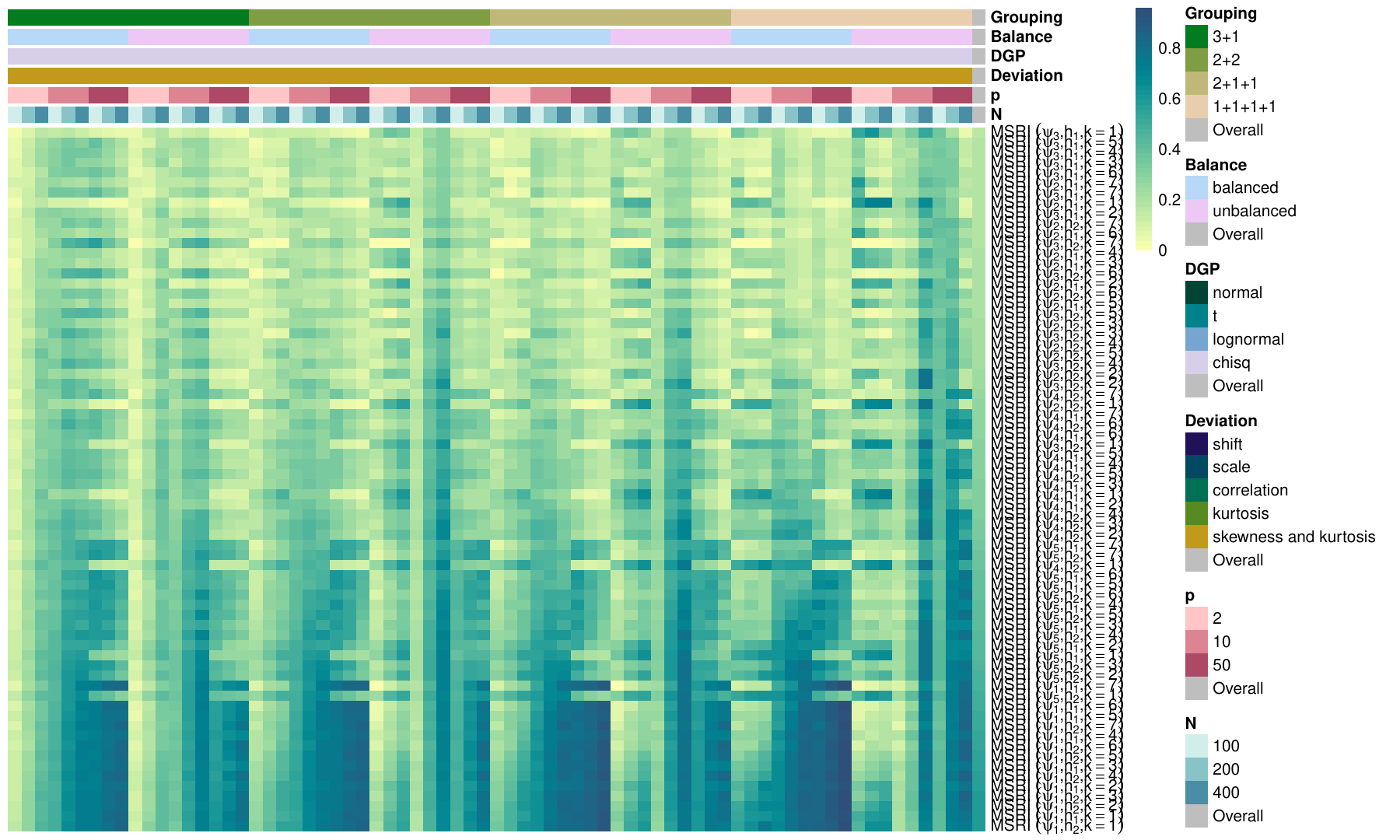}
		\caption{Mean PESR difference to best method per scenario for the skewness and kurtosis alternative and $k = 4$ datasets for all variants of the multiscale RI test \textcite{paul_clustering-based_2022}. Selected variant: MSRI~($\psi_3$,$h_1$,$k=1$).}
	\end{figure}
	
	\begin{figure}[!h]
		\centering
		\includegraphics[width=\linewidth]{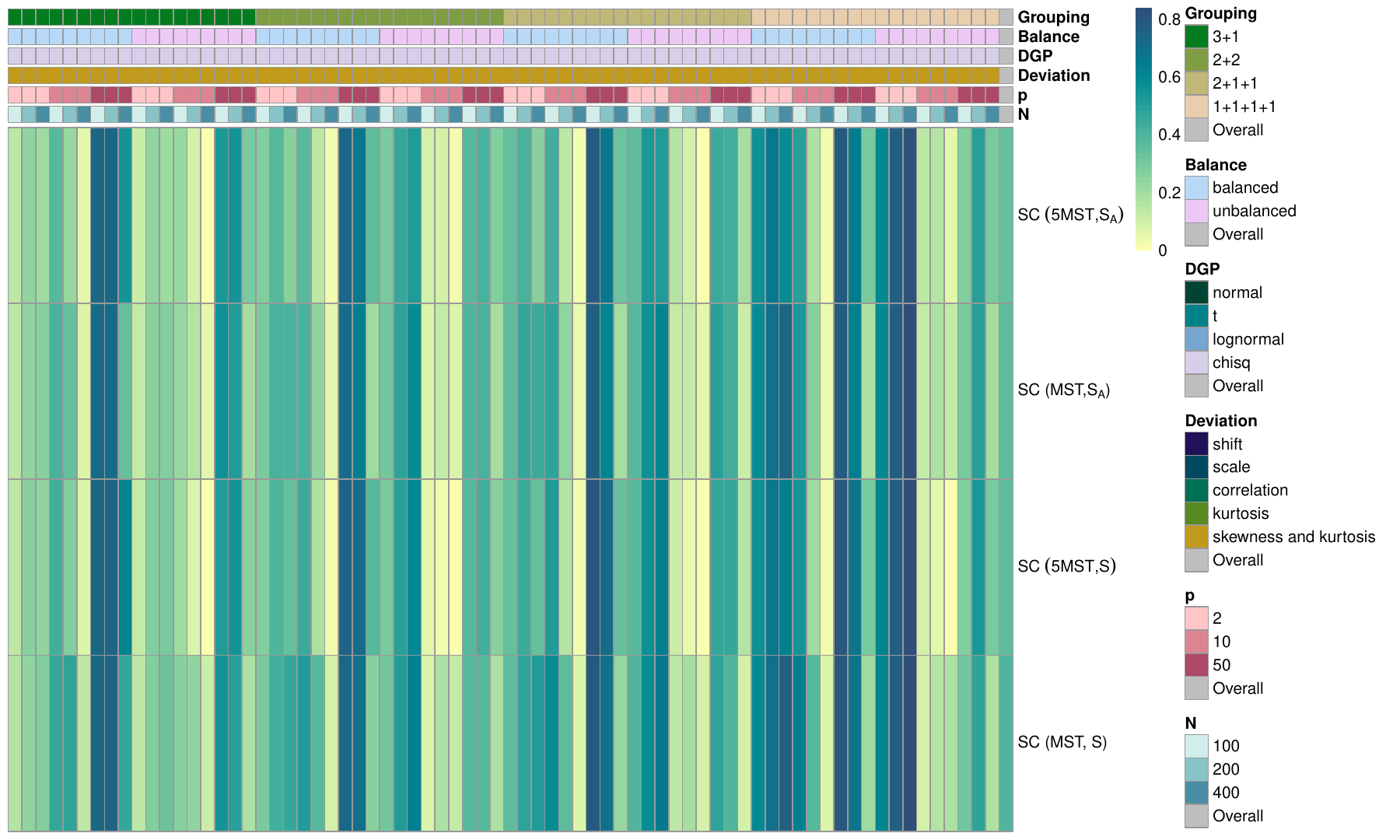}
		\caption{Mean PESR difference to best method per scenario for the skewness and kurtosis alternative and $k = 4$ datasets for all variants of the SC test \textcite{song_generalized_2021}. Selected variant: SC~(5MST,S$_{\text{A}}$).}
	\end{figure}	
	\clearpage

	\subsection{Method Rankings by Deviation}\label{app:diff.best.dev}
	\subsubsection[k = 2]{$k = 2$}
	\begin{figure}[!h]
		\centering
		\includegraphics[width=\linewidth]{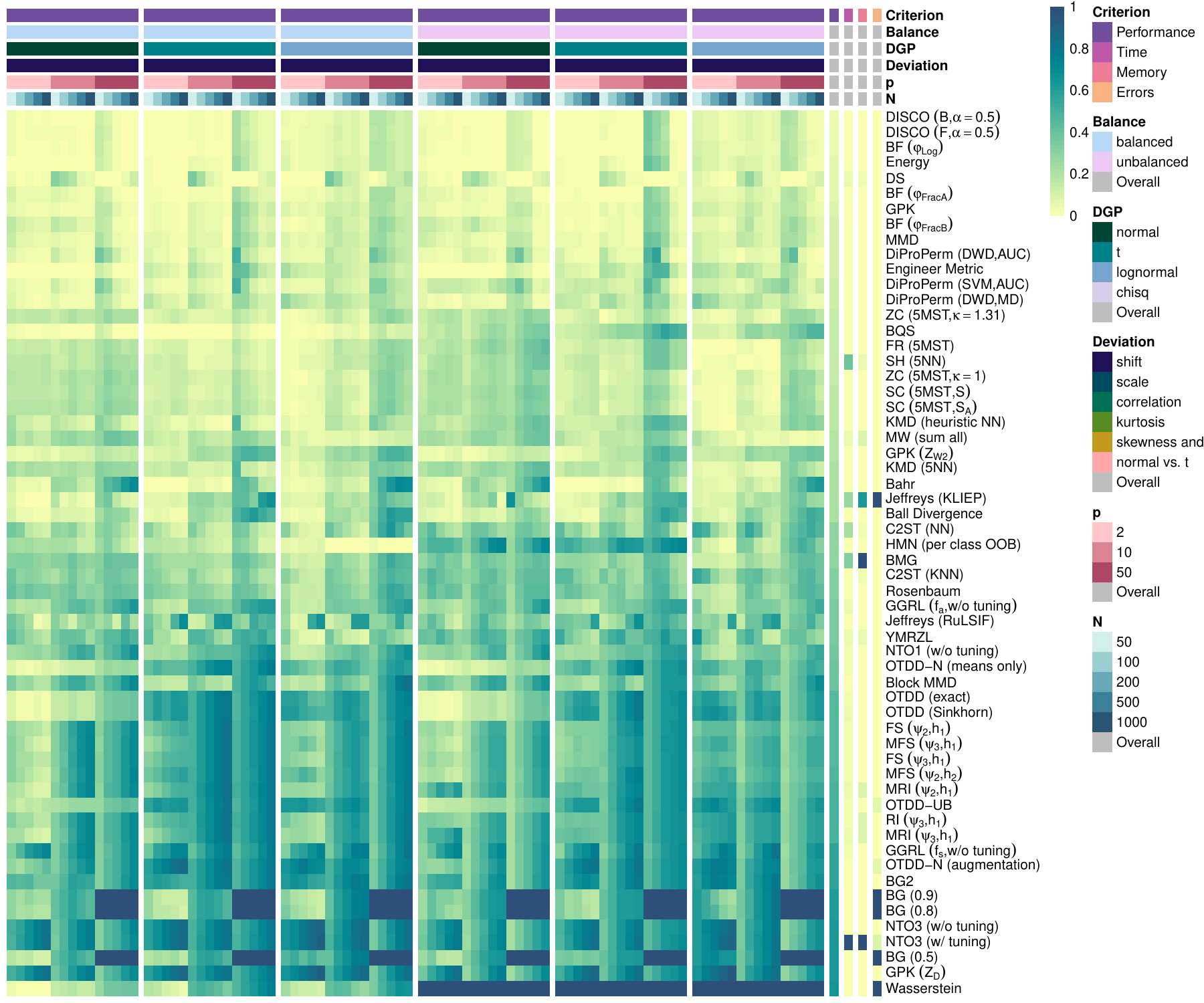}
		\caption{Mean difference to best method for shift scenarios. Methods are ordered from top to bottom by increasing mean difference over all scenarios (``Overall Performance'') and alphabetically in case of ties.}\label{fig:heat.diff.best.two.shift}
	\end{figure}
	
	\begin{figure}[!h]
		\centering
		\includegraphics[width=\linewidth]{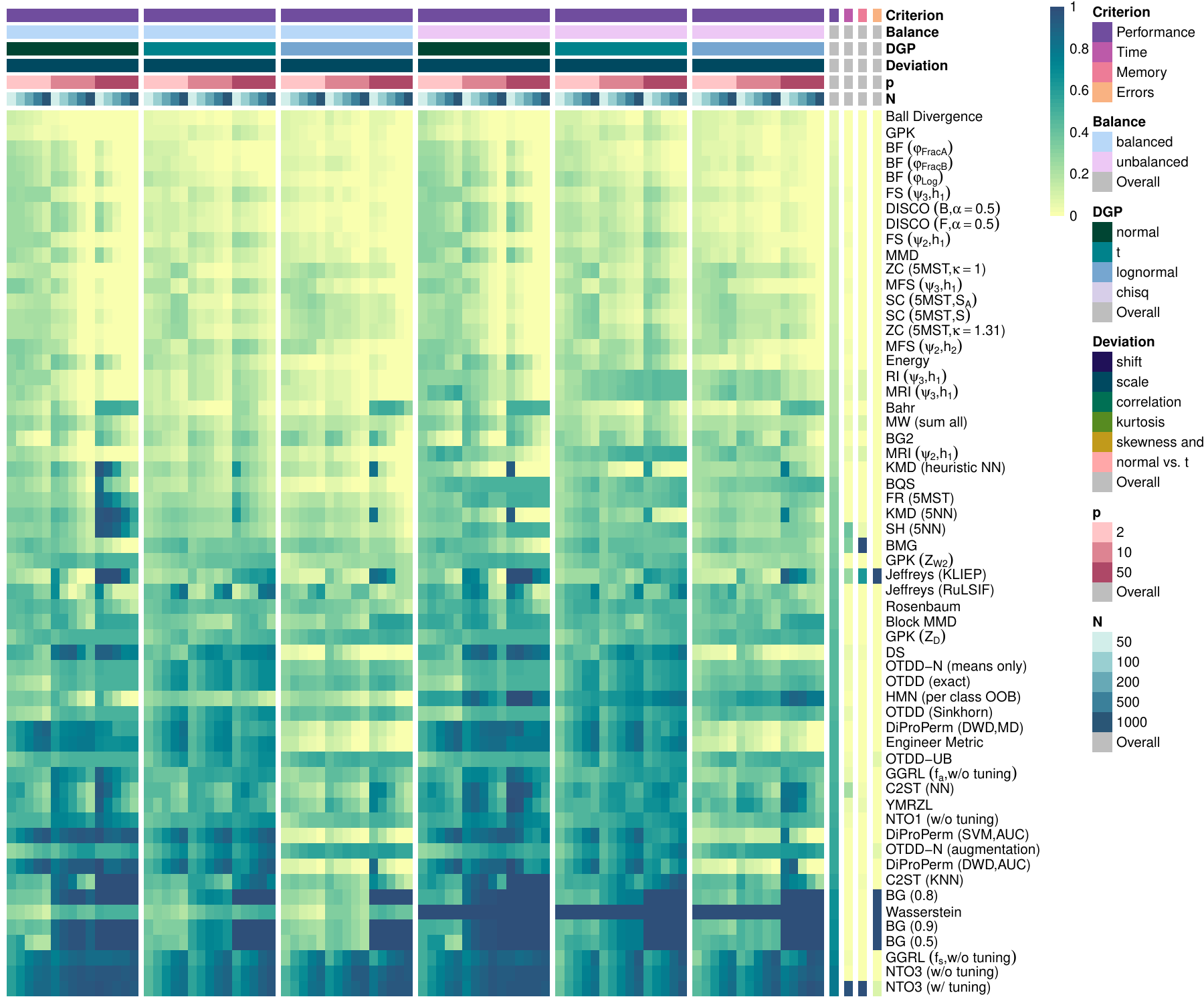}
		\caption{Mean difference to best method for scale scenarios. Methods are ordered from top to bottom by increasing mean difference over all scenarios (``Overall Performance'') and alphabetically in case of ties.}\label{fig:heat.diff.best.two.scale}
	\end{figure}
	
	\begin{figure}[!h]
		\centering
		\includegraphics[width=\linewidth]{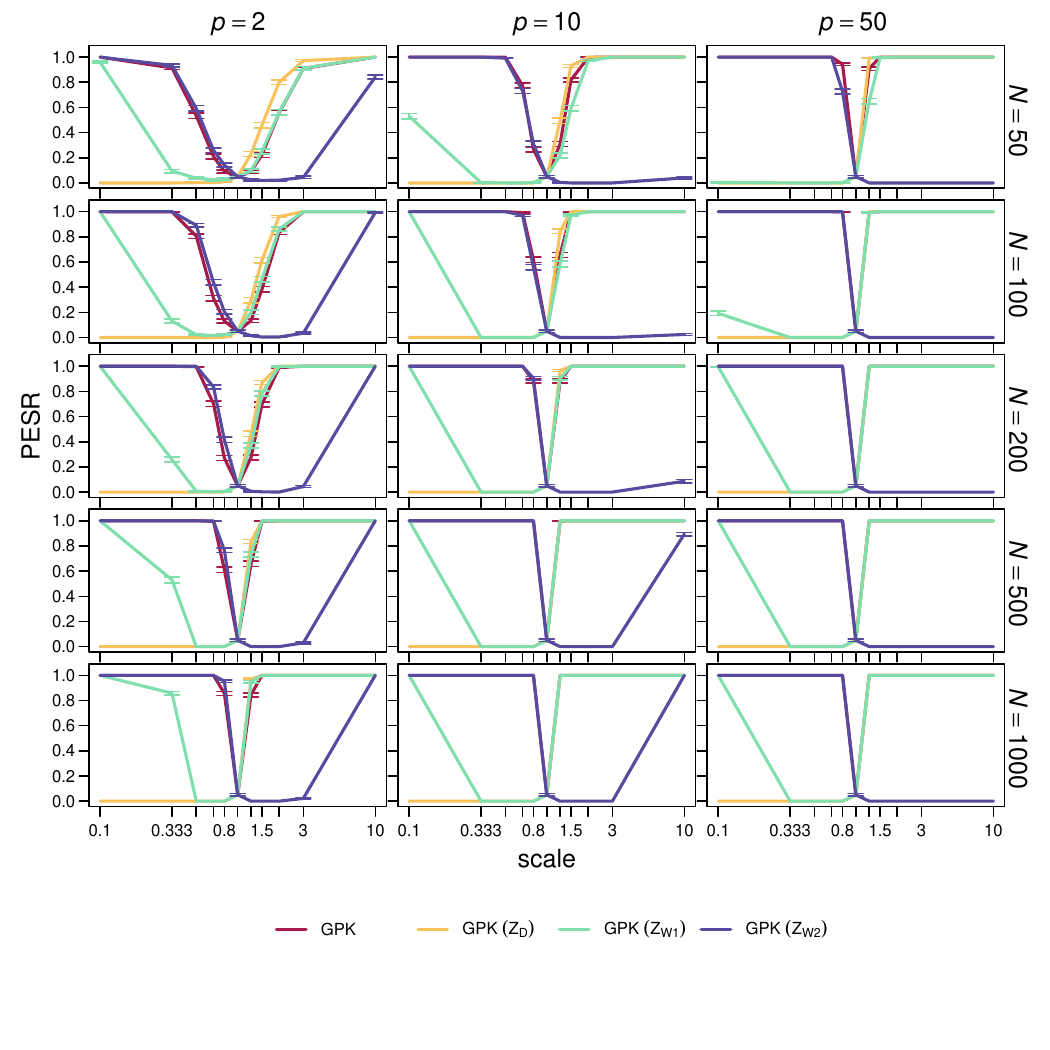}
		\caption{PESR curves for GPK variants for scale alternatives in normal datasets with equal sample sizes.}\label{fig:pesr.curve.gpk.normal.balanced.scale}
	\end{figure}
	
	\begin{figure}[!h]
		\centering
		\includegraphics[width=\linewidth]{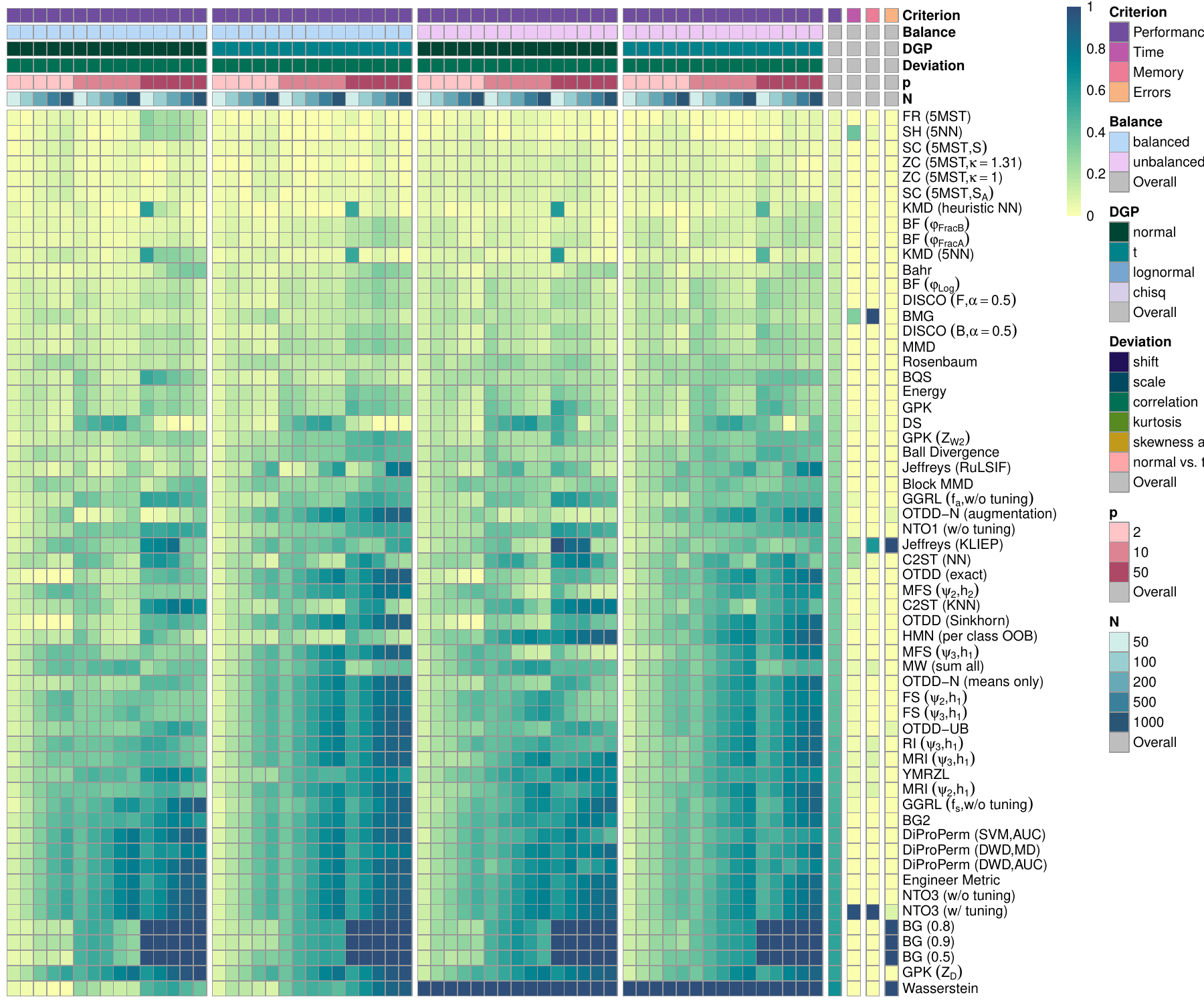}
		\caption{Mean difference to best method for correlation scenarios. Methods are ordered from top to bottom by increasing mean difference over all scenarios (``Overall Performance'') and alphabetically in case of ties.}\label{fig:heat.diff.best.two.cor}
	\end{figure}
	
	\begin{figure}[!h]
		\centering
		\includegraphics[width=\linewidth]{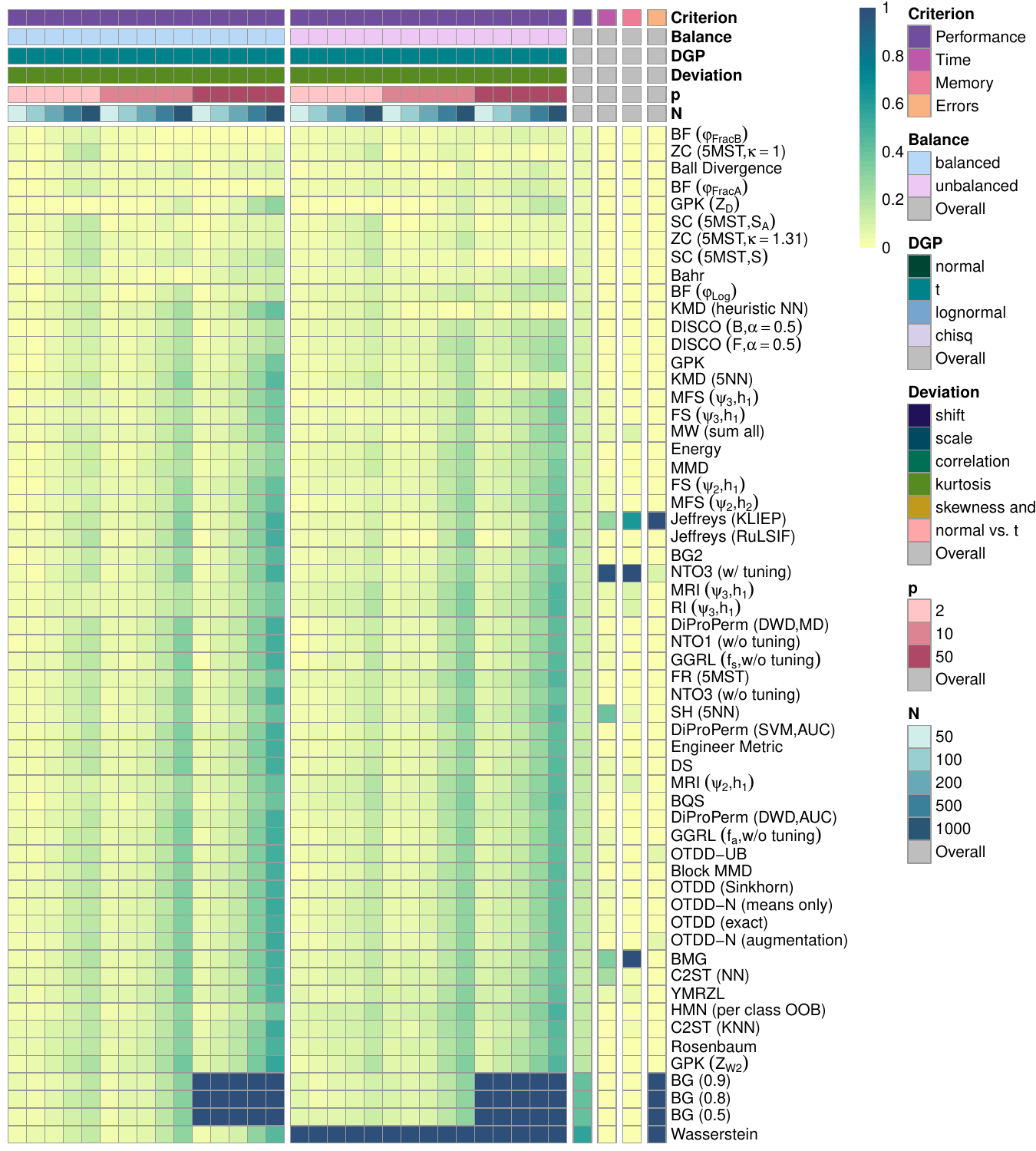}
		\caption{Mean difference to best method for kurtosis scenarios. Methods are ordered from top to bottom by increasing mean difference over all scenarios (``Overall Performance'') and alphabetically in case of ties.}\label{fig:heat.diff.best.two.kurtosis}
	\end{figure}
	
	\begin{figure}[!h]
		\centering
		\includegraphics[width=\linewidth]{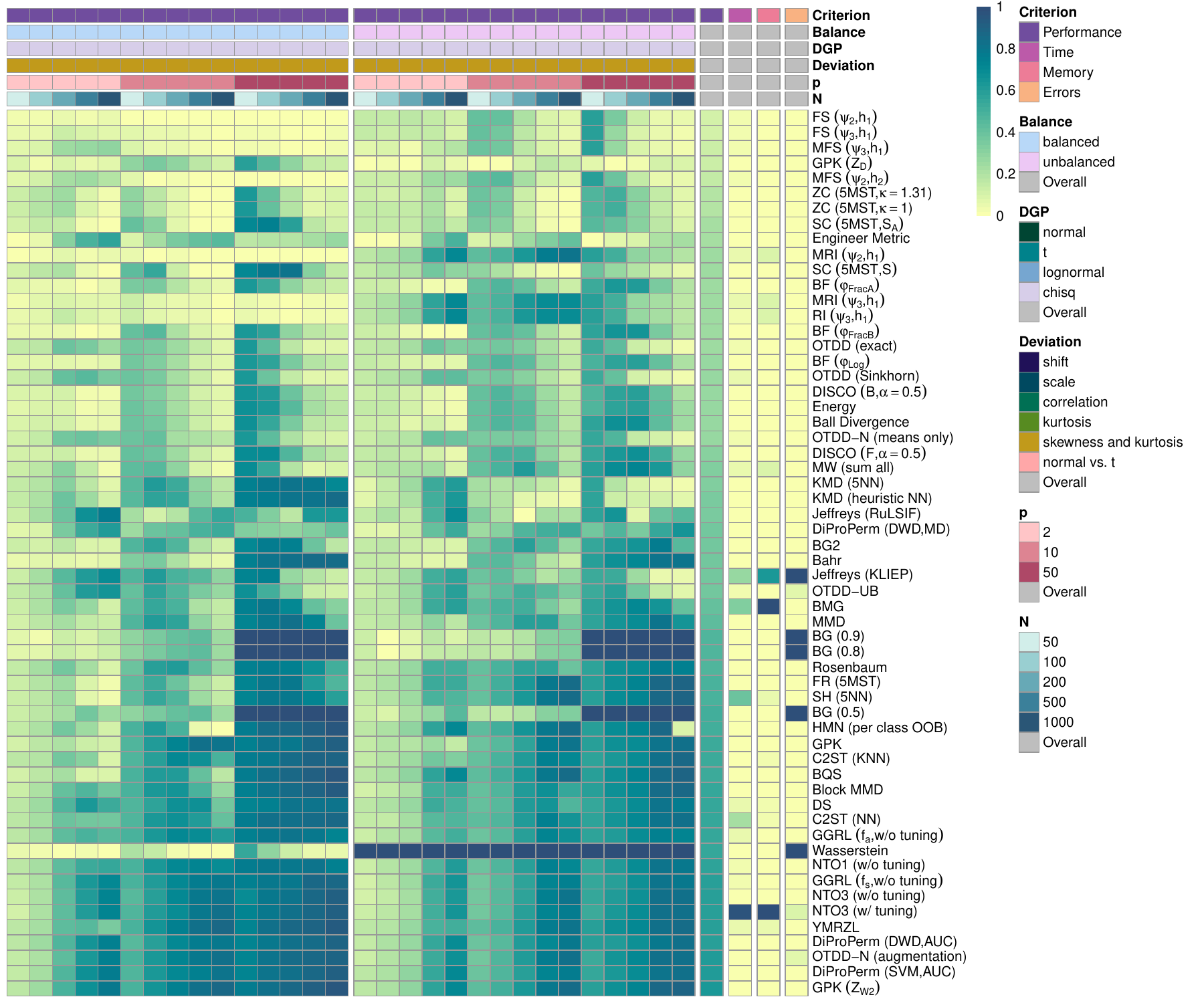}
		\caption{Mean difference to best method for skewness and kurtosis scenarios. Methods are ordered from top to bottom by increasing mean difference over all scenarios (``Overall Performance'') and alphabetically in case of ties.}\label{fig:heat.diff.best.two.skew.kurtosis}
	\end{figure}
	
	\begin{figure}[!h]
		\centering
		\includegraphics[width=\linewidth]{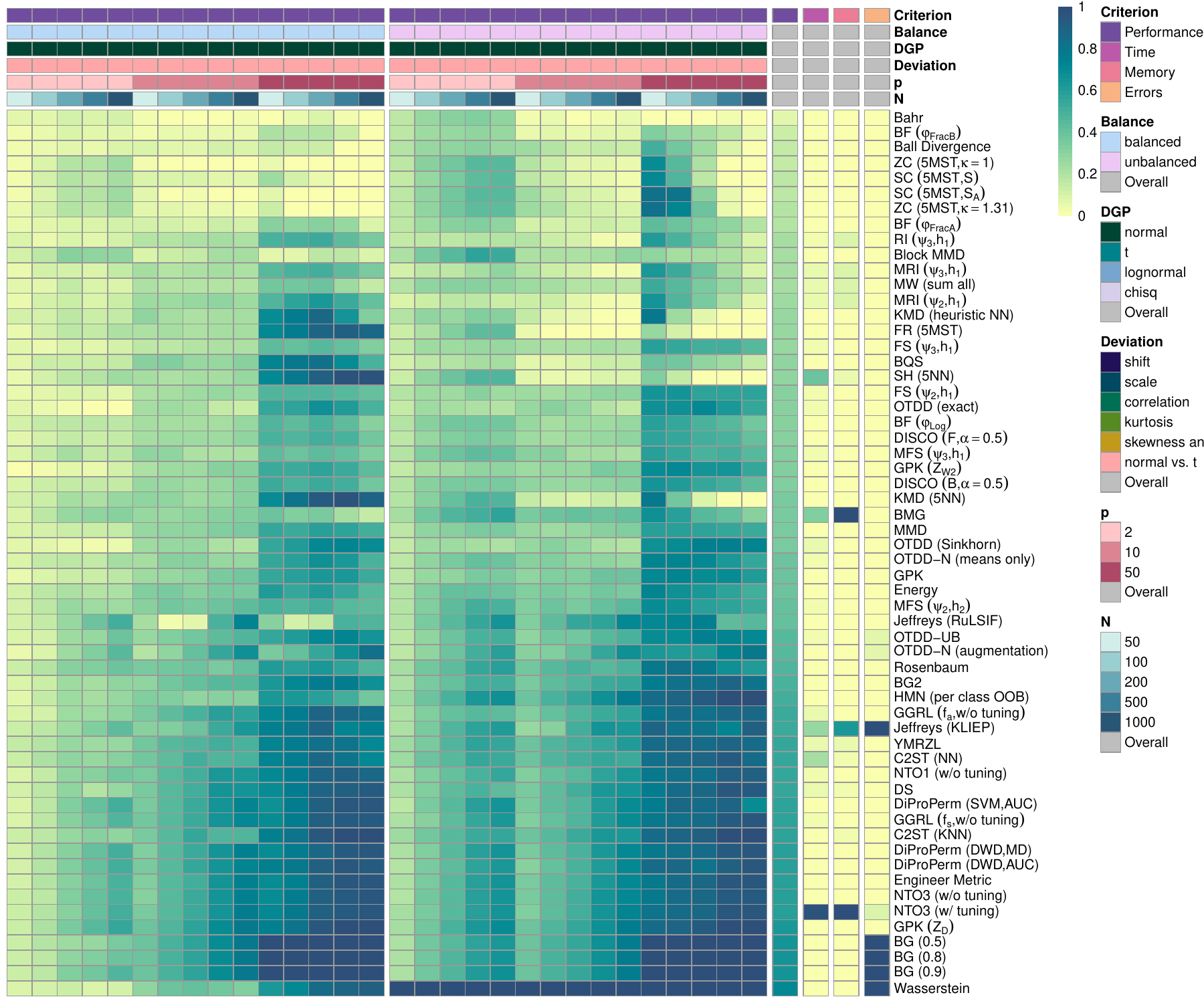}
		\caption{Mean difference to best method for normal vs.\ $t$ scenarios. Methods are ordered from top to bottom by increasing mean difference over all scenarios (``Overall Performance'') and alphabetically in case of ties.}\label{fig:heat.diff.best.two.normalt}
	\end{figure}
	
	\begin{figure}[!h]
		\centering
		\includegraphics[width=\linewidth]{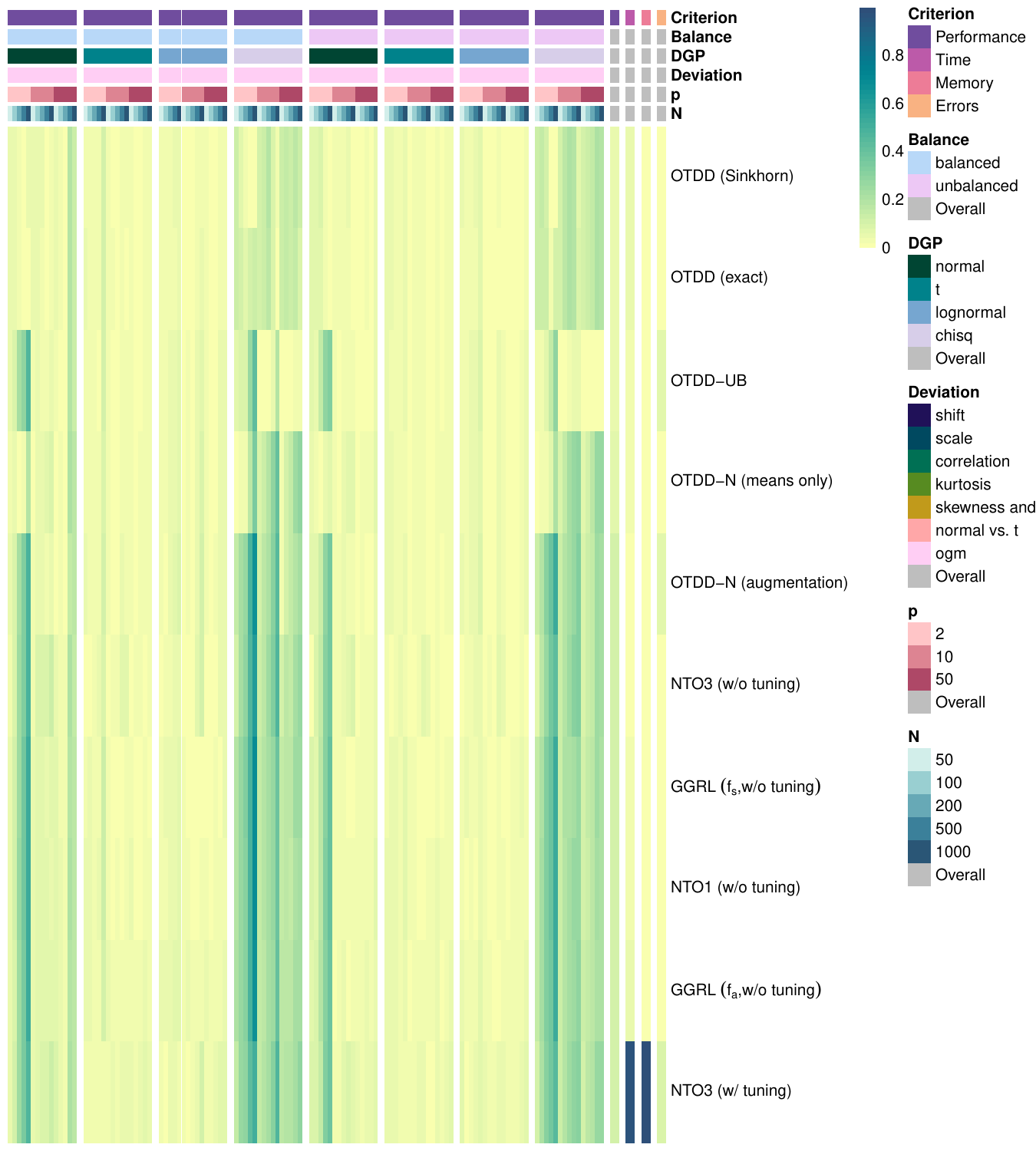}
		\caption{Mean difference to best method for OGM scenarios. Only methods that consider a target variable are shown. Methods are ordered from top to bottom by increasing mean difference over all scenarios (``Overall Performance'') and alphabetically in case of ties.}\label{fig:heat.diff.best.two.ogm}
	\end{figure}
	\clearpage
	
	\subsubsection[k = 4]{$k = 4$}

	\begin{figure}[!h]
		\centering
		\includegraphics[width=\linewidth]{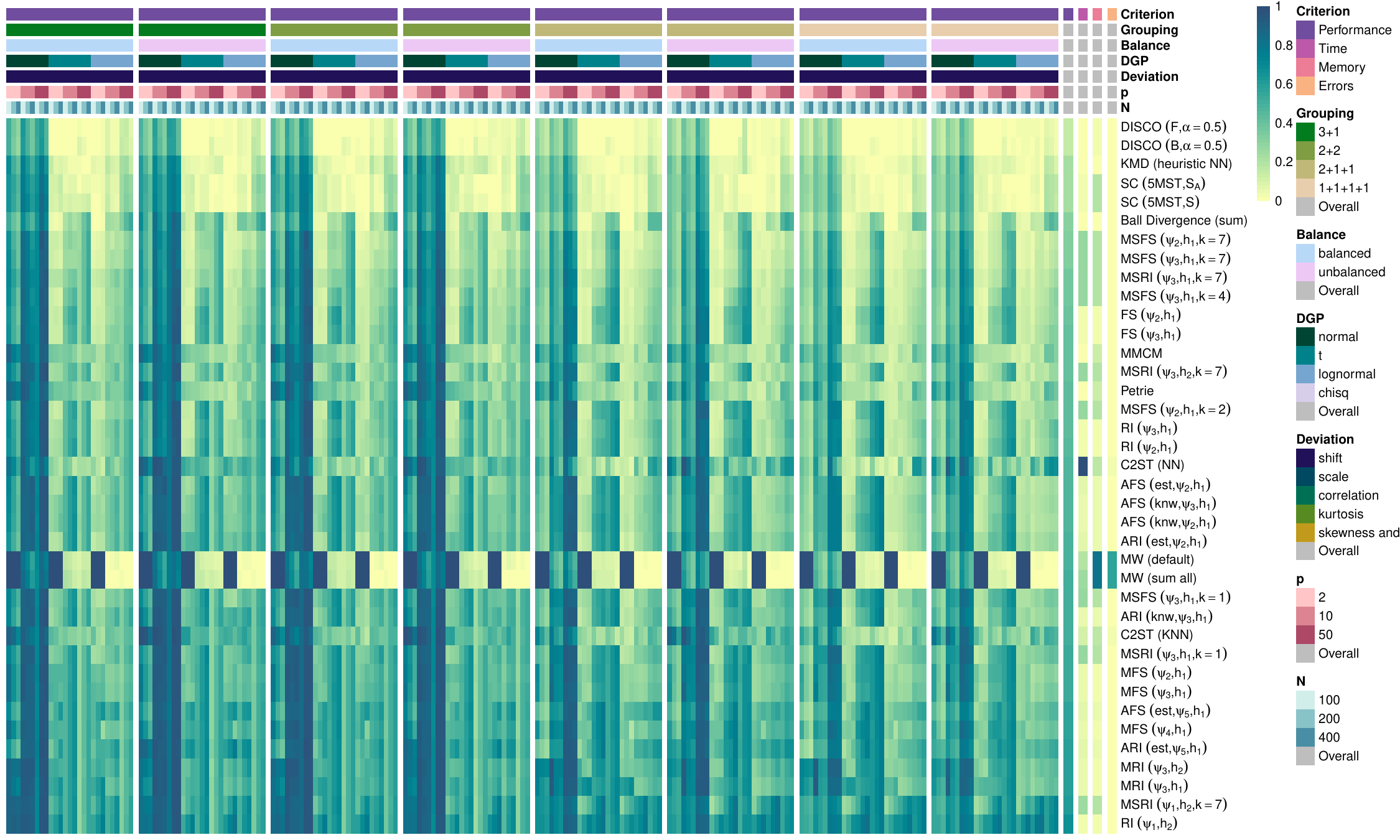}
		\caption{Mean difference to best method for shift scenarios. Methods are ordered from top to bottom by increasing mean difference over all scenarios (``Overall Performance'') and alphabetically in case of ties.}\label{fig:heat.diff.best.multi.shift}
	\end{figure}
	
	\begin{figure}[!h]
		\centering
		\includegraphics[width=\linewidth]{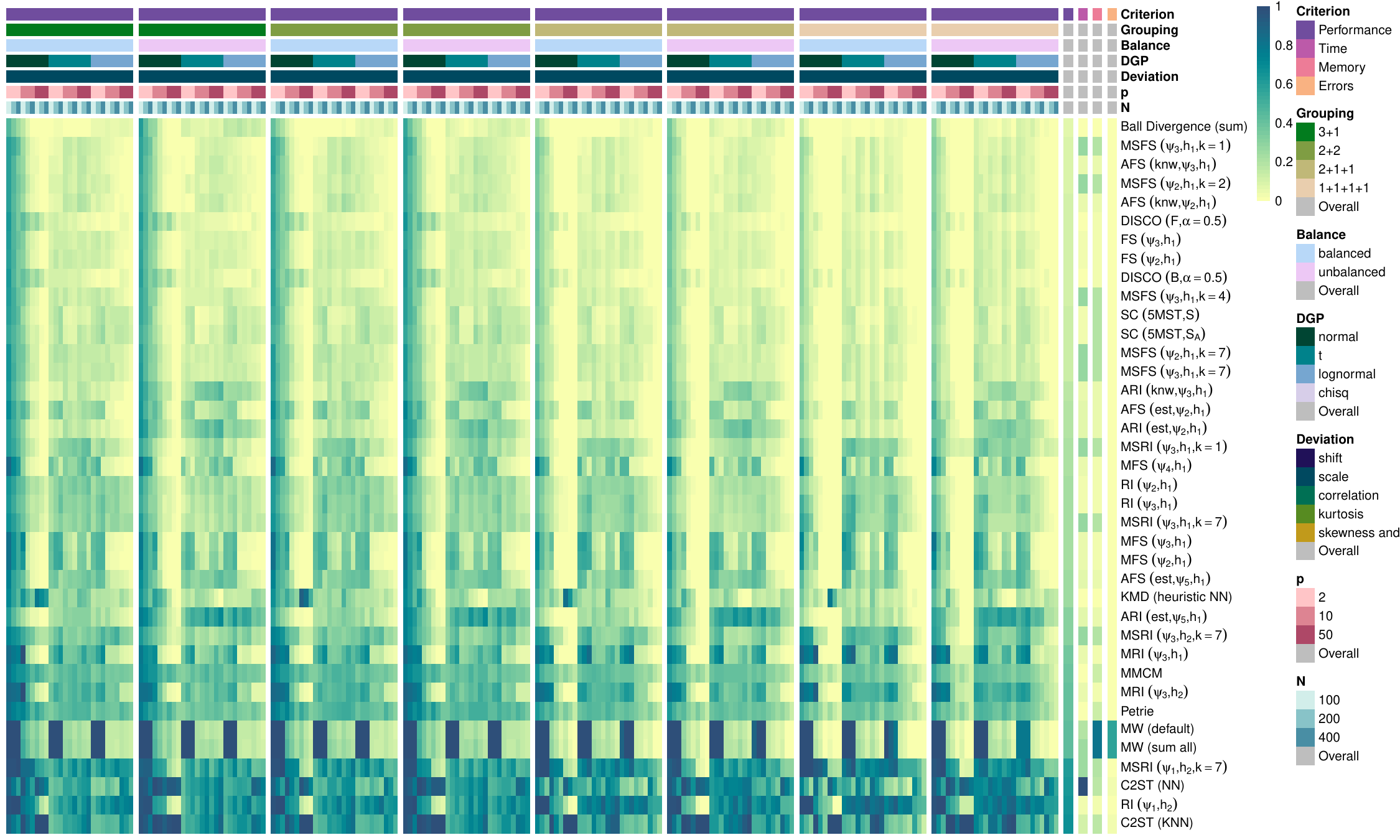}
		\caption{Mean difference to best method for scale scenarios. Methods are ordered from top to bottom by increasing mean difference over all scenarios (``Overall Performance'') and alphabetically in case of ties.}\label{fig:heat.diff.best.multi.scale}
	\end{figure}

	\begin{figure}[!h]
		\centering
		\includegraphics[width=\linewidth]{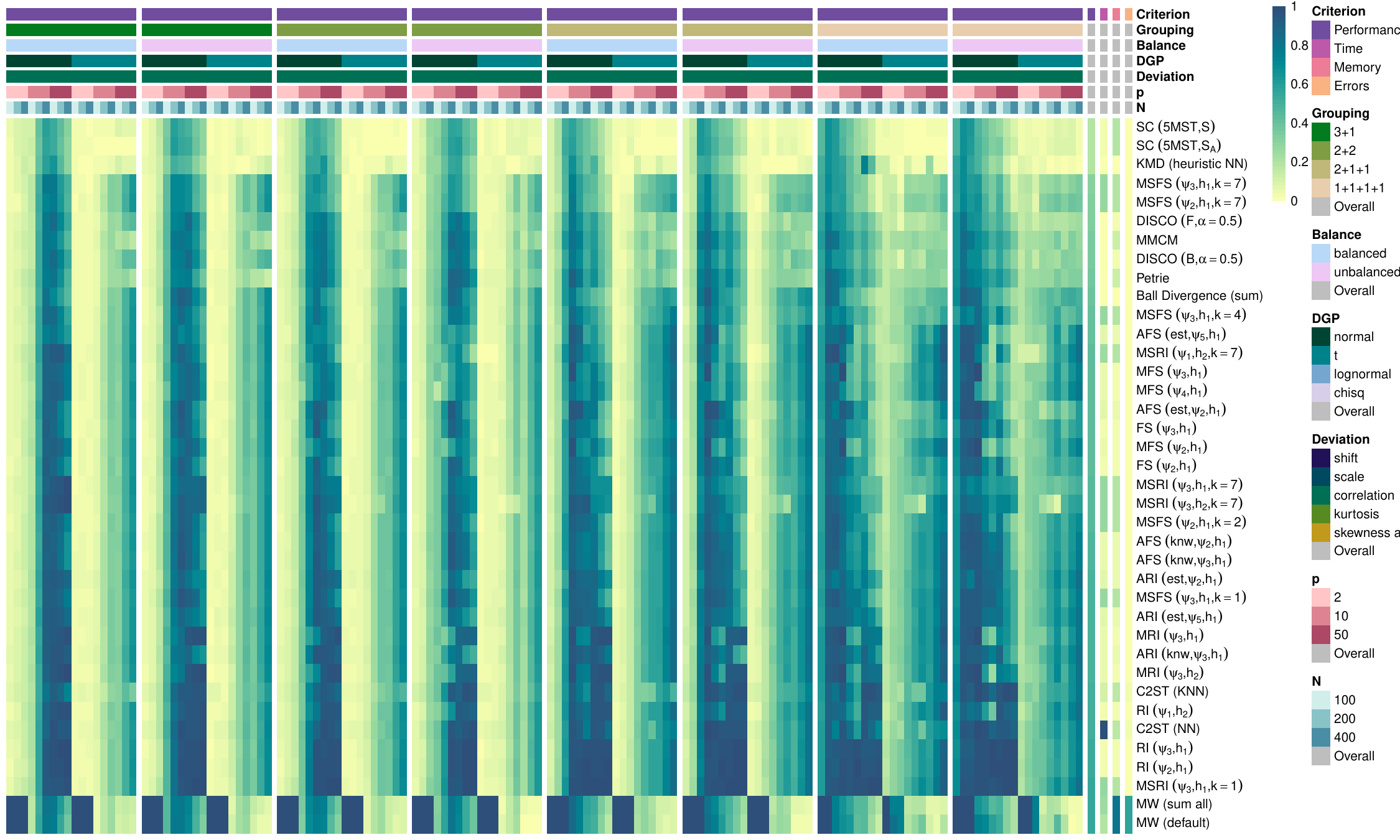}
		\caption{Mean difference to best method for correlation scenarios. Methods are ordered from top to bottom by increasing mean difference over all scenarios (``Overall Performance'') and alphabetically in case of ties.}\label{fig:heat.diff.best.multi.cor}
	\end{figure}
	
	\begin{figure}[!h]
		\centering
		\includegraphics[width=\linewidth]{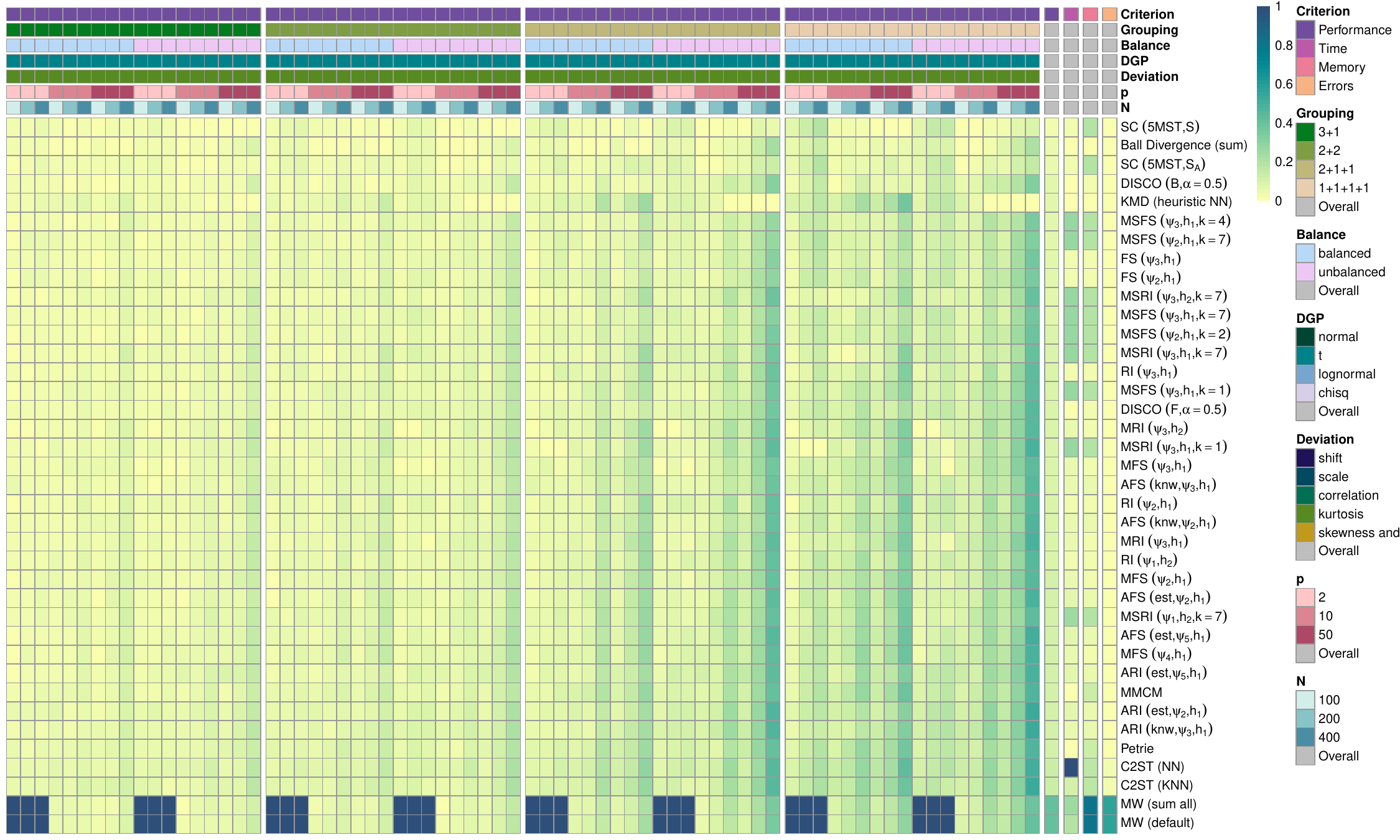}
		\caption{Mean difference to best method for kurtosis scenarios. Methods are ordered from top to bottom by increasing mean difference over all scenarios (``Overall Performance'') and alphabetically in case of ties.}\label{fig:heat.diff.best.multi.kurtosis}
	\end{figure}
	
	\begin{figure}[!h]
		\centering
		\includegraphics[width=\linewidth]{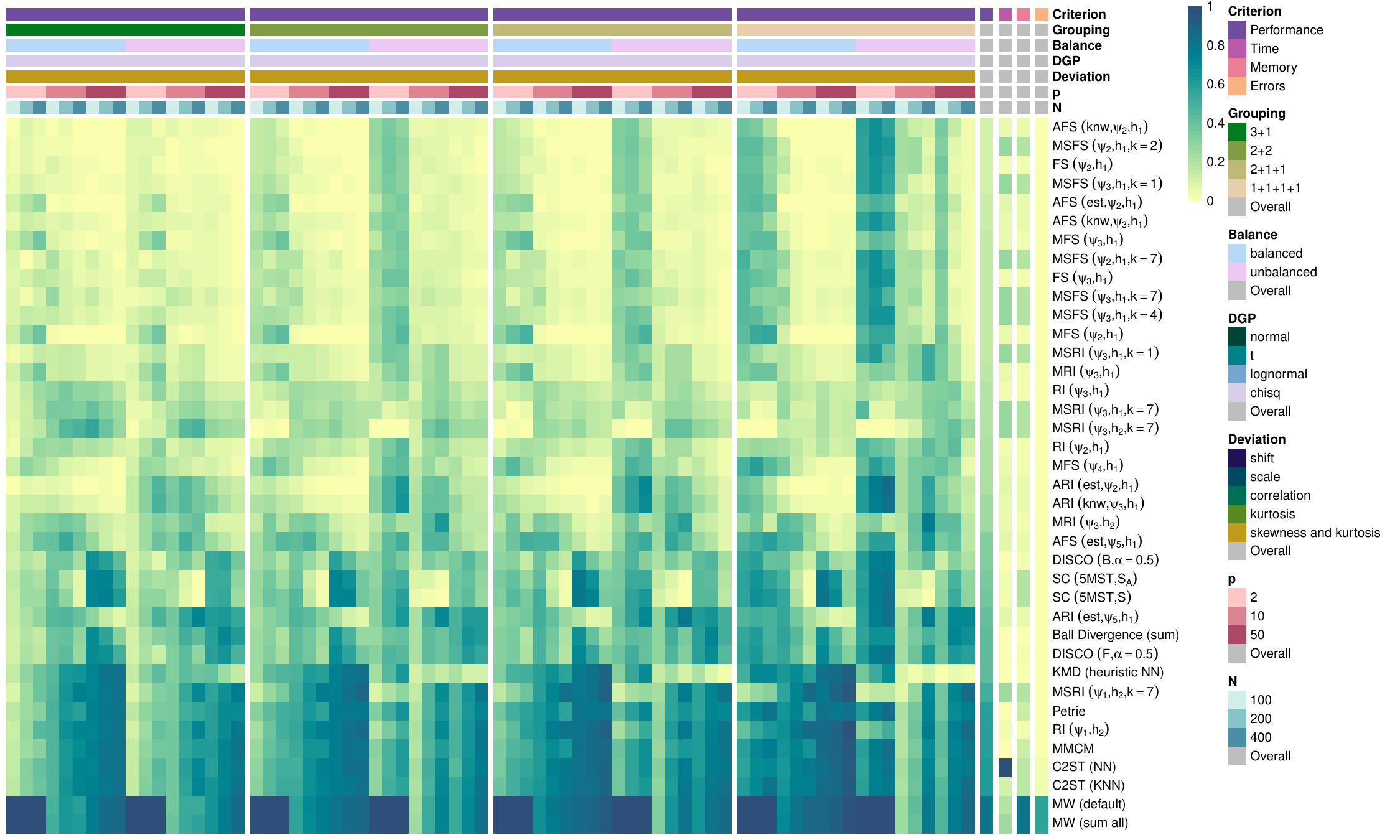}
		\caption{Mean difference to best method for skewness and kurtosis scenarios. Methods are ordered from top to bottom by increasing mean difference over all scenarios (``Overall Performance'') and alphabetically in case of ties.}\label{fig:heat.diff.best.multi.skew.kurtosis}
	\end{figure}
	\clearpage
		
	\subsection{Ranking Comparison}\label{app:comp.rank}
	\begin{figure}[!h]
		\centering
		\includegraphics[width=\linewidth]{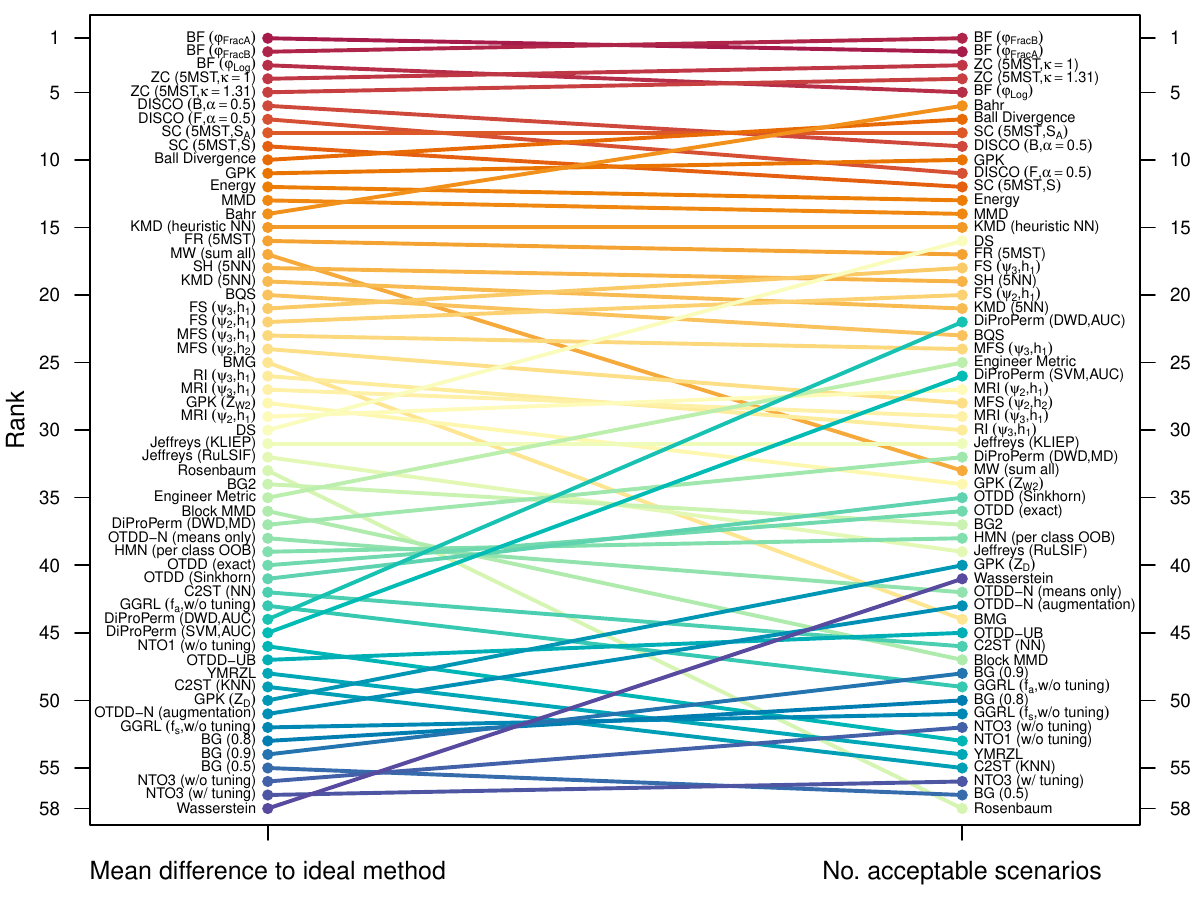}
		\caption{Rankings according to the overall mean difference in PESR to the ideal method and the proportion of scenarios for which the respective method is deemed acceptable for the two-sample case. Lower ranks correspond to better performance, i.e.\ lower mean difference in PESR or higher proportion of acceptable scenarios, respectively.}\label{fig:comp.rank.two}
	\end{figure}
	
	\begin{figure}[!h]
		\centering
		\includegraphics[width=\linewidth]{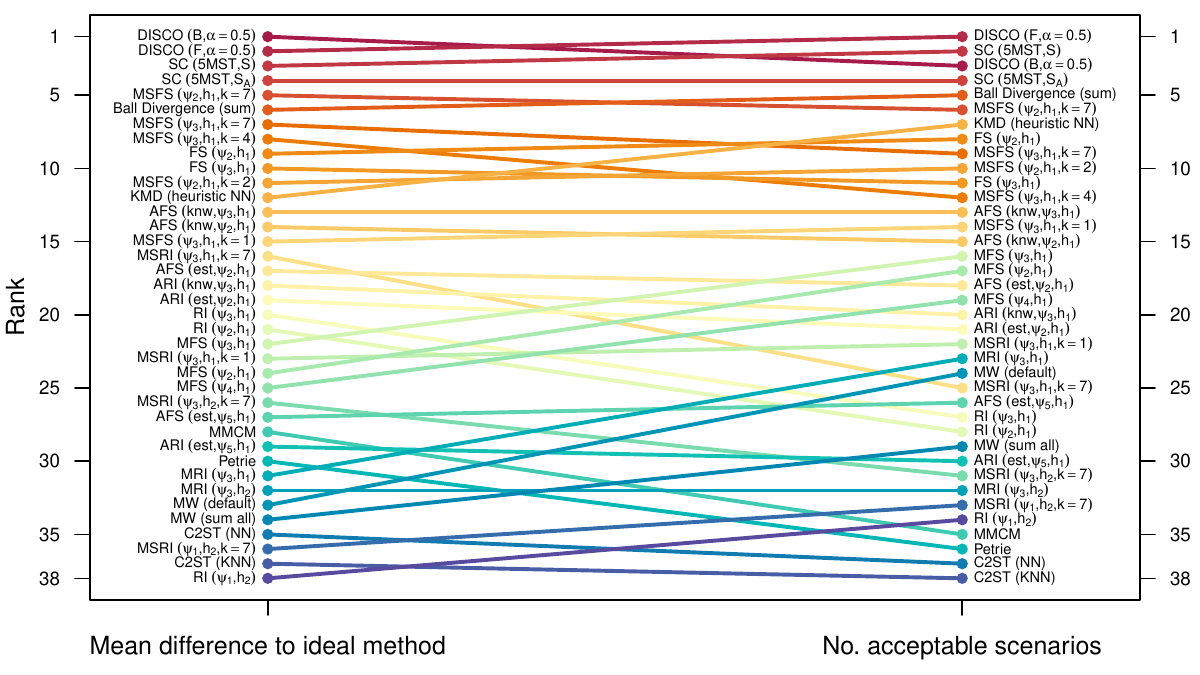}
		\caption{Rankings according to the overall mean difference in PESR to the ideal method and the proportion of scenarios for which the respective method is deemed acceptable for the multi-sample case. Lower ranks correspond to better performance, i.e.\ lower mean difference in PESR or higher proportion of acceptable scenarios, respectively.}\label{fig:comp.rank.multi}
	\end{figure}
	\clearpage
	
	\subsection{Decision Trees for Method Choice}\label{app:dec.trees}
	\subsubsection[k = 2]{$k = 2$}
	\paragraph{Overall}\mbox{}
	
	\begin{figure}[!h]
		\centering
		\includegraphics[width=\linewidth]{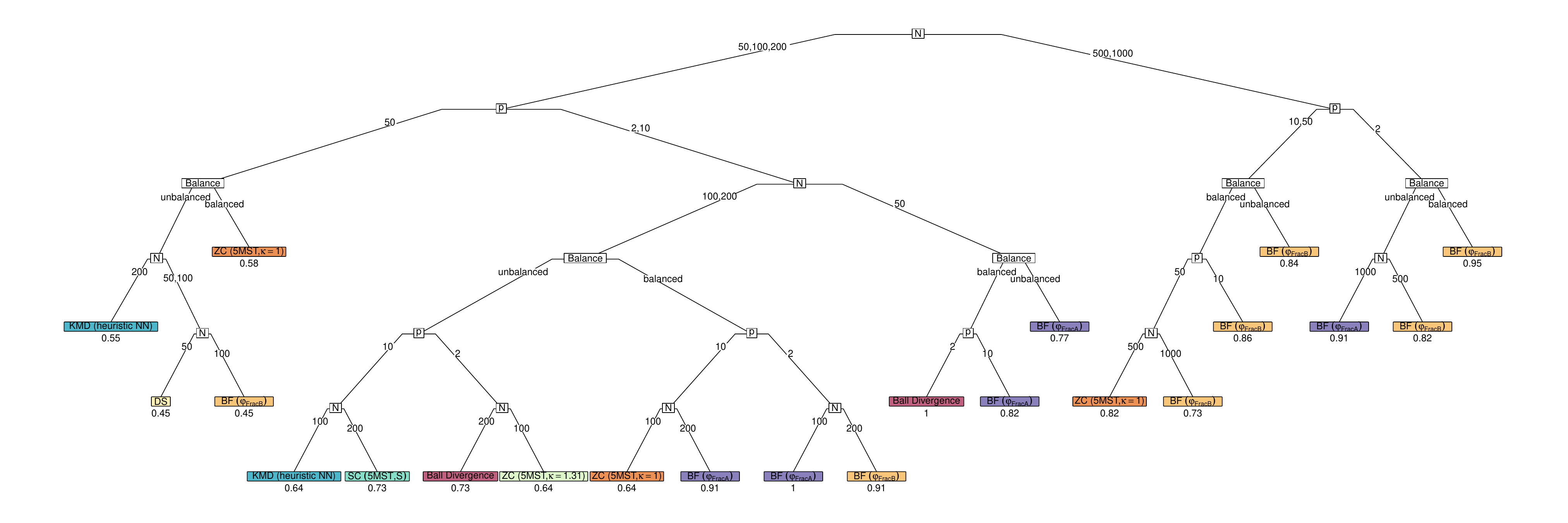}
		\caption{Decision rules for choosing a method that covers most scenarios for two numeric datasets given only the information about the dataset dimensions. Numbers underneath the leaf nodes give the proportion of scenarios in the respective leaf node that are covered by the chosen method.}\label{fig:tree.num.two}
	\end{figure}
	
	\paragraph{By DGP}\mbox{}
	
	\begin{figure}[!h]
		\centering
		\includegraphics[width=\linewidth]{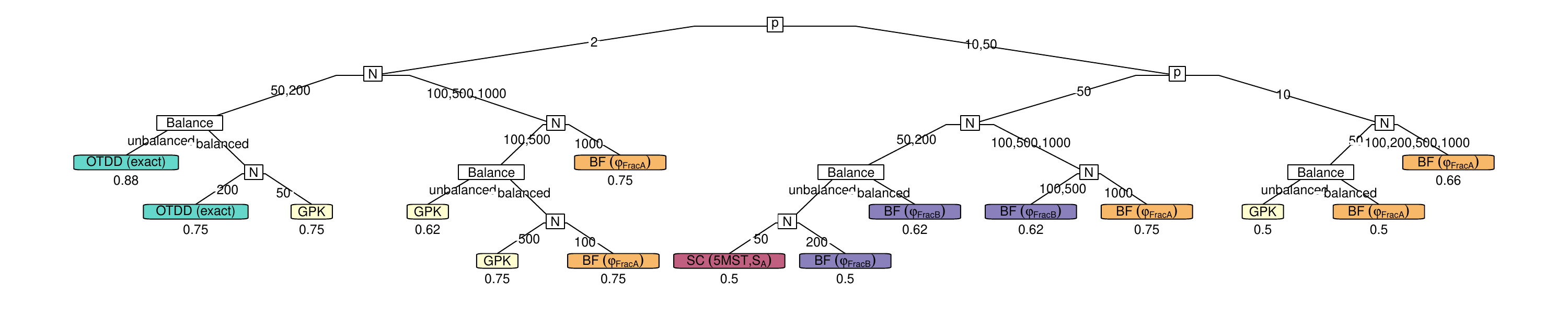}
		\caption{Decision rules for choosing a method that covers most scenarios for two normally distributed datasets given only the information about the dataset dimensions. Numbers underneath the leaf nodes give the proportion of scenarios in the respective leaf node that are covered by the chosen method. }\label{fig:tree.normal.two}
	\end{figure}
	
	\begin{figure}[!h]
		\centering
		\includegraphics[width=\linewidth]{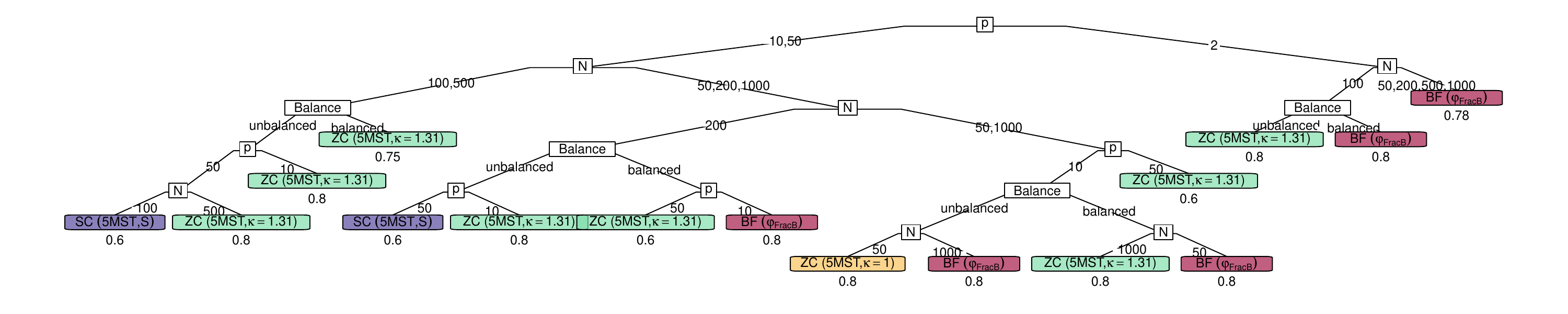}
		\caption{Decision rules for choosing a method that covers most scenarios for two $t$-distributed datasets given only the information about the dataset dimensions. Numbers underneath the leaf nodes give the proportion of scenarios in the respective leaf node that are covered by the chosen method.}\label{fig:tree.t.two}
	\end{figure}
	
	\begin{figure}[!h]
		\centering
		\includegraphics[width=\linewidth]{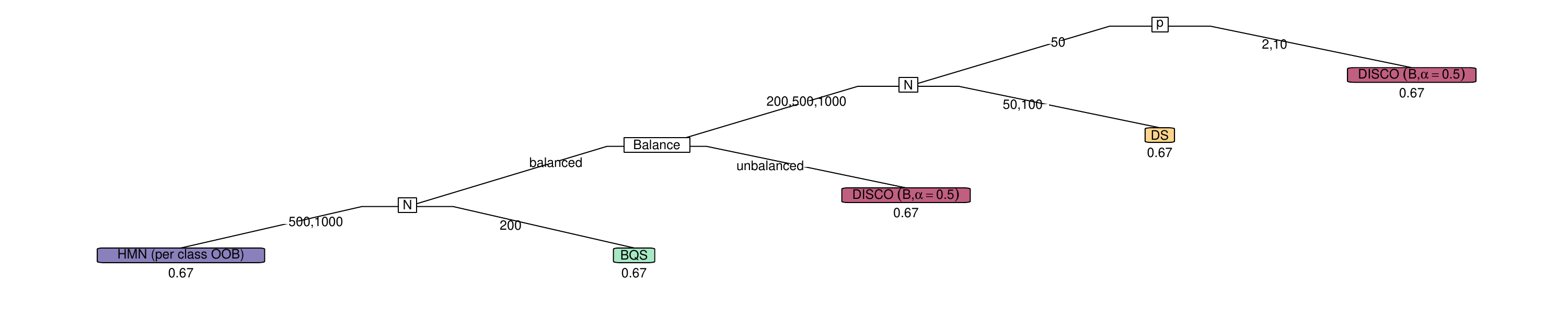}
		\caption{Decision rules for choosing a method that covers most scenarios for two log-normally distributed datasets given only the information about the dataset dimensions. Numbers underneath the leaf nodes give the proportion of scenarios in the respective leaf node that are covered by the chosen method. }\label{fig:tree.log-normal.two}
	\end{figure}
	
	\begin{figure}[!h]
		\centering
		\includegraphics[width=\linewidth]{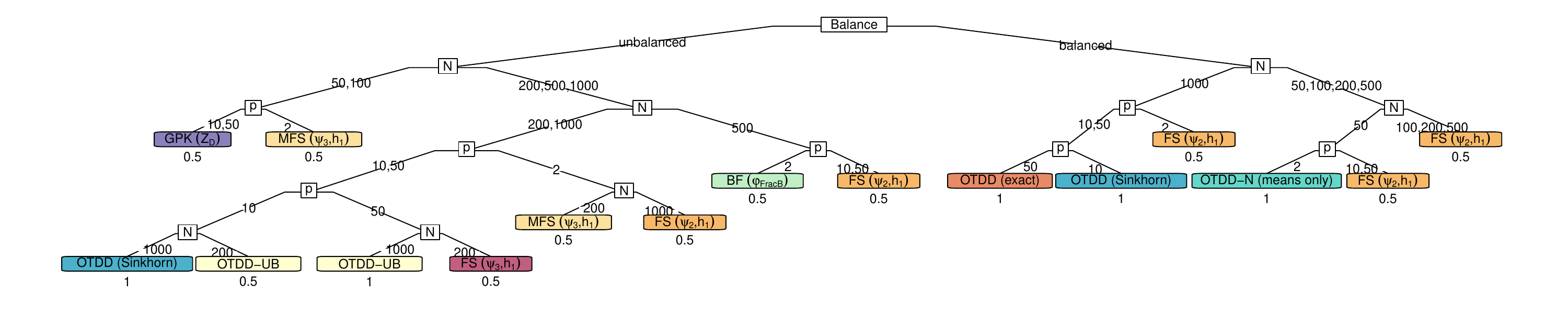}
		\caption{Decision rules for choosing a method that covers most scenarios for two $\chi^2$-distributed datasets given only the information about the dataset dimensions. Numbers underneath the leaf nodes give the proportion of scenarios in the respective leaf node that are covered by the chosen method. }\label{fig:tree.chisq.two}
	\end{figure}
	\clearpage

	\paragraph{By DGP and Deviation}\mbox{}

	\subparagraph{Normal distribution}\mbox{}

	\begin{figure}[!h]
		\centering
		\includegraphics[width=\linewidth]{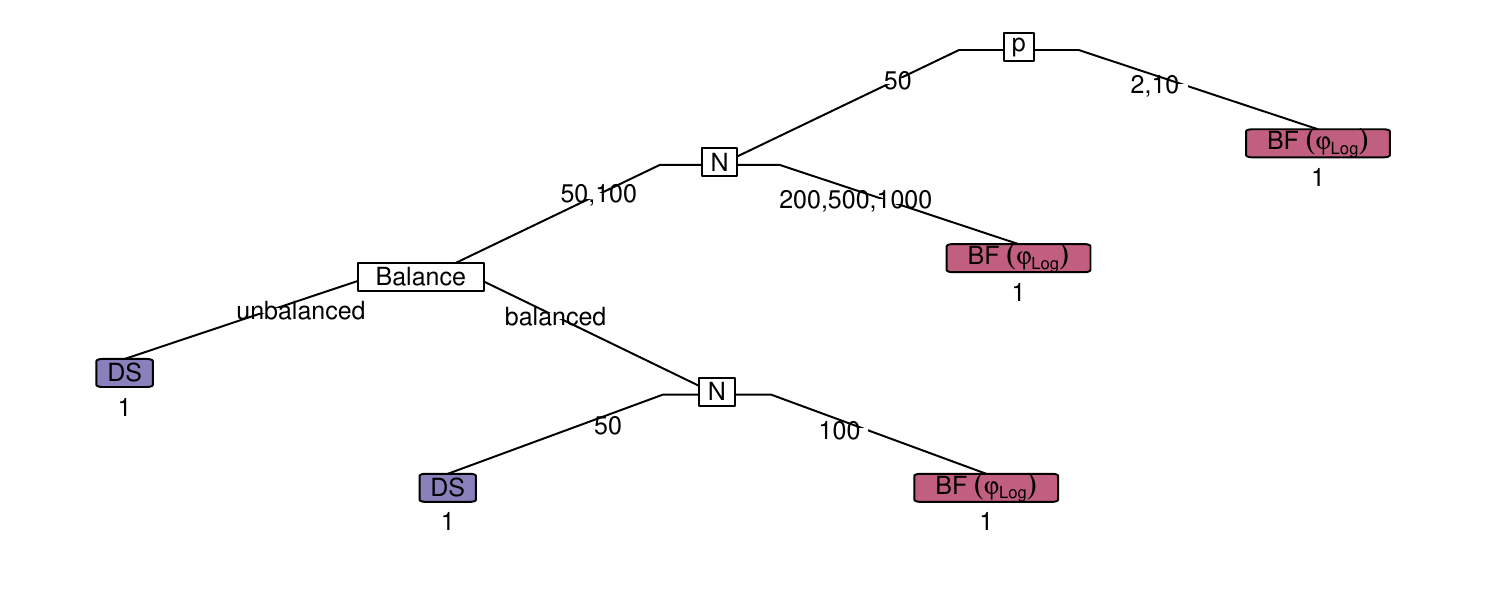}
		\caption{Decision rules for choosing a method that covers most scenarios for two normally distributed datasets and shift alternatives given only the information about the dataset dimensions. Numbers underneath the leaf nodes give the proportion of scenarios in the respective leaf node that are covered by the chosen method. }\label{fig:tree.normal.shift.two}
	\end{figure}
	

	
	For scale alternatives, for two normally distributed datasets, the Ball divergence covers most scenarios, regardless of $N$, $p$, and the sample size balance.
	
	\begin{figure}[!h]
		\centering
		\includegraphics[width=\linewidth]{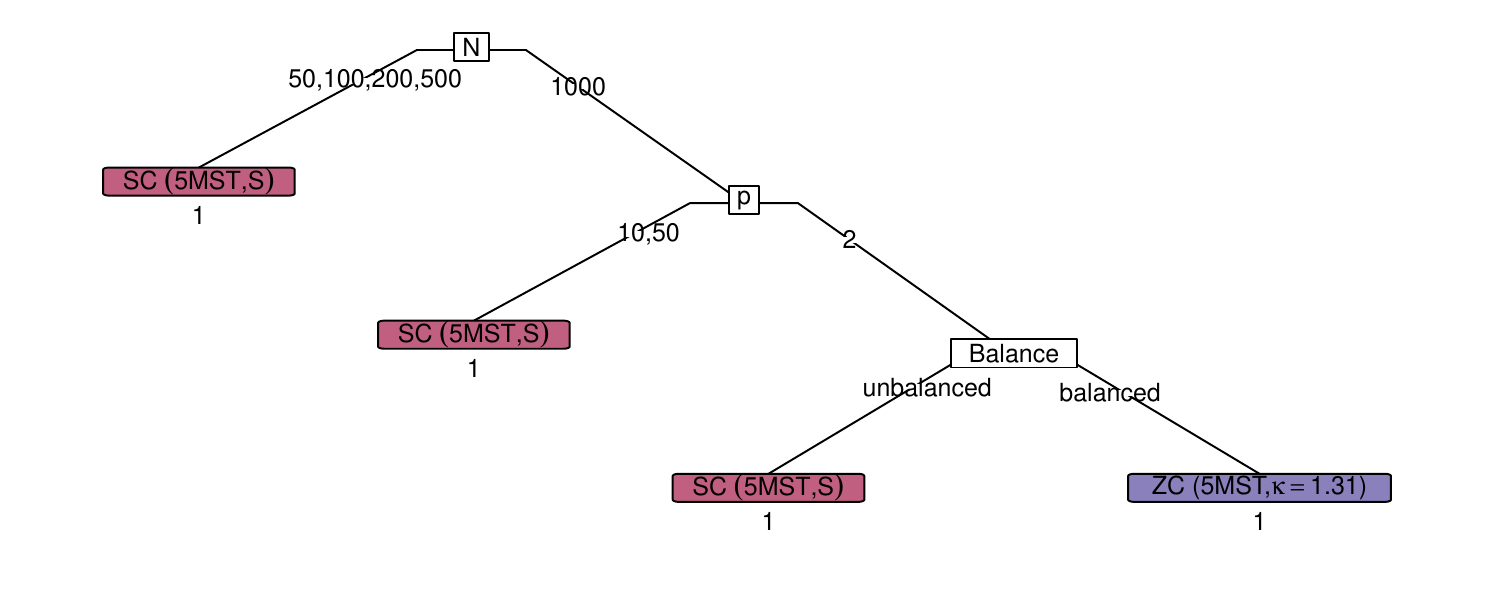}
		\caption{Decision rules for choosing a method that covers most scenarios for two normally distributed datasets and correlation alternatives given only the information about the dataset dimensions. Numbers underneath the leaf nodes give the proportion of scenarios in the respective leaf node that are covered by the chosen method. }\label{fig:tree.normal.cor.two}
	\end{figure}
	
	
	For OGM alternatives, for two normally distributed datasets, the OTDD~(exact) covers most scenarios, regardless of $N$, $p$, and the sample size balance.

	
	\subparagraph[t distribution]{$t$ distribution}\mbox{}
	
	\begin{figure}[!h]
		\centering
		\includegraphics[width=\linewidth]{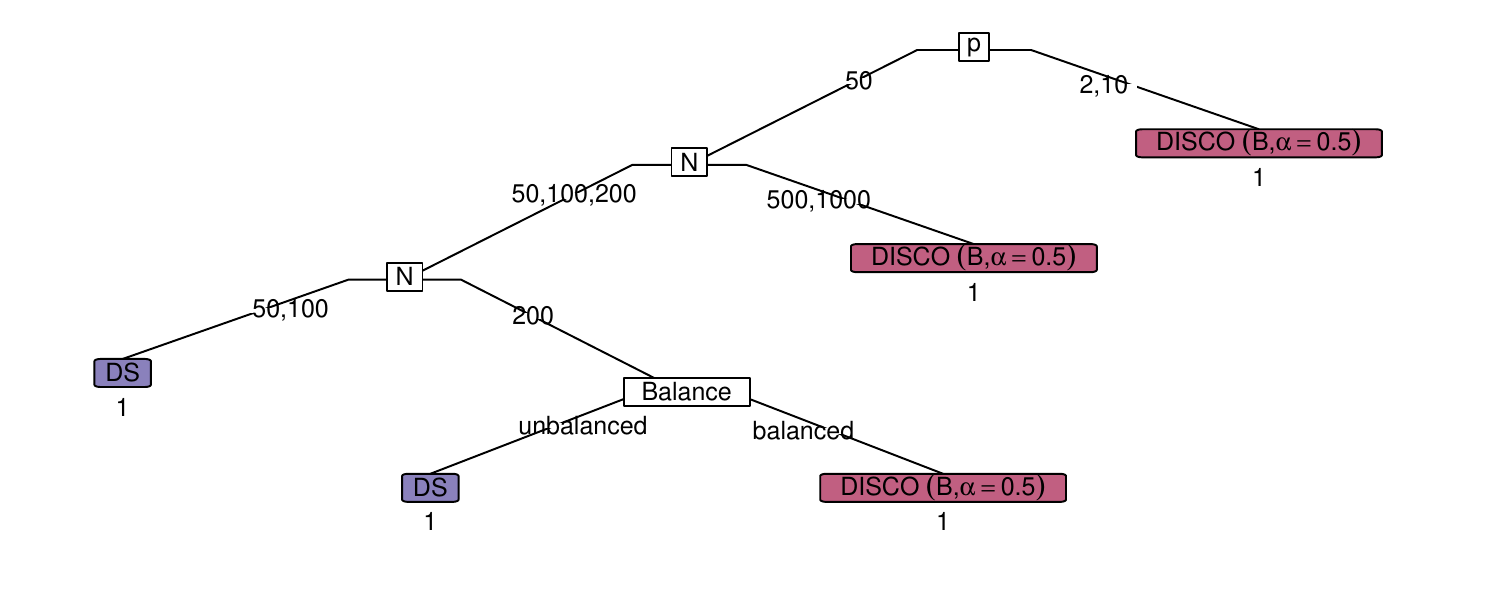}
		\caption{Decision rules for choosing a method that covers most scenarios for two $t$-distributed datasets and shift alternatives given only the information about the dataset dimensions. Numbers underneath the leaf nodes give the proportion of scenarios in the respective leaf node that are covered by the chosen method. }\label{fig:tree.t.shift.two}
	\end{figure}
	

	For scale alternatives, for two $t$-distributed datasets, the Ball divergence covers most scenarios, regardless of $N$, $p$, and the sample size balance.

%

	For correlation alternatives, for two $t$-distributed datasets, the FR~(5MST) covers most scenarios, regardless of $N$, $p$, and the sample size balance.

%

	For kurtosis alternatives, for two $t$-distributed datasets, the BF~($\varphi_{\text{FracB}}$) covers most scenarios, regardless of $N$, $p$ and the sample size balance.

%

	For OGM alternatives, for two $t$-distributed datasets, the GGRL~($f_a$,w/o tuning) covers most scenarios, regardless of $N$, $p$, and the sample size balance.
	\clearpage
	
	\subparagraph{Log-normal distribution}\mbox{}
	
	\begin{figure}[!h]
		\centering
		\includegraphics[width=\linewidth]{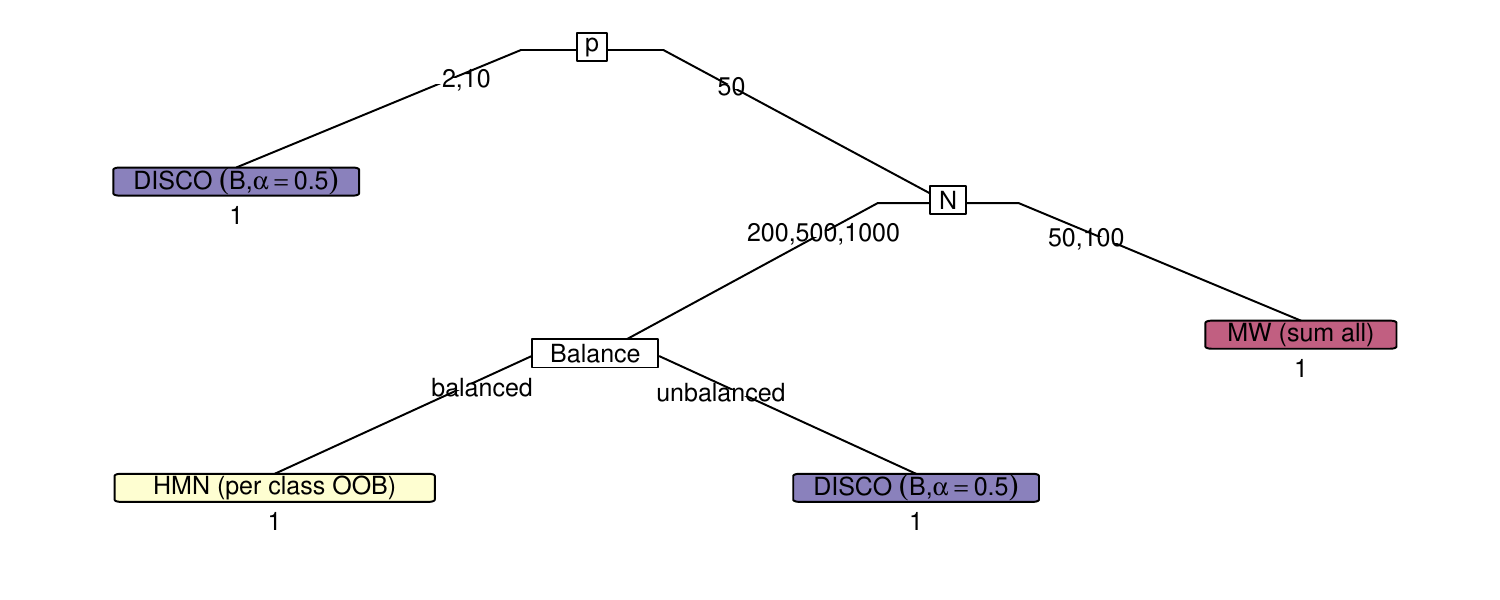}
		\caption{Decision rules for choosing a method that covers most scenarios for two log-normally distributed datasets and shift alternatives given only the information about the dataset dimensions. Numbers underneath the leaf nodes give the proportion of scenarios in the respective leaf node that are covered by the chosen method. }\label{fig:tree.log-normal.shift.two}
	\end{figure}
	
	
	
	For scale alternatives, for two log-normally distributed datasets, the Ball divergence covers most scenarios, regardless of $N$, $p$, and the sample size balance.
	
%


	For OGM alternatives, for two log-normally distributed datasets, the GGRL~($f_a$, w/o tuning) covers most scenarios, regardless of $N$, $p$, and the sample size balance.
	\clearpage	
	
	\subparagraph[Chi-squared distribution]{$\chi^2$ distribution}\mbox{}
	
	\begin{figure}[!h]
		\centering
		\includegraphics[width=\linewidth]{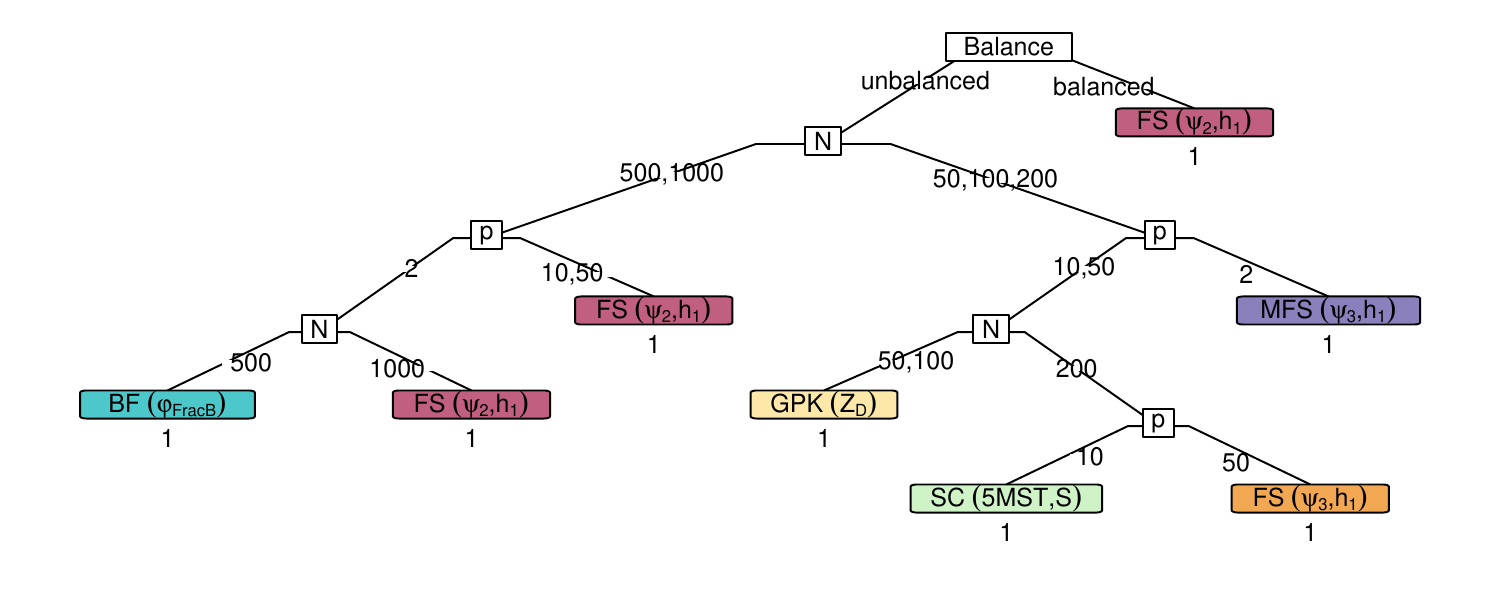}
		\caption{Decision rules for choosing a method that covers most scenarios for two $\chi^2$-distributed datasets and skewness and kurtosis alternatives given only the information about the dataset dimensions. Numbers underneath the leaf nodes give the proportion of scenarios in the respective leaf node that are covered by the chosen method. }\label{fig:tree.chisq.skew.kurtosis.two}
	\end{figure}
	
	\begin{figure}[!h]
		\centering
		\includegraphics[width=\linewidth]{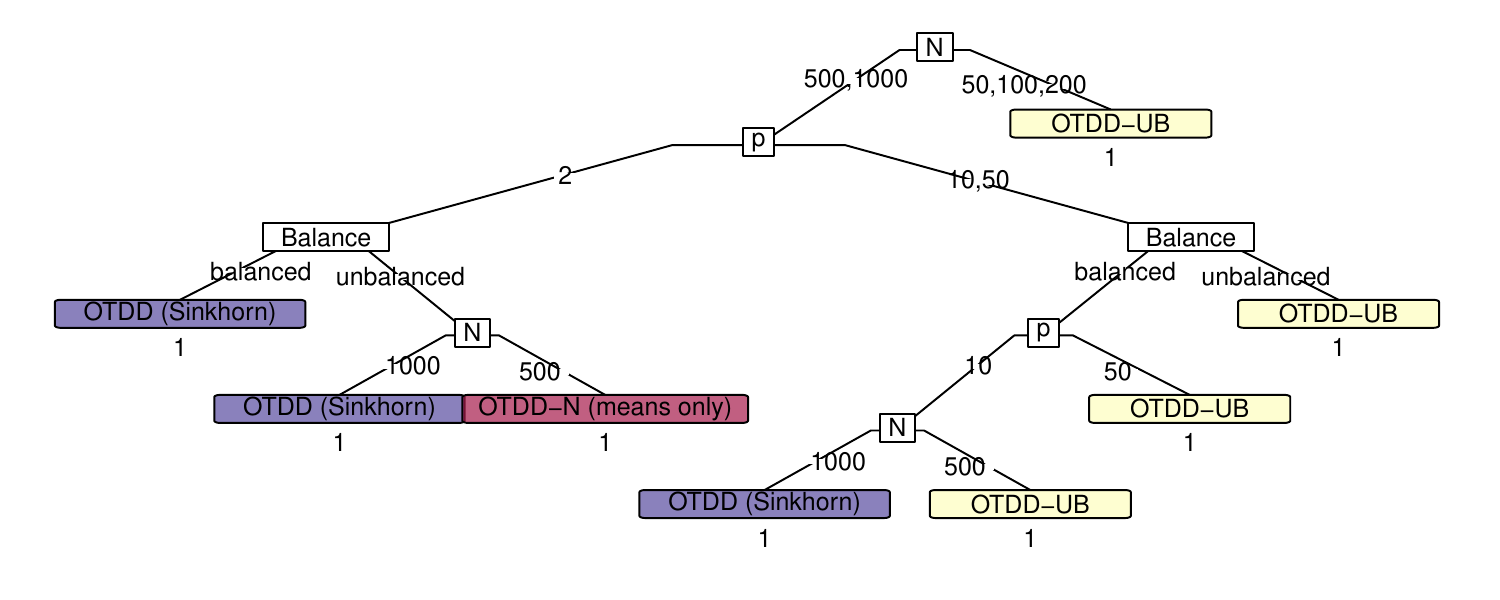}
		\caption{Decision rules for choosing a method that covers most scenarios for two $\chi^2$-distributed datasets and OGM alternatives given only the information about the dataset dimensions. Numbers underneath the leaf nodes give the proportion of scenarios in the respective leaf node that are covered by the chosen method. }\label{fig:tree.chisq.ogm.two}
	\end{figure}
	\clearpage
	
	\subsubsection[k = 4]{$k = 4$}
	
	\paragraph{By DGP}\mbox{}
	
	\begin{figure}[!h]
		\centering
		\includegraphics[width=\linewidth]{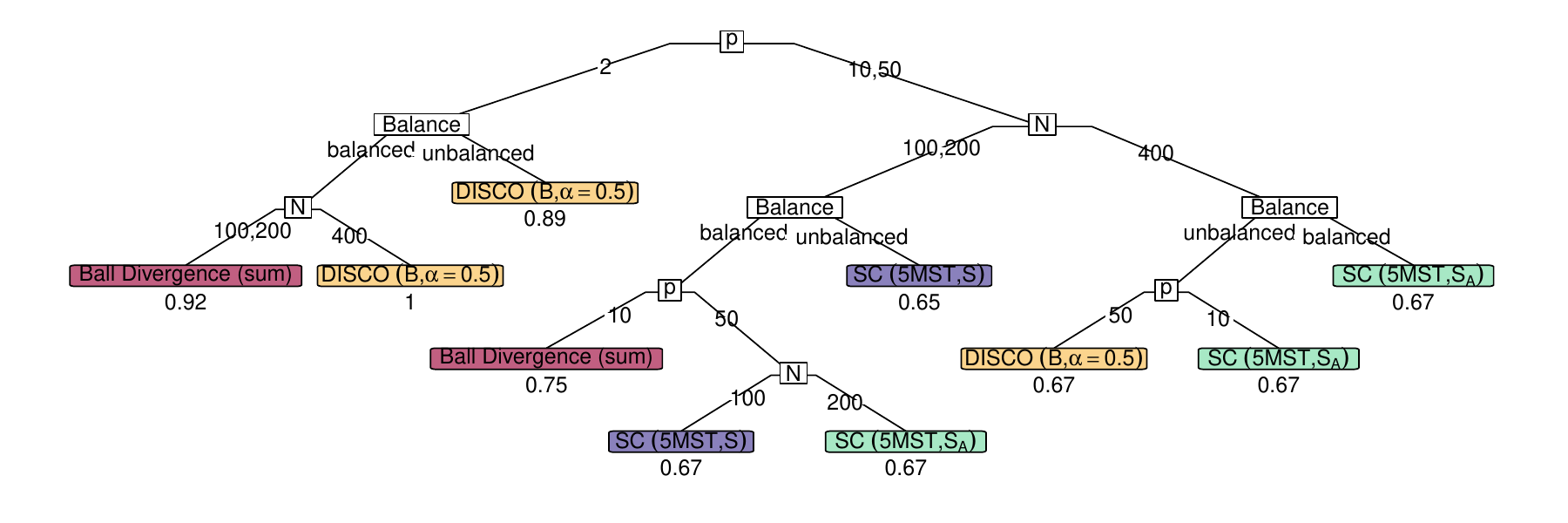}
		\caption{Decision rules for choosing a method that covers most scenarios for multiple normally distributed datasets given only the information about the dataset dimensions. Numbers underneath the leaf nodes give the proportion of scenarios in the respective leaf node that are covered by the chosen method. }\label{fig:tree.normal.multi}
	\end{figure}
	
	\begin{figure}[!h]
		\centering
		\includegraphics[width=\linewidth]{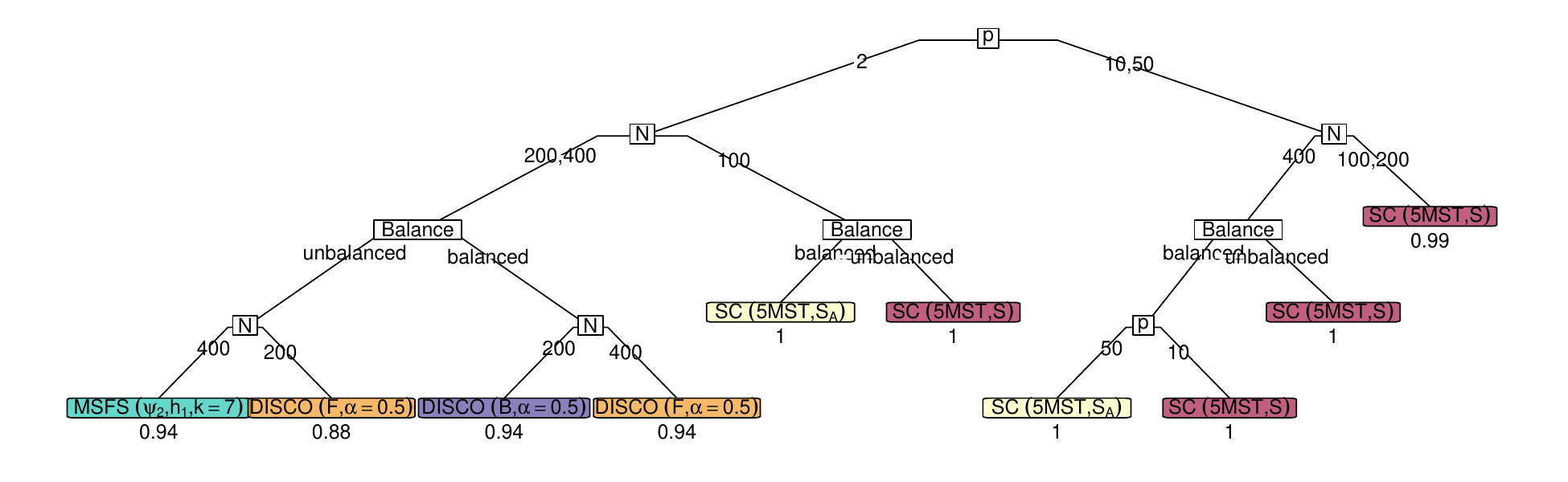}
		\caption{Decision rules for choosing a method that covers most scenarios for multiple $t$-distributed datasets given only the information about the dataset dimensions. Numbers underneath the leaf nodes give the proportion of scenarios in the respective leaf node that are covered by the chosen method.}\label{fig:tree.t.multi}
	\end{figure}

    For four log-normally distributed datasets, the DISCO~(F,$\alpha = 0.5$) covers most scenarios, regardless of $N$, $p$, and the sample size balance.
	

	
	\begin{figure}[!h]
		\centering
		\includegraphics[width=\linewidth]{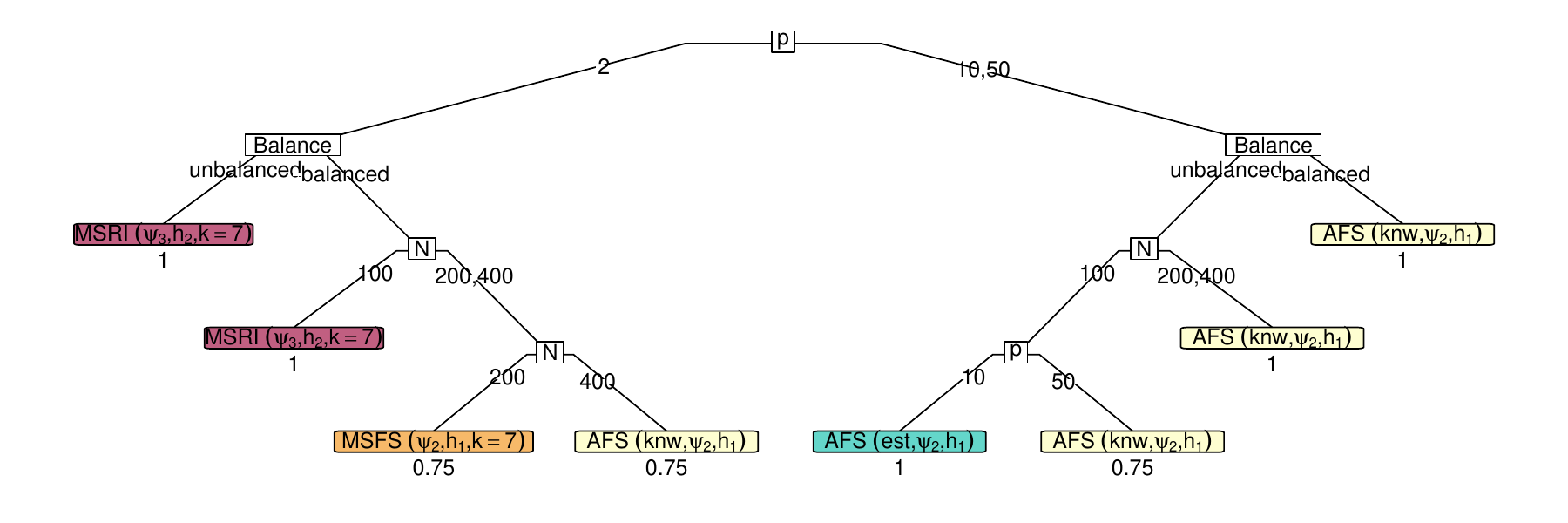}
		\caption{Decision rules for choosing a method that covers most scenarios for multiple $\chi^2$-distributed datasets given only the information about the dataset dimensions. Numbers underneath the leaf nodes give the proportion of scenarios in the respective leaf node that are covered by the chosen method. }\label{fig:tree.chisq.multi}
	\end{figure}
	\clearpage
	
	\paragraph{By DGP and Deviation}\mbox{}
	\subparagraph{Normal distribution}\mbox{}

    For shift alternatives on four normally distributed datasets, the DISCO~(B,$\alpha = 0.5$) covers most scenarios, regardless of $N$, $p$, and the sample size balance.
	

    For scale alternatives on four normally distributed datasets, the Ball divergence~(sum) covers most scenarios, regardless of $N$, $p$, and the sample size balance.
	

	
	\begin{figure}[!h]
		\centering
		\includegraphics[width=\linewidth]{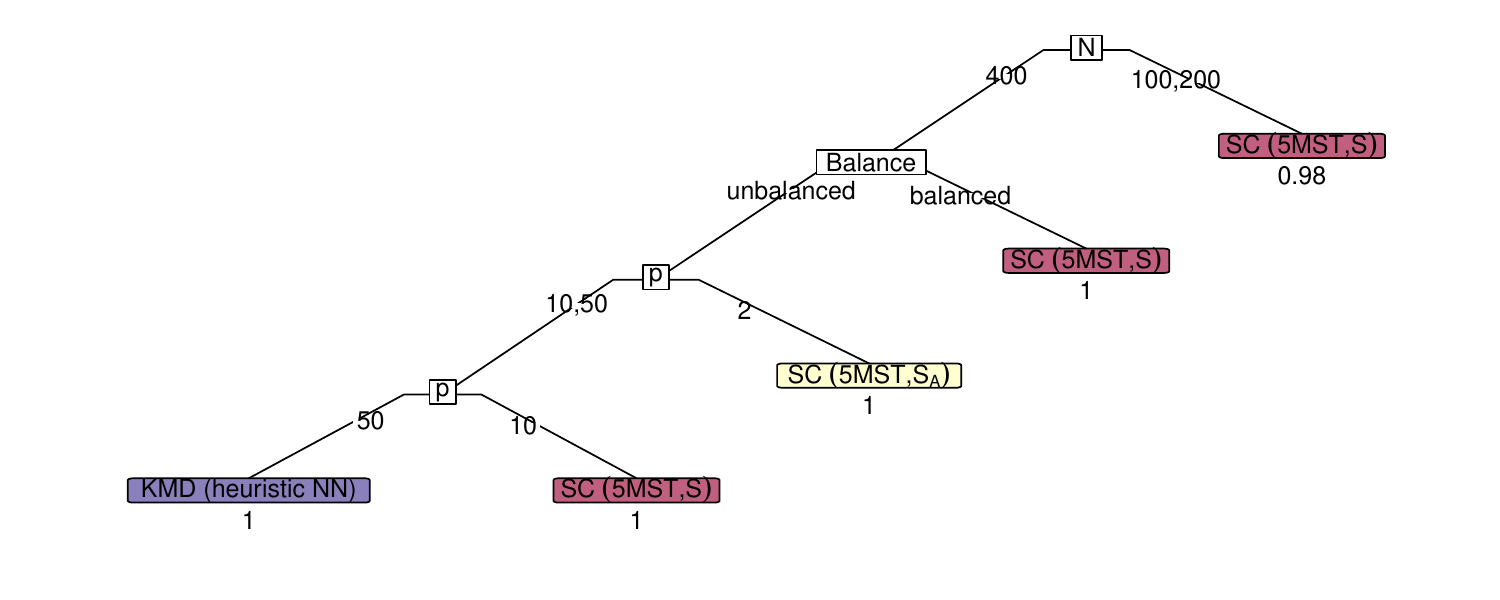}
		\caption{Decision rules for choosing a method that covers most scenarios for multiple normally distributed datasets and correlation alternatives given only the information about the dataset dimensions. Numbers underneath the leaf nodes give the proportion of scenarios in the respective leaf node that are covered by the chosen method. }\label{fig:tree.normal.cor.multi}
	\end{figure}

	\subparagraph[t distribution]{$t$ distribution}\mbox{}
	
	\begin{figure}[!h]
		\centering
		\includegraphics[width=\linewidth]{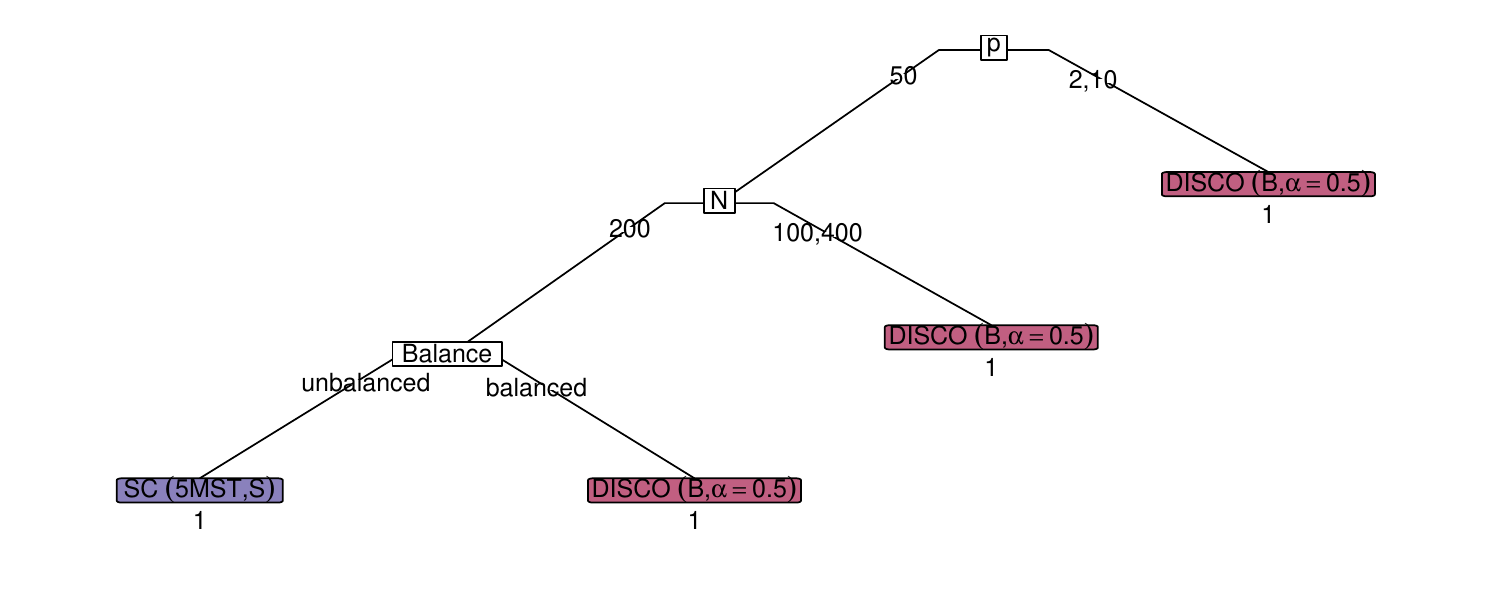}
		\caption{Decision rules for choosing a method that covers most scenarios for multiple $t$-distributed datasets and shift alternatives given only the information about the dataset dimensions. Numbers underneath the leaf nodes give the proportion of scenarios in the respective leaf node that are covered by the chosen method. }\label{fig:tree.t.shift.multi}
	\end{figure}
	
	For scale alternatives on four $t$-distributed datasets, the Ball divergence~(sum) covers most scenarios, regardless of $N$, $p$, and the sample size balance.

    For correlation alternatives on four $t$-distributed datasets, the SC~(5MST,S) covers most scenarios, regardless of $N$, $p$, and the sample size balance.
%

		
	\begin{figure}[!h]
		\centering
		\includegraphics[width=\linewidth]{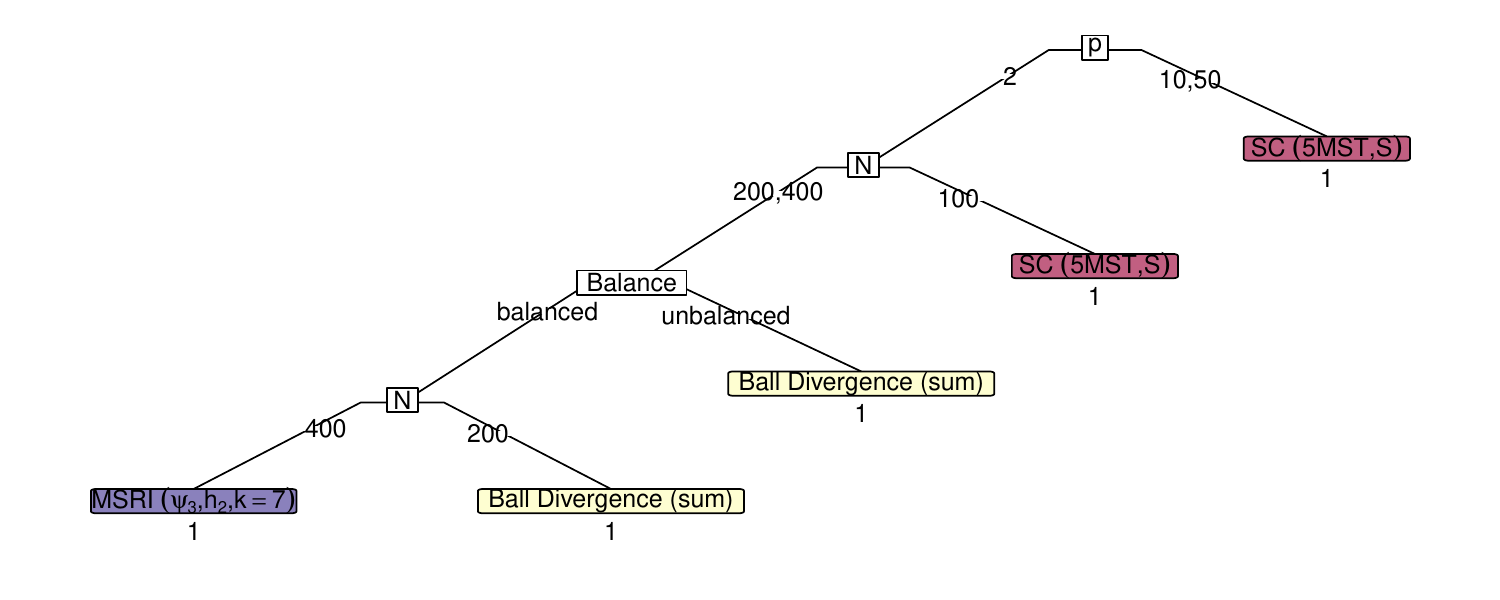}
		\caption{Decision rules for choosing a method that covers most scenarios for multiple $t$-distributed datasets and correlation alternatives given only the information about the dataset dimensions. Numbers underneath the leaf nodes give the proportion of scenarios in the respective leaf node that are covered by the chosen method. }\label{fig:tree.t.kurtosis.multi}
	\end{figure}
	\clearpage
	
	\subparagraph{Log-normal distribution}\mbox{}
	
	\begin{figure}[!h]
		\centering
		\includegraphics[width=\linewidth]{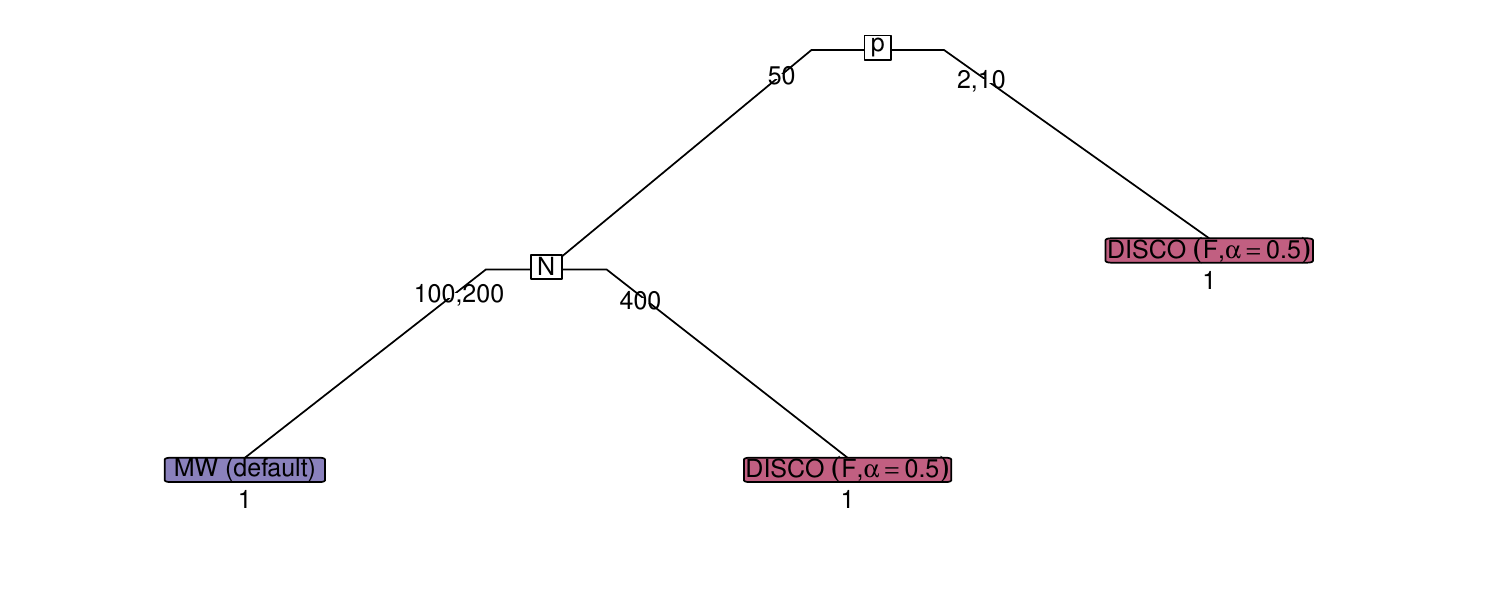}
		\caption{Decision rules for choosing a method that covers most scenarios for multiple log-normally distributed datasets and shift alternatives given only the information about the dataset dimensions. Numbers underneath the leaf nodes give the proportion of scenarios in the respective leaf node that are covered by the chosen method. }\label{fig:tree.log-normal.shift.multi}
	\end{figure}

    For scale alternatives on four log-normally distributed datasets, the AFS~(knw,$\psi_3$,$h_1$) covers most scenarios, regardless of $N$, $p$, and the sample size balance.


	\subparagraph[Chi-squared distribution]{$\chi^2$ distribution}\mbox{}
	
	\begin{figure}[!h]
		\centering
		\includegraphics[width=\linewidth]{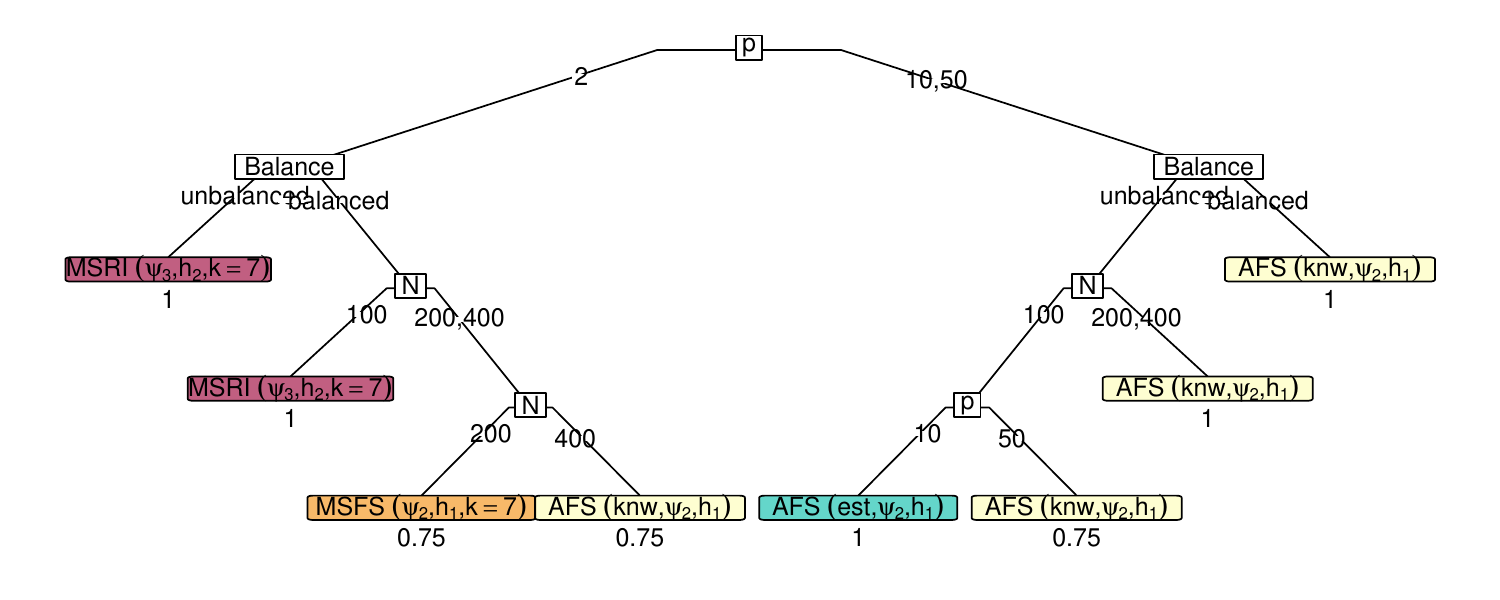}
		\caption{Decision rules for choosing a method that covers most scenarios for multiple $\chi^2$-distributed datasets and skewness and kurtosis alternatives given only the information about the dataset dimensions. Numbers underneath the leaf nodes give the proportion of scenarios in the respective leaf node that are covered by the chosen method. }\label{fig:tree.chisq.skew.kurtosis.multi}
	\end{figure}
	\clearpage
	
	\section{Comparison of Best Method Combinations According to Greedy and Exhaustive Search}\label{app:comp.greedy.exhaustive}
		\begin{figure}[!h]
		\centering
		\begin{subfigure}[t]{0.49\linewidth}
			\centering
			\includegraphics[width=\linewidth]{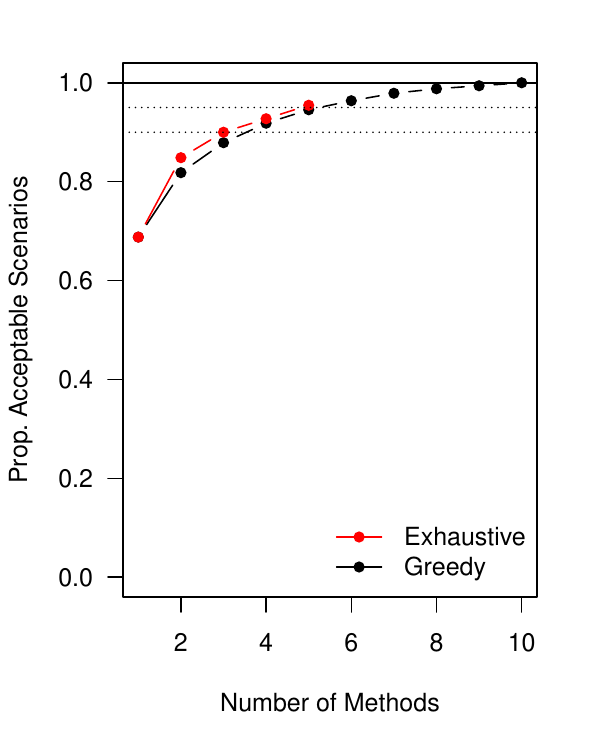}
			\caption{Two-sample Case}
			\label{fig:meth.sel.two}
		\end{subfigure}
		\hfill
		\begin{subfigure}[t]{0.49\linewidth}
			\centering
			\includegraphics[width=\linewidth]{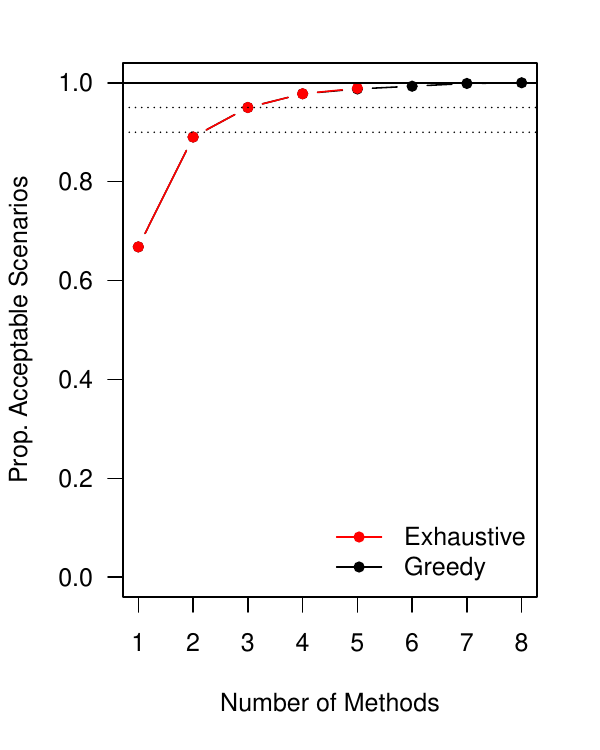}
			\caption{Multi-sample Case}
			\label{fig:meth.sel.multi}
		\end{subfigure}
		\caption{Proportions of scenarios that are covered by the best method combination found by greedy and exhaustive search. Exhaustive search for combinations consisting of more methods.}
		\label{fig:meth.sel}
	\end{figure} 
	\clearpage
	
	\section{Scenarios not Covered by Best Method Combinations}\label{app:scens.uncovered}
	\subsection[k=2]{$k=2$}

	\begin{table}[!h]
	\centering
	\begin{tabular}{rllrrll}
	  \toprule
	 & DGP & Deviation & $N$ & $p$ & Grouping & Balance \\ 
	  \midrule
	1 & log-normal & shift & 100 & 10 & 1+1 & balanced \\ 
	  2 & log-normal & shift & 50 & 10 & 1+1 & balanced \\ 
	  3 & log-normal & shift & 1000 & 50 & 1+1 & balanced \\ 
	  4 & log-normal & shift & 200 & 50 & 1+1 & balanced \\ 
	  5 & log-normal & shift & 500 & 50 & 1+1 & balanced \\ 
	  6 & $\chi^2$ & skewness\_and\_kurtosis & 100 & 10 & 1+1 & balanced \\ 
	  7 & $\chi^2$ & skewness and kurtosis & 50 & 10 & 1+1 & balanced \\ 
	  8 & $\chi^2$ & skewness and kurtosis & 100 & 50 & 1+1 & balanced \\ 
	  9 & $\chi^2$ & skewness and kurtosis & 200 & 50 & 1+1 & balanced \\ 
	  10 & $\chi^2$ & skewness and kurtosis & 50 & 50 & 1+1 & balanced \\ 
	  11 & normal & normal vs.\ $t$ & 100 & 2 & 1+1 & unbalanced \\ 
	  12 & normal & normal vs.\ $t$ & 1000 & 2 & 1+1 & unbalanced \\ 
	  13 & normal & normal vs.\ $t$ & 200 & 2 & 1+1 & unbalanced \\ 
	  14 & normal & normal vs.\ $t$ & 500 & 2 & 1+1 & unbalanced \\ 
	  15 & normal & normal vs.\ $t$ & 100 & 50 & 1+1 & unbalanced \\ 
	  16 & normal & normal vs.\ $t$ & 200 & 50 & 1+1 & unbalanced \\ 
	  17 & normal & normal vs.\ $t$ & 50 & 50 & 1+1 & unbalanced \\ 
	  18 & log-normal & shift & 100 & 10 & 1+1 & unbalanced \\ 
	  19 & $\chi^2$ & skewness and kurtosis & 100 & 10 & 1+1 & unbalanced \\ 
	  20 & $\chi^2$ & skewness and kurtosis & 200 & 10 & 1+1 & unbalanced \\ 
	  21 & $\chi^2$ & skewness and kurtosis & 50 & 10 & 1+1 & unbalanced \\ 
	  22 & $\chi^2$ & skewness and kurtosis & 100 & 2 & 1+1 & unbalanced \\ 
	  23 & $\chi^2$ & skewness and kurtosis & 50 & 2 & 1+1 & unbalanced \\ 
	  24 & $\chi^2$ & skewness and kurtosis & 100 & 50 & 1+1 & unbalanced \\ 
	  25 & $\chi^2$ & skewness and kurtosis & 200 & 50 & 1+1 & unbalanced \\ 
	  26 & $\chi^2$ & skewness and kurtosis & 50 & 50 & 1+1 & unbalanced \\ 
	  27 & $\chi^2$ & skewness and kurtosis & 500 & 50 & 1+1 & unbalanced \\ 
	   \bottomrule
	\end{tabular}
	\caption{List of scenarios that are not covered by the best combination of four methods for two datasets.}\label{tab:scens.not.covered.top4.two}
	\end{table}
	
	\begin{table}[!h]
	\centering
	\begin{tabular}{rllrrll}
	  \toprule
	 & DGP & Deviation & $N$ & $p$ & Grouping & Balance \\ 
	  \midrule
	1 & log-normal & shift & 100 & 10 & 1+1 & balanced \\ 
	  2 & log-normal & shift & 50 & 10 & 1+1 & balanced \\ 
	  3 & log-normal & shift & 1000 & 50 & 1+1 & balanced \\ 
	  4 & log-normal & shift & 200 & 50 & 1+1 & balanced \\ 
	  5 & log-normal & shift & 500 & 50 & 1+1 & balanced \\ 
	  6 & normal & normal vs.\ $t$ & 100 & 2 & 1+1 & unbalanced \\ 
	  7 & normal & normal vs.\ $t$ & 1000 & 2 & 1+1 & unbalanced \\ 
	  8 & normal & normal vs.\ $t$ & 100 & 50 & 1+1 & unbalanced \\ 
	  9 & normal & normal vs.\ $t$ & 200 & 50 & 1+1 & unbalanced \\ 
	  10 & normal & normal vs.\ $t$ & 50 & 50 & 1+1 & unbalanced \\ 
	  11 & log-normal & shift & 100 & 10 & 1+1 & unbalanced \\ 
	  12 & $\chi^2$ & skewness and kurtosis & 200 & 10 & 1+1 & unbalanced \\ 
	   \bottomrule
	\end{tabular}
	\caption{List of scenarios that are not covered by the best combination of six methods for two datasets.}\label{tab:scens.not.covered.top6.two}
	\end{table}
	
	\subsection[k=4]{$k=4$}

	\begin{longtable}{rllrrll}
	    \toprule
		& DGP & Deviation & $N$ & $p$ & Grouping & Balance \\
		\midrule
		\endhead
		\bottomrule
		\endfoot
		\bottomrule
		\caption{List of scenarios that are not covered by the best combination of two methods for multiple datasets.}\label{tab:scens.not.covered.top2.multi}\\
	\endlastfoot
	  1 & $t$ & kurtosis & 200 & 2 & 1+1+1+1 & balanced \\ 
	  2 & $t$ & kurtosis & 400 & 2 & 1+1+1+1 & balanced \\ 
	  3 & $\chi^2$ & skewness and kurtosis & 100 & 10 & 1+1+1+1 & balanced \\ 
	  4 & $\chi^2$ & skewness and kurtosis & 200 & 10 & 1+1+1+1 & balanced \\ 
	  5 & $\chi^2$ & skewness and kurtosis & 100 & 2 & 1+1+1+1 & balanced \\ 
	  6 & $\chi^2$ & skewness and kurtosis & 200 & 2 & 1+1+1+1 & balanced \\ 
	  7 & $\chi^2$ & skewness and kurtosis & 400 & 2 & 1+1+1+1 & balanced \\ 
	  8 & $\chi^2$ & skewness and kurtosis & 100 & 50 & 1+1+1+1 & balanced \\ 
	  9 & $\chi^2$ & skewness and kurtosis & 200 & 50 & 1+1+1+1 & balanced \\ 
	  10 & $\chi^2$ & skewness and kurtosis & 400 & 50 & 1+1+1+1 & balanced \\ 
	  11 & $t$ & kurtosis & 200 & 2 & 1+1+1+1 & unbalanced \\ 
	  12 & $t$ & kurtosis & 400 & 2 & 1+1+1+1 & unbalanced \\ 
	  13 & $\chi^2$ & skewness and kurtosis & 100 & 2 & 1+1+1+1 & unbalanced \\ 
	  14 & $\chi^2$ & skewness and kurtosis & 200 & 2 & 1+1+1+1 & unbalanced \\ 
	  15 & $\chi^2$ & skewness and kurtosis & 400 & 2 & 1+1+1+1 & unbalanced \\ 
	  16 & $\chi^2$ & skewness and kurtosis & 100 & 50 & 1+1+1+1 & unbalanced \\ 
	  17 & $\chi^2$ & skewness and kurtosis & 200 & 50 & 1+1+1+1 & unbalanced \\ 
	  18 & $\chi^2$ & skewness and kurtosis & 400 & 50 & 1+1+1+1 & unbalanced \\ 
	  19 & normal & scale & 100 & 2 & 2+1+1 & balanced \\ 
	  20 & normal & scale & 200 & 2 & 2+1+1 & balanced \\ 
	  21 & $\chi^2$ & skewness and kurtosis & 100 & 10 & 2+1+1 & balanced \\ 
	  22 & $\chi^2$ & skewness and kurtosis & 200 & 10 & 2+1+1 & balanced \\ 
	  23 & $\chi^2$ & skewness and kurtosis & 100 & 2 & 2+1+1 & balanced \\ 
	  24 & $\chi^2$ & skewness and kurtosis & 200 & 2 & 2+1+1 & balanced \\ 
	  25 & $\chi^2$ & skewness and kurtosis & 400 & 2 & 2+1+1 & balanced \\ 
	  26 & $\chi^2$ & skewness and kurtosis & 100 & 50 & 2+1+1 & balanced \\ 
	  27 & $\chi^2$ & skewness and kurtosis & 200 & 50 & 2+1+1 & balanced \\ 
	  28 & $\chi^2$ & skewness and kurtosis & 400 & 50 & 2+1+1 & balanced \\ 
	  29 & normal & correlation & 400 & 2 & 2+1+1 & unbalanced \\ 
	  30 & $\chi^2$ & skewness and kurtosis & 100 & 2 & 2+1+1 & unbalanced \\ 
	  31 & $\chi^2$ & skewness and kurtosis & 200 & 2 & 2+1+1 & unbalanced \\ 
	  32 & $\chi^2$ & skewness and kurtosis & 400 & 2 & 2+1+1 & unbalanced \\ 
	  33 & $\chi^2$ & skewness and kurtosis & 100 & 50 & 2+1+1 & unbalanced \\ 
	  34 & $\chi^2$ & skewness and kurtosis & 200 & 50 & 2+1+1 & unbalanced \\ 
	  35 & $\chi^2$ & skewness and kurtosis & 400 & 50 & 2+1+1 & unbalanced \\ 
	  36 & normal & scale & 100 & 2 & 2+2 & balanced \\ 
	  37 & normal & scale & 200 & 2 & 2+2 & balanced \\ 
	  38 & normal & scale & 400 & 2 & 2+2 & balanced \\ 
	  39 & log-normal & shift & 100 & 50 & 2+2 & balanced \\ 
	  40 & log-normal & shift & 200 & 50 & 2+2 & balanced \\ 
	  41 & $\chi^2$ & skewness and kurtosis & 100 & 10 & 2+2 & balanced \\ 
	  42 & $\chi^2$ & skewness and kurtosis & 200 & 10 & 2+2 & balanced \\ 
	  43 & $\chi^2$ & skewness and kurtosis & 100 & 2 & 2+2 & balanced \\ 
	  44 & $\chi^2$ & skewness and kurtosis & 200 & 2 & 2+2 & balanced \\ 
	  45 & $\chi^2$ & skewness and kurtosis & 400 & 2 & 2+2 & balanced \\ 
	  46 & $\chi^2$ & skewness and kurtosis & 100 & 50 & 2+2 & balanced \\ 
	  47 & $\chi^2$ & skewness and kurtosis & 200 & 50 & 2+2 & balanced \\ 
	  48 & $\chi^2$ & skewness and kurtosis & 400 & 50 & 2+2 & balanced \\ 
	  49 & normal & correlation & 100 & 10 & 2+2 & unbalanced \\ 
	  50 & normal & correlation & 400 & 50 & 2+2 & unbalanced \\ 
	  51 & log-normal & shift & 200 & 50 & 2+2 & unbalanced \\ 
	  52 & $\chi^2$ & skewness and kurtosis & 100 & 2 & 2+2 & unbalanced \\ 
	  53 & $\chi^2$ & skewness and kurtosis & 200 & 2 & 2+2 & unbalanced \\ 
	  54 & $\chi^2$ & skewness and kurtosis & 400 & 2 & 2+2 & unbalanced \\ 
	  55 & $\chi^2$ & skewness and kurtosis & 100 & 50 & 2+2 & unbalanced \\ 
	  56 & $\chi^2$ & skewness and kurtosis & 200 & 50 & 2+2 & unbalanced \\ 
	  57 & $\chi^2$ & skewness and kurtosis & 400 & 50 & 2+2 & unbalanced \\ 
	  58 & normal & scale & 100 & 2 & 3+1 & balanced \\ 
	  59 & normal & scale & 200 & 2 & 3+1 & balanced \\ 
	  60 & normal & scale & 400 & 2 & 3+1 & balanced \\ 
	  61 & $t$ & scale & 200 & 2 & 3+1 & balanced \\ 
	  62 & log-normal & shift & 100 & 50 & 3+1 & balanced \\ 
	  63 & log-normal & shift & 200 & 50 & 3+1 & balanced \\ 
	  64 & $\chi^2$ & skewness and kurtosis & 100 & 10 & 3+1 & balanced \\ 
	  65 & $\chi^2$ & skewness and kurtosis & 200 & 10 & 3+1 & balanced \\ 
	  66 & $\chi^2$ & skewness and kurtosis & 100 & 2 & 3+1 & balanced \\ 
	  67 & $\chi^2$ & skewness and kurtosis & 200 & 2 & 3+1 & balanced \\ 
	  68 & $\chi^2$ & skewness and kurtosis & 400 & 2 & 3+1 & balanced \\ 
	  69 & $\chi^2$ & skewness and kurtosis & 100 & 50 & 3+1 & balanced \\ 
	  70 & $\chi^2$ & skewness and kurtosis & 200 & 50 & 3+1 & balanced \\ 
	  71 & $\chi^2$ & skewness and kurtosis & 400 & 50 & 3+1 & balanced \\ 
	  72 & log-normal & shift & 100 & 50 & 3+1 & unbalanced \\ 
	  73 & log-normal & shift & 200 & 50 & 3+1 & unbalanced \\ 
	  74 & $\chi^2$ & skewness and kurtosis & 100 & 10 & 3+1 & unbalanced \\ 
	  75 & $\chi^2$ & skewness and kurtosis & 100 & 2 & 3+1 & unbalanced \\ 
	  76 & $\chi^2$ & skewness and kurtosis & 200 & 2 & 3+1 & unbalanced \\ 
	  77 & $\chi^2$ & skewness and kurtosis & 100 & 50 & 3+1 & unbalanced \\ 
	  78 & $\chi^2$ & skewness and kurtosis & 200 & 50 & 3+1 & unbalanced \\ 
	  79 & $\chi^2$ & skewness and kurtosis & 400 & 50 & 3+1 & unbalanced \\ 
	\end{longtable}	
	
	\begin{table}[!h]
	\centering
	\begin{tabular}{rllrrll}
	  \toprule
	 & DGP & Deviation & $N$ & $p$ & Grouping & Balance \\ 
	  \midrule
	1 & $t$ & kurtosis & 200 & 2 & 1+1+1+1 & balanced \\ 
	  2 & $t$ & kurtosis & 400 & 2 & 1+1+1+1 & balanced \\ 
	  3 & $\chi^2$ & skewness and kurtosis & 100 & 2 & 1+1+1+1 & balanced \\ 
	  4 & $\chi^2$ & skewness and kurtosis & 200 & 2 & 1+1+1+1 & balanced \\ 
	  5 & $\chi^2$ & skewness and kurtosis & 400 & 2 & 1+1+1+1 & balanced \\ 
	  6 & $t$ & kurtosis & 200 & 2 & 1+1+1+1 & unbalanced \\ 
	  7 & $t$ & kurtosis & 400 & 2 & 1+1+1+1 & unbalanced \\ 
	  8 & $\chi^2$ & skewness and kurtosis & 100 & 2 & 1+1+1+1 & unbalanced \\ 
	  9 & $\chi^2$ & skewness and kurtosis & 200 & 2 & 1+1+1+1 & unbalanced \\ 
	  10 & $\chi^2$ & skewness and kurtosis & 400 & 2 & 1+1+1+1 & unbalanced \\ 
	  11 & $\chi^2$ & skewness and kurtosis & 100 & 50 & 1+1+1+1 & unbalanced \\ 
	  12 & $\chi^2$ & skewness and kurtosis & 200 & 50 & 1+1+1+1 & unbalanced \\ 
	  13 & $\chi^2$ & skewness and kurtosis & 100 & 2 & 2+1+1 & balanced \\ 
	  14 & $\chi^2$ & skewness and kurtosis & 200 & 2 & 2+1+1 & balanced \\ 
	  15 & normal & correlation & 400 & 2 & 2+1+1 & unbalanced \\ 
	  16 & $\chi^2$ & skewness and kurtosis & 100 & 2 & 2+1+1 & unbalanced \\ 
	  17 & $\chi^2$ & skewness and kurtosis & 200 & 2 & 2+1+1 & unbalanced \\ 
	  18 & $\chi^2$ & skewness and kurtosis & 400 & 2 & 2+1+1 & unbalanced \\ 
	  19 & normal & scale & 100 & 2 & 2+2 & balanced \\ 
	  20 & normal & scale & 200 & 2 & 2+2 & balanced \\ 
	  21 & normal & scale & 400 & 2 & 2+2 & balanced \\ 
	  22 & log-normal & shift & 100 & 50 & 2+2 & balanced \\ 
	  23 & log-normal & shift & 200 & 50 & 2+2 & balanced \\ 
	  24 & $\chi^2$ & skewness and kurtosis & 100 & 2 & 2+2 & balanced \\ 
	  25 & $\chi^2$ & skewness and kurtosis & 200 & 2 & 2+2 & balanced \\ 
	  26 & normal & correlation & 100 & 10 & 2+2 & unbalanced \\ 
	  27 & normal & correlation & 400 & 50 & 2+2 & unbalanced \\ 
	  28 & log-normal & shift & 200 & 50 & 2+2 & unbalanced \\ 
	  29 & $\chi^2$ & skewness and kurtosis & 100 & 2 & 2+2 & unbalanced \\ 
	  30 & $\chi^2$ & skewness and kurtosis & 200 & 2 & 2+2 & unbalanced \\ 
	  31 & $\chi^2$ & skewness and kurtosis & 400 & 2 & 2+2 & unbalanced \\ 
	  32 & normal & scale & 100 & 2 & 3+1 & balanced \\ 
	  33 & log-normal & shift & 100 & 50 & 3+1 & balanced \\ 
	  34 & log-normal & shift & 200 & 50 & 3+1 & balanced \\ 
	  35 & log-normal & shift & 100 & 50 & 3+1 & unbalanced \\ 
	  36 & log-normal & shift & 200 & 50 & 3+1 & unbalanced \\ 
	   \bottomrule
	\end{tabular}
	\caption{List of scenarios that are not covered by the best combination of three methods for multiple datasets.}\label{tab:scens.not.covered.top3.multi}
	\end{table}

\end{document}